\newif\ifcomment
\newif\ifdraft
\newif\ifflush
\flushtrue
\newif\iflatexdiff
\def\revisionNumber{v1.5}
\documentclass[ALICE,manyauthors,12pt]{cernphprep}
\ifdraft
\usepackage[pdfencoding=auto, psdextra, hyperindex=false, linkcolor=NavyBlue, citecolor=LimeGreen, urlcolor=NavyBlue, colorlinks=true, backref=page, linktocpage]{hyperref}
\usepackage{draftwatermark}  
\else
\usepackage{hyperref}[citecolor=LimeGreen]
\fi
\usepackage[comma,square,numbers,sort&compress]{natbib}
\usepackage{hyperref}
\usepackage{graphicx}  
\usepackage{caption}
\usepackage{sidecap}
\usepackage{dcolumn}   
\usepackage{multirow}
\usepackage{bm}        
\usepackage{amssymb}   
\usepackage{amsfonts}
\usepackage{booktabs}  
\usepackage{graphics}
\usepackage[acronym]{glossaries}  

\usepackage{grffile}   
\usepackage{epsfig}
\usepackage{xargs}     
\usepackage[loose]{units}
\usepackage[usenames,dvipsnames]{xcolor}
\usepackage[normalem]{ulem} 
\usepackage{color}
\usepackage[utf8]{inputenc}
\usepackage[T1]{fontenc}
\usepackage{subfigure}
\usepackage{parskip}
\usepackage{placeins}  
\usepackage[us,12hr]{datetime}
\usepackage{pdfpages}
\usepackage{orcidlink}
\iflatexdiff
\RequirePackage{color}\definecolor{RED}{rgb}{1,0,0}\definecolor{BLUE}{rgb}{0,0,1}

\fi
\newcommand{\piz}{\ensuremath{\pi^{0}}}
\newcommand{\pt}{\ensuremath{p_{\rm T}}}
\newcommand{\ptt}{\ensuremath{p_{\rm T}^{\rm track}}}
\newcommand{\ptj}{\ensuremath{p_{\rm T, jet}}}

\newcommand{\pT}{\pt}
\newcommand{\eg}{e.g.}
\newcommand{\mom}{\mbox{\rm  GeV$\kern-0.15em /\kern-0.12em c$}}
\newcommand{\GeVc}{\ensuremath{\mathrm{GeV}/c}}
\newcommand{\GeVcs}{\ensuremath{\mathrm{GeV}/c^2}}
\newcommand{\Rn}{\ensuremath{R^{n^{\rm w}_{\rm cell}}_{\sigma_{\rm long}}}}
\newcommand{\Rg}{\ensuremath{R_{\rm{g}}}}
\newcommand{\PbPb}{\textnormal{Pb--Pb}}

\newcommand{\pPb}{\textnormal{p--Pb}}
\newcommand{\pp}{\mbox{pp}}
\newcommand{\s}{$\sqrt{s}$}
\newcommand{\snn}{\ensuremath{\sqrt{s_{_{\mathrm{NN}}}} }}

\newcommand{\sfive}{\s~=~5.02\,TeV}
\newcommand{\sseven}{\s~=~7\,TeV}
\newcommand{\seight}{\s~=~8\,TeV}
\newcommand{\stwo}{\s~=~2.76\,TeV}
\newcommand{\sthirteen}{\s~=~13\,TeV}
\newcommand{\stwolead}{\snn~=~2.76\,TeV}
\newcommand{\sfivelead}{\snn~=~5.02\,TeV}

\newcommand{\seightplead}{\snn~=~8.16\,TeV}
\newcommand{\shshlo}{\ensuremath{\sigma_\mathrm{long}^{2}}}
\newcommand{\shshloX}{\ensuremath{\sigma_\mathrm{long, 5x5}^{2}}}
\newcommand{\shshsh}{\ensuremath{\sigma_\mathrm{short}^{2}}}
 
\newcommandx{\Figure}[2][2=]{\hyperref[#1]{Figure~\ref*{#1}#2}}
\newcommandx{\Fig}[2][2=]{\hyperref[#1]{Fig.~\ref*{#1}#2}}
\newcommandx{\Figures}[2]{\hyperref[#1]{Figure~\ref*{#1}} and \hyperref[#2]{\ref*{#2}}}
\newcommandx{\Figs}[2]{\hyperref[#1]{Figs.~\ref*{#1}} and \hyperref[#2]{\ref*{#2}}}
\newcommandx{\Section}[2][2=]{\hyperref[#1]{Section~\ref*{#1}#2}}
\newcommandx{\Sec}[2][2=]{\hyperref[#1]{Sec.~\ref*{#1}#2}}
\newcommandx{\Sections}[2]{\hyperref[#1]{Sections~\ref*{#1}} and \hyperref[#2]{\ref*{#2}}}
\newcommandx{\Chapter}[2][2=]{\hyperref[#1]{Chapter~\ref*{#1}#2}}
\newcommandx{\App}[2][2=]{\hyperref[#1]{App.~\ref*{#1}#2}}
\newcommandx{\Appendix}[2][2=]{\hyperref[#1]{Appendix~\ref*{#1}#2}}
\newcommandx{\Table}[2][2=]{\hyperref[#1]{Table~\ref*{#1}#2}}
\newcommandx{\Tab}[2][2=]{\hyperref[#1]{Tab.~\ref*{#1}#2}}
\newcommandx{\Tables}[2]{\hyperref[#1]{Table~\ref*{#1}} and \hyperref[#2]{\ref*{#2}}}
\newcommandx{\TablesT}[2]{\hyperref[#1]{Tables~\ref*{#1}}-{\ref*{#2}}}
\newcommandx{\Equation}[2][2=]{\hyperref[#1]{Equation~\ref*{#1}#2}}
\newcommandx{\Eq}[2][2=]{\hyperref[#1]{Eq.~\ref*{#1}#2}}
\newcommandx{\Eqs}[2]{\hyperref[#1]{Eq.~\ref*{#1}} and \hyperref[#2]{\ref*{#2}}}
\newcommand{\ie}{i.e.}

\newcommand{\com}[1]{\relax}
\newcommand{\vectoralig}[1]{\mathbf{v}_{\rm #1}}
\iflatexdiff

\renewcommand{\xout}[1]{\textcolor{red}{\sout{#1}}}
 
\newcommand{\old}[1]{\textcolor{red}{\sout{#1}}}

\else

\renewcommand{\xout}[1]{}
\newcommand{\old}[1]{\relax}

\fi

\newacronym{ADC}{ADC}{Analog-to-Digital Converter}
\newacronym{ALICE}{ALICE}{A Large Ion Collider Experiment}
\newacronym{ALTRO}{ALTRO}{ALICE TPC Readout}
\newacronym{APD}{APD}{Avalanche Photo Diode}
\newacronym{BC}{BC}{Bunch Crossing}
\newacronym{BR}{BR}{Branching Ratio}
\newacronym{CCRF}{CCRF}{Conv-Calo ratio fit}
\newacronym{CCMF}{CCMF}{Conv-Calo mass fit}
\newacronym{CERN}{CERN}{European Organization for Nuclear Research}
\newacronym{CMF}{CMF}{Calo mass fit}
\newacronym{CMS}{CMS}{Compact Muon Solenoid}
\newacronym{CPS}{CPS}{Charge Sensitive Preamplifier}
\newacronym{CPV}{CPV}{charged particle veto}
\newacronym{CRF}{CRF}{Calo ratio fit}
\newacronym{CSP}{CSP}{Charge Sensitive Preamplifier}
\newacronym{CTP}{CTP}{Central Trigger Processor}
\newacronym{DAQ}{DAQ}{Data Aquisition}
\newacronym{DCal}{DCal}{Di-Jet Calorimeter}
\newacronym{EM}{EM}{Electromagnetic}
\newacronym{EMC}{EMC}{EMCal method}
\newacronym{EMCal}{EMCal}{Electro Magnetic Calorimeter}
\newacronym{FEC}{FEC}{Front End Card}
\newacronym{FEE}{FEE}{Front End Electronics}
\newacronym{GEANT}{GEANT}{GEometry ANd Tracking}
\newacronym{HLT}{HLT}{High Level Trigger}
\newacronym{IP}{IP}{Interaction Point}
\newacronym{IRC}{IRC}{InfraRed and Collinear}
\newacronym{ITS}{ITS}{Inner tracking system}
\newacronym{JER}{JER}{Jet energy resolution}
\newacronym{JES}{JES}{Jet energy scale}
\newacronym{L0}{L0}{Level-0}
\newacronym{L1}{L1}{Level-1}
\newacronym{LED}{LED}{Light-emitting diode}
\newacronym{LHC}{LHC}{Large Hadron Collider}
\newacronym{MB}{MB}{Minimum Bias}
\newacronym{MC}{MC}{Monte Carlo}
\newacronym{MEB}{MEB}{Multi-Event Buffer}
\newacronym{mEMC}{mEMC}{Merged Cluster Technique}
\newacronym{MIP}{MIP}{Minimum ionizing particle}
\newacronym{MWPC}{MWPC}{Multi-Wire Proportional Chambers}
\newacronym{NEF}{NEF}{Neutral Energy Fraction}
\newacronym{NLL}{NLL}{Next-to-Leading Log}
\newacronym{NLO}{NLO}{Next-to-Leading Order}
\newacronym{NNLO}{NNLO}{Next-Next-to-Leading Order}
\newacronym{OPAL}{OPAL}{Omni-Purpose Apparatus at LEP}
\newacronym{Overwatch}{Overwatch}{Online Visualization of Emerging tRends and Web Accessible deTector Conditions using the HLT}
\newacronym{PAR}{PAR}{Pause and Reset}
\newacronym{PCA}{PCA}{Point of Closest Approach}
\newacronym{PCM}{PCM}{Photon Conversion Method}
\newacronym{PCM-EMC}{PCM-EMC}{hybrid method}
\newacronym{PDG}{PDG}{Particle Data Group}
\newacronym{PID}{PID}{Particle Identification}
\newacronym{PHENIX}{PHENIX}{Pioneering High Energy Nuclear Interaction eXperiment}
\newacronym{PHOS}{PHOS}{PHOton Spectrometer}
\newacronym{pQCD}{pQCD}{perturbative Quantum Chromodynamics}
\newacronym{PYTHIA}{PYTHIA}{---}
\newacronym{PS}{PS}{Proton Synchroton}
\newacronym{QA}{QA}{Quality Assurance}
\newacronym{QCD}{QCD}{Quantum Chromo Dynamics}
\newacronym{QGP}{QGP}{Quark Gluon-Plasma}
\newacronym{RHIC}{RHIC}{Relativistic Heavy Ion Collider}
\newacronym{RF}{RF}{Rejection Factor}
\newacronym{SF}{SF}{Scale Factor}
\newacronym{SM}{SM}{Super Module}
\newacronym{SPD}{SPD}{Silicon Pixel Detector}
\newacronym{SPS}{SPS}{Super Proton Synchroton}
\newacronym{SRU}{SRU}{Scalable Readout Unit}
\newacronym{STU}{STU}{Summary Trigger Unit}
\newacronym{TB}{TB}{Test Beam}
\newacronym{TOF}{TOF}{Time Of Flight}
\newacronym{TPC}{TPC}{Time Projection Chamber}
\newacronym{TRU}{TRU}{Trigger Region Unit}
\newacronym{TRD}{TRD}{Transition Radiation Detector}
\newacronym{UE}{UE}{Underlying Event}
\newacronym{V0}{V0}{V0 detector}
\newacronym{WLS}{WLS}{Wavelength Shifting}

\makenoidxglossaries
\ifdraft
\SetWatermarkScale{1.8}
\usepackage{lineno}
\linenumbers
\modulolinenumbers[1]
\usepackage{fancyhdr}
\fi
\begin{document}
\includepdf{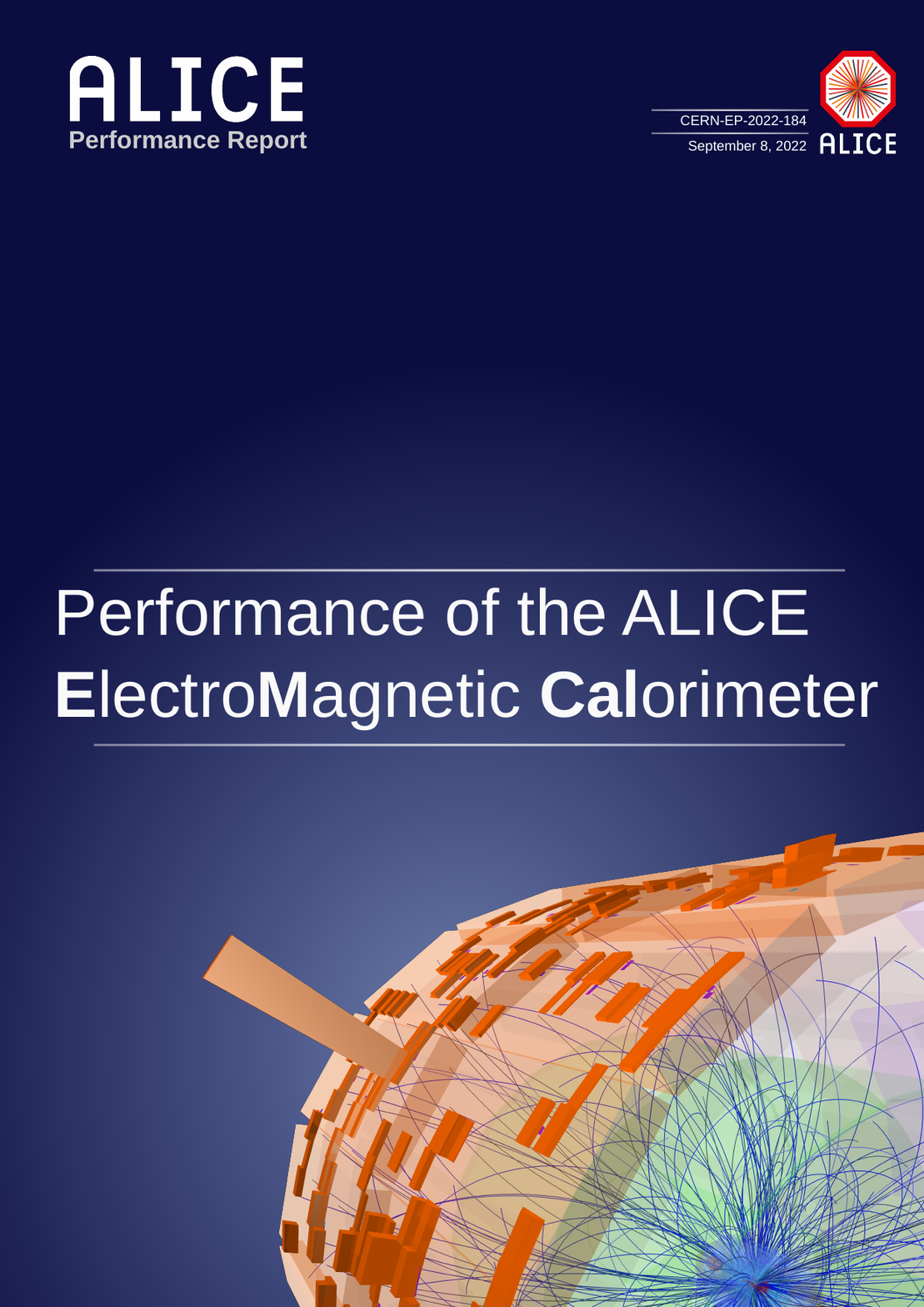}

\begin{titlepage}
\PHyear{2022}
\PHnumber{184} 
\PHdate{30 August}   
\title{Performance of the ALICE Electromagnetic Calorimeter}
\ShortTitle{Performance of the ALICE Electromagnetic Calorimeter} 
\Collaboration{ALICE Collaboration%
         \thanks{See Appendix~\ref{app:collab} for the list of collaboration members}
              }
\ShortAuthor{ALICE Collaboration} 
\begin{center}
\ifdraft
\today \\ 
\color{blue} Revision: \revisionNumber $\color{white}:$ \color{black}\vspace{0.3cm}
\fi
\end{center}
\begin{abstract}
The performance of the electromagnetic calorimeter of the ALICE experiment during operation in 2010--2018 at the Large Hadron Collider is presented.
After a short introduction into the design, readout, and trigger capabilities of the detector, the procedures for data taking, reconstruction, and validation are explained.
The methods used for the calibration and various derived corrections are presented in detail.
Subsequently, the capabilities of the calorimeter to reconstruct and measure photons, light mesons, electrons and jets are discussed.
The performance of the calorimeter is illustrated mainly with data obtained with test beams at the Proton Synchrotron and Super Proton Synchrotron or in proton--proton collisions at \sthirteen, and compared to simulations. 
\end{abstract}
\vspace{1.5cm}
\begin{center}
\begin{minipage}[h]{0.2\textwidth}
\includegraphics[width=3cm]{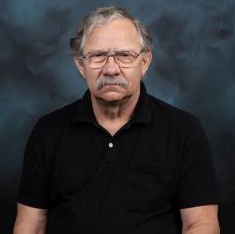}
\end{minipage}
\begin{minipage}[h]{0.3\textwidth}
This publication is dedicated to the memory of our colleague Dr.\ Thomas Cormier who lead the development of the EMCal and recently passed away.
\end{minipage}
\end{center}
\end{titlepage}
\setcounter{page}{2}
\tableofcontents
\newpage
\section{Introduction}
\label{sec:intro}
\gls{ALICE} is a multipurpose experiment at the \gls{LHC} at the \gls{CERN} consisting of multiple detector systems~\cite{Aamodt:2008zz}.
Its primary goal is to investigate the properties of hot quark–gluon matter created in ultrarelativistic heavy-ion collisions. 
This is accomplished by exploiting the unique features of the \gls{ALICE} detector systems, namely to be able to measure and identify both soft particles and hard probes, i.e.\ jets, heavy-flavor hadrons and quarkonia, as well as direct photons.
The \gls{ALICE} experiment incorporates detectors based on a variety of different particle reconstruction and identification techniques. 
The central tracking system covering a pseudorapidity of $|\eta|<0.9$ consists of the \gls{ITS}, \gls{TPC}, \gls{TRD} and \gls{TOF} and is able to detect and identify relatively soft charged particles with transverse momenta \pT~$>50$--$100$~MeV/$c$ in complex, high-multiplicity events~(see~\cite{Abelev:2014ffa} for details on the achieved performance).
All central barrel detectors are placed inside a large solenoid magnet with a maximum field of 0.5~T parallel to the beam direction.
%
In the central region, \gls{ALICE} includes two electromagnetic calorimeter systems: the \gls{PHOS} and the \gls{EMCal}.
The \gls{PHOS} calorimeter was designed to measure spectra and correlations of thermal and direct photons, and of neutral mesons via their decay into photon pairs. 
This requires high granularity as well as excellent energy and position resolution~\cite{Acharya:2019rum}. 
The \gls{EMCal}~\cite{Cortese:2008zza} was designed for the measurements of electrons from heavy-flavor hadron decays, the electromagnetic component of jets, and spectra of direct photons and neutral mesons. 
Compared to \gls{PHOS}, this requires a larger acceptance but less stringent requirements on the energy and position resolution. 
Both calorimeters are trigger detectors, they select collisions when there is a high energy deposition, typically a few GeV, from photons or electrons. The \gls{EMCal} also provides a dedicated trigger for jets. 
Such triggers are needed to sample most of the \gls{LHC} delivered luminosity, since not all the collisions that pass the minimum-bias~(MB) trigger condition~\cite{Abelev:2012sea} 
can be recorded due to \gls{TPC} readout time limitations.  

This document discusses the capabilities, data processing, calibration methods, and performance of the \gls{EMCal} with a focus on the experience and performance from the \gls{LHC} Run~2 from 2015--2018.
It constitutes the first complete description of the achieved performance since the report on the expected physics performance in 2009~\cite{Abeysekara:2010ze} before collision data were taken with the \gls{EMCal}, and since the general summary of the \gls{ALICE} performance from Run~1 with collision data in 2014~\cite{Abelev:2014ffa}.

\begin{figure}[t!]
\begin{center}
\begin{minipage}[t]{0.42\textwidth}
\includegraphics[width=0.99\textwidth]{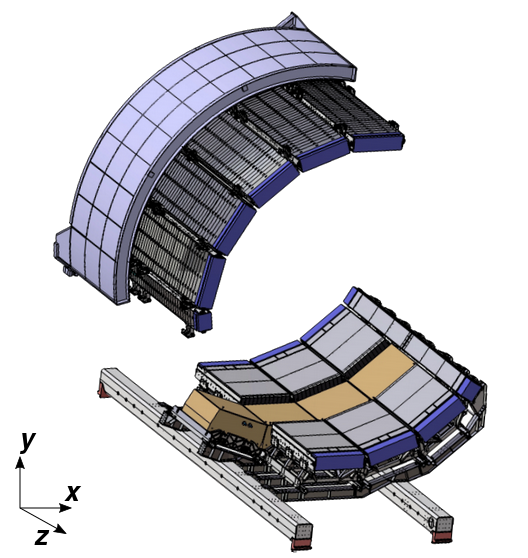}
\end{minipage}
\hspace{0.5cm}
\begin{minipage}[t]{0.42\textwidth}
\includegraphics[width=0.99\textwidth]{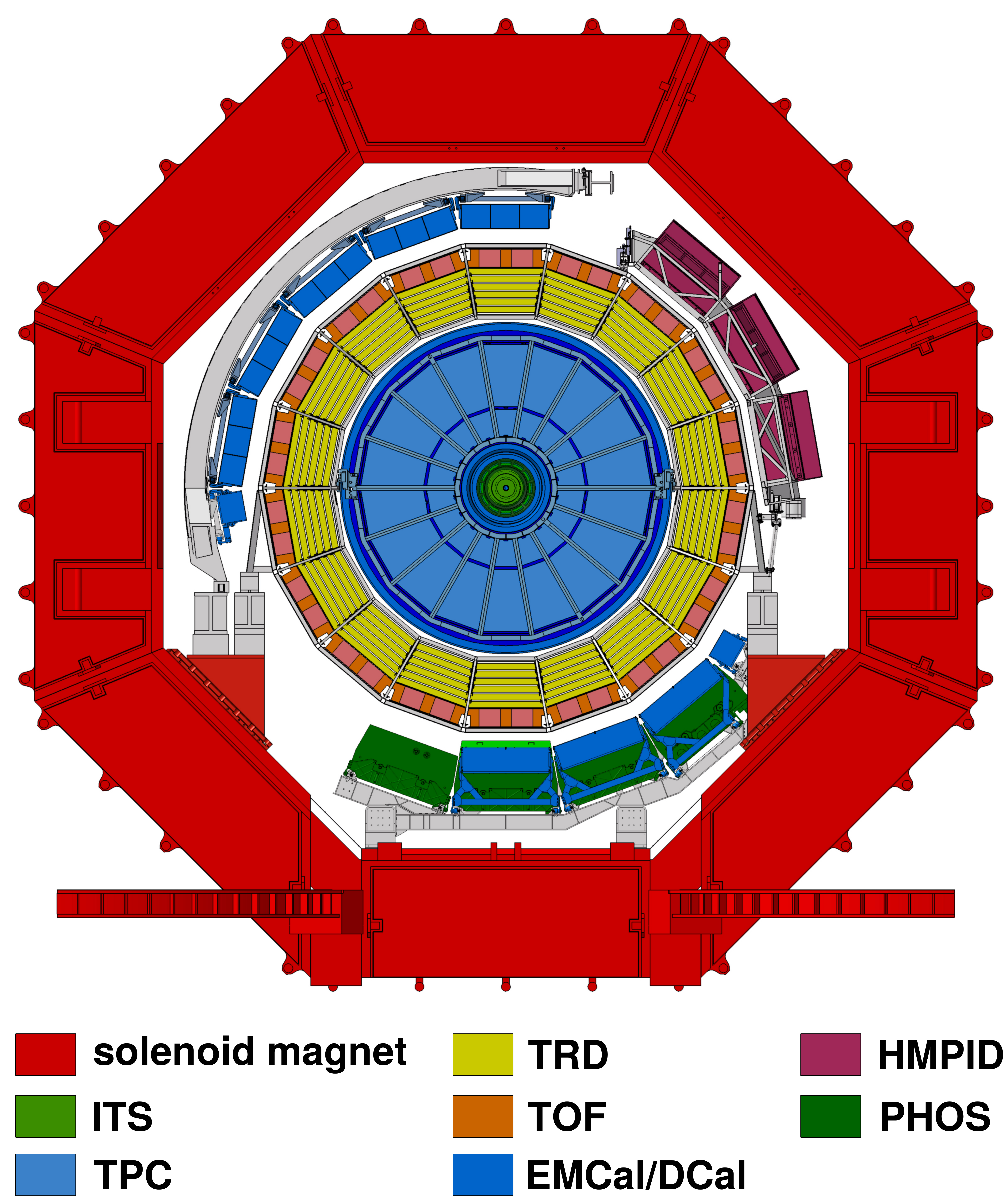}
\end{minipage}
\caption{Schematic view of the \gls{EMCal} (left) illustrating the module position on two approximately opposite locations in azimuth. The \gls{PHOS} calorimeter inside the \gls{DCal} is indicated in brown. The right figure shows a cross section of the \gls{ALICE} barrel detectors.}
\label{fig:0-Intr-Schematics}
\end{center}
\end{figure}

The \gls{EMCal} is a layered lead~(Pb)-scintillator~(Scint) sampling calorimeter with wavelength shifting fibers that run longitudinally through the Pb/Scint stack providing light collection (Shashlik)~\cite{Cortese:2008zza}.
It is located 4.5~m in radial distance from the beam pipe and covers two separate ranges in azimuth, as illustrated in \Fig{fig:0-Intr-Schematics}.
The \gls{EMCal} is structured in so-called \textit{\glspl{SM}}, as described in \Sec{sec:hardware-supermoduledesign}.
It was installed into \gls{ALICE} in several campaigns: 
in 2009, the first four \gls{SM}~(SM0--3, see \Fig{fig:1-HW-EMCal_overview_etaphi}) were installed; in 2011 the supermodules 4--11 were added. 
The rest of \gls{EMCal} supermodules~(SM12--19) were installed in 2014. 
Since these last supermodules are located about 180$^\circ$ opposite in azimuthal angle from the other \glspl{SM}, this second part of the \gls{EMCal} is often referred to as \gls{DCal}, highlighting its purpose to be able to measure dijets~\cite{Allen:2010stl}. 
We usually will refer to the complete detector as \gls{EMCal}~(SM0--19) and only treat the \gls{DCal} as an independent detector when we intend to point out or illustrate a difference. 
An additional change of setup of the \gls{EMCal} was introduced by the successive installation of the \gls{TRD} detector modules~\cite{Acharya:2017lco}, which are in front of the \gls{EMCal}. 
The \gls{TRD} modules provide additional material decreasing the yield of photons in some regions of the \gls{EMCal} due to the additional photon conversions occurring in the \gls{TRD} modules.

The ALICE coordinate system, used throughout the article, is a right-handed orthogonal Cartesian system with its origin at the \gls{LHC} \gls{IP}2. 
The $z$ axis is parallel to the mean beam direction at \gls{IP}2 and points along the \gls{LHC} beam~2~(i.e.\ \gls{LHC} anticlockwise). 
The $x$ axis is horizontal and points approximately towards the center of the \gls{LHC}. 
The $y$ axis, consequently, is approximately vertical and points upwards.
Other \gls{ALICE} or \gls{LHC} specific nomenclature that will be used throughout the document including beam-related properties such as bunch spacing and crossing and orbit are detailed in Refs.~\cite{Aamodt:2008zz,Evans:2008zzb}.

In the following, the data collected so far with the \gls{EMCal} are summarized. 
The data taking at the \gls{LHC} is organized in various units of time. 
There are \gls{LHC} run periods, which last several years and are followed by long shutdown periods. 
To date, \gls{ALICE} recorded data during two \gls{LHC} runs, Run~1 in 2009--2013 and Run~2 in 2015--2018. 
A single data-taking year is subdivided into different periods\footnote{Usually, a period advances from one \gls{LHC} technical stop to the next, but it can switch in between, if something changed in the detector configuration (eg.\ magnetic field) or the beam conditions (eg.\ collision system or energy).
Periods are denoted as \gls{LHC}{\it yyl}, where {\it yy} denotes the year and {\it l} corresponds to a letter to separate different periods, for example "LHC12c". 
Some of the figures in this article contain such period tags.}, which typically last several weeks to months for standard physics data-taking periods, or just days or hours, if special tests are done. 
Within these periods, \gls{ALICE} takes data during single ``runs'', which usually last a few hours and are tagged by a unique number~(run number). 
Tables~\ref{tab:DataSetspp}--\ref{tab:DataSetsHI} summarize the recorded 
datasets with the \gls{EMCal}. 
The \gls{EMCal} recorded events are reported in terms of integrated luminosity $L_{\rm int} = N_{\rm evt} {\rm RF} / \sigma_{\rm MB}$ where $N_{\rm evt}$ is the number of triggered collisions, \gls{RF} is the average number of rejected events per triggered event and $\sigma_{\rm MB}$ is the minimum-bias cross section~\cite{dEnterria:2020dwq,ALICE:2014gvw,ALICE:2012aa,ALICE:2012fjm}. 

\begin{table}[t!]
\centering
\caption{Data collected in \pp{} collisions for different center-of-mass energies (\s{}) with minimum-bias and \gls{EMCal} triggers.}
\begin{tabular}{rrrrr}
$\sqrt{s}$  & Year & MB events (x$10^6$) & $L_{\rm int}$ MB & $L_{\rm int}$ \gls{EMCal} \\
\toprule
0.9 TeV  & 2010 & 5.8 & 0.12 nb${}^{-1}$ & - \\ \midrule
2.76 TeV & 2011 & 26.4 & 0.55 nb${}^{-1}$ & 1.2 nb${}^{-1}$\\ 
         & 2013 & 15.6 & 0.33 nb${}^{-1}$ &47.1 nb${}^{-1}$\\ \midrule
5.02 TeV & 2015 & 100 & 1.95 nb${}^{-1}$ & 0.075 pb${}^{-1}$\\ 
         & 2017 & 1129 & 22.05 nb${}^{-1}$ & 0.435 pb${}^{-1}$ \\ \midrule
7 TeV    & 2010 & 358 & 5.74 nb${}^{-1}$ & - \\
         & 2011 & 1.8 & 0.03 nb${}^{-1}$ & 0.47 pb${}^{-1}$ \\ \midrule
8 TeV    & 2012 & 108 & 1.93 nb${}^{-1}$ & 0.62 pb${}^{-1}$ \\ \midrule
13 TeV   
		& 2016	& 382 & 6.61 nb${}^{-1}$ & 2.44 pb${}^{-1}$\\
		& 2017	& 519 & 8.97 nb${}^{-1}$ & 3.74 pb${}^{-1}$\\
		& 2018	& 615 & 10.64 nb${}^{-1}$ & 3.23 pb${}^{-1}$\\
\bottomrule
\end{tabular}
\label{tab:DataSetspp}
\end{table}

\begin{table}[t!]
    \centering
    \caption{Data collected in \pPb\ collisions for different nucleon--nucleon center-of-mass energies (\snn) with minimum~bias and \gls{EMCal} triggers. The numbers in parentheses denote the corresponding data set with only the triggering and vertexing detectors in the readout.}
    \begin{tabular}{rrrrr}
        \snn\ & year & MB events (x$10^6$) & $L_{\rm int}$ MB & $L_{\rm int}$ \gls{EMCal} \\
        \toprule
        5.02 TeV  & 2013 & 94    & 0.045 nb${}^{-1}$ & 7.38 nb${}^{-1}$\\
                  & 2016 & 490   & 0.235 nb${}^{-1}$  & - \\ \midrule
        8.16 TeV & 2016 & 38(83) & 0.018 (0.041) nb${}^{-1}$ & 1.42 (5.67) nb${}^{-1}$  \\ \bottomrule
        \end{tabular}
    \label{tab:DataSetspPb}
\end{table}

\begin{table}[t!]
    \centering
    \caption{Data collected in heavy-ion collisions at different nucleon--nucleon collision center-of-mass energies (\snn)  using minimum~bias, centrality~\cite{Abelev:2013qoq} or \gls{EMCal} triggers. Centrality triggers are defined as minimum bias (0-100\%), central (0-10\%), and mid-central (0-50\% for 2011 and 30-50\% for 2018). The corresponding cross sections are given in $\mu$b$^{-1}$ in the brackets.}
    \begin{tabular}{rrrrrr}
        &  & \multicolumn{3}{c}{centrality triggered events (x$10^6$)} & $L_{\rm int}$ \gls{EMCal} \\
        System & Year & minimum bias & central & mid-central &  \\
        \toprule
        Xe, 5.44 TeV  & 2017 & 1.7  (0.3) & -  & - & -\\ \midrule
        Pb, 2.76 TeV  & 2010 & 14.8  (2.0)& - & - & - \\
                      & 2011 & 1.6  (0.2) & 11.5  (15.3) & 10.6  (4.7) & 25.3 $\mu$b${}^{-1}$\\ \midrule
        Pb, 5.02 TeV  & 2015 & 74  (9.7) & - & - & 51.2 $\mu$b${}^{-1}$\\
                      & 2018 & 159  (20.9) & 133  (174.5) & 73  (48.1) & 116.9
                      $\mu$b${}^{-1}$ \\ \bottomrule
    \end{tabular}
    \label{tab:DataSetsHI}
\end{table}

The remaining part of the article is structured as follows. 
Details about the design and readout of the detector are given in \Sec{sec:hardware}.
Details on data taking, reconstruction and validation are given in \Sec{sec:dat}. 
In order to characterize the \gls{EMCal} in a controlled environment, a prototype was tested with beams of known energy in 2010.
The details of the setup and analysis of the beam test are documented in \Sec{sec:testeam}. 
To use the data of the \gls{EMCal}, several iterations of calibration and selections have to be applied. 
The technicalities of the related procedures are given in \Sec{sec:calibration}. 
The \gls{EMCal} performance is typically illustrated using data and simulations from \pp\ collisions at a center-of-mass energy of $\sqrt{s}=13$~TeV, and from \PbPb\ collisions at a center-of-mass energy per nucleon--nucleon of $\snn=5.02$~TeV.
To reconstruct objects like photons, light mesons, electrons, and jets, specific algorithms are applied to the calibrated data. 
The performance of the \gls{EMCal} regarding these reconstructed objects are discussed in \Sec{sec:physics}.
\enlargethispage{1cm}
\ifflush
\clearpage   
\fi
\section{Detector description}
\label{sec:hardware}
The main characteristics of the detector and the \gls{FEE} are briefly described below. They are relevant for understanding the physics performance discussed in the subsequent sections. For a complete description see~\cite{Cortese:2008zza, Allen:2010stl}. 

\subsection{Module design}
\label{sec:hardware-moduledesign}
\begin{figure}[b!]
\begin{center}
\includegraphics[scale=0.6]{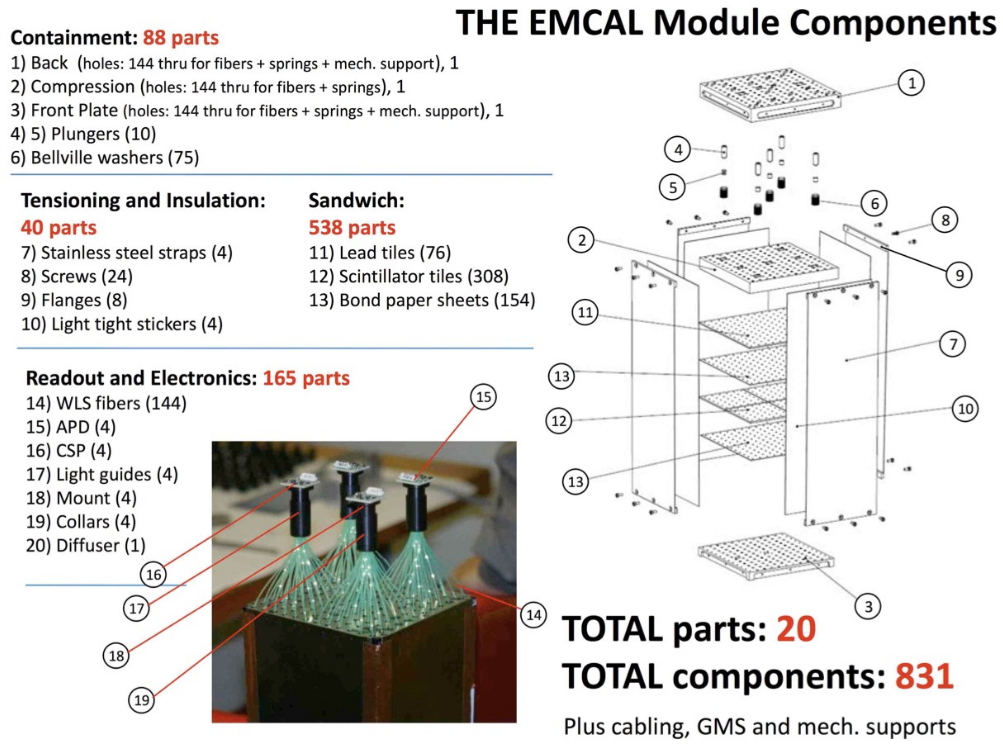}
\caption{
Photo and drawing of \gls{EMCal} module showing all components.} \label{fig:1-HW-ALICE_EMCAL_Module}
\end{center}
\end{figure}

The basic building block of the \gls{EMCal} is a Module, which consists of $2\times2$ optically isolated towers as illustrated in \Fig{fig:1-HW-ALICE_EMCAL_Module}. 

\begin{figure}[ht!]
\begin{center}
\includegraphics[scale=0.6]{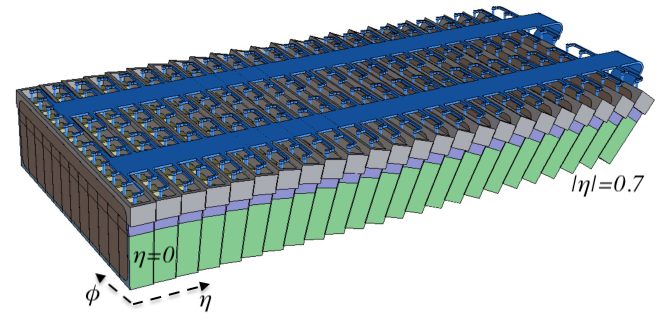}
\caption{Schematic view of the \gls{EMCal} full-size super modules (\glspl{SM}) illustrating the strip structure made of 24 strips.} \label{fig:1-HW-ALICE_EMCAL_SM_strips}
\end{center}
\end{figure}

Each tower is read out individually and spans a region of $\Delta \eta \times \Delta \varphi \simeq 0.0143 \times 0.0143$.
The tower lines are referred to as columns in the $\eta$ direction and as rows in the $\varphi$ direction.
Each module has a fixed width in the $\varphi$ direction and a tapered width in the $\eta$ direction with an angle of $1.5^{\circ}$. 
The resulting assembly of stacked strips made of 12 identical modules (along $\varphi$) is approximately projective in $\eta$ (see \Fig{fig:1-HW-ALICE_EMCAL_SM_strips}) with an average angle of incidence at the front face of a module of less than $2^{\circ}$ in $\eta$ and less than $5^{\circ}$ in $\varphi$.

The physical characteristics of the \gls{EMCal} modules are summarized in \Tab{Table-1}.
White, acid-free, bond paper serves as a diffuse reflector on the scintillator surfaces and provides friction between 76 layers of lead and 77 layers of scintillator. 
The scintillator edges are treated with TiO$_2$ loaded reflector to improve the transverse optical uniformity within a single tower and to provide tower-to-tower optical isolation better than $99\%$. 

\begin{table}[ht]
\begin{center}
\caption{\gls{EMCal} module physical parameters.} 
\begin{tabular}{ll}
Parameter & Value \\ \toprule
Tower Size (on front face) & 6.0 $\times$ 6.0 $\times$ 24.6 cm$^3$ \\  
Tower Size (at $\eta$=0 ) & $\Delta  \eta \times \Delta  \varphi \simeq 0.0143 \times 0.0143$ \\ 
Sampling Ratio & 1.44 mm Pb / 1.76 mm Scint. \\ 
Layers & 77 \\  
Scintillator & Polystyrene (BASF143E $+$ \\
             & 1.5\%pTP $+$ 0.04\%POPOP)  \\ 
Absorber  & natural lead \\
Effective radiation length $X_0$ & 12.3 mm \\ 
Effective Moli\`ere radius $R_{\rm M}$ & 3.20 cm \\  
Effective Density & 5.68 g/cm$^3$ \\  
Sampling Fraction & 1/10.5           \\  
No. of radiation lengths & 20.1 \\  \bottomrule
\end{tabular}
\label{Table-1}
\end{center}
\end{table}

Scintillation photons produced in each tower are captured by an array of 36
Kuraray Y-11 (200 M), double clad, \gls{WLS} fibers.
Each fiber within a given tower terminates in an aluminized mirror at the front face of the module and is integrated into a polished, circular group of 36 fibers at the photosensor end at the back of
the module.
 \glspl{APD} of $5\times5$ mm$^2$ size (Hamamatsu S8664-55 or Perkin Elmer C30739ECERH-l) are used as active photosensors for operation in the high 0.5~T field inside the \gls{ALICE} magnet.
The \glspl{APD} are operated at moderate gain for low noise and high gain stability in order to maximize energy and timing resolution. 
The number of primary electrons generated in the \gls{APD} from an electromagnetic shower is $\approx 4.4$ photoelectrons/MeV.   
The reverse bias voltage of the \glspl{APD} are individually controlled to provide an electron multiplication factor of $\sim$30, resulting in a charge output of $\approx 132$ electrons/MeV from the \glspl{APD}. The main characteristics of the \gls{EMCal} \gls{FEE} are summarized in \Tab{Table_Elpar}

\begin{table}[ht]
\begin{center}
\caption{\gls{EMCal} \gls{FEE} main characteristics.} 
\begin{tabular}{ll}
Parameter & Value \\ \toprule
High gain range & 15.3 MeV to 15.6 GeV\\
Low gain range  &  248 MeV to 250 GeV\\
Time integration window & 1.5 $\mu$s\\
Digitization sampling rate & 10 MHz\\
Light yield at APD  & $\approx 4.4$ photoelectrons/MeV\\
APD gain            & $\approx 30$ \\
Shaping time        & $\approx 235$~ns
\\  \bottomrule
\end{tabular}
\label{Table_Elpar}
\end{center}
\end{table}

The \gls{APD} is connected directly to a \gls{CSP} with a short rise time of $\approx 10$ ns and a long decay time of $\approx 130~\mu{\rm s}$, i.e., approximately a step pulse.
The amplitude of the step pulse is proportional to the number of integrated electrons from the \gls{APD} and therefore  proportional to the energy of the incident particle.

\begin{figure}[ht!]
\begin{center}
\includegraphics[scale=0.25]{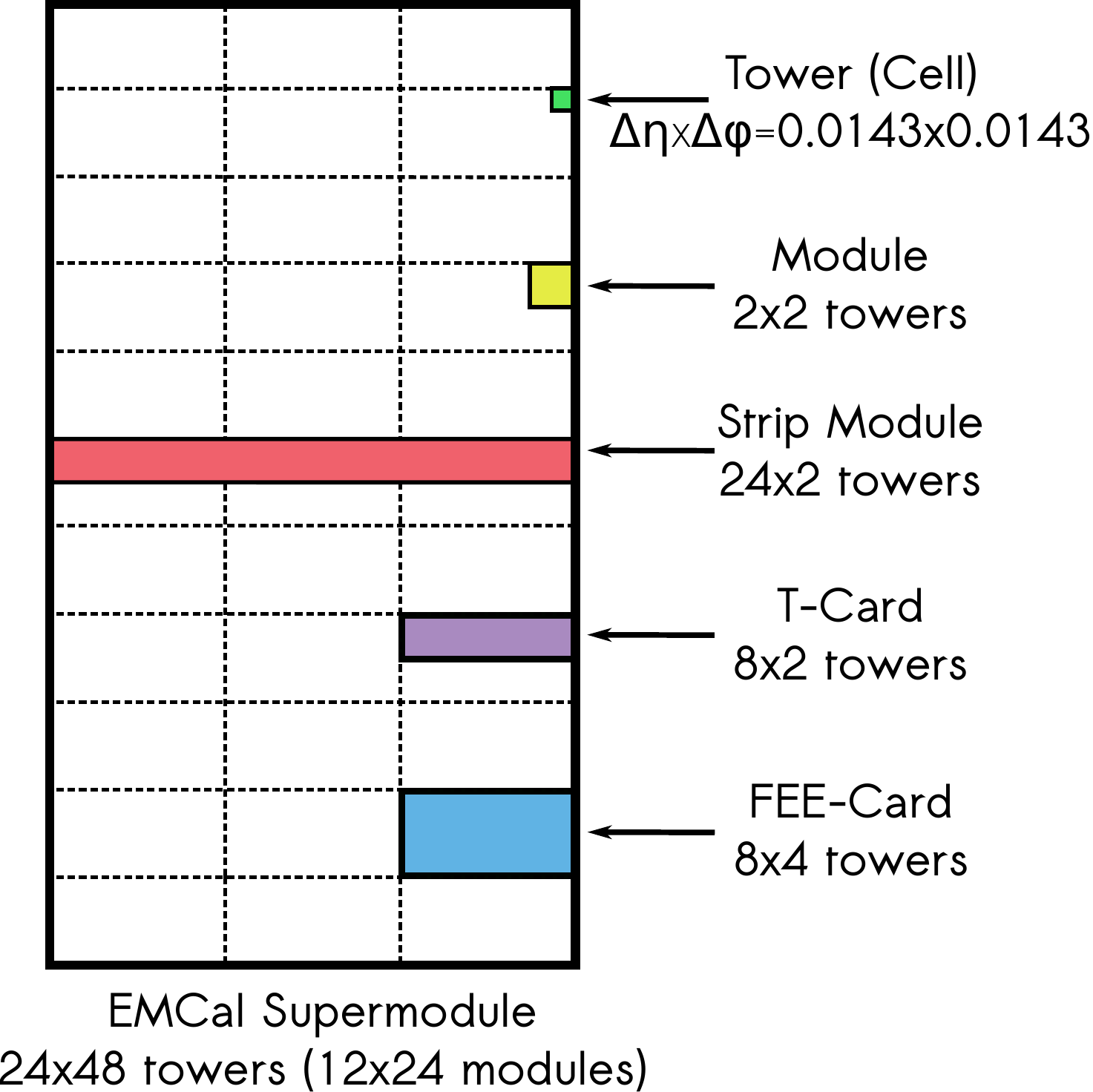}
\caption{(Color online) \gls{EMCal} full-size (\gls{SM}) in the $(\eta,\varphi)$ plane including visualizations of sub-components and their tower coverage.} \label{fig:1-HW-EMCal_supermodule_subparts}
\end{center}
\end{figure}

The \glspl{CSP} of $2\times8$ adjacent towers are connected to an adaptor called as T-Card, and the analog signals arrive via a ribbon cable to the \gls{FEE} cards located at the end of the \gls{SM}.  
\Figure{fig:1-HW-EMCal_supermodule_subparts} illustrates the acceptances covered by a tower, module, strip module, T- and \gls{FEE}- cards.

\subsection{Supermodule design}
\label{sec:hardware-supermoduledesign}
\begin{figure}[ht!]
\begin{center}
\includegraphics[scale=0.3]{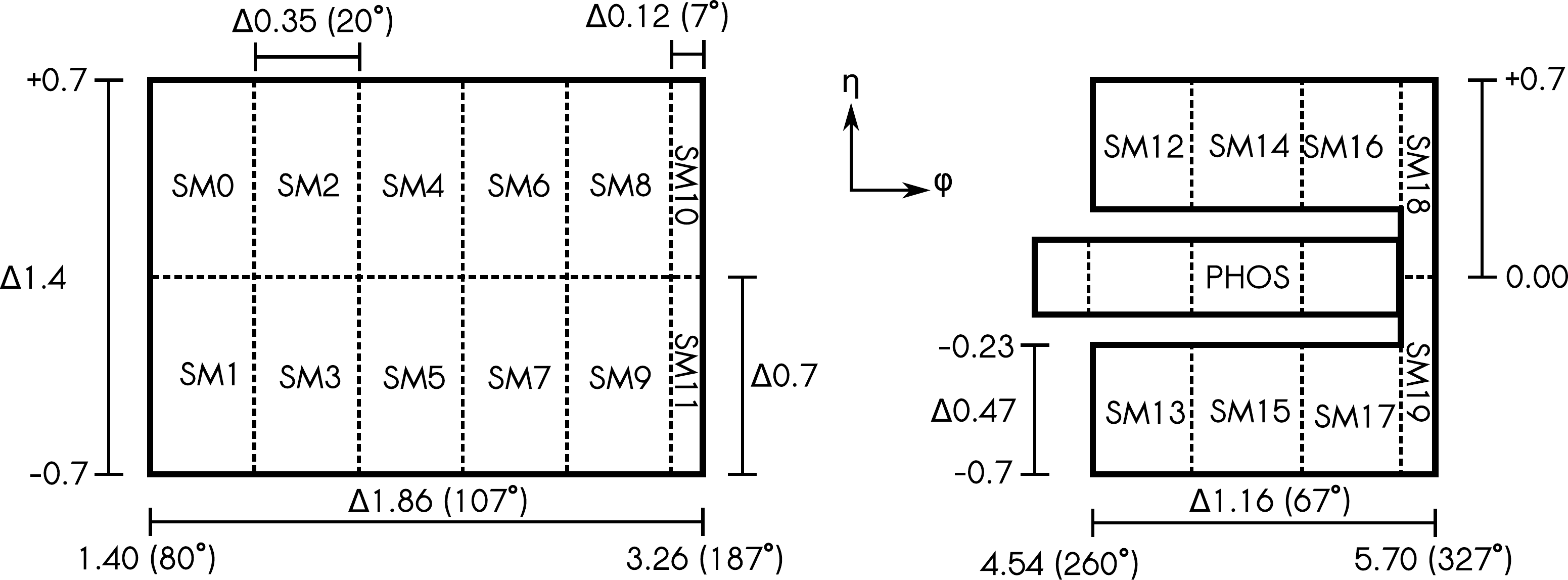}
\caption{Geometric overview of the \gls{EMCal} and \gls{DCal}  detectors in the $\eta$-$\varphi$ plane. The drawing outlines the full \gls{LHC} Run 2 setup with all 20 \glspl{SM} as well as the \gls{PHOS} detector in the \gls{DCal}  gap.} \label{fig:1-HW-EMCal_overview_etaphi}
\end{center}
\end{figure}

The overall design of the calorimeter is heavily influenced by its integration within the \gls{ALICE}~\cite{Aamodt:2008zz} setup and \glspl{SM} of 3 different sizes are used: full-, 2/3- and 1/3- size. 
Each full-size \gls{SM} is assembled from $12\times24= 288$ modules arranged in 24 strip modules of $12\times1$ modules each. Each 2/3-size \gls{SM} is assembled from $12\times16=192$ modules, and 
each one-third size \gls{SM} is assembled from $4\times24=96$ modules.
The \gls{EMCal} is made of 10 full-size \glspl{SM} and 2 1/3-size \glspl{SM} covering $|\eta|<0.7$ in pseudorapidity  and $80^{\circ} < \varphi < 187^{\circ}$ in azimuth.  
\gls{DCal} is made of 6 2/3-size \glspl{SM} and 2 1/3-size \glspl{SM} with acceptance:  $0.22 < |\eta| < 0.7, 260^{\circ} < \varphi < 320^{\circ} $ and $|\eta| < 0.7, 320^{\circ} < \varphi < 327^{\circ}$. 
A schematic view of the acceptance and the \gls{SM} numbering scheme used in the offline software is illustrated in \Fig{fig:1-HW-EMCal_overview_etaphi}. 
The \gls{EMCal} has 12288 towers and \gls{DCal} has 5376 towers. 
The \glspl{SM} are installed in the experimental cavern such that the radial distance of the \gls{SM} front face from the \gls{IP} is  
$R\simeq 4.3$ m, see \Fig{fig:0-Intr-Schematics}.

\subsection{Readout} 
\label{sec:readout}
A single \gls{FEE} card~\cite{Muller:2006jr} provides readout of 32 towers~(2 adjacent T-Cards) and splits the individual preamplifier signal into energy and trigger shaper channel. 
The energy signals are shaped with $\approx200$~ns shaping time in dual shaper channels differing by a factor of 16 in gain, sampled at 10 MHz with the 10-bit \gls{ALTRO} chip ~\cite{EsteveBosch:2003bj}. 
The shapers are designed such that for an \gls{APD} gain factor of 30, the range fills the full input range of 1 V of the \gls{ALTRO} chip, for each of the high ($\times16$) and low gain channels. The signal $V(t)$\com{ has an semi-Gaussian shape and} is characterised by the formula in \Eq{HW_Eq_Gauss2}:
\begin{equation}
V(t) = \frac{2Q  A^2}{C_{f}}\left[\frac{t-t_{0}}{\tau}\right]^2 \exp\left\{{-2\frac{t-t_0}{\tau}}\right\}
\label{HW_Eq_Gauss2}
\end{equation}
where $Q$ is the charge on \gls{APD}, $A$ the preamplifier gain, $C_f$ the capacitance in the preamplifier, $t_{0}$ the peak time, and $\tau$ the shaping time.

The dynamic range of the calorimeter  is defined by the requirement of 250~GeV maximum single channel energy. 
Given the 16:1 gain ratio of the high and low gain channels, the two overlapping 10-bit \gls{ADC} ranges correspond to an effective 14-bit dynamic range over the interval from $\sim$~16~MeV to 250~GeV, resulting in a Least Significant Bit on the low gain range of 250~MeV and on the high gain range of 16~MeV.

Since the digitization by the \gls{ALTRO} chip of the two parallel streams corresponding to low and high gain channels is  performed with 10~MHz (100~ns) sampling frequency, information about the signal time phase is required, as there are 4 possible phases of sampling in respect to trigger signal of 40~MHz granularity.
During LHC Run 1~(years 2009 to 2011), the phases were found online by the readout firmware, while during LHC Run 2~(years 2015 to 2018) they were found with an offline analysis of data as described in \Sec{sec:timeCalib}.

The \gls{ALTRO} chip copies a defined amount of samples from a rotating buffer to a \gls{MEB} on reception of the \gls{L0} trigger from the \gls{CTP}~\cite{Fabjan:684651}. 
Reception of the \gls{L1} signal validates the data in the multi-event buffer, so that this can not be rewritten by the following \gls{L0} trigger. 
Readout of the multi-event buffer can be performed after each \gls{L1} (no \gls{MEB} operation) or after up to 4 triggers.
The \gls{ALTRO} chips are configured to record fifteen 10-bit time (pre-)samples per readout channel per event, corresponding to 1.5~$\mu$s time integration window. 
They are compressed by discarding samples close to the reference level (pedestal) that contain no useful information (``zero suppressed''), reducing substantially the data volume provided to readout. 
The pedestals are obtained from special runs with no pre-programmed pedestal or signal present.

\subsection{Trigger} 
\label{sec:hardware-trigger}
Both \gls{EMCal} and \gls{DCal}  provide inputs to the \gls{L0} and \gls{L1}  trigger decisions in \gls{ALICE}. The trigger subsystem resides in specific hardware boards.
The analog signals of $2\times2$ groups of adjacent towers are summed in the \gls{FEE} boards and transmitted to a local \gls{TRU} board where the 
$2\times2$ tower sums from 12 \gls{FEE} cards (96, $2\times2$ sums) are digitized at the LHC clock frequency of 40 MHz~\cite{Kral:2012ae}. The digitized $2\times2$ tower sums are summed over time samples 
with pre-sample pedestal subtraction to provide an integral energy measurement, referred to as ``timesums''. 
Finally, overlapping $4\times4$ tower digital sums, called {\it trigger patches}, are formed within each \gls{TRU} and a peak-finding algorithm is used to find a signal peak.  
Each $4\times4$ sum signal peak amplitude is then compared against a threshold to provide a \gls{L0} trigger output that indicates the presence of a high energy shower in the \gls{TRU} region (1 \gls{TRU} covers 1/3 of the area for a full-size \gls{SM}). 
The \gls{L0} trigger decision from each \gls{TRU} is passed to a \gls{STU} that performs the logical OR of the \gls{L0} outputs from all TRUs to provide a single \gls{L0} input to the \gls{ALICE} \gls{CTP} within 800 ns after the interaction.

Upon receipt of an accepted \gls{L0} trigger from the \gls{CTP}, the digitized time-summed $2\times2$ tower sums from each \gls{TRU} are passed to the \gls{STU}. 
In the \gls{STU} the $4\times4$ overlapping tower sums are formed again, but across \gls{TRU} boundaries over the full acceptance to provide an improved \gls{L1} high-energy shower trigger referred to as \gls{L1}-$\gamma$ trigger~\cite{Bourrion:2012vn}. 
At the same time, tower sums over a large $8\times8$ trigger-channel window ($16\times16$ towers) and a $16\times16$ trigger-channel window ($32\times32$ towers) are also formed to provide a \gls{L1} jet trigger.
Both \gls{L1} triggers allow the definion of two thresholds for the event selection, referred to as high and low threshold.

In order to reduce the bias due to multiplicity fluctuations in heavy-ion collisions, there is a direct  communication between  \gls{EMCal} and \gls{DCal} \glspl{STU} for considering the underlying event background in the online \gls{L1} trigger decision.
For \gls{EMCal}, the background is estimated  based on the median of the energies deposited in  $8\times8$ trigger channel ($16\times16$ towers) windows in \gls{DCal}, and vice versa for \gls{DCal}.
The background is subtracted from the signal amplitude and then compared against a threshold to provide \gls{L1} triggers. 
During Run~1, when only the \gls{EMCal} was installed, the background for \gls{EMCal} was estimated based on the multiplicity measured by the V0 detector.
The settings and thresholds used for different data-taking periods are given in \Sec{sec:trigger}.

\subsection{APD pre-calibration and gain monitoring}
\label{sec:APD}
Prior to installation, each \gls{APD} was tested to verify its basic functionality and properties~\cite{Badala:2009zz}. 
In particular, the voltage needed for the \gls{APD} to obtain gain $M=30$ was required to be lower than 400~V due to the limitation in the \gls{EMCal} \gls{FEE}.  
The gain $M(V)$ is defined as the ratio between the amplitude at the voltage $V$ and the amplitude in the plateau at fixed temperature, and is parameterised with an exponential function  plus a constant~\cite{Badala:2008zzd}:
\begin{equation} 
M(V) = p_0 + p_1 \, e^{\,p_2\,V}\,.
\label{eqn-2-gain-gainVsHV} 
\end{equation}

The individual gain curves for most of the \glspl{APD} were determined and used to set the initial biases for each readout channel and equalize the gain for all towers. 
This allowed the equalisation of the initial tower calibrations to better than $\approx20\%$ in laboratory conditions.

Then, each \gls{SM} was calibrated prior to installation, by the use of the peak of the energy deposit spectrum of atmospheric muons traversing the calorimeter (see \Sec{sec-Energy-pre-calibration}).
The final relative tower-by-tower energy calibrations are obtained from measurements of the $\pi^0$ mass peak in the two-photon invariant mass spectrum as discussed in \Sec{sec-Energy-calibration-with-the-LHC-data}.
The absolute energy calibration is obtained by comparing the $\pi^0$ mass peak  positions from data and \gls{MC} as described in \Sec{sec:EMCalEnergyPositionCalib}.

The \gls{APD} gain also strongly depends on the temperature: since the avalanche multiplication depends on the mean free path of electrons between ionizing collisions, which is temperature dependent, the \gls{APD} gain decreases with temperature. 
The temperature across the \glspl{SM} is monitored by temperature sensors: eight sensors in the full-size and 2/3-size \glspl{SM} and four in the 1/3-size modules.
In parallel to periodically recording the measurements by temperature sensors during data taking, 
the temperature/time dependence of the \gls{APD} gain is monitored using a custom built \gls{LED} system triggered by an avalanche pulser system to provide an intense light pulse of several ns duration~\cite{Muller:2006jr, Cortese:2008zza}. 

While physics data-taking is ongoing, the events triggered by calibration triggers are taken during the long gap of about 2.97~$\mu$s at the time when the orbit reset from the LHC machine is sent, to avoid overlaps of ``physics'' pulses due to bunch--bunch crossings with the one generated by the \gls{LED} system. 
A calibration pre-pulse is provided to trigger the \gls{LED} pulser system by the \gls{CTP} followed by an L0 calibration event trigger with about 1.9~$\mu$s latency. 
At the reception of the calibration pre-pulse, up to $10^9$ photons of 470~nm wavelength are generated by a single ultra-bright blue \gls{LED}~(part no.~E7113UVC by eLED) and are transmitted to each strip module via an optical fiber. 
At the strip module, the optical fiber is split into twelve fibers that bring the light to a hole between the four towers at the back of each module. 
A diffuser reflects the \gls{LED} light back into the \gls{WLS} of each tower.  
The \glspl{LED}  are  monitored by photodiodes that are read out with an extra \gls{FEE} card installed in the \gls{FEE} crate close to the \gls{LED} system.
The variation of the \gls{EMCal} response to the \gls{LED} signal with time and temperature is presented in \Sec{tempCalib}. 

\ifflush
\clearpage
\fi
\section{Data processing}
\label{sec:dat}
\label{sec:DataReco}

In this section the basic steps of the \gls{EMCal} data reconstruction are discussed: raw data fitting, clusterization, cluster characterization, and data quality control.
Further, it summarizes all information regarding the \gls{EMCal} trigger performance for several collision systems and reconstructed objects.

\subsection{Raw data processing}
\label{sec:data-raw}
\begin{figure}[t]
\begin{center}
\includegraphics[width=0.5\textwidth]{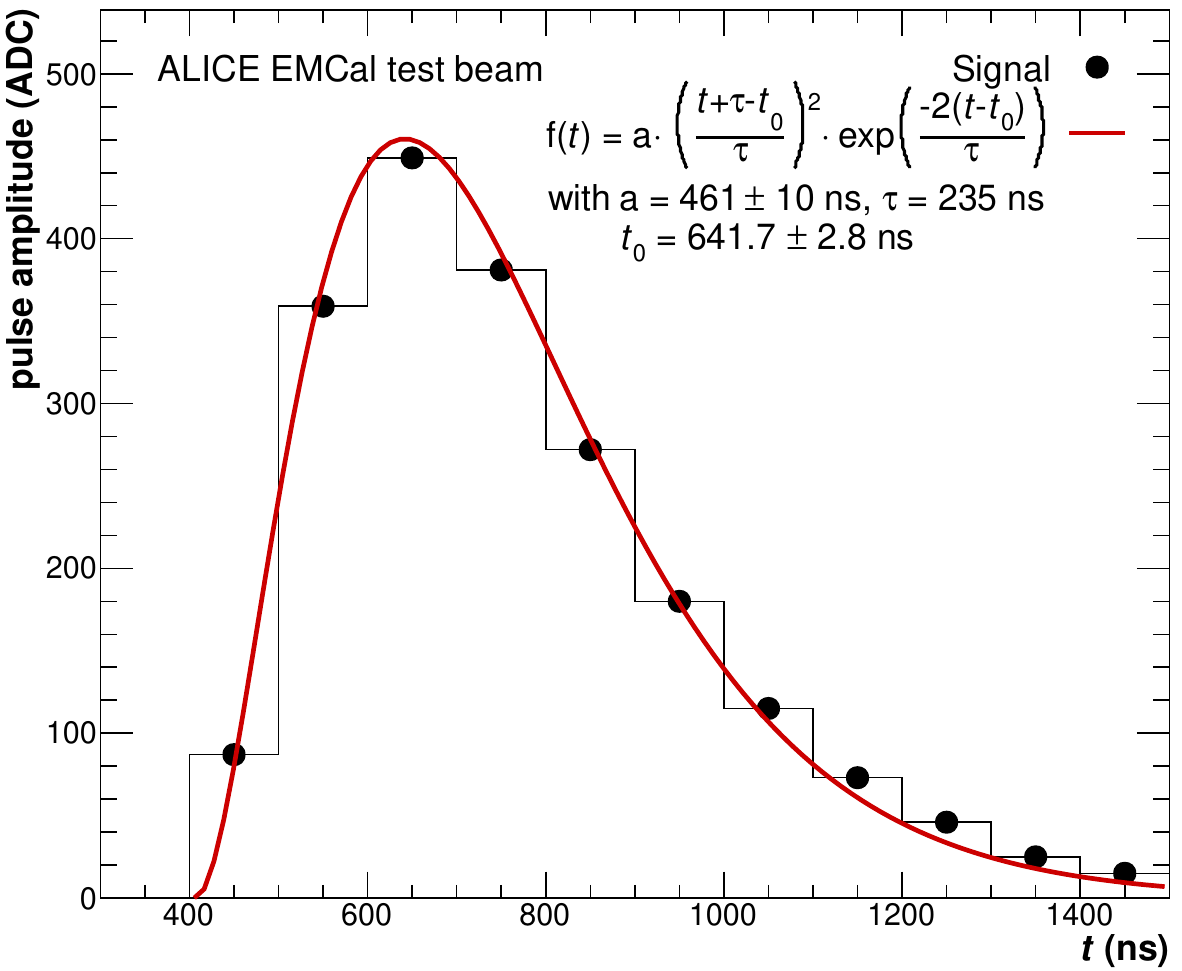}
\caption{Example of a fit to a digitized pulse from electron test beam data using the function defined in \Eq{HW_Eq_Gauss2} with a fixed shaping time parameter $\tau = 235~$ns. } 
\label{fig:3-Raw-PulseFit}
\end{center}
\end{figure}

As described in \Sec{sec:readout}, the electronics signal is digitized with a 10~MHz \gls{ADC} within the 1.5~$\mu$s time integration window. 
For each tower the pedestal subtraction and zero suppression are already applied in the electronics.
The amplitude of the digitized signal is lower than the amplitude of the input analog signal because a simple comparator is used in the \gls{ADC} and the difference depends on the sampling phase. 
In order to correct for this effect and measure the signal peak time accurately, the digitized-signal distribution is fitted with the function defined in \Eq{HW_Eq_Gauss2} during the offline reconstruction.
\Figure{fig:3-Raw-PulseFit} shows an example of the signal distribution for electron test beam data to illustrate the quality of the fit.
The fit was done with a fixed shaping time parameter $\tau = 235~$ns. 

From the fit, the signal {\it amplitude} (pulse maximum amplitude) and arrival {\it time} (time bin of the pulse maximum) are extracted per cell and stored in the reconstructed data. 
The amplitude is provided in \gls{ADC} counts, where an \gls{ADC} count corresponds to approximately 16~MeV in the high gain region and 250~MeV in the low gain region.
The exact correspondence varies from channel to channel and is determined in the energy calibration procedure (see \Sec{sec:e-calibchapter}).
The peak time is around $600~$ns due to cable lengths and delays, which are corrected for by the time calibration procedure described in \Sec{tempCalib}

\subsection{Detector response in simulations}
\label{sec:MC-setup}
The \gls{EMCal} simulation relies on two main steps: 
particle transport (particle interactions in the detector material), 
and digitization (transformation of the deposited particle energy into 
the equivalent quantities obtained from raw data after the raw signal fitting).
The transport of particles through the detector is handled with the \gls{GEANT} v3~\cite{Brun:1987ma} package for most of the simulations corresponding to Run~1 and~2 data. 
The \gls{GEANT} v4~\cite{Agostinelli:2002} package was used in a few instances for performance comparisons with the v3 version.
\gls{GEANT} v3 was preferred by \gls{ALICE} since it described the full detector system better and is faster in heavy-ion simulations with the software package used in Run 1 and 2.
In both transport models, particles are required to have a minimum kinetic energy of 1~MeV in order to be transported through the \gls{EMCal}.
The different processes that were typically activated are listed in \App{sec:GEANTConf} for the two transport models.
\gls{GEANT}4 relies on predefined ``physics lists'' that configure different processes. 
The one used in \gls{ALICE} is ``FTFP\_BERT\_EMV+optical+biasing'', which enforces fast simulation of electromagnetic processes via the Urban Special Model~\cite{Urban:2005} but does not describe the electromagnetic shower development in the calorimeter.
For the \gls{EMCal}, we ensure a detailed electromagnetic process modelling based on the \gls{GEANT}4 option ``EmStandard\_opt0''~\cite{Ivantchenko}.
For each particle entering the ``sensitive'' \gls{EMCal} material~(scintillator, polystyrene), the deposited energy of the particle ($\Delta E_{\rm deposited}$) and its shower daughters are recorded.
This energy is corrected for the diminished light output due to the ionization produced by the preceding particles~(saturation).
This is done by rescaling the energy deposited per particle using Birk's law~\cite{Birks:1951}, \Eq{equation:Birks}, as in \gls{GEANT}3's \texttt{G3BRIRK} routine:  
\begin{eqnarray}
{\rm Light\; yield} & = & 
 \frac{\Delta E_{\rm deposited}}{1+C_1 \delta + C_2 \delta^2  } \rm ,\: with
\label{equation:Birks} \\
\delta & = & 
 \frac{1}{\rho}\frac{dE}{dx}\; \left[\frac{{\rm MeV}\: {\rm cm}^{2}}{\rm g}\right]\nonumber , \\
C_{1} & = & \left\{ \begin{array}{ll}
                    0.013\: \left[\frac{\rm g}{{\rm MeV}\: {\rm cm}^{2}}\right] & Z=1 \\
                    0.00743\; \left[\frac{\rm g}{{\rm MeV}\: {\rm cm}^{2}}\right] & 
                                         Z>1 \end{array}\right.\nonumber \rm ,\: and \\
C_{2} & = & 9.6\times 10^{-6}\; 
                  \left[\frac{\rm g^{2}}{{\rm MeV}^{2}\: {\rm cm}^{4}}\right]\,. \nonumber
\end{eqnarray}
The polystyrene density is set to $\rho$ = 1.032 g/cm$^{3}$.
An additional sampling calibration factor depending on the transport model is needed to match the deposited energy and the particle energy. 
The sampling calibration factor is 10.87 for \gls{GEANT}3 and 8.85 for \gls{GEANT}4. 
Those factors were found by checking the mean reconstructed energy of photons generated at fixed energies.
At the digitization step, all the deposited energy in a cell, $E_{\rm cell}$, is summed. 
In order to include fluctuations in the \gls{APD} gains, $E_{\rm cell}$ is smeared using Poisson fluctuations. 
The fluctuations were parameterized according to a Poisson distribution with the variance:
\begin{eqnarray}
\label{equation:APDFluc} 
\lambda_{E}  =  E_{\rm cell}~\mu_{\gamma-e} / \sigma_{\rm \rm APD}\,,  
\end{eqnarray}
where the average number of photoelectrons $\mu_{\gamma-e} =4.4$ photoelectrons/MeV (see \Sec{sec:readout}) and  the \gls{APD} gain fluctuations $\sigma_{APD}=15$ were tuned using the \com{last \gls{SPS}}electron test beam measurements.   
Finally, electronic noise is assigned to each cell, via a Gaussian distribution centred at 0 with a width of 12 MeV.
The total cell energy is then transformed into \gls{ADC} counts\com{~(1~ADC $\approx$ 16~MeV in the low gain region)} to emulate the output of the raw data fitting, using the calibration factors from data reconstruction which can vary from cell to cell. 
If the total cell energy exceeds or is equal to 3~\gls{ADC} counts, corresponding to $\approx$ 50~MeV, the cell energy is accepted.
The current \gls{EMCal} simulation includes neither pileup events, whose contribution is considered negligible for Run~1 and Run~2 (see \Sec{sec:anaparam}), nor the direct interaction of particles~(in particular neutrons) with the \gls{APD}.
The latter is the most likely cause of the ``exotic clusters''~(see \Sec{sec:exotics}).
\subsection{Clusterization}
\label{sec:clusterization}
A particle interacting with the cell material produces a shower spreading its energy over neighbouring cells. 
The electromagnetic (or hadronic) shower can spread more than its Moli\`ere radius, which corresponds to about half an \gls{EMCal} cell size (see \Tab{Table-1}). 
A calorimeter cluster, an aggregate of adjacent calorimeter cells with energy above the noise threshold, is the main object delivered by the reconstruction software.
Ideally, such a cluster of cells is produced by a single particle hitting the detector, although depending on the particle energy, detector granularity, particle density, clusterization algorithm, and event type, a cluster can have contributions from different particles. 
For particles depositing their full energy in the calorimeter, like photons and electrons, the reconstructed cluster energy is approximately the particle energy.
For mesons that predominantly decay into two photons, like \piz\ and $\eta$ mesons, the energy is detected in the calorimeter as two separate clusters or a merged one, depending on their transverse momentum as well as on the granularity of the calorimeter.
These particles are commonly reconstructed as an excess in the di-photon invariant mass ($M_{\gamma_{1}\gamma_{2}}$) distributions:
\begin{equation}
M_{\gamma_{1}\gamma_{2}}=\sqrt{2E_{1}E_{2}\left(1-\cos\theta_{12}\right)},\label{eq:KineIMgg}\end{equation}
where $\theta_{12}$ is the relative opening angle between the photons in the laboratory frame, with
\begin{equation}
  \cos{\theta_{12}}=\frac{\gamma^{2}\beta_{\piz(\eta)}^{2}-\gamma^{2}\alpha_{\piz(\eta)}^{2}-1}{\gamma^{2}(1-\alpha_{\piz(\eta)}^{2})} \hspace{1.5cm} \label{eq:KineOpenAng}
\end{equation}
where
\begin{equation}
  \alpha_{\piz}=\frac{\left|E_{1}-E_{2}\right|}{E_{1}+E_{2}} \label{eq:KineAsym}
\end{equation}
is the decay asymmetry for the respective meson, $\beta_{\piz(\eta)}={p}/E$ and $\gamma=1/\sqrt{1-\beta_{\piz(\eta)}^{2}}$. 
The opening angle $\theta_{12}$ becomes smaller with increasing meson energy and is minimal at $\alpha_{\piz(\eta)}=0$. 
Asymmetric pairs ($\alpha_{\piz(\eta)}\approx1$) have a larger $\theta_{12}$, thus for high-energy mesons only asymmetric decays can be separated.

\begin{figure}[t]
    \centering
    \includegraphics[width=0.48\textwidth]{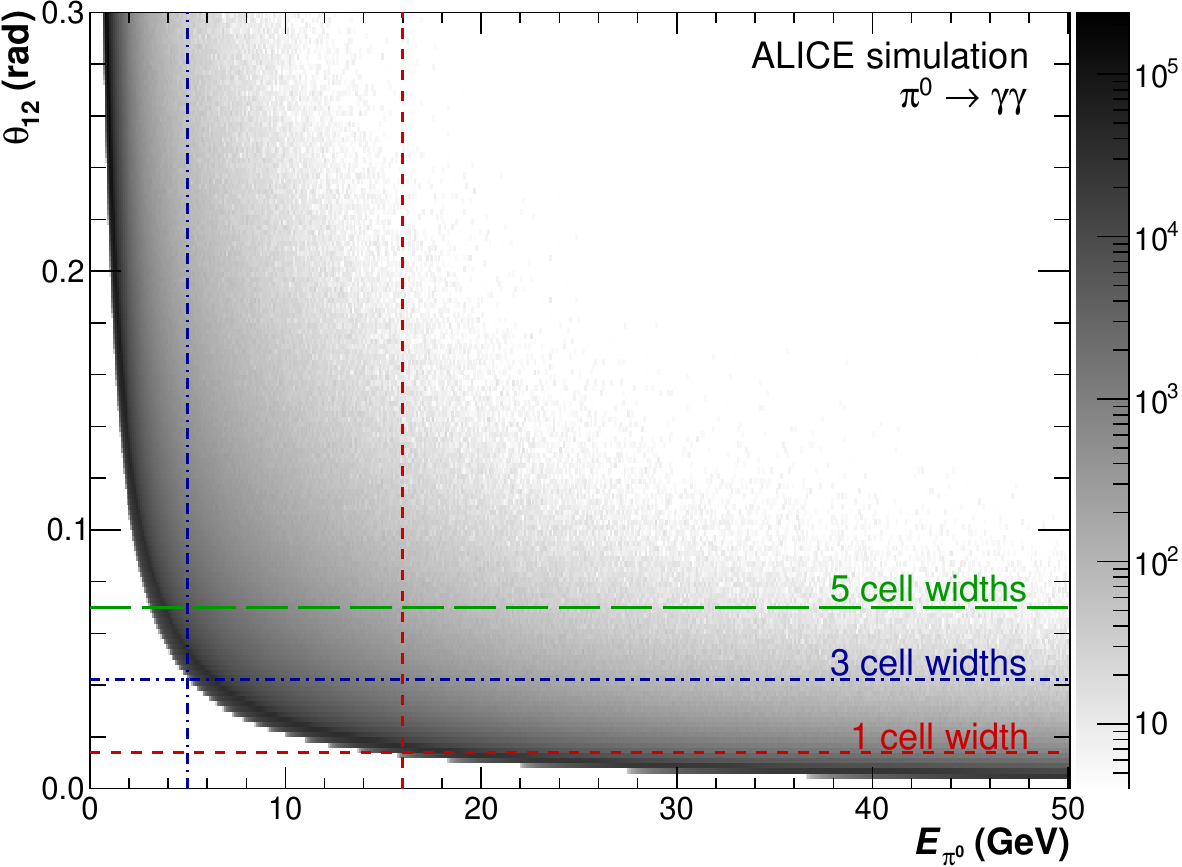}
    \hspace{0.2cm}
    \includegraphics[width=0.48\textwidth]{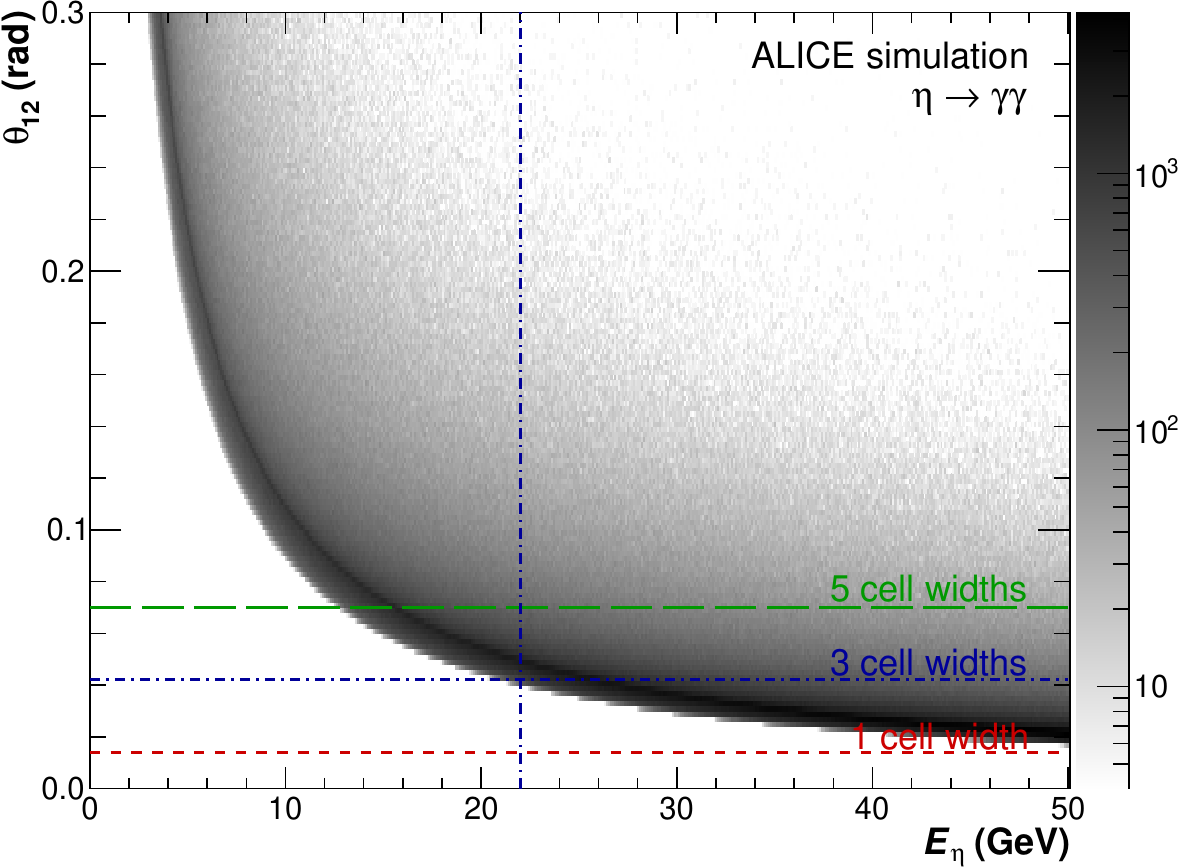}
    \caption{
          (Color online) Distribution of the opening angle $\theta_{12}$ of two decay photons from \piz\ (left) and $\eta$ (right) mesons decays as a function of the meson energy obtained at generator level from a \gls{MC} simulation. 
          The horizontal lines indicate the opening angle corresponding to a width of approximately 1, 3 and 5 cells separating the two photons.
          Two photons completely merge into one cluster if they fall below the one cell distance, while they start merging at a width of approximately 3 cells, depending on the clusterizer.
          The colored vertical lines correspond to the energy limits where two photons could still be fully resolved using the V1 (blue) and V2 (red) clusterizers.
        }
    \label{fig:OpeningAngle}
\end{figure}
%
In \Fig{fig:OpeningAngle}, the decay kinematics for \piz\ and $\eta$ mesons are compared to \gls{EMCal} cell sizes. 
For opening angles close to or smaller than 0.0143 rad, the two decay photons are too close to be resolved.
For $\piz$ and $\eta$ mesons with energies larger than 5~GeV and 22~GeV, respectively, the opening-angle distribution is such that the outer cells of the photon clusters can overlap, and above 16~GeV for \piz mesons and 60~GeV for $\eta$ mesons the minimum opening angle is close to or smaller than the cell size. 

\begin{figure}[t!]
\centering
\includegraphics[width=0.40\textwidth]{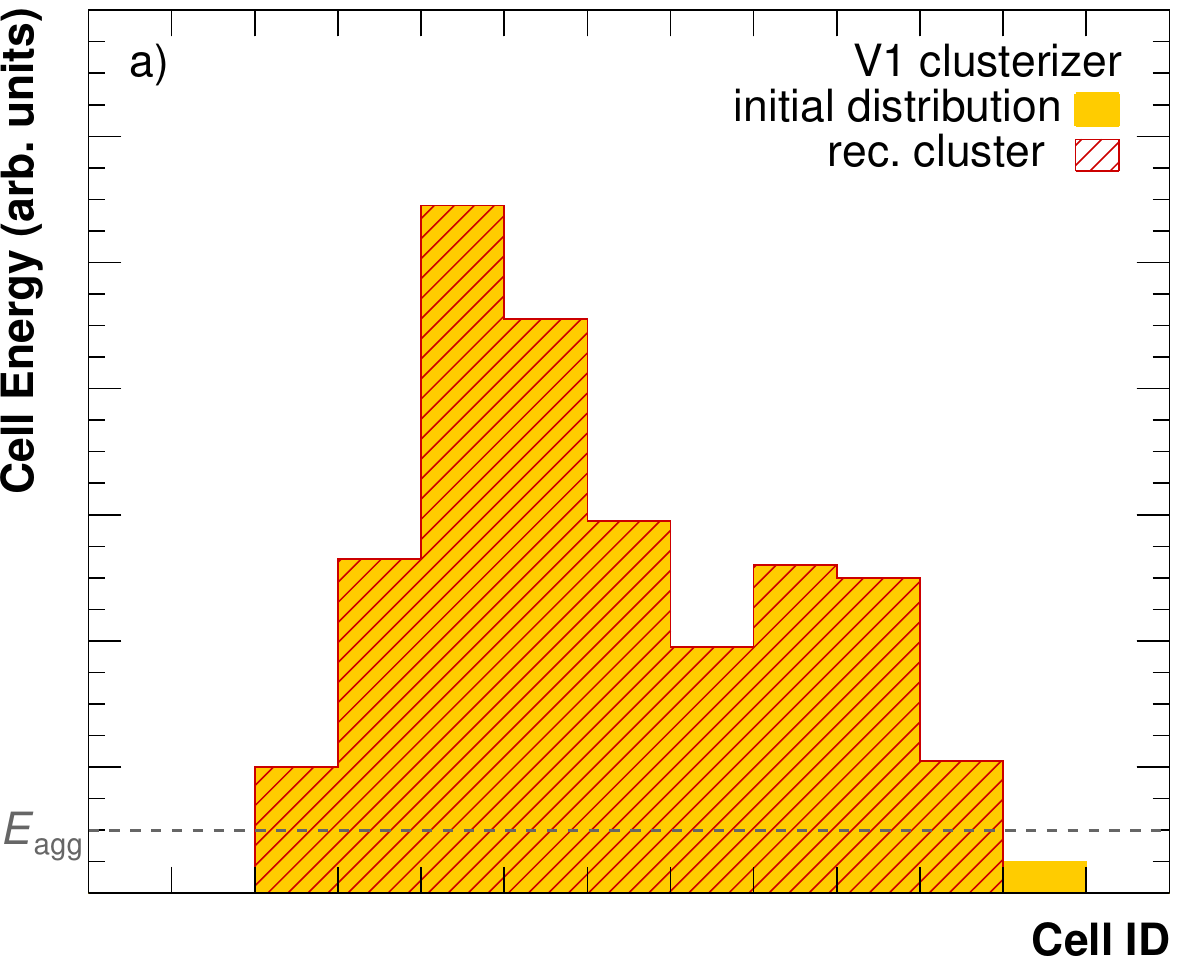} \hspace{0.4cm}
\includegraphics[width=0.40\textwidth]{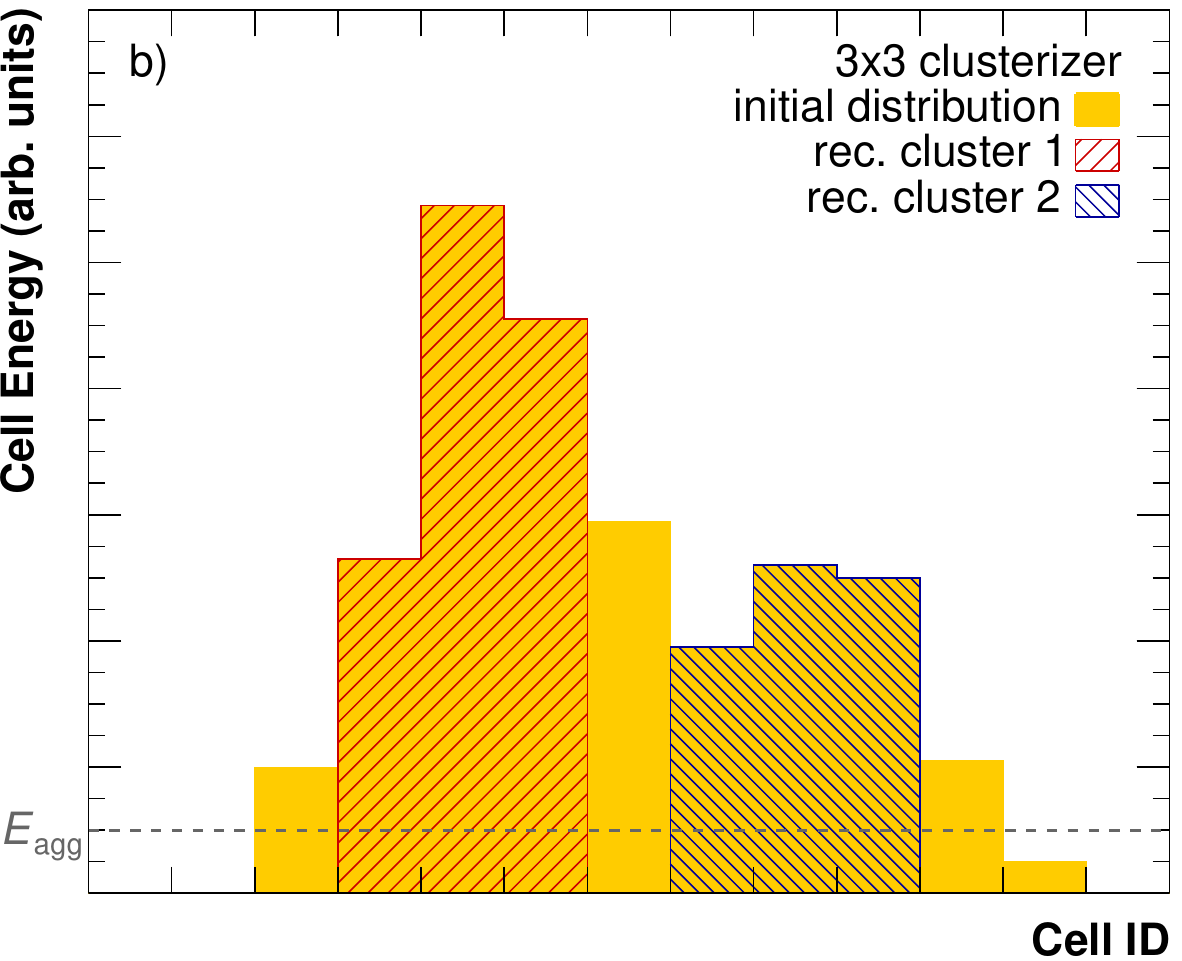}\\ 
\includegraphics[width=0.40\textwidth]{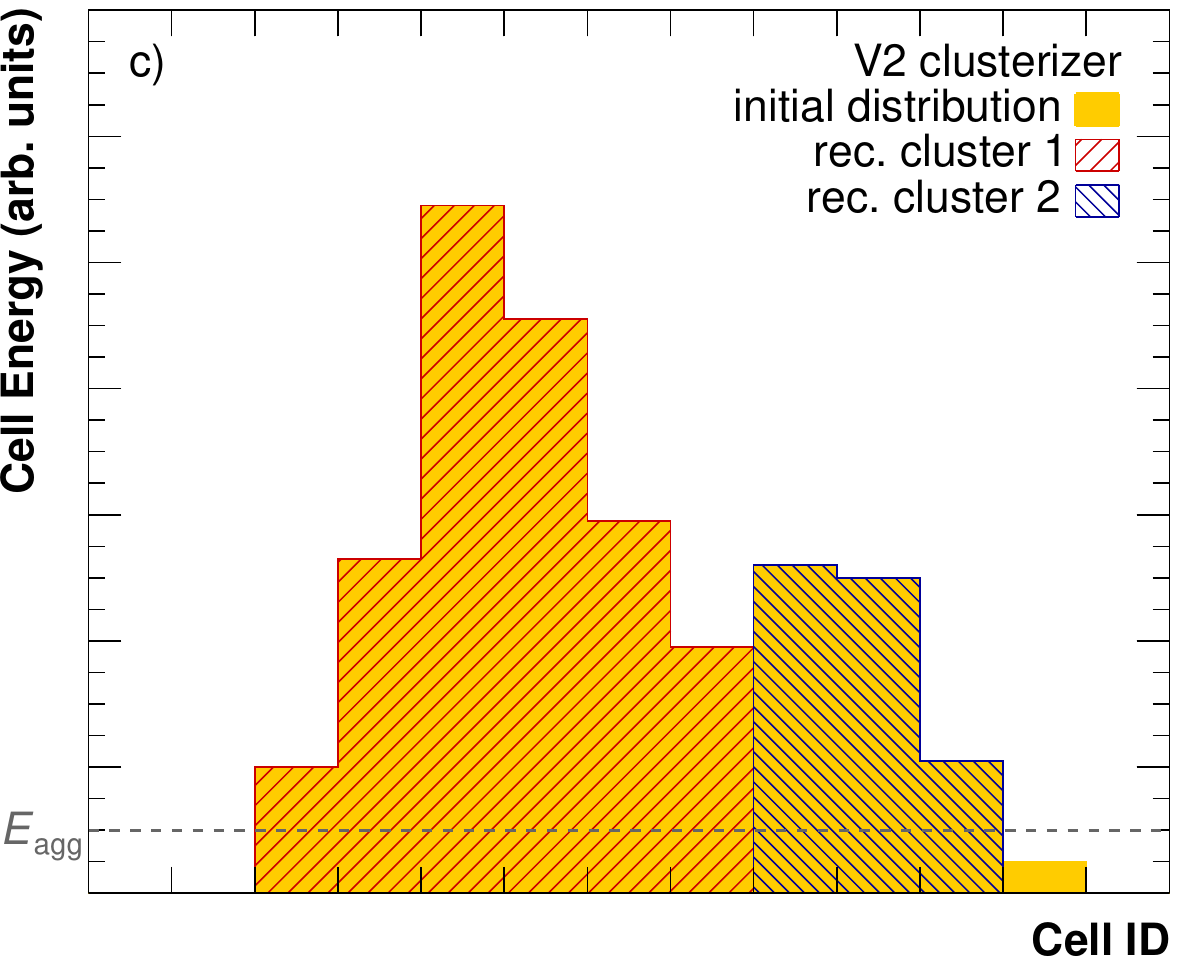} \hspace{0.4cm}
\includegraphics[width=0.40\textwidth]{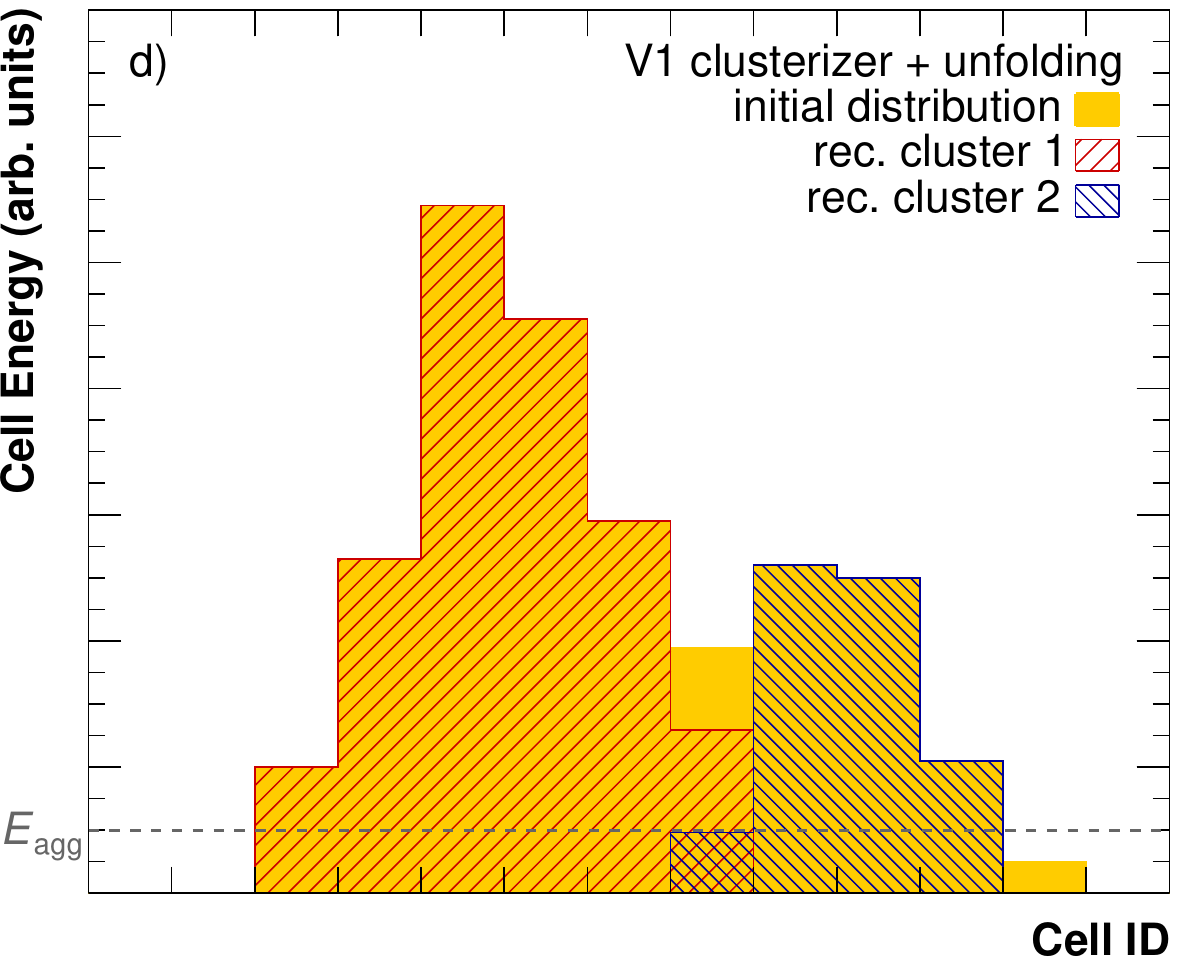} 
\caption{(Color online) Schematic comparison of different clusterization algorithms. 
Only one dimension is shown for simplicity.
Yellow boxes represent the energy in each cell. $E_\mathrm{agg}$ is the clusterization threshold as defined in the text. The different clusters are indicated by blue and red hatched areas.
Each panel represents a clusterization algorithm: a) V1, b) $3 \times 3$, c) V2, d) V1+unfolding.}
\label{fig:clusterization} 
\end{figure}

Several clusterization methods have been developed for the \gls{EMCal} which can separate these decay products up to different incident particle energies. 
The selection of the clusterization method depends on the goal of the analysis.
\Figure{fig:clusterization} schematically explains the differences between the clusterization methods. 
All clusterizing methods build clusters starting from the highest energetic cell in a region, referred to as seed cell, and associate cells in the vicinity that share a common side with cells already in the cluster. 
Cells are aggregated into the cluster in case their energy is above the aggregation threshold $E_{\rm agg}$ (typically $100$~MeV). 
The seed threshold depends on the probe, ranging from 300 MeV for jet measurements, where a high efficiency on the energy reconstruction is desired, to 500 MeV for particle identification, which is more sensitive to noise. 
The simplest clusterization algorithm, called ``V1'' clusterizer, aggregates cells as long as the conditions specified above are fulfilled.
For the ``V2'' clusterizer, only cells with an energy smaller than neighboring cells that are already part of the cluster are aggregated to the cluster. 
This additional requirement makes the clusterization algorithm more robust against shower overlaps, in particular in high particle density environments.

Besides these most frequently-utilized methods, other algorithms were implemented, like the  ``$N \times M$'' clusterizer or the ``V1+unfolding'' clusterizer.
The $N \times M$ clusterizer restricts the size of the cluster to $N \times M$ cells in $\eta$ and $\varphi$ direction around the seed cell, starting with the cell with highest energy in the event.
The V1+unfolding clusterizer splits V1 clusters into several sub-clusters by fitting the cell energy profile, allowing a cell to be present in two clusters, with a different fraction of energy in the cell assigned to each cluster. 
This clusterizer is expected to provide a better performance for separating overlapping showers and cluster energies than the other clusterizers but is significantly slower. 

\begin{figure}[t!]
    \centering
    \includegraphics[width=0.49\textwidth]{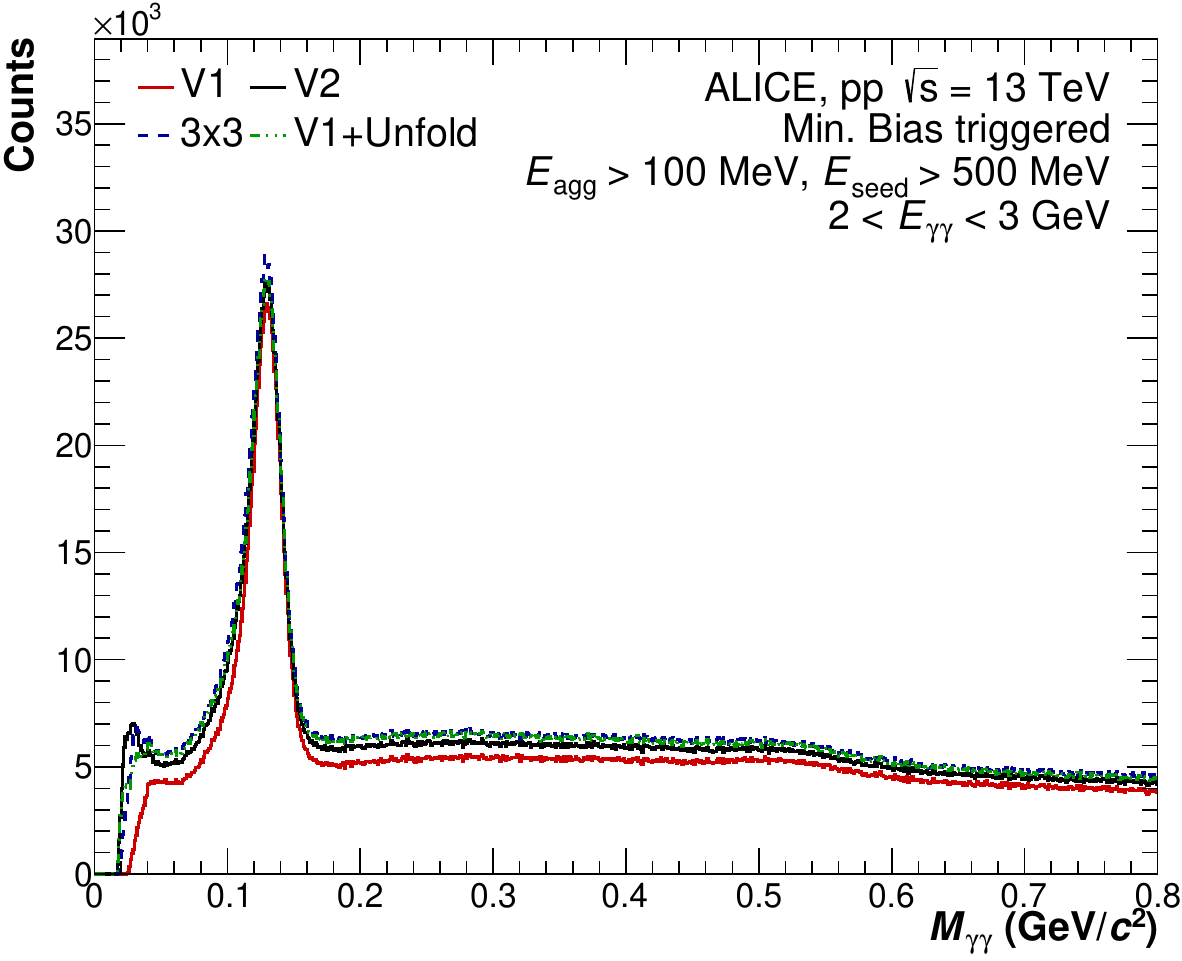} 
    \includegraphics[width=0.49\textwidth]{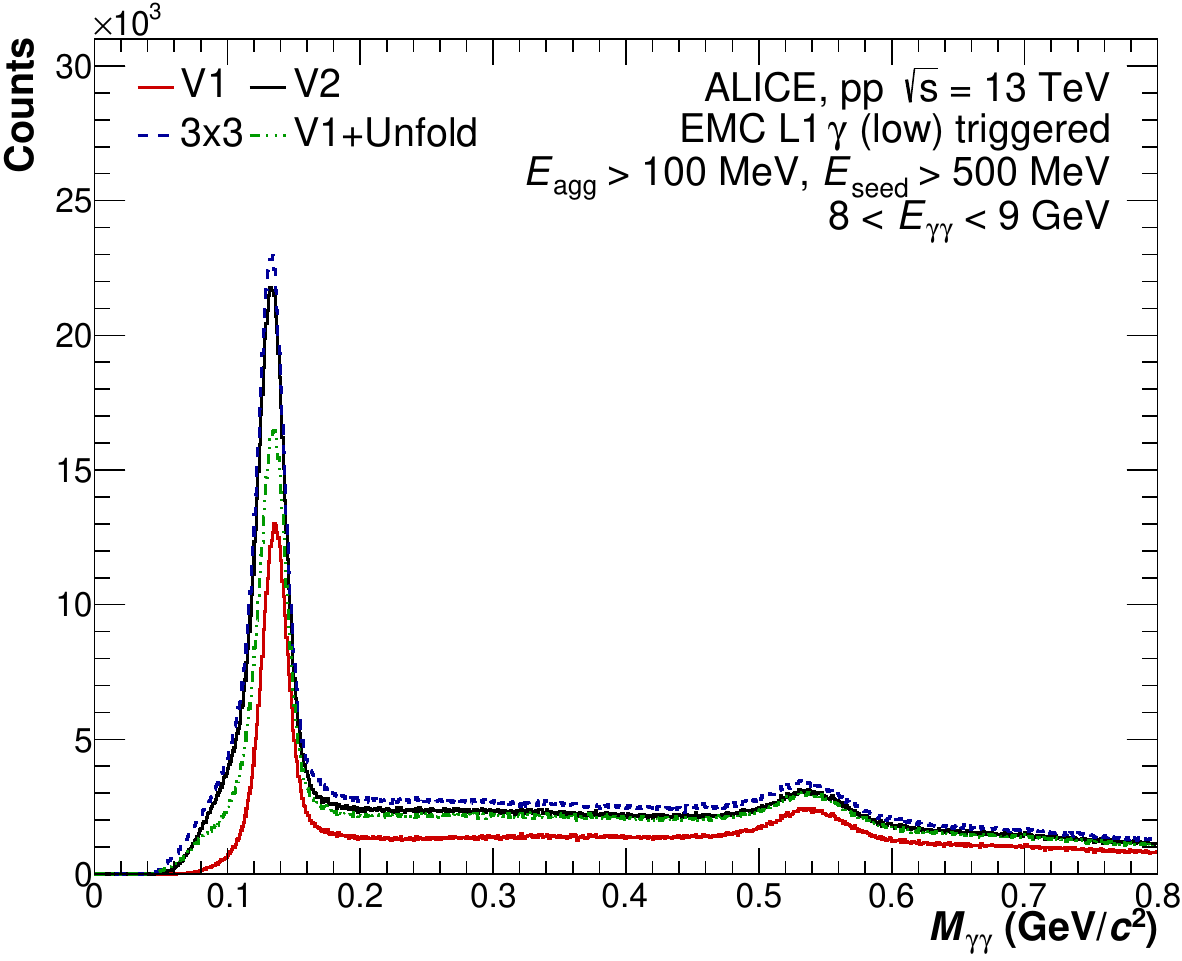} \\
    \includegraphics[width=0.49\textwidth]{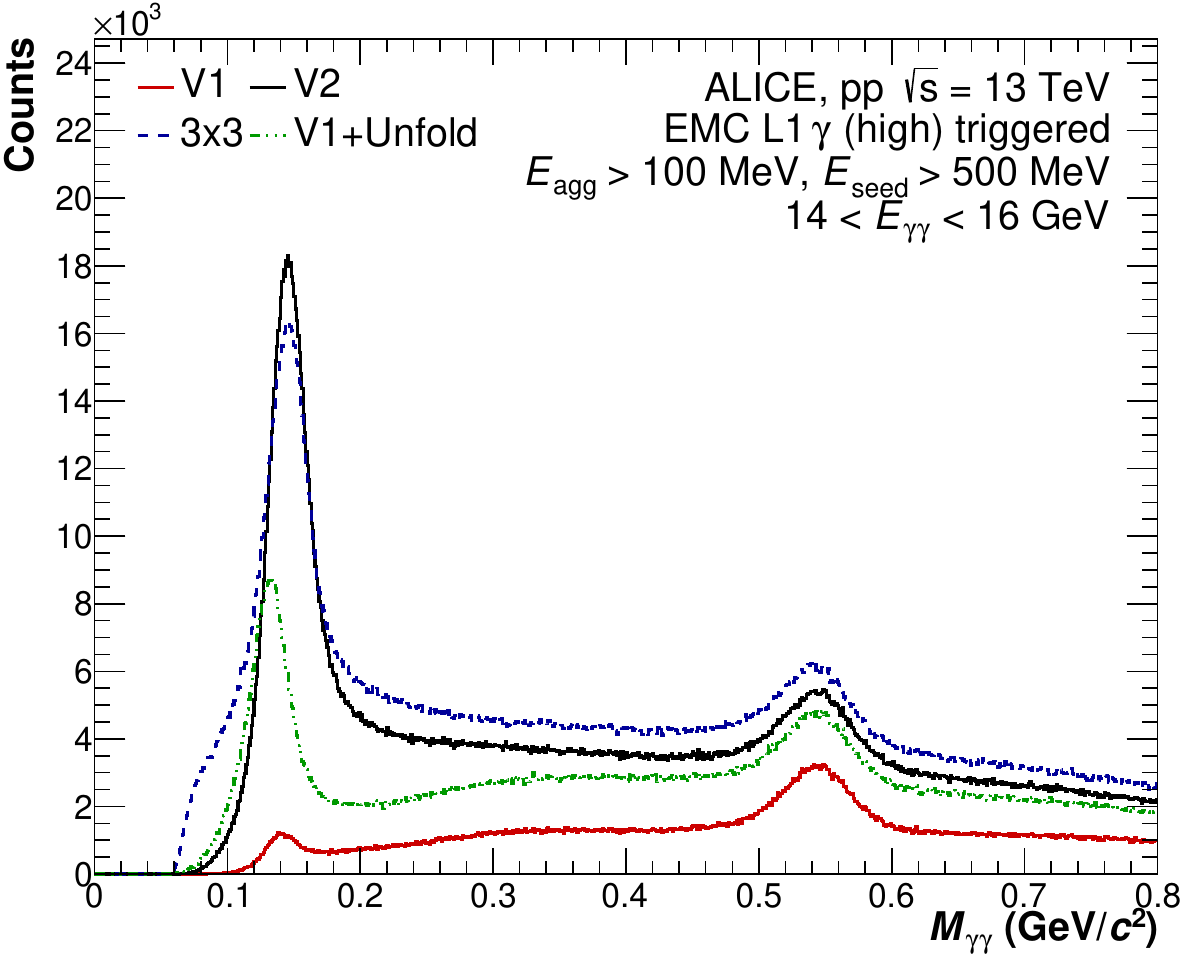}
    \includegraphics[width=0.49\textwidth]{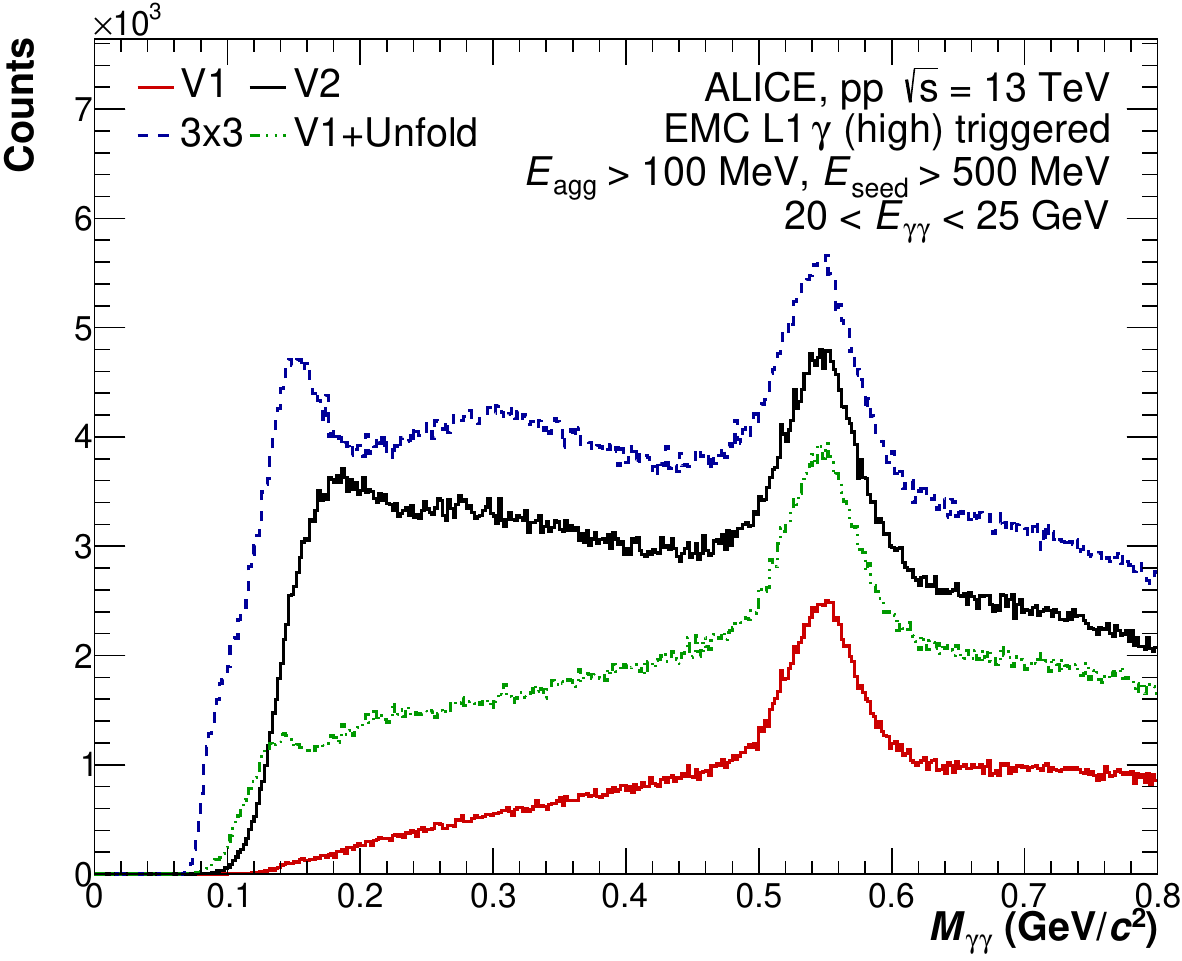}\\
    \includegraphics[width=0.49\textwidth]{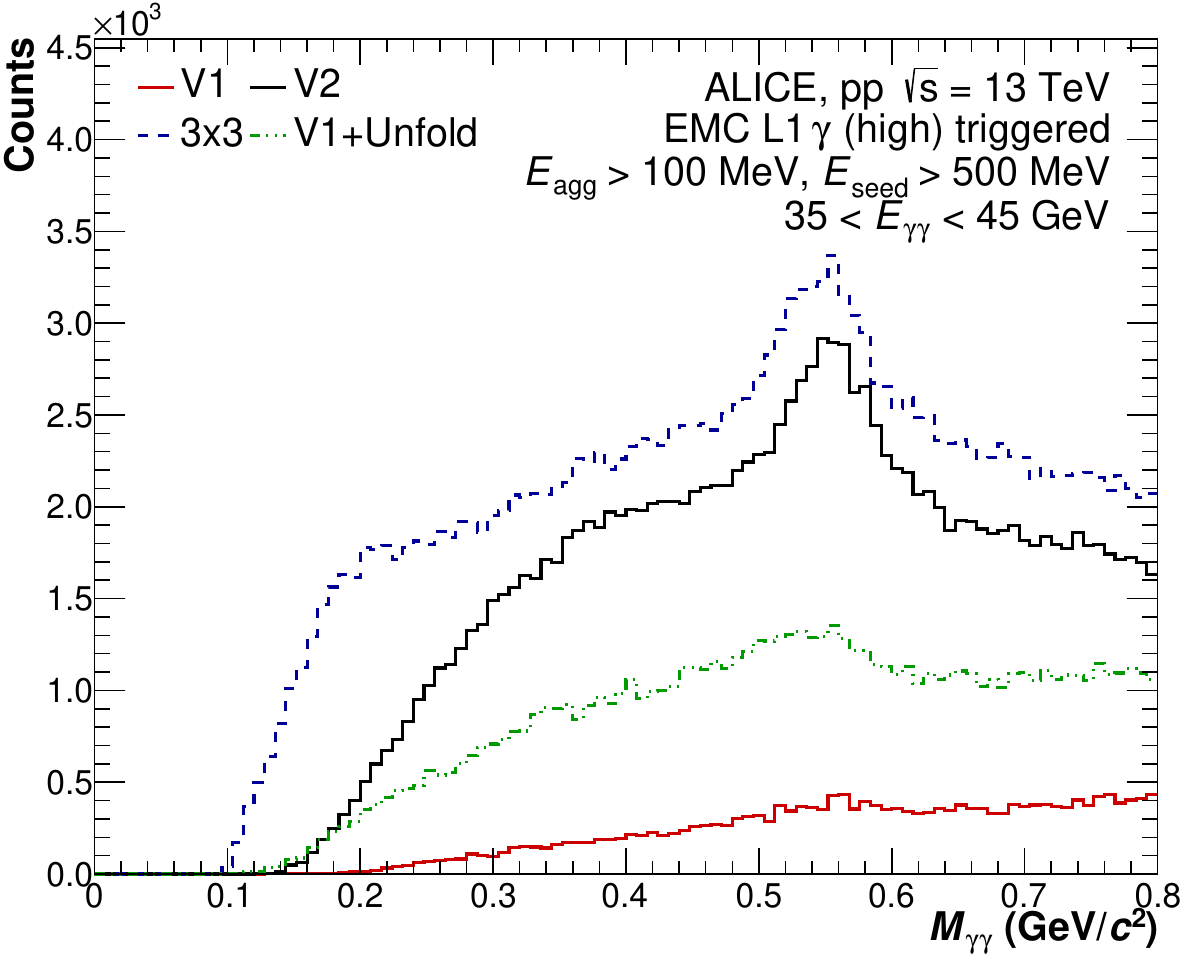}
    \includegraphics[width=0.49\textwidth]{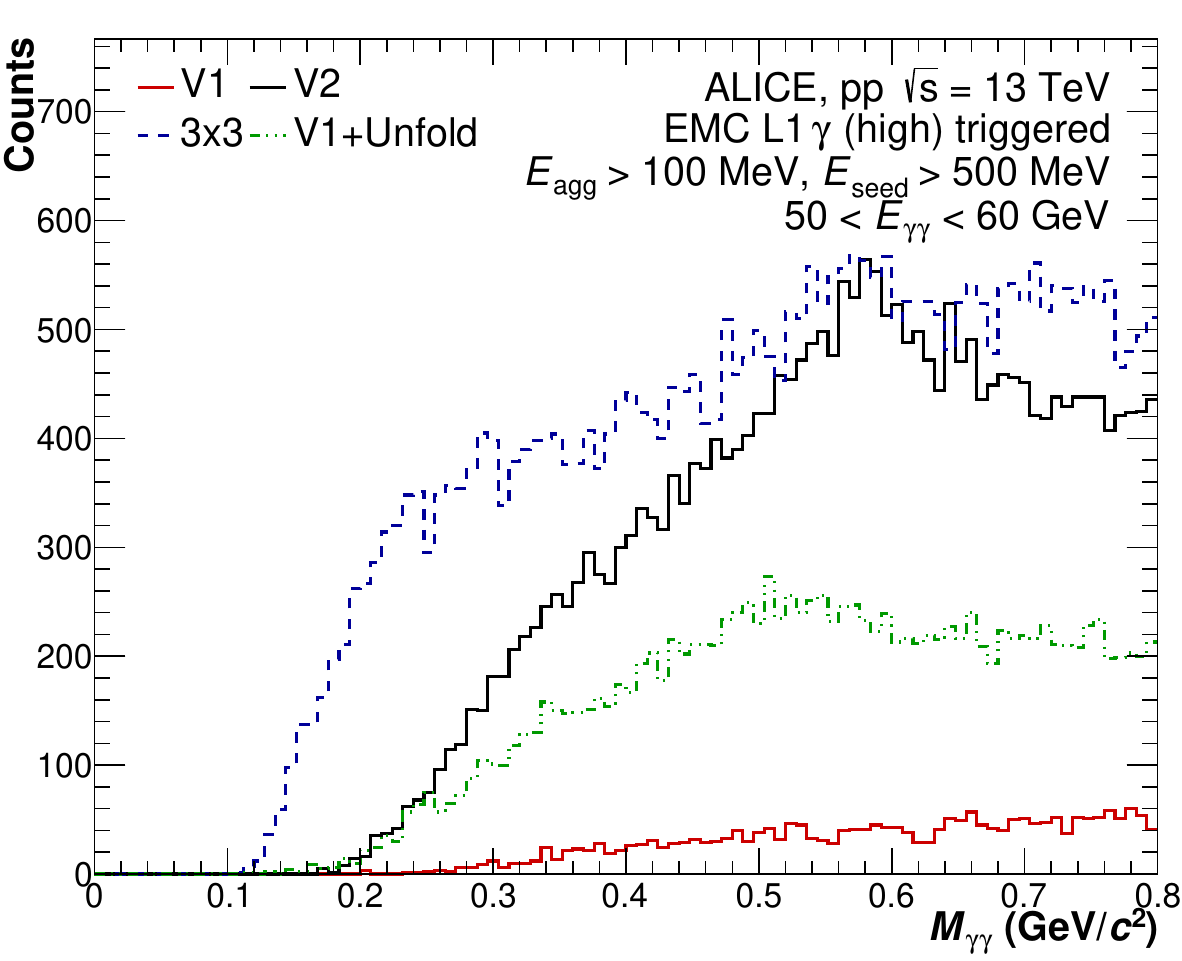}
    \caption{(Color online) Invariant mass distribution of cluster pairs in pp collisions at \sthirteen\ for different intervals of pair energy. The differently colored lines correspond to different clusterizer types, using the same aggregation $E_\mathrm{agg}=100$~MeV and seed $E_\mathrm{seed} = 500$~MeV thresholds. 
              The lowest bin in energy uses the data sample with minimum bias trigger, while the others are obtained from the \gls{EMCal} \gls{L1} triggered data with thresholds at about $E \approx 4$ and $9$~GeV, respectively. }
    \label{fig:InvMassEBins} 
\end{figure}

The performance of each clusterizer was evaluated for different physics objects at the beginning of \gls{LHC} Run~1. 
In general, it was found that while the V1 clusterizer is suitable for analyses that heavily rely on separating the signal and background through \gls{PID} selections (i.e. shower shape, see \Sec{sec:showershape}), it is not optimal for high-particle-density environments.
In particular, within a strongly collimated jet or in heavy-ion collisions, the other three clusterization techniques (V2, NxM and V1-unfolding) perform better, resulting in clusters which are more closely related to the incident particles in terms of energy and position.
While the V1-unfolding algorithm has promising features, its improved performance is devalued by its higher computational time.
Thus, the simpler V2 algorithm is typically used.

\Figure{fig:InvMassEBins} demonstrates the performance of the different clusterization methods on the neutral mesons invariant mass analysis of pp collisions at \sthirteen. 
While all of these algorithms perform similarly well below $E_{\gamma\gamma} = 10$~GeV, significant differences can be seen above $E_{\gamma\gamma} = 12$~GeV. 
Starting from $6$~GeV the V1 clusterizer will start to absorb the second photon into the higher energetic cluster due to their small opening angle
depleting the lower invariant masses in the cluster-pair distribution until no \piz\ peak is visible any longer around $E_{\gamma\gamma} = 15$~GeV.
Using the V1-unfolding algorithm, the \piz\ can be reconstructed up to $E_{\gamma\gamma} = 20$ GeV through two-dimensional template fits to the cell-energy distribution within the V1 cluster and subsequent splitting of the cluster.
Nonetheless, the V2 and 3x3 clusterizers have proven to be better suited to reconstruct \piz\ and $\eta$ mesons with small opening angles as they are able to split even very asymmetric decays.
Consequently, they are more performant at high \pT\ in an invariant-mass analysis.
However, their cluster centers and energy sharing will be slightly biased towards the higher energetic photon leading to a shift of the neutral pion peak position towards higher invariant masses for very small opening angles. 
Moreover, the signal-to-background ratio will be slightly worse due to the on average larger number of clusters found per event in comparison to the V1-unfolding algorithm.

\begin{figure}[t]
    \includegraphics[width=0.49\textwidth]{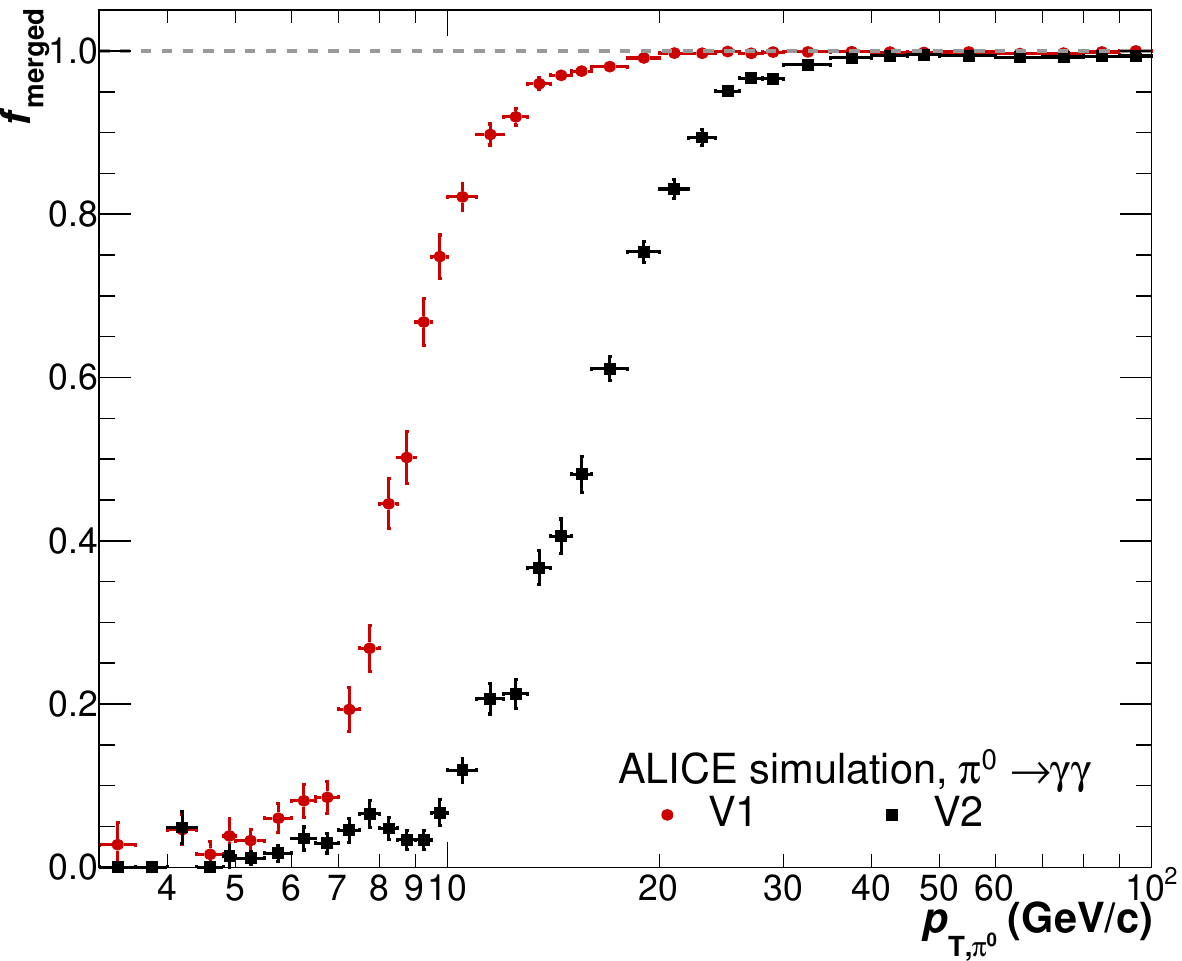}
    \includegraphics[width=0.49\textwidth]{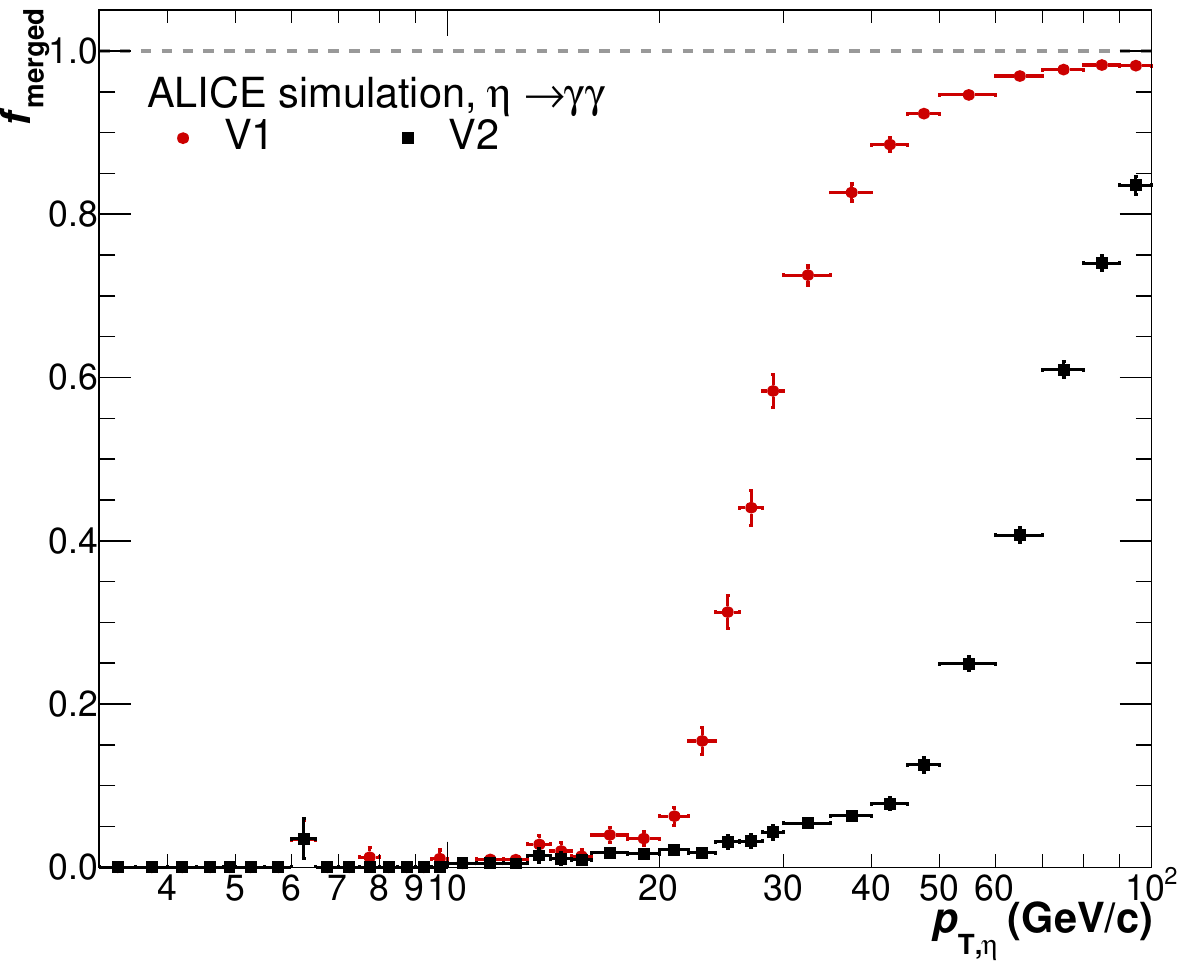}
    \caption{(Color online) 
    Fraction of neutral pions (left) and eta mesons (right) for which the showers from their decay photons are merged into a single cluster and can not be reconstructed using an invariant-mass analysis.}
    \label{fig:fractionClusterizer}
\end{figure}

\Figure{fig:fractionClusterizer} shows the fraction of \piz\ and $\eta$ mesons for which the two photons were reconstructed in one cluster for different clusterizers. 
The choice of the optimal clusterizer for a given measurement highly depends on the transverse momentum range under study. 
Each clusterizer has to be evaluated based on purity, momentum resolution, and efficiency considerations. 
For example, it can be seen in \Fig{fig:fractionClusterizer}{ (left)}, that the merging of the clusters arises at lower \piz\ meson energy for the V1 clusterizer (around 6~GeV), while the V2 clusterizer can split the cluster also at higher momenta based on the two maxima in the found object.
For \piz\ meson energies above 20~GeV, neither the V1 nor the V2 clusterizers are able to resolve the majority of their decay photons. 
In this range, the so-called merged pion analysis exploits the features of merged clusters to reconstruct neutral pions as explained in \Sec{sec:MesonShowerShape}.
Since the V1 and V2 clusterizers were found to be the most suitable for most of the analyses, only the performance of these two algorithms will be discussed in the remainder of this document.
Their advantages and disadvantages for different analysis types and physics objects are addressed in more detail in \Sec{sec:physics}.
Once the cluster is formed, the cluster energy $E$ is calculated as the sum of the energies of the cluster cells ($E_{{\rm cell},~i}$), 

\begin{equation}
\label{eq:e_cluster}
E = \sum_i E_{{\rm cell},~i},\\
\end{equation}
where $i$ indicates a cell that belongs to the cluster. 
The cluster position (centroid)  
in the ALICE global coordinate system 
is obtained by a weighted average of the cells position ($x_i,y_i,z_i$)
\begin{equation}
\label{eq:xyz_centroid}
\langle x \rangle= \sum_i \frac{w_i x_i}{w_\mathrm{tot}}, \quad
\langle y \rangle= \sum_i \frac{w_i y_i}{w_\mathrm{tot}}, \quad
\langle z \rangle= \sum_i \frac{w_i z_i}{w_\mathrm{tot}},
\end{equation}
where the weights $w_i$ depend logarithmically on the cell energies,
\begin{equation}
\label{eq:w_i}
w_i = {\rm Maximum}(0,w_{\rm max}+\ln(E_{{\rm cell},~i}/E))
\end{equation}
and
\begin{equation}
\label{eq:w_tot}
w_\mathrm{tot} = \sum_i w_i,\\
\end{equation}
with $w_{\rm max}$ set to 4.5~\cite{Awes:1992yp} to exclude cells with energy smaller than 1.1\% of the cluster energy.
 
\subsection{Cluster selection}
\label{sec:anaparam}
Only cells which are calibrated and fulfill the quality assurance criteria described in \Sec{sec:calibration} are clusterized. 
For single-particle analysis, which often rely on a good understanding of the purity~($P$), the electronic noise in data has to be reduced to a minimum and therefore a cut on the number of cells of $n_{\rm cells} > 1$ is strictly required. 
For invariant-mass ($M_{\gamma\gamma}$) based analyses this cut can be released in order to maximize the efficiency of the single-photon reconstruction, as random electronic noise merely increases the combinatorial-background contribution to the invariant-mass distribution by a few percent.
In order to suppress distortions to the clusters from masked cells or edge effects, some analyses require in addition a minimum distance of up to two cells from the highest energy cell in the cluster to the masked cells ($d_{\rm masked}$) or the borders of the \glspl{SM}~($d_{\rm edge}$), except for the \glspl{SM} borders at $\eta=0$ where there is no separation between \glspl{SM} located at the same azimuthal angle, see Fig.~\ref{fig:1-HW-EMCal_overview_etaphi}.
In addition, cells at the edge of the \glspl{SM} tend to be calibrated with less precision due to a lack of statistics and, thus, might perform worse than the other cells. 
The requirement of a minimum distance to dead detector areas can increase the quality of the cluster sample, however, by reducing the acceptance.
Thus, these requirements are only imposed for analyses needing a very high purity of the cluster sample and a high quality of its cluster properties, like those involving isolated photons or electrons from semi-leptonic heavy-flavour hadron decays.

Due to the rather wide time integration window of $1.5$~$\mu$s of the \gls{EMCal}, multiple collisions besides the triggering one are recorded for each event, also referred to as pileup. 
To select the main triggering collision, a cut on the arrival time of the leading cell for each cluster, referred to as cluster time, after time calibration is performed. 
The selection criteria depend on the bunch spacing within the \gls{LHC} and are chosen such that only the primary bunch crossing is selected, while still being as open as possible in order not to introduce effects from the limited time resolution at lower transverse momenta (see \Sec{sec:timeCalib}). 
The efficiency loss for signal clusters was found to be negligible for timing cuts $|t| > 25$~ns. 
For tighter selections, a small efficiency loss of up to $5\%$ at low energies is expected.
Thus, the cluster timing selection window ranges from $\pm25$~ns to $\pm250$~ns, depending on the bunch spacing of the data-taking period. 
In order to reject background and to discriminate between different particle species hitting the \gls{EMCal}, the clusters can be further distinguished by 
association of clusters with tracks propagated to the \gls{EMCal}, shower shape discrimination, and cluster exoticity, which will be described in detail in the next sections.

\Table{tab:clusterbasiccuts}, summarizes the basic cuts and recommended values at the cluster level to select good quality clusters from what is discussed in the previous and next paragraphs.

\begin{table}[t!]
\centering
\caption{Basic cluster cuts, depending on the particle multiplicity of the collision, and the section or equation where more information can be found. The quoted time cut is the tightest one applied. It can be relaxed, if collision pileup is small and/or the bunch spacing is large. For more specific analysis selections, see \Tab{tab:photonbasiccuts}.}
\begin{tabular}{lrrr}
Parameter & \multicolumn{2}{c}{Multiplicity} & Section/ Equation \\
          & low  & high &  \\
\toprule
$n_{\rm cells}$        & \multicolumn{2}{c}{ }     & \ref{sec:anaparam}, \ref{sec:NCellEffi}\\ 
\hspace{0.2cm} $M_{\gamma\gamma}$ based analysis  & \multicolumn{2}{c}{$\geq1$}     & \\ 
\hspace{0.2cm} $P$ based analysis  & \multicolumn{2}{c}{$>1$}     & \\ 
$|t|$ (ns) & \multicolumn{2}{c}{$<25$}   & \ref{sec:anaparam}\\ 
\shshlo                & \multicolumn{2}{c}{$>0.1$}  & \ref{sec:showershape}, \ref{sec:exotics}, \Eq{eq:ss1}\\ 
$F_{+}$                       & $<0.97$ & $<0.95$ & \ref{sec:exotics}, \Eq{eq:exoticity}\\ 
\bottomrule
\end{tabular}
\label{tab:clusterbasiccuts}
\end{table}


\subsubsection{Association of clusters and tracks}
\label{sec:trackmatch}
The \gls{EMCal} is designed to measure the energy of particles that interact electromagnetically with the material of the \gls{EMCal}, \ie\ photons and electrons. 
However, hadrons can also deposit energy in the \gls{EMCal}, charged hadrons most commonly via ionization, but also via nuclear interactions generating hadronic showers. 
In the measurements where the distinction between showers originating from charged and neutral particles is required, clusters are associated to charged-particle tracks. \\
In most cases, cluster-track association is used to veto clusters with a significant contribution from charged particles in order to avoid double counting of the energy in case of jet reconstruction or contamination from hadrons and electrons in the photon sample. 
However, it can also be used to select clusters originating mainly from electrons. 

\begin{figure}[t]
    \centering
    \includegraphics[width=0.329\textwidth]{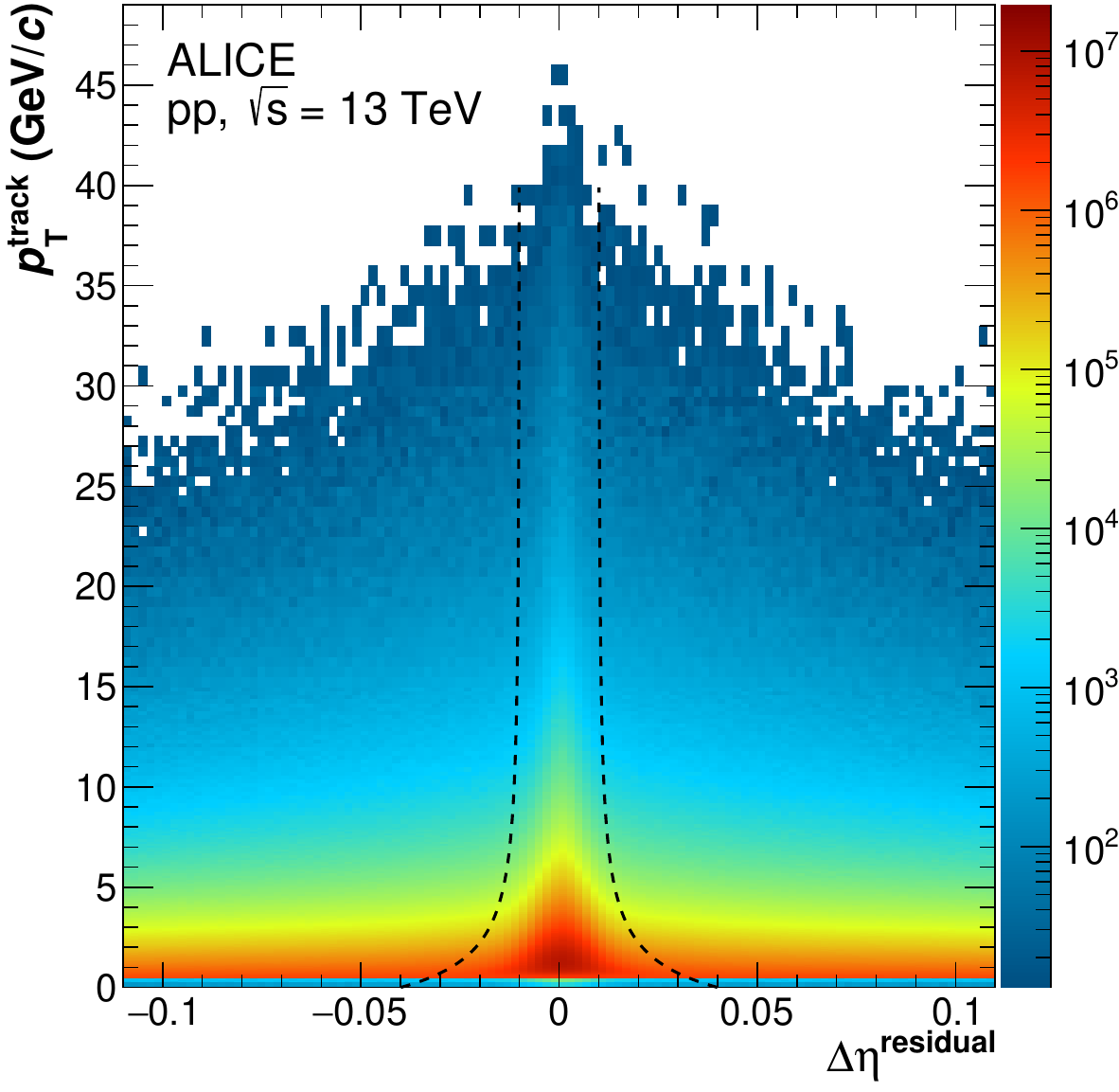}
    \includegraphics[width=0.329\textwidth]{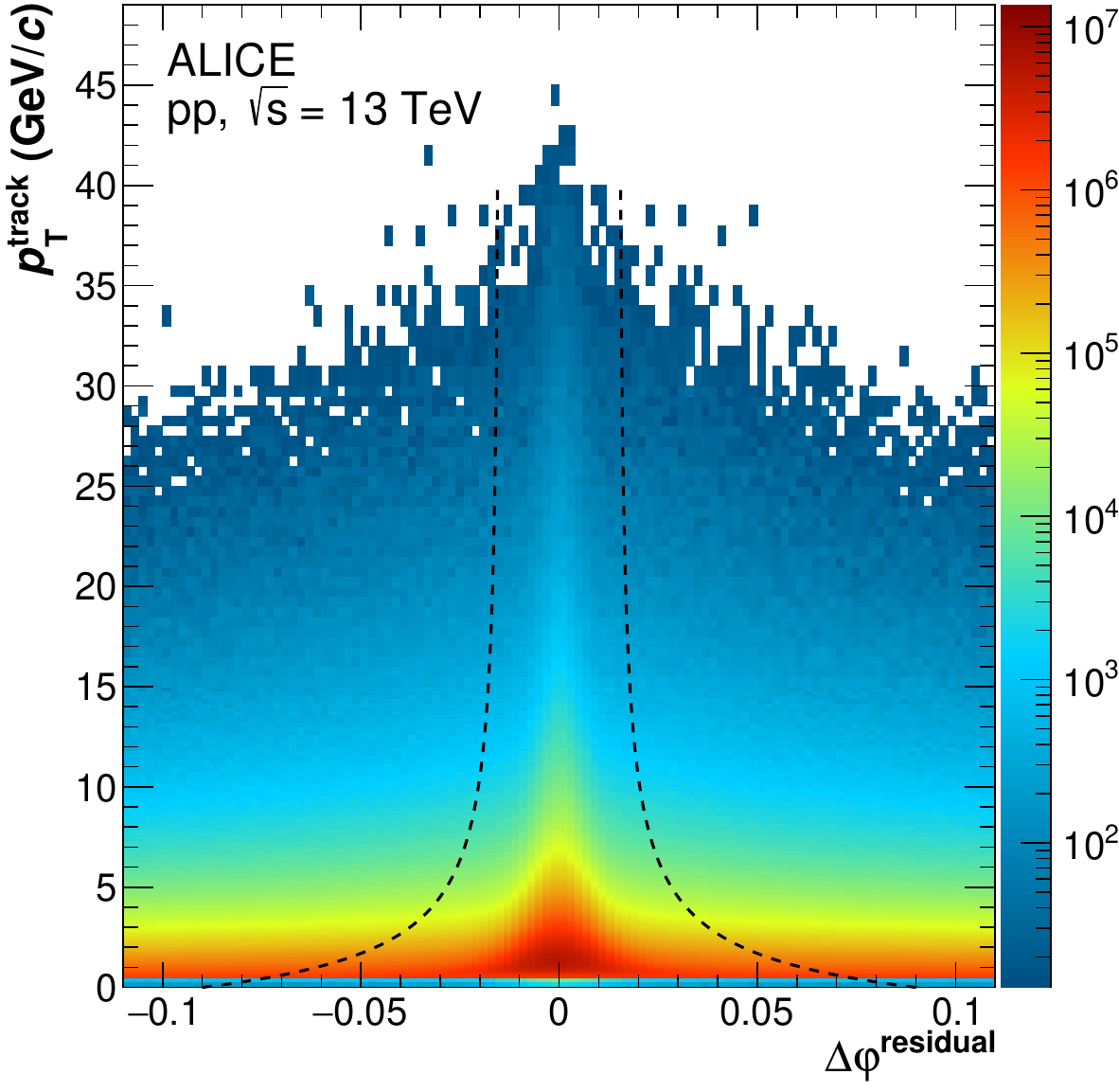}
    \includegraphics[width=0.329\textwidth]{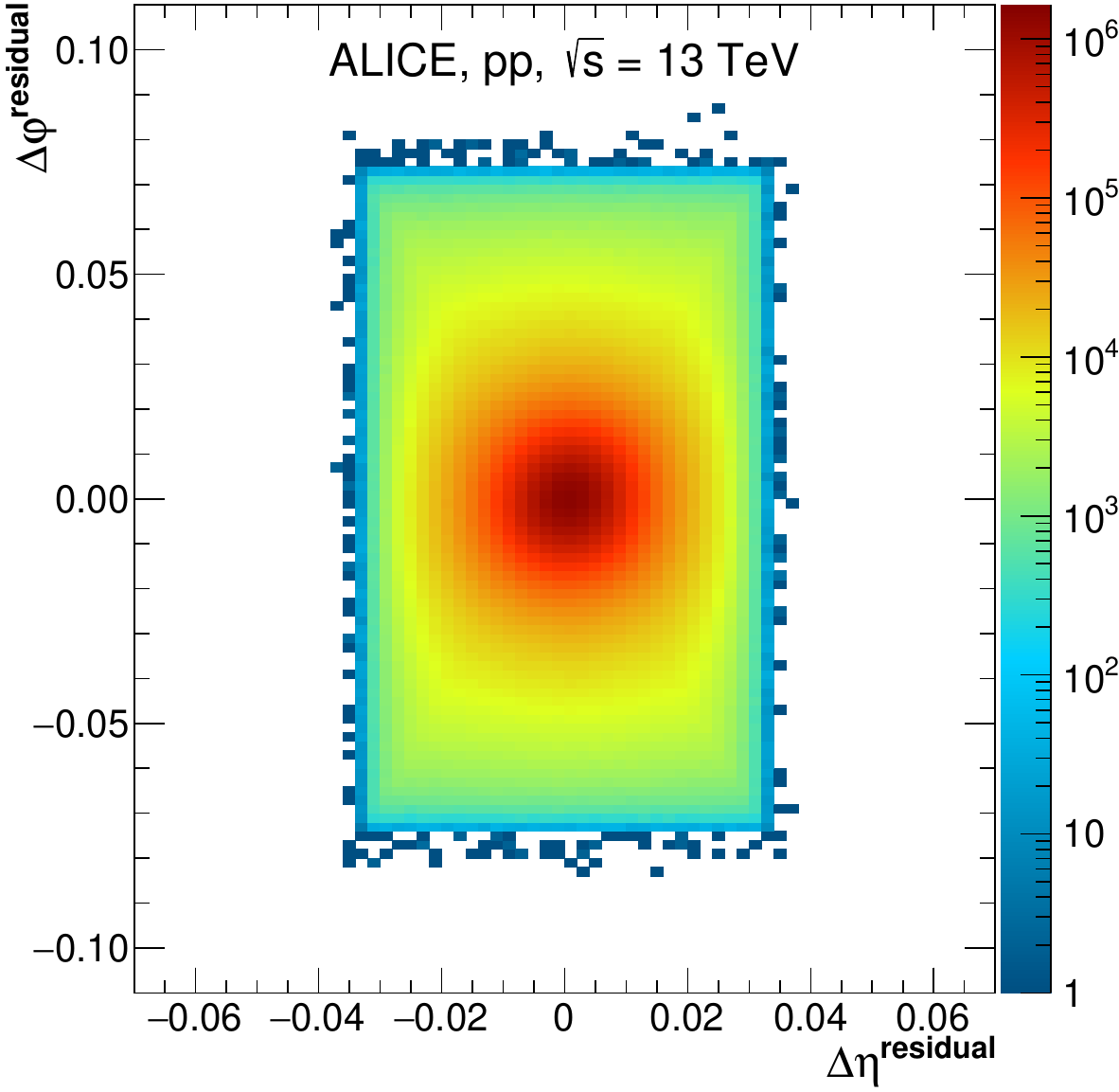}
    \caption{ 
        (Color online) Distance between a cluster and the closest projected track in $\eta$~(left) or $\varphi$ angle~(middle) versus the track momentum and for matched track-cluster pairs~(right) in pp collisions at \sthirteen\ collected with the minimum bias trigger. 
        Clusters are reconstructed using the V2 clusterizer. 
        The black lines in the left and middle panels indicate the suggested selection criteria expressed in \Eq{Eq:TMVetoCriteria}.
    }
    \label{fig:trmatch}
\end{figure}

Within \gls{ALICE}, charged particles are most commonly reconstructed using combined \gls{ITS} and \gls{TPC} tracking.
In order to determine whether a charged particle points to a reconstructed \gls{EMCal} cluster, tracks are further extrapolated to the \gls{EMCal}, taking into account the energy loss of the particle when it traverses the detector materials. 
The matching algorithm works as follows, for each track: 
As a first step, the trajectories are extrapolated to a fixed depth of 450 cm radial distance from the beam axis. 
This distance correspond to the average depth of the cluster energy deposition . 
By default, they are propagated with a step size of 20 cm in order to account for the average energy loss in the material. 
Afterwards, for every track and cluster pair with an angular distance smaller than $0.2$ rad, the track is extrapolated to the exact radial distance of the cluster with a step size of 5~cm.  For these pairs the residuals in $\varphi$ an $\eta$ are calculated. \\
In case several tracks fulfill the matching criterion for a given cluster, only the closest track is considered as the associated track. 
Clusters or tracks can be used as a target for association. 
For photon reconstruction, it is preferable to associate tracks to clusters, while for electron identification, clusters need to be associated with tracks.
The distributions of the residuals of clusters to the closest track are displayed in \Fig{fig:trmatch} in $\Delta\eta$~(left) and $\Delta\phi$~(middle) as a function of track $\pt$. 
The $\Delta\varphi$ distributions are significantly wider, mainly due to the orientation of the magnetic field, and thus different values should be chosen for selecting the matches in $\Delta\eta$ and $\Delta\varphi$. 
In addition, the $\Delta\varphi$ residuals are asymmetric when selecting only tracks with positive or negative charge.
The direction of the tail depends on the charge of the incident particle and the polarity of the magnetic field.
As the width of the distribution depends on the transverse momentum of the particle, a \pT-dependent window in the $\Delta\eta$-$\Delta\varphi$ plane is used to select cluster-track pairs.
\begin{figure}[t]
    \centering
    \includegraphics[width=0.49\textwidth]{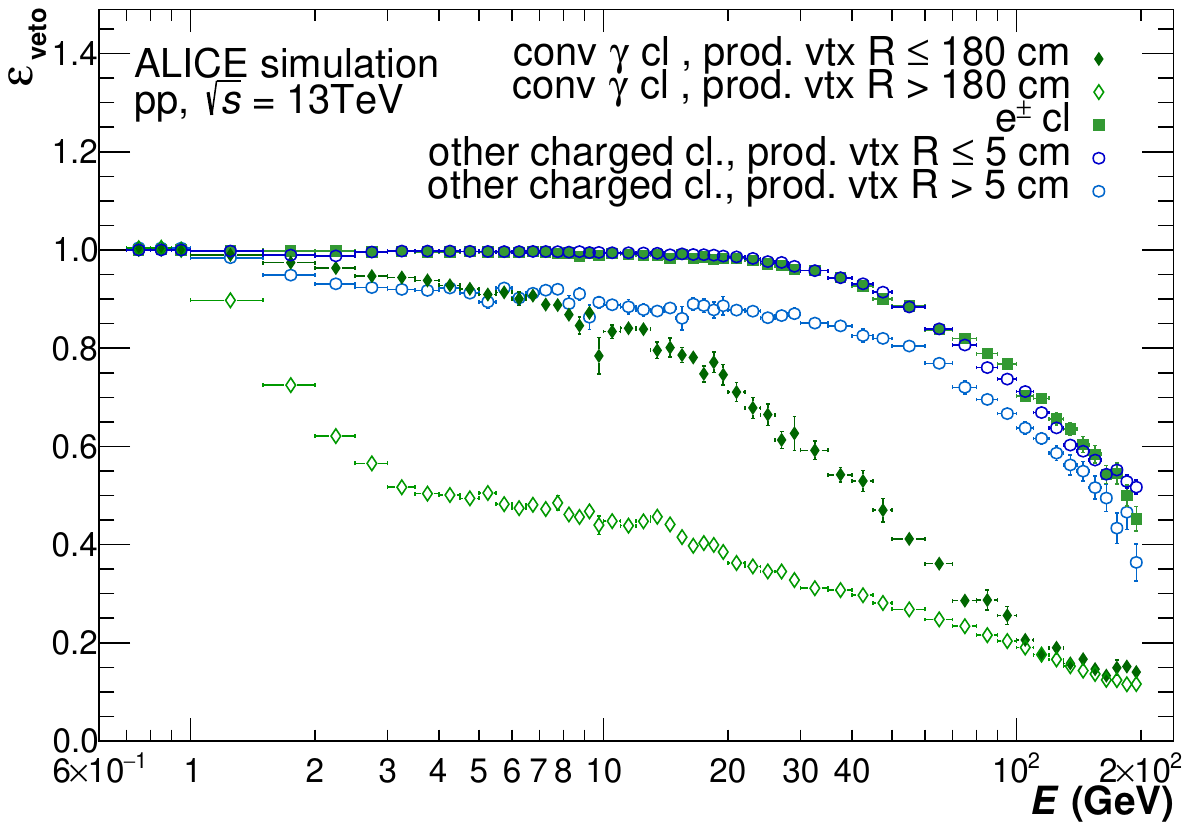}
    \includegraphics[width=0.49\textwidth]{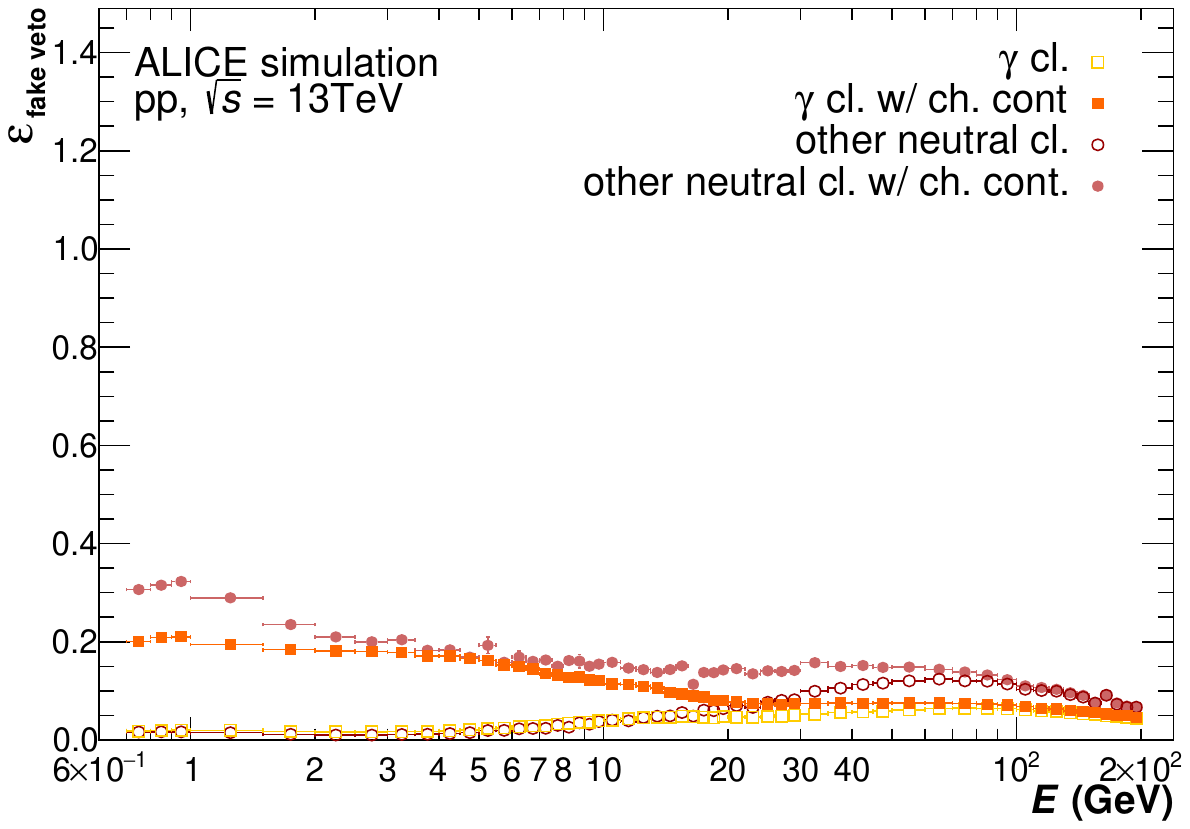}
    \caption{
        (Color online) Left: cluster-veto efficiency for primary particles (dark blue open circles) and electrons (green squares) as well as conversion electrons with a production vertex below 180~cm (green closed diamonds) and above 180~cm (green open diamonds) and other secondary particles (cyan open circles) as obtained from simulations of pp collisions at \sthirteen . 
        Right: fraction of fake track-to-cluster matches for clusters originating from photons (yellow open squares) and other neutral particles (red open circles). 
        Additionally, the same categories are shown for clusters that have additional charged particle contributions for photon clusters (orange squares) and other neutral particles (light red circles). 
    }
    \label{fig:trmatcheff}
\end{figure}
After simultaneous optimization of the photon purity and efficiency, the following conditions were used to tag  a cluster as ``neutral'' in most of the photon and neutral meson analyses:
\begin{equation}
    |\Delta \eta^\text{residual}| > 0.010 +(\pt^\text{track} +4.07)^{-2.5}\ \mathrm{and }\ |\Delta \varphi^\text{residual}| > 0.015 +(\pt^\text{track} +3.65)^{-2} ~\mathrm{rad,}
    \label{Eq:TMVetoCriteria}
\end{equation}
where $\Delta \varphi^\text{residual}= \varphi^\text{track}-\varphi^\text{cluster}$, 
$\Delta \eta^\text{residual}= \eta^\text{track}-\eta^\text{cluster}$ and the track transverse momentum ($\pt^\text{track}$) is in \GeVc\ units as detailed in Ref.~\cite{Acharya:2017tlv}. 
The selection window is approximately one \gls{EMCal} cell size at high \ptt\ and few cell sizes below 1~GeV/$c$.
The \pT\ integrated $\Delta\eta$-$\Delta\varphi$ distribution is illustrated in \Fig{fig:trmatch}~(right) for track-matched clusters. 

As most of the charged hadrons will not deposit their full energy in the calorimeter, the ratio of the cluster energy to the track momentum ($E/p$) can be used to discriminate between charged hadrons and electrons, as described in \Sec{sec:electrons}. 
To reduce the number of fake vetos~(clusters accidentally matched to tracks by the matching procedure) for cluster energies above 10~GeV the ratio of the cluster energy to the track momentum ($E/p$) can be required to be small to match the cluster and the track. 
A value of $(E/p)_\mathrm{max}=1.7$ was determined to provide the best purity and reconstruction efficiency for photon analyses.
The requirement additionally reduces the probability for a cluster to be wrongly matched to a track due to a charged particle depositing only minimum ionization energy in the cluster.\\
The cluster-veto efficiency as a function of the cluster energy for \pp\ collisions at \sthirteen\ is shown in \Fig{fig:trmatcheff}~(left) for different charged particles types. 
The matching criteria were optimized such that fake matches are kept at a minimum, as can be seen in \Fig{fig:trmatcheff}~(right), while still maintaining a high track-veto efficiency in particular at high cluster energies.
Clusters containing a contribution from a particle corresponding to a reconstructed track can be vetoed with an efficiency of 100\% for leading primary charged particles~(dark blue open circles) or leading primary electrons~(green squares). 
At higher cluster energies the $E/p$ criterion becomes relevant, leading to a decrease in the matching efficiency.\\
Most of the electron-positron pairs from photon conversions cannot be rejected using track matching as they occur in the material between the \gls{TPC} and the \gls{EMCal}, however, those which occur at a radius between $5$ and $180$ cm from the \gls{IP} can be rejected with a similar efficiency as secondary tracks.\\
While for the photon and jet analyses it is key to keep the fake matches to a minimum, for the primary electron analyses the matching efficiency as a function of the track \pT\ should be maximized, hence the $E/p$-matching veto is not applied.  
For electrons, a different selection on the ratio of the cluster energy to the track momentum ($E/p$) is used to discriminate between charged hadrons and electrons, as described in \Sec{sec:electrons}, and increase the purity. 

\subsubsection{Distribution of energy within a cluster: shower shape}
\label{sec:showershape}
\begin{figure}[t!]
 \begin{center}
    \includegraphics[width=0.3\textwidth]{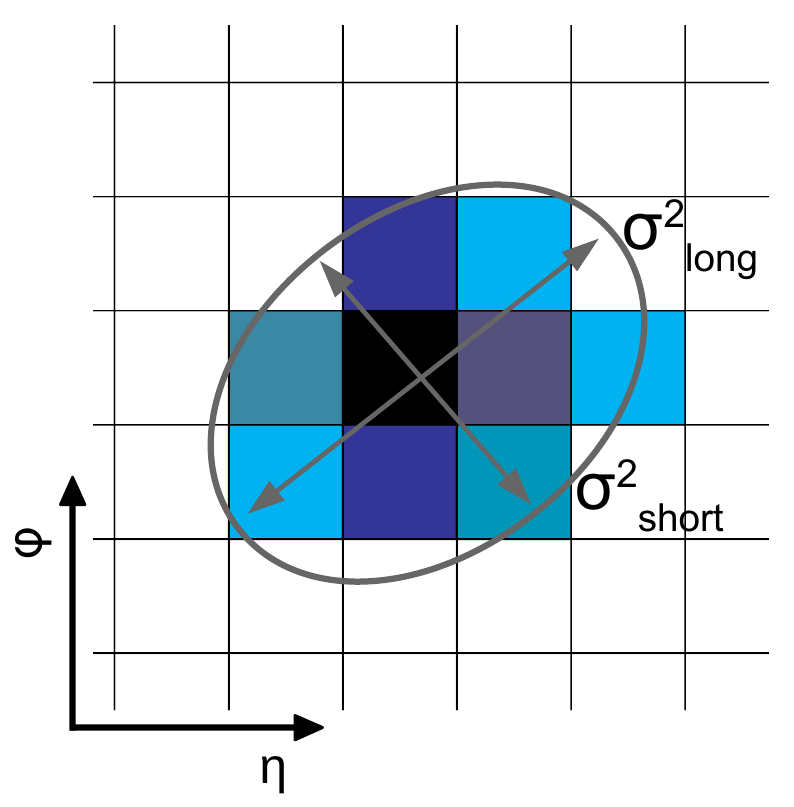}
    \hspace{1cm}
    \raisebox{+15mm}{%
    \begin{minipage}[b]{7.cm}
    \caption{\label{fig:ShowerSchema}(Color online) Schematic representation of the shower shape and the ellipse axes. The different colors indicate the amount of energy deposited in each cell, the darker the more energy.}
    \end{minipage}}
 \end{center}
\end{figure}
The distribution of energy within a cluster, referred to as ``shower shape'', is described using
a parametrization of the shower surface ellipse axes~\cite{Alessandro:2006yt, Abeysekara:2010ze}. 
The shower surface is defined by the intersection of the cone containing the shower with the front plane of the calorimeter, as displayed schematically in \Fig{fig:ShowerSchema}. 
The energy distribution along the $\eta$ and $\varphi$ directions is represented by a covariance matrix with terms ${\sigma^2_{\varphi\varphi}}$, ${\sigma^2_{\eta\eta}}$ and ${\sigma^2_{\varphi\eta}}$, which are calculated using logarithmic energy weights $w_i$~(see Eq.~\ref{eq:w_i} and~\ref{eq:w_tot}),
\begin{equation}
\sigma^{2}_{\alpha\beta} = \sum_i \frac{w_i\alpha_i\beta_i}{w_\mathrm{tot}}-\sum_i \frac{w_i\alpha_i}{w_\mathrm{tot}}\sum_i \frac{w_i\beta_i}{w_\mathrm{tot}}\,,
\label{eq:ss_centroid}
\end{equation}
where $\alpha_{i}$ and $\beta_{i}$ are the cell indices in the  $\eta$ or $\varphi$ direction. 

The shower shape parameters \shshlo\ (long axis) and \shshsh\ (short axis) are defined as the eigenvalues of the covariance matrix, and are calculated as
\begin{figure}[t!]
\begin{center}
\includegraphics[width=0.45\textwidth]{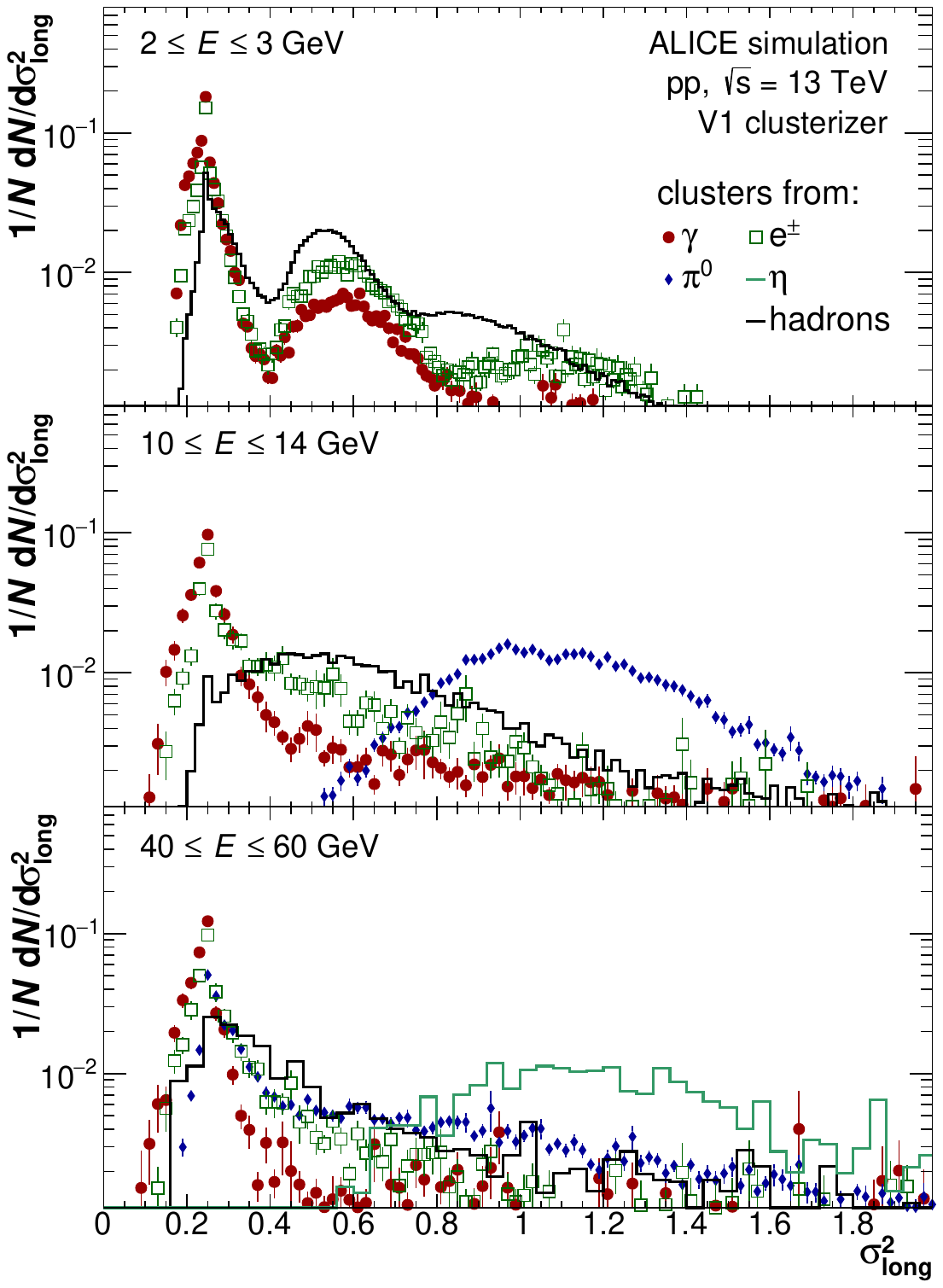}
\hspace{0.3cm}
\includegraphics[width=0.45\textwidth]{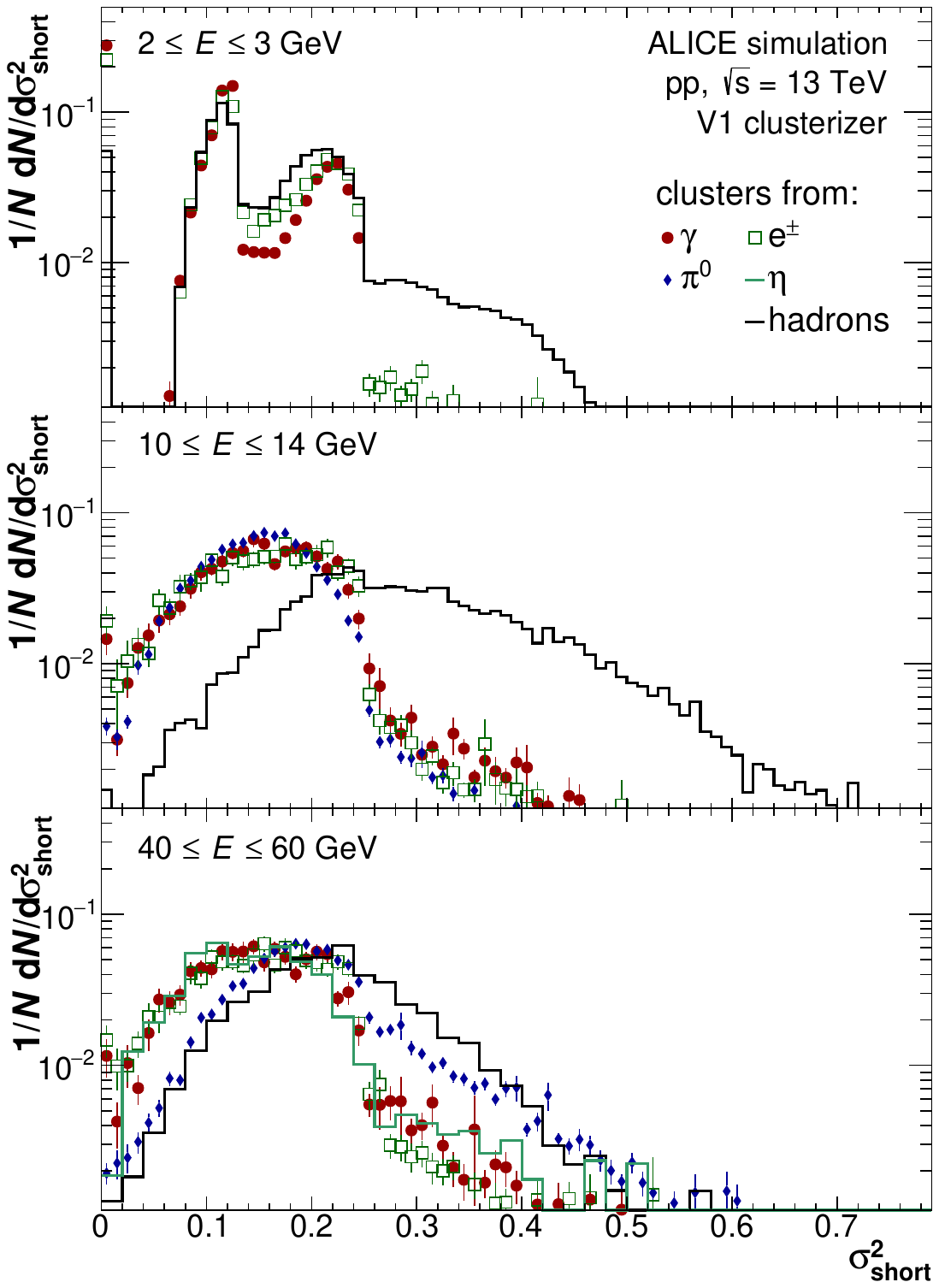}
\end{center}
\vspace{-0.2cm}
\caption{\label{fig:SSPhotonPi0MC}(Color online) \shshlo\ (left) and \shshsh\ (right) distributions in three energy intervals for photons, electrons, hadrons, \piz\ and $\eta$ mesons. The distributions are obtained using the V1 clusterizer from a simulation of pp collisions at \sthirteen\ performed with the \gls{PYTHIA} event generator, in which events are required to contain either two jets or a jet and a high-energy direct photon. Each distribution is normalized to its integral. A model simulating the effect of the cross talk was applied as discussed in \Sec{sec:crosstalk}. 
}
\vspace{0.2cm}

\begin{center}
\includegraphics[width=0.45\textwidth]{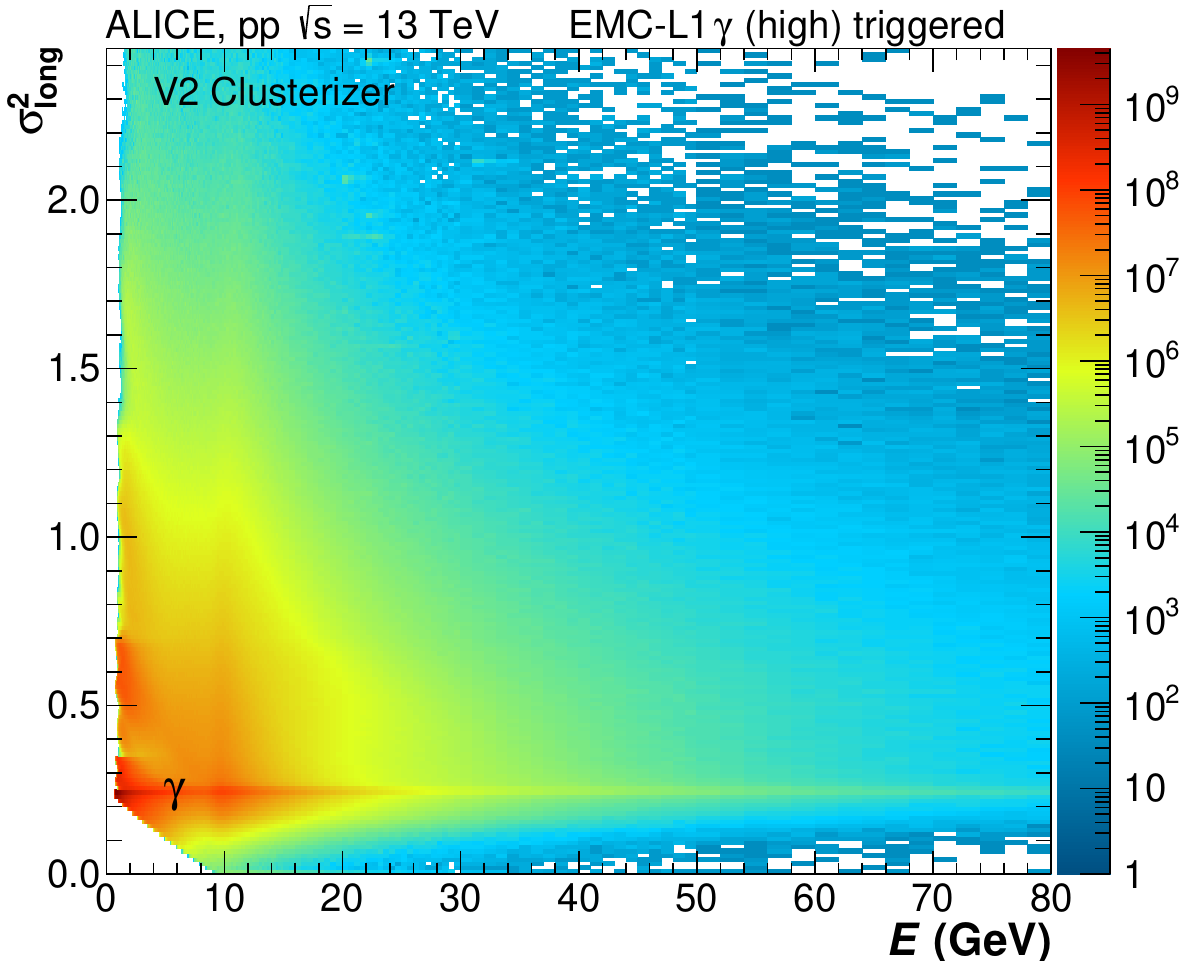}
\hspace{0.3cm}
\includegraphics[width=0.45\textwidth]{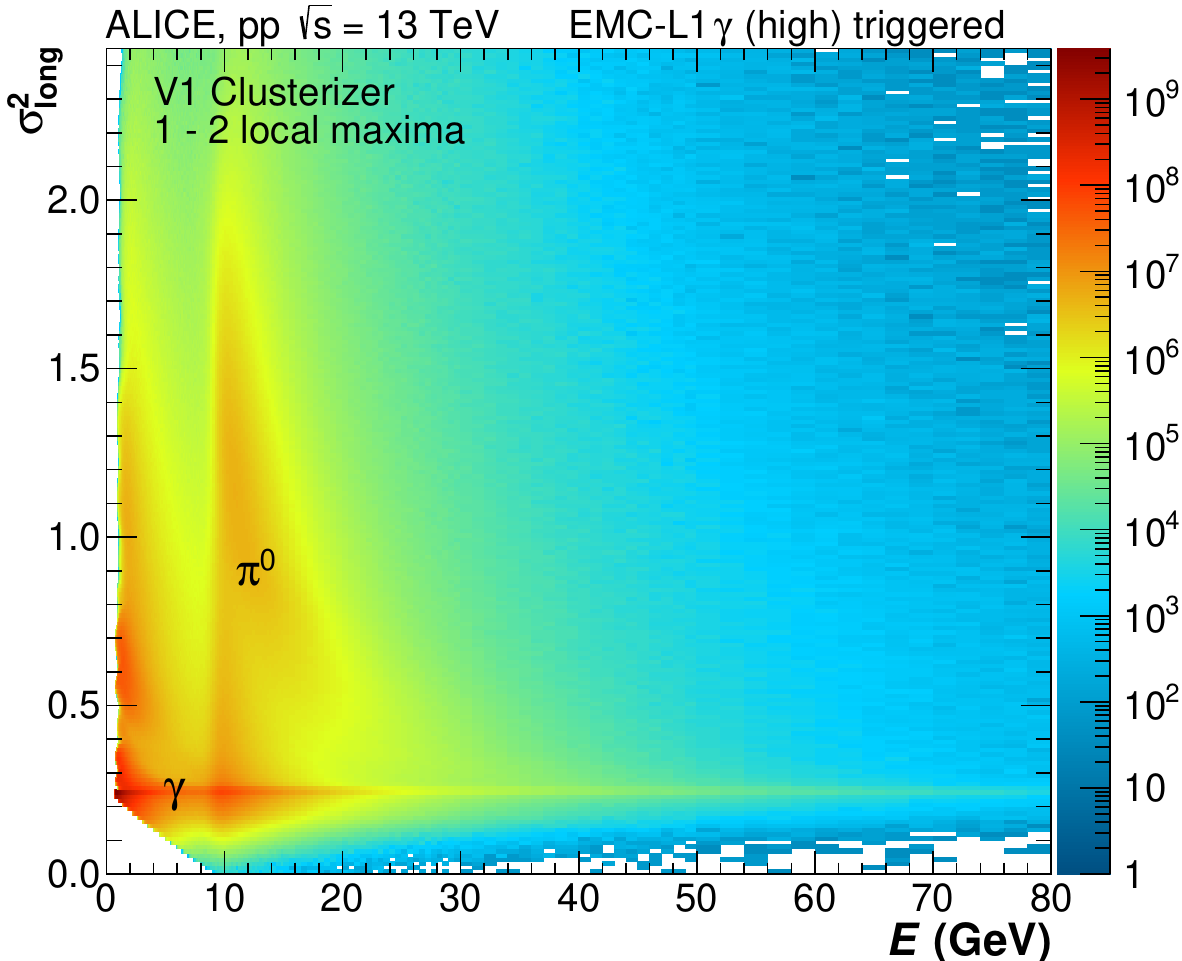}
\end{center}
\caption{\label{fig:SSPhotonPi0Data}(Color online)  Distributions of \shshlo\ versus cluster energy in pp collisions at \sthirteen\ triggered by the \gls{EMCal} \gls{L1} at approximately 9~GeV for the V2 (left) and V1 clusterizer (right). Each energy bin is normalized to its integral and exotic clusters were rejected (\Sec{sec:exotics}). 
}
\end{figure}
\begin{eqnarray}
\shshlo = 0.5(\sigma^{2}_{\varphi\varphi}+\sigma^{2}_{\eta\eta})+\sqrt{0.25(\sigma^{2}_{\varphi\varphi}-\sigma^{2}_{\eta\eta})^2+\sigma^{2}_{\eta\varphi}}, \label{eq:ss1}\\
\shshsh = 0.5(\sigma^{2}_{\varphi\varphi}+\sigma^{2}_{\eta\eta})-\sqrt{0.25(\sigma^{2}_{\varphi\varphi}-\sigma^{2}_{\eta\eta})^2+\sigma^{2}_{\eta\varphi}} \label{eq:ss2},
\end{eqnarray}
The particle shower spread measured with these parameters can be used to distinguish symmetric electromagnetic showers (small spread) originating from photons or electrons from non-symmetric showers caused by hadronic interactions of, for instance, neutrons, protons or charged pions.
In particular at low \pT~($\pT < 0.5$ GeV), the shower shape of charged particles can also be elongated by the angle of incidence.
Furthermore, the merging of showers from electromagnetic processes, \ie\ ${\rm e^{+}e^{-}}$ pairs from conversions within a close distance to the \gls{EMCal} or photons from neutral meson decays with high transverse momenta, also lead to asymmetric shower shapes, as described in \Sec{sec:clusterization}.
Additionally, several particles from a collimated jet can contribute to the same cluster, or random overlaps can occur. 
The latter is particularly important in high particle-density environments like heavy-ion collisions, but can also be relevant in \pp\ events due to pileup of multiple \pp\ collisions.

\Figure{fig:SSPhotonPi0MC} shows the normalized \shshlo\ and \shshsh\ distributions for different particle types in three different energy intervals using the V1 clusterizer, obtained from \gls{PYTHIA} simulations of \pp\ collisions at \sthirteen.
While the distributions of photons and electrons clearly peak around $\shshlo = 0.25$ within a narrow range $0.1<\shshlo<0.3$ independently of the energy, hadrons exhibit a much wider distribution with large tails towards higher shower shape values. 
At low cluster energies, up to few GeV, the distributions show other less prominent peaks than that at 0.25. These are due to cases in which a low number of cells in specific configurations~(two or three cells aligned or L-shaped) contributes to the shower shape calculation, giving rise to a non-monotonous distribution.
For electrons, the tail of the \shshlo\ distribution extends to larger values than for photons in particular at low cluster energy due to the magnetic field and rescattering in the detector material.
The \shshlo\ distribution for the \piz\ and $\eta$ mesons\com{ merged decay photon clusters} changes significantly with the meson energy and its mean value decreases rapidly towards higher energies, coming closer to the photon distribution with increasing meson energy.

As there is no clear discriminating power in the \shshsh\ distributions for neutral particles, the most sensitive parameter to study the particle composition for neutral particles appears to be \shshlo. 
\Figure{fig:SSPhotonPi0Data} shows the \shshlo\ for the V2~(left) and V1~(right) clusterizer as a function of the cluster energy for \gls{L1} triggered events in \pp\ collisions at \sthirteen. 
The dominant regions for photons and merged \piz\ mesons are indicated by the corresponding labels. 
For intermediate energies~($6<E<20$~GeV), a better separation between the two components can be obtained by using the V1 clusterizer. 
Above 20~GeV, however, the distributions appear to be similar as expected.

\subsubsection{Rejection of anomalous clusters with high energy deposit in a single cell}
\label{sec:exotics}
In electromagnetic calorimeters, in addition to the electromagnetic response, hadrons can be detected.
Most of these hadrons, except slow neutrons, on average only deposit the \gls{MIP} energy in the calorimeter.
Typical energy depositions from slow neutrons arise from hits that occur in the \gls{APD} of the cells.
These hits produce large signals, which are reconstructed as highly energetic clusters localized in single channels as opposed to a spread across multiple channels as expected from purely electromagnetic showers. 

Such characteristic clusters are called {\it exotic clusters}, and are observed in both calorimeters, \gls{EMCal} and \gls{PHOS}. 
While they are known to be present in the data, the corresponding response is not implemented in the simulation, mainly due to the missing description of the \gls{APD} behavior.
\begin{figure}[t]
    \centering
    \includegraphics[width=0.49\textwidth]{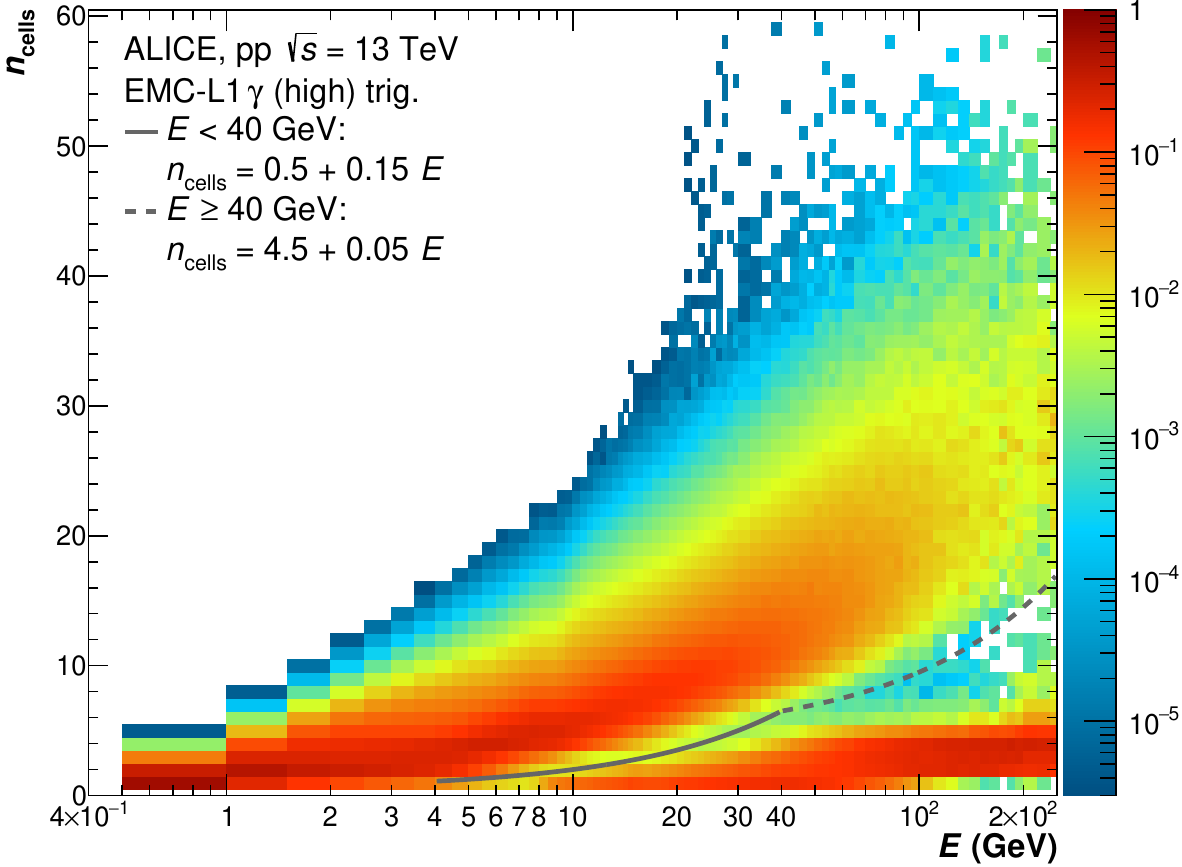}
    \includegraphics[width=0.49\textwidth]{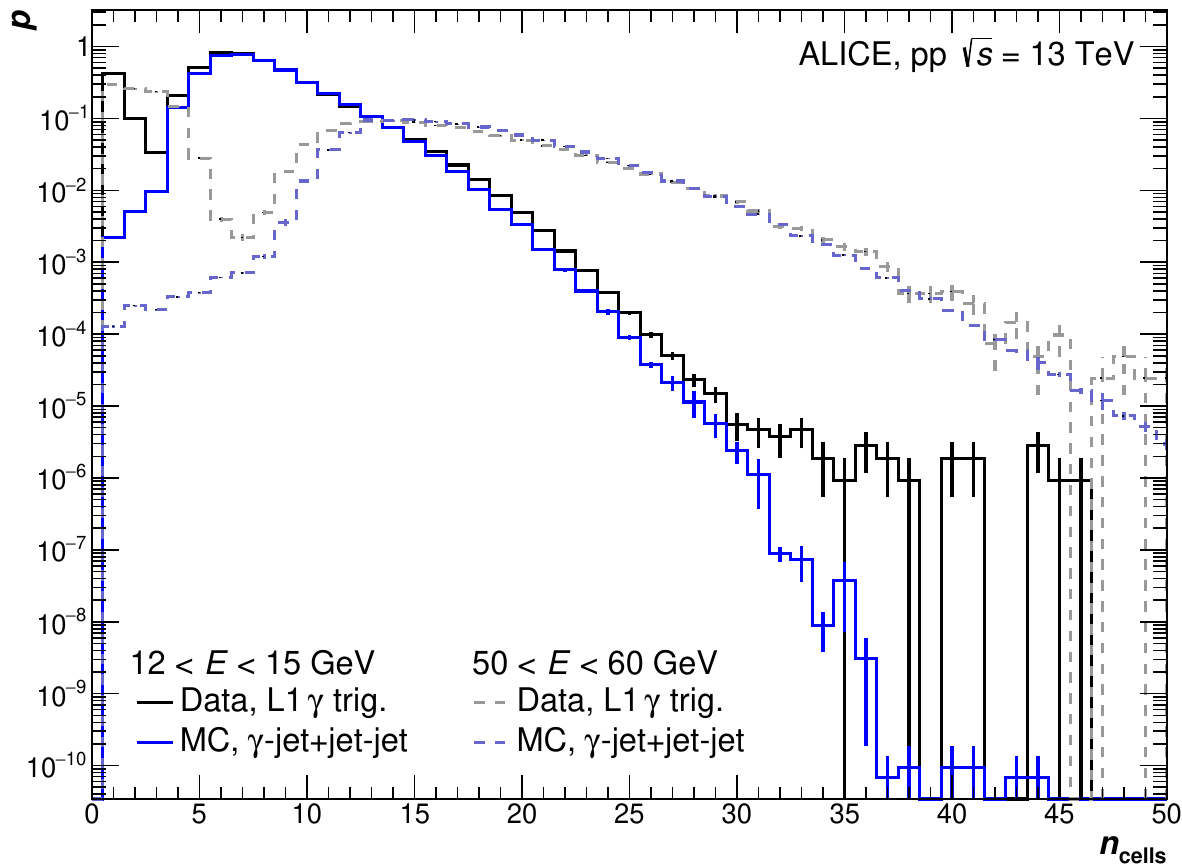}
    \caption{
        \label{fig:ncells_exotic}(Color online) Left: number of cells as a function of the cluster energy found with the V2 clusterizer in \pp\ collisions at \sthirteen\ using the \gls{EMCal} high threshold \gls{L1} $\gamma$ trigger.
        The region below the lines is populated by exotic clusters.
        The distribution for each energy bin is normalized to its integral. 
        Right: comparison of $n_{\text{cells}}$ probability distributions for measured data (black), projection of the left plot, to simulated (blue) collisions for two different cluster energy bins. 
        Each distribution is normalized by the integral of the distribution for $n_{\rm cells}$ > 10. 
    }
\end{figure}
Exotic clusters can be easily identified as they typically have a low number of cells in the cluster despite a large energy, as illustrated in \Fig{fig:ncells_exotic}{ (left)}. 
This effect is not observed in simulated events as shown in \Fig{fig:ncells_exotic}{ (right)} where cluster size distributions in data and simulation for different cluster energies are compared.
Similar distributions are observed for minimum bias and triggered data in \pp, \pPb\ and \PbPb\ collisions at all center-of-mass energies.

\begin{figure}[t]
    \centering
    \includegraphics[width=0.49\textwidth]{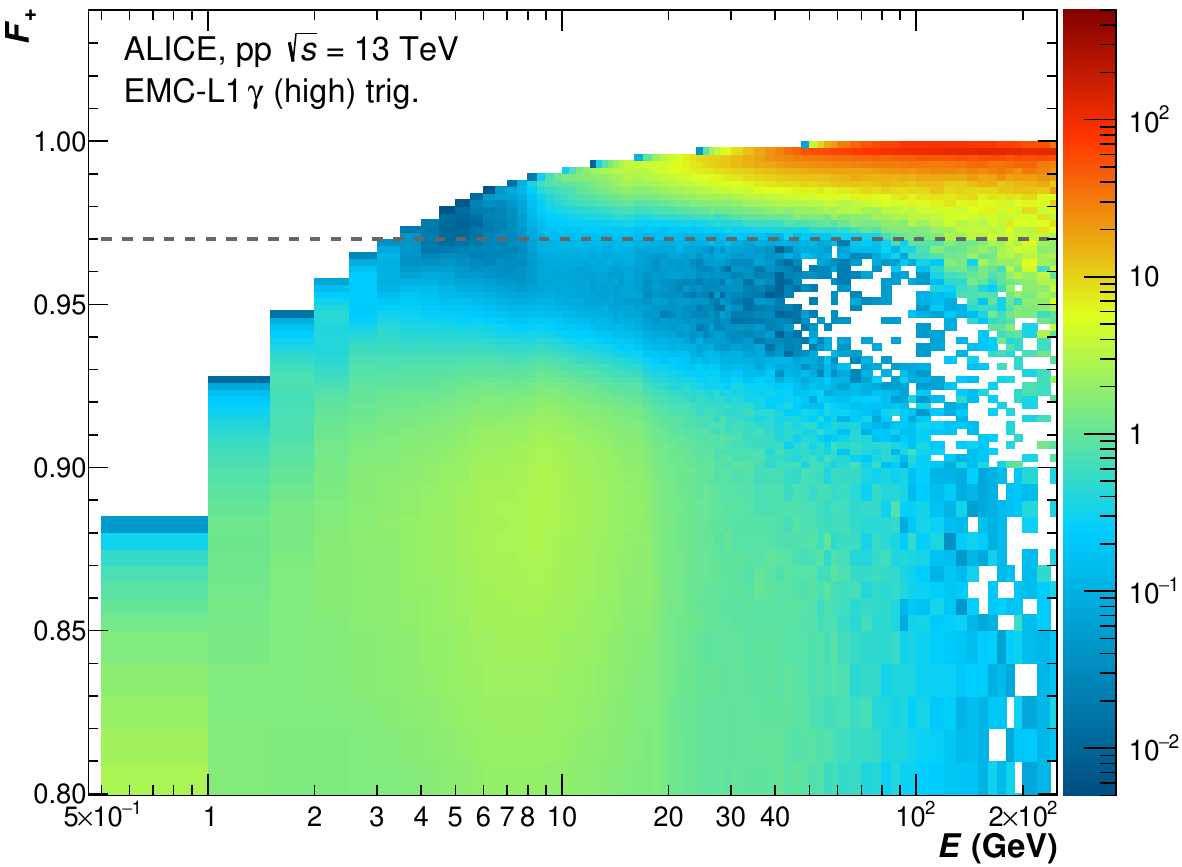}
    \includegraphics[width=0.49\textwidth]{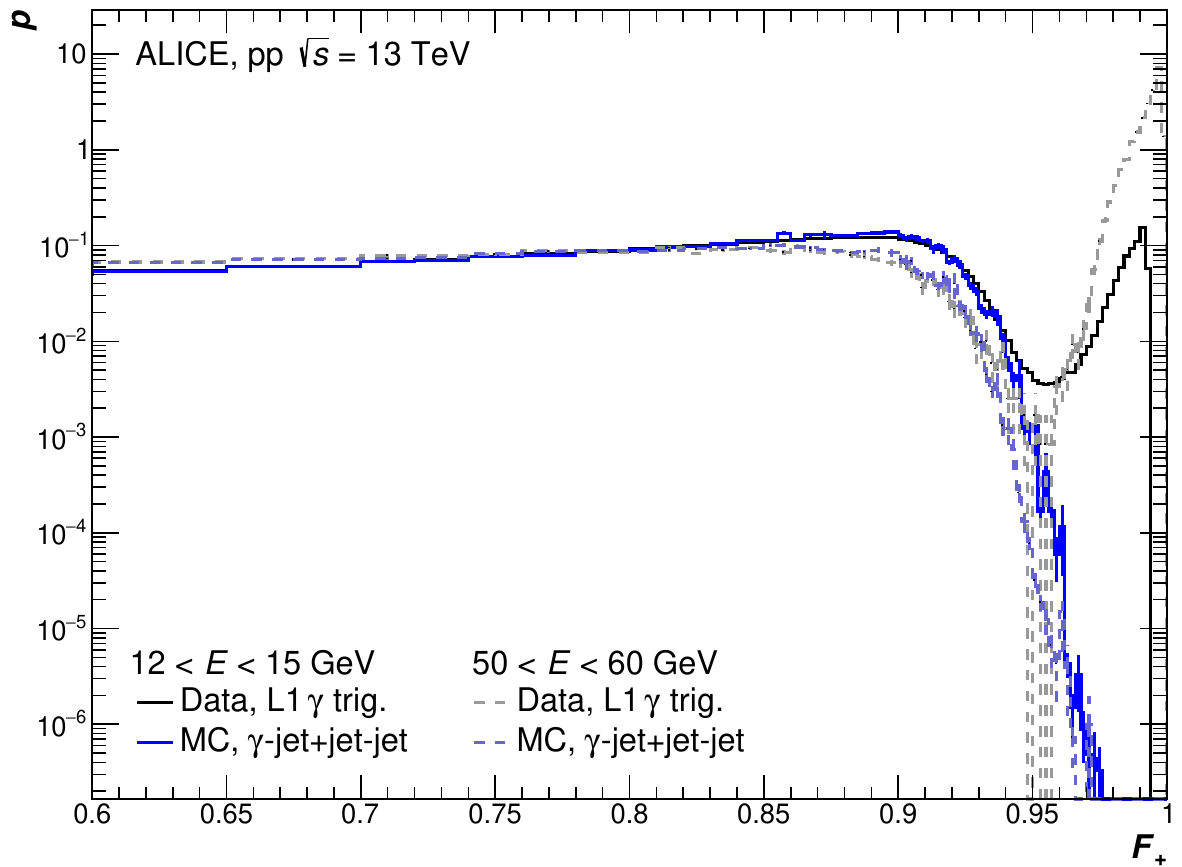}
    \caption{
        \label{fig:fcross_exotic} (Color online) Left: ``exoticity'' ($F_{\rm +}$) as function of the cluster energy found with the V2 clusterizer  in \pp\ collisions at \sthirteen\ using the \gls{EMCal} high threshold \gls{L1} $\gamma$ trigger.
        The region above the line is populated by exotic clusters.
        The distributions are normalized to have an integral of unity for each energy bin. 
        Right: comparison of $F_{+}$ probability distributions for measured data (black), projection of the left panel, and simulated (blue) collisions for two different cluster-energy intervals. 
        Each distribution is normalized by the integral of the distribution for $F_{+}<0.85$.
    }
\end{figure}
Most of these exotic clusters can be rejected by requiring a minimum number of cells, \ie\ $n_{\text{cells}} \geq 2$, within the cluster. 
However, some of them still enter the reconstructed cluster sample.
Exotic clusters typically exhibit a characteristic high energy tower neighbored by very low energy depositions. 
Those low energy cells are likely due to cross-talk between readout channels, as discussed in \Sec{sec:crosstalk}, or random overlaps with the underlying event.
Therefore, a topological cut on $F_{+}$, called the exoticity, can be used to remove such clusters, similar to the method developed by the \gls{CMS} collaboration~\cite{Bialas:2013wra}. 

The exoticity parameter is defined as:
\begin{equation}
    F_{+}=1-E_{+}/E^{\rm max}_{\rm cell}
    \label{eq:exoticity}
\end{equation} 
where $E^{\rm max}_{\rm cell}$ is the energy of the cell with highest energy and $E_{+}$ is the sum of the energy of the four cells sharing an edge to the maximum energy cell.  
The exoticity describes the degree of the homogeneity of the energy partition within the cluster, allowing to reject clusters with a dominant contribution from a single cell. 

\Figure{fig:fcross_exotic}~(left) shows the dependence of $F_{+}$ on the cluster energy. 
Above 10~GeV, the value of $F_{+}$ exhibits a spike at unity not observed in simulation in \Fig{fig:fcross_exotic}~(right). 
A minimum is observed between $0.95 < F_{+} < 0.97$ in all collision systems for cluster energies below 50 GeV, and the low $F_{+}$ values dominantly originate from physical clusters.
While the fraction of exotic clusters within the cluster sample is negligible below $\sim$ $5$~GeV it rises continuously for higher cluster energies until it plateaus at about 90\% for cluster energies above $60$~GeV.
For $E>$~30~GeV, about half the clusters have $F_{+} > 0.97$.
For heavy-ion collisions, the $F_{+}$ distribution of good clusters widens due to the additional energy input from the underlying event. 
This results in an increase of  the fraction of good clusters with $F_{+}>0.97$ and a shift of the minimum towards $F_{+}=0.95$ with increasing event centrality.

\begin{figure}[t]
  \centering
    \includegraphics[width=0.49\textwidth]{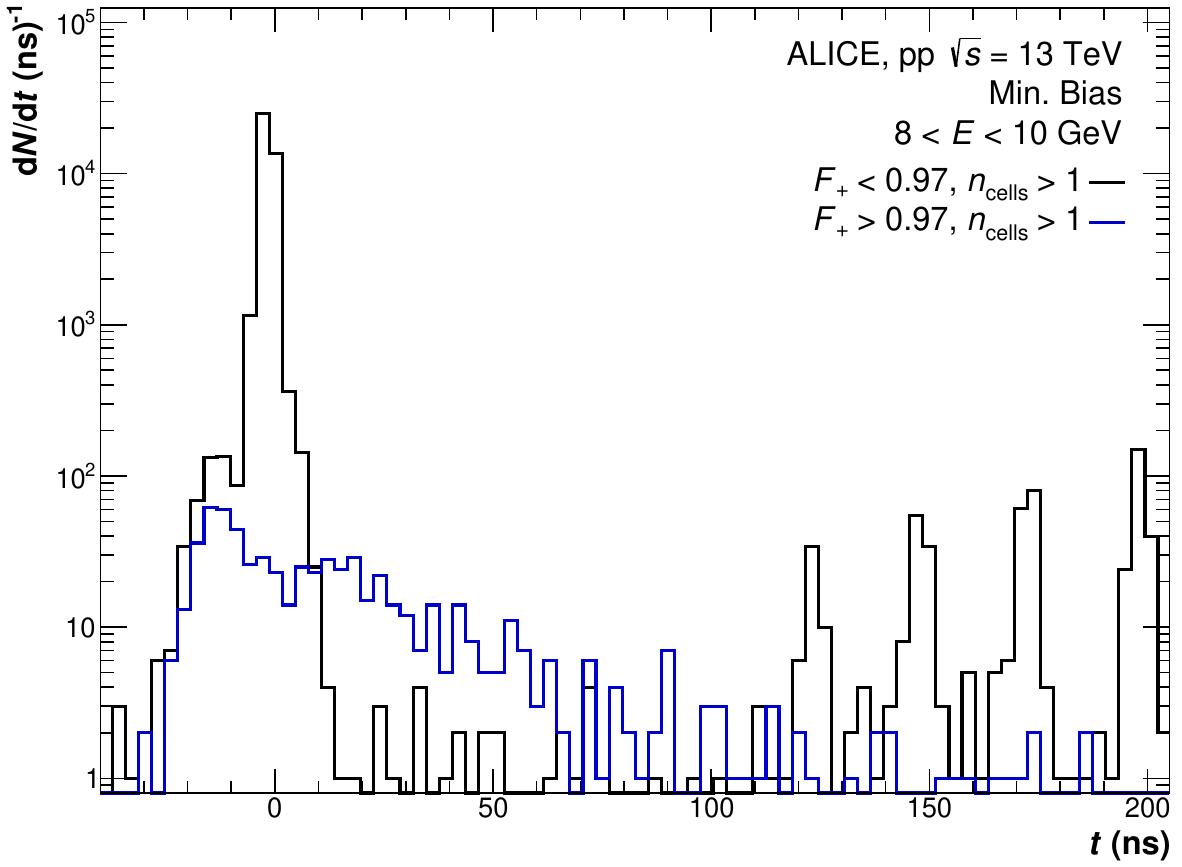}
    \includegraphics[width=0.49\textwidth]{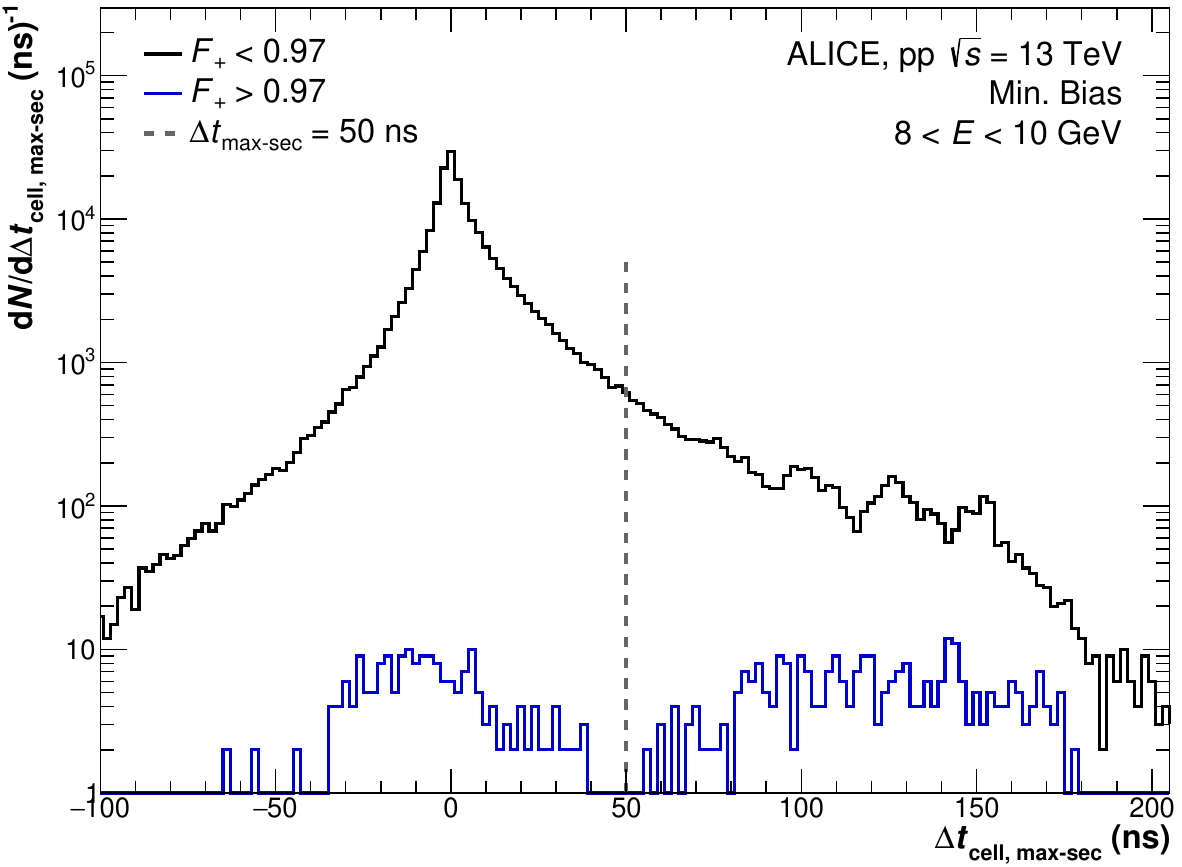}\\
    
    \caption{
        \label{fig:exoTime}(Color online) Left: cluster time for exotic ($F_{+}>$ 0.97 and non-exotic ($F_{+}<$ 0.97) clusters. 
        The additional peaks in the time distribution beyond $100$ ns arise from additional bunch crossings, which could not be rejected by the online Past-Future protection using the \gls{V0} detector.
        Right: difference in time of the most and second most energetic cell in the cluster for exotic and non exotic clusters. 
        Both distributions are obtained for V2 clusters with energy in the interval $8 < E < 10$~GeV from data taken from pp collisions at \sthirteen.
    }
\end{figure}
\begin{figure}[t]
    \includegraphics[width=0.49\textwidth]{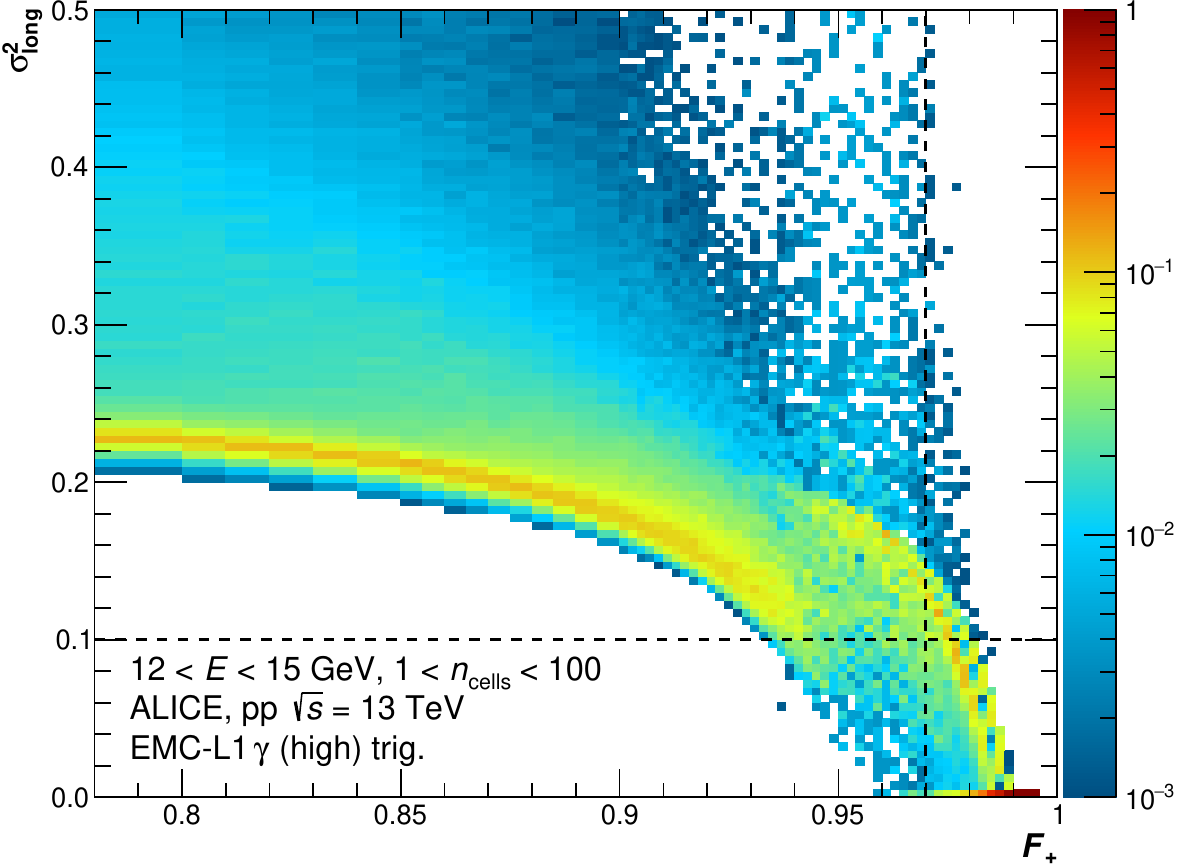} 
    \includegraphics[width=0.49\textwidth]{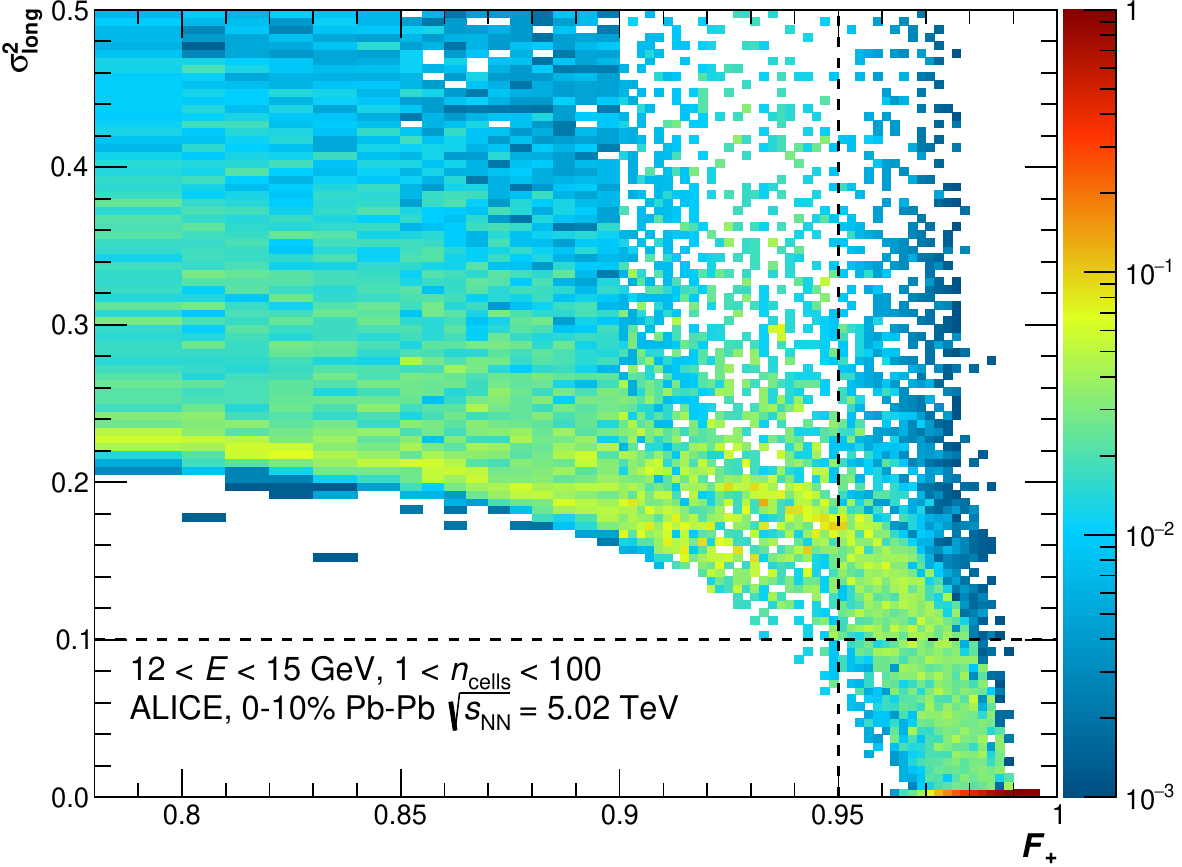} 
    \caption{
        \label{fig:exoM02}(Color online) \shshlo\ as a function of the exoticity parameter $F_{+}$ obtained from V2 clusters with $12<E<15$~GeV. The distributions are shown for pp collisions at \sthirteen\ using the \gls{EMCal} \gls{L1} $\gamma$ (high) trigger at $E^{\rm L1-trig}$~=~9~GeV (left) and \PbPb\ collisions at \sfivelead\ for 0-10\% central collisions.
    }
\end{figure}
Differences in the cluster time distribution for exotic clusters with respect to non-exotic clusters indicate that exotic clusters have no physical origin and are thus safe to remove.
\Figure{fig:exoTime}~(left) shows a broad and almost flat timing distribution for exotic cluster candidates (blue) with almost no peak at $t=0$~ns. 
A similar timing distribution was observed for clusters with only one cell (not shown), while those with $F_{+} < 0.97$ and more than one cell (black) show the typical expected cluster time distribution.
The presence of long tails in the timing distribution hints at a random time association for those clusters within the readout time window.
This is supported by the time difference between the leading and subleading cell times $\Delta t_{\rm max-sec}$ shown in \Fig{fig:exoTime}~(right).
For non-exotic clusters~(black), we observe a peaked and rather symmetric distribution around 0~ns difference. 
For exotic clusters, two regimes can be observed:
a)~timing within the expected one but without a clear peak at 0~ns and the average shifted to negative timing, most likely random cell associations with the underlying event or noise; 
b)~timing shifted beyond 50~ns, likely cross-talk induced cells.
%

The exoticity variable $F_{+}$ defined here is strongly correlated with the shower shape variable \shshlo{} defined in the previous section. 
\Figure{fig:exoM02} shows the correlation of the width of the shower along the long axis $\sigma^{2}_{\rm long}$ and $F_{+}$.
Clusters with $\sigma^{2}_{\rm long}<0.2$ also satisfy the $F_{+}$ cut at about 0.95. 
Since the region of $0.1<\sigma^{2}_{\rm long}<0.2$ contains a substantial fraction of clusters attributed to physical origin, a combination of cuts on $\sigma^{2}_{\rm long}<0.1$ and $F_{+}>0.97$ is recommended to safely remove exotic clusters.
 
\subsection{Trigger performance}
\label{sec:trigger}
\begin{table}[t]
    \centering
    \caption{Area covered by \gls{EMCal} towers, FastORs and trigger patches 
    of various sizes. For trigger patches the closest resolution parameter for
    jets fully contained within the trigger patch is listed for comparison.}
    \begin{tabular}{l rrl}
         Object & Area & Approx. $R$ & Usage \\
         \toprule
         Tower       & 0.0143$\times$0.0143 & \\
         FastOR      & 0.0286$\times$0.0286 & \\
         \midrule
         $2\times2$ patch   & 0.0572$\times$0.0572 & 0.025 & \gls{L0}, \gls{L1}$\gamma$ \\
         $8\times8$ patch   & 0.2288$\times$0.2288 & 0.1 & \gls{L1}jet \PbPb{}, \gls{DCal} \gls{L1}jet\\
         $16\times16$ patch & 0.4576$\times$0.4576 & 0.2 & \gls{EMCal} \gls{L1}jet \pp{} and \pPb{} \\ \bottomrule
    \end{tabular}
    \label{tab:trigger::area}
\end{table}
To utilize the \gls{EMCal} trigger functionality, it is imperative to characterize the trigger settings and performance.
Different patch sizes were used for the various triggers, covering areas
in the $\eta$-$\phi$ space as listed in \Tab{tab:trigger::area}.
The trigger parameters varied with time according to evolving expected data-taking conditions, and are described in further detail below. 
To ensure effective operation, each \gls{LHC} run began with a trigger commissioning period.
\Table{tab::trigger::SetupRun} lists the trigger configuration for datasets collected in \pp{}, \pPb{}, and \PbPb{} collisions during \gls{LHC} Run~1~(2009--2013) and Run~2~(2015--2018) at a variety of \s. 

\begin{figure}
    \centering
    \includegraphics[width=0.6\textwidth]{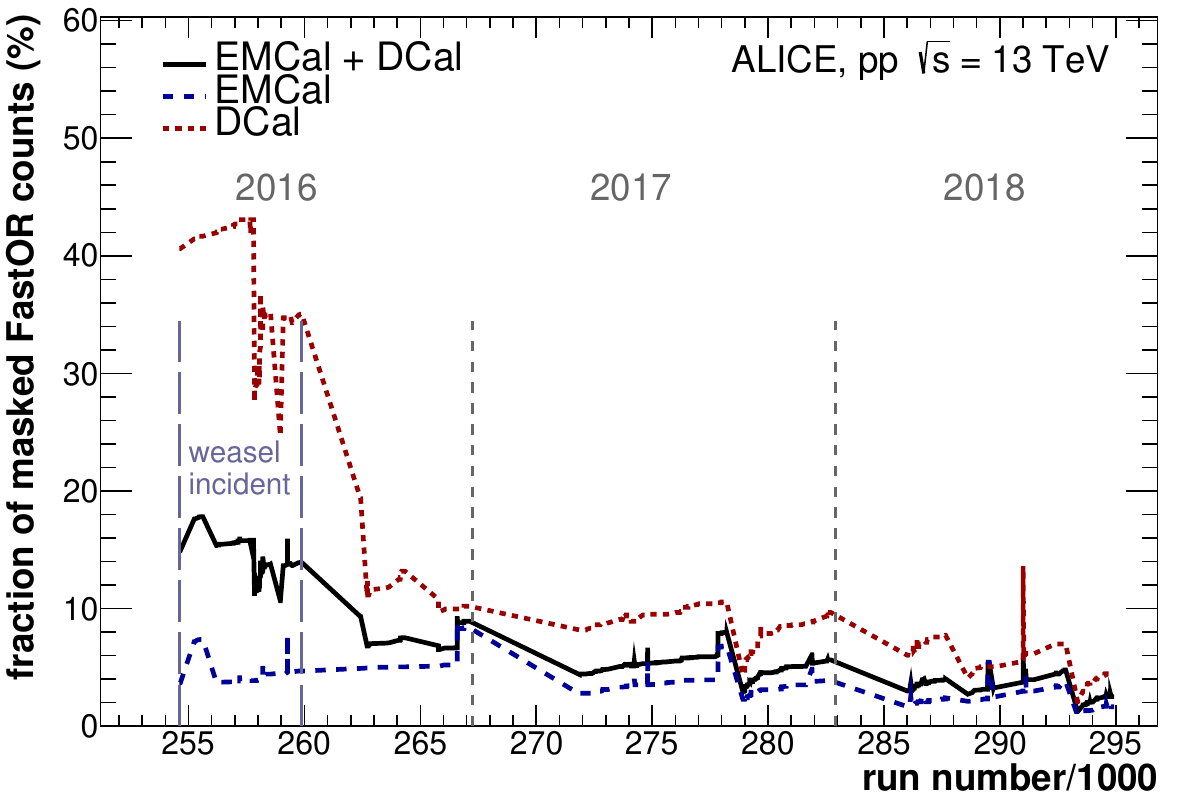}
    \caption{Fraction of masked FastORs as a function of the run number}
    \label{fig::trigger::masktrending}
\end{figure}

\Figure{fig::trigger::masktrending} shows the fraction of masked channels as a function of the run number for the pp data taking at \sthirteen{}. 
Masked channels consist of channels that are dead or suffering from a substantial noise contribution and are therefore excluded from the trigger
electronics, resulting in a reduction of the acceptance at trigger level. 
The fraction of masked channels decreases with time due to the maintenance and repair of problematic hardware. 
In addition, during the first part of the data taking in 2016 the \gls{DCal} suffered from hardware damage created by a power cut 
resulting in a temporary loss of two 1/3-size supermodules in the \gls{DCal}, corresponding to two \glspl{TRU}. 
The affected supermodules were re-included after the hardware repair starting from the last runs in 2016.

\begin{figure}[b]
    \centering
    \includegraphics[width=0.58\textwidth]{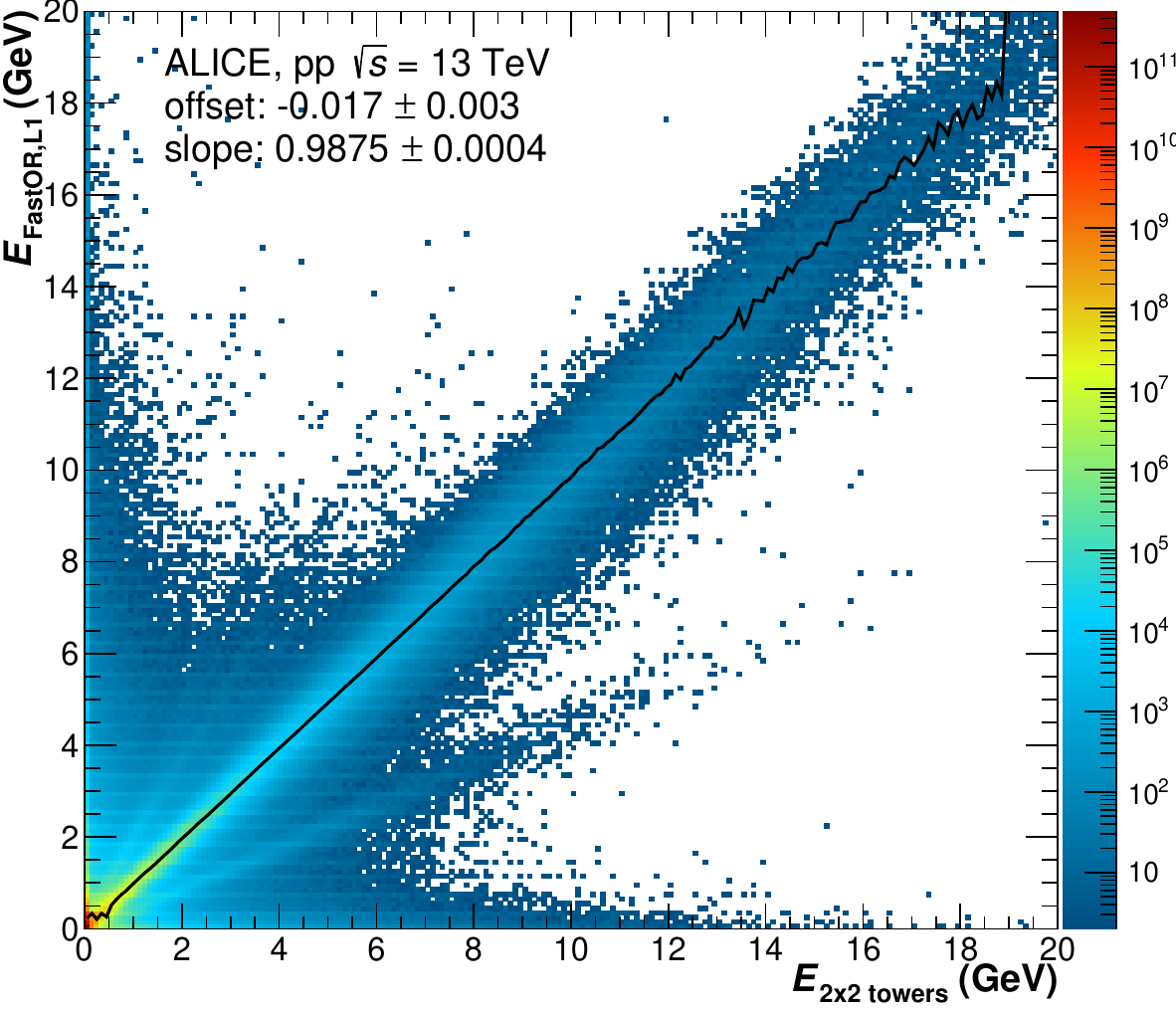}
    \caption{(Color online) Correlation between energy from the \gls{FEE} and FastOR readout based on the \gls{L1} \glspl{ADC} for towers corresponding to the same FastOR module. 
    The black line indicates the mean energy in the FastOR for a given energy interval at \gls{FEE} level.}
    \label{fig:trigger::comparisonEnergy}
\end{figure}

In order to assess the quality of the energy calibration applied in the trigger electronics, we compared the energy from $2\times2$ towers belonging to the same module from the front-end readout and the FastOR \gls{L1} timesums serving as \gls{L1} \gls{ADC} values.
\Figure{fig:trigger::comparisonEnergy} shows this comparison for all modules. 
Energy from towers that are masked at \gls{FEE} level are not included in the sum of the $2\times2$ tower energies. 
A main correlation band with an approximate slope of $1$ is visible. 
The spread around the mean leads to a smearing of the rise~(also called ``turn-on'') of the trigger efficiency or rejection in the vicinity of the nominal trigger threshold.  
Additional bands are associated with modules where a fraction of the channels is masked.

\begin{figure}[t]
    \centering
    \includegraphics[width=0.48\textwidth]{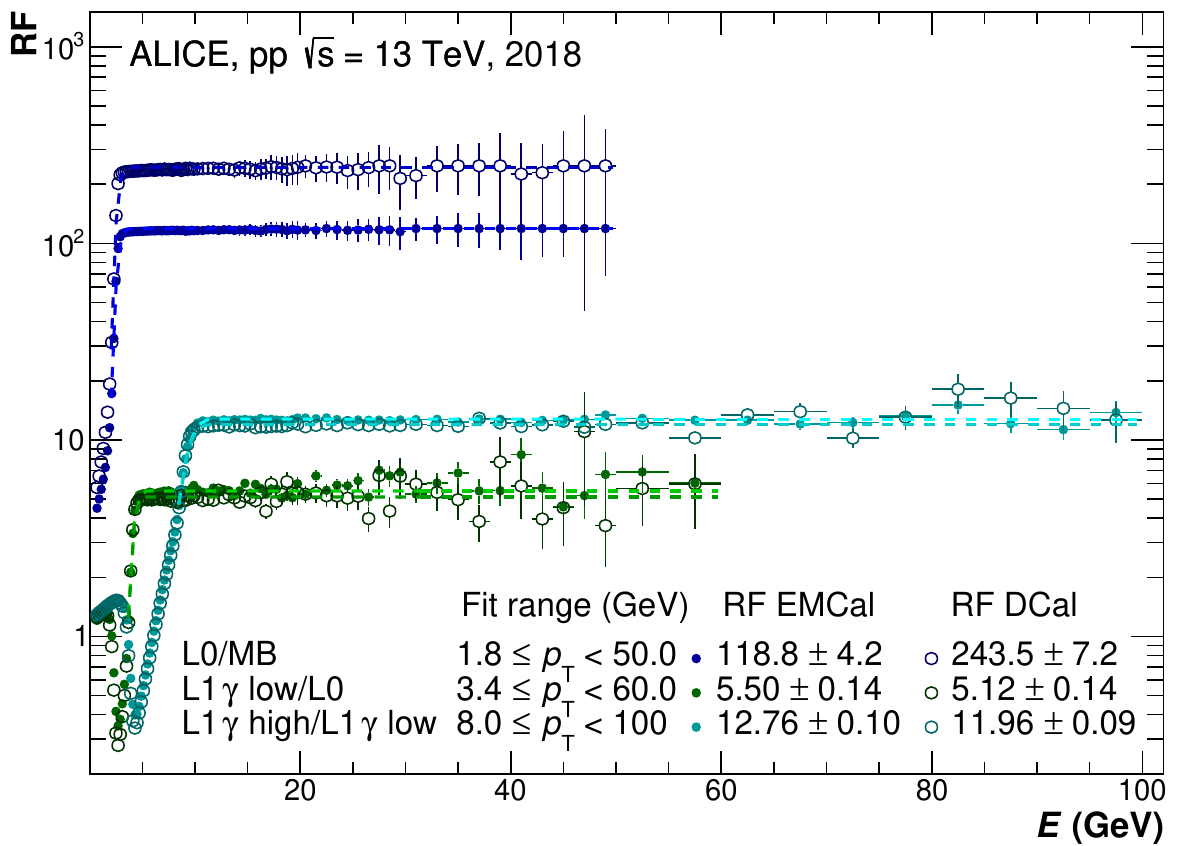}
    \includegraphics[width=0.48\textwidth]{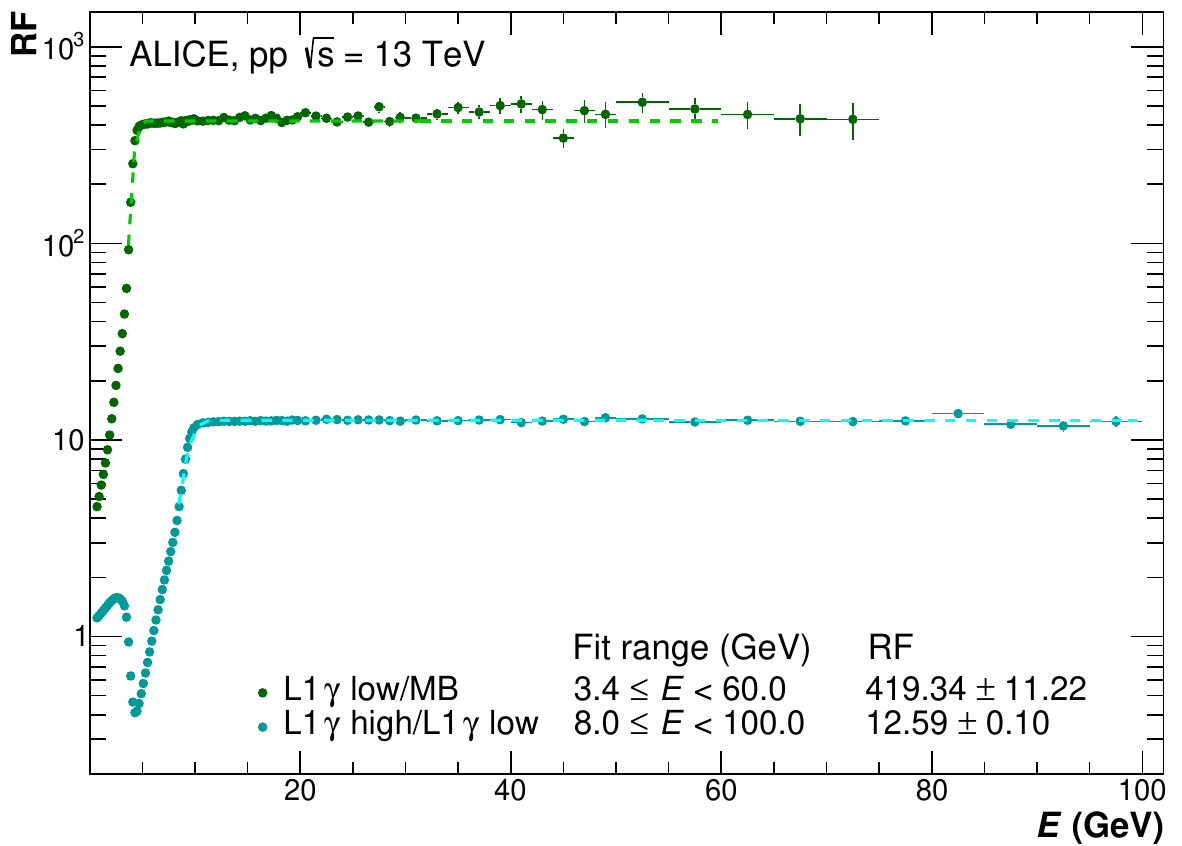}
    \caption{(Color online) Trigger rejection factor (\gls{RF}) for the single-shower trigger in \pp{} collisions at 
    \sthirteen{} for the single samples for 2018 (left) and the combined \gls{EMCal} and \gls{DCal} triggers (right).}
    \label{fig::trigger::rejectionEGApp}
\end{figure}

\begin{table}[t]
    \centering
    \caption{Setup of the \gls{EMCal} triggers in various collision systems in \gls{LHC} Run 1 (2009-2013) and Run 2 (2015-2018). 
            Thresholds for \PbPb{} collisions at \stwolead{} (2011) depend on the online multiplicity measured by the V0 detector.
            During Run 2, equivalent \gls{DCal} triggers were enabled using the same thresholds. 
            The \gls{L0} and \gls{L1}-$\gamma$ triggers are based on $2\times2$ FastORs.
            The patch size of the \gls{L1}-jet triggers for \gls{EMCal} is $16\times16$ FastORs, while \gls{DCal} uses $8\times8$ FastORs in \pp\ and \pPb\ collisions.
            In \PbPb\ collisions the \gls{L1}-jet trigger patch size is $8\times8$ FastORs for both parts of the calorimeter.
            }
    \begin{tabular}{llrrrrrr}
       System & Energy & Year &\multicolumn{5}{c}{Energy threshold (GeV)}  \\  
         && & \gls{L0} & $\gamma$ High & $\gamma$ Low & Jet high & Jet Low \\
         \toprule
         \pp{} & \stwo{}   & 2011 & 3.4 & - & - & - & - \\ 
               &           & 2013 & 2 & 6 & 4 & 10 & 7 \\
               &\sfive{}  & 2015 & 2.5 & - & - & - & - \\
               &           & 2017 & 2.5 &  - &4  & - & 16\\
               & \sseven{} & 2011 & 5.5 & - & - & - & - \\
               & \seight{} & 2012 & 2 & 10 & - & 16 & - \\
               & \sthirteen{} & 2015-2018 & 2.5 & 9 & 4  & 20 & 16 \\
         \midrule
         \pPb{} & \sfivelead{} & 2013 & 3 & 11 & 7 & 20 & 10 \\
                & \seightplead{} & 2016 & 3  & 8 & 5.5 & 23 & 18 \\
         \midrule
         \PbPb{} & \stwolead{} & 2011 & 1 & 5 & - & 10 & - \\
                 & \sfivelead{}  & 2015 & - & 10 & - & 20 & - \\
                 &               & 2018 & 2.5 & 10 & 5 & 20 & - \\
         \bottomrule
    \end{tabular}
    \label{tab::trigger::SetupRun}
\end{table}
%
The trigger performance was studied using the ratios of the event-normalized energy spectra of \gls{EMCal} clusters measured in triggered events to the spectra measured in minimum-bias events.
An example of such a comparison is shown in \Fig{fig::trigger::rejectionEGApp}.
For cluster energies beyond the trigger threshold, an approximately constant plateau region can be observed, corresponding to maximum efficiency of the trigger. 
The value of the plateau region is referred to as trigger \acrfull{RF}, which is used to estimate the integrated luminosity inspected by the trigger. 
The trigger \gls{RF} can also be calculated by using cluster energy spectra measured with different trigger thresholds.
The ratio is approximately constant for cluster energies larger than the largest trigger threshold.
This method is more robust for the \gls{L1} triggers, as the statistical uncertainties on the spectra are negligible for both trigger thresholds.

\begin{table}[t]
    \centering
    \caption{Trigger rejection factor (\gls{RF}) of different single-shower triggers in \pp{} and \pPb{} collisions at various center-of-mass energies. In addition to the RF values the relative statistical uncertainties are given.}
    \begin{tabular}{llrrrr}
        \toprule
        System & Collision & Year & \multicolumn{3}{c}{\gls{RF}} \\
               & energy  &  & \gls{L0} & \gls{L1} $\gamma$ (low) & \gls{L1} $\gamma$ (high)\\ \midrule
        \pp{}  & \stwo~\cite{ALICE:2017nce}   & 2011 & $1217 \pm 5.5\%$ & - & -\\
               &        & 2013 & $126 \pm 3.4\%$ & $1959 \pm 6.7\%$ & $7743 \pm 8.8\%$\\   
               & \sfive & 2015 & $1976 \pm 3.6\%$ & - & -\\
               &        & 2017 & & $848 \pm 1.7\%$\\
               & \sseven~\cite{ALICE:2019rtd}& 2011 & $2941 \pm 5.9\%$  & - & - \\
               & \seight~\cite{ALICE:2017ryd}& 2012 & $ 65 \pm 1.6\%$ & - & $14712 \pm 3.8\%$  \\
               & \sthirteen & 2016--2018 &  & $419.34 \pm 2.7\%$ & $5279 \pm 2.8\%$\\ \midrule
        \pPb{} & \sfivelead & 2013 & $90 \pm 3.5$\% & $1759 \pm 7.6\%$ & $7211 \ \pm 7.6\%$ \\
                              & \seightplead~\cite{ALICE:2021est}& 2016 &  & $288 \pm 2.8\%$ & $991 \pm 3.0\%$\\ 
        \bottomrule
    \end{tabular}
    \label{tab::trigger::rejectionGamma}
\end{table}

The resulting \glspl{RF} are listed in \Tab{tab::trigger::rejectionGamma}, where the quoted relative uncertainties are obtained from variations of the cluster energy range used in the plateau region fit. 
As visible in \Fig{fig::trigger::rejectionEGApp} (left), only the \gls{RF} of the \gls{L0} trigger differs between \gls{EMCal} and \gls{DCal}, mainly due to their different acceptances, while the rejection at \gls{L1} is the same for both subdetectors.
However, the \gls{L0} trigger was used for this data set only as a pre-trigger for the \gls{L1} triggers and a control sample was taken, which was synchronously scaled down together with the minimum bias data.
The right side of \Fig{fig::trigger::rejectionEGApp} shows the \gls{RF} obtained from events triggered by the \gls{L1} \gls{EMCal} or \gls{DCal} $\gamma$ triggers for the full data sample collected in pp collisions at \sthirteen{}.
The trigger rejection obtained in this way does not necessarily coincide with the rejection factors obtained for each detector independently, due to a small but non-negligible overlap of events for which both triggers fire. 

\begin{figure}[t]
    \centering
    \includegraphics[width=0.8\textwidth]{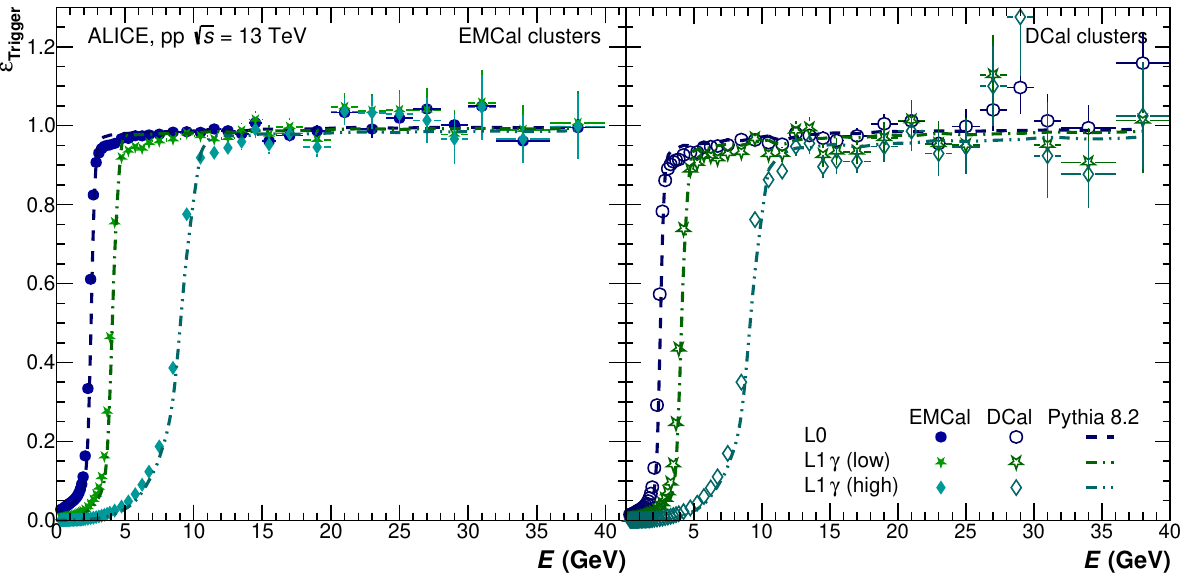}
    \caption{(Color online) Trigger efficiency for \gls{EMCal} (left) and \gls{DCal} (right) in \pp{} collisions at \sthirteen{}.}
    \label{fig::trigger::efficiencyData}
\end{figure}

The trigger efficiency was obtained by comparing the cluster energy spectra after normalizing by the luminosity inspected by the trigger. 
The luminosity is determined using the number of minimum bias triggers before prescaling. 
The trigger efficiency for the single shower triggers was obtained from the value of the ratio of the cluster energy spectrum for a given \gls{EMCal} trigger, \ie\ from the \gls{L0} trigger, and the corresponding spectrum in minimum bias collisions in the plateau region.  
\Figure{fig::trigger::efficiencyData} shows the trigger efficiency for \gls{EMCal} and \gls{DCal} single shower triggers obtained in \pp{} collisions at \sthirteen. 
The trigger efficiency is 99\% for \gls{EMCal} and 97\% for \gls{DCal} triggers and is well reproduced by the simulations.
The reduction of the trigger efficiency is due to FastOR channels which are masked at trigger level whereas the corresponding towers are not masked in the Front-End readout. 
FastORs in the trigger electronics are masked in the \gls{TRU} at \gls{L0}, therefore the trigger acceptance is the same at \gls{L0} and \gls{L1}.

\begin{table}[t]
    \centering
    \caption{Purity of the single-shower triggers in \pp{} collisions at \sthirteen{}}
    \begin{tabular}{lrrr}
        Detector & \gls{L0} trigger & low threshold & high threshold \\
        \toprule
        \gls{EMCal} & 76.2\% & 78.0\% & 65.4\% \\
        \gls{DCal} & 68.8\% & 66.5\% & 52.1\% \\
        \bottomrule
    \end{tabular}
    \label{tab::trigger::puritySingleShower}
\end{table}

The trigger purity of the single-shower trigger is determined by counting the fraction of triggered events with a cluster passing the standard cluster selection, defined in \Sec{sec:anaparam}, with an energy above the nominal threshold of the trigger in the corresponding subdetector firing the trigger. 
For single shower triggers, clusters with an energy above the trigger threshold are considered as physics signal correlated with the trigger signal as can be derived from \Fig{fig::trigger::efficiencyData}. 
Consequently events lacking clusters with a sufficient energy can be considered as random correlations of noise sources with the interaction trigger leading to a trigger selection in the detectors that needs to be considered as impurity of the trigger.
The purity values for the various single shower triggers are listed in \Tab{tab::trigger::puritySingleShower}. 
The purity reaches $\approx~80\%$ for the \gls{L0}- and low-threshold trigger in the \gls{EMCal}.
The impurity is driven by exotic clusters, by regions in the detector that were masked in the cluster reconstruction but were active in the trigger system, and by residual energy decalibration for the trigger energy estimates. 
The difference in purity between \gls{EMCal} and \gls{DCal} can be attributed to a larger noise contribution in the \gls{DCal}.

The trigger performance of the jet trigger is studied using jets reconstructed from calorimeter clusters. 
Due to the presence of the \gls{PHOS}\com{ hole} on the \gls{DCal} side (see Figs.~\ref{fig:0-Intr-Schematics} and~\ref{fig:1-HW-EMCal_overview_etaphi}), the performance is only characterized in the \gls{EMCal} acceptance. 
As the size of the 16x16 FastOR jet patch used in pp and p--Pb collisions corresponds roughly to the size of $R$ = 0.2 jets, those jets are used for the performance evaluation.
\begin{figure}[t]
    \centering
    \includegraphics[height=6.2cm]{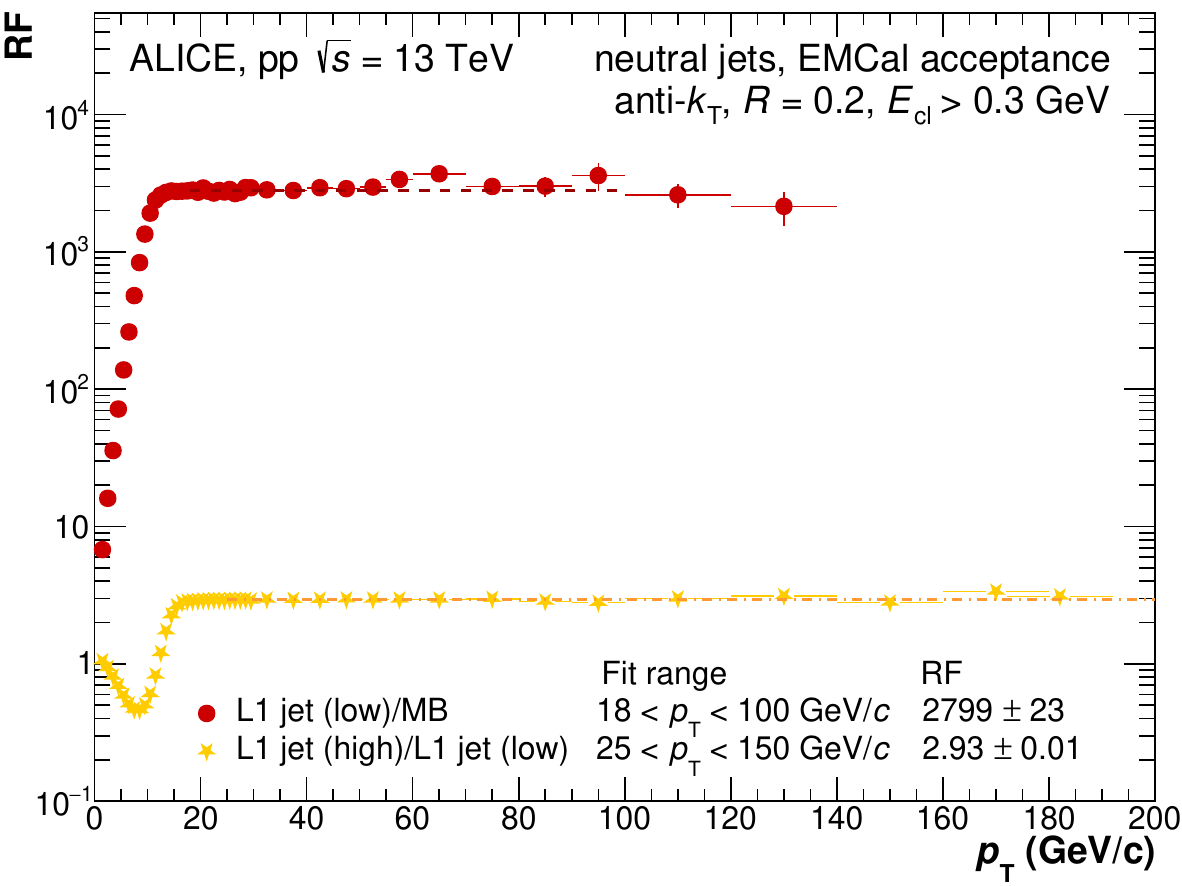}
    \includegraphics[height=6.2cm]{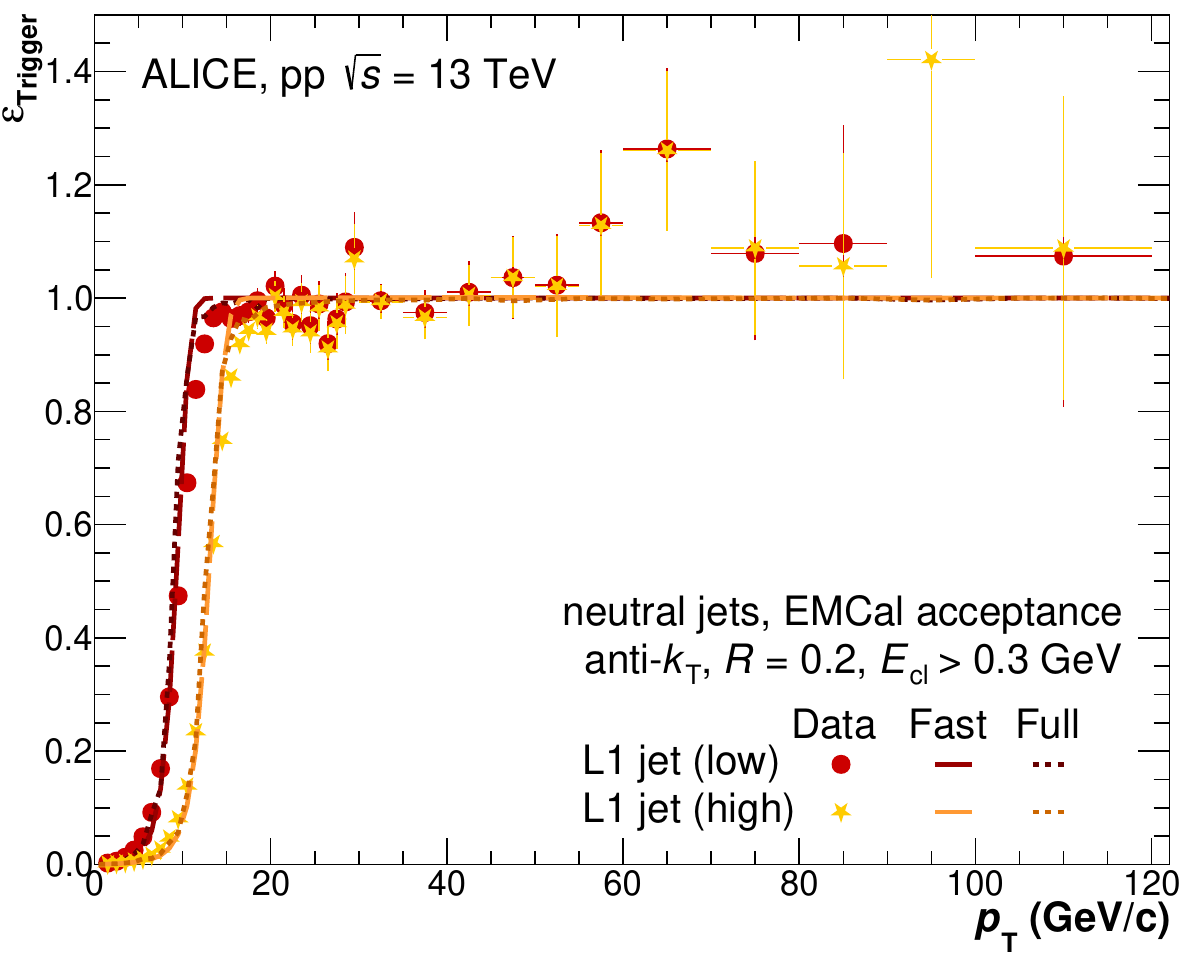}
    \caption{(Color online) Left: trigger rejection of the jet triggers obtained from calorimeter-based jets with $R$ = 0.2 in \pp{} collisions at \sthirteen{} collected in 2017 and 2018. Ratios are with respect to minimum-bias events (red) or to events triggered by the low-threshold jet trigger (yellow). Right: corresponding trigger efficiency of the jet triggers in \pp{} collisions at \sthirteen{} obtained with fast simulations on cell level and with full simulations including the trigger response.}
    \label{fig::trigger::turnonCaloJetsPP13TeV}
\end{figure}
\begin{table}[t]
    \centering
    \caption{Trigger \gls{RF} of different jet triggers in \pp{} and \pPb{} collisions at various center-of-mass energies.}
    \begin{tabular}{llrrr}
        \toprule
        System & Collision & Year & \multicolumn{2}{c}{\gls{RF}} \\
               & energy    & & \gls{L1} jet (low) & \gls{L1} jet (high)\\ \midrule
        \pp{} & \seight & 2012 & - & $5144 \pm 5\%$\\ 
              &  \sthirteen & 2017-2018 & $2799 \pm 0.8\%$ & $8201 \pm 0.9\%$\\ \midrule
        \pPb{} & \sfivelead & 2013 & $269 \pm 1.9\%$ & $6358 \pm 2\%$ \\
               & \seightplead & 2016 & $1621 \pm 2.7\%$ & $3568 \pm 2.7\%$  \\
        \bottomrule
    \end{tabular}
    \label{tab::trigger::rejectionJet}
\end{table}
\Figure{fig::trigger::turnonCaloJetsPP13TeV}~(left) shows the trigger rejection for the jet triggers in \pp{} collisions at \sthirteen{}. 
The $R$ = 0.2 jets were reconstructed with the anti-${k}_{\rm t}$ algorithm from FastJet~\cite{Cacciari:2011ma, Cacciari:2008gp}, using only calorimeter clusters in the \gls{EMCal} acceptance as jet constituents. 
Jets are required to be fully contained within the \gls{EMCal} fiducial acceptance.
Similar to clusters in the case of the single-shower triggers, the jet trigger rejection becomes approximately constant for \pt{} above the threshold. 
The ratio between the low- and the high-threshold trigger can be used to determine the trigger rejection factor of the high-threshold trigger with a better precision. 
The resulting trigger \glspl{RF} are listed in \Tab{tab::trigger::rejectionJet}.

\begin{figure}[t]
    \centering
    \includegraphics[width=0.7\textwidth]{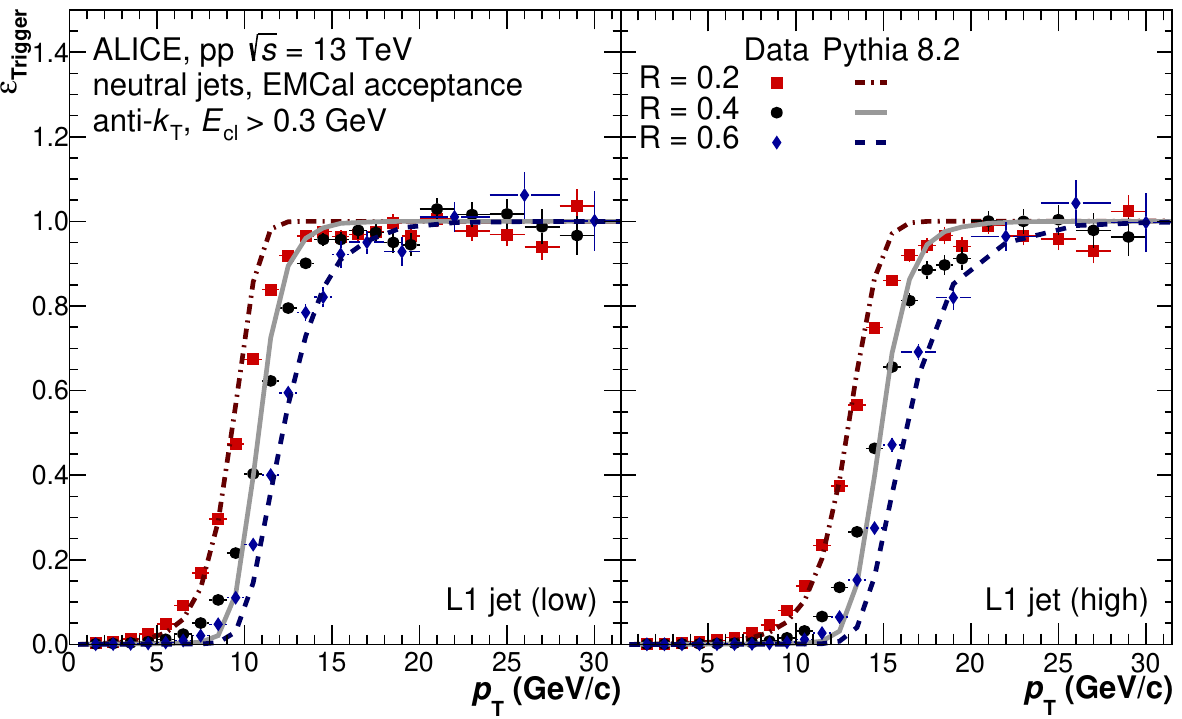}
    \caption{(Color online)  Trigger efficiency for different jet resolution parameters for the low-threshold (left panel) and high-threshold (right panel) trigger for calorimeter based jets as a function of their transverse momentum. }
    \label{fig::trigger::efficiencyCaloJetsPP13TeV}
\end{figure}

In order to determine the trigger efficiency, the jet spectra and the number of triggers are corrected by the pre-scaling of the trigger in the same way as for the single shower triggers. \Figure{fig::trigger::turnonCaloJetsPP13TeV} (right) shows the trigger efficiency for calorimeter-based jets with a jet resolution parameter $R$ = 0.2 for $0~<~p_{\rm{T}}~<~120~{\rm GeV}/c$. 
Due to the large size of the jet patch, acceptance losses due to dead channels play a minor role at sufficiently high \pt{} and mostly lead to a broadening in the turn-on region, leading to a trigger efficiency converging at 1 at high \pt{}. 
Good agreement with simulation is observed in a wide range of \pt{}.

In the turn-on region a sensitivity of the trigger efficiency on the jet resolution parameter is expected from the energy distribution within a jet. 
Due to the fixed patch size the jet trigger only measures a fraction of the jet energy for jets with $R > 0.2$. 
Correspondingly, the jet \pt{} at which jets are fully efficiently selected by the trigger increases with increasing $R$.
\Figure{fig::trigger::efficiencyCaloJetsPP13TeV} shows the trigger efficiency for calorimeter-based jets with different resolution parameters ranging from $R=0.2$ to $R=0.6$ for the low~(left panel) and high~(right panel) thresholds. 
The trigger efficiency is approximately 1 beyond 10 GeV for the low threshold and 15 GeV for the high threshold for $R$ = 0.2. 
Those are significantly lower than the thresholds applied in hardware at 20 (16) GeV for the high (low) threshold. 
Due to the large patch size, the jet trigger is more sensitive to noise in the trigger system. 
Therefore, a small noise contribution can lead to a sizable shift of the turn-on. 
The trigger efficiency is described by simulation in which we assume a random noise component for each FastOR with a width of 50~MeV, corresponding to 1~ADC count.

\begin{figure}[t]
    \centering
    \includegraphics[width=0.6\textwidth]{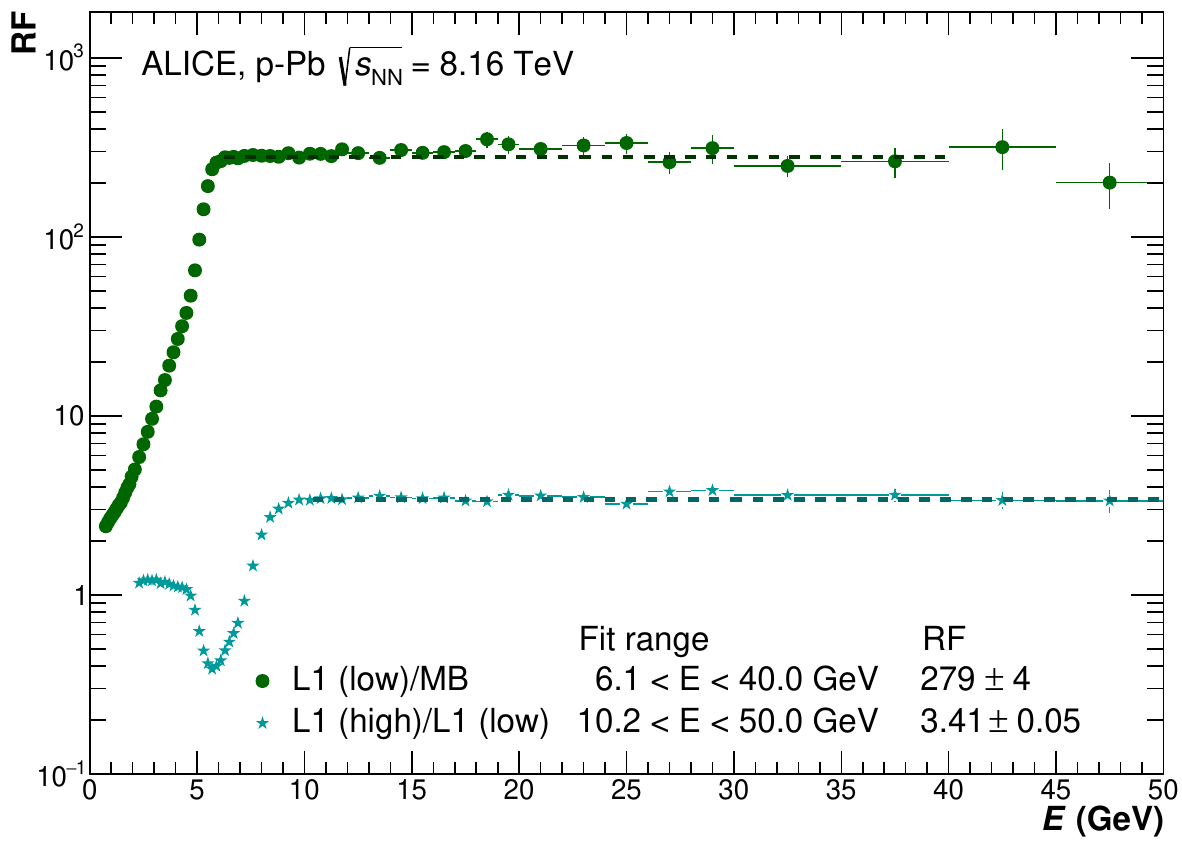}
    \caption{(Color online) Trigger rejection for single-shower triggers in \pPb\ collisions at \seightplead, obtained from cluster energy spectra.}
    \label{fig::trigger::rejectionpPb}
\end{figure}

\Figure{fig::trigger::rejectionpPb} shows the trigger \gls{RF} for single shower triggers obtained from the cluster energy spectra in \pPb{} collisions at \seightplead{}. 
Due to the larger event activity, in \pPb\ collisions the rejection is almost a factor of 2 smaller than in \pp\ collisions for similar thresholds.

\begin{figure}[t]
    \centering
    \includegraphics[width=0.6\textwidth]{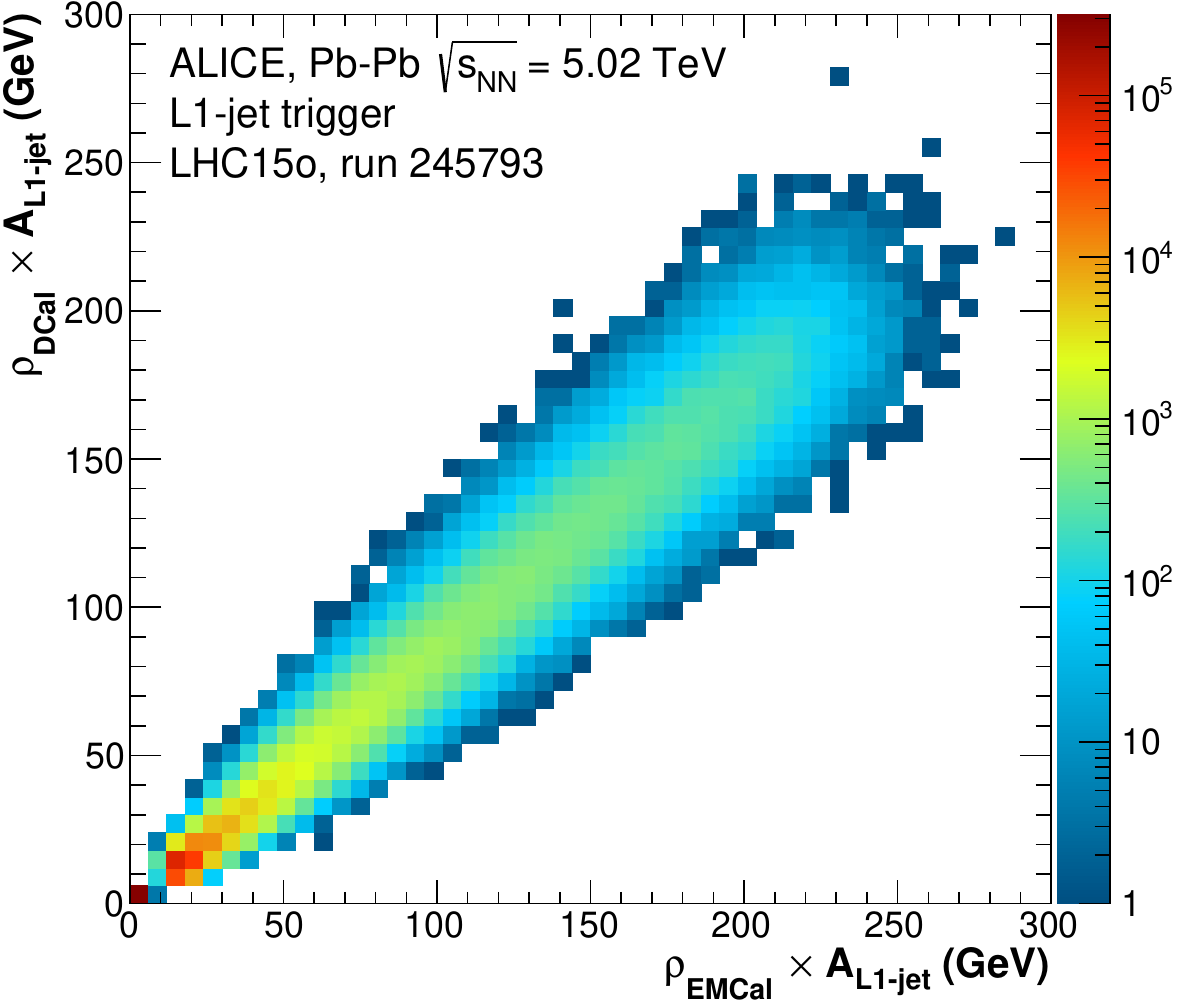}
    \caption{Correlation of the average energy density $\rho$ in \gls{EMCal} and \gls{DCal} scaled with the area of the L1-jet patch, determined with the L1 trigger electronics}
    \label{fig::trigger::PbPb_centrality}
\end{figure}

\begin{figure}[t]
  \centering
  \includegraphics[height=6.1cm]{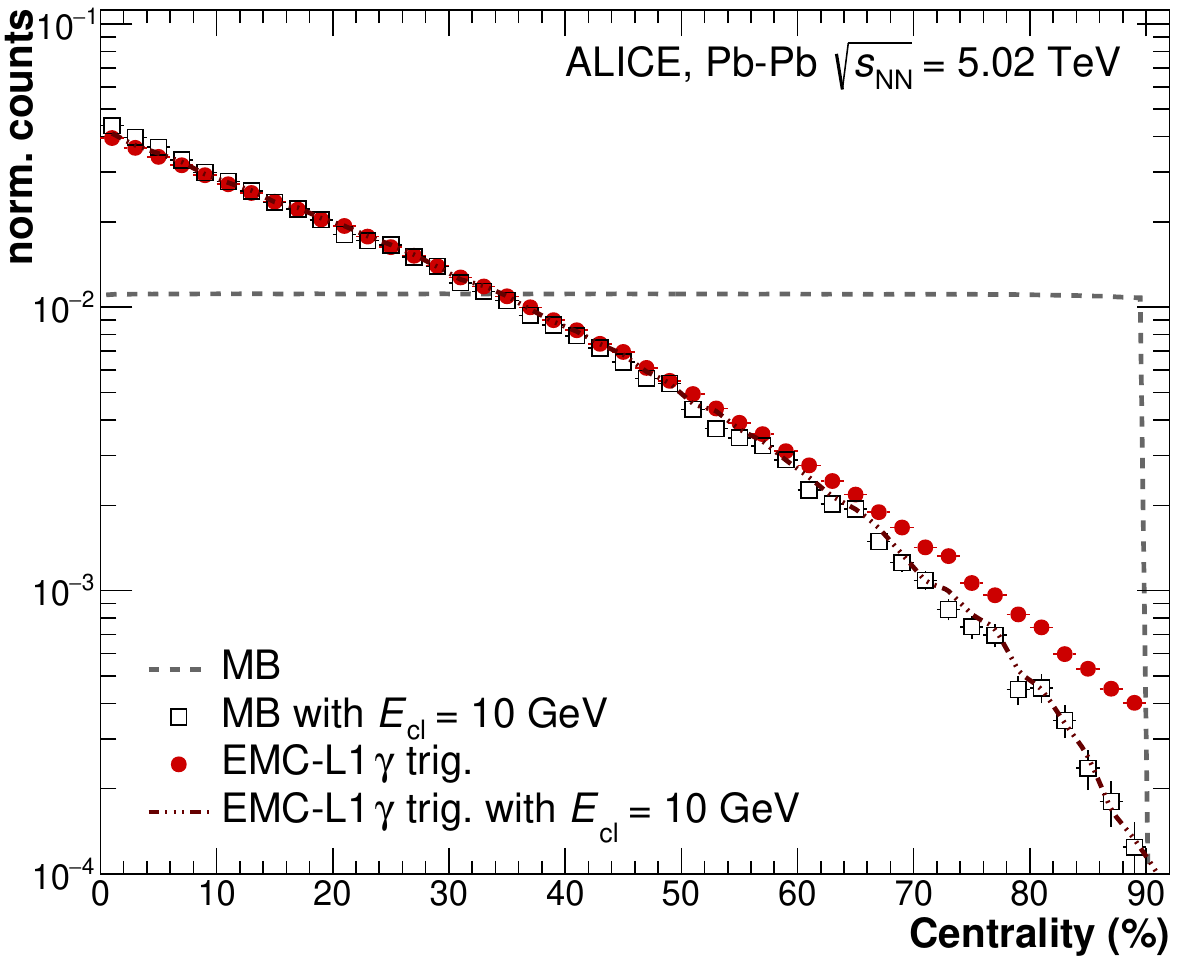}
  \includegraphics[height=6.1cm]{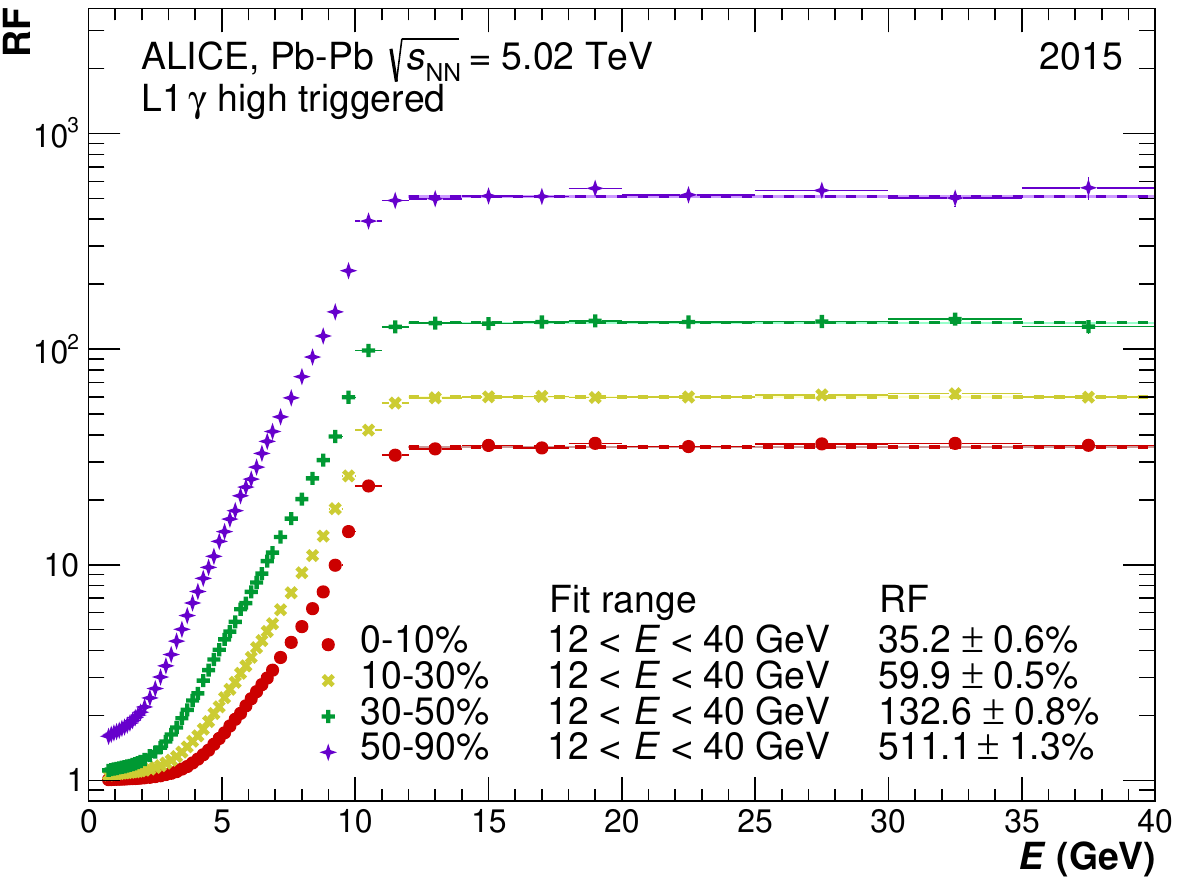} 
  \caption{(Color online) Left: Centrality percentile distribution of \gls{EMCal} \gls{L1} $\gamma$ triggered events (red) in comparison to the pure minimum bias distribution and minimum bias triggered events with a 10 GeV cluster in the event. 
                          Right: Trigger \glspl{RF} for the \gls{EMCal} or \gls{DCal} \gls{L1} $\gamma$ triggers in different centrality classes for \PbPb\ collisions at $\snn= 5.02$~TeV. 
                          Only statistical uncertainties of the trigger \gls{RF} are given in the legend.}
  \label{fig::trigger::PbPb_GA}
\end{figure}


\Table{tab::trigger::SetupRun} also lists the trigger setup used during the \PbPb{} data taking. 
In the 2018 \PbPb{} run, the low threshold $\gamma$ trigger was additionally applied to peripheral events as selected by the V0 centrality trigger. 
In heavy-ion collisions, a background subtraction algorithm based on the average energy density in calorimeter on the opposite side of the triggering subdetector was implemented in Run~2 in order to reduce the sensitivity of the trigger on the energy from the underlying event. \Figure{fig::trigger::PbPb_centrality} shows the correlation of the average energy density $\rho$ measured with the two subdetectors, \gls{EMCal} and \gls{DCal}, scaled by the area of the L1-jet patch, which is used for the background subtraction. 

The $\gamma$ and jet triggers are expected to bias the centrality distribution of the triggered events by selecting preferentially more central events because the hard processes producing high-energy triggers scale with the number of binary collisions.
The effect can be demonstrated by comparing the centrality distribution from events triggered by the \gls{L1} $\gamma$ trigger to the centrality distribution in events triggered by the minimum-bias trigger, which are required to have at least one cluster with an energy above the trigger threshold.
This comparison is shown \Fig{fig::trigger::PbPb_GA}~(left). 
Without the cluster requirement, the centrality percentile distribution is approximately constant in minimum-bias events.  
When requiring at least one cluster over the trigger threshold, the centrality distribution is in qualitative agreement with that obtained from \gls{EMCal} \gls{L1} $\gamma$-triggered events.
Remaining differences in the centrality distributions from pure \gls{L1}-triggered events and minimum-bias events including the cluster requirement are found for the most peripheral events, dominantly in the 60-90\% centrality class. 
Impurities in the trigger resulting from noise or exotic clusters as discussed above have a larger effect in more peripheral collisions with lower multiplicities. 
Therefore, a larger event count is seen in triggered events for peripheral collisions than what is expected from minimum-bias events with the cluster requirement. 
After rejecting impurities by applying the same cluster condition also in triggered events (red dashed line) the expected centrality distribution is obtained.

\Figure{fig::trigger::PbPb_GA} (right) shows the \gls{RF} of the combined \gls{L1} $\gamma$ trigger in different bins of centrality. 
The rejection factor is the smallest for the 10\% most central collisions and increases towards more peripheral events.
\begin{figure}[t!]
  \includegraphics[width=0.48\textwidth]{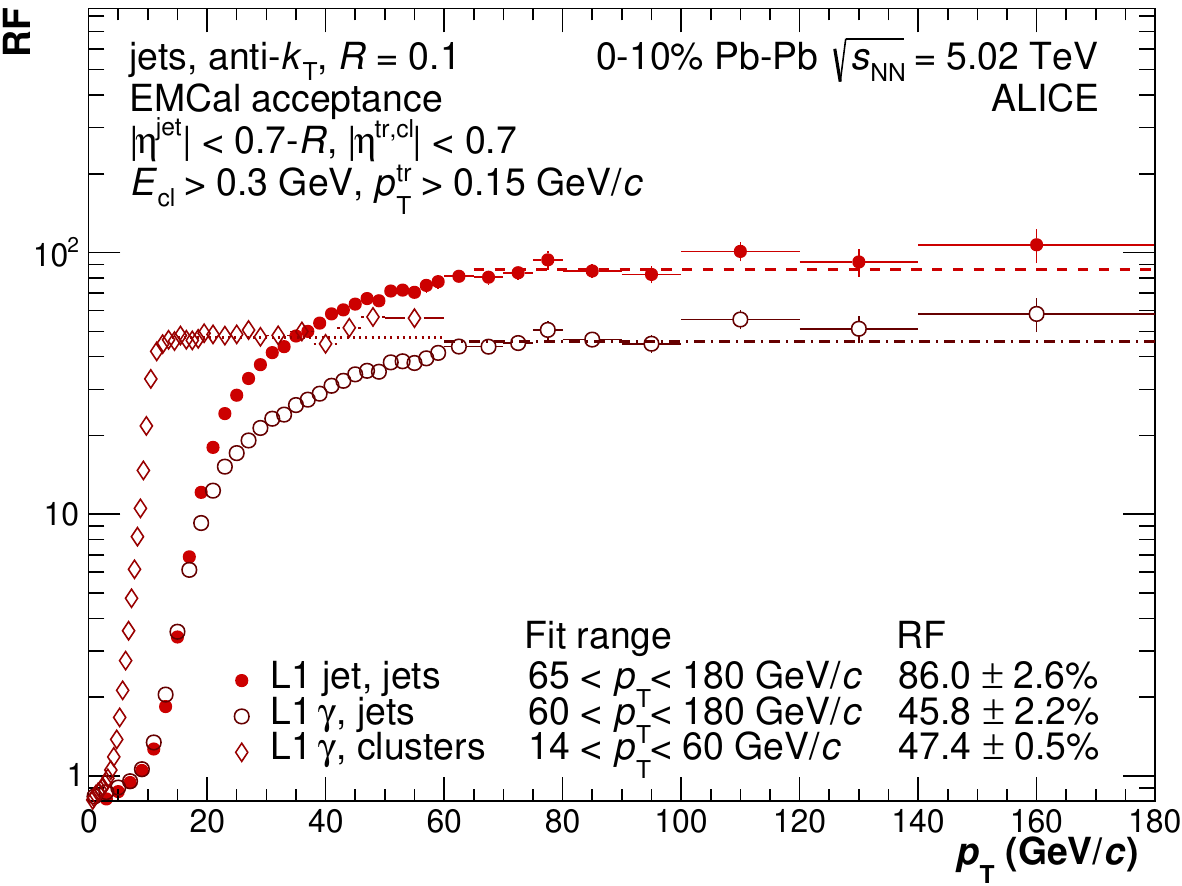}
  \hspace{0.2cm}
  \includegraphics[width=0.48\textwidth]{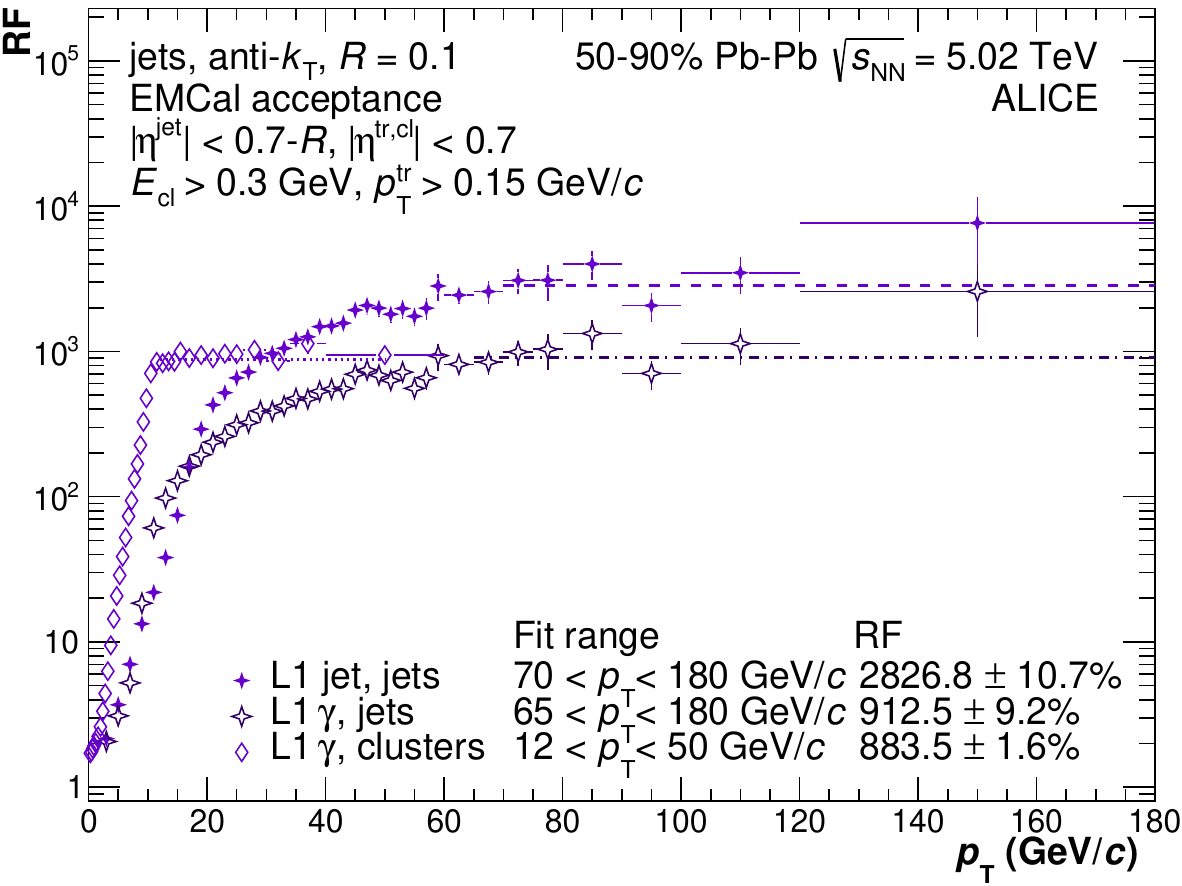}
  \caption{(Color online) Comparison of the trigger \glspl{RF} based on clusters and full jets for the \gls{EMCal} \gls{L1} $\gamma$ and jet trigger in 0--10\% (left) and 50--90\%~(right) central \PbPb\ collisions at \sfivelead. 
            Only statistical uncertainties of the trigger \gls{RF} are given in the legend. } 
  \label{fig::trigger::PbPb_JE}
\end{figure}
In addition to the \gls{L1} $\gamma$ trigger the \gls{L1} jet trigger was used in \PbPb\ collisions at \sfivelead\ as well.
A comparison of the resulting \glspl{RF} for reconstructed jets between the \gls{L1} $\gamma$ and \gls{L1} jet trigger can be seen in \Fig{fig::trigger::PbPb_JE} for central and peripheral collisions.
The clusters and jets were reconstructed in the \gls{EMCal} acceptance with a jet radius of $R = 0.1$, which is close to the area of the patch size of 8x8 FastORs used in Pb--Pb collisions, in order to have the least bias from fluctuations of the underlying event.
Both triggers appear to be fully efficient above jet momenta of about 70--80~GeV/$c$ with the \gls{L1} $\gamma$ reaching its maximum efficiency slightly earlier.
The \gls{L1} jet trigger is, however, more effective and thus has a higher \gls{RF}.
For nearly all \gls{L1} jet-triggered events, a coincidence with the \gls{L1} $\gamma$ trigger on the same side was observed, indicating that the \gls{L1} jet trigger is more selective in heavy-ion collisions. 
For the \gls{L1} $\gamma$ trigger the \glspl{RF} obtained from the cluster and jet spectra agree within their statistical uncertainties, as expected, when restricting the cluster acceptance to the \gls{EMCal} as well. 
This means that most of the jets beyond \pT\ = 80~GeV/$c$ contain at least one cluster above the trigger threshold of about $10$~GeV.
 
\subsection{Data quality assurance}
\label{sec:dataQA}
The \gls{EMCal} offline 
\gls{QA} tools are integrated into the general \gls{ALICE} offline \gls{QA} framework.
The goal of the \gls{QA} process is to provide immediate feedback on the data quality, enabling the determination of good run lists for analyzers, and detecting and classifying anomalies. 
If anomalies are detected, a dedicated calibration may be necessary.

A fast reconstruction of the \gls{EMCal} data is done immediately after data taking~(within a few hours). 
These first data quality checks are particularly important, as issues discovered at this stage can be fixed during data taking.
The \gls{QA} is based on automatic post-processing of detector specific data produced by algorithms running at the end of the reconstruction.  
The post processing produces run-by-run energy distributions of each cell, clusters, trigger information, and correlations to other detectors. 

\begin{figure}[t]
\centering
\includegraphics[width=0.49\textwidth]{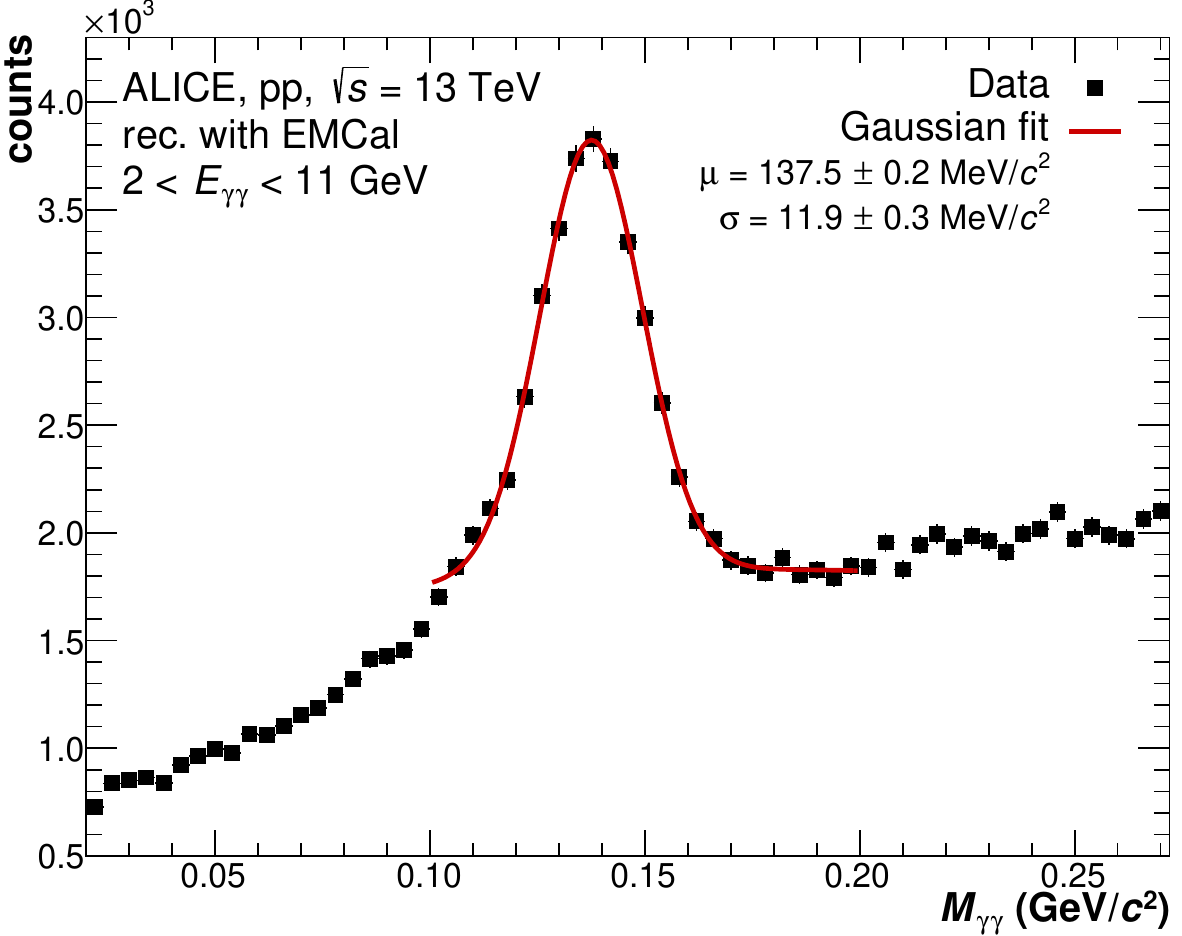}
\includegraphics[width=0.49\textwidth]{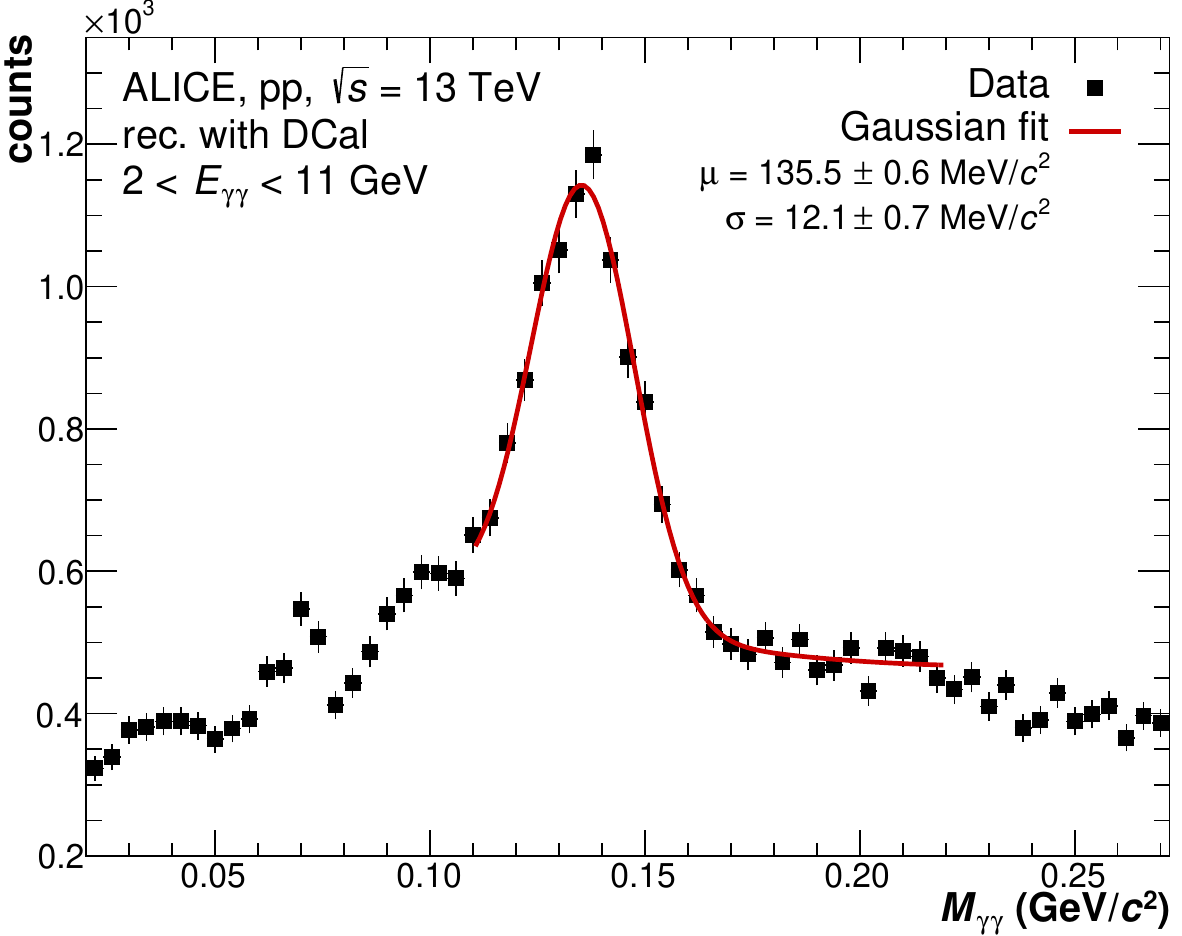}
\caption{\label{fig:2-Dat-QA_OnlinePi0} (Color online) Raw distribution of the invariant mass of cluster pairs in \gls{EMCal} (left) and \gls{DCal} (right) for one run of 2018 data taking obtained during the \gls{QA} process. 
        The red line corresponds to a fit to the invariant mass distribution with a Gaussian function for the \piz\ signal and a second-order polynomial for the background. 
        The fit parameters are used to monitor the performance of the reconstruction and of the detectors. In the displayed run, 819535 events were collected.
        }
\end{figure}

The invariant mass distribution of cluster pairs is shown in \Fig{fig:2-Dat-QA_OnlinePi0} as a concrete example of this type of \gls{QA}, with the \gls{EMCal} on the left and the \gls{DCal} on the right.
The \piz\ meson mass peak is fit for both distributions, allowing for straightforward assessment of the data quality by observing the stability of the peak position and width values.
Most of these quantities are collected per \gls{SM} to enable a more detailed characterization.
All the results are automatically posted to a web-based repository. 

The stability of the \gls{EMCal} and \gls{DCal} performance is monitored over longer periods covering several runs, from a few tens to a few hundred,  with the same detector conditions to check for overall deviations from normal operation.
These checks are done on average quantities extracted from the run-by-run \gls{QA}:  the mean values and dispersion of the number of cells with signal per event, the number of clusters per event,  the mean number of cells per cluster, the position of the \piz\ invariant mass peak.
The number of charged tracks reconstructed in the \gls{ITS} and \gls{TPC} which are associated to a cluster is also monitored for stability.
All trending plots are systematically inspected for both minimum-bias and \gls{EMCal}-triggered data at each reconstruction step and for each data-taking period to identify outliers. 
These checks are particularly important to identify run ranges for which different calibrations or bad channel determinations are needed. 

\begin{figure}[t]
    \centering
    \includegraphics[width=0.65\textwidth]{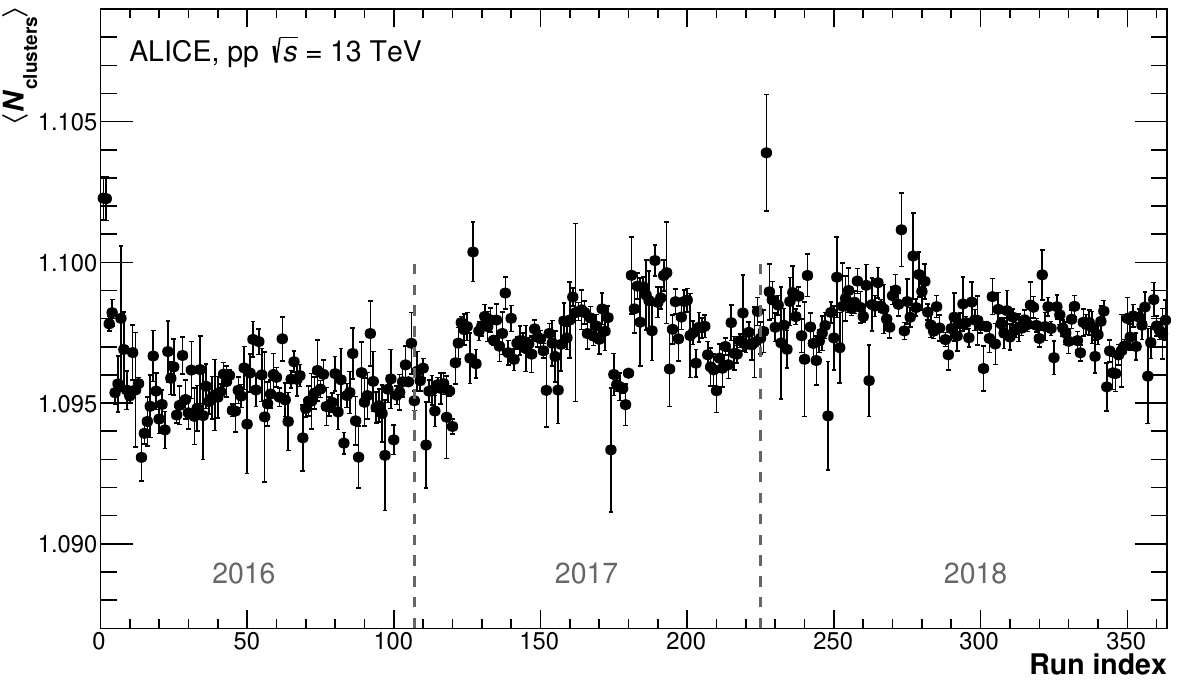}
    \caption{
        \label{fig:3-Dat-QA_trendingClusterNb} Mean number of clusters per event as a function of the run index for \pp{} collisions at \sthirteen{}. Example runs with similar data taking conditions  are  displayed.
        Only clusters with energy above 0.5 GeV were used for the mean estimation. 
        The gray vertical lines correspond to the start of different data-taking years.}
    \centering
    \includegraphics[width=0.49\textwidth]{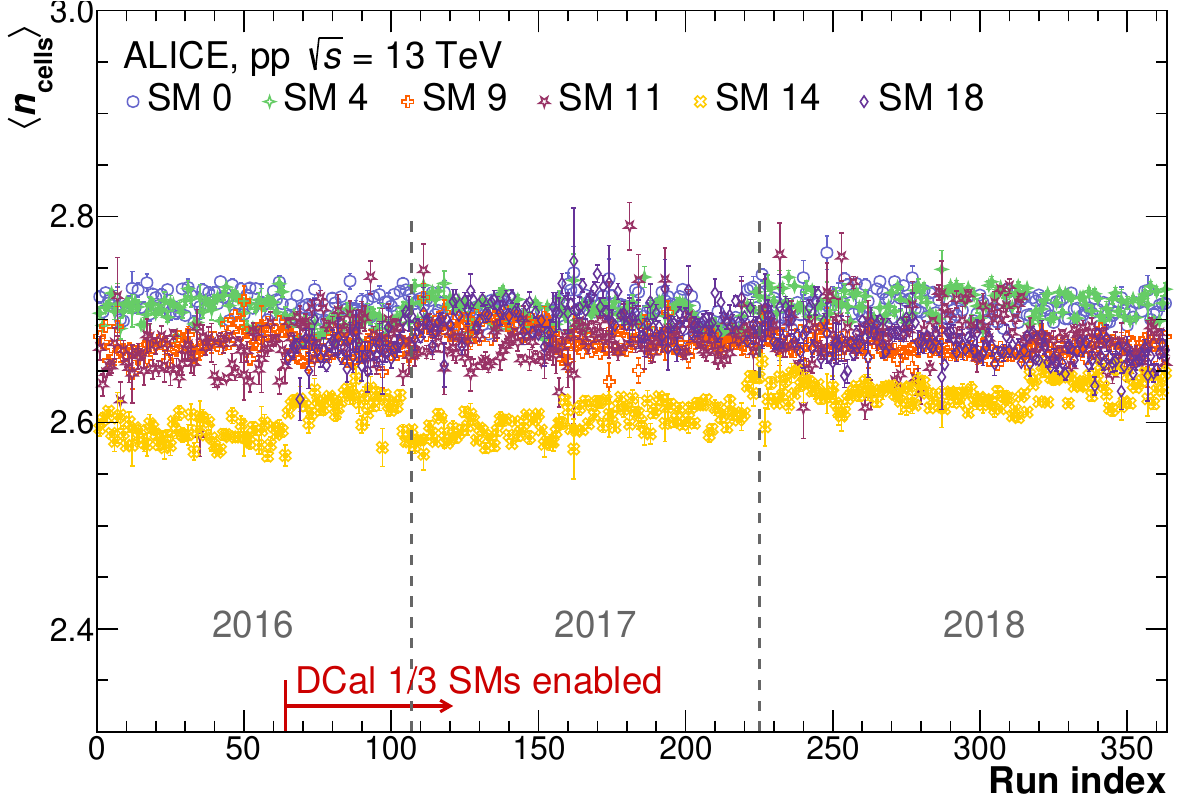}
    \includegraphics[width=0.49\textwidth]{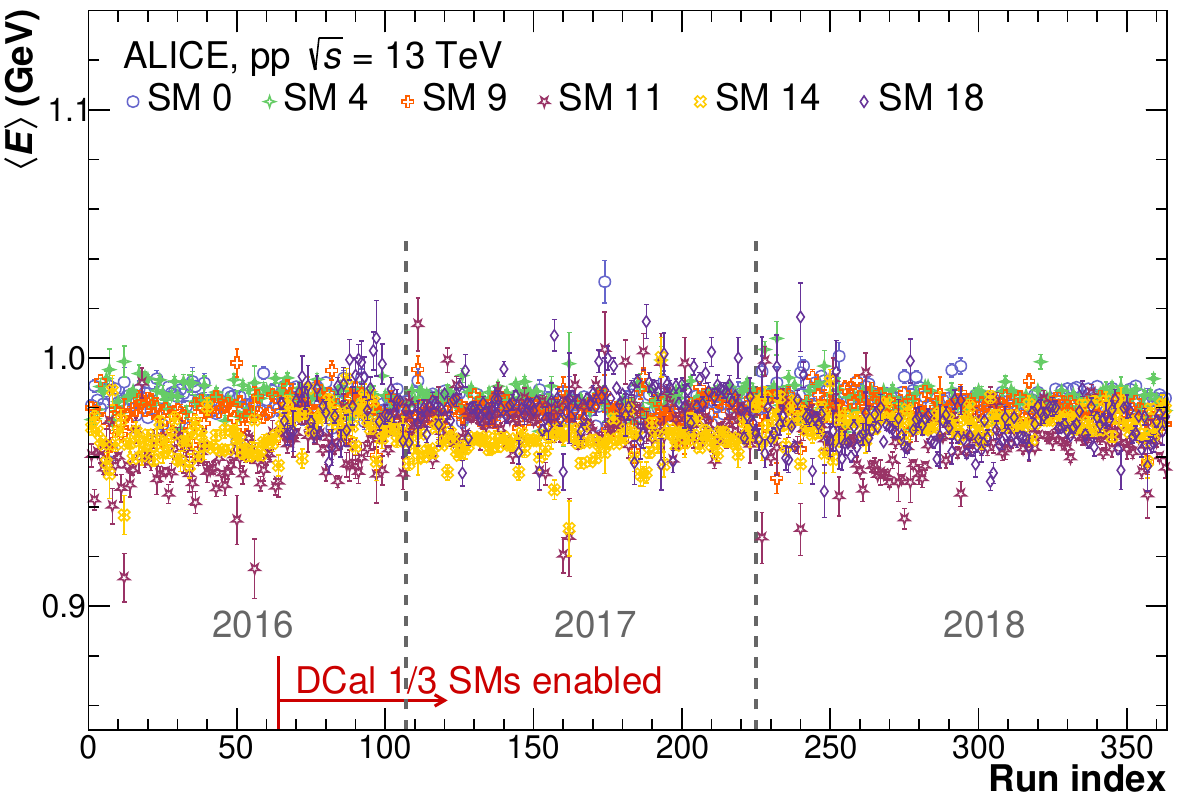}
    \caption{\label{fig:4-Dat-QA_trendingClusterCells} \label{fig:5-Dat-QA_trendingClusterE} (Color online) Mean number of cells per cluster (left) and mean cluster energy (right) for a selection of \glspl{SM} as a function of the run index in \pp{} collisions at \sthirteen{}. Example runs with similar data taking conditions  are  displayed. Only clusters with energy above 0.5 GeV were used for the mean estimation. The vertical red line indicates the run at which the 2 last \gls{DCal}~\glspl{SM} were inserted into the readout. 
    The vertical gray lines correspond to the start of different data-taking years.}
\end{figure}

To illustrate the \gls{QA} process, the \gls{EMCal} and \gls{DCal} performances for \pp{} collisions at \sthirteen\ collected in 2016, 2017, and 2018 are presented in the following.
The average number of clusters per event as a function of run index is shown in \Fig{fig:3-Dat-QA_trendingClusterNb}. 
The mean was calculated using clusters with energy larger than 0.5 GeV to select signal-like clusters.  
For each data-taking year, this metric was generally stable within less than 1\%.

The mean number of cells per cluster and the mean cluster energy in minimum-bias events are presented on the left and right of \Fig{fig:4-Dat-QA_trendingClusterCells}, respectively.  
These quantities are presented for a selection of \glspl{SM}.
Although there is some variation in the mean number of cells per cluster among \glspl{SM}, the run-by-run performance for a single \gls{SM} is constant over time.
The mean cluster energy is consistent between \glspl{SM} and run-by-run.
At the beginning of the 2016 data taking (runs before the red vertical line), two \glspl{SM} from \gls{DCal}~were not included in the readout and the corresponding values (cells/cluster and mean energy/cluster) for \glspl{SM} 18 and 19 are zero. 

Collectively, these trending plots illustrate the stability of cluster performance over the 3 years of data taking for Run 2. 
For \pp{} collisions at \sthirteen{}, less than 10 \% of the data collected with the \gls{EMCal} was affected by anomalies spotted during the quality assurance process.
Most of the anomalies, corresponding to about 8\% of the collected data sample, are due to issues in the cell-time distributions.
These anomalies can be recovered via a specific time calibration, as described in \Sec{sec:timeCalib}.
The remaining issues, corresponding to less than 2\% of the collected sample, are mostly due to pedestal subtraction malfunction, or to cases in which large parts of the detector were disabled during data taking. 
 
\subsection{Online data-quality monitoring on the high-level trigger}
In an effort to further monitor the \gls{EMCal} performance in real time a monitoring system based on the capabilities of the \gls{HLT}~\cite{Fabjan:684651}, a large computing system performing immediate reconstruction during the data-taking process, was developed and deployed during Run 2.

In order to be able to operate on the \gls{HLT} it was necessary to develop an of interconnected synchronous and asynchronous processing components to handle \gls{EMCal} data reconstruction and processing~\cite{ALICE:2019jgt}.
These components consisted of a new reconstruction chain, separate from that used for standard offline reconstruction, to convert raw data into digits, clusters, and triggers.  
In order to meet the strict performance requirements, these components were purposely built for the \gls{HLT}. 
In particular, data were handled as flat structures minimizing the overhead in copy processes. 
Within the context of monitoring, the trigger reconstruction and clusterization components were especially important to enable monitoring of higher level and more complex information than would be otherwise possible. 
Dedicated quality assurance components ran asynchronously, extracting derived information from all the steps of the reconstruction to characterize the detector performance. 
Examples of monitored quantities are cluster spectra, trigger rates, and the comparison of the median energy of the trigger patch measured in the \gls{EMCal} versus \gls{DCal}.

The \gls{HLT} distributes events to be reconstructed to the components in a round-robin manner. 
The components process the data and then return the results to an \gls{HLT} merger component that then makes the data available for further processing and display, as described below. 
This design provides a time resolution on the order of minutes, with the time resolution scaling with the computation time, which itself is proportional to the event size. 
For \PbPb{} data-taking conditions, the \gls{EMCal} components could process up to 6000~events/s~\cite{ALICE:2019jgt}, resulting in the full bandwidth being inspected.

\begin{figure}[t]
    \begin{center}
    \includegraphics[width=0.7\textwidth]{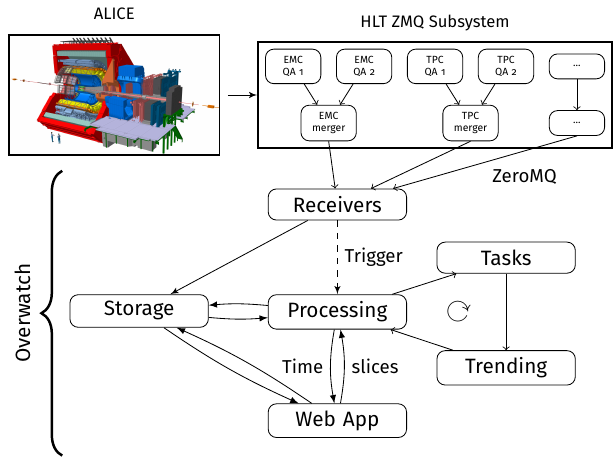}
    \end{center}
    \caption{(Color online) Data flow and architecture from the raw-data collection by \gls{ALICE} to the eventual graphical representation in the \gls{Overwatch} web application~\cite{Ehlers:2018ihn}.\label{fig:3-HLT-OverwatchArchitecture}}
\end{figure}
\begin{figure}[t!]
    \begin{center}
    \includegraphics[width=0.6\textwidth]{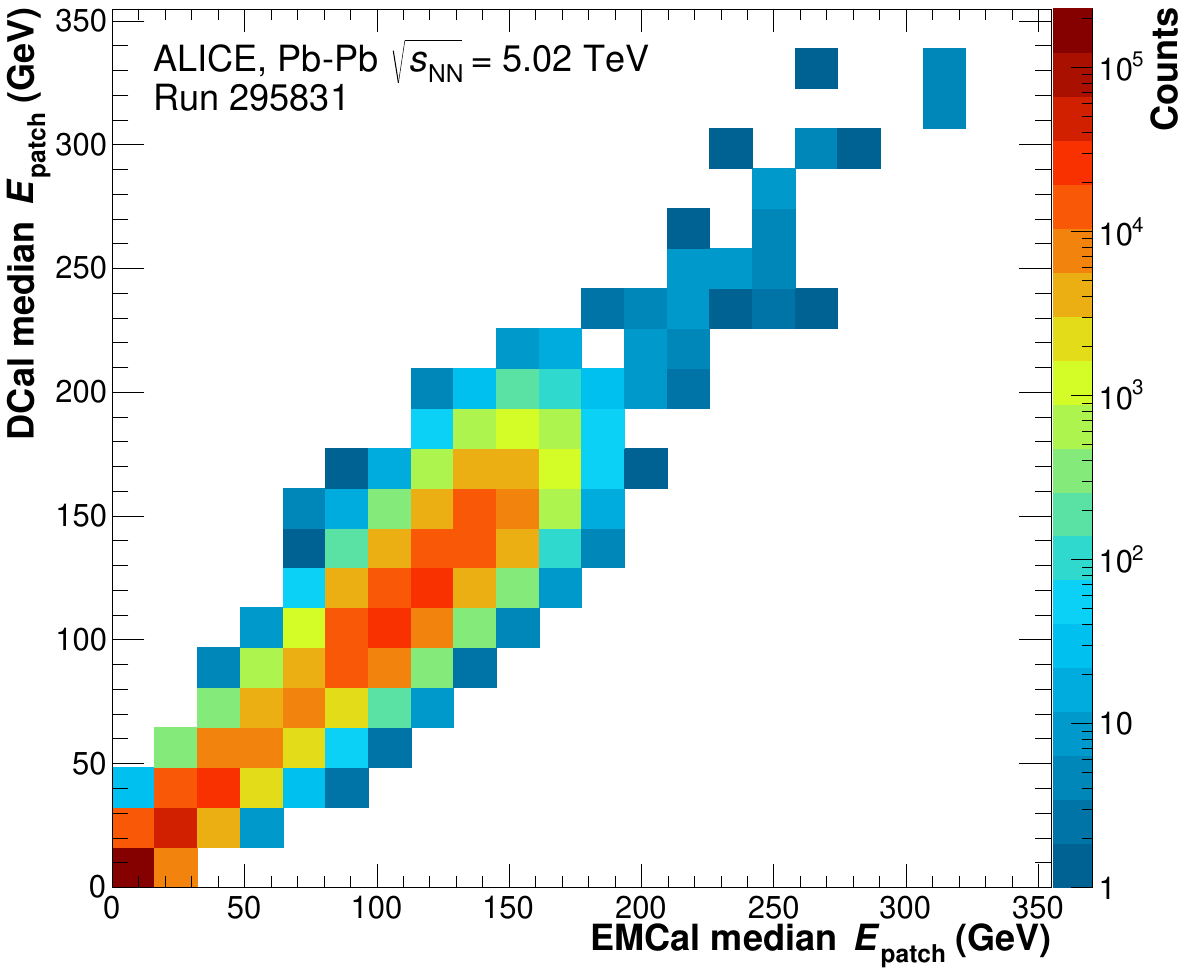}
    \end{center}
    \caption{(Color online) Comparison of \gls{EMCal} and \gls{DCal} median energies of the trigger patch from the online \gls{QA} on the \gls{HLT}. 
    The linear correlation indicates that both subdetectors are measuring similar event activity.\label{fig:3-HLT-MedianComparsion}}
\end{figure}

On top of the output from the \gls{HLT} merger, the \gls{Overwatch} project was developed and utilized to further process and display the \gls{QA} information~\cite{Ehlers:2018ihn}. 
It provided the capabilities to store and further contextualize the \gls{QA} data, extracting and trending values, projecting and plotting histograms, and adding additional information for graphical representation, such as locations of the \glspl{SM}. 
The data flow from collection by \gls{ALICE} through display in \gls{Overwatch} is represented in \Fig{fig:3-HLT-OverwatchArchitecture}. 
Information provided via the \gls{HLT} is stored long term, processed according to the specifications provided by the \gls{EMCal} specific module, and then displayed via the web application. 
Of particular interest for quality assurance is the ability for the user to select a particular time range for reprocessing via the web application, allowing time-dependent investigation of observed phenomena.

The flow of data from the \gls{EMCal} components on the \gls{HLT} through the merger to \gls{Overwatch} forms the same structure as a device within the $\mathrm{O}^{2}$~\cite{Buncic:2015ari} framework. 
The experience gained in developing this project aided in both the ongoing development of the \gls{EMCal} $\mathrm{O}^{2}$ framework, as well as in the broader development of quality control efforts for Run 3.
One example of the \gls{QA} provided through \gls{Overwatch} is the comparison of the \gls{EMCal}-\gls{DCal} median energy in the trigger patch shown in \Fig{fig:3-HLT-MedianComparsion}. 
The histogram, collected during the 2018 \PbPb{} data-taking period, shows a strong correlation between the median tower energy in both regions, indicating that both are measuring similar event activity, reflecting the event centrality.

\ifflush
\clearpage
\fi
\section{Test beam}
\label{sec:testeam}
\label{sec:TestBeam} 
\subsection{Test beam setup}
\label{sec:testbeamdetails}
An \gls{EMCal} mini-module of $8 \times 8$ towers built according to the final design of the production modules was tested in the summer of 2010 at the \gls{CERN} \gls{PS} and \gls{SPS} facilities using the same detector configuration as installed in \gls{ALICE}. 
Production versions of the \gls{EMCal} \gls{FEE} boards and the final \gls{LED} monitoring system (\Sec{sec:hardware}) were used during these tests.
Beam tests were performed also with earlier (not final) versions of the \gls{FEE}  at FNAL in 2005 and at \gls{CERN} in 2007, and the results are summarised in Ref.~\cite{Allen:2009aa}.\\
%
All towers were scanned with beams of electrons, muons and hadrons in order to investigate the response of the \gls{EMCal} to these particles. 
As discussed in \Sec{sec:hardware-moduledesign}, the modules in \gls{ALICE} have a $1.5^{\rm {o}}$ taper in the $\eta$ direction to provide an approximately projective geometry.
During the scan, the \gls{EMCal} mini-module was placed on a movable table, allowing to select the position of incidence of the beam on the mini-module. The scan of each tower was performed such that the beam hit the tower surface perpendicularly.

\begin{figure}[t!]
\begin{center}
\includegraphics[width=14cm]{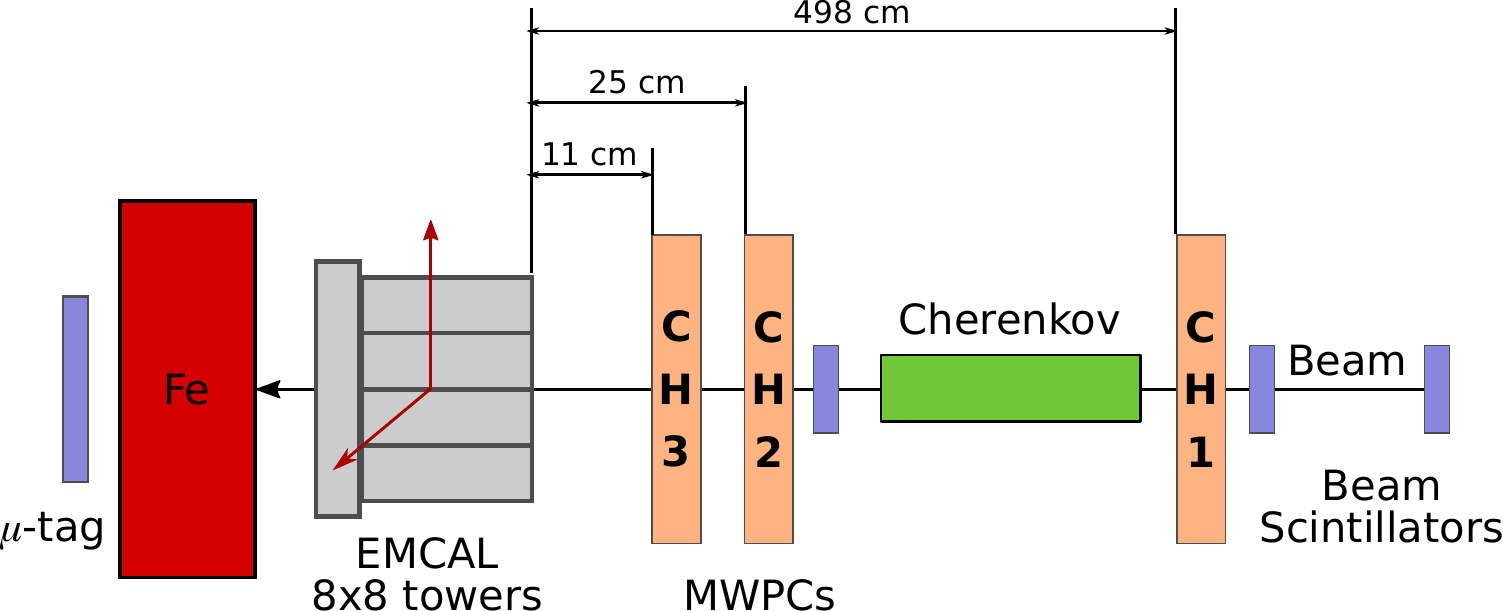}
\caption{(Color online) Schematic view of the \gls{ALICE} \gls{EMCal} mini-module at the \gls{PS} T10 beam line. 
                        The beam enters from the right. 
                        The Cherenkov detector was used for identification of the beam particle. 
                        The mini-module could be moved in the directions indicated by the red arrows in order to scan different towers.}\label{fig:5-Tes-psSetup}
\end{center}
\end{figure}

The \gls{PS} accelerator at \gls{CERN} accelerates protons up to 25~GeV. 
The protons impinge on a production target to produce electrons and hadrons with an approximately exponential energy distribution. 
The momentum of the produced particle is selected by a system of magnets and collimators. 
In this study, the final momentum selection collimator was typically set to achieve a $0.5\%$ selection on the momentum. 
The test beam was not sharply focused spatially in order to investigate a large area of the \gls{EMCal} for each position setting of the \gls{EMCal}  mini-module. 
For each configuration, typically about  $8$--$10$ adjacent towers gave signals. 
Electrons and hadrons within an energy range from 0.5 to 6~GeV were studied during the \gls{PS} beam period. 
The setup used at the T10 beam line of \gls{PS} is shown in \Fig{fig:5-Tes-psSetup}.
A threshold Cherenkov detector was used in the \gls{PS} experimental setup in order to discriminate between electrons and hadrons~(mostly pions) in the mixed beams of the \gls{PS}.
The T10 beam line at the \gls{PS} was recommended for operation with a minimum beam momentum of 1 \GeVc.
In order to operate at lower beam momenta of 0.75 \GeVc, the magnet settings were extrapolated.
Thus, for the lower energy, a 1\% uncertainty was assigned because of this extrapolation. 

\begin{figure}[t!]
\begin{center}
\includegraphics[width=13cm]{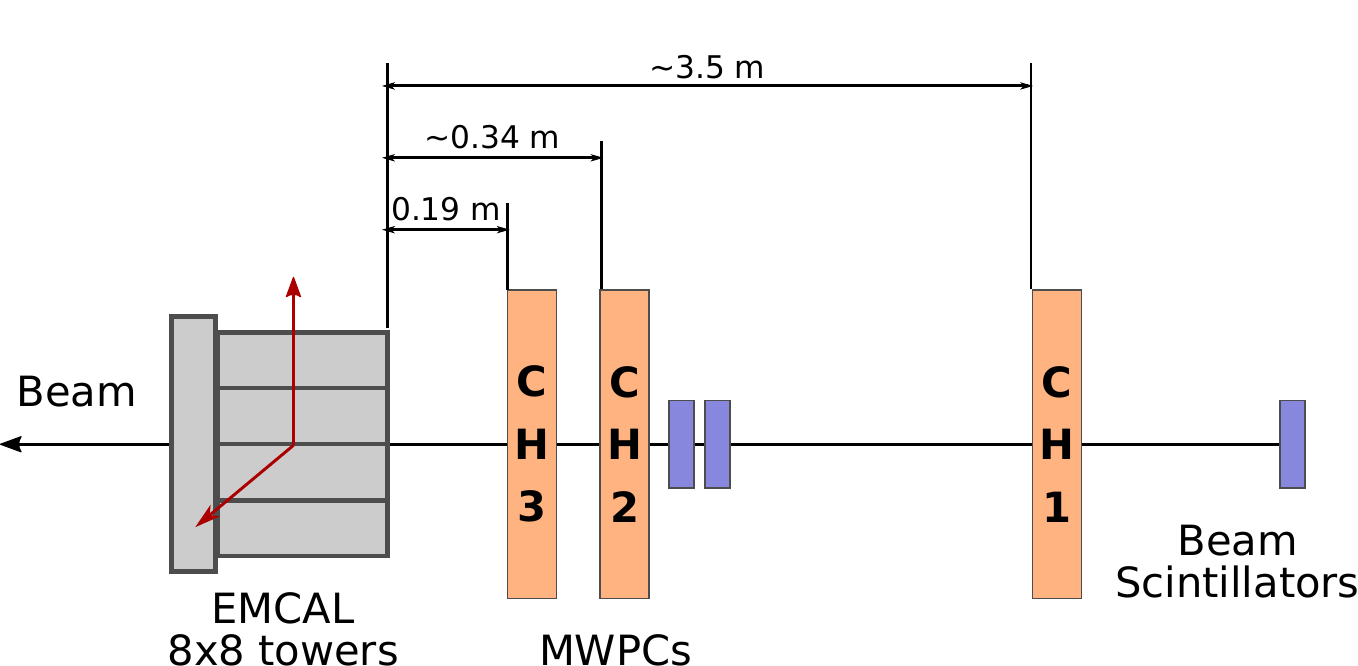}
\caption{(Color online) Schematic view of the \gls{ALICE} \gls{EMCal} mini-module at the \gls{SPS} H4 beam line. 
                        The beam enters from the right. 
                        The mini-module could be moved in the directions indicated by the red arrows in order to scan different towers.}\label{fig:5-Tes-spsSetup}
\end{center}
\end{figure}

In addition, data were taken with the same mini-module setup in the H4 beam line at the \gls{SPS}. 
The \gls{SPS} operates at a maximum beam energy of 450~GeV, allowing higher energy particles to be used. The electron and hadron energies studied with the mini-module at the \gls{SPS} were in the range of $6$--$225$~GeV.  
In the H4 beam-line, clean electron beams were provided as a tertiary beam after photons (the secondary beam) from neutral pion decays, produced in the production target, were converted to electrons in a lead converter. 
The setup used during the \gls{SPS} data-taking is shown in \Fig{fig:5-Tes-spsSetup}.
For both test beam periods, scintillator counters provided the beam trigger. 
Three \gls{MWPC}~(indicated as CH1, CH2, CH3 in the figures) ensured that only a single track per event was registered and provided the position of incidence of the beam particle during the offline analysis.  


\subsection{Calibration and corrections} 
\label{sec:shaperNL}
The relative calibrations of the towers were determined with initial runs in which the movable platform was moved to scan all towers with the test beam. 
For the \gls{PS} period, the calibration runs were taken with 6~GeV electrons, while for the \gls{SPS} period the calibration runs were taken with 10~GeV electrons. 
For each tower, enough calibration data was taken to provide a statistical uncertainty of less than 1$\%$ on the reconstructed energy peak. 
Corrections for the temperature dependence of the \gls{APD} gains for each tower were applied on a run-by-run basis in the offline analysis. 
Because the electromagnetic showers spread over several towers, and the sharing of the energy between towers depends on the position of incidence on the tower, the reconstructed energy ($E_{\rm rec}$) must be determined as a sum over several towers with signals. 
As a result, the calibration factor for each tower must be determined by comparing the sum of energy deposits in several towers with the incident energy ($E_{\rm beam}$). 
To this purpose, the events were grouped according to the position of the tower with the highest energy deposit~(leading cell) and for each group a $\chi^2$ function was defined as
\begin{equation}
\chi_{k}^{2} = \frac{1}{N_{\rm event}} \sum_{\rm event} \left[ E_{\rm beam} - \sum_{i=1}^{\rm towers} \alpha_{i}\,{\rm ADC}_{i} - \beta E_{\rm agg}\right]^{2}, \label{TB_Chi2_PerTower}
\end{equation}
where $N_{\rm event}$ is the number of events in the given group, $\alpha_{i}$ is the energy calibration coefficient for the $i$-th tower, and $ADC_{i}$ is the measured pulse amplitude. 
The last term, introduced as a linear function of the minimum aggregation energy threshold per tower~($E_{\rm agg}$), was included for fixing the absolute energy scale. 
Both the $\alpha_{i}$ and the $\beta$ coefficients were found by minimising the (global) $\chi^2$ defined as a sum of $\chi_{k}^{2}$ for different leading cell positions and for several values of minimum aggregation energy threshold: 
 \begin{equation}
\chi{^2} = \sum_{{\rm LeadingTower},\, E_{\rm agg}} \chi_{k}^{2} \,.
\label{TB_Chi2_glob}
\end{equation}
In order to avoid the edge effects and to ensure that the electromagnetic shower is fully captured by the mini-module, it was required that the maximum energy deposit is in one of the four central towers (conditionally labeled by symbols A, B, C and D).
The events with a maximum tower energy of less than $E_\mathrm{seed} = 500$~MeV were discarded from the analysis in order to be compatible with the cluster selection used for analyses of \gls{ALICE} physics data, where the clusterizer is only applied for the so-called {\it seed} towers with an energy above this threshold (\Sec{sec:clusterization}).
Since this threshold is higher than the energy deposited by a \gls{MIP}, it results in better energy and position resolutions but affects the low energy clusters. 
Additional towers were used during the clusterization only if they had energies above $E_{\rm agg} \geq 50$~MeV. 
This energy corresponds to approximately 3~\gls{ADC} counts.
In addition, clusters made of single towers were excluded from the study in order to be consistent with the analysis of \gls{ALICE} data, where such clusters are rejected to suppress the noise and "exotic" clusters~(see \Sec{sec:exotics}). 
During the calibration procedure, $E_{\rm agg} = $ 50, 100 and 150~MeV values were used in \Eq{TB_Chi2_PerTower}, and the $\beta$ coefficients determined from \gls{PS} 6~GeV runs and from \gls{SPS} 10~GeV runs were found to be compatible with $\beta \simeq$ 3.4. 
Thus, in the case of requiring the minimum aggregation energy per tower to be at least $\sim$100~MeV, 
a deficit of about 3.4\% of the ``measured energy'' is expected for electrons with incident energy of $\sim$1~GeV. 

Since the beam energies used for energy calibrations are well below 16~GeV, in the operational region of high-gain regime, the $\alpha_{i}$ coefficients were only determined for the high-gain channels.
The calibration of low-gain channels was accomplished on a channel-by-channel basis by comparing the pulses in high- and low-gain channels from higher energy electron beam data. 
The high- and low-gain amplitudes were found to be well correlated with an average gain ratio of 16.3 and an RMS of 0.15.
 
\begin{figure}[t]
  \centering
  \includegraphics[height=6.5cm]{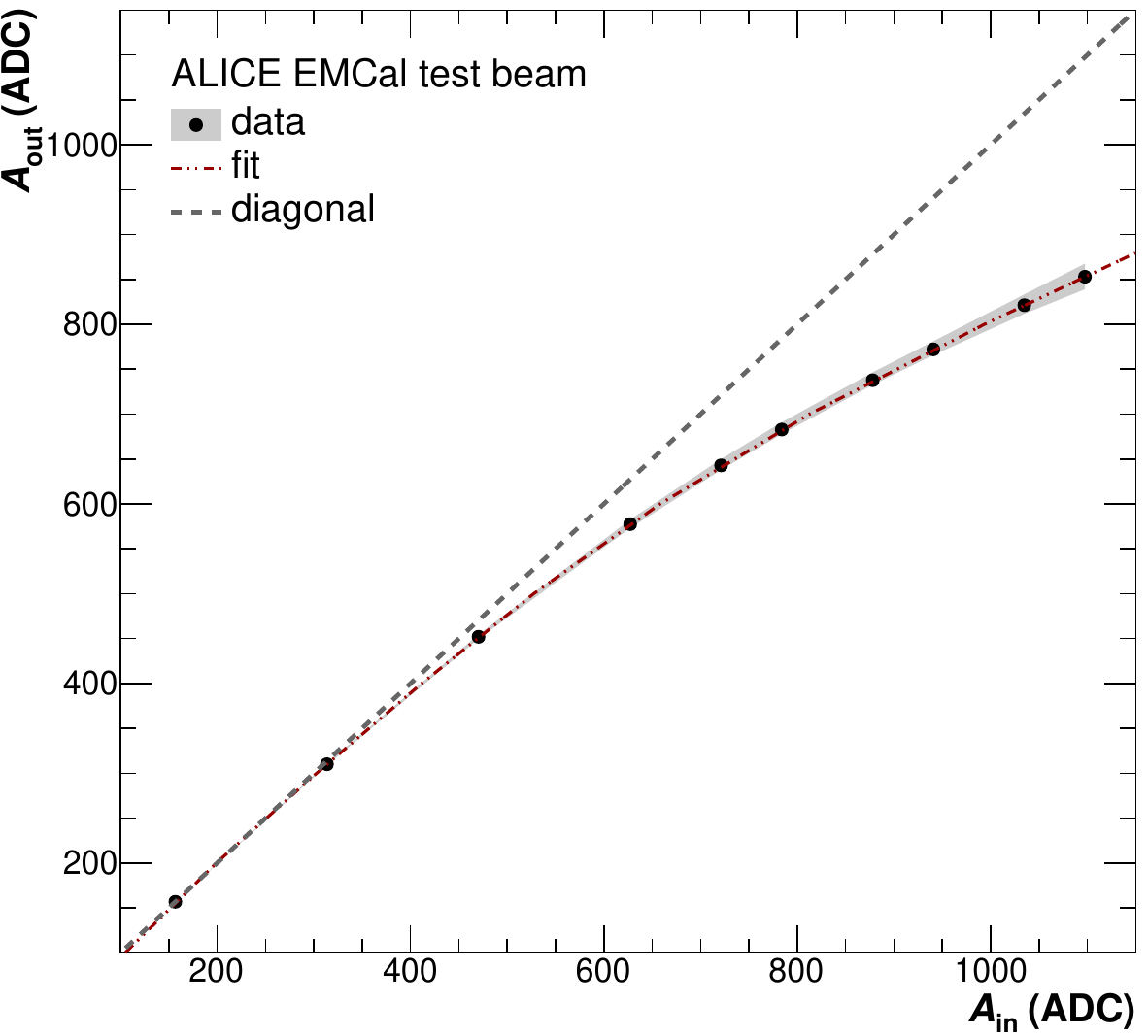}
  \includegraphics[height=6.5cm]{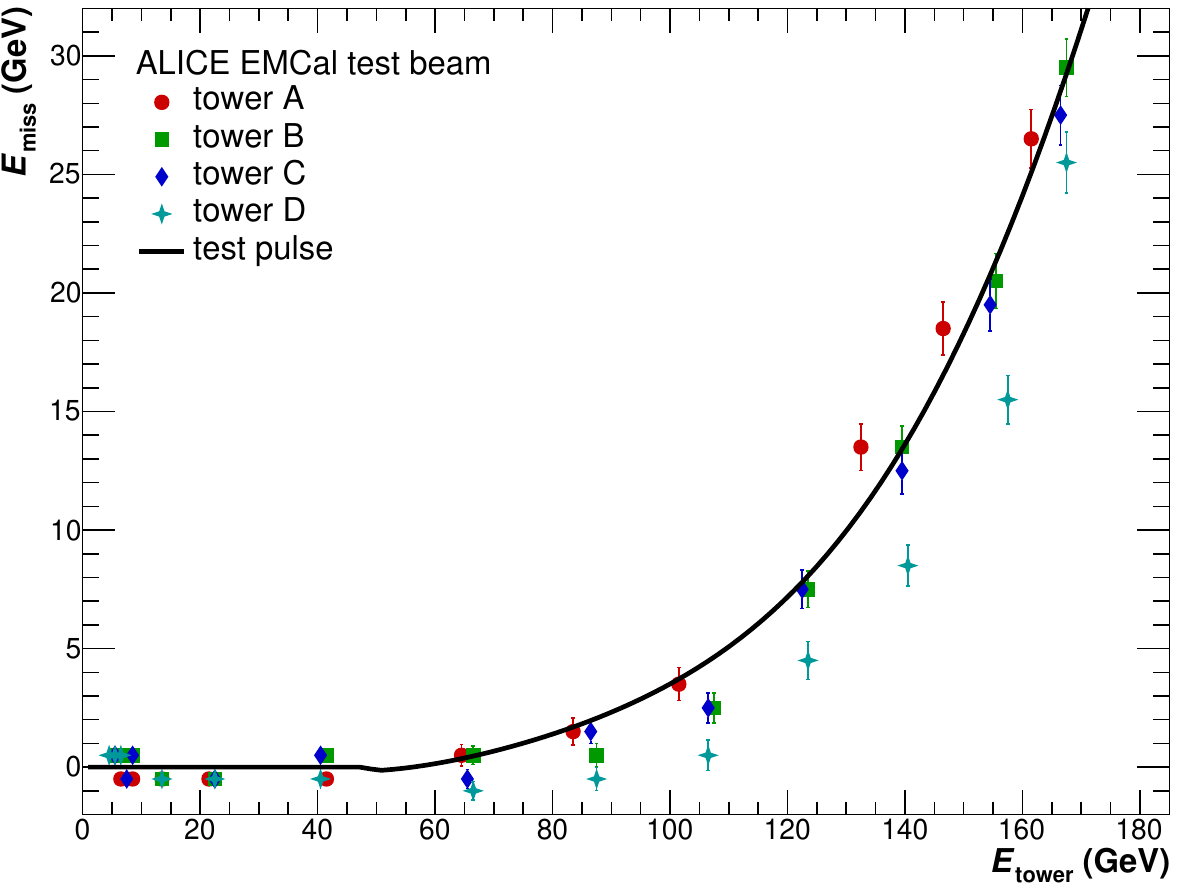}
  \caption{(Color online) Left: Measured pulse amplitude~($A_{\rm out}$) as a function of input pulse amplitude obtained from laboratory measurements.
                            The dashed gray line indicates the case of a linear shaper. 
                            Right: Comparison of laboratory measurements with the \gls{TB} data on missing energy~($E_{\rm miss}$) as a function of the measured energy.}  
  \label{fig:ShaperNonLin}

\end{figure}

During the analysis of high energy beam data, a significant energy nonlinearity was observed for beam energies $\gtrsim 100$~GeV.
Specifically, an energy deficit increasing with the beam energy was found, which is not described by \gls{MC} simulations, likely a possible consequence of electromagnetic shower leakage. 
With laboratory measurements it was confirmed that the nonlinearity arises from the \gls{FEE} response, namely due to the buffer of the shaper used in the \gls{FEE} card. 
The \gls{FEE} (shaper) nonlinearity was studied in detail and  parameterized by correlating the measured pulse amplitude to an input pulse amplitude, injected from an external pulse generator TG5011 and from a dedicated light generator described in Ref.~\cite{Rusu:2019pvk}. 
The correlation for 20 channels is displayed in \Fig{fig:ShaperNonLin} (left). 
The mean of the measured pulse amplitude as a function of leading tower energy for various towers is well described by a $6^{\rm th}$-order polynomial function. 
Based on the parametrization from laboratory measurements, the  missing energy as a function of measured energy was calculated and compared with test-beam data in \Fig{fig:ShaperNonLin}~(right). 
In the latter figure, the dependence of the difference between the beam and reconstructed energies ($E_{\rm miss}$) is shown as a function of leading tower energy for various beam energies and for various positions of the leading tower. 
This comparison must be considered as a qualitative comparison, because the missing energy cannot be accurately measured from data. The dispersion of the data points per tower is mostly due to the variation of the \gls{FEE} (shaper) nonlinearity per channel (the gray band in the left panel), the variation of energy calibration coefficients used to convert from \gls{ADC} to GeV units, and the cross-talk (see \Sec{sec:crosstalk}).

For both the test beam and LHC \gls{ALICE} data reconstructions, the \gls{FEE} (shaper) nonlinearity correction is absorbed in the tower-level energy calibration. 

\subsection{Data analysis and results} \label{sec:TBanalysis}
\begin{figure}[t]
  \centering
  \includegraphics[width=0.49\textwidth]{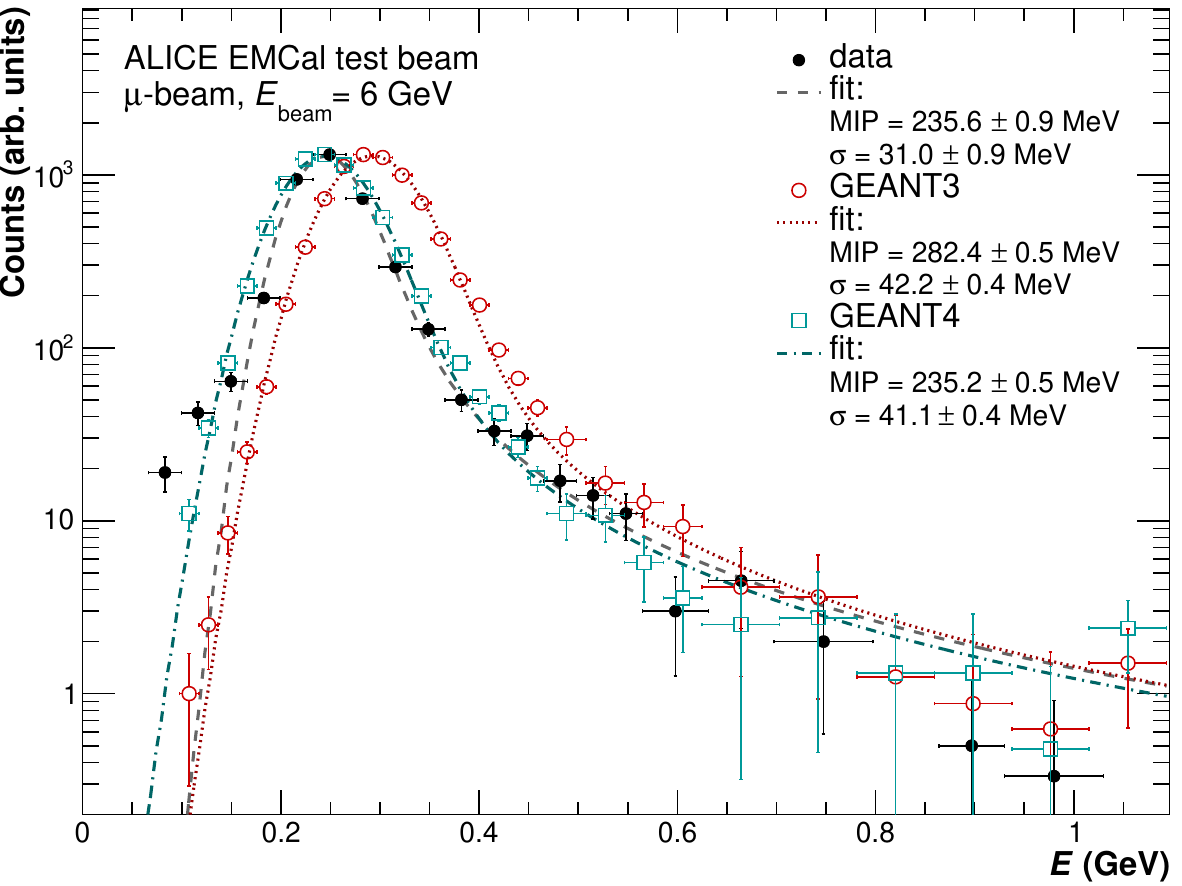}
  \includegraphics[width=0.49\textwidth]{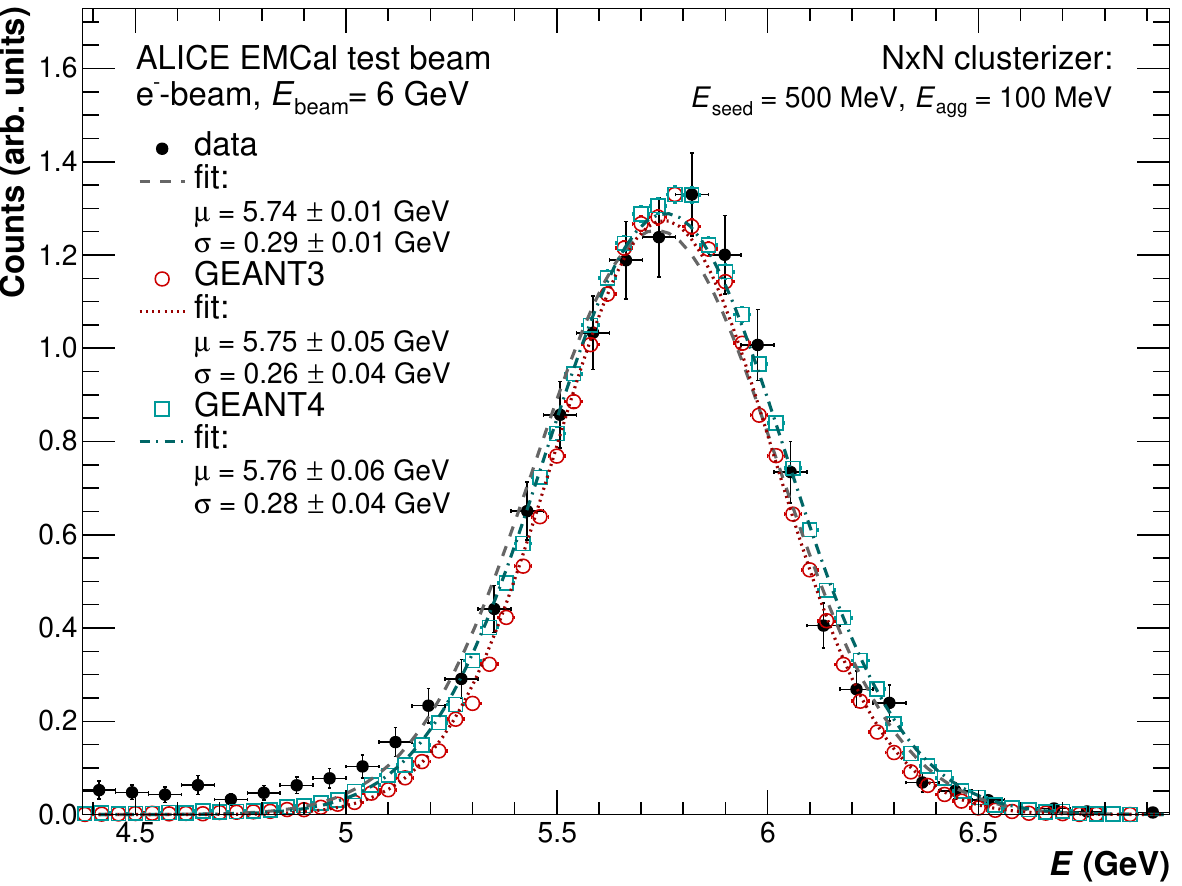}
  \caption{(Color online) Left: energy distribution of single cell clusters obtained from scans with a 6 GeV muon-beam. 
            Right: energy distribution of clusters obtained from scans with a 6~GeV electron beam. 
            For both cases the data are shown with black markers and compared with the predictions from \gls{MC} simulations with \gls{GEANT}3 and \gls{GEANT}4 transport codes.}
  \label{fig:MIPposition}
\end{figure}

Even though the clusters with leading cell energy less than 500~MeV were excluded from the energy-calibration procedure, special attention was paid to lower-energy single-cell clusters from the data taken with muon- and hadron- beams to determine the \gls{MIP} energy. 
Muon- and hadron- beams showed identical results. 
The measured energy distribution with a 6 GeV energy muon beam is shown in \Fig{fig:MIPposition} (left). 
A fit with a Landau–Gaussian convolution achieves a good description of the data and yields approximately 236~MeV for the \gls{MIP} energy.
The \gls{MC} prediction with the \gls{GEANT}4 transport code is in good agreement with data, whereas the one with \gls{GEANT}3 overestimates the \gls{MIP} energy by about 50~MeV. 

In \Fig{fig:MIPposition} (right), the reconstructed energy distribution measured from data with a 6~GeV electron beam is presented and compared with the predictions from \gls{MC} simulations with \gls{GEANT}3 and \gls{GEANT}4 transport codes. 
Both predictions are in good agreement with the data.

\subsubsection{Response at low energies} \label{sec:lowEeffi}

The granularity of \gls{EMCal} towers is such that the tower size is about twice the Moli\`ere radius~(see \Tab{Table-1}). 
This is similar to the calorimeters of the \gls{PHENIX}~\cite{OPAL:1990yff} and \gls{OPAL}~\cite{PHENIX:2003fvo} experiment, whereas for many  other calorimeters the tower size is comparable to the Moli\`ere radius. 
Such segmentation of \gls{EMCal} results in a uniform response across towers.
For incident electrons with energy $E \lesssim 4$~GeV, the transverse size of the \gls{EMCal} shower is compatible or smaller than the size of one tower.
Thus, when the incident particle hits the center of a tower, the electromagnetic shower is fully contained in a single tower. 
However, when the particle hits the tower near its edge, the shower is split among two or more towers. 
Therefore, the probability of finding clusters with at least two cells is smaller when the low energy particles hit near the center of towers. In addition, due to the non-linear energy response vs shower energy, in case the  shower is split in several towers due to hitting by a particle near the tower edge, the reconstructed energy is smaller than it would be in case the particle would hit the center of a tower and the whole shower would contain in a single tower.  
Thus, at low energies the detector response, and as a consequence the energy nonlinearity, depend both on the energy and on the hit position due to the way the shower splits among towers and on the light accumulation and  propagation properties. 

\begin{figure}[t]
  \centering
  \includegraphics[width = 0.49\textwidth]{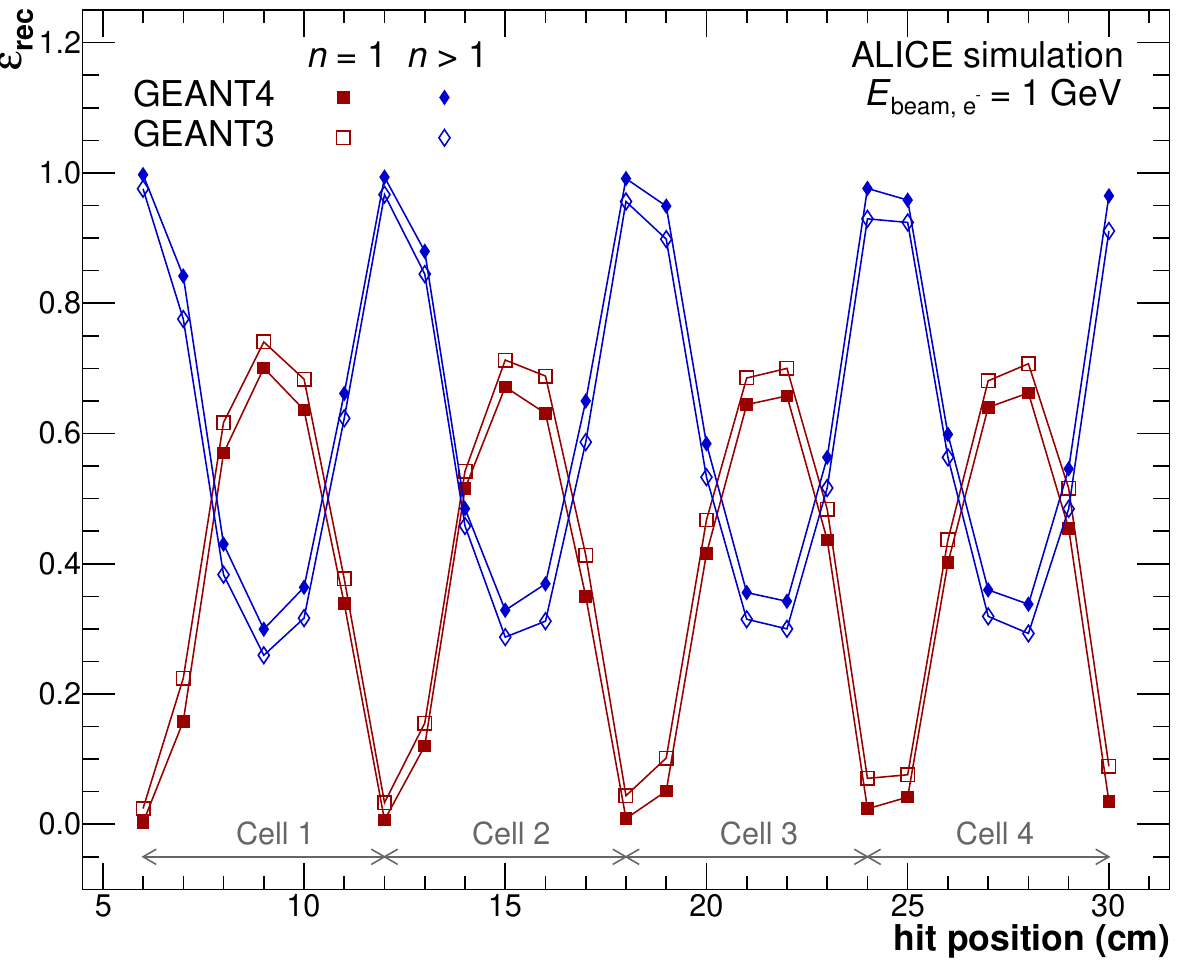}
  \includegraphics[width = 0.49\textwidth]{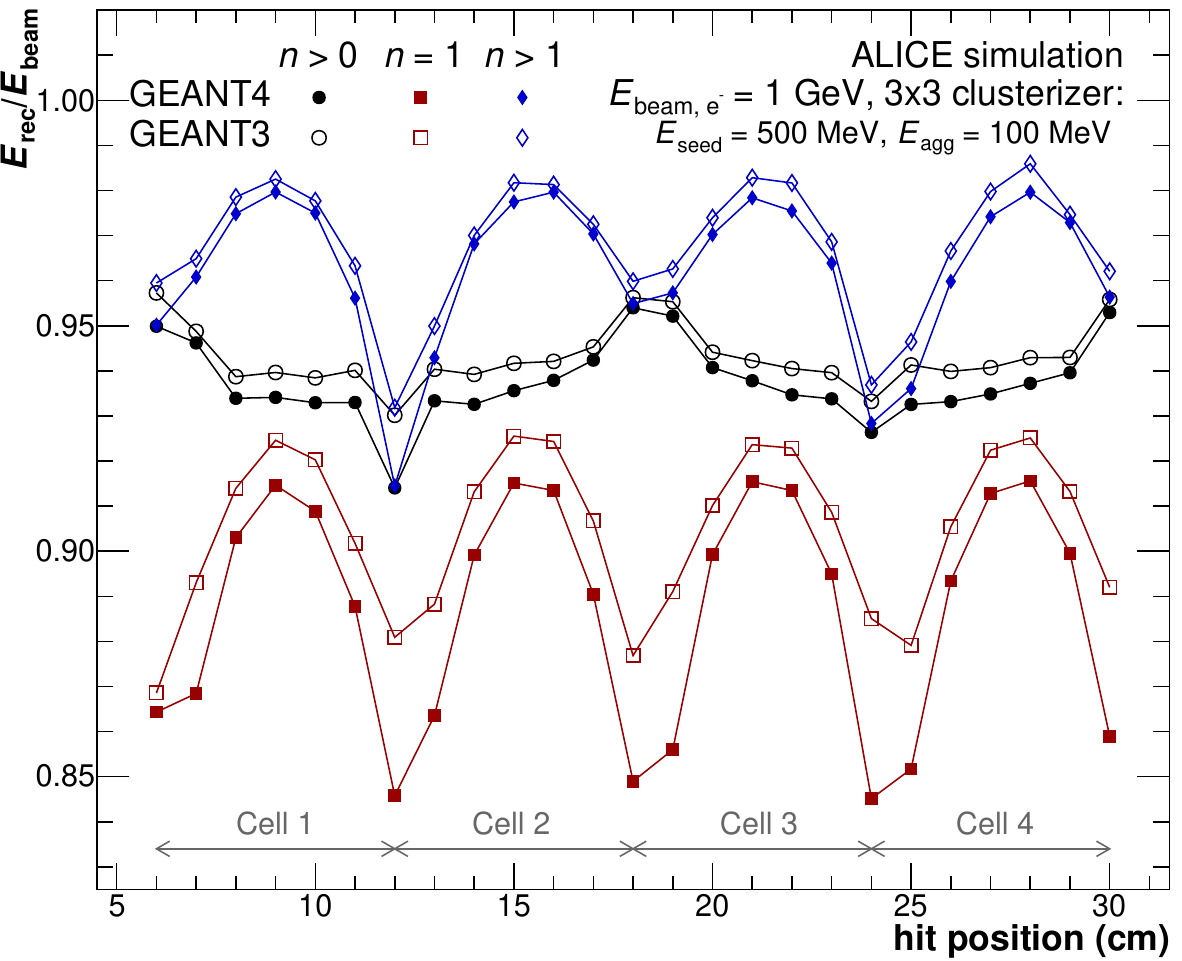}
  \caption{Cluster reconstruction/finding efficiency (left) and energy nonlinearity (right) as a function of hit position obtained from \gls{MC} simulations for 1~GeV electrons. Red markers stand for single cell clusters ($n=1$), blue makers stand for the clusters made of at least two cells ($n>1$), and the black markers stand for the clusters with any number of cells ($n>0$).}
  \label{fig:5-TB-RecEffic_1GeV_MC}
\end{figure}

The cluster-reconstruction and cluster-finding efficiency ($\varepsilon_{\text{rec}}$) as well as the energy response as a function of hit position obtained from \gls{MC} simulations are shown in \Fig{fig:5-TB-RecEffic_1GeV_MC}. 
The latter quantity is quantified as the ratio between the reconstructed energy and the incident energy and also referred to as energy nonlinearity.
The scan was performed by varying the $x$-coordinate of the hit position at fixed $y=0$ (at the center of tower in the perpendicular direction). 
The cases when the cluster is made of a single tower and at least two towers are shown separately in \Fig{fig:5-TB-RecEffic_1GeV_MC}.
In the analysis of the LHC collision data, at least two towers in the cluster are required to suppress the noise (\Sec{sec:clusterization}), resulting in the cluster finding and reconstruction uniformity across towers.

\begin{figure}[t]
  \centering
  \includegraphics[width = 0.49\textwidth]{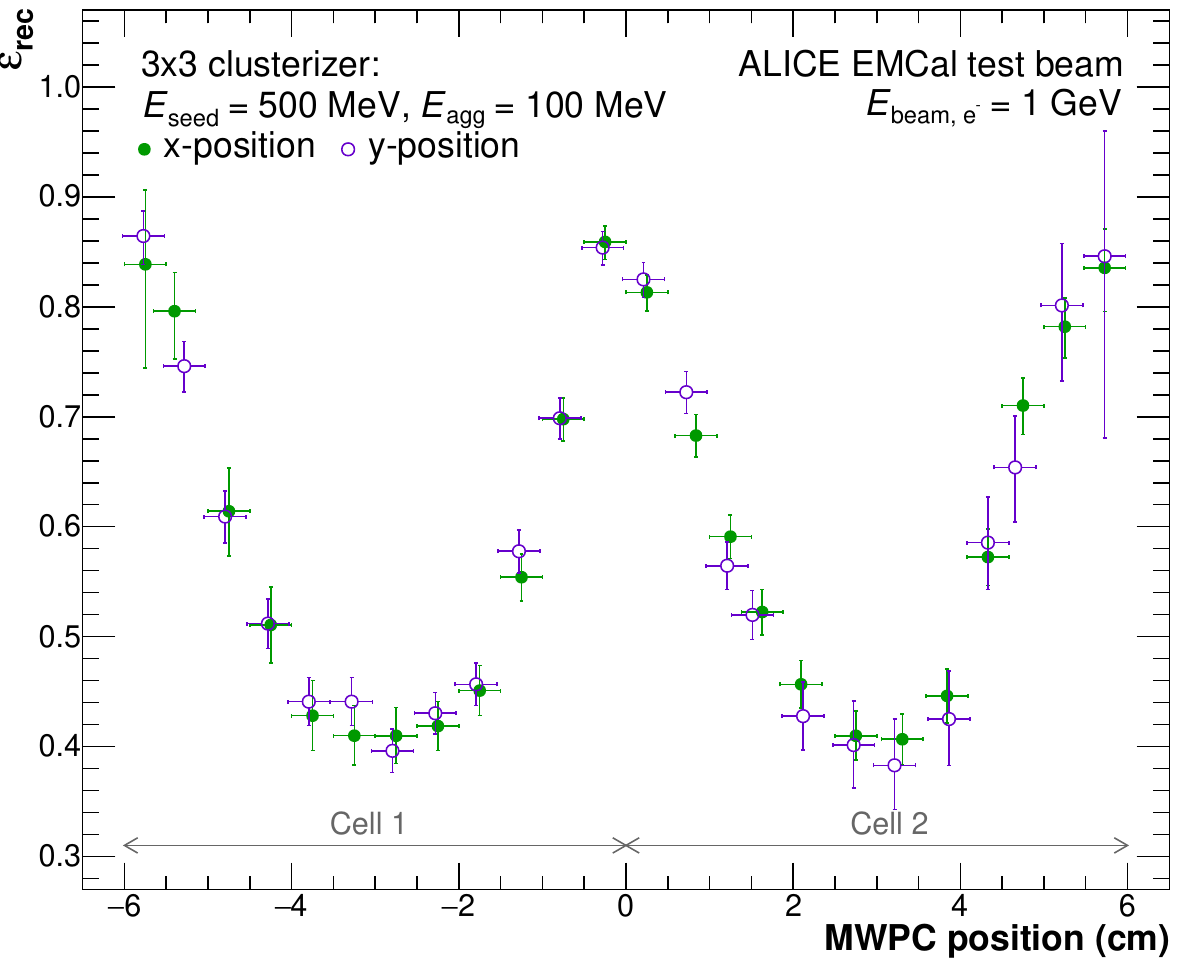}
  \includegraphics[width = 0.49\textwidth]{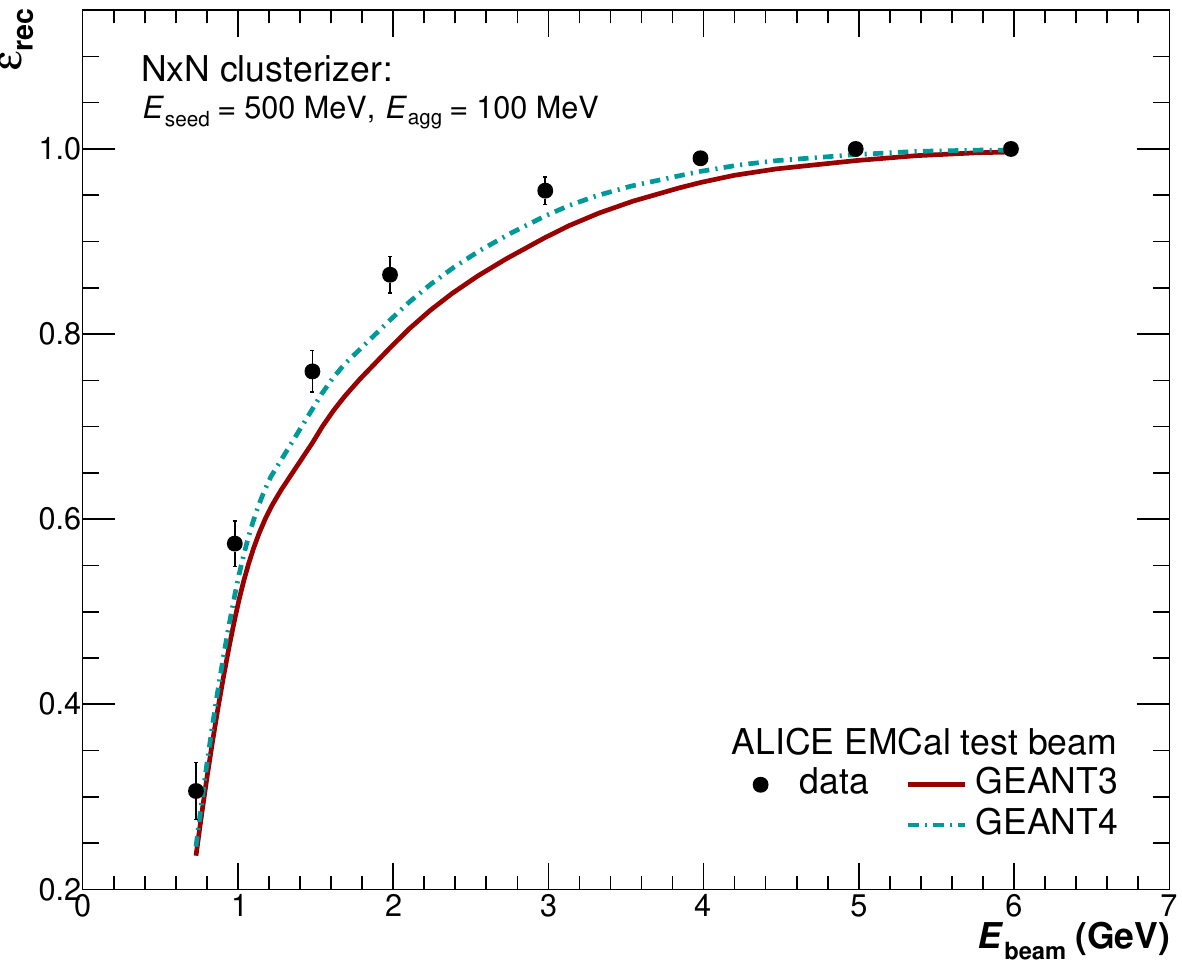}
  \caption{(Color online) The reconstruction efficiency for the clusters made of at least two cells and for 1 GeV electrons as a function of hit position measured using the \glspl{MWPC} (left) and as a function of the incoming particle energy (right).}
  \label{fig:5-TB-RecEffic_1GeV_data}
  \label{fig:5-TB-RecEfficVsEbeam}
\end{figure}

\Figure{fig:5-TB-RecEffic_1GeV_data}{ (left)} shows the cluster $\varepsilon_{\text{rec}}$ for clusters with at least two towers as a function of the incident hit position measured from \gls{TB} data.
Due to the lack of statistics it was not possible to make a $x$-scan ($y$-scan) in slices of $y$~($x$). 
Instead, the integral over all hit positions in the perpendicular direction was done. 
For the same reason, the energy nonlinearity dependence on the hit position could not be measured with a reliable precision. 
We expect a few percent uncertainty for the average energy nonlinearity determination depending on the beam-center position and the beam profile. 
The mean $\varepsilon_{\text{rec}}$ (for a uniform distribution of incident particles across towers) as a function of incident particle energy found from data and \gls{MC} simulations is shown in \Fig{fig:5-TB-RecEfficVsEbeam}~(right). 
The efficiency is close to 100\% for electrons with energies larger than 4~GeV in both data and \gls{MC}. 
However, for lower energies the \gls{MC} simulations systematically underestimate the data, indicating that the simulations predict more collimated electromagnetic showers. 

\subsubsection{Energy nonlinearity and resolution} \label{sec:TBNonLin}
\begin{figure}[t]
  \centering
  \includegraphics[width=0.75\textwidth]{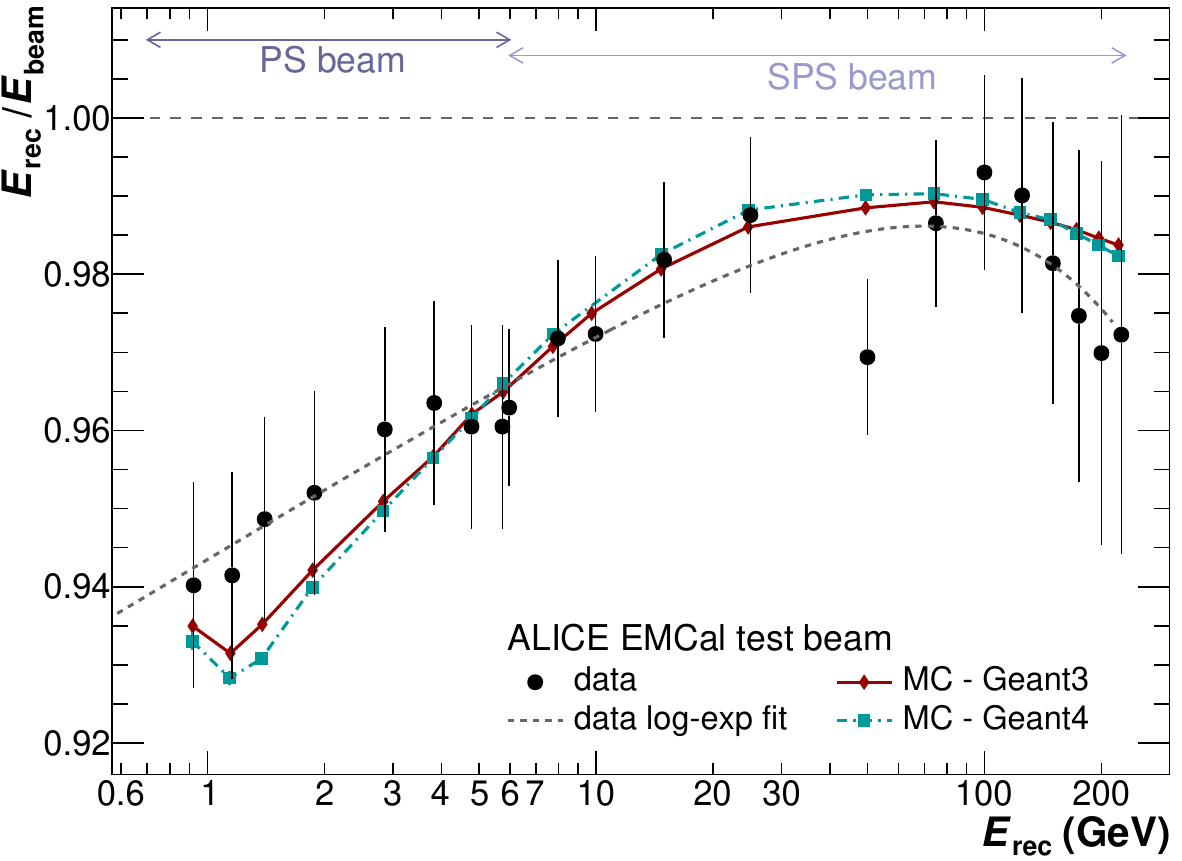}
  \caption{(Color online) Energy nonlinearity correction ($E_{\rm rec}/E_{\rm beam}$) as a function of reconstructed energy for electrons obtained from \gls{TB} data (black points), and from \gls{MC} simulations with \gls{GEANT}3 (red points) and \gls{GEANT}4 (cyan points) transport codes.}
  \label{fig:5-TB-EnergyNonLinearity}
\end{figure}

\Figure{fig:5-TB-EnergyNonLinearity} shows the ratio of reconstructed energy over true beam energy ($E_{\rm rec}/E_{\rm beam}$) as a function of $E_{\rm rec}$ measured in the test-beam data and obtained from \gls{MC} simulations. 
This dependence presents the  energy nonlinearity of the detector response.
For the data the channel-by-channel shaper nonlinearity correction (see \Fig{fig:ShaperNonLin} left) is already applied as described above.

A reasonable agreement between data and \gls{MC} predictions was achieved.
The energy nonlinearity can be parameterized as
\begin{equation}
  f(E_{\rm rec}) = \frac{p_{0}+p_{1}\ln(E_{\rm rec})}{1+p_{2}\exp(E_{\rm rec}/p_{3}) }\,,
\end{equation}
with parameters: $p_0 = 4.3 \pm  0.6$, $p_1 = 0.06   \pm 0.02$, $p_2 =  3.5 \pm  0.6$, $p_3 = 4172 \pm   2276$~(energy in units of GeV).

The energy dependence of the energy resolution of an electromagnetic calorimeter is parameterized as
\begin{equation}
\sigma(E)/E = a \oplus b/\sqrt{E} \oplus c/E,
\label{eqn:5-TB-energyResolution}
\end{equation}
where $E$ is the incident energy~(in units of GeV).
The intrinsic resolution is characterized by the parameter $b$ that arises from stochastic fluctuations due to intrinsic detector effects such as energy deposition, energy sampling and light-collection efficiency. 
The constant term, $a$, originates from systematic effects, such as shower leakage, detector nonuniformity or channel-by-channel calibration errors.
The third term, $c$, is due to electronic noise summed over the towers of the cluster used to reconstruct the electromagnetic shower. 
The three resolution contributions are added  in quadrature.

\begin{figure}[t]
  \centering
  \includegraphics[width = 0.49\textwidth]{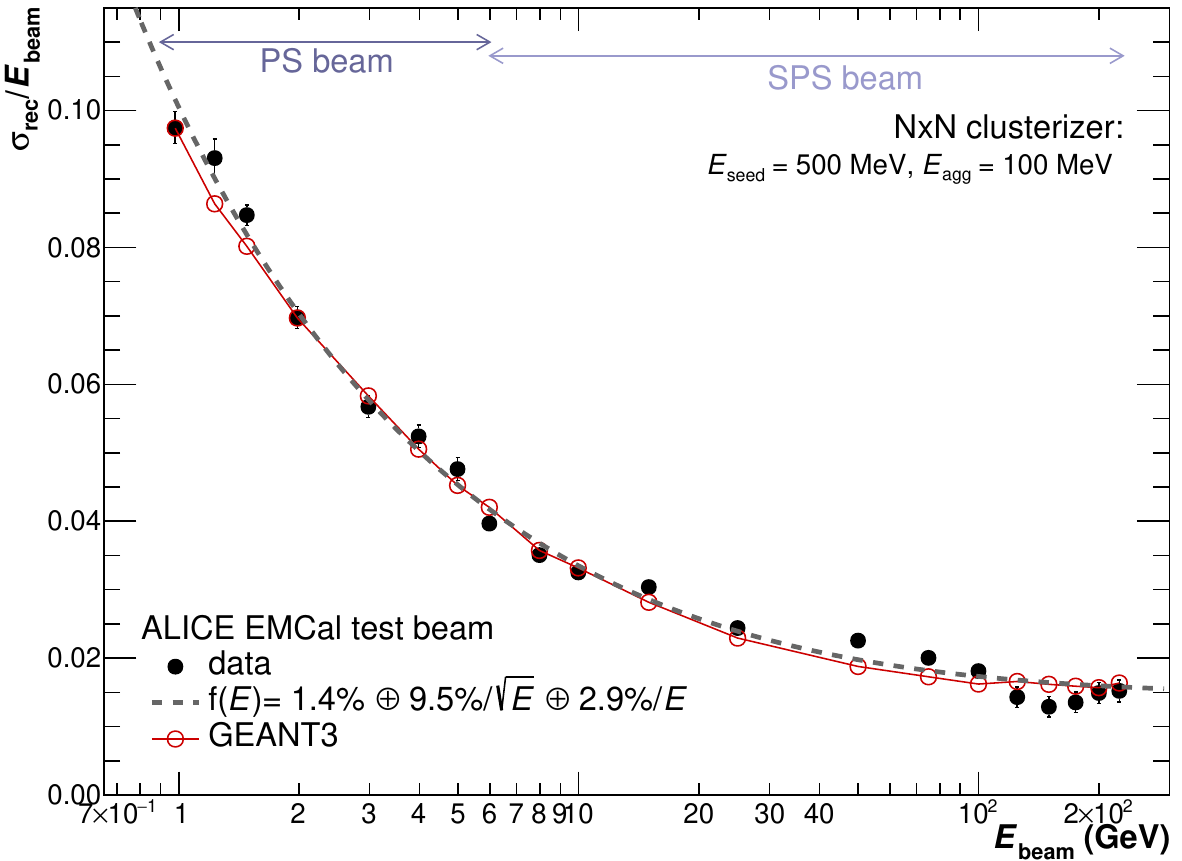}
  \includegraphics[width = 0.49\textwidth]{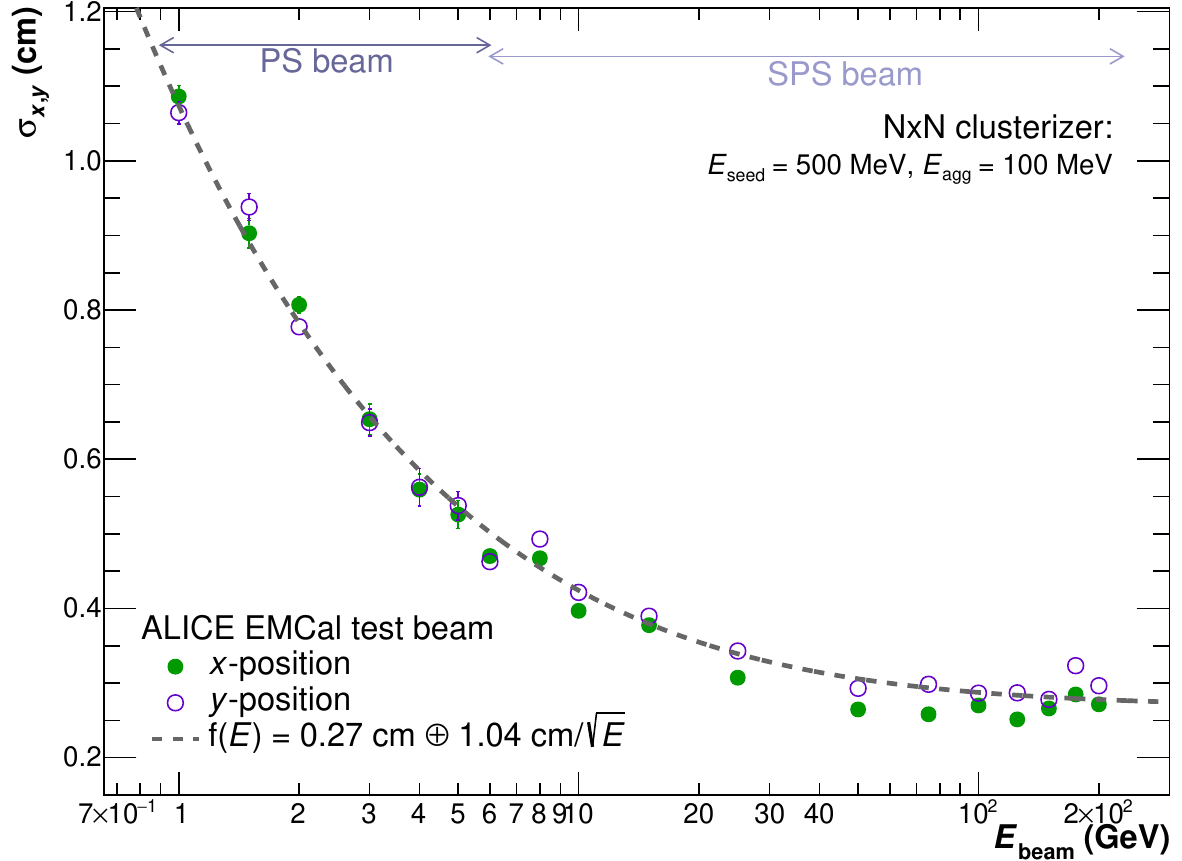}
  \caption{(Color online) Left: relative energy resolution as a function of beam energy. Right: cluster-position resolution as a function of beam energy.}
  \label{fig:5-TB-energy_resolution}
  \label{fig:5-TB-position_resolution}
\end{figure}

The energy resolutions obtained from test beam data and \gls{MC} simulations are shown in \Fig{fig:5-TB-energy_resolution}.  The following numerical values were found from a fit to data for the energy resolution parameters: $a = 1.4 \pm 0.1 $, $b = 9.5 \pm 0.2$~GeV$^{1/2}$, $c = 2.9 \pm 0.9$~GeV.

The energy response was also studied for different positions corresponding to the modules as installed in \gls{ALICE}. 
Most of the test beam data were taken with a configuration where the beam hits the \gls{EMCal} modules perpendicularly. 
Data were also taken with configurations where the modules were tilted in $\varphi$ by $6^\circ$ or $9^\circ$ at different surface positions. 
The response at such tilted configurations is consistent  with the energy nonlinearity and the average resolution as a function of energy presented in \Figs{fig:5-TB-EnergyNonLinearity}{fig:5-TB-energy_resolution} (left). 
No significant deviations from the averages at $0^\circ$ were observed.

\subsubsection{Position resolution} \label{sec:PosResol}
The segmentation of the calorimeter allows one to obtain the hit position from the energy distribution inside a cluster with an accuracy better than the tower size. 
The $x$ and $y$ coordinate locations are calculated using a logarithmic weighting of the tower energy deposits as described in \Sec{sec:showershape}. 
The \glspl{MWPC} used in the test beam data provided a measurement of the reference position with an uncertainty smaller than 2~mm.
\Figure{fig:5-TB-position_resolution}{ (right)} shows the $x$ and $y$ position resolution as a function of the energy deposit for electrons. 
As expected, no significant difference in the resolution in the $x$ and $y$ directions is observed, and the position resolution is significantly smaller than the tower size even for $\sim700$~MeV clusters.
The electromagnetic shower position resolution can be described by $a \oplus b/\sqrt{E}$, where the two contributions are added in quadrature, with the following numerical values for the parameters found from the fit to data: $a = 0.268 \pm 0.001$~cm, $b = 1.042 \pm 0.003$~cm/GeV$^{1/2}$.

\ifflush
\clearpage
\fi
\section{Calibrations and corrections}
\label{sec:calibration}
\subsection{Survey alignment}
\label{sec:alignment}
The geometry of the \gls{EMCal} was initially implemented in the software following the design of the \gls{ALICE} experiment, and this geometry is referred to as the ``ideal geometry'' in this section. 
However, during the installation of the detector in the cavern, the detector was positioned in a slightly different location, which is referred here as the ``actual geometry''. 
This results in a mismatch between the geometry implemented in the software and the actual position of the detectors. 
In the \gls{ALICE} reconstruction framework, this is corrected by the introduction of alignment matrices, which describe the rotation and translation of the detectors from the ideal to the actual positions. 
The \gls{EMCal} \glspl{SM} were installed individually in the cavern, and therefore they are considered to be independent when calculating the alignment matrices. 
In this setup, each module inside an \gls{SM} was built and mounted with few hundred micrometer precision, and has no freedom to move, which further justifies the use of each \gls{SM}'s position as the degrees of freedom in the alignment. 

\begin{figure}[b]
  \centering
  \includegraphics[width=0.5\textwidth]{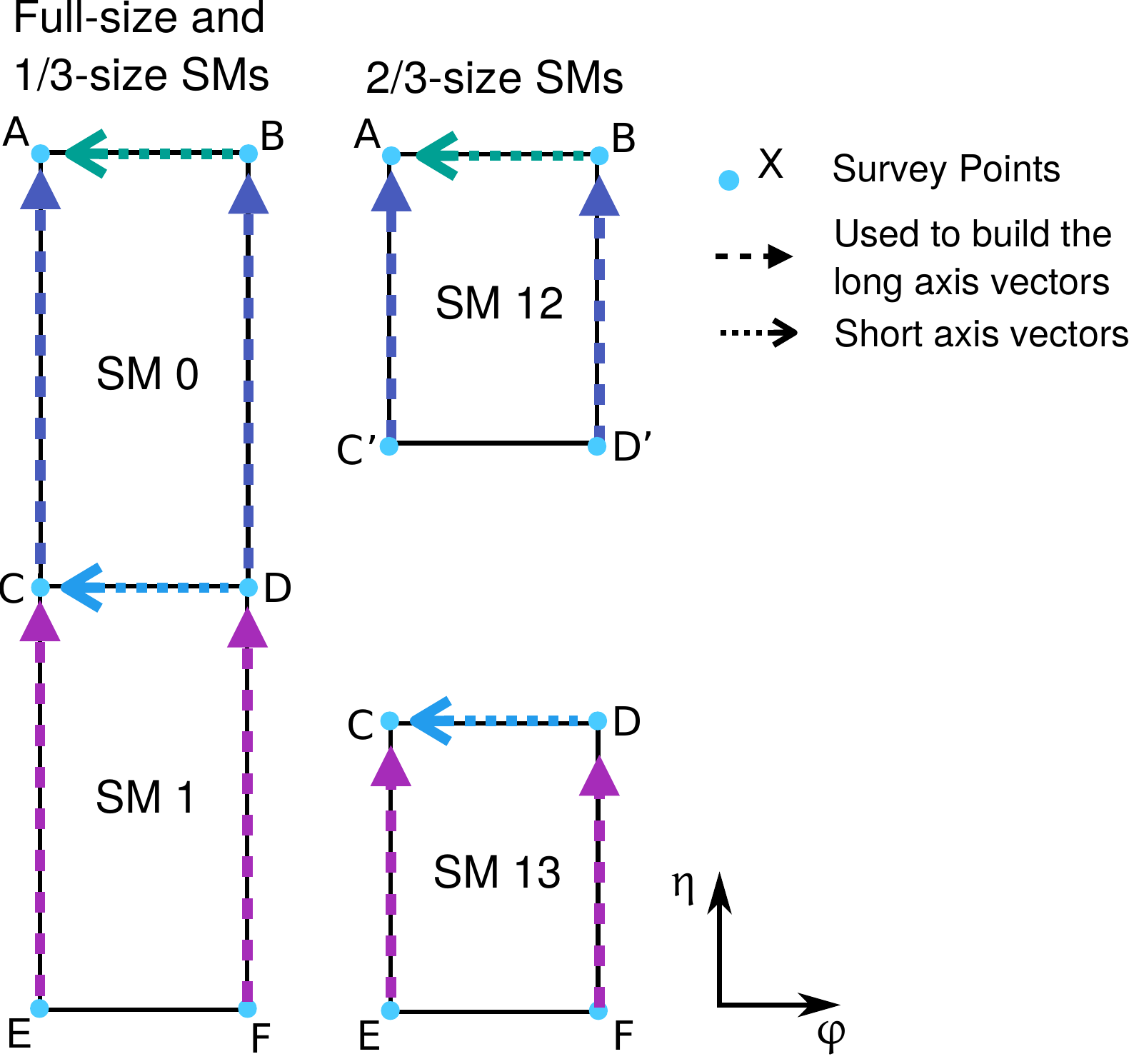}
  \caption{Schematic overview of the positions of the survey points and how the vectors used for the alignment were constructed.}
  \label{fig:calib_alig_markers}
\end{figure}

The initial alignment was based on the measurement of the position of reference markers attached to the corners of the \glspl{SM}, as illustrated in \Fig{fig:calib_alig_markers}.
Using these positions, the alignment was performed in the following steps. 
At first, two vectors along the edges of each \gls{SM} were used to span the plane of the surface of the calorimeter. 
The first vector, called short-axis vector ($\vectoralig{S}$), which is approximately in the $\varphi$--direction, was built using the two points of the survey in the shortest length of the \gls{SM}. 
The positions used were the closest to $\eta = 0$ for the \glspl{SM} in the $\eta <0$ region (C and D in \Fig{fig:calib_alig_markers}) and the closest points to $\eta \approx 0.7$ for the \glspl{SM} in the $\eta >0$ region (A and B in \Fig{fig:calib_alig_markers}). 
The second vector, called long-axis vector ($\vectoralig{L}$), which is approximately in the $\eta$--direction, was built along the longest dimension of the \gls{SM}. 
It was constructed by taking the average of the two vectors in the longest direction of each \gls{SM} (see \Fig{fig:calib_alig_markers}). 
For example, $\vectoralig{L}$ was calculated using the following expression for \gls{SM}0:

\begin{equation}
  \vectoralig{L} = \frac{\vectoralig{AC}+ \vectoralig{BD}}{2},
\end{equation}
where $\vectoralig{XY}$ is the vector that connects the point $Y$ to the point $X$.
Then, a vector orthogonal to the actual surface of the detector ($\vectoralig{O}$) was computed using the cross product of the long- and short-axis vectors:

\begin{equation}
  \vectoralig{O} = \vectoralig{L} \times \vectoralig{S}.
\end{equation}

\begin{figure}
  \centering
  \includegraphics[width=0.49\textwidth]{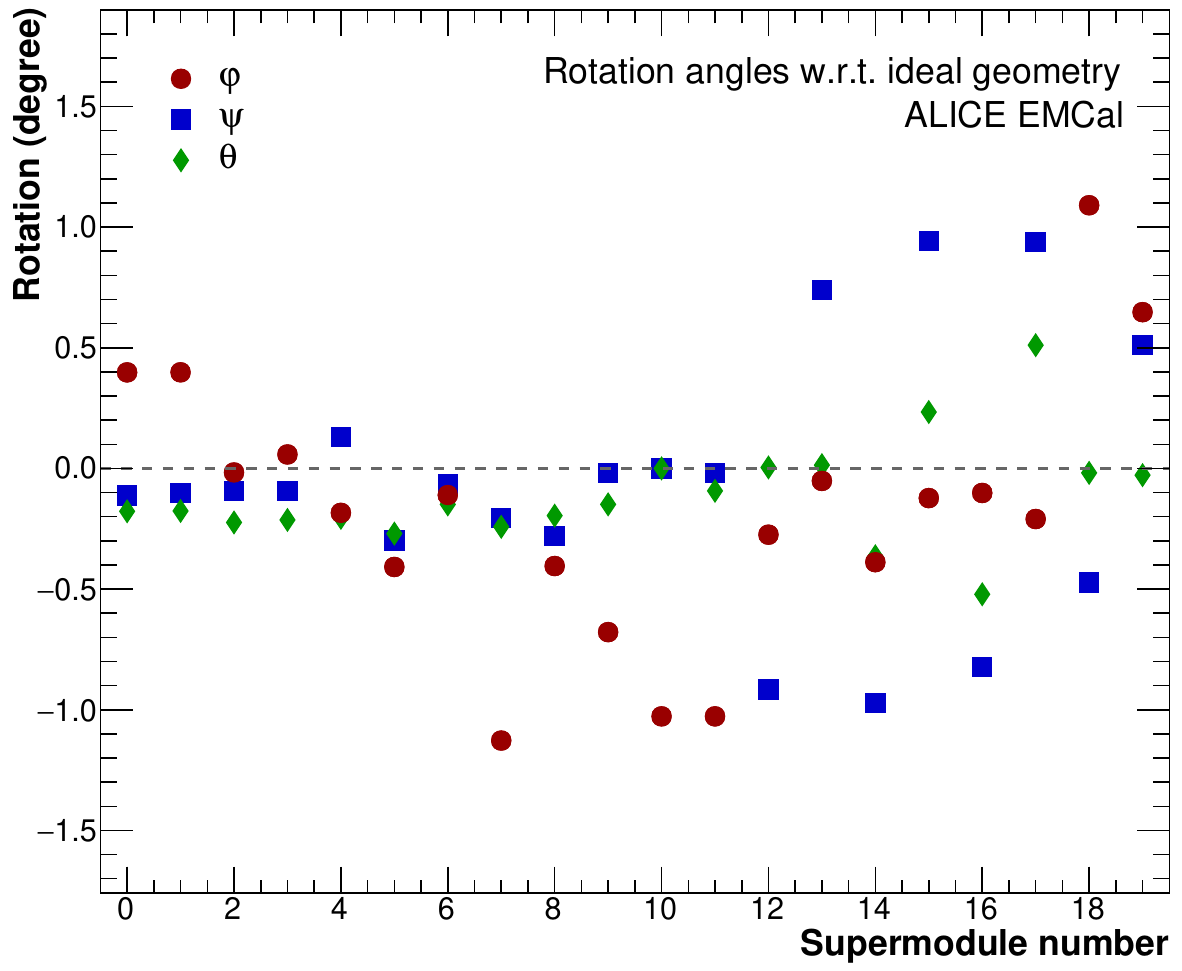}
  \includegraphics[width=0.49\textwidth]{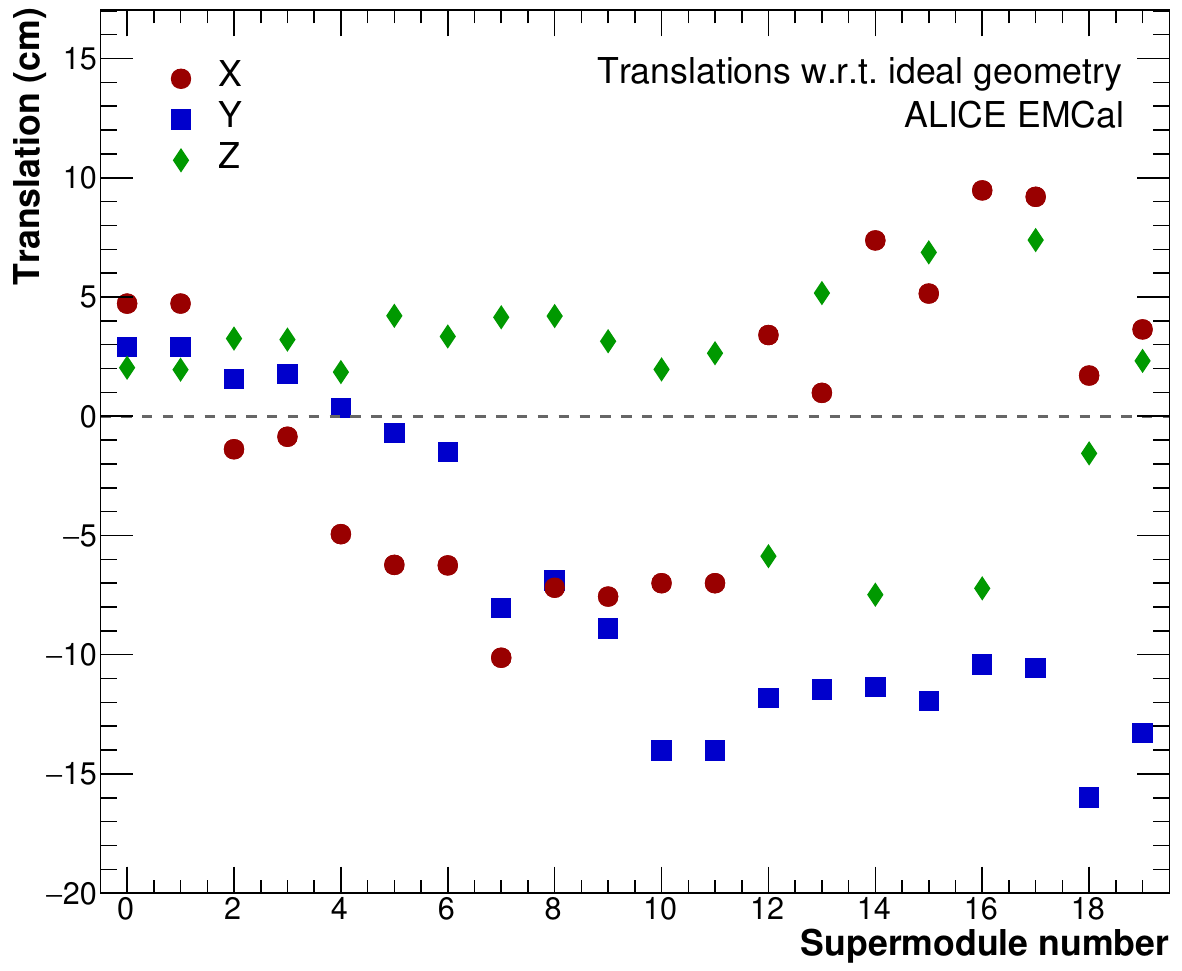}
  \caption{ (Color online)  Left:  Angular rotation of each \gls{SM} with respect to the ideal geometry.  Right: Translation of each \gls{SM} with respect to the ideal geometry.}
  \label{fig:calib_alig_angles}
  \label{fig:calib_alig_translation}
\end{figure}

This orthogonal vector was compared to the one in the ideal geometry, and the angular difference between them was calculated, in the reference system defined at the center of the \gls{SM}, in terms of roll-pitch-yaw  angles $\phi, \psi$ and $\theta$ standing for the rotations around the axes $x, y$ and $z$, respectively ~\cite{miller2010euler}. 
The angular differences for each \gls{SM} are presented in \Fig{fig:calib_alig_angles} (left). 
The differences are smaller than 1 degree for most cases.
The ideal geometry was then rotated by these angles, and the translations in the $x$, $y$ and $z$ directions between the rotated ideal geometry and the actual position were calculated. 
They are reported in \Fig{fig:calib_alig_translation} (right).
This took into account an additional constraint applied for the full-size and 1/3 size \glspl{SM}, to ensure a smooth transition at $\eta \approx 0$. 
The translations are of the order of a few centimetres, and a global shift of approximately 11 cm in the $y$-direction is found for the \gls{DCal} \glspl{SM} (SM 12--19). 
In the end, the total alignment matrix was calculated by combining the translations in the $x$, $y$, and $z$ axes, and the three rotation angles calculated in the previous step. 

\begin{figure}[t]
  \centering
  \includegraphics[width=0.49\textwidth]{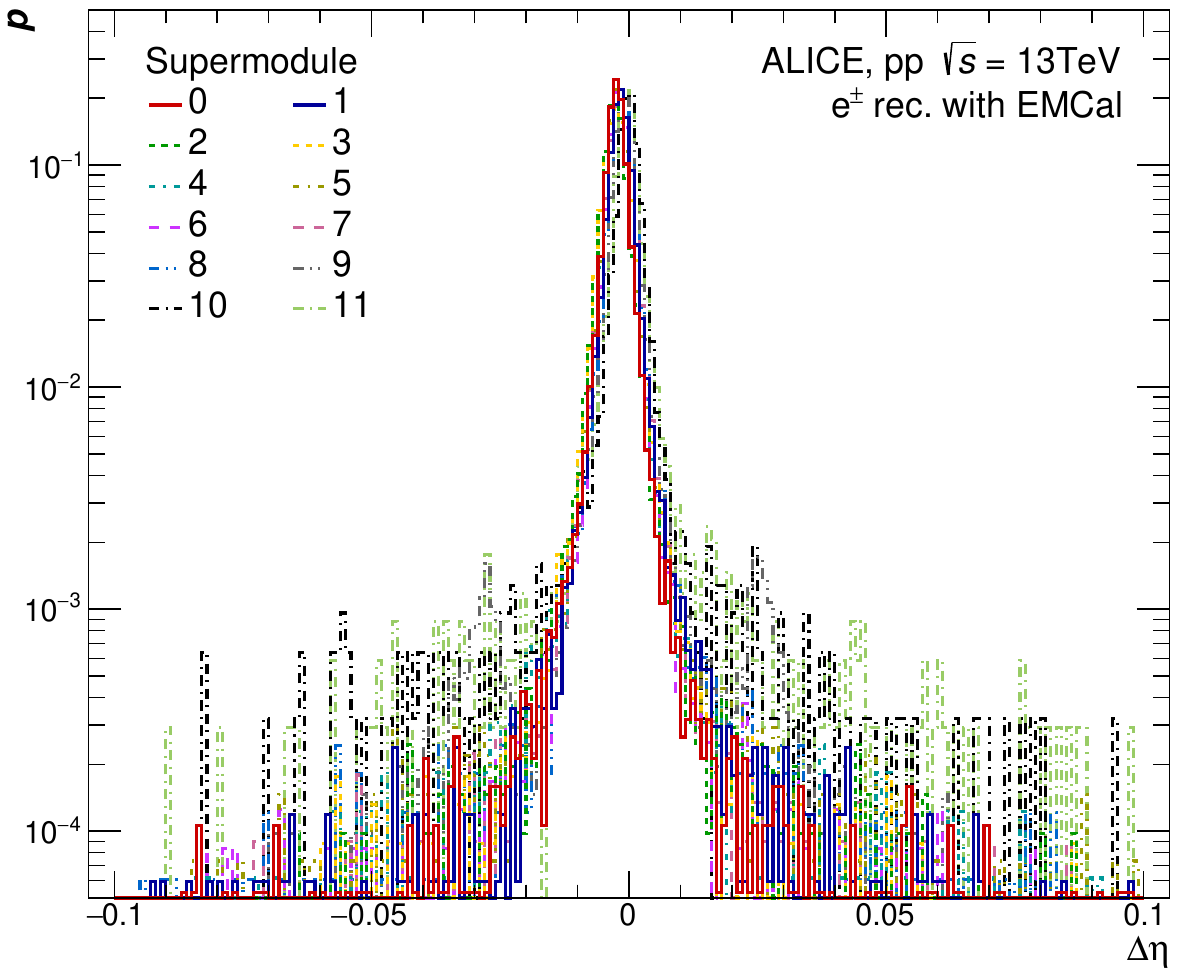}
  \includegraphics[width=0.49\textwidth]{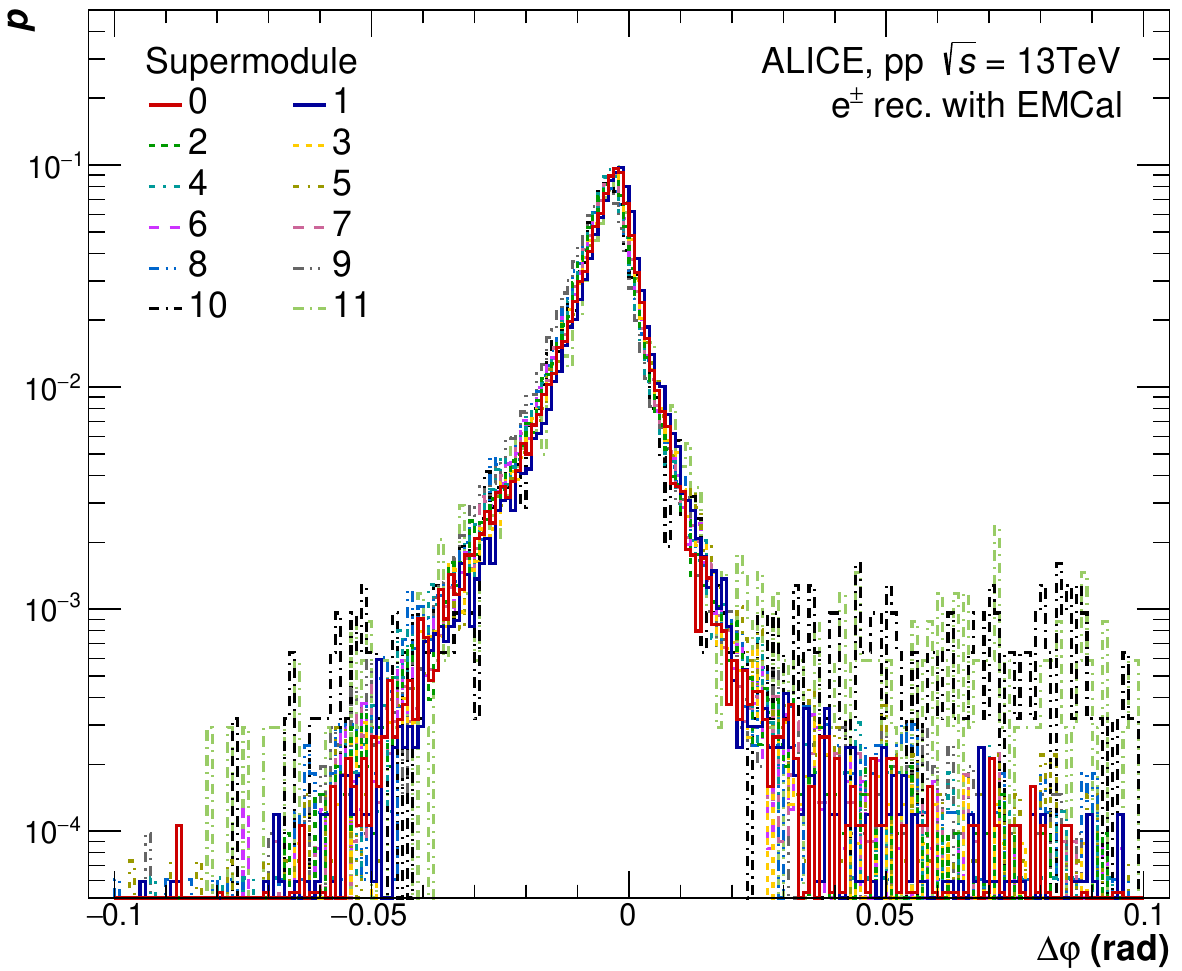}
  \caption{(Color online) The probability distribution of the position of electron tracks propagated to the \gls{EMCal} \glspl{SM} surface and their associated cluster in $\eta$-direction (left) and $\varphi$-direction (right). The distributions are normalized by their integral.}
  \label{fig:calib_alig_eta_phi_diff_dcal}
  \includegraphics[width=0.49\textwidth]{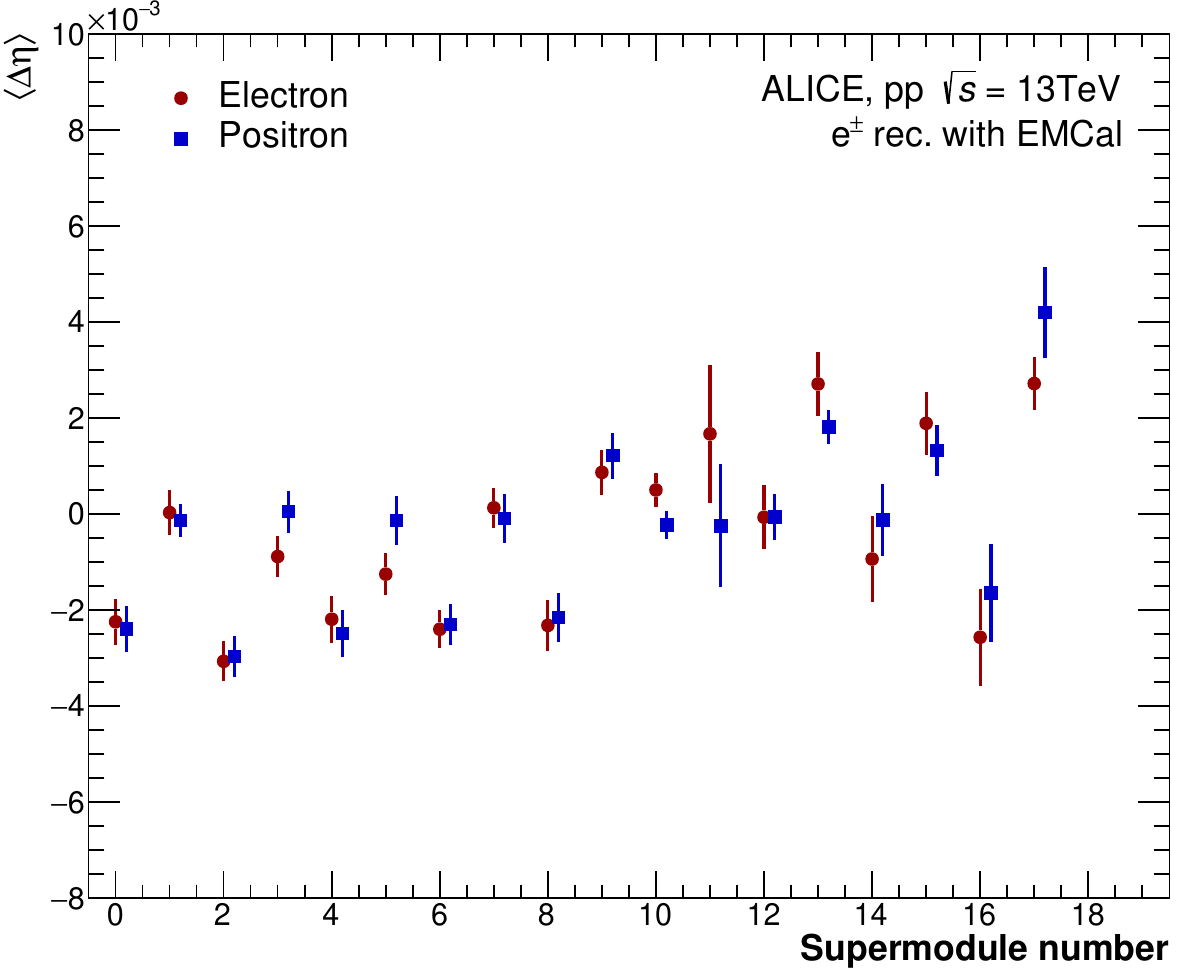}
  \includegraphics[width=0.49\textwidth]{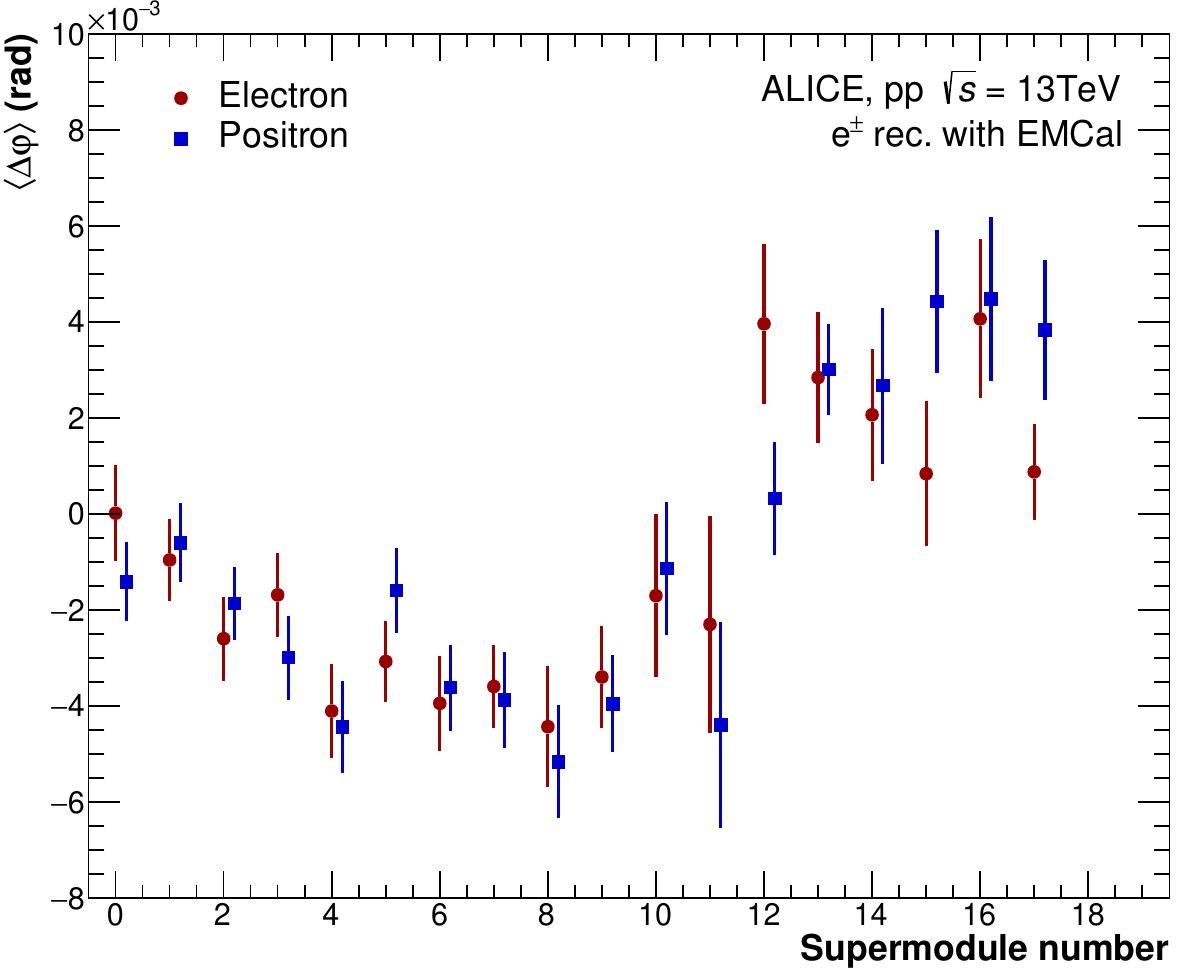}
  \caption{(Color online) Mean differences between the position in $\eta$ (left) and $\varphi$ (right) of clusters and electron tracks propagated to the \gls{EMCal} / \gls{DCal} as function of the \gls{SM} number. The colors represent electrons (red) and positrons (blue).}
  \label{fig:calib_alig_eta_phi_diff_mean}
\end{figure}

To check the accuracy of the alignment and possibly correct for remaining misalignment, a study of the matched tracks to the \gls{EMCal} clusters in pp collisions was performed. 
The strategy was to use the differences between the position of the tracks and clusters as a benchmark of the quality of the alignment, where the cluster position is given by the weighted average position according to \Eq{eq:xyz_centroid}.
 Since electrons deposit their full energy in the electromagnetic calorimeter, electrons with \pT $>$~2~GeV/$c$ and $|\eta| < $ 0.7 were chosen for this study. 
They were identified using the charged tracks reconstructed in the \gls{ITS} and the \gls{TPC}, and propagated to the calorimeter's surface (\Sec{sec:trackmatch}). 
Each track was propagated to the position of the \gls{EMCal} and was associated with a cluster if the corresponding difference in $\varphi$ and $\eta$ (of the track and the cluster) was less than $0.1$. 
When being matched, the tracks were propagated to the position of the cluster. 
The particle identification used the specific ionisation energy loss in the \gls{TPC} with a cut of $-1 < n_\sigma^{\rm{TPC}} < 3$, where $n_\sigma$ is the difference between the measured and expected detector response signals for electrons normalised to the response resolution. The purity was further improved by applying a selection in the ratio of the energy of the cluster over momentum of the track ($0.8 < E/p < 1.2$). 
An example of the result of such a study is shown in \Fig{fig:calib_alig_eta_phi_diff_dcal}, where the differences in $\varphi$ and $\eta$ for the \gls{EMCal} \glspl{SM} are reported.
The distributions show a clear peak around 0, showcasing the good performance of the alignment. Nevertheless, it is still possible to see a small remaining misalignment, most visible in the asymmetry of the residuals in the $\varphi$-angle. 
A compilation of all the mean values for each \gls{SM} is shown in \Fig{fig:calib_alig_eta_phi_diff_mean}. 

For the \gls{EMCal}, the track matching studies indicated a residual misalignment in the beam direction (Z-axis) which were added to the alignment matrices. No further alignment was applied in the \gls{DCal}, since the remaining misalignment was smaller than the size of a cell.


\subsection{Cell energy calibration}
\label{sec:e-calibchapter}
The goal of the energy calibration is to obtain for each cell $i$ a coefficient $c_i$ such that when multiplied by the actual cell responses, all cells give the same value to an identical stimulation. 

\subsubsection{Energy pre-calibration} \label{sec-Energy-pre-calibration}
The testing and relative pre-calibration of each \gls{SM} were performed upon assembly in the laboratory using the response to cosmic muons as a \gls{MIP}. \glspl{SM} were calibrated in three sections consisting of 8 strip modules per test.  
The trigger was defined by 16 scintillator paddles positioned in pairs above and below the strip modules.
The size, position and orientation of the paddles were such that a particle that crossed both paddles of a pair only crossed the associated strip module. Both ends of a paddle were instrumented with photomultiplier tubes. When all four photomultipliers of a pair of top and bottom scintillator paddles received a signal a trigger was issued and the data from the entire 1/3 of the SM was read out. The triggering paddle pair was identified, and the times when the signal was observed by each of the four photomultipliers were registered. In the final analysis the timing information allowed reconstruction of the muon crossing position along the top and bottom paddles with $\sim3$~cm precision. With an additional isolation cut (with no signals in neighboring modules) events were selected for use in the calibration. 
This procedure allowed us to obtain an initial tower relative energy calibration with $\approx2\%$ precision, see~\cite{Faivre:2011zz} for details.

\subsubsection{Energy calibration with the LHC data} \label{sec-Energy-calibration-with-the-LHC-data}
Since the collected data does not contain enough $Z^0$ boson $\rightarrow $e$^+$e$^-$ or even $\eta$ meson $\rightarrow \gamma\gamma$ decays~\cite{Acharya:2019rum}, the cells were calibrated by using the \piz\ meson $\rightarrow \gamma\gamma$ decays. 
In this section, ``miscalibration'' will be used for convenience to denote the difference with respect to the experimentally unreachable perfect calibration, regardless of the fact that the cells were or were not calibrated. 
``Residual miscalibration'' will be used when the cells are calibrated.

The procedure consists of measuring the invariant mass distribution of \piz\ meson candidates for which one of the two decay photons has the cluster centroid (see \Eq{eq:ss_centroid}) located in the considered cell. 
This distribution is then fit by the sum of a Gaussian (for the \piz\ meson peak) and a second-order polynomial (for the combinatorial background). The coefficient $c_i$ is obtained from the ratio of the known \piz\ meson mass $M_{\piz}^{\rm PDG}$ to the mean of the fitted Gaussian $\mu_i^{\rm fit}$: $c_i = \left(M_{\piz}^{\rm PDG}/\mu_i^{\rm fit}\right)^n$, where $n$ is a coefficient chosen between 1 and 2. 
While \Eq{eq:KineIMgg} suggests to choose $n=2$ for a random set of not calibrated cells, studies showed that convergence is faster with values around $n=1.5$~\cite{Acharya:2019rum}. 
Since the total shower energy, distributed in several cells, and the energy deposited by the second decay photon are required in the estimate of the invariant mass, the calibration procedure is carried out iteratively.
The cells located on the edges of the \glspl{SM} cannot be calibrated this way, because part of the electromagnetic shower is lost. 

The calibration data sets were obtained in \pp\ collisions using a lower threshold for the \gls{L0} trigger than that used in the data sets for physics analyses, in order to trigger on decay photons from \piz\ meson decays that lead to well-separated clusters in the calorimeter.
These data sets were  taken in 2012, 2015 and 2018, with a \unit[2.5]{GeV} \gls{L0}-trigger threshold, and contain an average number of 17, 41, and 18 thousand events per cell~(\gls{SM} edges excluded), respectively. 
A cell energy threshold of \unit[50]{MeV} and a minimal seed energy of \unit[100]{MeV} were used for the cluster reconstruction, and the nonlinearity of the energy response of the calorimeter was corrected (see~\Sec{sec:EMCalEnergyPositionCalib}). 
To reconstruct \piz\ meson candidates, only approximately circular clusters~($\shshlo < 0.5$) that were not matched to a track were used. 
They also were required to have an associated time less than \unit[20]{ns} relative to the collision time, and a time difference between the two clusters less than \unit[20]{ns}. 
In addition, the cluster energy was required to be larger than \unit[0.7]{GeV} and lower than \unit[10]{GeV}.
Finally, both clusters used to reconstruct the \piz\ meson were required to be in the same \gls{SM}, in order not to be affected by residual misalignment.

Since the cluster energies used for energy calibration are well below 16~GeV, in the operational region of high gain regime~(see \Sec{sec:readout}), the $c_i$ coefficients implicitly refer to high gain channels. 
The calibration of low gain channels is obtained using the test beam measurements where a factor 16.3~(with RMS of 0.15) was found for the low over high gain ratio~(see \Sec{sec:shaperNL}).


\subsubsection{Target energy resolution} \label{sec-7-targ}
\begin{figure}
\centering
\includegraphics[width=9cm]{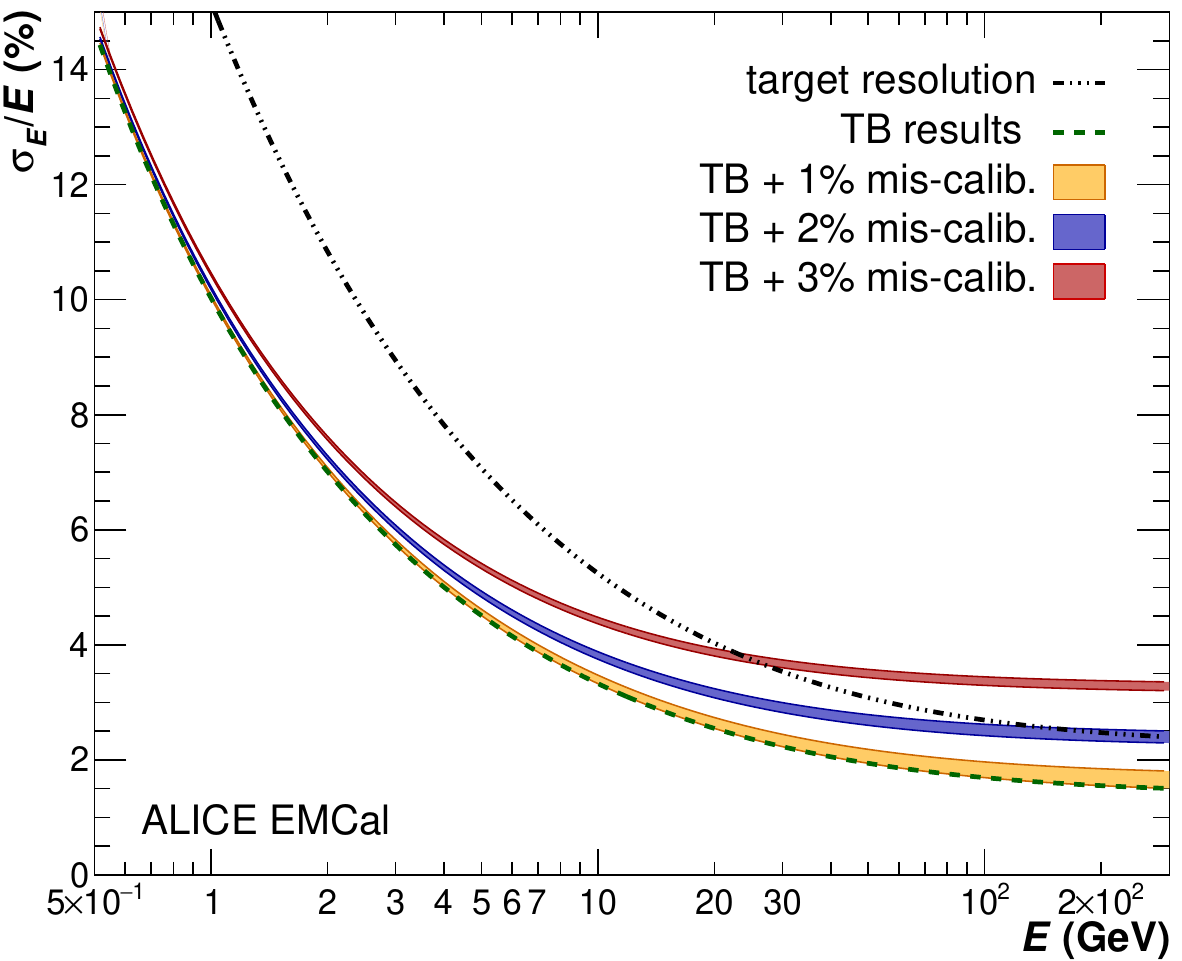}
\hfill
\raisebox{+10mm}{%
\begin{minipage}[b]{6.5cm}
\caption{(Color online) Relative energy resolution as a function of the energy of the incident particle. 
         Displayed are the target resolution (black, dashed-dotted) and the measured energy resolution from the test beam (green, dashed),  \Sec{sec:TBNonLin}.
         Additionally, three different bands are added showing the intrisic resolution with added 1, 2 and 3\% miscalibration, considering the residual miscalibration during the test beam to be between 0\% (upper band limit) and 1\% (lower band limit).}
\label{fig-7-targ-Eresolution}
\end{minipage}
}
\end{figure}

The calorimeter intrinsic energy resolution $\sigma_E^{\rm intrinsic}$ was measured in beam tests, with a relative energy miscalibration estimated to be smaller than 1\%~(\Sec{sec:shaperNL}).
The relative uncertainty $\sigma_E^{\rm calib}/E$ due to the relative miscalibration of the cells adds up quadratically to the constant term $a$ of \Eq{eqn:5-TB-energyResolution}. 
The energy calibration aims at reducing $\sigma_E^{\rm calib}$ down to a level, which makes the total energy resolution compatible with the physics program of the experiment. 
Studies showed that an energy resolution $\sigma_E^{\rm tot} = \sigma_E^{\rm intrinsic} \oplus \sigma_E^{\rm calib}$ of about $(2\% \oplus 15\% /\sqrt{E}) \oplus 1\%$ is sufficient to achieve the physics goals of \gls{EMCal}~\cite{Cortese:2008zza,Allen:2009aa}. 
\Figure{fig-7-targ-Eresolution} shows that a residual miscalibration of 2\% allows to cover the full energy range with the desired resolution, while a residual miscalibration of 3\% would only affect electron and photon measurements above 30~GeV, an energy domain 
where the statistical uncertainties exceed those arising from miscalibration.

Since what is actually measured are reconstructed masses of \piz\ mesons, a relationship has to be drawn between the energy miscalibration and the spread of the \piz\ mesons masses measured in the various cells.

For different instances of actual energy deposited, $E^{\rm true}$, the energy measured in the cell is denoted by $E_{i,k}^{\rm meas}$ where $i$ is an index for the cell, and $k$ corresponds to each measurement being different due to the intrinsic calorimeter resolution. Using $\alpha_i$ to characterise the effect of the non perfect calibration of cell $i$ and $\beta_k$ for the calorimeter resolution, the measured energy can be written as:
\begin{equation}
E_{i,k}^{\rm meas} = E^{\rm true}(1+\alpha_i)(1+\beta_k) \simeq E^{\rm true}(1 + \alpha_i + \beta_k), 
\end{equation}
where $\alpha_i$ and $\beta_k$ both follow Gaussian distributions centred at 0 and of widths $\sigma_E^{\rm calib}/E^{\rm true}$ and $\sigma_E^{\rm intrinsic}/E^{\rm true}$, respectively.

Let us first consider the simple case of a \piz\ meson decay where the energy of each decay photon is fully contained in a pair of single cells $i$ and $j$.
In this case the reconstructed \piz\ meson invariant mass can be expressed as
\begin{equation}
M_{\gamma\gamma}^{\rm reco} = M_{\piz}^{\rm PDG} \sqrt{(1 + \alpha_i + \beta_1)(1 + \alpha_j + \beta_2)} \;\simeq\; M_{\piz}^{\rm PDG} \left(1 + \frac{1}{2} (\alpha_i + \alpha_j + \beta_1 + \beta_2)\right),
\end{equation}
where $\beta_1$ and $\beta_2$ are associated to the energy deposits in cells $i$ and $j$, respectively.

Using index $n$ to number the  \piz\ mesons reconstructed with a photon hitting a cell $i$, the average of the measured invariant masses can be written as
\begin{equation}
\frac{1}{N} \sum_{n = 1}^N M_{\gamma\gamma, n}^{\rm reco} = 
M_{\piz}^{\rm PDG} \left(1 + \frac{1}{2N} \left(\sum_n \alpha_i + \sum_n \alpha_j + \sum_n \beta_{k_1} + \sum_n \beta_{k_2}\right)\right).
\end{equation}
The indices $j$, $k_1$ and $k_2$ depend on $n$, so the $\beta$ terms (average shift induced by the intrinsic energy resolution) and the $\alpha_j$ term (average miscalibration of many cells) vanish.
Therefore, assuming that the fit made on the invariant mass distribution is a good estimate of the average reconstructed mass, $\mu_i^{\rm fit} \simeq M_{\piz}^{\rm PDG} \left(1 + \frac{\alpha_i}{2}\right)$.

The distribution of $\mu_i^{\rm fit}/M_{\piz}^{\rm PDG}$ is therefore expected to follow 
a Gaussian with half of the width of the corresponding
distribution of $\alpha_i$.
A \com{toy-}simulation confirmed this result, and therefore the uncertainty values reported below on the measurement of \piz\ meson masses will be multiplied by 2 to obtain the corresponding uncertainty on an energy measurement.

\subsubsection{Consequences of the upstream material budget} 
\label{sec-7-mate}
As the \piz\ meson life time is very short with $\tau_{\pi^0}\approx8.5\times10^{-17}$s, the decay photons must cross the material of all the central \gls{ALICE} detectors before reaching the calorimeter. 
Conversions happening before the first half of the \gls{TPC} are efficiently rejected by the veto applied when the cluster is matched to a charged-particle track, see \Fig{fig:trmatcheff}. 
Conversions in \gls{TRD} and  \gls{TOF} are partially removed by the requirement on the shower shape because the positron and electron are separated by the magnetic field. 
However, the remaining clusters suffer from an energy loss in the material crossed after the conversion. 
This results in a slightly more pronounced tail of the \piz\ meson peak at low invariant masses (see \Fig{fig:Pi0ClustersandDecompMassPeaksFine}). 
The limited number of  \piz\ mesons in the most affected cells does not allow for adding more free parameters to the fit function in order to describe 
the asymmetry in the invariant-mass peak.\\
Columns 10--13 and 38--43 (0 being towards midrapidity) were defined as zones with more material, as the space frame and \gls{TRD} module borders induce a very significant widening of the \piz\ meson peak width and a loss of the number of collected \piz\ mesons up to a factor of 2. 
Calibrating these cells therefore requires more data than the other areas, and they are treated differently as explained in \Sec{sec-7-shap}.\\
In addition to the more pronounced tail, the average invariant mass is also shifted to lower values, which needs to be accounted for in the absolute energy calibration.
The corresponding correction depends slightly on the geometry of the super-module as well as on the average material installed in front of it.  
Thus, it ranges from 0.6\% for the elongated \glspl{SM} to up to 1.4\% for the 2/3 \glspl{SM}.  

\subsubsection{Statistical uncertainty on the measured $\pi^{0}$ mass}
\label{sec-7-stat}
The statistical uncertainty on the estimation of the reconstructed \piz\ meson mass $\mu_i^{\rm fit}$ depends on the accumulated number $N_{\piz,i}$ of \piz\ mesons in the considered cell $i$. 
This uncertainty is therefore estimated in different bins of $N_{\piz}$, and is obtained by comparing the mass fitted in two independent event samples, {\tt S1} and {\tt S2}, for each cell.
The data used for this study is the large 2015 calibration data set.

In order to minimize the influence on the result of possible miscalibration among the cells of both photon clusters, the sample {\tt S2} was finely calibrated, until the width of a Gaussian fit on the distribution of the fitted masses $\mu_i^{\rm fit}$ reached 0.1\%. 
The width of the same distribution for the sample {\tt S1} remained almost constant during the convergence process, which confirms that it is dominated by statistical variations and not residual miscalibration.

\begin{figure}
\centering
\includegraphics[width=0.49\textwidth]{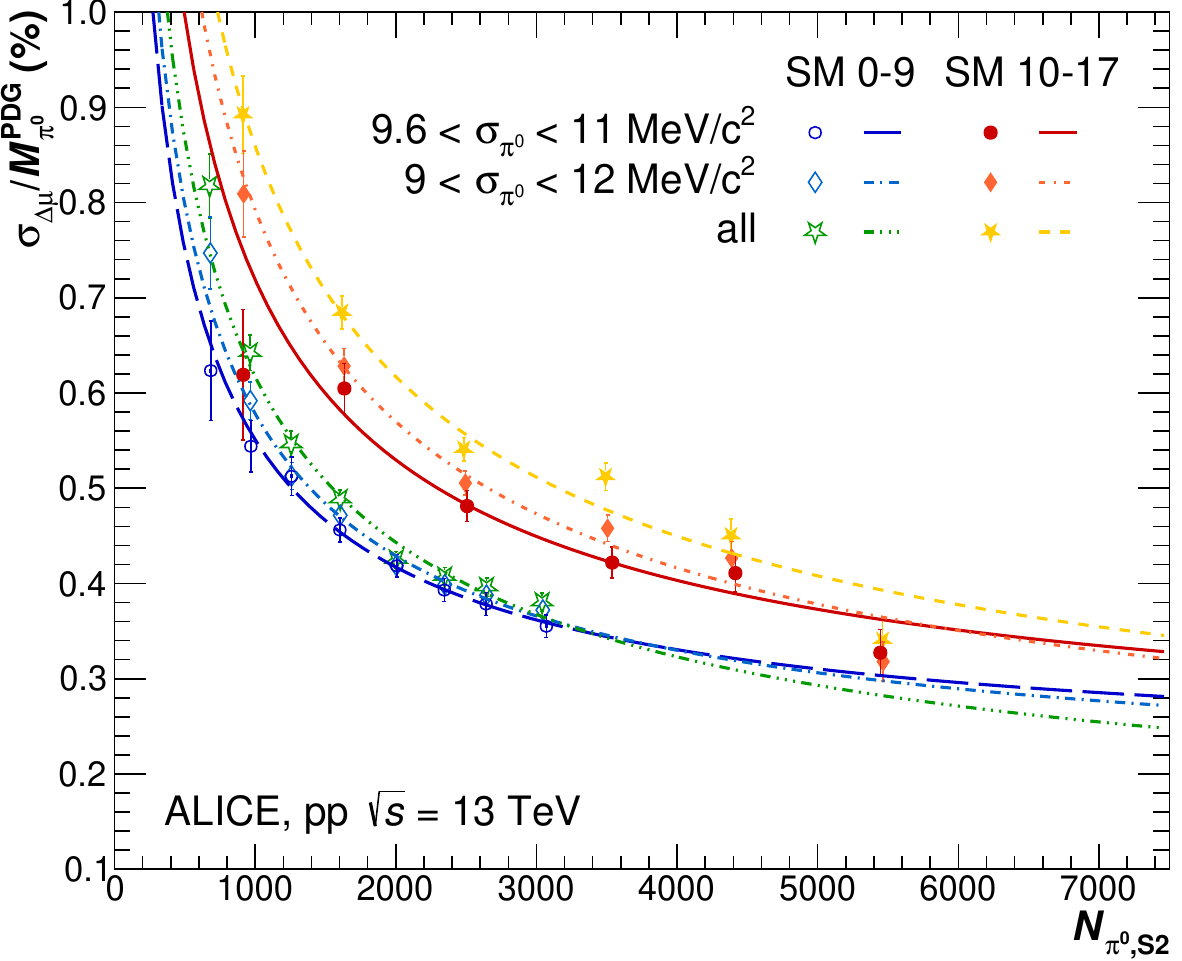}
\includegraphics[width=0.49\textwidth]{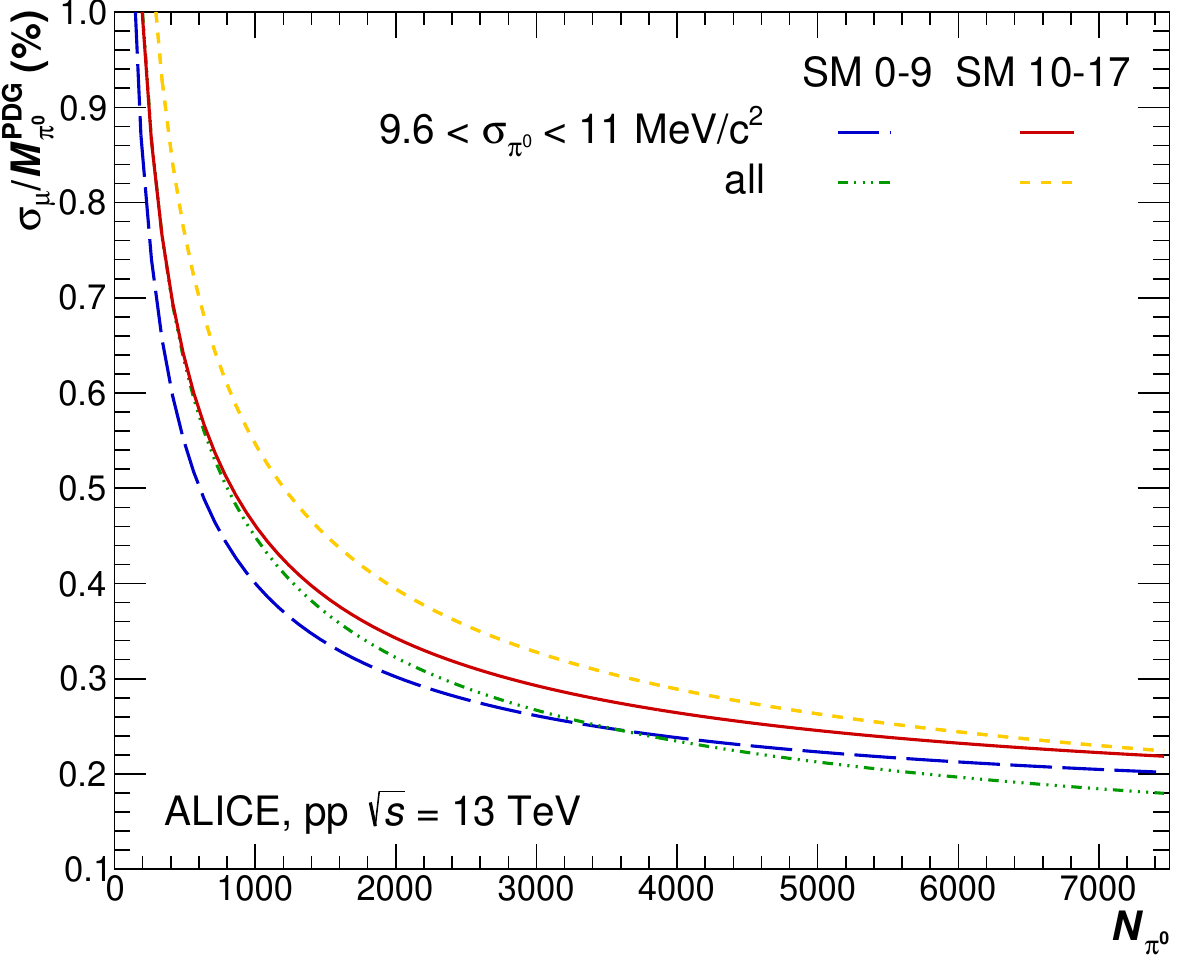}
\caption{(Color online) Left: Width of the distribution of $\Delta \mu_i$ (see text) as a function of the number of  \piz\ mesons in sample {\tt S2}, for \glspl{SM} 0-9 (blue, green, cyan) and \glspl{SM} 10-17 (red, orange, yellow), for various $\sigma_{\piz}$ cell selection: $9.6 < \sigma_{\piz} < 11.0$~MeV/$c^2$ (circles), $9.0 < \sigma_{\piz} < 12.0$~MeV/$c^2$ (diamonds) and none (stars). Right:~Statistical uncertainty on the \piz\ meson mass as a function of the number of  \piz\ mesons collected in the cell, for \glspl{SM} 0--9 with the tightest (blue) and without (green) selection on $\sigma_{\piz}$ as well as for \glspl{SM} 10--17 with the tightest (red) and without (yellow) selection on $\sigma_{\piz}$.}
\label{fig-7-stat-widthVsStatisticsLargerSigmasStatUncert}
\label{fig-7-stat-statisticalUncertaintiesSigmasStatUncert}
\end{figure}

The uncertainty on the measurement of the mass difference in cell $i$, $\Delta \mu_i = \mu_{i,{\text{\tt S1}}}^{\rm fit} - \mu_{i,{\text{\tt S2}}}^{\rm fit}$, is $\sigma_{\Delta \mu, i} = \sqrt{\sigma_{\mu,i,\text{\tt S1}}^2 + \sigma_{\mu,i,\text{\tt S2}}^2}$. 
Since the analysis is performed independently on classes of cells with similar number of  \piz\ mesons, it is reasonable to assume that the uncertainty on the mass measurement is the same for all cells belonging to the considered class.
The width $\sigma_{\Delta\mu,i}$ can be written as $\sigma_{\Delta\mu} = \sqrt{\sigma_{\mu,\text{\tt S1}}^2 + \sigma_{\mu,\text{\tt S2}}^2}$ and the $\Delta \mu_i$ distribution for a class is expected to be a Gaussian centered at 0, of width $\sigma_{\Delta \mu}$. 
The width obtained by a Gaussian fit for each statistics interval $N_{\piz,\text{\tt S2}}$, is reported in \Fig{fig-7-stat-widthVsStatisticsLargerSigmasStatUncert}~(left).

To obtain the statistical uncertainty on the fitted \piz\ meson mass $\mu_i^{\rm fit}$ as a function of the number of \piz\ mesons collected, we require a functional shape $\sigma_\mu (N_{\piz})$ and a relation between $\sigma_{\mu,\text{\tt S1}}$ and $\sigma_{\mu,\text{\tt S2}}$. 
The latter can be achieved by introducing the ratio $k = N_{\piz,\text{\tt S2}} / N_{\piz,\text{\tt S1}}$. 
This ratio was measured in each cell, and its distribution differs for \glspl{SM} 0--9 and 10--17 due to varying data-taking conditions (see summary \Tab{tab-7-stat-fitParametersStatUncert}); the study was thus made separately on the two sets of \glspl{SM}. 
For each of them, the profile of the 2-dimensional distribution of $(N_{\piz,\text{\tt S1}} ; N_{\piz,\text{\tt S2}})$ appeared to be linear over the full range of number of  \piz\ mesons, and was fitted with the function $N_{\piz,\text{\tt S2}} = k \times N_{\piz,\text{\tt S1}}$. 
To describe the $\sigma_\mu (N_{\piz})$ behavior, the function $\sigma_\mu (N_{\piz}) =  a / \sqrt{N_{\piz}} \oplus b$ was assumed. 
The resulting uncertainty on $\Delta \mu$ as a function of the number of  \piz\ mesons in sample {\tt S2} can be written as
\begin{equation}
\sigma_{\Delta \mu} (N_{\piz,\text{\tt S2}}) \;\simeq\; \sigma_\mu (N_{\piz,\text{\tt S2}}/k) \oplus \sigma_\mu (N_{\piz,\text{\tt S2}}) \;=\; \sqrt{ \frac{a^2}{N_{\piz,\text{\tt S2}}} (k+1) + 2 b^2} \label{eq:statisticalTerm}
\end{equation}

\begin{table}
\centering
\caption{Values of the ratio $k$ and of parameters $a$ and $b$ of the function used to fit the data using \Eq{eq:statisticalTerm}, for the two sets of \glspl{SM} and 3 selections of $\sigma_{\piz}$.}
\begin{tabular}{rl|lll}
  & $\sigma_{\piz}$ interval (\unit{MeV/$c^2$}) & $k$     & $a$ (\unit{MeV/$c^2$}) & $b$ (\unit{MeV/$c^2$}) \\ \toprule
                  & [9.6; 11.0]         & 0.878   & $15.8 \pm 0.8$  & $0.20 \pm 0.029$ \\
  \gls{SM} 0--9   & [9.0 ; 12.0]        & 0.883   & $17.1 \pm 0.6$  & $0.18 \pm 0.027$ \\
                  & None                & 0.885   & $18.9 \pm 0.5$  & $0.11 \pm 0.04$ \\ \midrule
                  & [9.6; 11.0]         & 1.482   & $18.6 \pm 1.2$  & $0.20 \pm 0.05$ \\
  \gls{SM} 10--17 & [9.0 ; 12.0]        & 1.486   & $21.1 \pm 0.8$  & $0.14 \pm 0.05$ \\
                  & None                & 1.488   & $22.9 \pm 0.8$  & $0.15 \pm 0.06$ \\ \bottomrule
  
\end{tabular}
\label{tab-7-stat-fitParametersStatUncert}
\end{table}

The results, shown in \Fig{fig-7-stat-widthVsStatisticsLargerSigmasStatUncert}~(left) and in \Tab{tab-7-stat-fitParametersStatUncert}, reveal a moderate increase of the uncertainty when the selected range of the width $\sigma_{\piz}$ of the fitted \piz\ meson peak is made wider. 
A larger uncertainty is also found for \glspl{SM} 10--17 with respect to \glspl{SM} 0-9. 
\Figure{fig-7-stat-statisticalUncertaintiesSigmasStatUncert}~(right) summarizes the behavior of the corresponding functions $\sigma_\mu (N_{\piz})$, over a typical range of number of \piz\ mesons for a typical energy calibration dataset. 
To be conservative, the final uncertainty chosen is that obtained with \glspl{SM} 10--17 and without any selection on the \piz\ meson peak width.

\subsubsection{Uncertainty on the $\pi^{0}$ mass due to the fit} \label{sec-7-shap}
\begin{figure}
\centering
\includegraphics[width=0.6\textwidth]{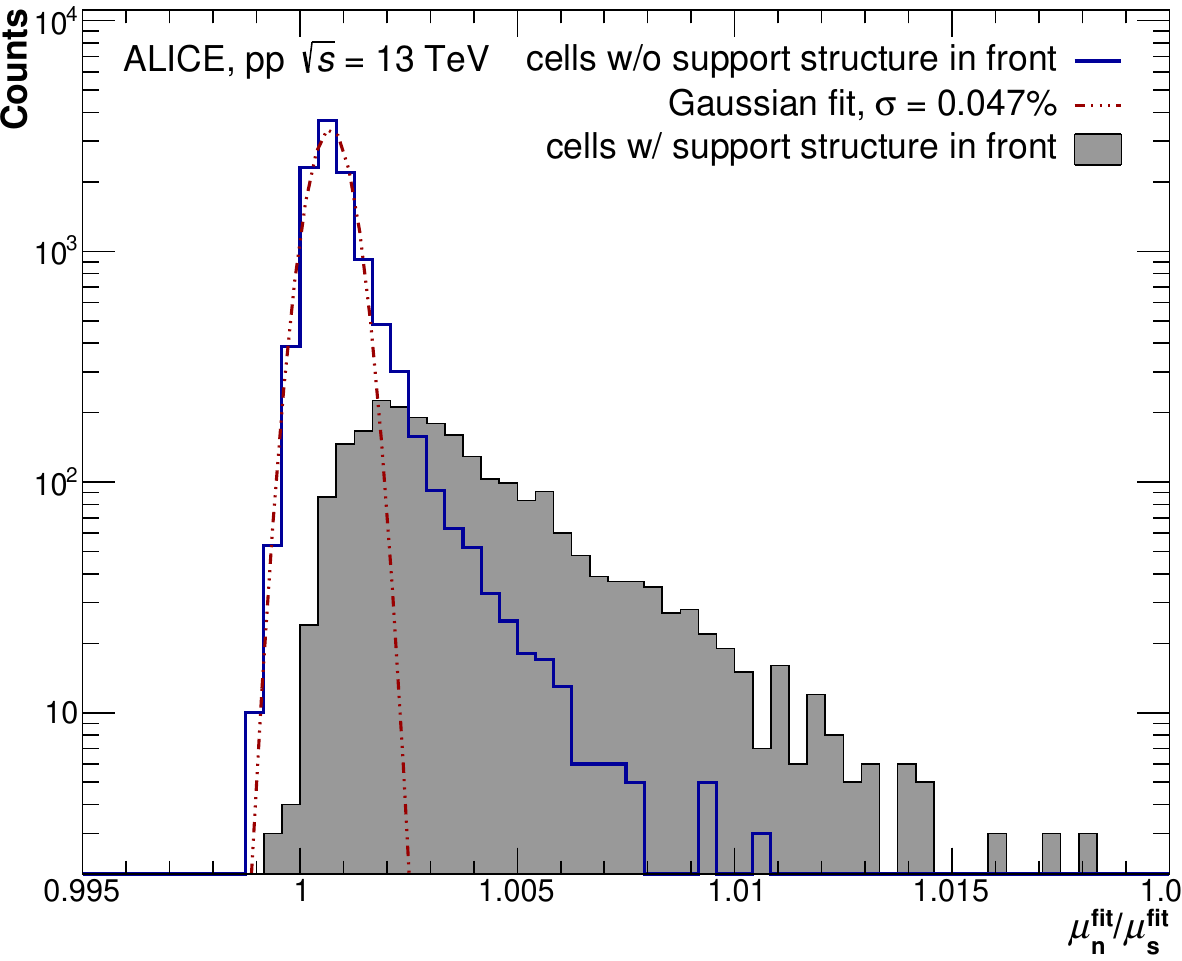}
\hfill
\raisebox{+20mm}{%
\begin{minipage}[b]{0.38\textwidth}
\caption{Distribution of the ratios of the \piz\ meson masses found by a fit in the narrow interval ($\mu^{\text{fit}}_n$) of $70 <\mu_\pi^{0} < 220$~MeV/$c^2$ with respect to the standard range ($\mu^{\text{fit}}_s$) of $50 <\mu_\pi^{0} < 300$~MeV/$c^2$, for cells located behind the zones with more material (filled distribution) and for the other cells (blue distributions) for \glspl{SM} 0--17. The latter histogram is fit with a Gaussian (red).}
\label{fig-7-shap-distribMassRatios}
\end{minipage}
}
\end{figure}

In order to ensure a good fit convergence for cells with low number of \piz\ mesons, a wide standard range of $50 \leq M_{\gamma\gamma} \leq \unit[300]{MeV}/c^2$ has been chosen for fitting the $\gamma\gamma$ invariant mass distributions.
The uncertainty due to the fit in the estimation of the \piz\ meson mass was estimated by reducing this range down to $ 70 \leq M_{\gamma\gamma} \leq \unit[220]{MeV}/c^2$.
When tightening either of the fit limits, the fitted \piz\ meson mass evolved monotonically to larger values. 
\Figure{fig-7-shap-distribMassRatios} shows the distribution of the ratios $\mu^{\rm fit}_{i,n}/\mu^{\rm fit}_{i,s}$, where $n$ and $s$ refer to the narrowest and standard intervals, which were used for fitting the \piz\ meson mass distribution. 
The cells were divided into two samples: those located behind the zones with more material and those with less material.
The distribution for the latter cells exhibits a bulk which follows a Gaussian, and a tail which extends up to a 1\% deviation. 
A Gaussian fit indicates that the relative width of the bulk, 0.047\%, is almost an order of magnitude lower than typical values of the statistical uncertainty of the \piz\ meson mass. 
The uncertainty due to the fit range can thus be neglected. 
This argument does not hold for the tail of this distribution. 
Yet, the cells that exhibited a change of the fitted \piz\ meson mass by more than 0.2\% were found to be located next to excluded cells or \glspl{FEE}. 
Since the analyses sensitive to such effects use clusters with a centroid located more than one cell away from excluded cells, such cells can only give a minor contribution to the total cluster energy. 
There is therefore no expected impact on the analyses.
Finally, the cells located behind the zones with more material exhibit a stronger sensitivity to the fit range, which was taken into account in the estimation of the global uncertainty~(\Sec{sec-7-resu}).

\subsubsection{Consequence of a front-end electronic card change on the tower gains} \label{sec-7-FECch}

Aging and damages induced by glitches on the electrical network require in some occassions to replace \gls{FEE} cards. 
The \gls{FEE}-cards control the bias voltages, and since the voltages that are actually applied to the \gls{APD} are not strictly equal to the desired set voltage, changing an \gls{FEE} card results in a measurable voltage modification, causing a change of the gain in the towers.
The actual output voltages of the channels of three \gls{FEE} cards were measured for three values of the voltage setting: 300, 335 and \unit[370]{V}. 
The distribution of the measured output voltages was fitted with a Gaussian. 
The width was found to be \unit[1.3]{V}, in the worst case, and without significant variations with the set voltage.
The transformation of these voltage variations into an uncertainty on the tower gains $M$ is done using \Eq{eqn-2-gain-gainVsHV}. 
For each tower of \glspl{SM} 0-17, a random voltage change $\Delta V$ is picked according to the \unit[1.3]{V} wide Gaussian.
The value of the voltage set $V_{\rm set}$ and the three Gaussian parameters $(p_0 ; p_1 ; p_2)$  are read from the databases, and the induced gain change is calculated as $M_{\rm new}/M_{\rm old} = M(V_{\rm set} + \Delta V)/M(V_{\rm set})$. 
This variable was found to follow a Gaussian distribution with relative width of 2.5\%, which is thus the gain spread to expect when an \gls{FEE} card is changed.

\subsubsection{Consequence of inactive calorimeter areas on the reconstructed \texorpdfstring{$\pi^{0}$ }\, mass}\label{sec-7-FECoff}

Due to a change of the acceptance available to build pairs of clusters, the invariant mass distributions of the clusters from towers close to an inactive or dead area of a \gls{SM} have a different combinatorial background shape, possibly affecting the fit result. 
Therefore, the 2015 calibration dataset was used to quantify the \piz\ meson mass displacement, which is induced by masking some areas of a \gls{SM} before reconstructing the data.
Various sizes and placements of the masked area were tested. 
In cells located in the same columns (respectively: rows) as the masked area, the \piz\ meson masses were found to be systematically displaced to lower (resp. larger) masses. 
Masking 2 to 5 T-Card wide areas (8 rows and 4-10 columns) leads to significantly larger displacements as the size of the masked area is strongly correlated to the shift in the mass peak position.

Masking a 4-T-Card wide area in the center of a \gls{SM} can lead to a shift in the \piz\ meson mass of up to 1\% in some towers. 
Masking T-Cards close to an edge~(rows 0 to 7) leads to similar results, except that the mass change in the same columns remained smaller than 0.3\% even for 4 masked T-Cards. 
When several full rows or several full columns were masked instead, the mass shifts obtained remained below 0.5\%.
From these results we conclude, that when an area of the calorimeter has no or little calibration data, the \piz\ meson mass found in cells with a distance of less than 4 columns (excluding those that share an edge with the missing area) differs by up to 1\% from what would have been obtained if that area had collected data. 
For cells located farther away, the bias is smaller than the statistical uncertainty.

\subsubsection{Energy calibration performance} 
\label{sec-7-resu}

The global uncertainty due to the relative calibration results from the uncertainties listed above, should be combined with the remaining difference between the achieved calibration level and perfect calibration of the data sample. 
This difference was quantified on the calibration data sample from the width the reconstructed \piz\ meson mass distribution in the various calorimeter cells at the end of the calibration procedure. 
As this width can be made arbitrarily small, the calibration process was continued until it reached a value, which could be neglected with respect to the statistical uncertainty.

\begin{figure}
\centering
\includegraphics[width=0.484\textwidth]{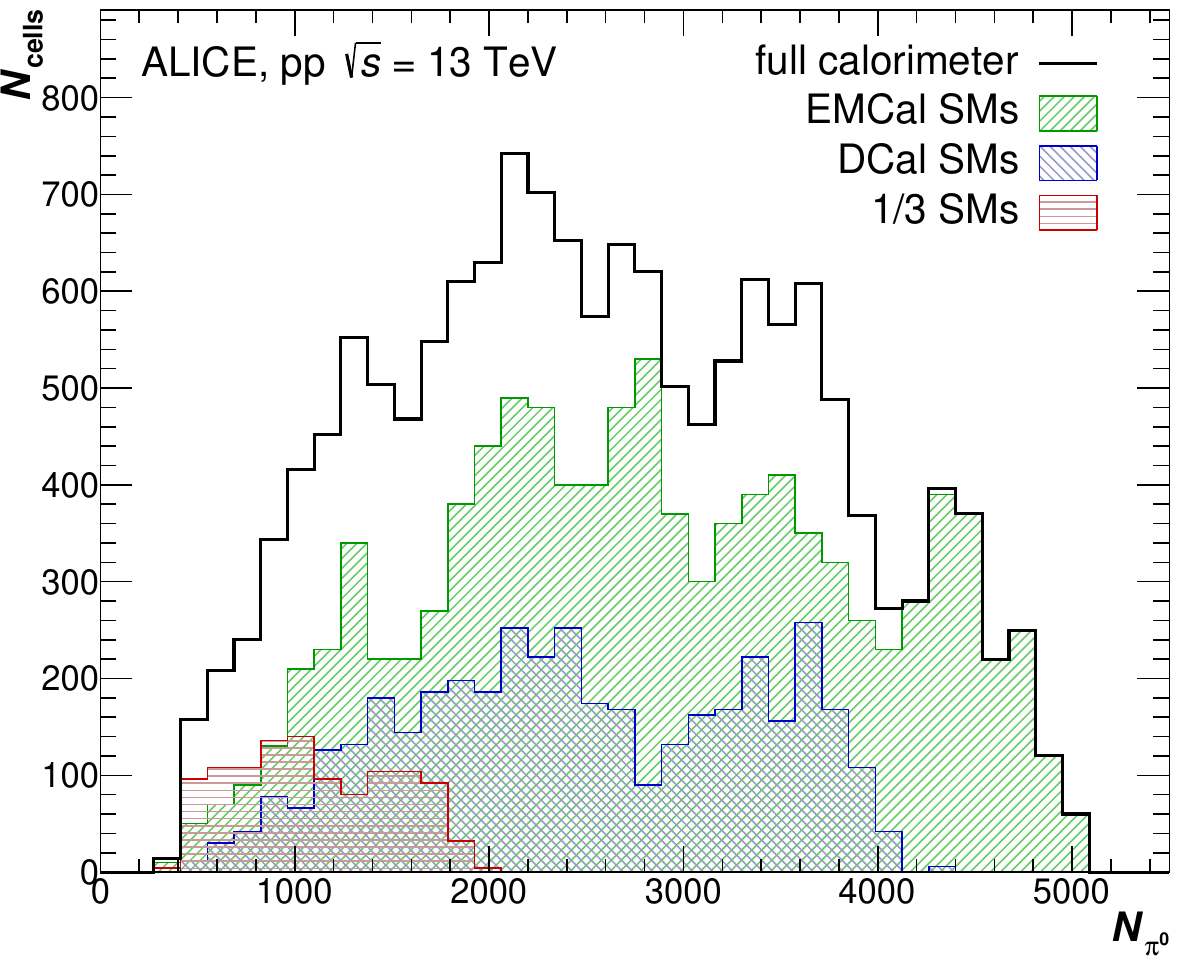}
\includegraphics[width=0.50\textwidth]{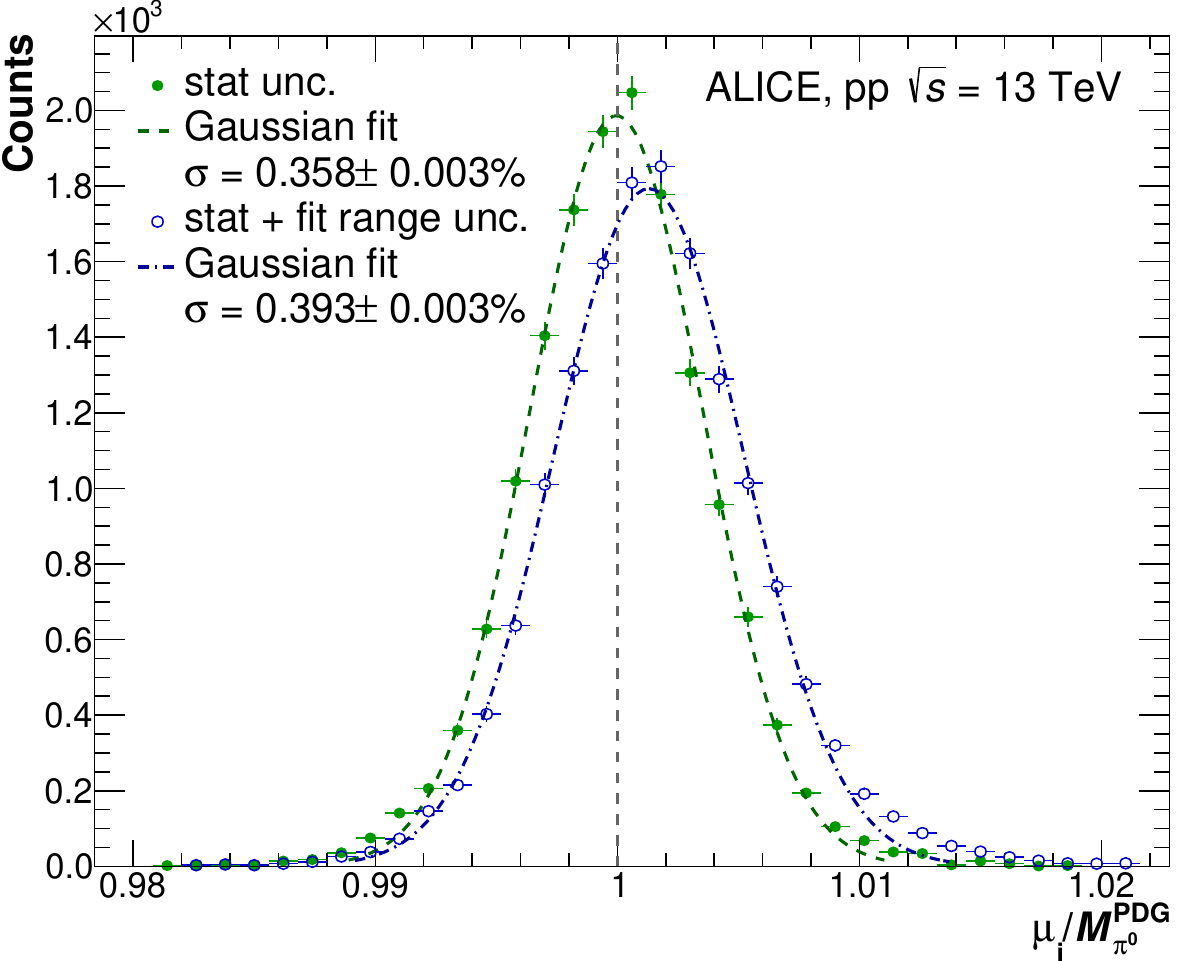}
\caption{(Color online) Left: Distribution of the number of  \piz\ mesons collected in an ideal calorimeter (black). 
        The respective contributions of the 10 \gls{EMCal}, 6 2/3-sized \gls{DCal} and 4 1/3-sized \glspl{SM} are displayed in different colors (see \Sec{sec:hardware-supermoduledesign}).
        Right: Distributions of the smeared \piz\ meson masses normalized by $M_{\piz}^{\rm PDG}$, when only the statistical uncertainty is applied (closed symbols), and when also the uncertainty due to the fit range is applied (opened symbols).}
\label{fig-7-resu-hDistrStatInSyntheticCalo}
\label{fig-7-resu-hDistrGlobalUncert}
\end{figure}

To estimate the global uncertainty on the calibration, a distribution of the average masses obtained in each cell was built by applying on the fixed $M_{\piz}^{\rm PDG}$ value a random coefficient that reproduces the various effects.
The statistical uncertainty was reproduced by picking for each cell a coefficient according to a Gaussian distribution of width equal to the value $\sigma_{\mu}(N_{\piz})$, where $\sigma_{\mu}(N_{\piz})$ is the function established in \Sec{sec-7-stat} and $N_{\piz}$ is the expected number of  \piz\ mesons in this cell according to its position. 
The collected number of  \piz\ mesons per cell is indeed not uniform, 
due on one hand to the requirement that both photons hit the same \gls{SM}, and on the other hand to the non-uniform upstream material budget.
An ideal distribution for the three types of \gls{SM} sizes was built from real distributions to mimic a \gls{SM} without dead cells and where all \gls{FEE} cards would have registered data from the same amount of collisions. 
\Figure{fig-7-resu-hDistrStatInSyntheticCalo}~(left) shows the resulting distribution for the number of  \piz\ mesons in the various cells for an ideal full calorimeter. 
The variations of the number of reconstructed $\piz$ mesons per cell arise from areas of the calorimeter with different material budget in front of them~(see \Sec{sec-7-mate}).

For the systematic uncertainty due to the fit range, the random coefficient was picked according to the relevant distribution of \Fig{fig-7-shap-distribMassRatios}, depending on whether the considered cell was or was not placed in a zone with more material.
\Figure{fig-7-resu-hDistrGlobalUncert}~(right) shows that the overall smearing of the \piz\ meson mass based purely on the statistical uncertainty amounts to 0.36\%. 
Considering in addition the uncertainty due to the fit range, the mass smearing rises to 0.39\%, with a tail at larger values.

Since the calibration uncertainty $\sigma_E^{\rm calib}$ is twice the uncertainty on the \piz\ meson mass~(see \Sec{sec-7-targ}), we concluded that $\sigma_E^{\rm calib}$ is lower than 1\% and therefore meets the $\sigma_E^{\rm calib} < 3\%$ requirement.

Since, as shown in \Sec{sec-7-FECoff}, missing parts of the calorimeter only induce mass shifts below 1\% and in a limited fraction of their \gls{SM}, such a result suggests that a satisfactory level of calibration could be achieved with a smaller number of collision events than have been taken in 2015. 
However, a larger size of the calibration sample is very useful to obtain a minimum of reconstructed \piz\ mesons  to guarantee fit convergence.
Moreover, it allows to check the calibration quality even for the cells with reduced entries, e.g. due to the non-uniformity of the reconstructable \piz\ mesons, to the larger \piz\ meson peak width and smaller signal-to-noise ratio in zones with more material, and to sometimes imperfectly working cells or \gls{FEE} cards.

Finally, a \gls{FEE} card change as studied in~\ref{sec-7-FECch} adds a 2.5\% spread on $\sigma_E^{\rm calib}$. 
Since this remains within the conditions established in \Sec{sec-7-targ}, and since only a fraction of the \gls{FEE} cards of the calorimeter are changed during the data-taking period to which the calibration applies, we concluded that the energy calibration is satisfactory.

\subsection{Bad channel masking procedure} 
\label{sec:badchannel}
\begin{figure}[t!]
\includegraphics[height=6.9cm]{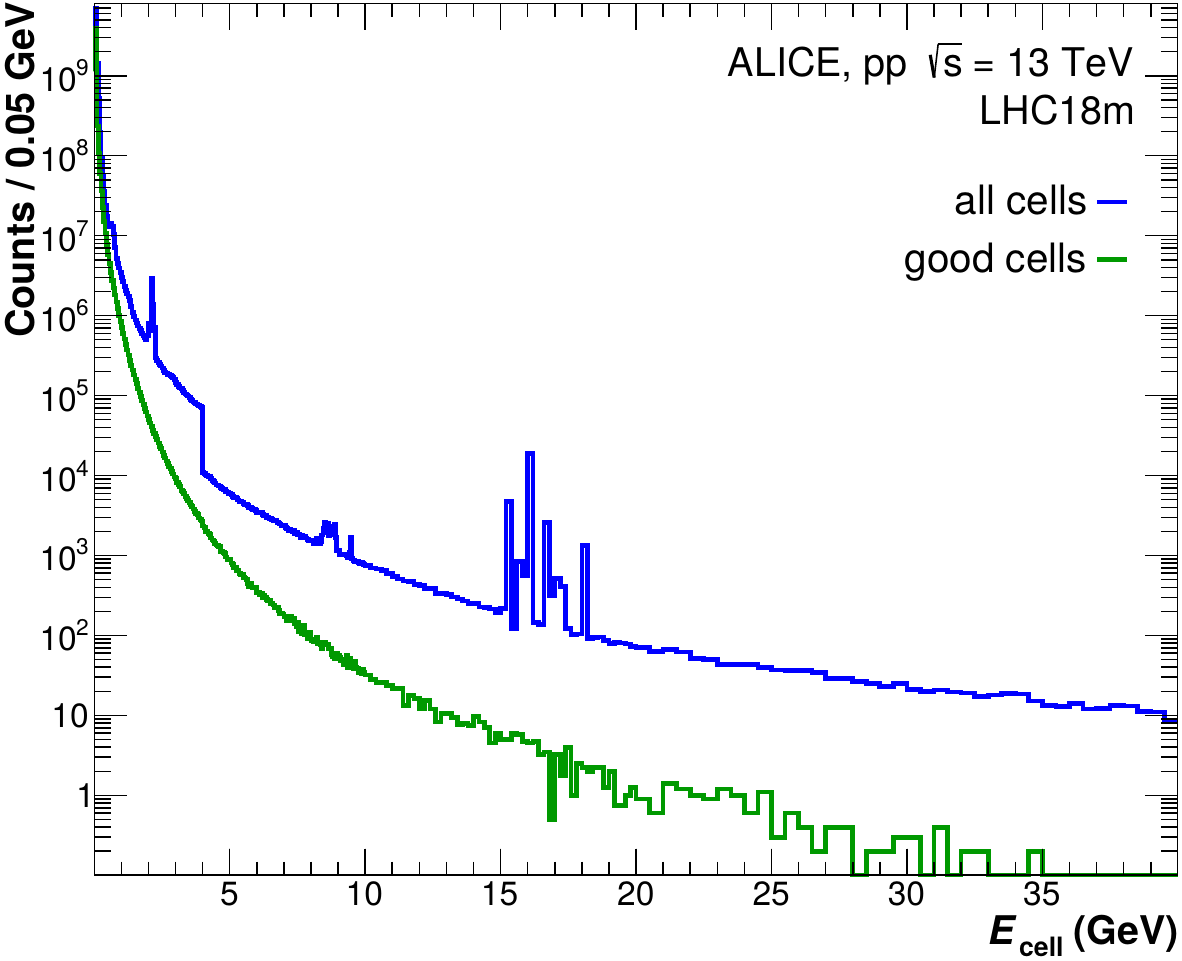}
\includegraphics[height=6.9cm]{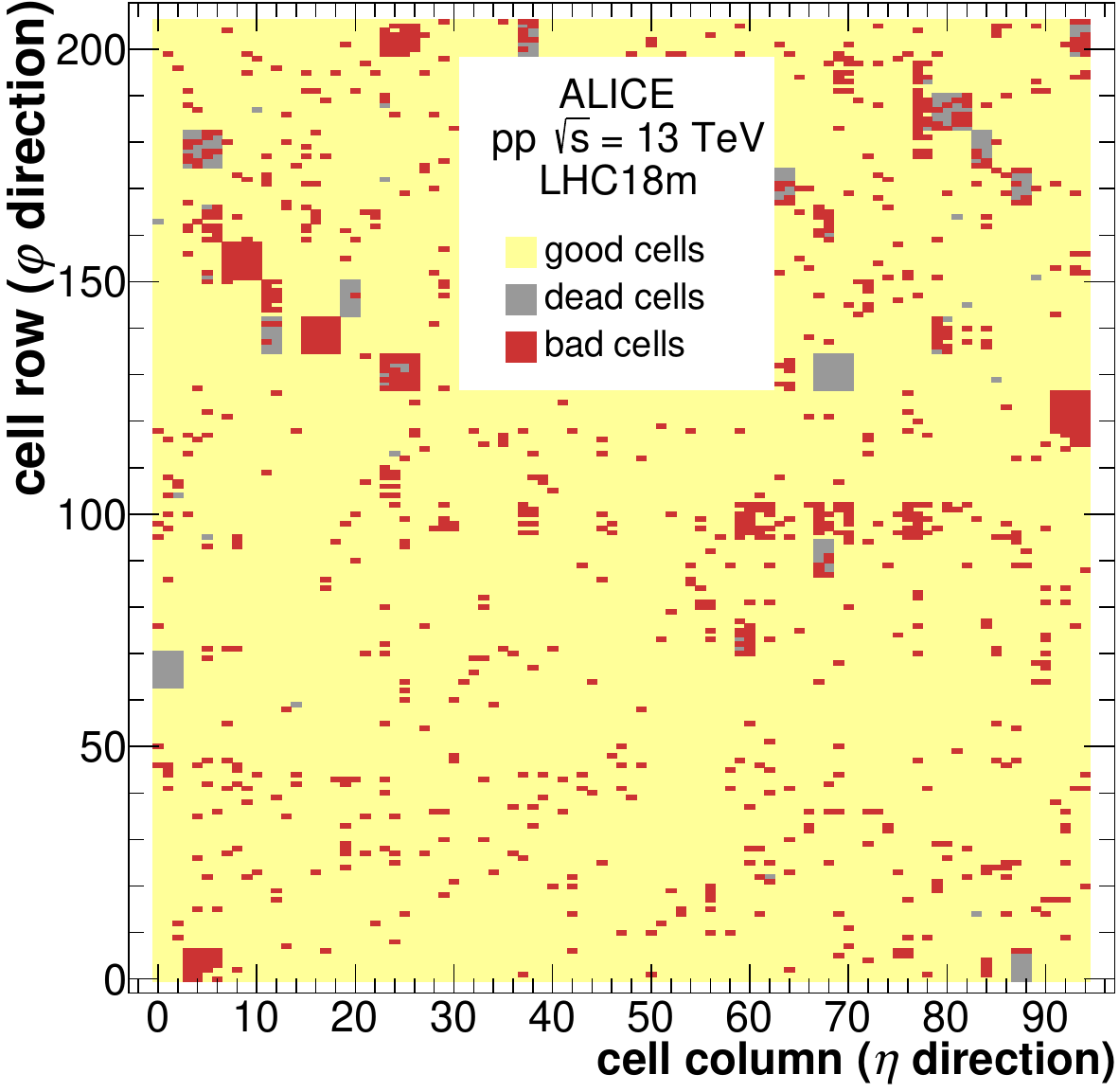}
\caption{(Color online) Left: Energy distribution of all cells (blue), and all good cells (green) of the \gls{EMCal}. 
	Right: Map of cells classified as good (yellow), bad (red) and dead (gray) for the part of the data taking period LHC18m as a function of row and  column.}
\label{fig:7-Bad-badChannel}
\end{figure}

Some channels in the calorimeter give an improper response to a hit, are noisy or have a discontinuous energy spectrum. 
These so-called {\it bad channels} need to be masked and excluded from any data analysis.
\Figure{fig:7-Bad-badChannel} shows the energy spectrum of all channels of \gls{EMCal} in blue and the energy spectrum without the bad channels in green. 
The former clearly shows kinks and spikes that do not originate from the measured energy of particles hitting the detector but most likely originate from bad channels.

To identify these channels, we used two observables: the mean energy per channel and the number of hits recorded by a channel.
Since some cells gave an improper response only over certain energy ranges, we evaluated these two quantities independently in several energy intervals $\Delta E$. 
This way, the analysis is sensitive to faulty behavior in all energy regions. 
The identification of the bad channels relies on the assumption that the number of hits per channel only differs due to statistical fluctuations.
Strictly speaking, this assumption is not valid for all channels of the \gls{EMCal}, mainly due to material in front of the calorimeter and due to the non-uniform rapidity distribution of produced particles. 
As mentioned in \Sec{sec-7-mate}, the support structures in front of the \gls{EMCal}  affect the abundance and energy of particles traversing the detector. 
Cells behind the \gls{TRD} support structures record significantly fewer hits than an average cell. 
The position of those cells can be parametrized as a function of $\varphi$ (row) and $\eta$ (column). 
Furthermore, the particle-rapidity distribution is not flat in $\eta$ and therefore also affects the number of hits in a cell as a function of the cells position in $\eta$ (column).

\begin{figure}[t!]
	\centering
  \includegraphics[width=0.7\textwidth]{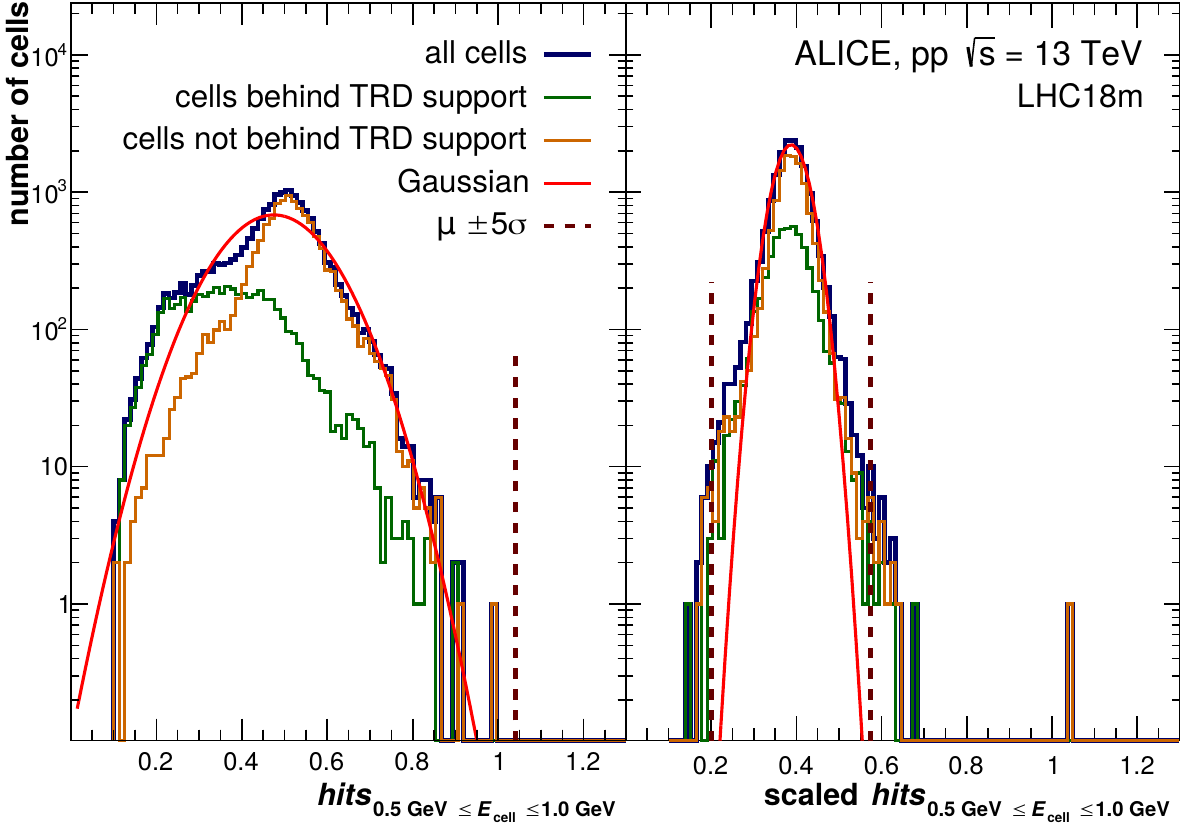}
  \hspace{0.1cm}
	\raisebox{+10mm}{%
    \begin{minipage}[b]{4.4cm}
    \caption{(Color online) Hit distribution for 0.5~GeV $ \leq E_{\text{cell}} < $ 1.0~GeV before (left) and after (right) the scaling procedure. All cells are shown in blue, cells behind the TRD support structure in green, and cells not behind the TRD support structure in orange.}
    \label{fig:7-Bad-badChannel2}
    \end{minipage}
  }
\end{figure}

To correct for these effects, we scaled the number of hits in a cell according to the mean number of hits per row and per column $\langle \text{hits} \rangle _{\rm row/column}$ in which the cell is positioned.
Since potential bad channels should not be included in the calculation of the scaling factor, we identified and excluded them with an algorithm described later. 
The corrected number of hits for a cell was determined by: $N_{\rm hits}^{\rm after} = N_{\rm hits}^{\rm before} \times  \frac{\langle \text{hits} \rangle _{\rm global}}{ \langle \text{hits} \rangle _{\rm row/column}}$. 
Here, $N_{\rm hits}^{\rm after}$ is the number of hits after the correction, $N_{\rm hits}^{\rm before}$ is the number of hits before the correction and $ \langle \text{hits} \rangle _{\rm global}$ is the mean number of hits of all channels of the calorimeter.
Since we applied the correction for the rows and for the columns separately, the total correction was determined in an iterative procedure, alternating between the correction for the row and for the column. 
After four iterations the mean number of hits for all rows and columns no longer changed. 
\Figure{fig:7-Bad-badChannel2} shows the hit distribution in an energy range of 0.5~GeV $ \leq E_{\text{cell}} < $ 1.0~GeV before the scaling procedure (left) and after (right) the scaling procedure. The cells behind the  \gls{TRD} support structures clearly differ from the other cells in the distribution.
The hit distribution before scaling is only shown to illustrate the importance of the scaling procedure and is not used to identify bad channels.
After this initial adaptation, the bad channel analysis can be performed.

The analysis distinguishes three classes of channels: good, bad, and dead channels.
If a channel has zero recorded hits, the analysis classifies this channel as dead. 
These are typically channels where an \gls{FEE} card was switched off for replacement.
The identification of the bad channels was done by selections on the mean energy distribution and the hit distribution in all selected energy intervals $\Delta E$. 
The distributions obtained after applying the scaling procedure described above, as shown in \Figure{fig:7-Bad-badChannel2} (right), were fitted with a Gaussian and the standard deviation $\sigma$ was used as a reference for the threshold to tag a channel as bad. 
We typically use $\mu \pm 5\sigma$ as a selection criterion although there can be slight variations, depending on the period. 
Channels outside this interval were declared as bad.

Furthermore, a cut was applied to reject cells that fire frequently outside the nominal expected hit time: 
cells which fired outside the time interval of 550 to 700~ns\footnote{Typically the first step in the bad channel masking is performed before the cell time calibration~(\Sec{sec:timeCalib}). 
Before the time calibration the cell time distribution peaks in the given time interval as can be seen in \Fig{fig:emcalTime}.} with a high frequency were classified as noisy and thus bad for physics analysis.

The bad channel masking used in the data analysis was not defined run-by-run.
Instead, the same bad channel masking was applied to multiple consecutive runs exhibiting similar detector conditions, so-called \textit{run blocks}. 
To identify these run blocks, a preliminary bad channel analysis was performed for each run individually. 
A channel that was declared as bad by a selection on one distribution, was declared as bad for the respective run. 
Only cells that were flagged as bad in more than 20\% of the runs were considered for the algorithm that determines the run blocks.
In a second step, the individual bad channels were compared among all the runs to find similar behavior. 
The algorithm itself minimizes the channels that need to be masked for the entire run block even though they only misbehaved for a part of the runs. 
It further takes into account that runs do not contain an equal number of events, thus, to minimize unnecessary masking, the number of events of the run is used as a weight in the definition of masked channels.
After the runs which get a common map were found by the algorithm, a new bad channel analysis was performed on the combined event sample for each run block.
When all other calibrations are applied, the masking procedure in each run block is repeated with the cell time taken into account as well as the simulated detector response for the corresponding cell.
In this step, typically 30 to 100 additional bad channels were found. 
Mostly, these channels were masked due to a bad time distribution. 
Those cannot be identified by the first masking procedure since at that point the time usually is not yet calibrated.
The total number of bad and dead channels is typically in a range of 1000 to 1600, which corresponds to 6\% to 9\% of all channels, depending on the run-block. On average, 300 to 400 channels of all problematic channels were dead.

   
\subsection{Cell time information calibration} 
\label{sec:timeCalib}
The main goal of the time calibration is to correct the cell time information by the average cell time over a period of data taking. 
The calibration corrects for the cable length (signal propagation in the cables takes $\sim 600$~ns), electronic response time, and time shifts due to clock phase differences.
The period can be as long as one \gls{ALICE} period, or the entire year. 
However, all the electronic changes and different trigger setups impact the calibration.
The \gls{BC} in pp collisions during \gls{LHC} Run 2 occurred every 25~ns\footnote{During \gls{LHC} Run 1  pp collisions occurred every 50~ns. 
During Pb--Pb data taking periods the bunch spacing was even larger, 150~ns or 75~ns.}. 
The \gls{ALTRO} clock of the \gls{EMCal} readout runs with a different frequency than the \gls{LHC} clock, and samples every 100 ns (see \Sec{sec:readout}).
During Run 1 the phase difference between \gls{ALTRO} and \gls{LHC} clocks was determined and corrected for by the readout firmware, whereas during Run 2 it was determined with an offline analysis on a run-by-run basis for each readout unit separately.  
In most cases, the clock phase difference was stable during a run. 
However, in runs where the \gls{FEE} required a reconfiguration because of hardware failures, a change of phase often occurred. 
For such cases, the phase differences are determined separately for the events collected before and after the reconfiguration.
After the run-by-run phase correction was applied, the average correction for each individual cell due to the different cable lengths and electronic response is obtained for a longer period of time. 
In principle, the time calibration should stay the same as long as no hardware modifications were performed.
The shift is determined from the average time of the main bunch crossing with respect to 0~ns for cells with energies above 0.4~GeV. 
As the response time is slightly different for high gain~(low energies) and low gain~(high energies), an additional shift for the latter was obtained for each channel for cell energies above $\sim 16$~GeV.
These time shifts were determined for each cell based on the whole pp data sample at \sthirteen, as they are expected to be constant and their evaluation needs large statistics.

\begin{figure}[t]
\includegraphics[width=0.49\textwidth]{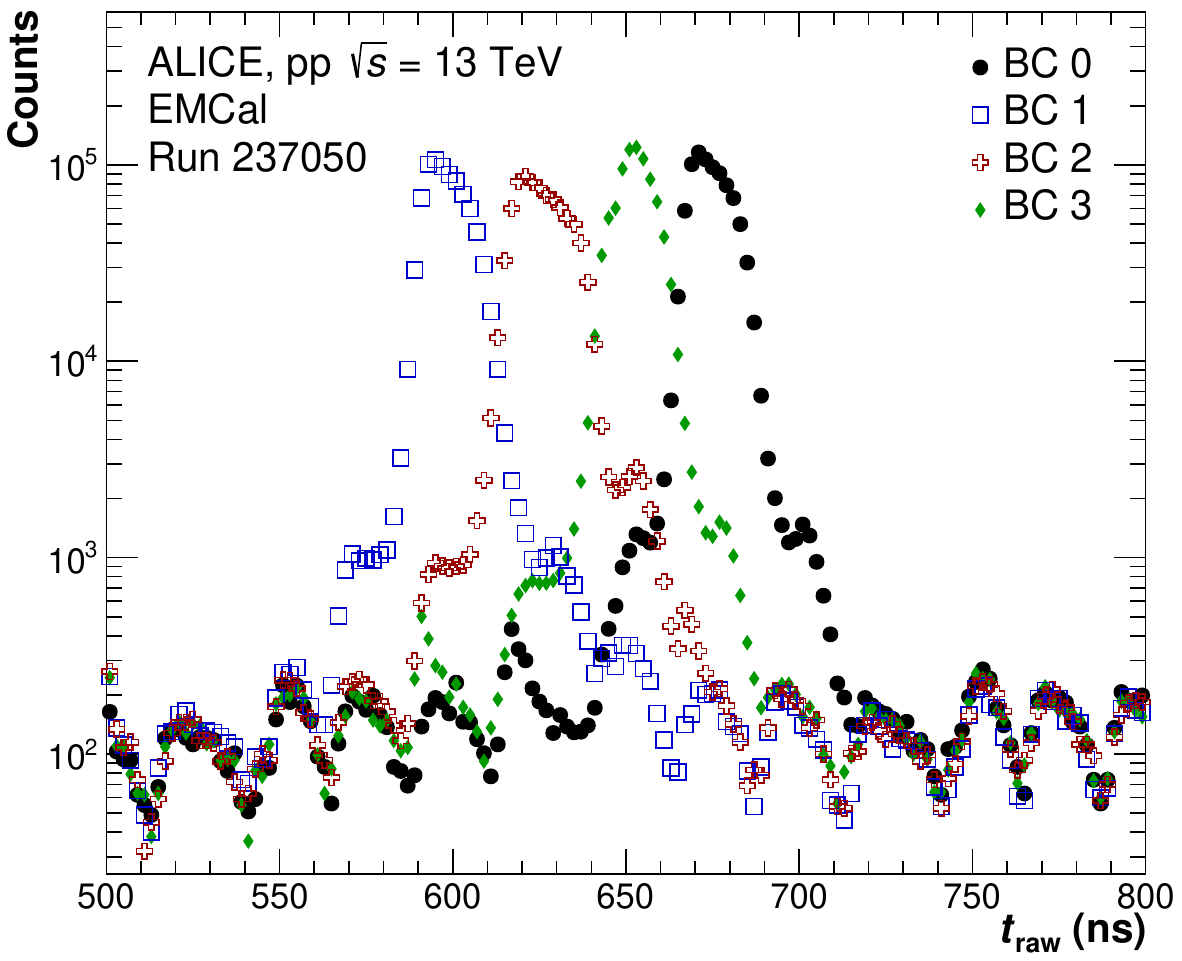}
\includegraphics[width=0.49\textwidth]{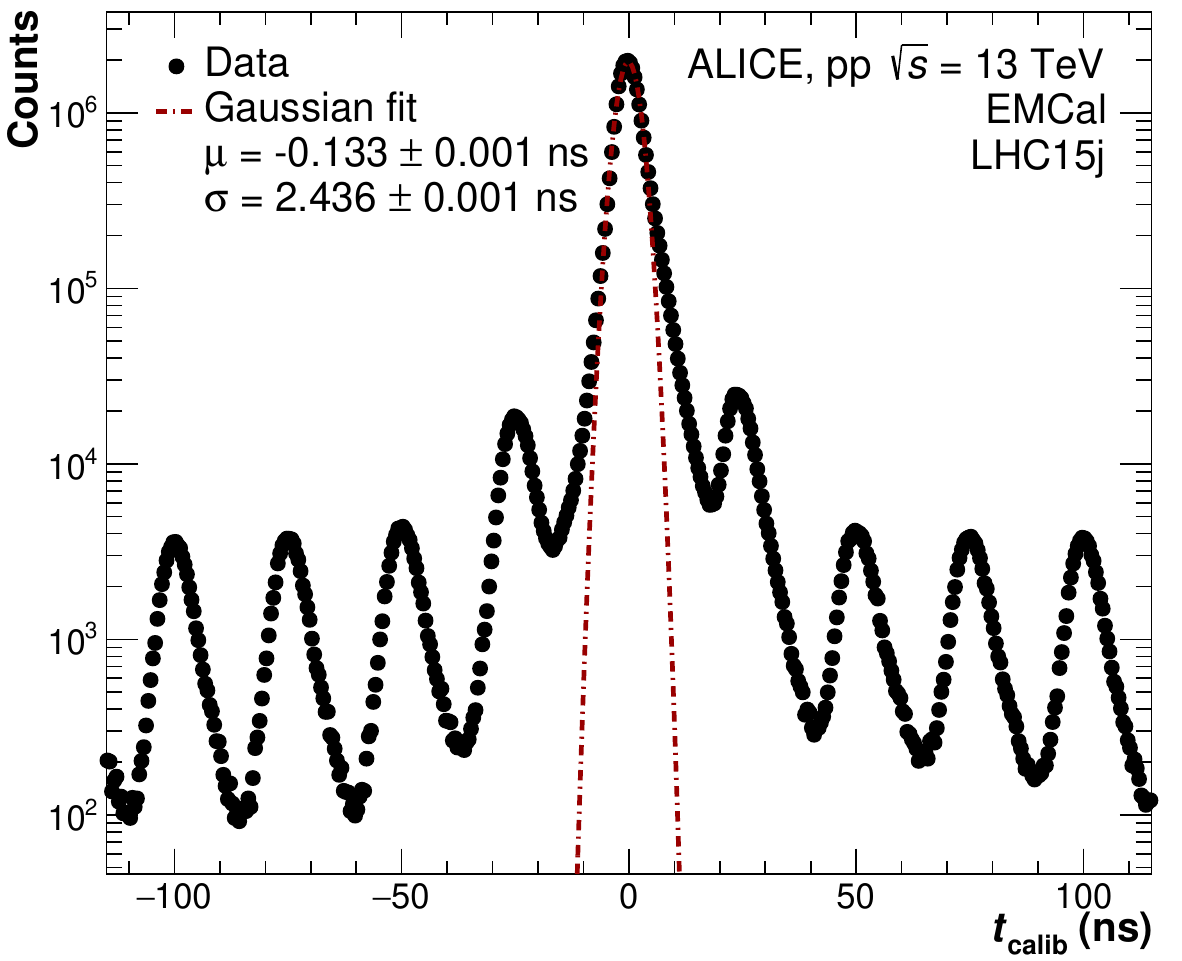}
\caption{(Color online) Left: Time distribution of \gls{EMCal} cells for different bunch crossings before the time calibration. Right: Aligned time distribution after the time calibration for cells with $E>$2 GeV.}
\label{fig:emcalTime}
\end{figure}
\begin{figure}[t]
\includegraphics[width=0.49\textwidth]{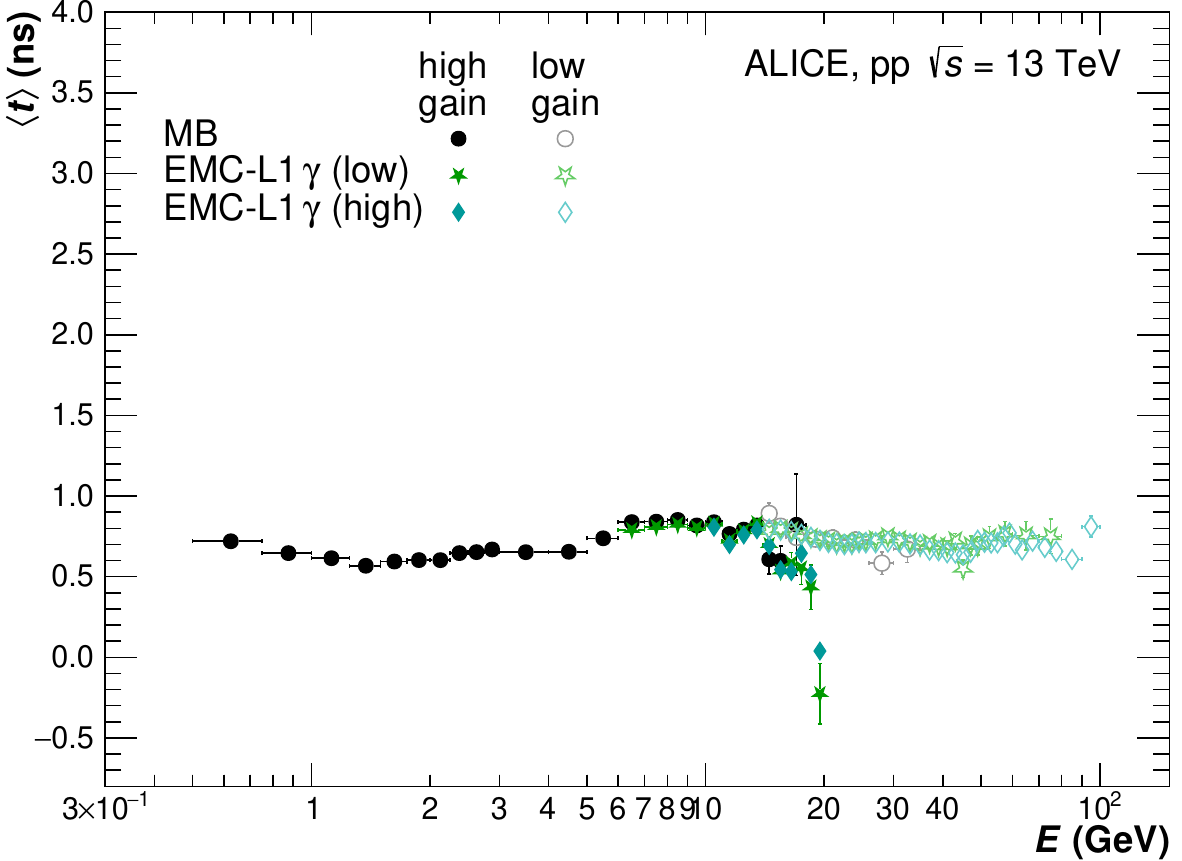}
\includegraphics[width=0.49\textwidth]{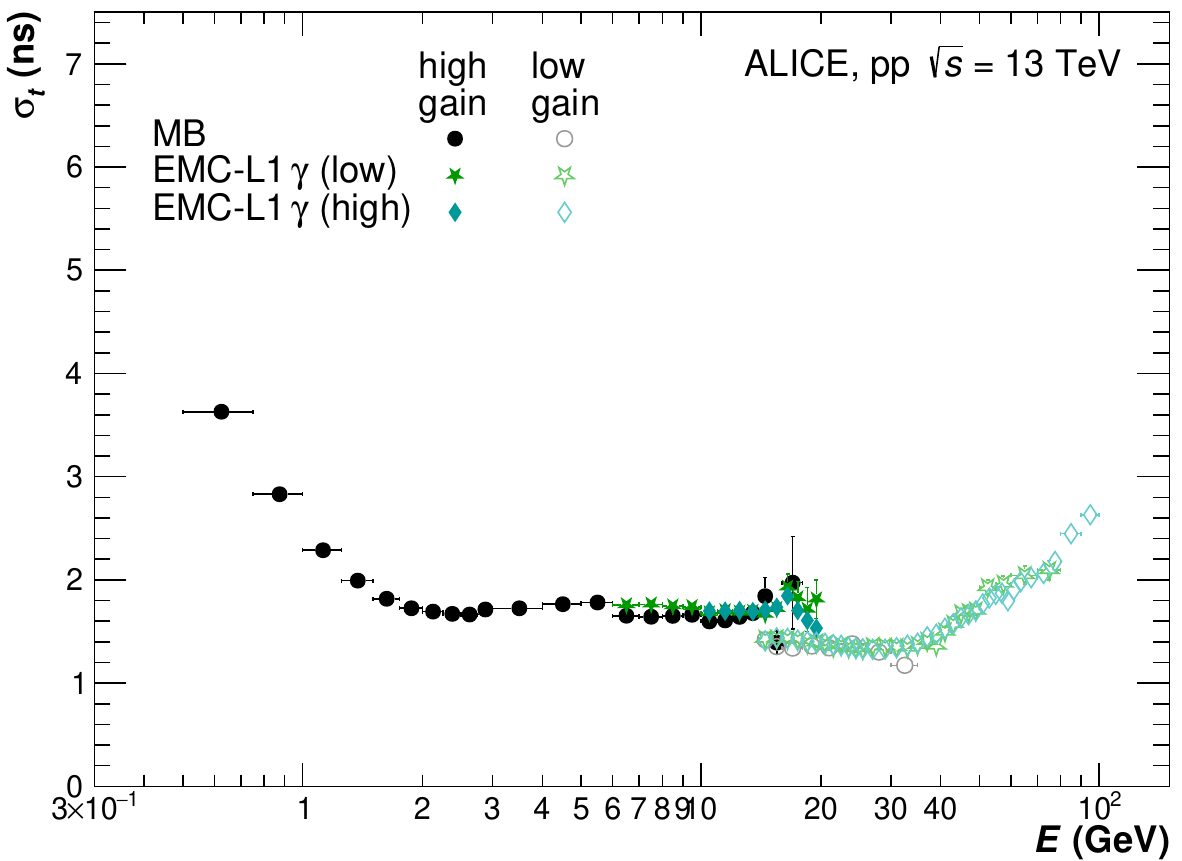}
\caption{(Color online) Gaussian mean (left) and width (right) of the time distribution of the main bunch crossing as a function of cluster energy. }
\label{fig:emcalTimeReso}
\end{figure}

\Figure{fig:emcalTime} shows the time distribution before (left) and after (right) the time calibration.
After the calibration, the cell time distribution is aligned at zero.
The neighbouring peaks, well visible on the plots, come from out-of-bunch pileup. 
The energy-dependence of the mean and width of the main peak in the timing distributions are shown in \Fig{fig:emcalTimeReso}.
A shift of about $0.8$~ns is visible, which can be attributed to the slightly asymmetric timing distribution with a tail towards later times.
The time resolution improves with increasing cluster energy from about $4$ ns at $0.8$ GeV to $1.4$ ns at $4$ GeV. 
For energies larger than $20$ GeV the time resolution worsens due to the limited statistics for the calibration of the low-gain sample.
The timing signal for energies at $\sim 100$ GeV is affected by the shaper nonlinearity discussed in  \Sec{sec:shaperNL}. 
However, as the magnitude of the time skewing is smaller than   the timing resolution, no special correction is implemented.
  
\subsection{Cell temperature calibration} 
\label{tempCalib}
As the gain of each \gls{APD} depends linearly on the temperature at which it is operating, a correction to the gain has to be applied offline in case the data were taken under varying conditions.
Thus, a three component monitoring and correction system was devised for the \gls{EMCal}, as described in \Sec{sec:APD} and~\cite{Cortese:2008zza}.
Its main components are the \gls{LED} system and temperature sensors, which sit close to the \glspl{APD} at various locations.
Special calibration triggers fire the \gls{LED} pulse and corresponding readout.
These events are available for online monitoring of the gains as well as offline calibration, but do not enter the physics data stream. \\
As the strength of the \gls{LED} pulse might vary as a function of time for each strip module~(48 cells), it is monitored using the back light collected from the light diffuser~(see \Sec{sec:hardware} for details on the hardware implementation).
These light yields are read out using a separate monitoring front-end card and averaged per run for stability. 
They are referred to as the \gls{LED} monitoring signal in the remainder of this section.
In addition, temperature sensors were installed close to the \glspl{APD} (eight for full and 2/3-size modules and four for 1/3 size modules).
Due to the fact that some of these sensors were, however, failing during the data taking and their absolute calibration is not known, only the average temperature per \gls{SM} of reliable sensors was used to calibrate the data.
As the temperature within one run does not change dramatically, these values were also averaged per run. This guarantees a reasonable stability of the corresponding reference values. 
Afterwards, the most likely temperature for each \gls{SM} was determined for the data taken during Run 2 of the \gls{LHC}. 
All cells within one \gls{SM} were calibrated to these reference temperatures. Due to the limited heat dissipation within the L3 magnet of \gls{ALICE}, the observed average as well as minimum and maximum temperatures are significantly different between the \gls{EMCal} and \gls{DCal}. 
While for the latter, the temperature rarely varies by more than $1.2$ $^\circ$C around $\sim20$ $^\circ$C, the variation for the top six \gls{EMCal} modules is up to 4 $^\circ$C with an average temperature around 24.5$^\circ$C. 
The analysis of the \gls{LED} events was performed for each cell and each signal per cell was normalized to the \gls{LED} monitoring signal in the corresponding strip module.
The \gls{LED} signal divided by the \gls{LED} monitoring-signal is further referred to as the normalized \gls{LED} signal.

\begin{figure}[t!]
\begin{minipage}[t]{0.4\textwidth}
\includegraphics[height=1.05\textwidth]{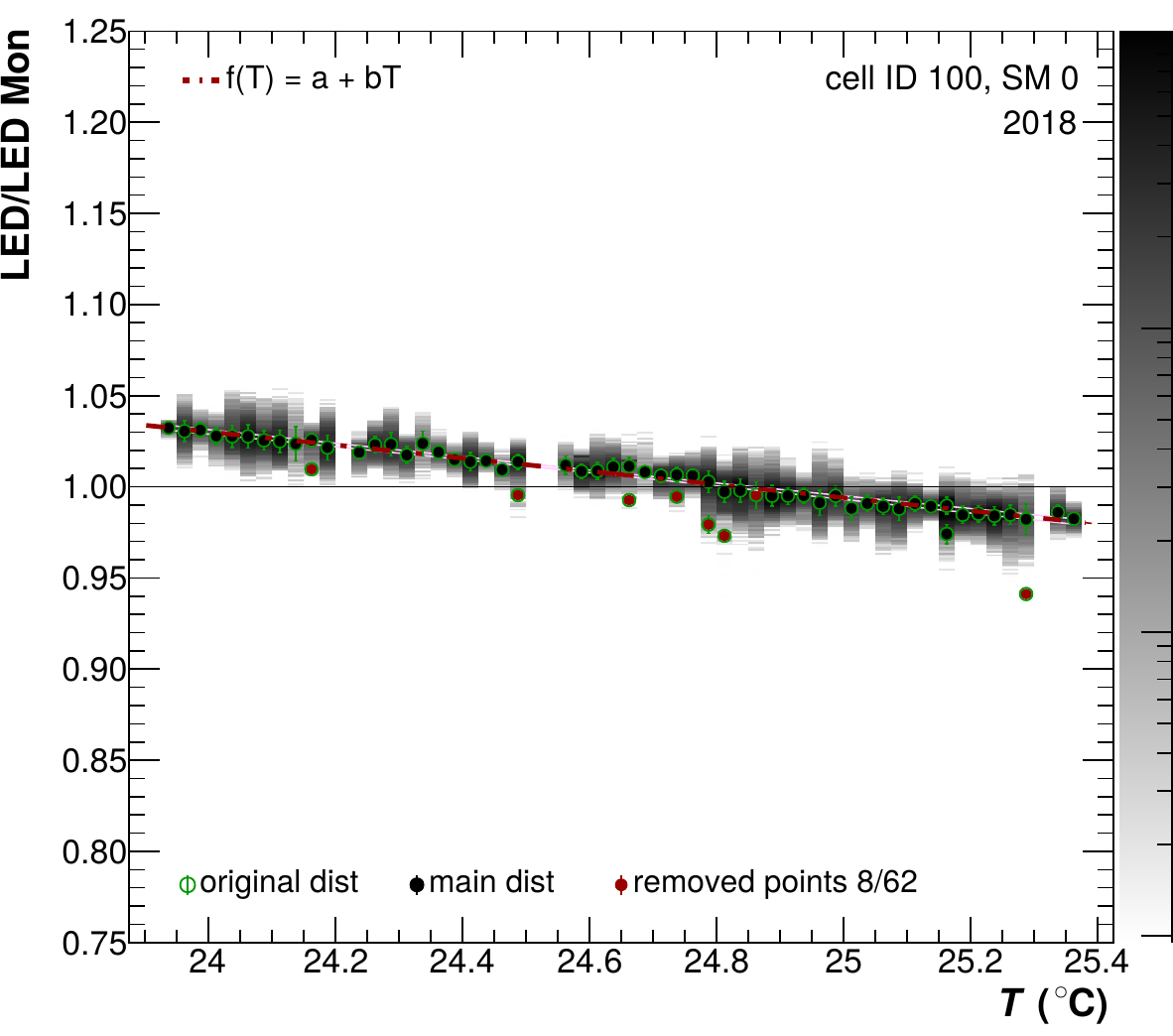}
\end{minipage}
\hspace{1.5cm}
\begin{minipage}[t]{0.4\textwidth}
\includegraphics[height=1.05\textwidth]{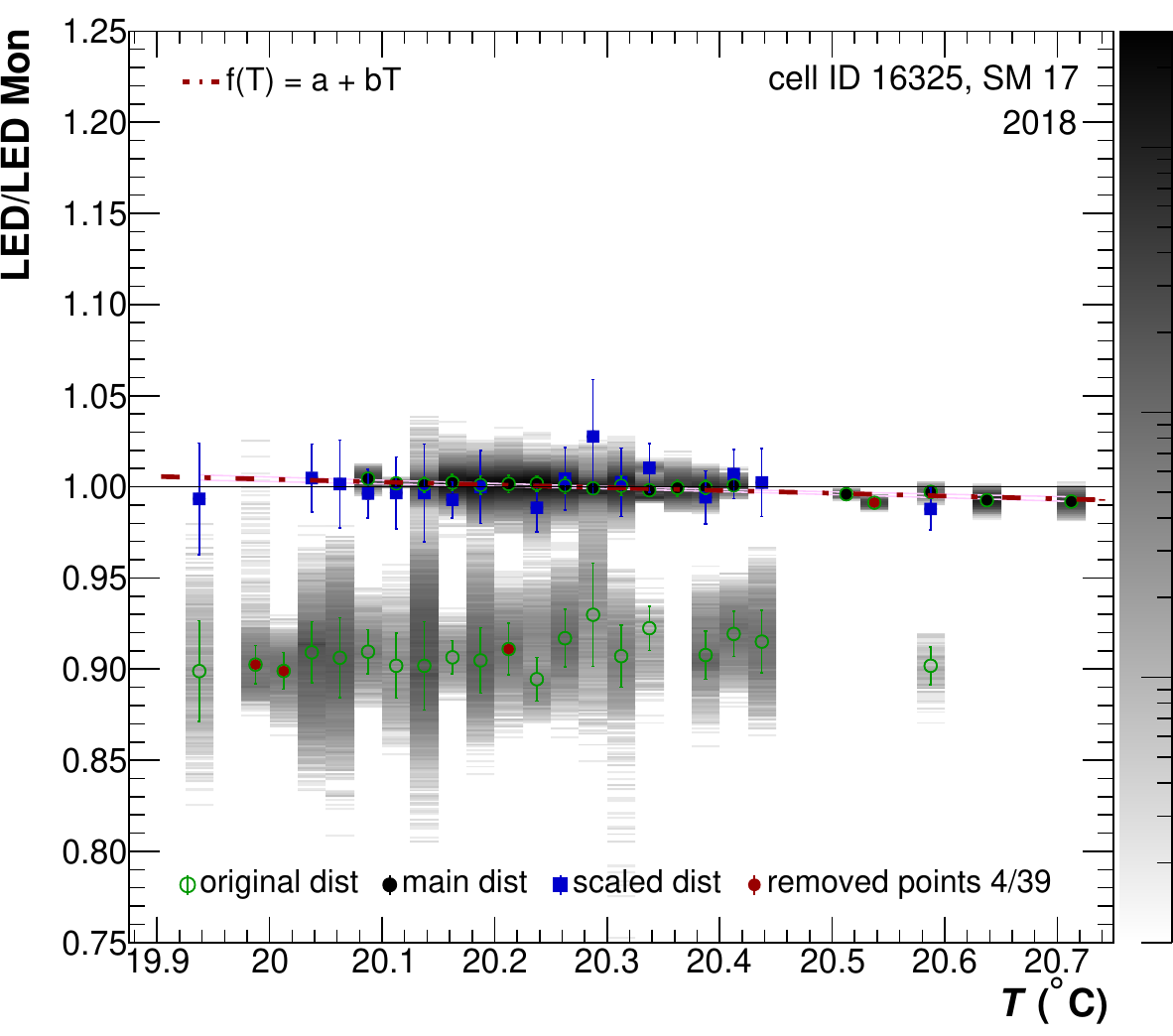}
\end{minipage} 
\caption{Illustration of the fitting procedure of the normalized \gls{LED} signal for a good cell in \gls{EMCal} (left) and a problematic one in \gls{DCal} (right). 
The raw distribution obtained for the full 2018 data sample is shown in gray scales in the background, while the maxima in each temperature slice are indicated by green open circles. 
As there might be multiple clusters of points (as seen on the right) the distribution that is considered as the dominant cluster is marked by black closed circles, while the blue squares represent the  shifted distributions after the correction for their offset is applied.
The final fit to the combined distribution of black and blue points is given as a dashed red line. 
Points marked in red were iteratively excluded from the fit as they were considered outliers. }
\label{fig:emcalTempExtr}
\end{figure}

\Figure{fig:emcalTempExtr} presents the resulting distributions for two example cells for the data collected in 2018. 
As the measured light yield in one cell not only depends on the temperature but also on other factors (\ie\ electronic noise in the front end card), it is possible that the normalized \gls{LED} signal at a given temperature varies with time. 
Consequently, the normalized \gls{LED} distribution in each temperature slice might have multiple maxima if more than one run was taken at the same average temperature (green empty circles). 
We assume, however, that the temperature dependence of the normalized \gls{LED} signal is identical for all runs recorded at the same conditions and only an absolute shift in the signal is observed for different conditions.
An example of such a split distribution is shown in \Fig{fig:emcalTempExtr}~(right). 
If more than one distinct cluster of points is identified, the clusters were shifted to a common baseline. For this, we select a main distribution (black circles) and determine the offset towards the secondary distribution. 
The secondary clusters are then shifted by this offset (blue squares). 
Afterwards, the combined distribution is fitted. Strong outliers from the combined distributions (red points) are iteratively removed from the fitting procedure.

\begin{figure}[t!]
\begin{minipage}[t]{0.4\textwidth}
\includegraphics[height=1.05\textwidth]{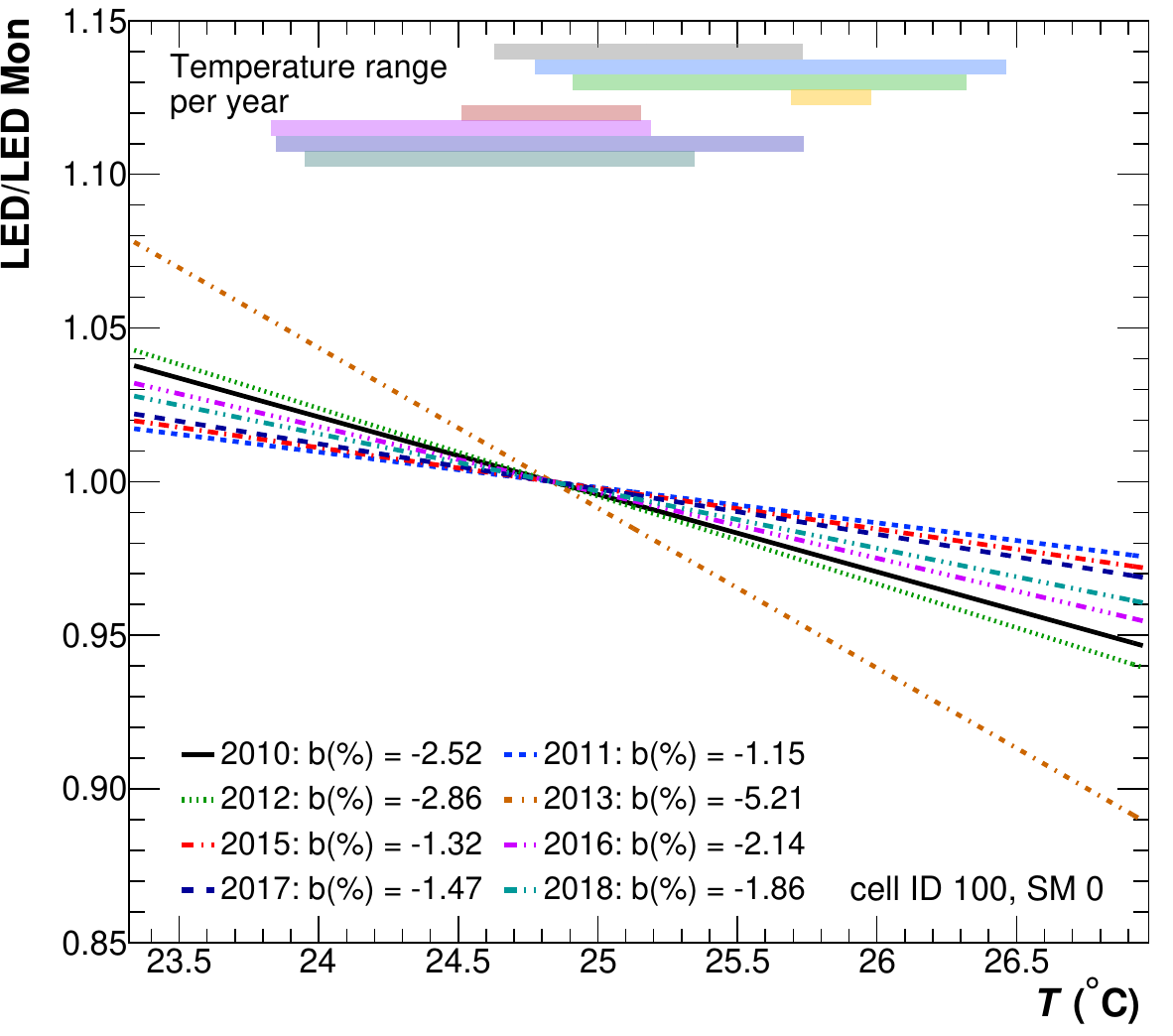}
\end{minipage}
\hspace{1.5cm}
\begin{minipage}[t]{0.4\textwidth}
\includegraphics[height=1.05\textwidth]{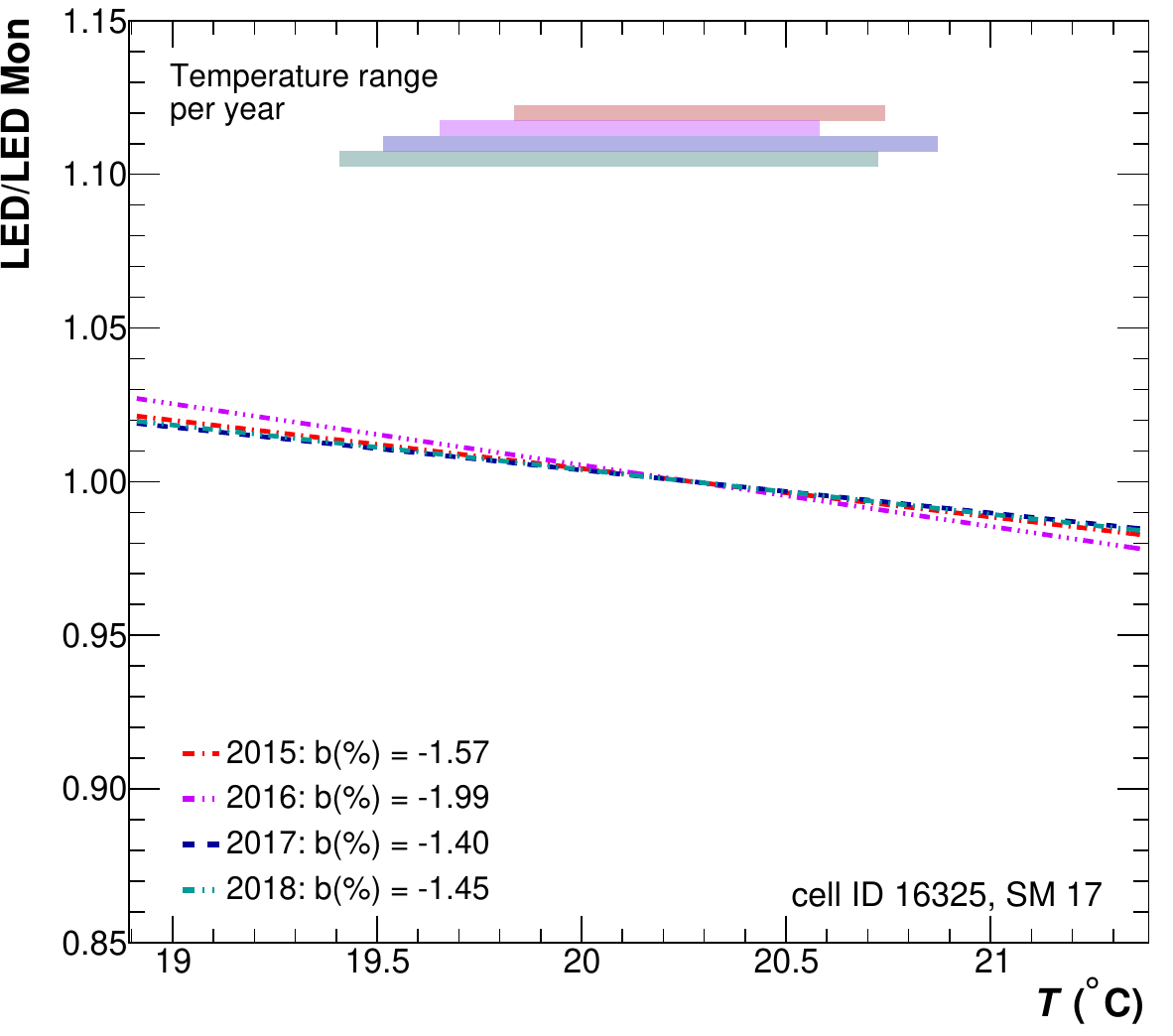}
\end{minipage} 
\caption{(Color online) Comparison of the obtained temperature calibration parameters in the \gls{EMCal} (left) and \gls{DCal} (right). 
The same cells were chosen as for \Fig{fig:emcalTempExtr}.
The calibration parameters were obtained separately for all years during which the corresponding \gls{SM} was installed and the cell considered good. 
The accessible temperature ranges for each year are indicated by the shaded areas in the same colors as the corresponding fits for the respective years.}
\label{fig:emcalTempYears}
\end{figure}

The determination of the calibration coefficients was performed for each year of running separately, taking into account only good cells and runs for data taking. 
A comparison of the obtained calibration coefficients for each year is presented in \Fig{fig:emcalTempYears}. 
The accuracy of this calibration strongly depends on the temperature range covered during the corresponding year, which is indicated as a shaded band in each of the comparison figures. 
For very short data-taking periods, like 2013, the outlined procedure often resulted in unreliable parameters due to the finite temperature resolution provided by the sensors. 
Thus, for the affected year, the parameters from the 2012 run were taken. 
In general no strong variation of the slope as a function of time is expected, except if the gain of the corresponding \gls{APD} has been changed, which is not done very often.
Consequently, the variation in the slope parameters can give an indication of the systematic uncertainty of the extraction procedure.
For most cells the slope parameter $b$ ranges between $-2.5\%$/$^\circ$C and $-1.2\%$/$^\circ$C with its most frequent value around $-1.8\%$/$^\circ$C. 
Similar values were estimated during the test beam campaign~\cite{Cortese:2008zza,Badala:2008zzd}, where only one type of the \glspl{APD} used in the \gls{EMCal} was tested. 
Based on these results we concluded that the fitting procedure of the normalized \gls{LED} signal failed if slopes below $-7\%$/$^\circ$C or above $-1.2\%$/$^\circ$C were found. 
For these cells the average slope of all cells for the corresponding year was set as the calibration parameter. 
Consequently, the gain correction factor does not exceed $\pm 5\%$ for most cells, even under extreme temperature variations. 
On average the temperature variations lead to a gain correction factor below 1\%, which is too small to affect the $\piz$ invariant mass peak width, which is dominated by the energy resolution at low energies and shower overlaps at high $\pt$. 
  
\subsection{Monte Carlo cluster energy fine tuning}
\label{sec:EMCalEnergyPositionCalib}
\begin{figure}[t]
    \centering
    \includegraphics[height=6.7cm]{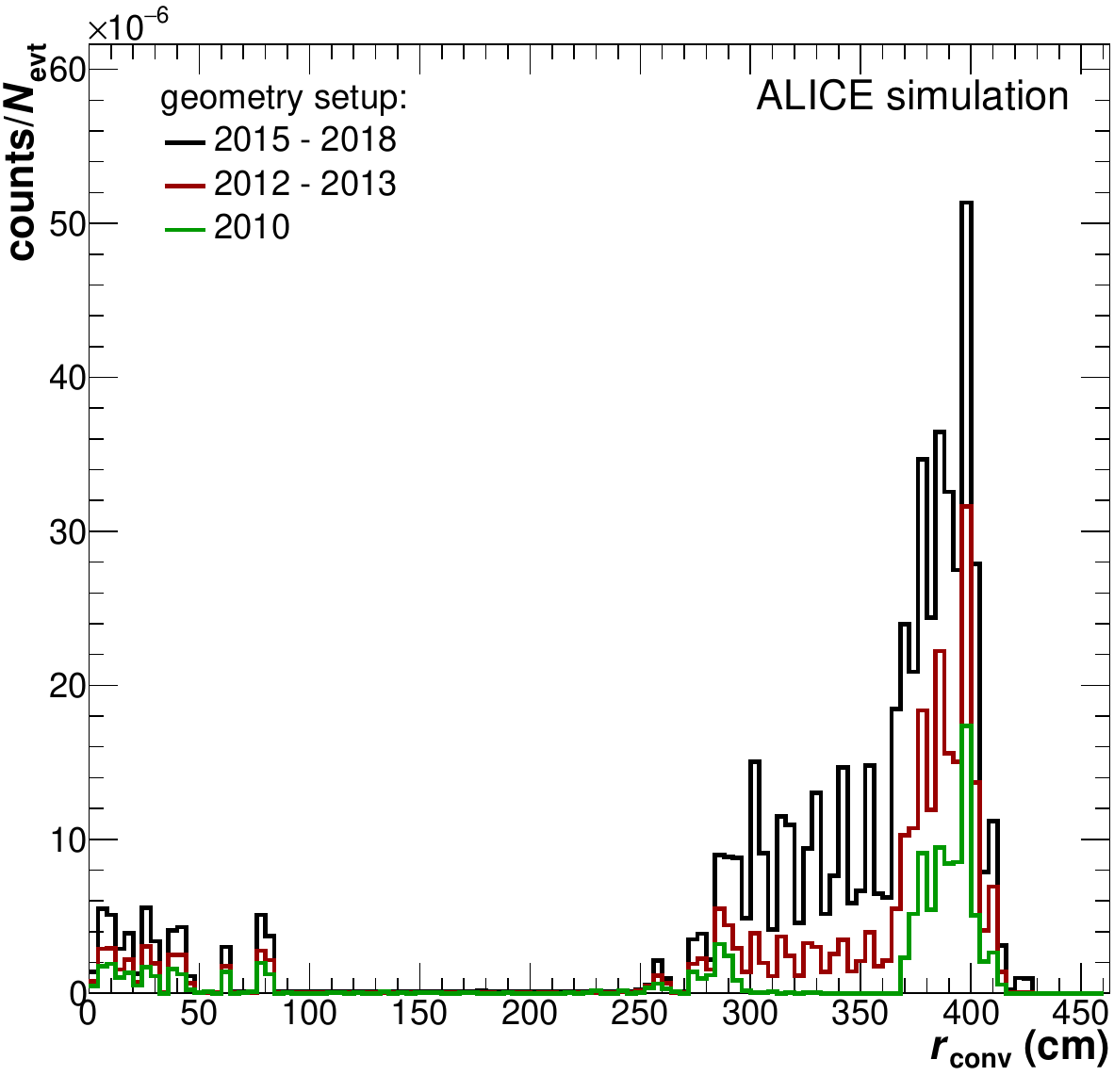}
    \includegraphics[height=6.5cm]{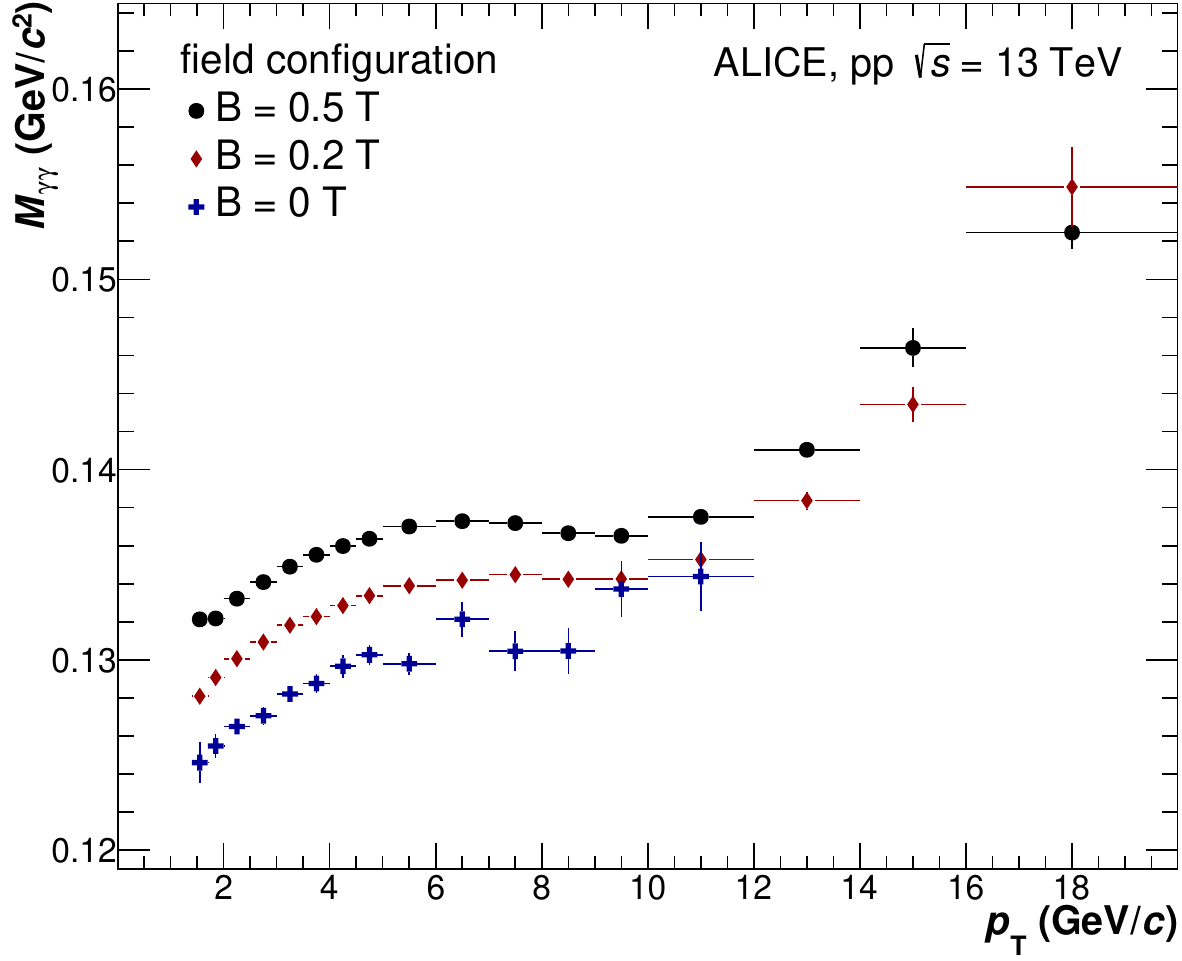}
    \caption{(Color online) Left: Radial distance from the \gls{IP} of photon conversions in the detector material for different detector configurations in 2010, 2012 and 2015-2018. 
    The distributions are obtained for \gls{PYTHIA}8 simulations and only for photons whose conversion products were reconstructed as clusters in \gls{EMCal} and formed, when paired with another cluster, a signal in the $\pi^0$ invariant mass window. 
    For 2011 the same number of super modules was installed as in 2012 - 2013, but two fewer modules had the \gls{TRD} installed in front of them.
    Right: Neutral pion invariant mass peak position as a function of \pT\ for different magnetic field configurations for \pp\ collisions at \sthirteen.}
    \label{fig:ConversionRadiusAndBField}
    \centering
    \includegraphics[width=1.\textwidth]{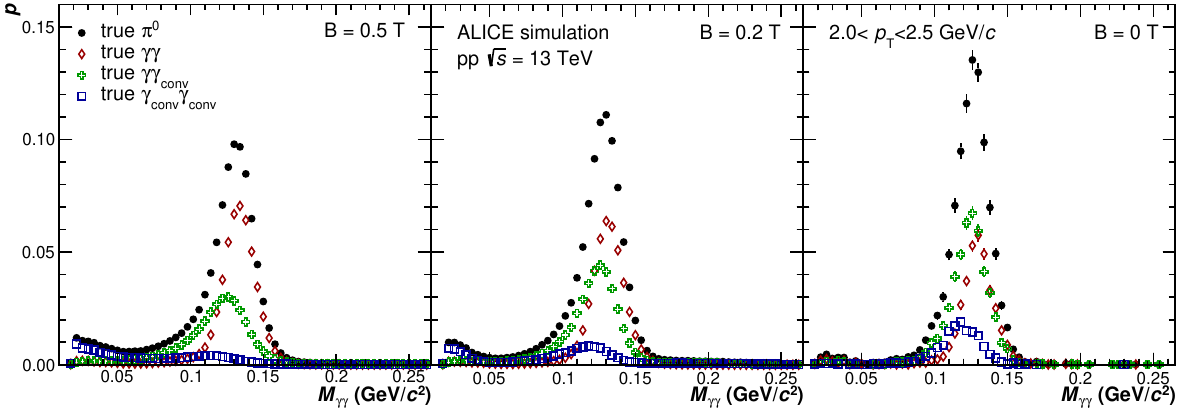}
    
    \caption{(Color online) Invariant mass distribution of reconstructed $\pi^0$ mesons in \gls{MC} simulations. 
    Contributions from pure photon pairs as well as from clusters which contain converted photon contributions are shown separately for $B = 0.5$~T (left), $B = 0.2$~T (middle) and $B = 0$~T (right) for pp collisions at \sthirteen.}

    \label{fig:Pi0ClustersandDecompMassPeaksFine}
\end{figure}

The nonlinearity correction based on test-beam data calibrates the cluster energies in data and simulation to an agreement within the percent level. 
An additional cluster energy correction is needed in order to provide a per-mil level agreement of the energy scale and therefore smaller systematic uncertainties on the analysis level. \\
The necessity of this correction arises from the detector material that is present in front of the  \gls{EMCal} within the central barrel of \gls{ALICE}. 
This material causes additional photon conversions, which influence the position of the $\pi^0$ mass peak.  
This effect was not present in the \gls{EMCal} test-beam data taking and thus, its consequences are not included in the test-beam nonlinearity.
\Figure{fig:ConversionRadiusAndBField} (left) shows the conversion radius of $\pi^0$ decay photons that were reconstructed as clusters in the \gls{EMCal} for different geometrical configurations of the \gls{TRD} and different number of \gls{EMCal} \glspl{SM}.
The material, especially the one in the \gls{TRD} and \gls{TOF} directly in front of the \gls{EMCal}, is responsible for a large fraction of photon conversions, which smear the cluster energies reconstructed in the \gls{EMCal}, especially at low energies.
These late conversions cannot be removed from the cluster sample via charged-particle track matching as the conversions appear at radii where track reconstruction is no longer possible within \gls{ALICE}.
The effect of late conversions can be seen in \Fig{fig:Pi0ClustersandDecompMassPeaksFine}~(left). 
There, the invariant mass distribution~(according to \Eq{eq:KineIMgg}) of two \gls{EMCal} clusters, which originate from a simulated $\pi^0$, is shown in the neutral pion mass region (events were simulated with \gls{PYTHIA}8 and subsequently propagated through \gls{ALICE} with \gls{GEANT}3).
The reconstructed $\pi^0$ mesons which contain contributions from conversions, shown in magenta and cyan, exhibit a shift of the peak to lower invariant masses and a pronounced tail compared to reconstructed $\pi^0$ mesons where both photons are not converted.
This smearing effect, combined with the conversion electron clusters being indistinguishable from real photon clusters due to the lack of track matching, is the motivation for the additional cluster energy correction presented in this chapter. 
The Monte Carlo fine tuning will thus also partially correct for any imperfection in the material implementation of the \gls{TRD} and \gls{TOF}.
The correction is based on a comparison of the measured $\pi^{0}\rightarrow\gamma\gamma$ invariant mass peak position in data and simulation.

However, the influence of the material in front of the \gls{EMCal} will not only depend on the actual material but also on the magnetic field configuration. 
\Figure{fig:ConversionRadiusAndBField} (right) shows the neutral pion invariant mass peak position for lower magnetic field configurations in comparison to the nominal field of $B = 0.5$~T.  
For lower magnetic fields, the probability increases to reconstruct late conversions within the same cluster, thus increasing the pion reconstruction efficiency, in particular at lower transverse momenta. 
Therefore, more $\pi^0$ mesons containing conversion contributions are reconstructed closer to their nominal invariant mass for $B = 0.2$~T and $B = 0$~T than in the nominal field case.
As presented in  \Fig{fig:Pi0ClustersandDecompMassPeaksFine}~(right), the tail towards lower invariant masses is less pronounced, while the average mass shifts by $2$--$3$~MeV/$c^2$ for $B = 0.2$~T and $5$--$6$~MeV/$c^2$ for $B = 0$~T and the width of the peak decreases with respect to the same neutral pion distribution for $B = 0.5$~T.\\
The correction procedure itself uses the peak position in the $\pi^0$ meson invariant mass distributions in data and simulation in order to obtain a cluster energy dependent correction that is applied on the reconstructed cluster energies in simulation.
The data is therefore taken as the ``truth'' in this calibration scheme after the energy calibration, described in \Sec{sec:e-calibchapter}.
For this, the $\pi^0$ mesons can be reconstructed with a \gls{PCM-EMC}~\cite{Acharya:2017hyu,Acharya:2017tlv} in which one of the $\pi^0$ decay photons is reconstructed as a cluster in \gls{EMCal} and the other decay photon is reconstructed from electron-positron pairs from photon conversions in the detector material, using only the \gls{ITS} and \gls{TPC} tracking detectors, called the \gls{PCM}~\cite{Abelev:2012cn}.
The reconstruction of conversion photons benefits from the high momentum resolution of the \gls{ALICE} central barrel tracking in order to enhance the precision of the correction.
Furthermore, they allow this fine tuning to go up to high cluster energies without risking the decay photons of the $\pi^0$ to merge into the same cluster.
For this method, based on the hybrid \gls{PCM-EMC} reconstruction, the reconstructed invariant mass of the $\pi^0$ candidate is plotted directly versus the cluster energy to determine the energy fine tuning.

The second method, called \gls{EMC} in the following, uses only \gls{EMCal} clusters with a cut on symmetric $\pi^0$ decays.

The general strategy of the energy fine tuning is the following:
First, the $\pi^0$ invariant mass peaks are fit in increasing $\pt$ bins in order to determine the mass positions in data and simulation.
Second, the ratios of mass positions in data and simulation are calculated.
Third, these ratios are parameterized to get the correction function, which can be used to perform the fine tuning of cluster energies in simulations.


\textbf{\gls{PCM-EMC} method}:
    Neutral mesons are reconstructed with the hybrid method, using one conversion photon and one \gls{EMCal} cluster.
    The invariant mass versus cluster energy is then filled into a 2-dimensional histogram.
    The same is done for pure background using an event mixing technique.
    Afterwards, these distributions are binned in energy, the mixed event background is normalized to the same-event distribution outside the \piz\ signal region and then subtracted. 
    This procedure is performed for data and simulation. 
    Subsequently, the peak position is determined for data and simulation using pure Gaussian fits and the ratio of these is calculated.
    The obtained values are squared to reflect the proportionality of $m^{2}_{\pi^{0}}\propto E$.
    These ratios are fitted with \Eq{eq:convcalo}, where $p_{0}$, $p_{1}$ and $p_{2}$ are the free parameters:
    \begin{equation}
        f(E)=p_{0}+\exp\left(p_{1}+p_{2}\times E\right).
        \label{eq:convcalo}
    \end{equation}
    This function, which is referred to as \gls{CCRF}, is then used to recalculate the cluster energies in simulation. 

\textbf{\gls{EMC} method}:
    Neutral mesons are reconstructed using two \gls{EMCal} clusters in this method (\gls{EMC}). 
    To select two photons, which approximately have the same energy in the laboratory rest frame, a strict cut on the asymmetry $|\alpha| < 0.1$, as defined in \Eq{eq:KineAsym}, is performed, simplifying \Eq{eq:KineIMgg} to $M^{2}_{\pi^{0}} = 2E^{2}_{\gamma}\left(1-\cos(\theta)\right)$ assuming $E_{\gamma_{1}}\approx E_{\gamma_{2}}:=E_{\gamma}$.
    By using $E_{\pi^{0}} \approx 2E_{\gamma}$, the invariant masses of pion candidates are  binned in cluster energy, similar to the \gls{PCM-EMC} method.
    From here on, the procedure is exactly the same as for the \gls{PCM-EMC} method, except the fit is called \gls{CRF} and the data-to-simulation ratio and the subsequent fit are performed using $m$ instead of $m^2$, to reflect that two photon candidates enter the calculation.

The \gls{PCM-EMC} method does not suffer that much from cluster merging at high transverse momentum, as compared to the \gls{EMC} method, which is limited to cluster energies below 8~GeV due to the asymmetry cut. 
If the clusters get closer to each other, the cluster splitting algorithms tend to distort the clusters in a manner that is not fully reproduced in the simulation.
Additionally, the \gls{PCM-EMC} method is more efficient at lower momentum, since there is only one cluster needed to reconstruct the meson and the reconstructed converted partner photon can have even lower energies.
Moreover, the \gls{PCM-EMC} method has a better mass resolution than the \gls{EMC} as it follows a mixture of the tracking and \gls{EMCal} resolutions.
The \gls{PCM-EMC} method was hence chosen as the default method for the fine-tuning, while the \gls{EMC} method was mainly used for systematic studies.

\begin{figure}[t]
    \centering
    \includegraphics[width=0.49\textwidth]{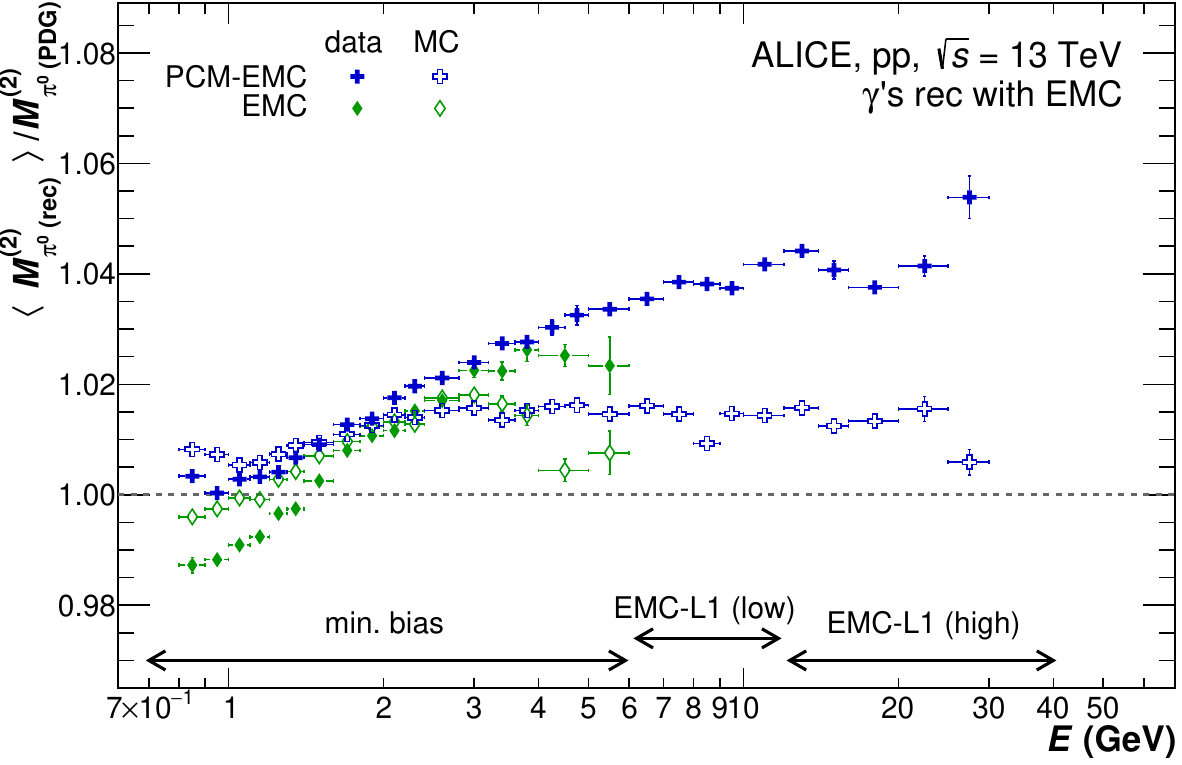}
    \includegraphics[width=0.49\textwidth]{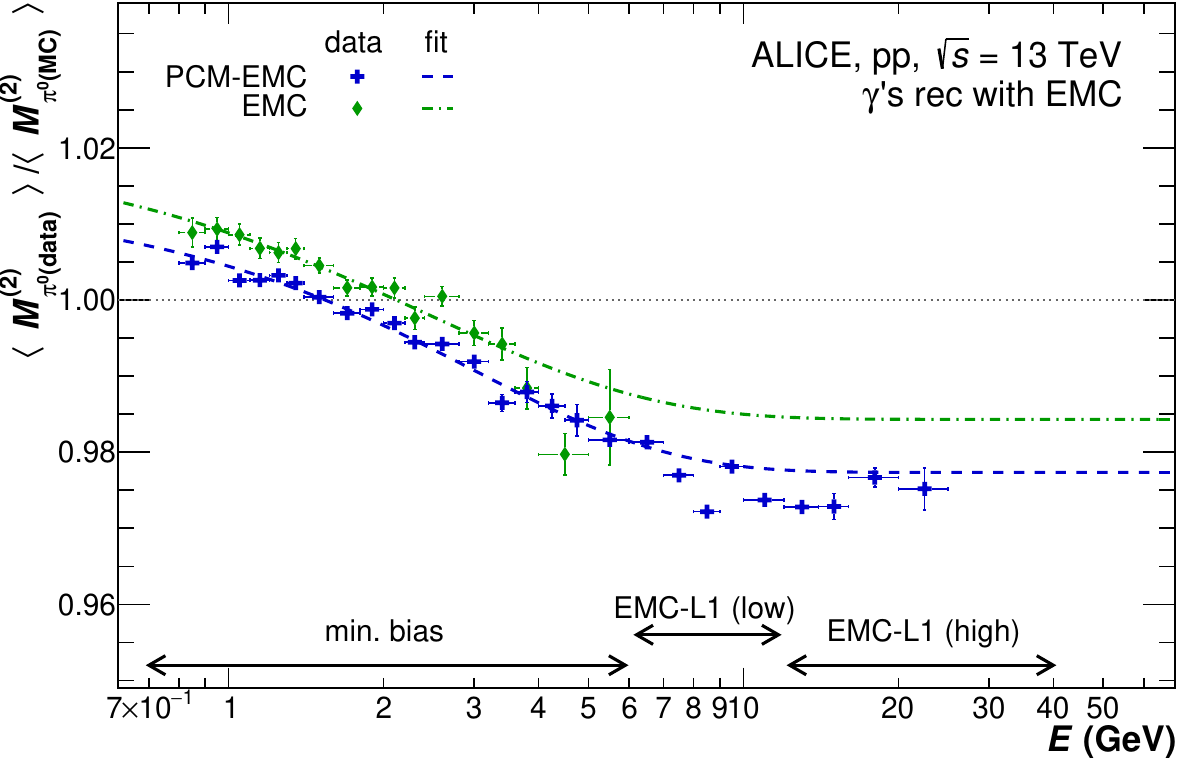}\\
    \caption{(Color online) Left: Mass positions for \gls{PCM-EMC}(blue) and \gls{EMC} (green) after applying the nonlinearity correction obtained from \Sec{sec:TBNonLin} \Figure{fig:5-TB-EnergyNonLinearity}. The mass positions are normalized to the neutral pion rest mass. In the case of \gls{PCM-EMC} the data points represent the squared mass position. Right: Ratio of mass positions in data and \gls{MC} for both techniques with their corresponding fits according to \Eq{eq:convcalo}.}
    \label{fig:MassRatios}
\end{figure}

\Figure{fig:MassRatios} shows the obtained invariant mass peak positions for data and \gls{PYTHIA}8 simulations (left) as well as the corresponding data-to-simulation ratios for the \gls{PCM-EMC} (blue) and \gls{EMC} (green) case (right).
The mass positions are normalized to the neutral pion rest mass of $0.13498~\GeVc^{2}$.
The ratios on the right are overlaid with the inverse of the correction function \gls{CCRF}(\gls{CRF}).
In order to enhance the number of \piz\ mesons at higher cluster energies, the triggered data~(\gls{EMC} \gls{L1} - low and high threshold) were used\com{ for higher cluster energies} for both methods in the cluster energy ranges indicated in the figure.

\begin{figure}[t!]
    \centering
    \includegraphics[width=0.6\textwidth]{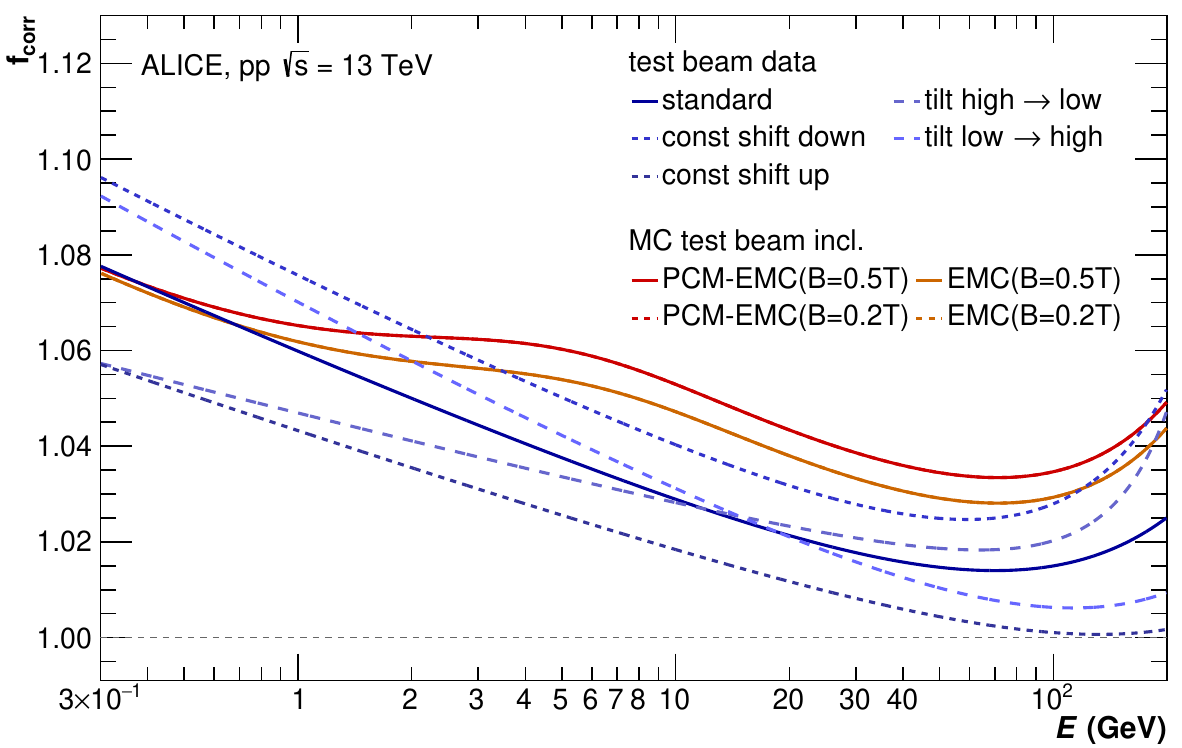}
    \caption{(Color online) Correction functions ($f_{\rm corr}$) for the default analysis cuts (see \Tab{tab:photonbasiccuts}) for \gls{CRF} and \gls{CCRF}, which are applied in \gls{MC}, together with the test-beam correction for data and its variation from \Sec{sec:TBNonLin}. 
    }
    \label{fig:ESTotalCorrection2}
\end{figure}

\begin{figure}[t!]
    \centering
    \includegraphics[width=0.48\textwidth]{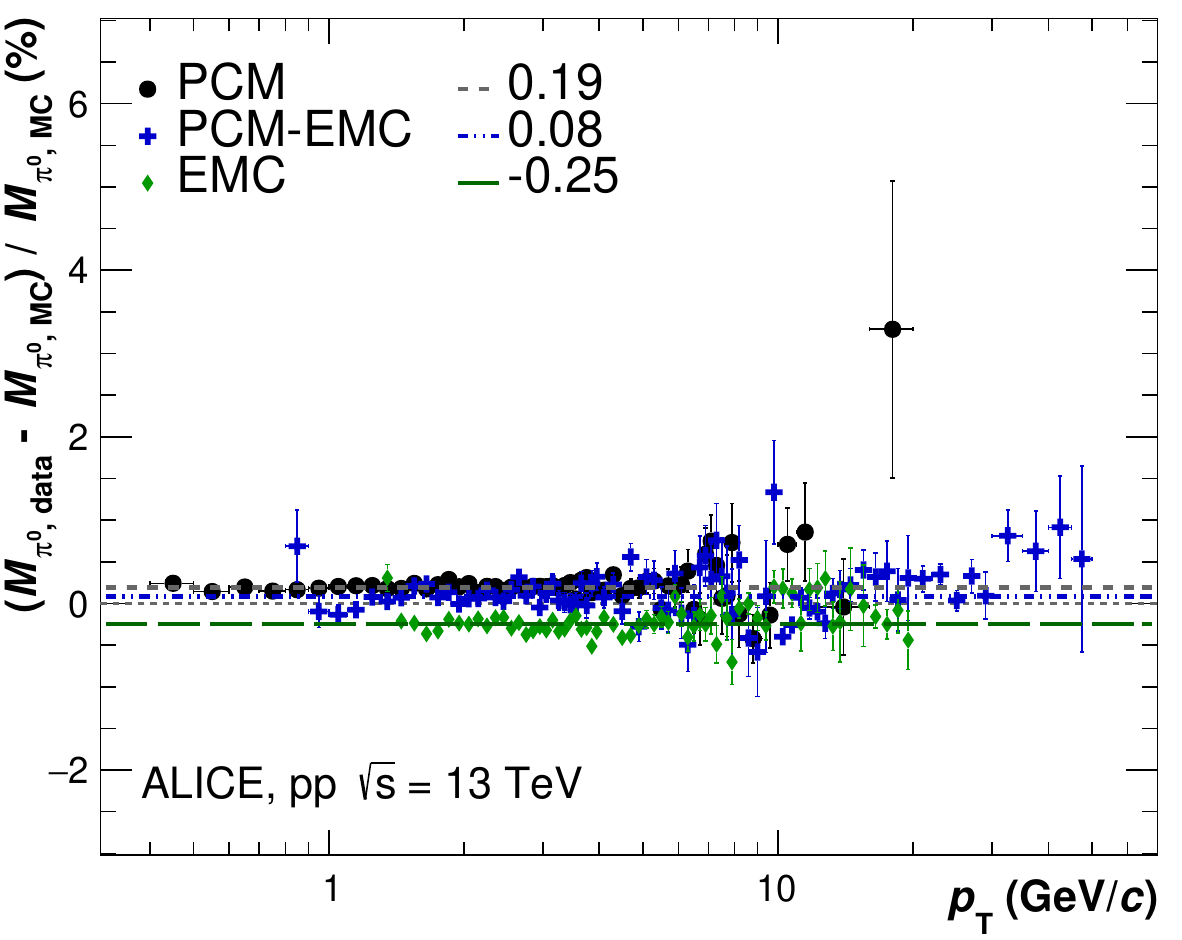}
    \includegraphics[width=0.48\textwidth]{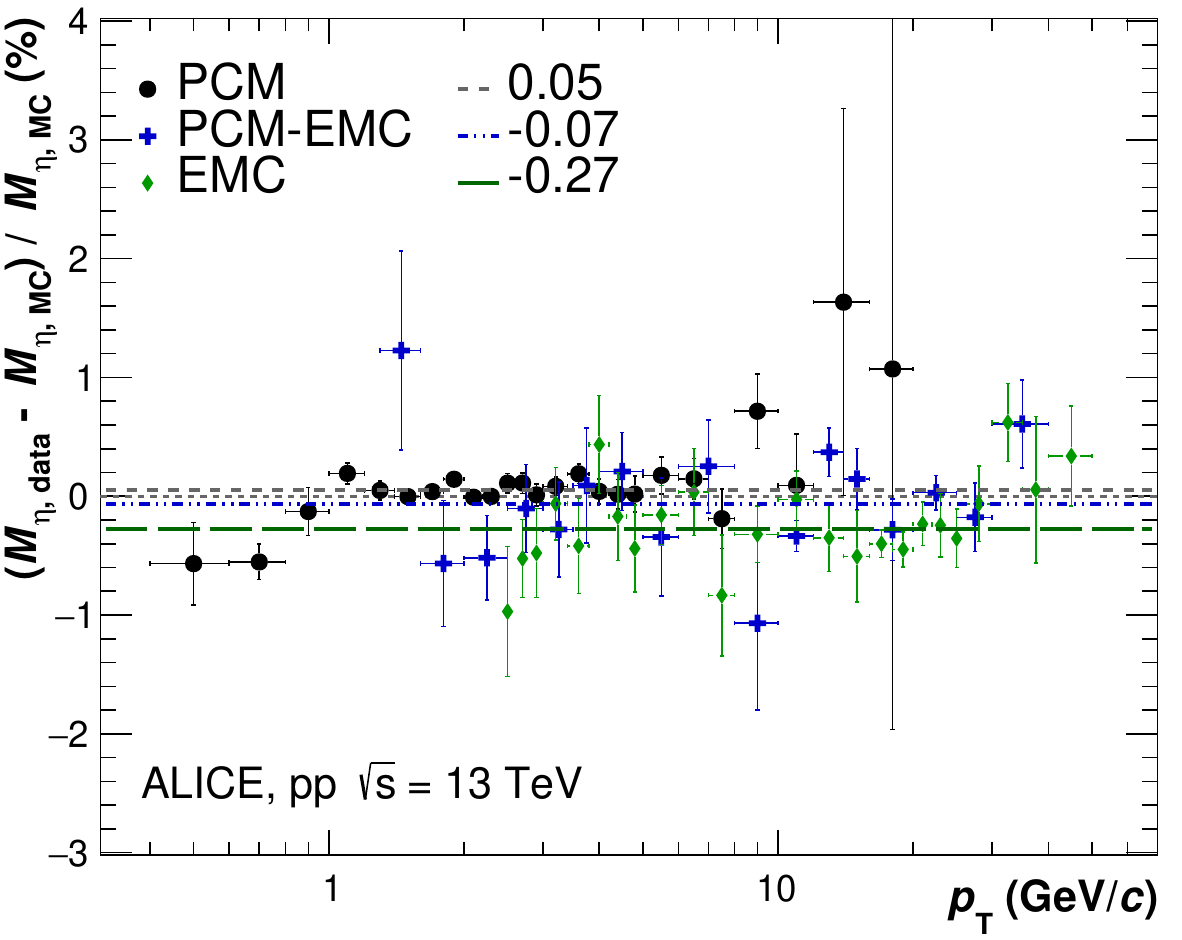}
    \caption{(Color online) Relative mass position difference between data and simulations for the neutral pion (left) and $\eta$ meson (right) for different reconstruction techniques in \pp\ collisions at \sthirteen. }
    \label{fig:ESConvCalo}
\end{figure}

\Figure{fig:ESTotalCorrection2} presents the correction function for the cluster energy\com{ after applying the formula from \Eq{eq:convcalo}. 
The corresponding correction for the data is shown} together with its variations necessary to correct for the different systematic parametrizations of the test-beam nonlinearity~(\Sec{sec:TBNonLin}, \Figure{fig:5-TB-EnergyNonLinearity}). 
To apply the nonlinearity and fine-tuning correction, the reconstructed cluster energies are directly multiplied by the function to obtain the corrected cluster energies: $E_{\rm corr} = E_{\rm rec} \times f_{\rm corr}$. Here, $f_{\rm corr}$ contains both the nonlinearity correction obtained from \Figure{fig:5-TB-EnergyNonLinearity} as well as the energy fine tuning presented in \Figure{fig:MassRatios}.
\Figure{fig:ESConvCalo} quantifies the \gls{MC}--data $\pi^0$ mass difference after full correction with \gls{CCRF}, where a near perfect agreement with unity in the ratio of data to simulation is observed for both \gls{EMCal} related reconstruction techniques.
As the \piz\ meson mass has been used for the calibration procedure this level of agreement was expected, while the $\eta$ meson serves as a cross-check for the applied \gls{MC} energy calibration fine tuning. 
For both mesons, the agreement of the peak positions is better than $0.3\%$ for both reconstruction techniques after subtracting the residual decalibration of 0.2\% caused by the \gls{PCM} photons.
A similar agreement at lower transverse momenta is obtained based on the \gls{CRF} method, albeit with less precision at higher transverse momenta.

\begin{table}[t]
    \hspace*{-0.4cm}
    \footnotesize
    \centering
    \caption{Summary of the parameters obtained for \gls{CCRF} and \gls{CRF} for \Eq{eq:convcalo}.}
    \begin{tabular}{cccccc} 
        magnetic field & \textbf{scheme} & $p_{0}$ & $p_{1}$ & $p_{2}$ (GeV$^{-1}$) \\
        \toprule
        \multirow{2}{*}{B = 0.5 T} & \gls{CCRF}  &  0.979235 & -3.17131 & -0.464198 \\
        & \gls{CRF}  &  0.984314 & -3.30941 & -0.399441  \\ \midrule
        \multirow{2}{*}{B = 0.2 T} & \gls{CCRF}  &  0.988503 & -3.10024 & -0.28337 \\
        & \gls{CRF}  &  0.99759 & -3.21271 & -0.363656  \\
        \bottomrule
    \end{tabular}
    \label{tab:energypos2}
    \label{tab:energypos}
\end{table}	

These results demonstrate that the energy calibration including the fine-tuning  scheme provides a precise energy calibration over the full momentum range for different triggers and meson species. 
\Table{tab:energypos2} summarizes the parameters for the \gls{CCRF} and \gls{CRF} fine-tuning functions for the nominal-field (B = 0.5 T) and low-field (B = 0.2 T) configurations.
The presented fine-tuning  correction is valid for all \pp\ and \pPb\ data taken during the \gls{LHC} Run 2.
An additional constant 1.3\% cluster energy increase in simulation is necessary when using \gls{LHC} Run 1 data from 2012 and 2013 where not all \gls{EMCal} modules were covered by \gls{TRD} modules, as seen in \Fig{fig:ConversionRadiusAndBField}{ (right)}.
This increase is due to a significantly smaller amount of photon conversions and therefore, a better cluster energy resolution combined with a limited precision of the detector material implementation in \gls{GEANT}3.
Similarly, the fine-tuning correction has to be adapted for low-field configurations in order to correct for the different influence from converted photons.
For \PbPb\ collisions, additional correction factors for high multiplicity events are needed in order to account for the high track density which further affects the energy scale.

The systematic uncertainty associated with this fine-tuning correction is obtained from separate fine-tuning  corrections for each systematic variation of the test-beam nonlinearity correction.
In addition, the difference between the \gls{CCRF} and \gls{CRF} fine-tuning  corrections can be used for the estimation of systematic uncertainties.
Studies of the effect of the fine-tuning systematic variations on $\pi^0$ meson transverse momentum spectra found the systematic uncertainty on light meson spectra measurements to be less than 2\% and $\pt$-independent.
     
\subsection{Cluster size correction}
\label{sec:NCellEffi}

\begin{figure}[t!]
\includegraphics[height=7.2cm]{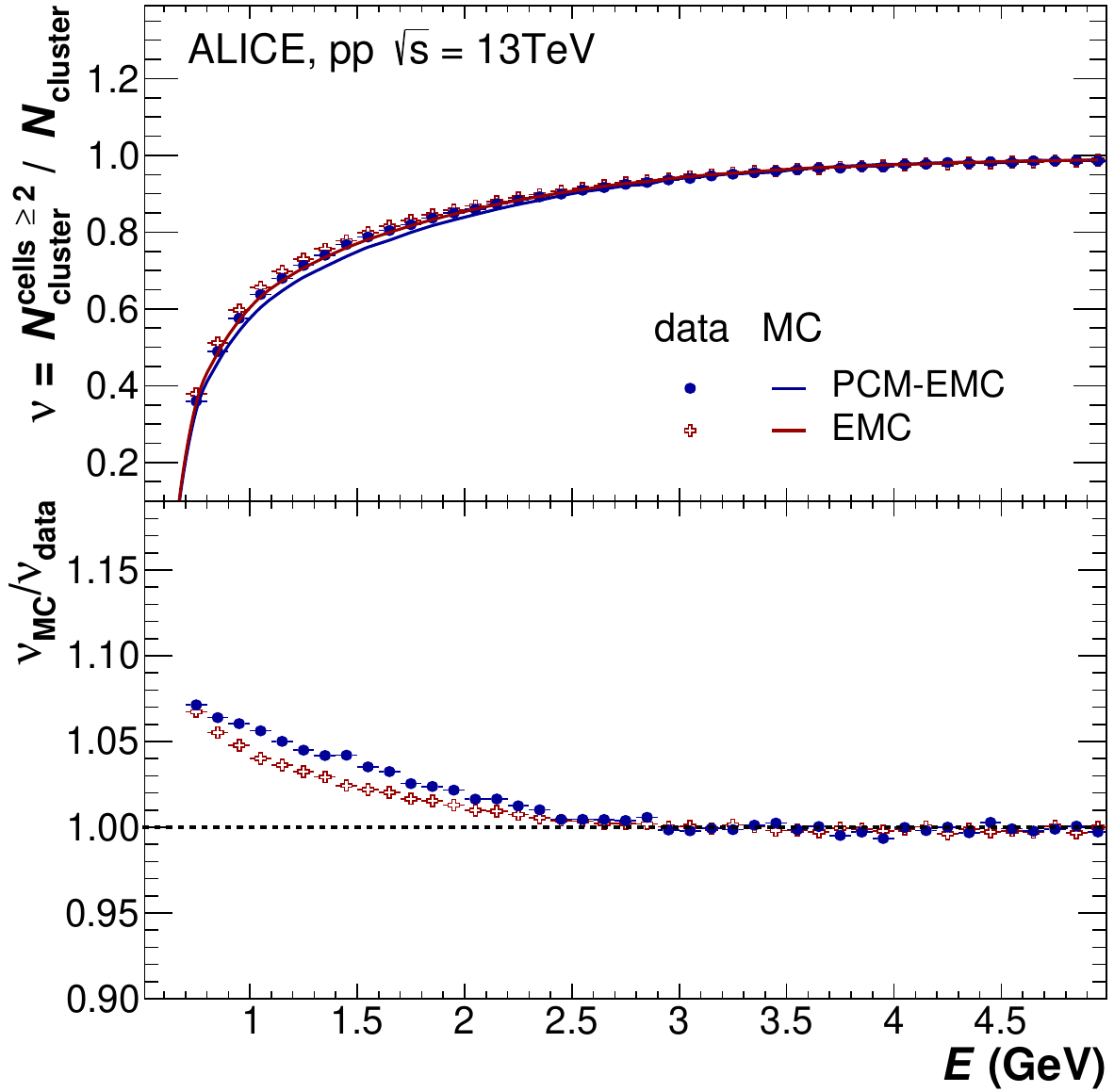}
\includegraphics[height=7.2cm]{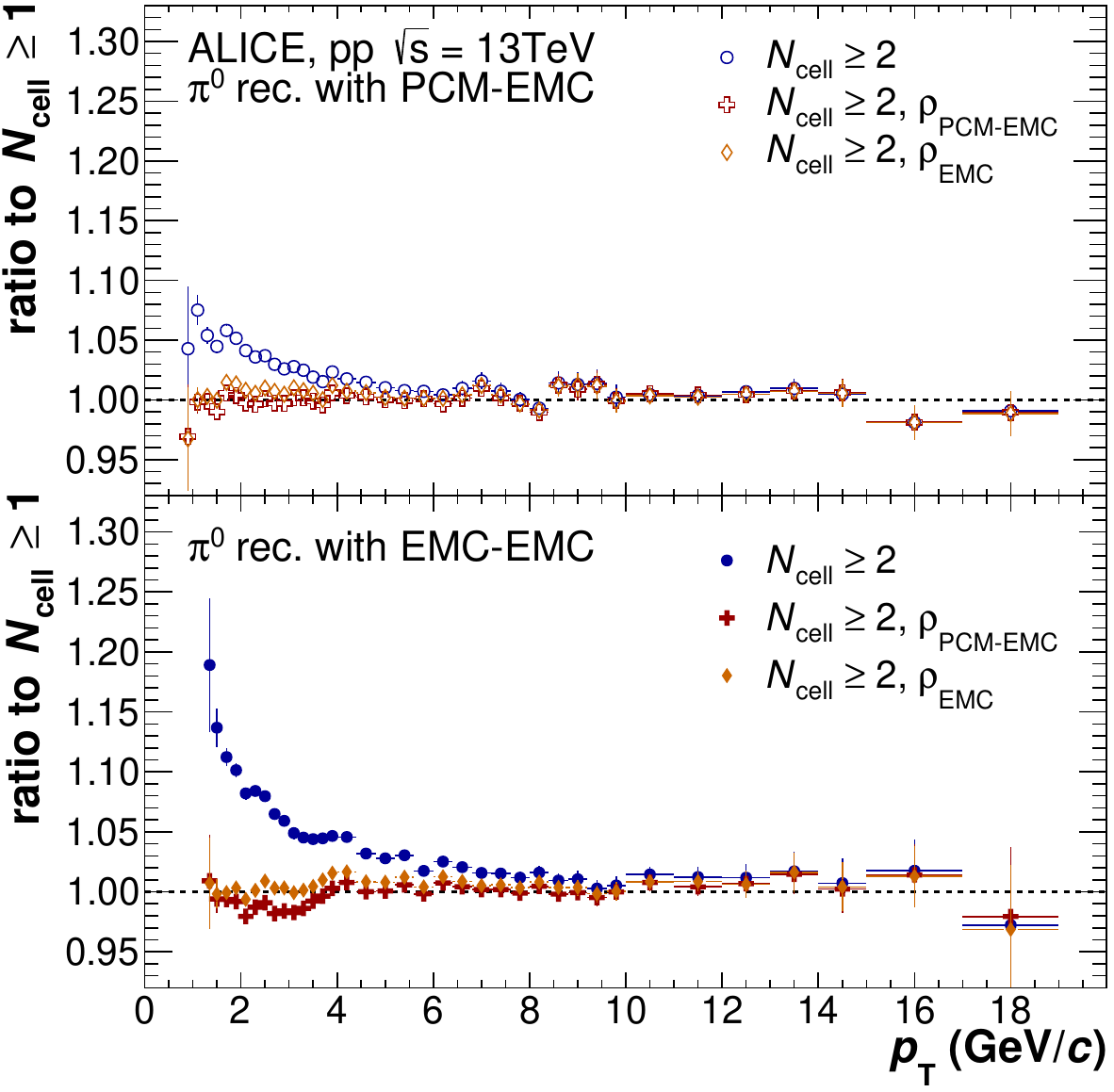}
\caption{(Color online) Left: Fraction of clusters with 2 or more cells for data and \gls{MC} for clusters selected with \gls{PCM-EMC} tagging and \gls{EMC} tagging. 
                        Right: Ratio of fully corrected \piz\ meson spectra obtained with \gls{PCM-EMC} (top) and \gls{EMC} (bottom) with $N_{\rm{cell}} \geq 2$ to the $N_{\rm{cell}} \geq 1$ with and without applying the cluster-size correction.}
\label{fig:NCellEffi}
\end{figure}
As shown in \Fig{fig:5-TB-RecEffic_1GeV_MC} and \Fig{fig:5-TB-RecEffic_1GeV_data}, the number of cells per cluster is not perfectly reproduced by the simulation. 
More single-cell clusters are observed in \gls{MC} than in data. 
For purity-based analyses a cut on single-cell clusters is essential to reject non-physical, exotic clusters (see \Sec{sec:exotics}). 
At low energies, exotic clusters typically have only one cell and are indistinguishable from physical clusters. 
However, a cut on the minimum number of cells results in an inaccurate reconstruction efficiency correction since a larger fraction of clusters are rejected in the \gls{MC} than in data.
To be able to quantify the difference of the fraction of the physical one cell clusters between data and MC, exotic clusters have to be removed from the cluster sample. 
This is accomplished by selecting clusters from cluster pairs which are likely to originate from \piz\ meson decays with $M_{\piz}^{\rm{PDG}} - 0.05 \leq M_{\gamma\gamma} < M_{\piz}^{\rm{PDG}} + 0.02$ (\Eq{eq:KineIMgg}).
The tagging technique can be performed using two clusters with the \gls{EMCal}, called \gls{EMC} tagging, or by pairing a conversion photon with an \gls{EMCal} cluster, called \gls{PCM-EMC} tagging. 
Details on the meson reconstruction technique can be found in \Sec{sec:mesons}. 
The latter is preferred due to a smaller fraction of overlapping clusters from meson decays.
As a result of the large signal to background ratio around the \piz\ meson mass in the $M_{\gamma\gamma}$ distribution, the obtained cluster sample mainly consists of $\gamma$ and $\gamma_{\rm conv.}$ clusters originating from \piz\ meson decays. 
These clusters are used in the following to calculate the difference in the number of cells per cluster between data and \gls{MC}.
The difference can be expressed using the fraction $\nu$ of clusters containing two or more cells 
\begin{equation}
\nu = N_{\rm{cluster}}^{\rm{cells} \geq 2}  / N_{\rm{cluster}} ~~, 
\end{equation}
where $N_{\rm{cluster}}^{\rm{cells} \geq 2}$ are all clusters with two or more cells and $N_{\rm{cluster}}$ are all selected clusters. 
\Figure{fig:NCellEffi}{ (left)} shows $\nu$ as a function of the cluster energy for the two tagging methods for data and MC. 
With decreasing energy, the fraction of single-cell clusters increases both for data and MC.
The ratio between \gls{MC} and data is shown in the bottom panel for both tagging techniques, where a disagreement between data and \gls{MC} of up to $\approx 7\%$ is observed at low energies. 
The difference between the \gls{EMC} tagged and \gls{PCM-EMC} tagged clusters is up to $2 \%$ depending on the cluster energy. 
To correct for the disagreement between data and \gls{MC}, a fraction $\rho(E)$ of the single-cell cluster sample will be treated as two-cell clusters in the simulation such that $\nu_{\rm{MC}} / \nu_{\rm{data}} = 1$, thus, artificially correcting the imbalance of one and more cell clusters in data and simulation.
This correction only impacts low energetic clusters and quickly approaches zero above $2$ GeV. \\
In contrast to purity-based measurements, invariant mass based analyses are not affected by exotic clusters or noise contributions as these clusters are removed by the background subtraction during the peak extraction. 
A cut on the number of cells is therefore not mandatory and thus the transverse momentum dependent invariant yield of the \piz\ meson without an $N_{\rm cell}$ selection criterion can be used as reference for the validation of this correction procedure. 
\Figure{fig:NCellEffi}{ (right)} shows the \piz\ meson invariant yield ratios for transverse momentum spectra obtained using only clusters with $N_{\rm{cell}} \geq 2$ with respect to the spectrum without such a cluster selection criterion.
In order to obtain these spectra, the cluster sample with a cut on the number of cells is either corrected with the correction factor based on the \gls{EMC} or \gls{PCM-EMC} method ($\rho_{\rm{EMC / PCM-EMC}}$). 
The ratios of the spectra without this additional correction are also shown for comparison.
The top panel shows the \piz\ meson spectra obtained with \gls{PCM-EMC} and the bottom panel shows the \piz\ meson spectra obtained with \gls{EMC}.
The corrected spectrum based on clusters with a cut on $N_{\rm{cell} \geq 2}$ (blue points) deviates up to $\approx$ 7 \% (15\%) for \gls{PCM-EMC} (\gls{EMC}) as a result of the incorrect description of the number of cells per cluster by the \gls{MC}. 
With the $\rho$ correction applied, the deviation is reduced to less than 1.5 \% (3\%) for \gls{PCM-EMC} (\gls{EMC}) with respect to the result without selection on $N_{\rm{cell}}$. 
The remaining difference after applying the correction with the two different functions is taken as systematic error.

\subsection{Cross-talk emulation}
\label{sec:crosstalk}
\begin{figure}[t]
    \centering
    \includegraphics[width=0.5\textwidth]{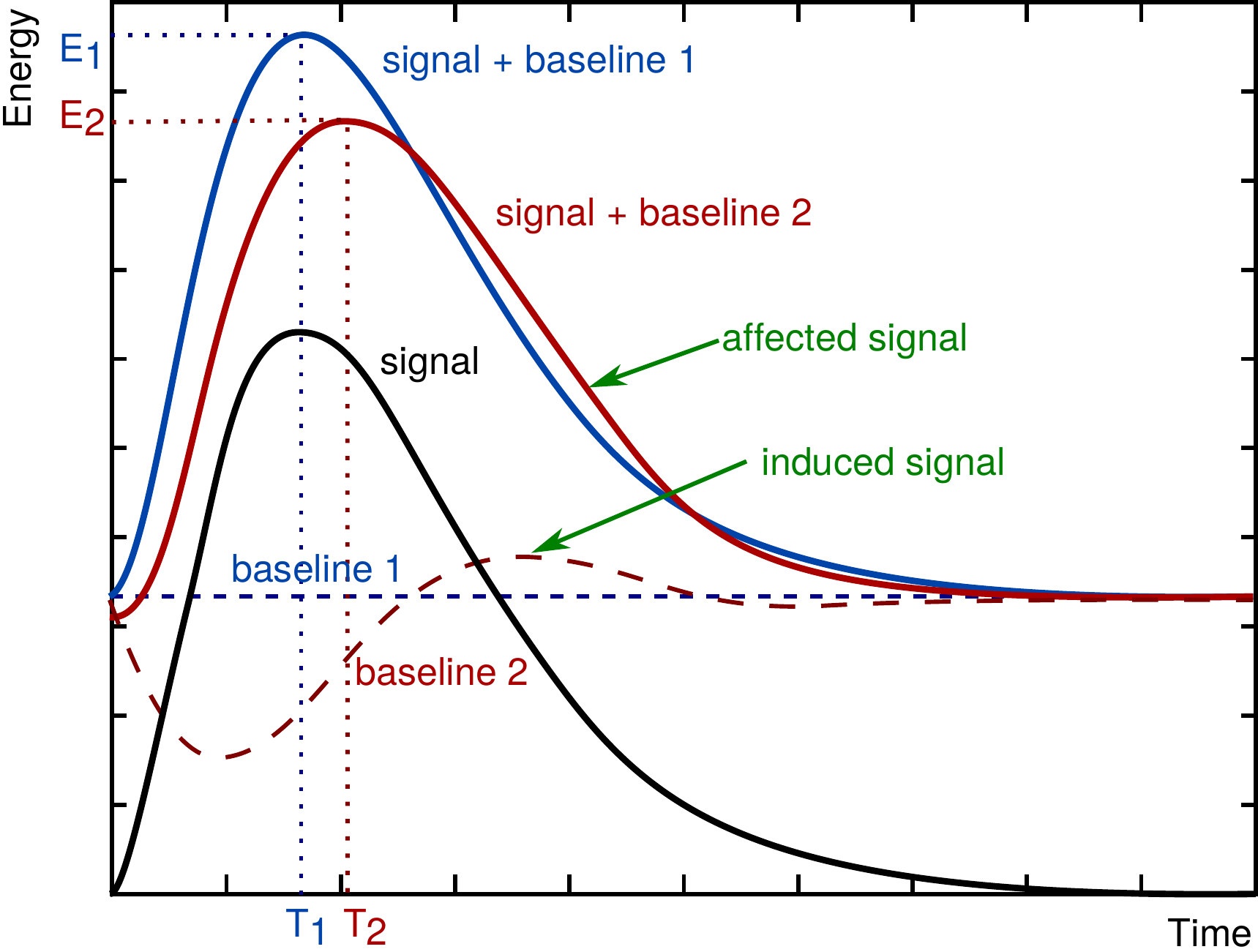} 
      \hspace{0.2cm}
	\raisebox{+18mm}{%
    \begin{minipage}[b]{7.4cm}
    \caption{\label{fig:ADCbaseline} (Color online) Schematic view of the measured \gls{ADC} time distribution shapes considering the contribution of the signal and a baseline. 
        The effect of a baseline modification on the final shape is shown. 
    }
    \end{minipage}
  }    
\end{figure}

\begin{figure}[t!]
    \centering
    \includegraphics[width=0.3284\textwidth]{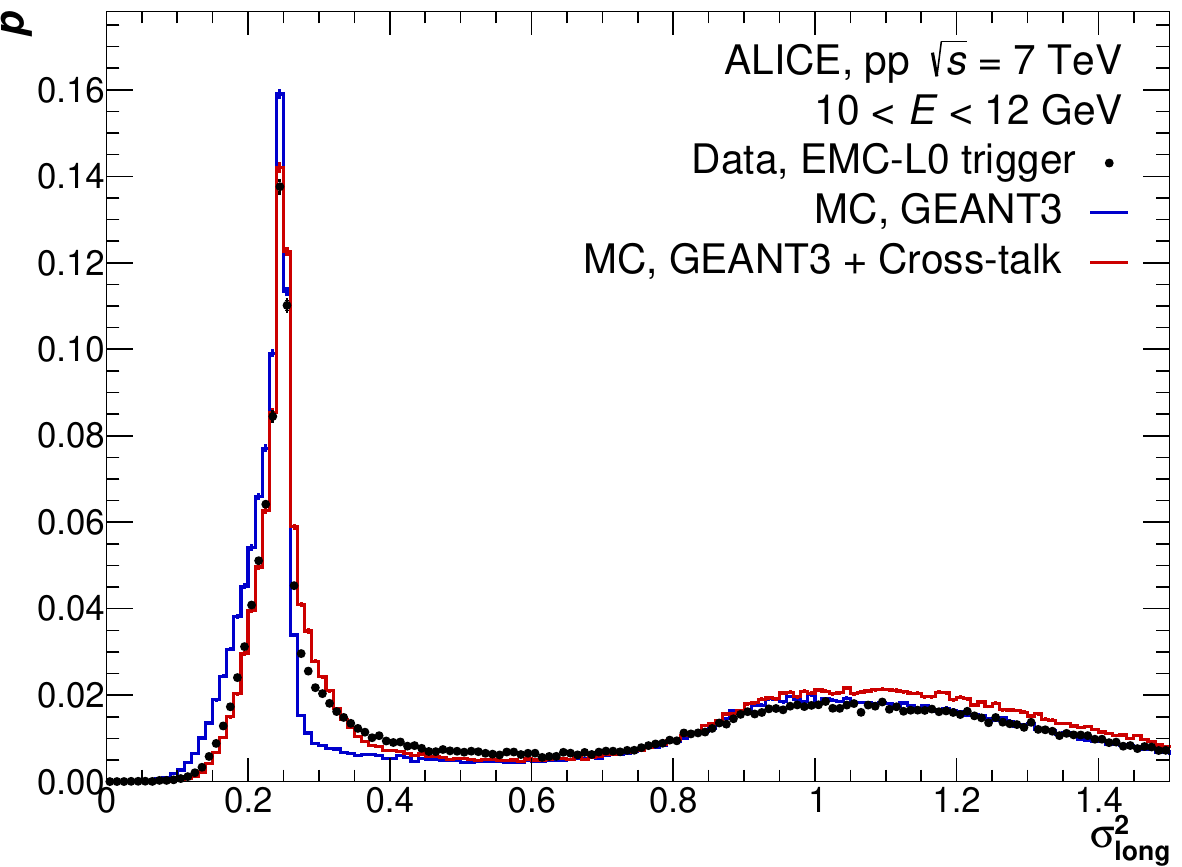}
    \includegraphics[width=0.3284\textwidth]{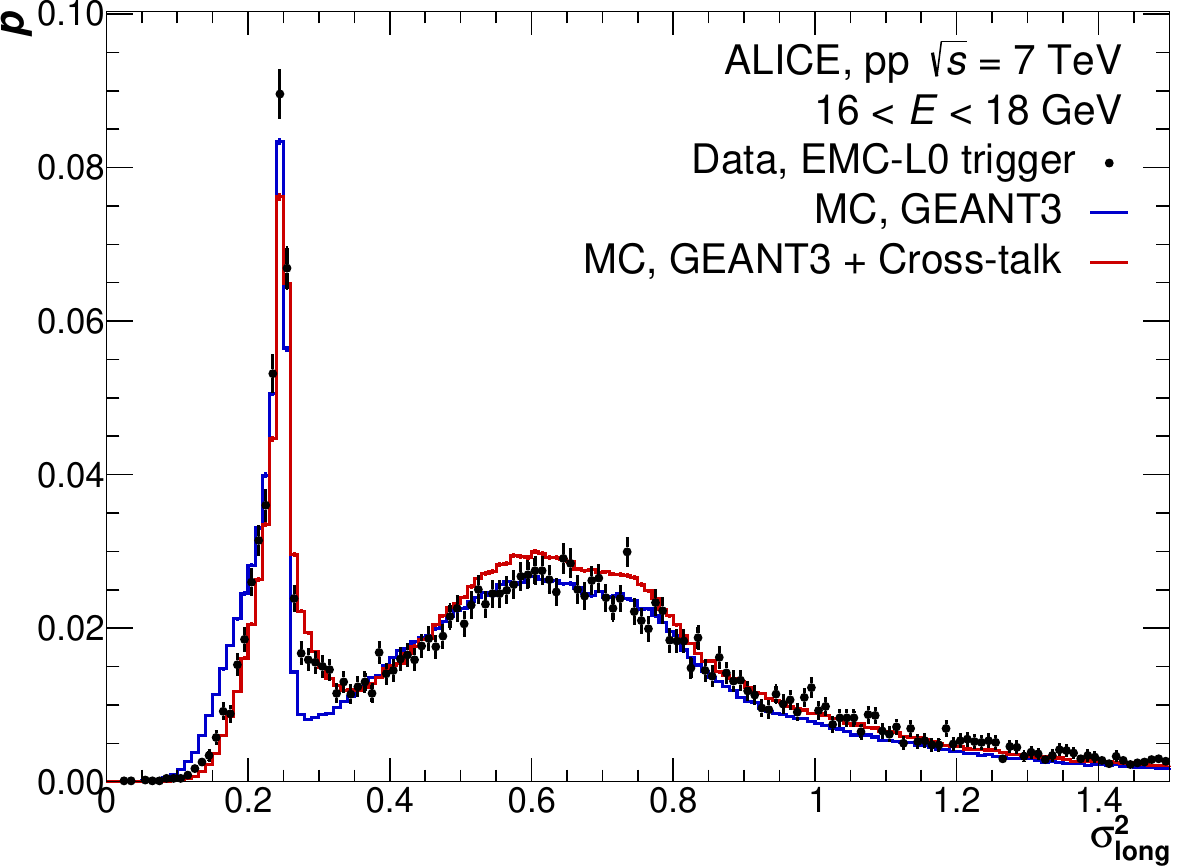}
    \includegraphics[width=0.3284\textwidth]{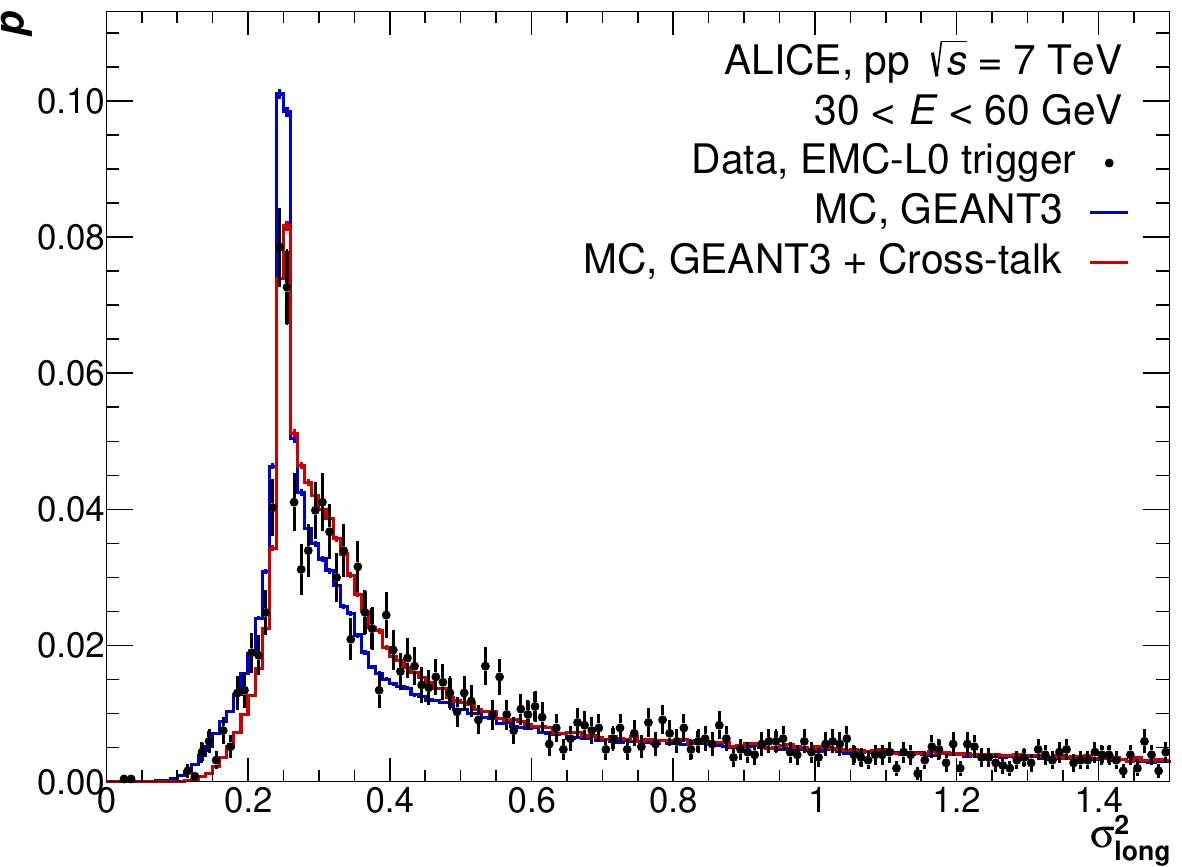}
    \caption{
        \label{fig:SigmaLongDataMCxTalk} (Color online) Probability distributions of the shower shape parameter $\shshlo$ of neutral clusters in data and simulations.
        The different panels show different neutral cluster energy intervals. All distributions are normalized to their integral.
        Data are shown as black histograms and simulations (\gls{PYTHIA}6 events with two jets or a direct photon and a jet in the final state, with \gls{GEANT}3 default settings) in blue. 
        For the red histograms the modelling of the cross-talk observed in the \gls{EMCal} electronics was included in the simulations.
    }
\end{figure}

Laboratory tests~\cite{Rusu:2019pvk} on the electronic cards revealed that channels induce cross-talk signal to other channels belonging to the same T-Card resulting in a nonconstant baseline described schematically in \Fig{fig:ADCbaseline}, affecting the measured cell energy and time.
The cross talk also modifies the energy and time distribution of the adjacent cells and thus the shower shape of the cluster.
Laboratory studies demonstrated that the cross-talk is mainly due to the ribbon cables used to connect the T-Cards with the \gls{FEE}-Cards and has a random nature. 
No analytical model could reproduce its effect in simulations.
Instead, we rely on data-driven techniques to quantify the cross-talk and to emulate it in the simulation.

In early analyses of the cluster shapes, a difference between the distribution of energy over the calorimeter cells between data and \gls{MC} simulation was found. An example of this difference is shown in \Fig{fig:SigmaLongDataMCxTalk}.
The real data distribution appears to be broader than the simulated data, in particular in the range $0.3 < \shshlo < 0.4$ on the right side of the {\it photon peak} ($0.1 < \shshlo < 0.3$, see \Fig{fig:SSPhotonPi0MC}).

While initially this effect was attributed to a mismatch in the calorimeter response in \gls{GEANT}3, in fact the largest contribution arises from problems on the hardware level. 
The observation of unexpected low mass peaks in the \piz\ meson energy calibration revealed correlations of non-physical origin between distant towers when one was noisy, which triggered the testing of the \gls{FEE} and T-Card~(see \Sec{sec:hardware-moduledesign}). 

Broader cluster shapes can also be caused by additional material present in the detector, which is not implemented in the simulation.
This expected broadening can be observed when looking at the cluster activity as a function of $\eta$ and $\varphi$ for clusters in the tail region $0.3<\sigma^{2}_{\rm long}<0.4$.
A clear enhancement in the regions corresponding to the support structures of the inner detectors at $|\eta|\approx 0.6$ emerges, both in data and simulation.
However, clusters measured in full \glspl{SM} 3 and 7 have broader cluster shape compared to the other \glspl{SM} in data but not in simulation.
This difference is enhanced by requiring a minimum number of cells selection in addition.
According to \Eq{eq:w_i} (see \Sec{sec:showershape}), only cells with $w_i>0$ and thus at least $\sim$1.1\% of the cluster energy, contribute to the shower shape calculation.
In the presence of cross-talk, some energy can leak from a sufficiently high-energy cell to neighbor cells with low energy, which can cause the neighbor cells to go over threshold. 

\begin{figure}[t!]
    \centering
    \includegraphics[width=0.49\textwidth]{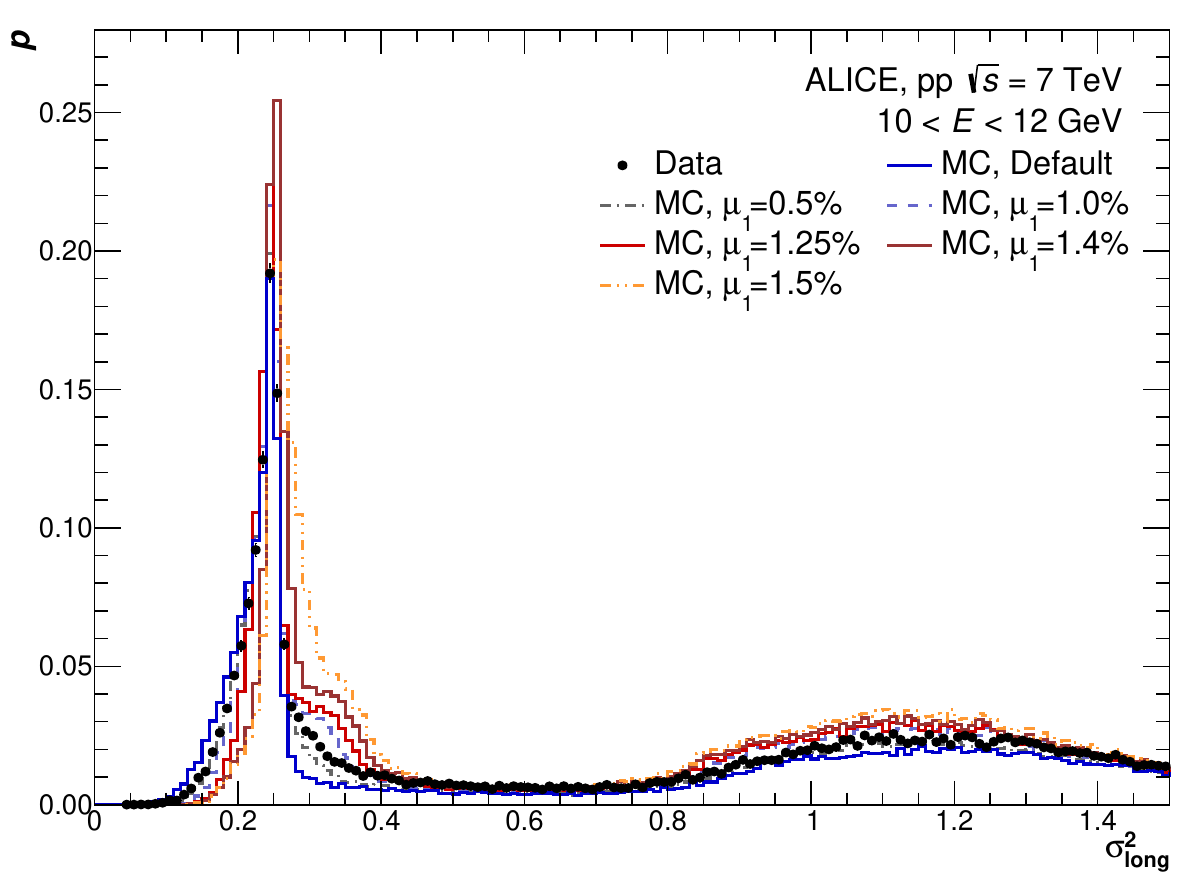} 
    \includegraphics[width=0.49\textwidth]{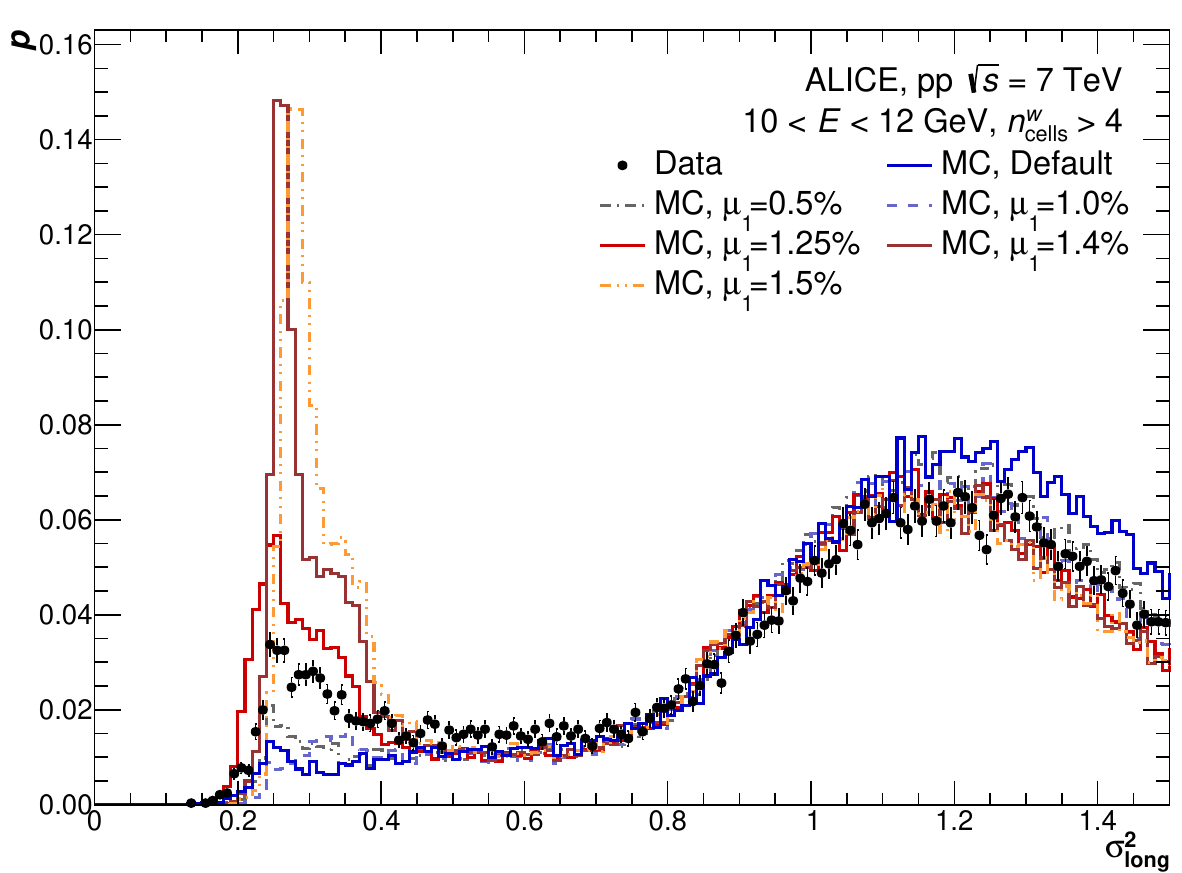} 
    \caption{
        \label{fig:shower_shape_8GeV_indE} (Color online) Comparison of the probability distribution of the shower shape parameter, $\shshlo$, of neutral clusters with $n^{\rm w}_{\rm cell}>1$~(left) and $>4$~(right) for different fractions of induced energy in the cross talk model (see \Eq{eq:crosstalk}), in pp collisions at \sseven\ \gls{EMCal} triggered data are compared to \gls{PYTHIA}6 simulated events with a direct photon and a jet or two jets in the final state, where one jet is triggered by a decay $\gamma$ on \gls{EMCal} acceptance with \pt$>$~3.5 GeV/$c$. 
        Data and default \gls{MC} (untuned simulation) are the same as in \Fig{fig:SigmaLongDataMCxTalk}.
    }
\end{figure}
\begin{figure}[t!]
    \centering
    \includegraphics[width=0.32\textwidth]{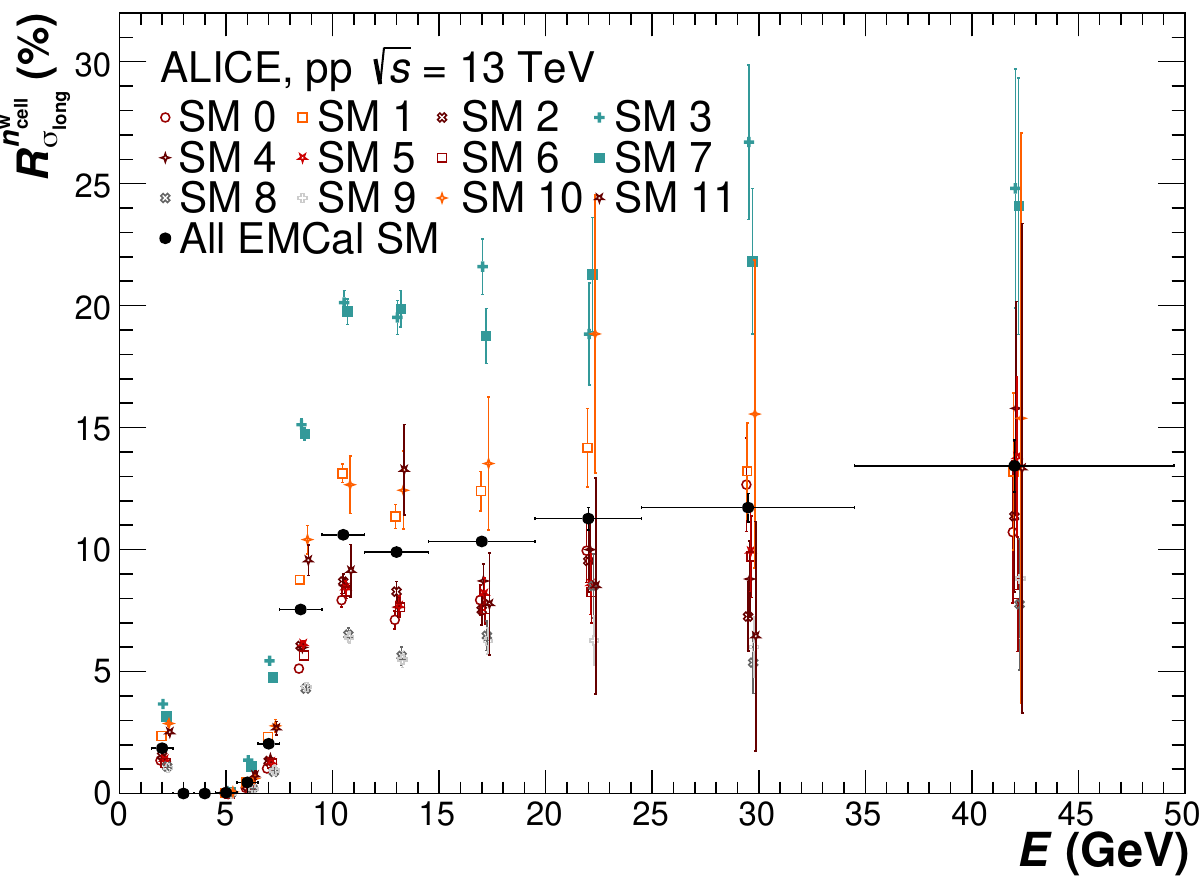} 
    \includegraphics[width=0.32\textwidth]{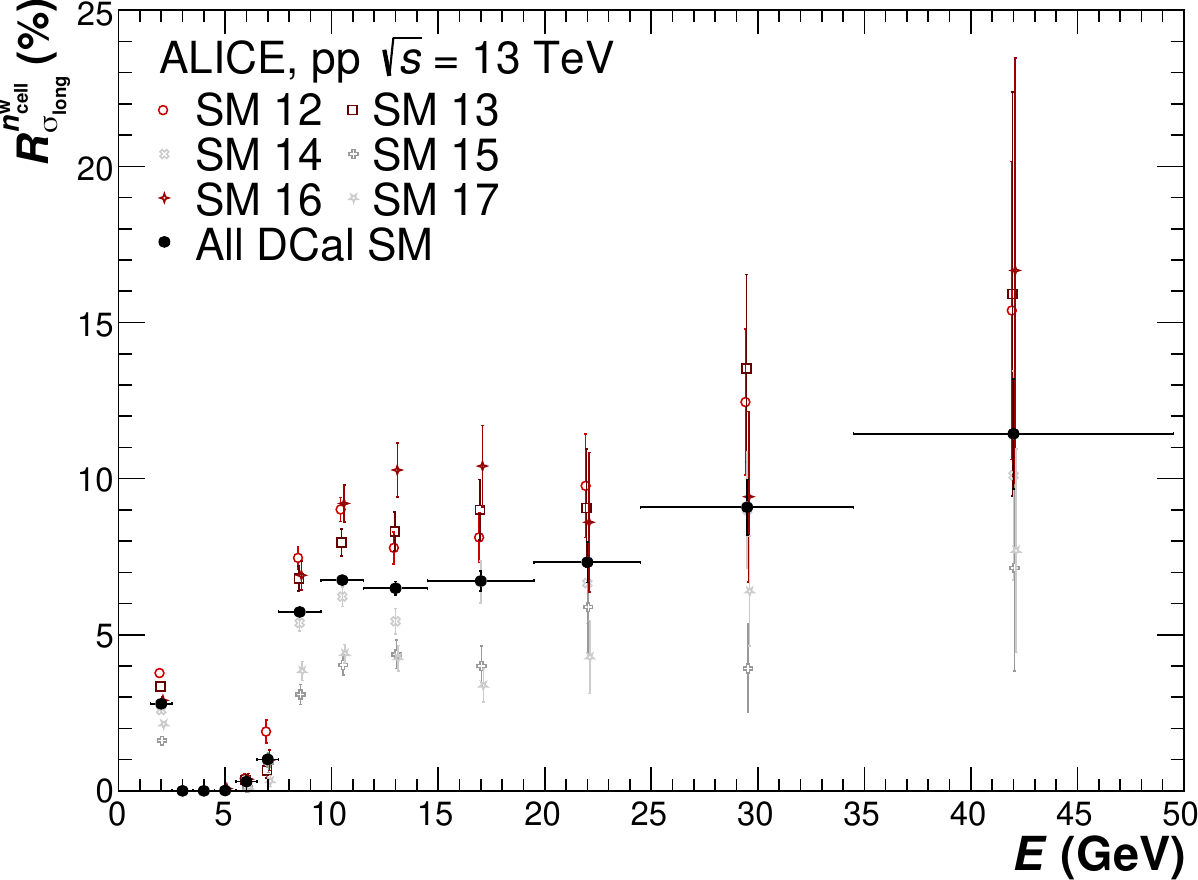} 
    \includegraphics[width=0.32\textwidth]{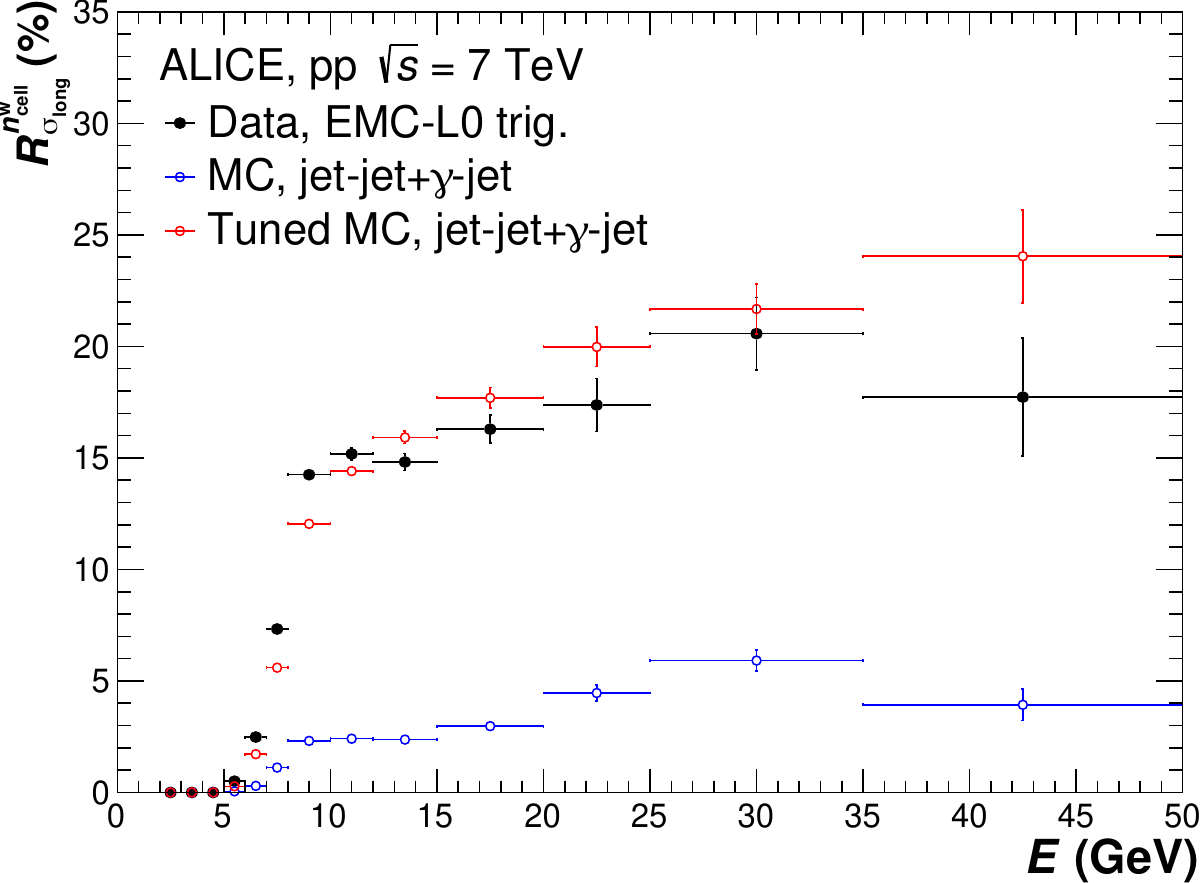}
    \caption{
        \label{fig:shower_shape_Incl_EvsNcut_Run2data_pp}(Color online) Fraction of clusters with $n^{\rm w}_{\rm cell}>4$ within the range $0.1<\sigma_{\rm long}^{2}<0.3$ per \gls{SM}. 
        Left and middle: pp \sthirteen\ \gls{L1} $\gamma$ triggered data. 
        Right: Clusters (enhanced merged decay population by few \%) in \gls{EMCal} for pp \sseven\ minimum bias and \gls{L0} trigger data (black marker), compared to simulations with 2 jets in the final state with different photon trigger thresholds, with (red marker) and without (blue marker) cross-talk tuning. 
    }
\end{figure}

\Figure{fig:shower_shape_8GeV_indE} shows the $\sigma^{2}_{\rm long}$ distribution for neutral clusters for which the number of cells in the cluster that contribute to the shower shape calculation, $n^{\rm w}_{\rm cell}$, is larger than 1 (left panel) and 4 (right panel). The distributions for $n^{\rm w}_{\rm cell}>4$ show a bump
in the photon region $0.1<\sigma^{2}_{\rm long}<0.3$ not observed in the simulation. 
The fraction \Rn of clusters in the photon region with $n^{\rm w}_{\rm cell} > 4$ with respect to all clusters in that region is calculated for each \gls{SM} and compared to simulations to quantify the effect per \gls{SM}.

\Figure{fig:shower_shape_Incl_EvsNcut_Run2data_pp} indicates that \Rn 
is rather constant for $E>5$~GeV but with a different value depending 
on the \gls{SM} number. For $E<5$~GeV, 
there is a suppression of clusters with a large number of cells due to the clusterization threshold $E_{\rm agg}$ and the energy threshold for inclusion in the shower shape calculation $w_{0}$. 
Different categories can be identified: the most affected \glspl{SM} are \gls{SM} 3 and 7, followed by \glspl{SM} 1 and 10, while the remaining \glspl{SM} of the \gls{EMCal} show significantly fewer clusters with $n^{\rm w}_{\rm cell} > 4$.
In the simulation, the plateau is lower~(\Rn$\sim2$\%) than for any of the \gls{SM} categories observed in data ($4 < $\Rn$ < 20$\%).

In order to emulate the observed features in the simulations, the following assumptions were made:
the problem is purely a cross-talk between cells inside a T-Card; all cells in the calorimeter may induce cross-talk; given a signal in a cell, the neighbouring cells are affected, and possibly those 2 rows away; it can depend on the cell energy; it can depend on the \gls{SM} number but not on rapidity or azimuth within the \gls{SM}. 
The assumptions on the rapidity independence are not entirely correct as, due to cable lengths, the region close to $\eta=0$ is more affected, but these assumptions were made to simplify the emulation process.\\
The proposed algorithm emulating the cross-talk at the simulation level works as follows: 
inspect all the cells with signal in the simulation, and for each cell with energy $E_{\rm cell}$, add to the surrounding 5 cells in the same T-Card an induced energy $E^{\rm ind}_{i,j}$ ($i$ and $j$ indicate the column and row with respect to the selected cell)
\begin{equation}
    E_{i,j}^{\rm ind} = F^{\rm ind} E_{\rm cell},~{\rm with}~F^{\rm ind} = \mu_{1}+\mu_{2}~E_{\rm cell},
    \label{eq:crosstalk}
\end{equation}
where $\mu_{1}$ and $\mu_{2}$ depend on the \gls{SM} and the location of the cells $(i,j)$. If $F^{\rm ind}$ is above or below a given value $F^{\rm ind}_{\rm max}$ or $F^{\rm ind}_{\rm min}$, respectively, those values are used instead.
Each $E_{i,j}^{\rm ind}$ is optionally smeared by a Gaussian random distribution with width $\sigma_{\rm ind}$.
Finally, the total induced energy after smearing in the nearby T-Card cells is subtracted from the main signal cell, 
$E_{\rm cell}^{ \rm final} = E_{ \rm cell}-\Sigma E_{i,j}^{\rm ind}$,
so that the energy scale is conserved. 
For simplicity, all five surrounding cells in the T-Card use the same $\mu_{1}$ and $\mu_{2}$ and a smaller $\mu_{1}$ is used for the cells two rows away from the reference signal cell. 
Additionally, the amount of induced energy is limited for lower energies in order not to provoke additional cluster nonlinearity by requiring:
$E_{i,j}+E_{i,j}^{\rm ind} > E^{\rm ind}_{ \rm min}$,
where $E^{\rm ind}_{ \rm min}$ is the same as the clusterization minimum cell energy $E_{ \rm agg}$. 
If the sum is smaller, the induced energy is not subtracted from the signal reference cell.
\begin{table}[b!]
    \centering
    \caption{
        \label{tab:EmulationParam} Emulation parameters used to calculate the $E^{\rm ind}_{i,j}$ in the 5 surrounding cells to a given cell with signal in the T-Card depending on the \gls{SM} number. 
        Note that we allow also the inducing energy in cells 2 rows away, but only with $F^{\rm ind}$=$F^{\rm ind}_{\rm min}$. 
    }
    \begin{tabular}{l*{6}{r}r}
                  & group 1 & group 2 & group 3 & group 4\\ 
    \toprule
    \glspl{SM} \gls{EMCal}   & 3, 7  & 1, 10 & 0, 2, 4, 5, 6, 11  & 8, 9  \\
    \glspl{SM} \gls{DCal}    & -     & -     & 12, 13, 16      & 14, 15, 17, 18, 19  \\
    $F^{\rm ind}_{\rm min}$  & 0.6\% & 0.5\% & 0.45\% & 0.35\%  \\
    $F^{\rm ind}_{\rm max}$ & 1.8\% & 1.6\% & 1.6\%  & 1.6\%   \\
    $\mu_{1}$         & 1.2\% & 1.2\% & 1.15\% & 0.8\%   \\
    $\mu_{2}$ (GeV$^{-1}$)         & -0.11\% & -0.11\% & -0.11\% & -0.11\%   \\
    $\sigma_{\rm ind}$  & 0.5\% & 0.5\% & 0.5\% & 0.5\%   \\
    $E^{\rm ind}_{\rm min}$ (GeV) & $E_{\rm agg}$ & $E_{\rm agg}$ & $E_{\rm agg}$ & $E_{\rm agg}$   \\ \bottomrule
    \end{tabular}
\end{table}

\Figure{fig:shower_shape_8GeV_indE} shows the effect on $\sigma_{\rm long}$ with and without the cut on $n^{\rm w}_{\rm cell}$ for a fixed $\mu_{1}$ value of a few percent, $\mu_{2}=0$, no smearing of the induced energy and no cut on the leaked energy, applied on the 5 surrounding cells in simulation and its comparison to data. 
A clear broadening and even a shift of the distribution develops when increasing $\mu_{1}$ from 0.5\% to 1.5\%, the
extent of which depends on the cluster energy. 
For the chosen cluster energy, a value of $\mu_{1} \approx 1.35$\% is favoured for \glspl{SM} 3 and 7, and $\mu_{1} \approx 1.2$\% is favoured for the other \glspl{SM}.
After studying the energy dependence, the optimum parameterization for $\mu_{1}$, $\mu_{2}$, $\sigma_{\rm ind}$, $F^{\rm ind}_{\rm min}$ and $F^{\rm ind}_{\rm max}$ as listed in \Tab{tab:EmulationParam} for different \glspl{SM} was found.\\
\begin{figure}[t]
    \centering
    \includegraphics[width=0.49\textwidth]{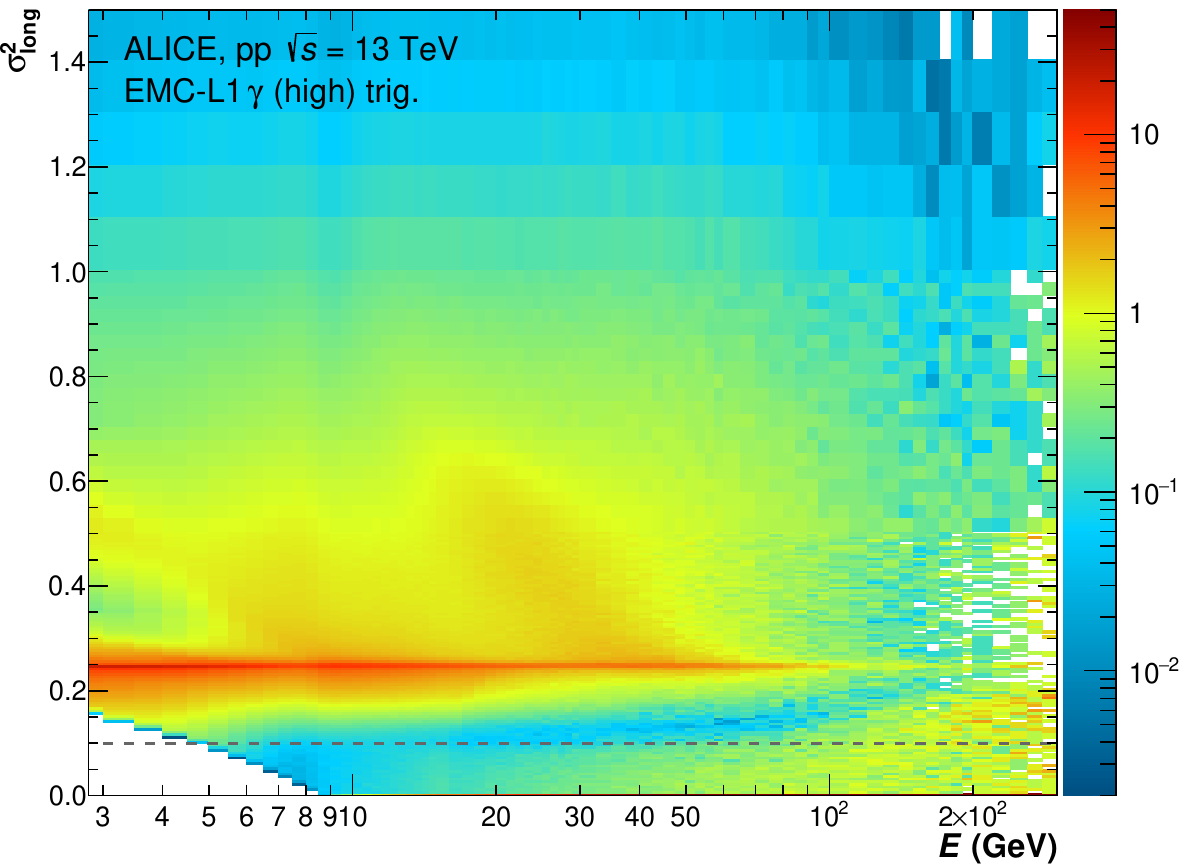}
    \includegraphics[width=0.49\textwidth]{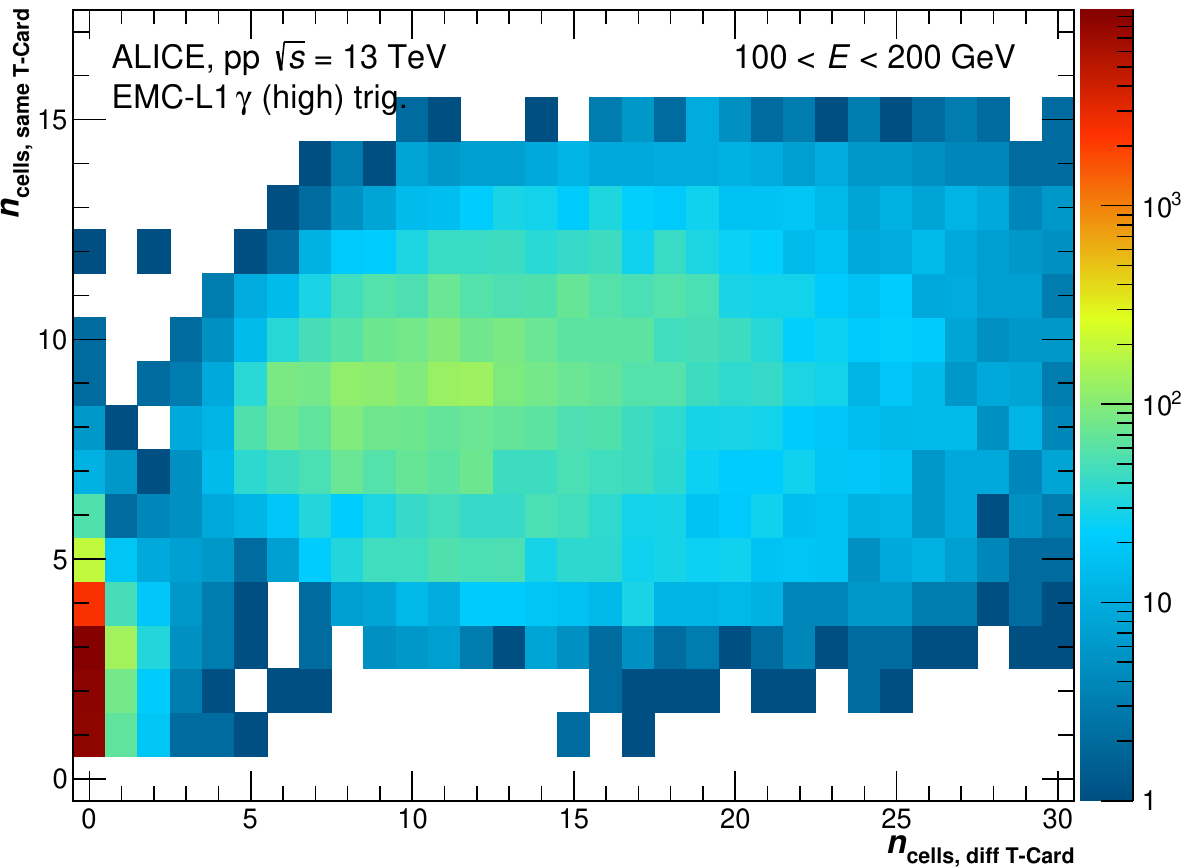}
    \caption{
        (Color online) Left: $\shshlo$ as a function of the cluster energy. 
        Data, pp collisions at \sthirteen\ triggered by the \gls{EMCal} and \gls{DCal} with \gls{L1}-$\gamma$ at $8.5$~GeV, V2 clusterizer.
        Right: Number of cells in the cluster in the same T-Card as the highest energy cell as a function of the number of cells in a different T-Card, for clusters between 100 and 200~GeV. 
    }
    \label{fig:HighEnCrossTalk}
    \label{fig:HighEnCrossTalkFractionCellsTCard}
\end{figure}
\Figures{fig:SigmaLongDataMCxTalk}{fig:shower_shape_Incl_EvsNcut_Run2data_pp} (right) show the final agreement between data and simulation for the $\sigma_{\rm long}^{2}$ distributions and the \Rn variable with and without the final cross-talk emulation.
The cross-talk emulation also improves the agreement between data and simulation for all other cluster properties, like $\sigma_{\rm short}^{2}$ and number of cells.
%
\begin{figure}[t!]
    \centering
    \includegraphics[height=3.8cm]{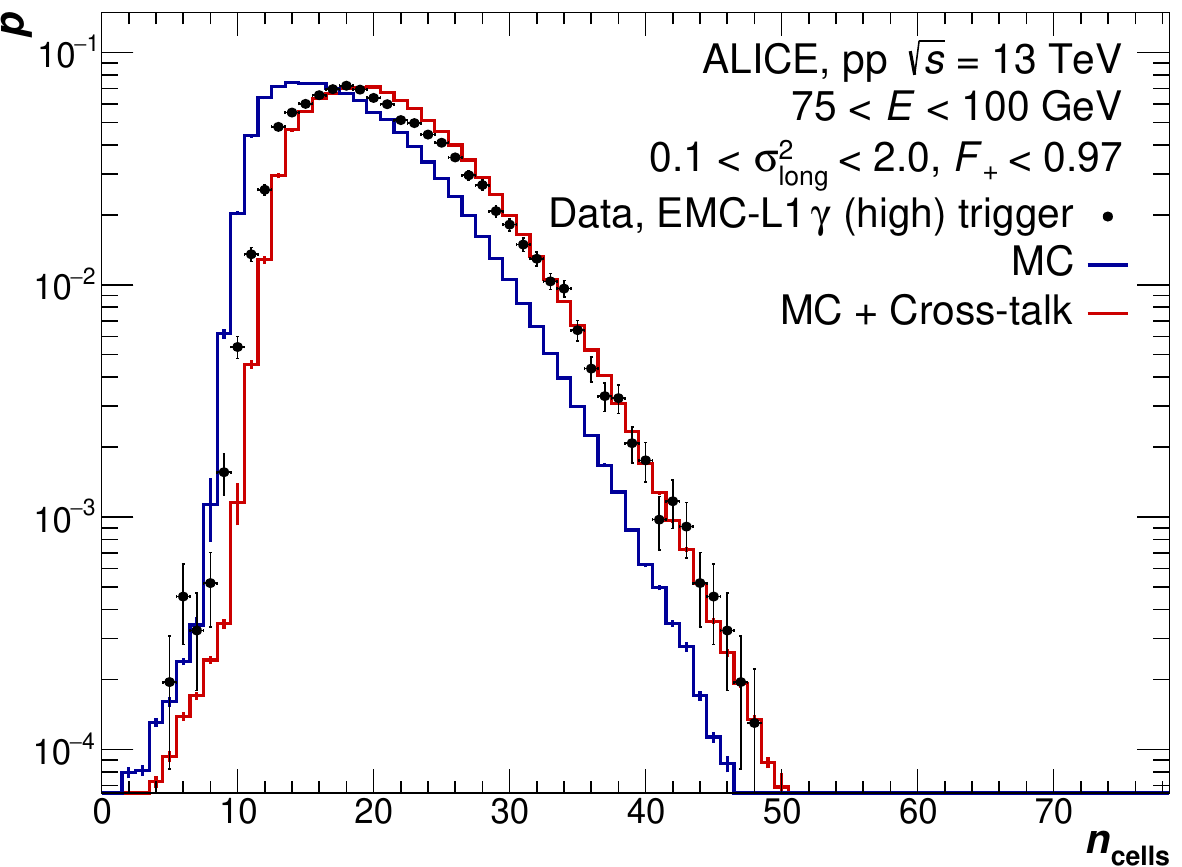}
    \includegraphics[height=3.8cm]{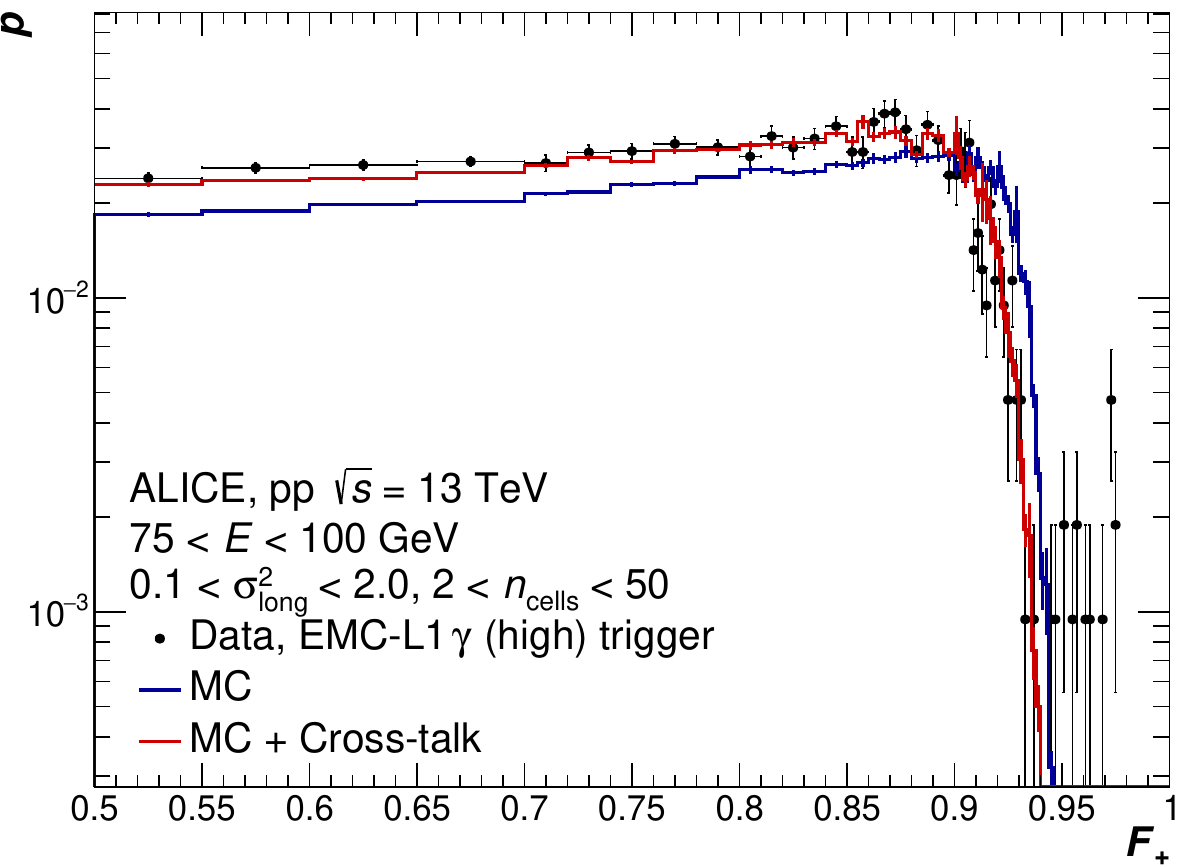}
    \includegraphics[height=3.9cm]{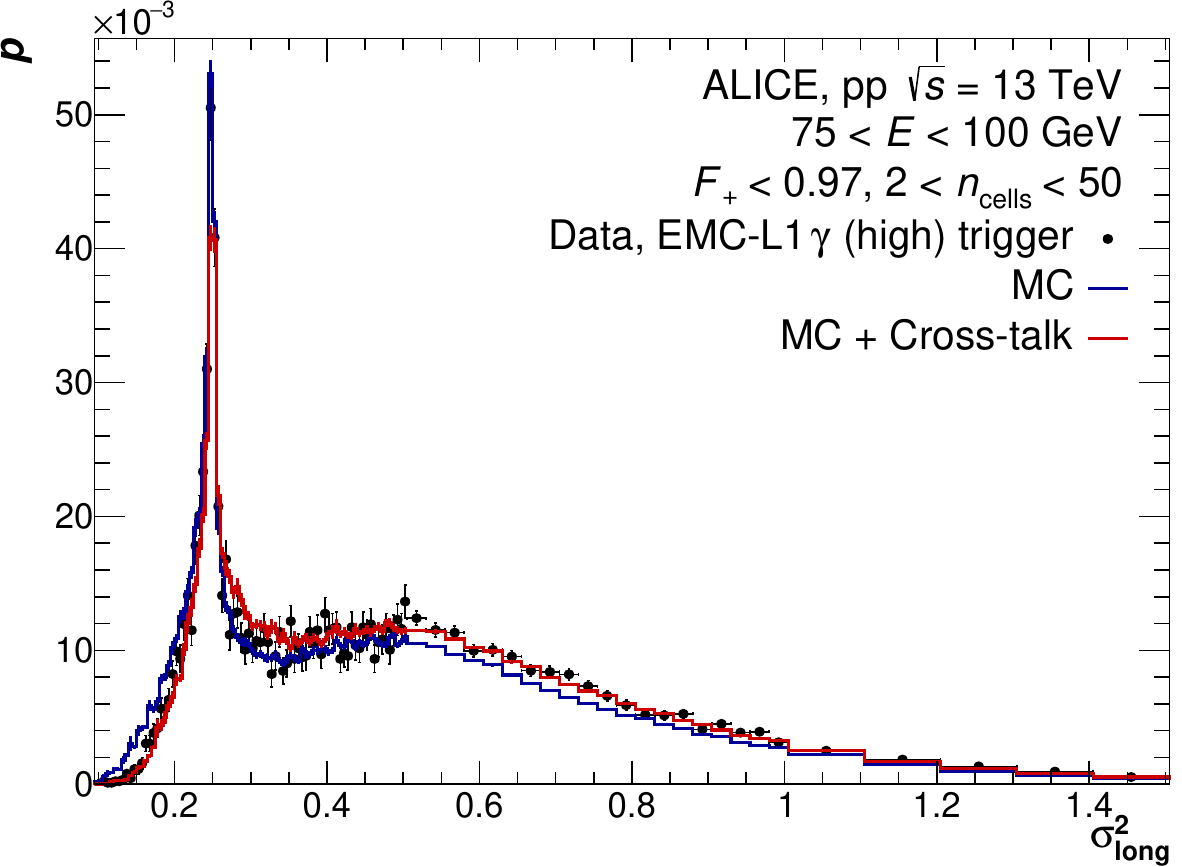}
    \caption{
        \label{fig:HighEnCrossTalkCutDataMC} (Color online) Comparison of the probability distributions between data and simulation for cluster $n_{\rm cells}$ (left), $F_{+}$ (middle) and $\shshlo$ (right) for V2 clusters with $75 < E < 100$~GeV. Selections applied: reject clusters with no cell in a different T-Card as the highest energy cell, $F_{+}>0.95$ and $\shshlo < 0.1$ where it applies.
        Data and simulation of pp collisions at \sthirteen, simulation of \gls{PYTHIA}8 events with two jets or a direct photon and a jet in the final state, and with cross talk emulation activated (red points) and not activated (blue points).
    }
\end{figure}

For very high cluster energies ($E > 50$~GeV), an enhancement of the neutral cluster population at low shower-shape values (between $0.1<\sigma^{2}_{\rm long}<0.25$) appears, which merges with the photon band above $150$~GeV as seen in \Fig{fig:HighEnCrossTalk} (left).
These clusters can be classified as exotic since they have a much lower average number of cells per cluster and very high exoticity (see~\Sec{sec:exotics}).
At such high energies, the leading cell appears to be leaking a significant fraction of its energy into the neighboring cells, which in turn contributes to the expectation of a relatively large value of shower shape parameters.
While the exotics appear to be present in all calorimeters using \glspl{APD} at the \gls{LHC} experiments, the cross-talk observed in the \gls{EMCal} enhances their presence in cluster samples with $N_{\text{cell}} \geq 2$.
This can be seen by correlating the number of cells in a high energy cluster belonging to the same or to a different T-Card of the highest energy cell for data and simulation. 
The corresponding distribution for V2 clusters with $100 < E < 250$~GeV is shown in \Fig{fig:HighEnCrossTalk} (right).
Two distinct structures are visible: A strongly peaked distribution with none or very few cells in a different T-Card, and a much wider spread distribution that has approximately equal amounts in the same and a different T-Card.
While the latter is present in simulations as well, the former is not, indicating that the former stems most likely from exotic signals that are not implemented in the simulation. Those signals in data induce energy by cross-talk on the neighboring cells that sit in the same T-Card as the signal.
Simulations show that above 50~GeV, less than one per mil of clusters cover just one T-Card, except those whose highest energy cell sits in a border, which are usually removed from most of the analyses. 
Consequently, requiring that clusters with energy above 50~GeV have at least one cell contributing to the cluster in a T-Card different than the highest energy cell, cleans the cluster sample of exotic clusters and the standard selection criteria used below 50~GeV can be used normally to further clean the sample from the few remaining exotic clusters. \\
\Figure{fig:HighEnCrossTalkCutDataMC} shows that the agreement between data and simulation at high energies is satisfactory when the selection on the number of cells in the T-Card is used together with the cross-talk emulation.


\ifflush
\clearpage
\fi
\section{Physics performance}
\label{sec:physics}
In this chapter, the performance of the \gls{EMCal} is demonstrated for the identification and reconstruction of different physics observables.
The main variables used to distinguish different particles using the \gls{EMCal} are the shower shape variable \shshlo~(see \Sec{sec:showershape}), as well as the matching of a track to the cluster and its $E/p$~(see \Sec{sec:trackmatch}). 
Combining the information from these sources allows to distinguish electrons, hadrons and photons creating a cluster in the \gls{EMCal}.
The identified photons can be used to reconstruct mesons or baryons with one or more photons as decay products, most prominently the $\pi^0$ and $\eta$ mesons.
Additionally, the \gls{EMCal} clusters are used to improve jet energy measurements in \gls{ALICE} by measuring their electromagnetic component. 
Different clusterizer algorithms are used to reconstruct \gls{EMCal} clusters depending on the requirement of the analysis.

In the following section, the general features of the photon reconstruction method will be explained, first by highlighting the different selection criteria for the inclusive statistical photon measurements (\Sec{sec:photonrec}) and for isolated photons at high transverse momenta (\Sec{sec:photoniso}).
Afterwards, the performance of reconstructing $\pi^0$, $\eta$ and $\eta'$ mesons using their di-photon decay channel will be described and a novel \gls{PID} analysis technique based on the shower shape for the neutral pions will be introduced (\Sec{sec:mesons}). 
Using the reconstructed $\pi^0$ and $\eta$ mesons, we explore the reconstruction performance for  $\omega$ and $\eta'$ mesons using their 3-body decay channels, presented at the end of this section. 
\Section{sec:electrons} is devoted to the performance of the electron identification, which can be used for reconstructing J$/\psi$ mesons via their di-electron decay channel, as well as open heavy-flavour hadrons via their semi-leptonic decay channels (c,b$\rightarrow$e). 
We also present the improvement of the sample size for the D-meson reconstruction via their hadronic decay channels, by taking advantage of the \gls{EMCal} as a trigger detector.
The description of the jet performance in \Sec{sec:jets} concludes the chapter.

\subsection{Photons}
\label{sec:photons}
\subsubsection{Identification of photons}
\label{sec:photonrec}
In order to reconstruct photons using the \gls{EMCal}, both clusterizers, V1 and V2, are used (see \Sec{sec:clusterization}).
For the decay photon reconstruction, the V2 clusterizer is more suitable as the cluster energy is closer to the actual photon energy. 
Additionally, the cluster properties relate better to those of single particles hitting the \gls{EMCal}, since the V2 clusterizer reduces the effects due to merging of clusters from different particles.
However, the performance is rather similar to the V1 clusterizer for momenta below 4~\GeVc. 
While both clusterizers have been used in various photon and meson analyses, the V2 clusterizer is generally preferred when trying to analyse simultaneously the direct photon and neutral pion spectra 
since an invariant mass analysis can be used up to higher \pt\ (see \Fig{fig:InvMassEBins} and \Fig{fig:fractionClusterizer}).
The first results using the \gls{EMCal} in pp collisions for such analyses were published in~\cite{Acharya:2018dqe}.
On the other hand, the V1 clusterizer is a better choice for the isolated photon reconstruction, as the number of maxima in the cluster and the shower shape can be used to discriminate between photons originating from a neutral pion decay and those being truly from a primary photon. 
In the following paragraphs, the reconstructed cluster selection criteria that are common to the different photon analyses will be discussed and the performance of the two clusterizers will be contrasted.

\begin{figure}[t]
    \includegraphics[width=0.49\textwidth]{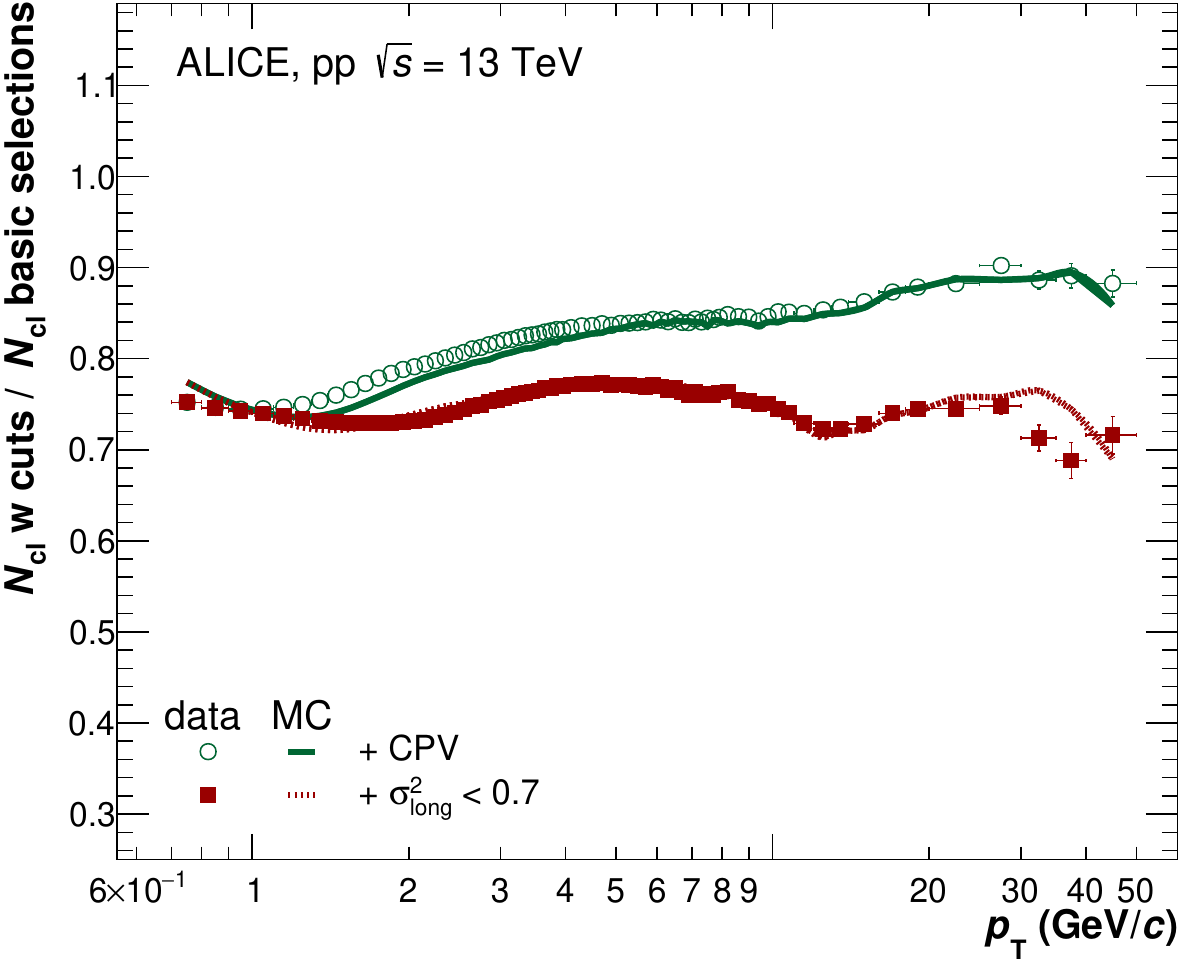}
    \includegraphics[width=0.49\textwidth]{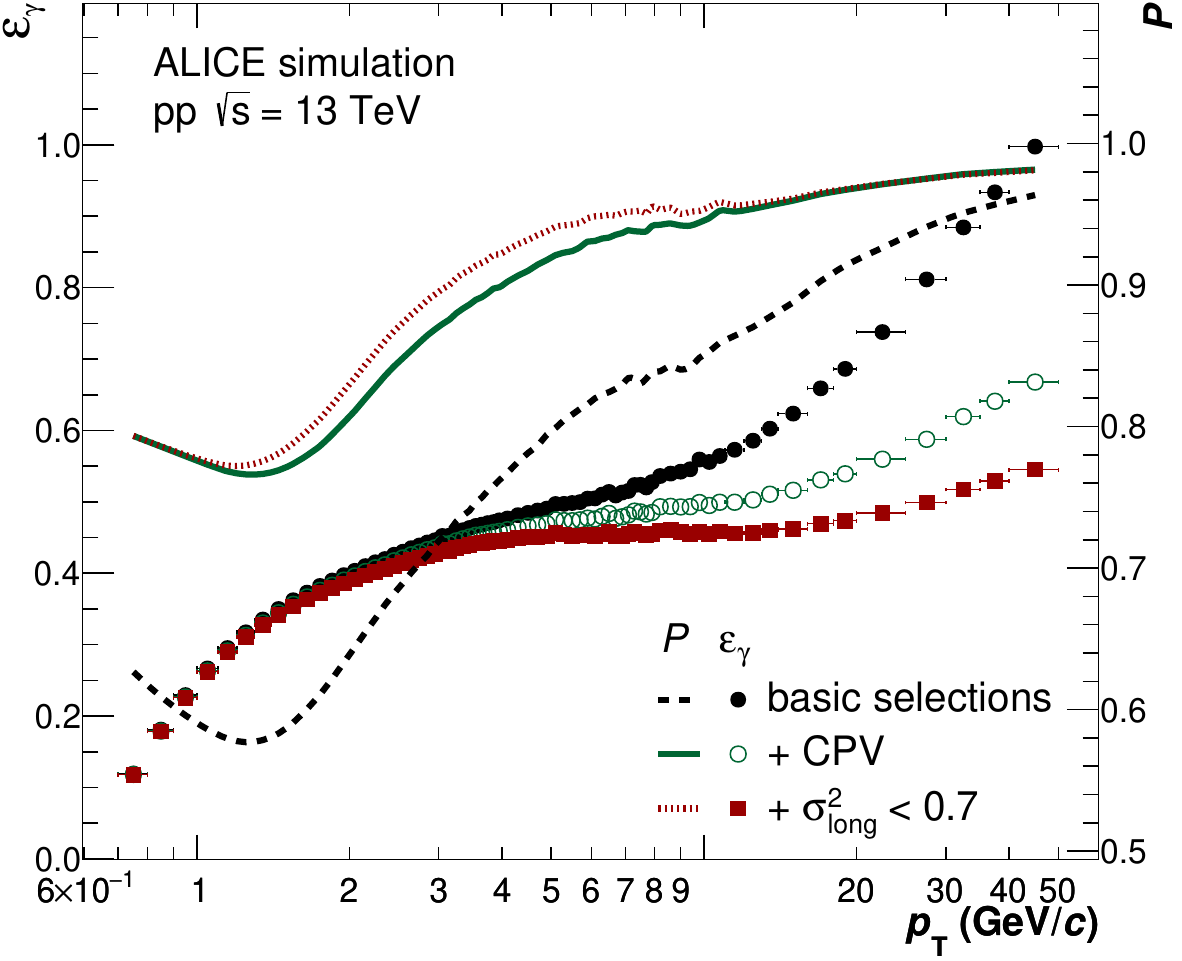}
    \caption{(Color online) Left: Ratio of the number of selected clusters after each applied selection with respect to basic cluster selections (see \Tab{tab:clusterbasiccuts}) as a function of the cluster momentum. 
                            The green markers indicate the use of the CPV requirement, while the red markers include the CPV and $\shshlo$ selection criteria. 
                            For each step, the data are represented by the points, while the lines indicate the same selection step in the simulations. 
                            Right: Inclusive photon purity $P$ and reconstruction efficiency $\epsilon_{\gamma}$ as defined in \Eqs{eq:photonPurity}{eq:photonEffi} after applying each cluster selection criterion. 
                            Both figures are done using V2 clusters as inputs.}
    \label{fig:photonRecCutFlow}
\end{figure}
\begin{figure}[t]
    \includegraphics[width=0.49\textwidth]{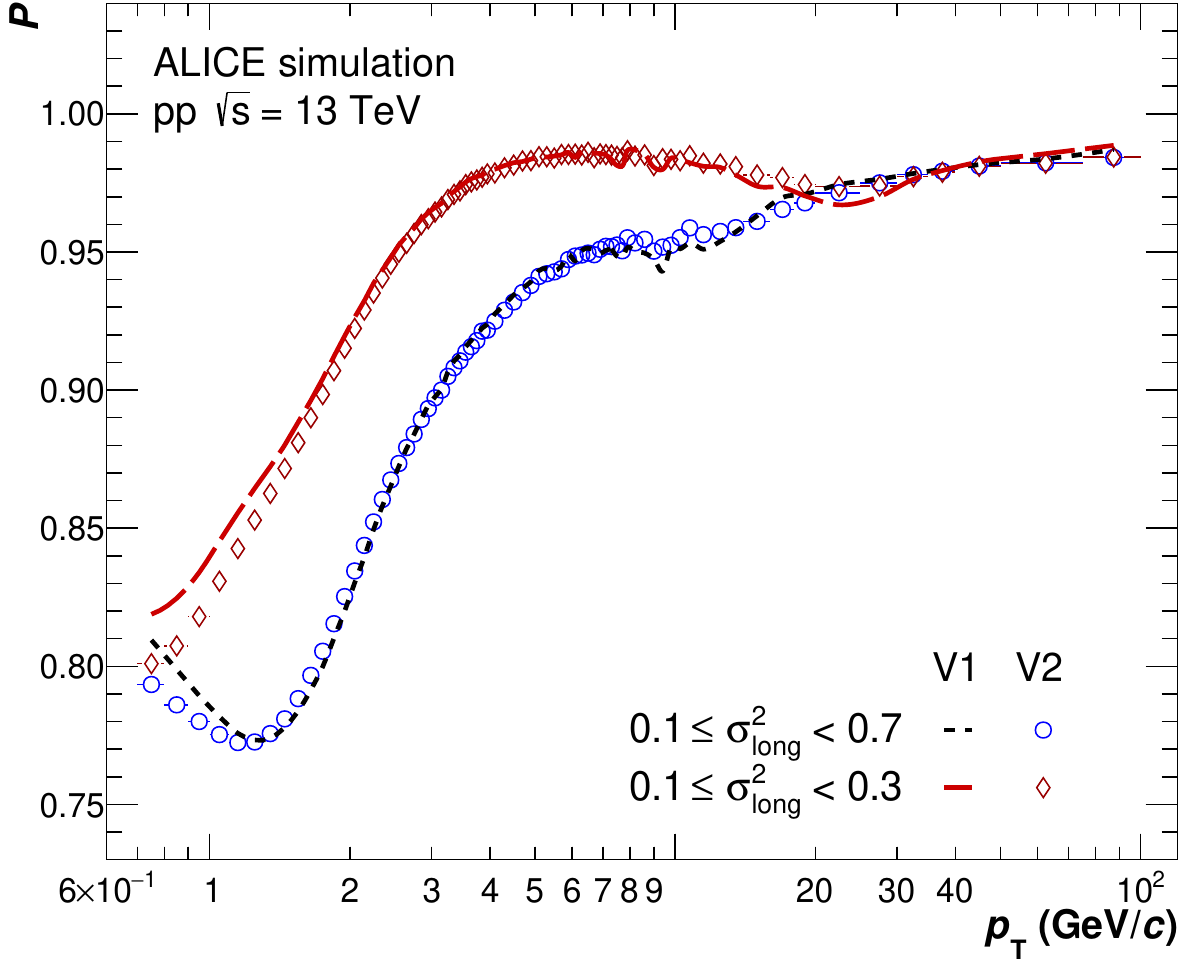}
    \includegraphics[width=0.49\textwidth]{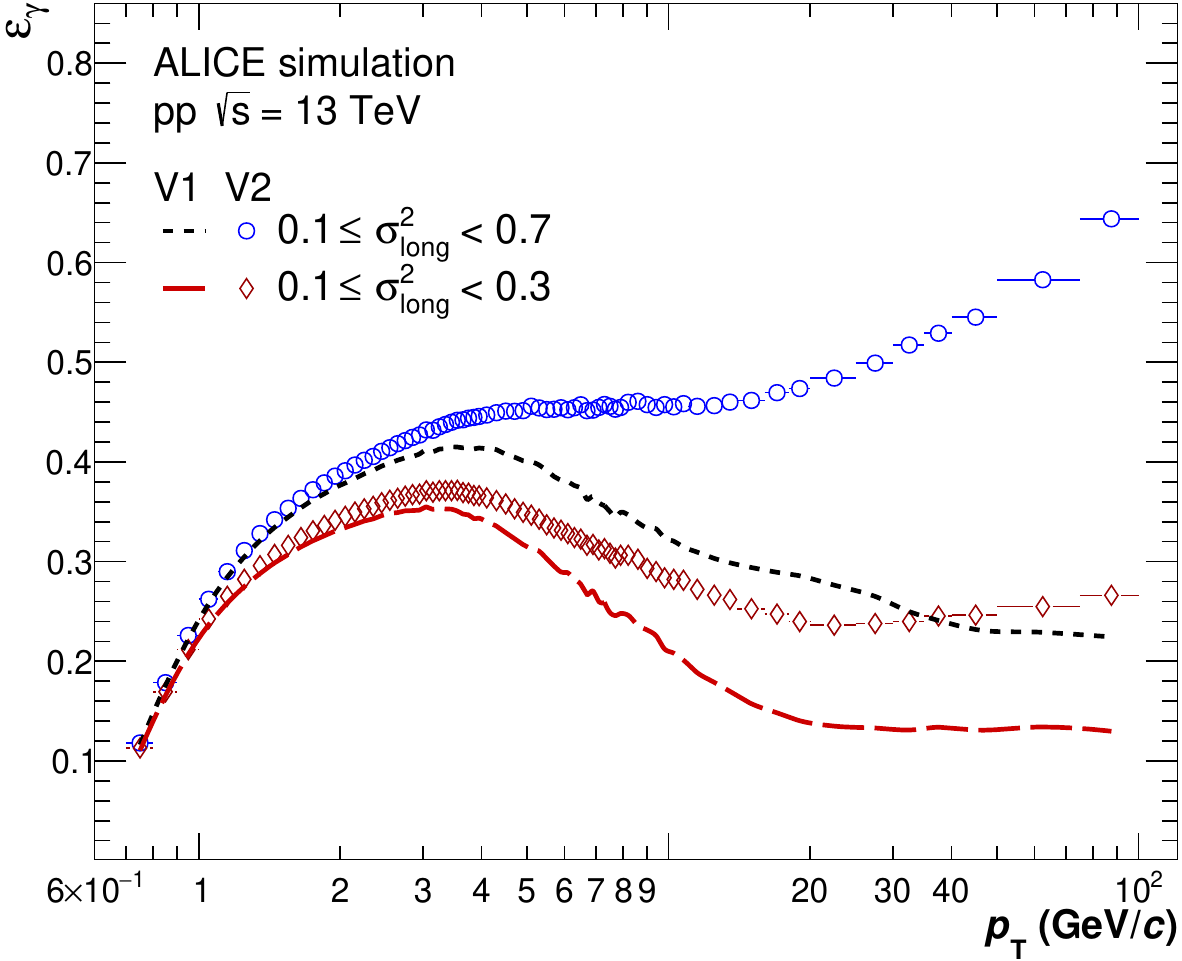}
    \caption{(Color online) Comparison of the purity~(left) and reconstruction efficiency~(right) for inclusive photons for the V1~(lines) and V2~(points) clusterizers. 
              The two quantities are shown for a tight (0.1 < \shshlo < 0.3, red) and loose (0.1 < \shshlo < 0.7, blue) shower shape selection.
    }
    \label{fig:photonRecProp}
\end{figure}

For the photon analysis, only fully calibrated and corrected clusters with a seed energy of $E_{\text{seed}} \geq 500$~MeV and an aggregation threshold of $E_{\text{agg}} = 100$~MeV are considered (see definition in~\Sec{sec:clusterization}). 
Ideally, only clusters from the main bunch crossing are chosen, thus the cluster-timing selection window ranges from $\pm25$~ns to $\pm250$~ns, depending on the data set, as discussed in \Sec{sec:anaparam}. 
Additionally, a minimum cluster energy threshold of $0.7$~GeV is imposed to reduce any effect of the nonlinearity at low cluster energies. 
In order to reduce the contamination from exotic clusters (as described in \Sec{sec:exotics}), a minimum shower shape value ($\shshlo>0.1$) and a minimum number of cells per cluster ($n_{\text{cell}}\geq 2$) are requested. 
The latter reduces not only the contribution from high-energy exotic clusters, but also removes a large fraction of low-momentum clusters which are due to noisy channels in the data. 
To further suppress these exotic clusters, we reject clusters in which one cell carries most of the cluster energy ($F_+ < 0.95-0.97$, see Eq.~\ref{eq:exoticity}).  
At higher cluster energies ($E > 50$~GeV), clusters formed with cells only in one T-Card are rejected since exotic cells induce energy only on surrounding cells in the same T-Card (see \Sec{sec:crosstalk}). 
To increase even further the inclusive photon purity of the cluster sample, clusters originating from charged hadrons or electrons can be rejected using a \gls{CPV} dependent on the track \pT\, as described in \Sec{sec:trackmatch}.
In order not to reject clusters where the hadron contributes only a small fraction of the energy, we apply the \gls{CPV} only for charged-particle track-cluster pairs for which $E/p_{\rm track} < 1.7$. 
Moreover, a mild maximum $\shshlo$ selection criterion is employed to suppress the contribution from converted electrons and hadrons at low transverse momenta. 
Most of the very  elongated clusters with $\shshlo > 0.7 (0.5)$ and $\pT<2$~\GeVc\ in \pPb\ or \pp\ (\PbPb) collisions originate either directly from electrons and hadrons or have a contribution from a second particle hitting the same cell.  
\Figure{fig:photonRecCutFlow}~(left) shows the effect of each photon selection criterion on the number of reconstructed clusters as a function of \pt\ with respect to the basic selections (see \Tab{tab:clusterbasiccuts}) for data and simulation. 
To judge the performance of the different cluster selections for the photon identification, the photon reconstruction efficiency ($\varepsilon_{\gamma}$) and purity ($P$) need to be evaluated simultaneously. They are defined as:
\begin{eqnarray}
  P (p_{\text{\tiny T, rec}}) &=& \frac{N_{\gamma, \rm  rec}(p_{\text{\tiny T, rec}})}{N_{\rm cl, rec} (p_{\text{\tiny T, rec}})} \label{eq:photonPurity}\\
  \varepsilon_{\gamma} (p_{\text{\tiny T, MC}}) &=& \frac{N_{\gamma, \rm rec} (p_{\text{\tiny T, MC}})}{N_{\gamma} (p_{\text{\tiny T, MC}})} \, \label{eq:photonEffi}
\end{eqnarray}
where $N_{\rm cl,rec}$ are the number of reconstructed clusters, $N_{\gamma, \rm rec}$ are all reconstructed clusters with a leading contribution from a photon and $N_{\gamma}$ are the number of photons within the \gls{EMCal} acceptance. Furthermore $p_{\text{\tiny T, MC}}$ is the \pT\ of the simulated photon and $p_{\text{\tiny T, rec}}$ is the \pT\ of the reconstructed cluster.
\Figure{fig:photonRecCutFlow}~(right) shows the effect of the dedicated photon selection criteria on the photon reconstruction efficiency and purity obtained in simulations.
Introducing the \gls{CPV} requirement increases the photon purity by more than 20\% for \pT\ below 10 \GeVc\ while reducing the photon reconstruction efficiency by about 2\% at 6 \GeVc\ and up to 30\% at 40 \GeVc. 
At the same point, since the contributions from cluster overlaps are significantly reduced, the necessary corrections to the cluster energy are smaller.
The contribution from hadrons and electrons are suppressed through the mild selection on the shower shape increasing the purity by 2-10\% at low momenta. 
Similarly, the contribution from merged pions is reduced in the \pt\ range 6 to 20~\GeVc\ by using such a mild shower-shape selection criterion without reducing the efficiency significantly.\\
\Figure{fig:photonRecProp} shows the purity~(left) and efficiency~(right) of the reconstructed inclusive photon sample in pp collisions at \sthirteen\ for the V1 (lines) and V2 (points) clusterizers for a strict (red) and a loose (blue) photon selection criterion regarding the shower shape. 
For these comparisons, all clusters with 1 or 2 local maxima ($n_{\text{LM}}$) are accepted for the V1 clusterizer to reduce the contribution from cluster overlaps.
It can be clearly seen that below 4~\GeVc\ the performance of the two clusterizers differs only by a few percent and that the purity steadily rises with increasing \pT\ from about 80\% to 97\%.
For estimating both $\varepsilon_{\gamma}$ and $P$, each cluster is counted once, and only the higher momentum photon is counted as reconstructed in the efficiency definition for clusters where the neutral pion merges into one cluster.
Additionally, these clusters would be treated as photons in the purity. 
The excess energy from other particles in the cluster is corrected later through unfolding or the effective resolution correction.
Thus, the V2 clusterizer outperforms the V1 clusterizer when using the same selection criteria, allowing for a more efficient reconstruction of individual decay photons up to 20~\GeVc.
For higher momenta, the fraction of merged pions in the cluster sample with respect to photons is too large and stricter selection criteria need to be applied as discussed in \Sec{sec:photoniso}.

\begin{figure}[t]
\centering
\includegraphics[width=0.5\textwidth]{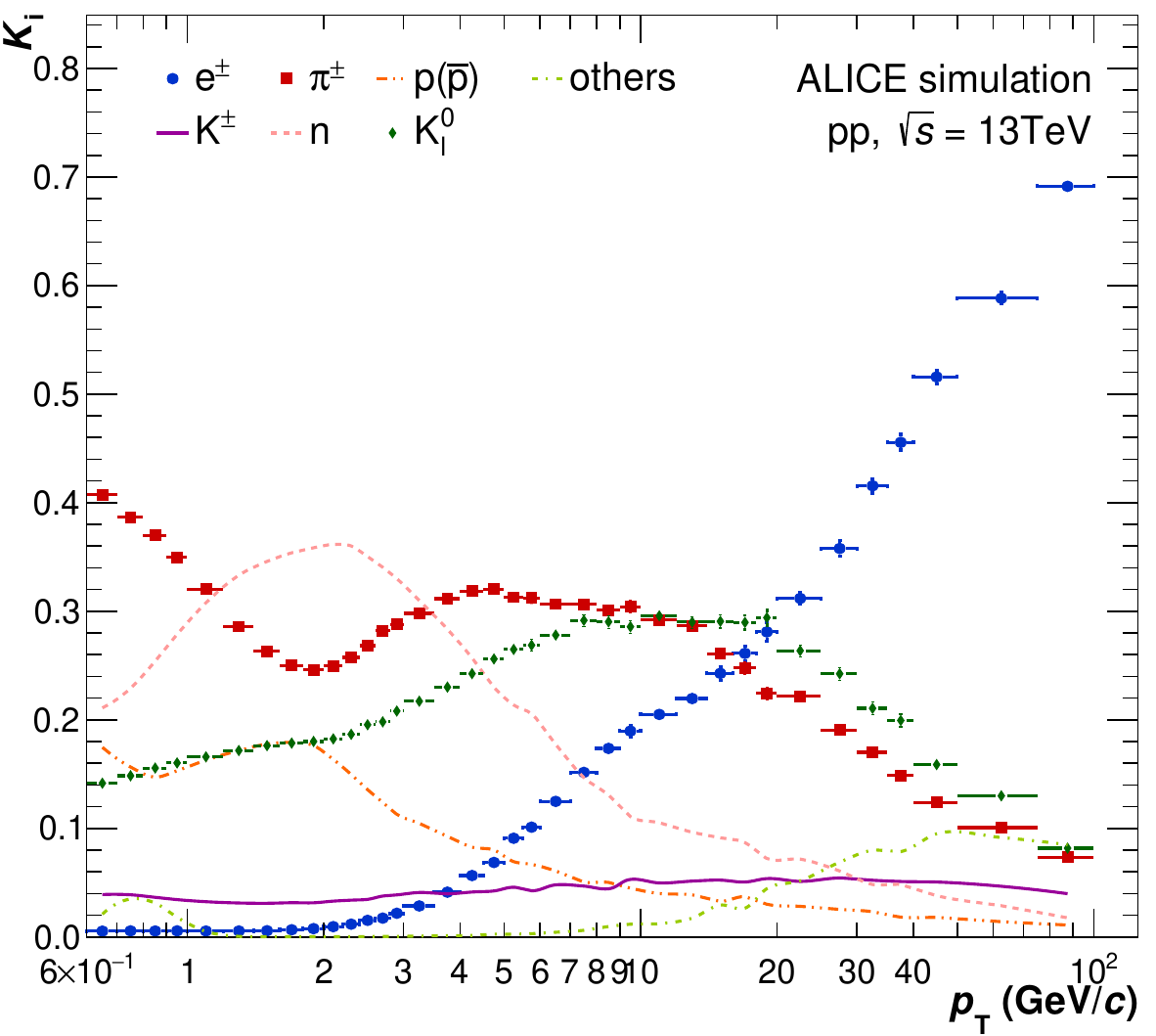}
\hspace{0.2cm}
\raisebox{+20mm}{%
\begin{minipage}[b]{7.5cm}
\caption{(Color online) Individual contributions ($\kappa_i$) to the remaining contamination of the photon sample when using the V2 clusterizer, as a function of the reconstructed \pT. }
\label{fig:BGcomp}
\end{minipage}}
\end{figure}

\Figure{fig:BGcomp} shows the decomposition of the remaining contamination of the selected cluster sample for the V2 clusterizer shown in  \Fig{fig:photonRecProp}. 
In the range  $4 < \pT < 20$ \GeVc, there is a 5\% overall contamination, which is mostly due to charged pions that constitute about 30\% of all the contributions to the contamination.
These charged pions could not be rejected using the track matching veto, as their track was most likely not reconstructed. 
Of similar magnitude is the contribution of neutrons between 1 and 3~\GeVc. 
Furthermore, about $15$--$30\%$ of the remaining background arises from K$^0_L$ directly hitting the \gls{EMCal} surface. 
At higher momenta, the electron contamination rises to the same level as the charged pion contamination, while still representing only about 2\% of the total photon sample.

\begin{figure}[t]
    \includegraphics[height=6.5cm]{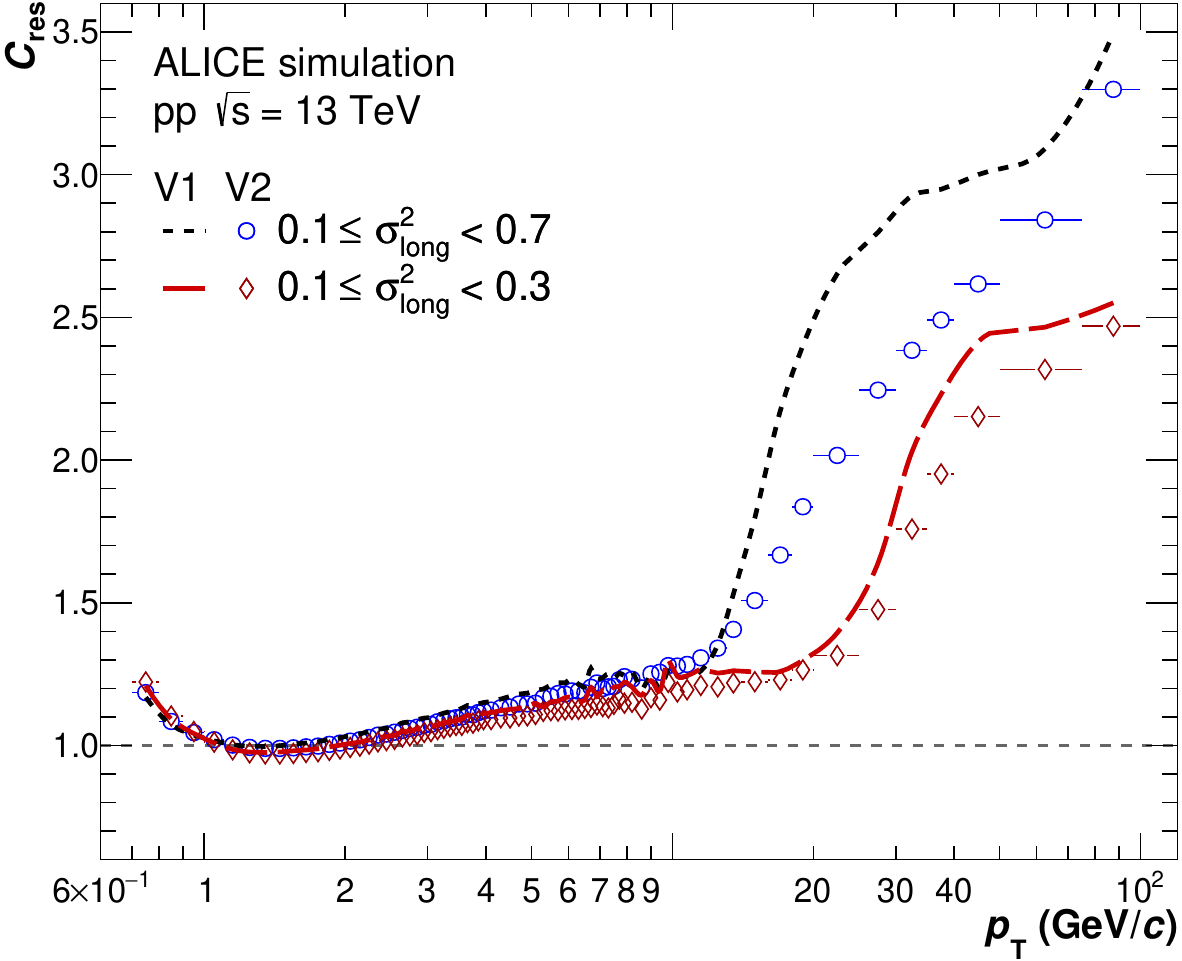}
    \includegraphics[height=6.5cm]{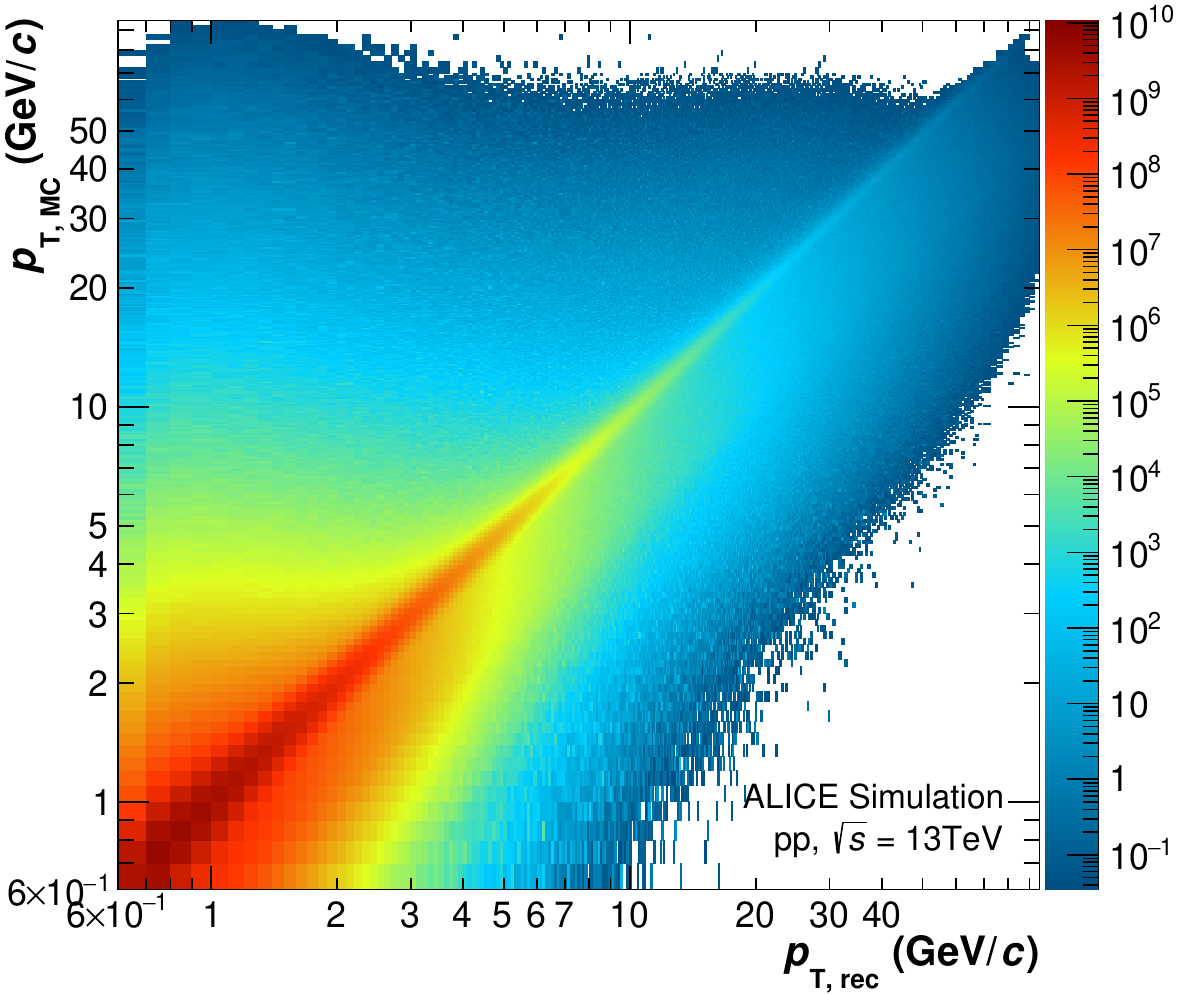}
    \caption{(Color online) Left: Comparison of the transverse momentum smearing correction ($C_{\rm res}$) contained in the photon efficiency for the V1 (lines) and V2 (points) clusterizers in pp collisions at $\sqrt{s} = 13$~TeV. The different colors indicate different shower shape selections.
    Right: Relation between reconstructed cluster and true photon momenta for photon candidates reconstructed with the V2 clusterizer.}
    \label{fig:photonUnfolding}
\end{figure}
Since the transverse momentum of the reconstructed cluster and of the incident photon are not the same,
a momentum smearing correction must be applied to recover the initial photon transverse momentum. 
This can either be done through unfolding of the reconstructed \pT\ distribution using the detector response matrix~(\Fig{fig:photonUnfolding}~right) or by incorporating this correction into the efficiency correction~(\Fig{fig:photonUnfolding} left), denoted as $C_{\text{res}} (p_{\text{\tiny T, rec}}) = \varepsilon_{\gamma}(p_{\text{\tiny T, rec}})/\varepsilon_{\gamma} (p_{\text{\tiny T, MC}})$.
However, the latter can only be applied if the spectral shape of the photons and their contaminations is well described by the simulations, otherwise an iterative reweighting of the initial spectra has to be performed. 
For pp collisions, these corrections are of the order of 5--25\% below 20 \GeVc\ and are on average smaller for the V2 clusterizer compared to the V1 clusterizer.
Above 20 \GeVc, this correction increases significantly due to the increasing overlap with other particles which makes the unfolding procedure unstable.
In this region, stronger $\shshlo$ selections are required to improve the energy resolution for the single photons as described in the next section.
In larger collision systems, the correction can reach up to 20--35\% below 20 \GeVc\ and the V1 clusterizer requires significantly larger corrections than the V2 clusterizer, in particular when considering clusters with more than one maximum.

\Table{tab:photonbasiccuts} summarizes the dedicated photon identification criteria that are applied in conjunction with the basic cluster cuts of \Tab{tab:clusterbasiccuts}.

\begin{table}[t!]
\centering
\caption{Dedicated cluster selection criteria for photon and electron analyses, depending on the multiplicity of the particles produced in the collision. 
        To be used with \Tab{tab:clusterbasiccuts}.}
\begin{tabular}{llcc}
& & \multicolumn{2}{c}{Multiplicity }  \\
Analysis type & Parameter &  low & \hspace{0.2cm}high  \\
\toprule
\multirow{7}{*}{statistical photon analysis}& clusterizer & \multicolumn{2}{c}{ V1/ V2}\\
& $n_{\rm LM}$, V1 clust.      & \multicolumn{2}{c}{ $\leq 2$}    \\ 
& \shshlo          & $<0.7$  & \hspace{0.2cm}$<0.5$  \\ 
& \gls{CPV}\\
& \hspace{0.2cm} - $\Delta\varphi^{\rm residual}$ (rad), $\Delta\eta^{\rm residual}$ &  \multicolumn{2}{c}{\Eq{Eq:TMVetoCriteria}} \\ 
& \hspace{0.2cm} - $E/p$ \gls{CPV} veto & $>1.7$ & -  \\  
& \hspace{0.2cm} - \gls{MIP} subtraction & no & yes \\  \midrule  
\multirow{8}{*}{isolated photon analysis} & clusterizer & \multicolumn{2}{c}{ V1}  \\
& $n_{\rm LM}$, V1 clust.       &  \multicolumn{2}{c}{ $\leq 2$}     \\ 
&\shshlo          &  \multicolumn{2}{c}{$<0.3-0.4$ (\pT  ~dep.)}    \\ 
& $d_{\rm mask} $ (cells)       & \multicolumn{2}{c}{$>1$}    \\  
&$d_{\rm edge} $ (cells)       & \multicolumn{2}{c}{$>1$}    \\  
& \gls{CPV}\\
& \hspace{0.2cm} - $\Delta\varphi^{\rm residual}$ (rad), $\Delta\eta^{\rm residual}$ & \multicolumn{2}{c}{\Eq{Eq:TMVetoCriteria}} \\ 
& \hspace{0.2cm} - $E/p$ \gls{CPV} veto   & \multicolumn{2}{c}{$>1.7$}  \\  \midrule 
\multirow{5}{*}{electron analysis} & \gls{TPC} d$E$/d$x$ & \multicolumn{2}{c}{$-1 \leq n\sigma^{\textrm{TPC}}_{e^\pm} \leq 3$} \\
& cluster track assoc. & \multicolumn{2}{c}{$\Delta\eta < 0.01, \Delta\varphi < 0.01$} \\
&\shshlo  \pT $< 15$ \GeVc   &  \multicolumn{2}{c}{ $0.05< \shshlo < 0.9$}    \\ 
&\hspace{0.2cm} \pT $\leq 15$ \GeVc                &  \multicolumn{2}{c}{ $0.05< \shshlo < 0.6$} \\ 
& $E/p$           &  \multicolumn{2}{c}{$0.8 < E/p < 1.2$} \\ 
\bottomrule 
\end{tabular}
\label{tab:photonbasiccuts}
\end{table}

\subsubsection{Isolated photon performance}
\label{sec:photoniso}
Direct photons from  $2 \rightarrow 2$ scattering processes are expected to appear isolated, as they are produced with no hadronic activity in their vicinity except for the underlying event of the collision, in contrast to other photon sources like decays of mesons that are likely generated by parton fragmentation~\cite{Ichou:2010wc} and have a high probability to be accompanied by other fragments. 
To increase the purity of the direct photons from  $2 \rightarrow 2$ processes in the sample, the direct photon candidates measured in the \gls{EMCal} are required to be isolated. 
This isolation technique will also reject the largest part of photons from fragmentation and is explained in more detail in Ref.~\cite{Acharya:2019jkx}.\\
The isolation criterion is based on the so-called \textit{isolation momentum} $p_{\rm T}^{\rm iso}$.
It is defined by considering the transverse momentum of all particles with an angular position ($\eta^{i}$,$\varphi^{i}$) measured inside a cone of radius 
\begin{equation}
\label{eq:rsize}
R = \sqrt{ (\eta^{i} - \eta^{\gamma})^2 + (\varphi^{i}-\varphi^{\gamma})^2}=0.4,
\end{equation}
around the candidate's angular position ($\eta^{\gamma}$,$\varphi^{\gamma}$).
The isolation momentum is calculated by adding the transverse momenta of all neutral clusters (clusters not matched to charged particles) 
in the calorimeter, excluding the \pt\ of the candidate photon cluster, and the transverse momenta of all charged-particle tracks that fall into the cone. The momentum sum is corrected by the collision \gls{UE} contribution, which is not correlated to the hard process at the origin of the direct photon:
\begin{equation}
\label{eq:etiso}
p_{\rm T}^{\rm iso}=\sum_{\rm in~cone -candidate} p_{\rm T}^{\rm neutral~cluster}+\sum_{\rm in~cone} p_{\rm T}^{\rm track} - \rho_{\rm UE}\pi R^{2},
\end{equation}
where $\rho_{\rm UE}$ is the \gls{UE} density in the cone.\\
For the analyses of pp collisions, the \gls{UE} contribution for charged particles in the cone is $\rho_{\rm UE} = 1.6$~\mom\ as determined in collisions at \sfive\ ~\cite{Acharya:2020sxs}.
This \gls{UE} contribution results in a small third term in \Eq{eq:etiso} and is therefore ignored in the $p_{\rm T}^{\rm iso}$ calculation in \pp\ analyses. 
The candidate photon is declared isolated if $p_{\rm T}^{\rm iso}<2$~\GeVc. 
This value was chosen after studying the efficiency and purity in \pp\ collisions and strongly depends on the analysis strategy. 
Alternatively, the isolation criteria can be based solely on the charged tracks within the cone.
In order to reach a similar isolation efficiency in this case, the isolation requirement must be reduced to $p_{\rm T}^{\rm iso} < 1.5$~\GeVc.
The latter allows to use the full acceptance of the \gls{EMCal} and \gls{DCal} in the analysis without needing corrections to the isolation energy when the cone is partially out of the calorimeter acceptance.
This technique was used for the measurement of isolated photon-hadron correlations in \pp\ and \pPb\ collisions at \sfivelead~\cite{Acharya:2020sxs}.

\begin{figure}[t]
\centering
\includegraphics[width=0.55\textwidth]{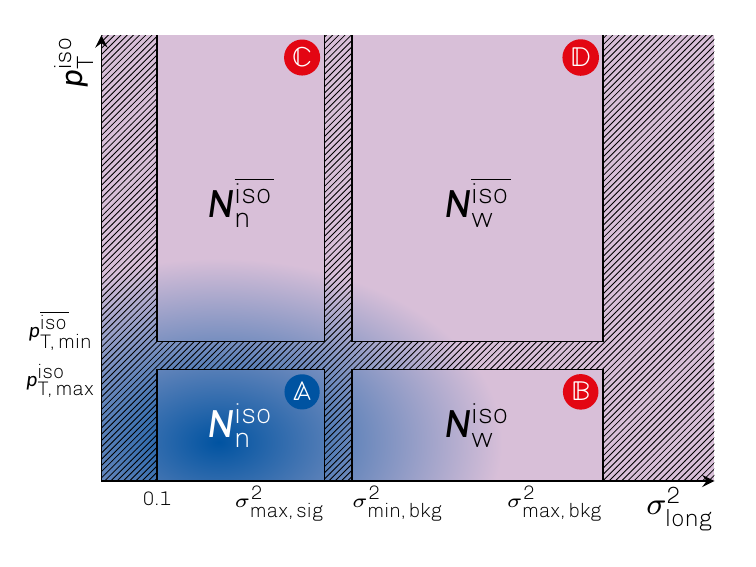}
\raisebox{+10mm}{%
\begin{minipage}[b]{7.cm}
\caption{(Color online) Illustration of the parameter-space of the photon $p_{\rm T}^{\rm iso}$ and \shshlo, 
used to estimate the background yield in the signal region ($\mathbb{A}$) from the observed yields in the three control regions 
($\mathbb{B}$, $\mathbb{C}$, $\mathbb{D}$). The red regions indicate areas dominated by background and the blue regions by the photon signal. 
The color gradient indicates mixture of signal and background. }
\label{fig:abcdCartoon}
\end{minipage}}
\end{figure}

In \PbPb\ and \pPb\ collisions, the \gls{UE} contributes to the $p_{\rm T}^{\rm iso}$, thereby biasing it to higher values.
The \gls{UE} contribution $\rho_{\rm UE}$ must therefore be estimated and subtracted event-by-event. 
One technique consists in using the $\sum p_{\rm T}^{\rm track}$ value measured in cones oriented 90 degrees in azimuth away from the direction of the isolated photon candidate.
Alternatively, the underlying event can be estimated using the FASTJET median area density method~\cite{Cacciari:2009dp,Cacciari:2011ma}. 
In \pPb\ collisions, both techniques yield very similar results.
When using only charged-particle tracks in the calculation of $p_{\rm T}^{\rm iso}$, the \gls{UE} contributes with an energy density of $\rho_{\rm UE} = 3.2$~\mom\ in the cone for \pPb\ collisions~\cite{Acharya:2020sxs}, which is about twice as high as for \pp\ collisions.
The underlying event contribution in \pPb\ collisions is however still  much smaller than its contribution in the most central heavy-ion collisions where it can reach few tens of \mom\ ~\cite{Aad:2015lcb,Chatrchyan:2012vq}.\\
Even after applying the isolation criteria, the photon candidate sample still has a non-negligible contribution from background clusters, mainly from meson decays. 
To estimate the background contamination, different classes of measured clusters can be used~\cite{Aad:2011tw,Acharya:2019jkx}:
(1) classes based on the shower shape \shshlo, i.e.\ \textit{wide} (\shshlo > 0.4, elongated clusters) and \textit{narrow} ($0.1<\shshlo < 0.3$), and 
(2) classes defined by the isolation momentum, i.e.\ \textit{isolated} ($p_{\rm T}^{\rm iso}<2$~\GeVc) and \textit{non-isolated} ($p_{\rm T}^{\rm iso} > 3$~\GeVc). 
The selection criteria reported above are illustrative; the exact values depend on the collision system and may vary with candidate \pt.
Such \pt-dependent values have been used in the analysis of \pp\ collisions at \sseven~\cite{Acharya:2019jkx}.
The different classes are denoted by sub- and superscripts, e.g. quantities related to isolated, narrow clusters are labeled $X_{\rm n}^{\rm iso}$ and non-isolated, wide clusters are $X_{\rm w}^{\overline{\rm iso}}$. 
The yield of isolated photon candidates in this nomenclature is $N_{\rm n}^{\rm iso}$. 
It consists of signal ($S$) and background ($B$) contributions: 
$N_{\rm n}^{\rm iso} = S_{\rm n}^{\rm iso} + B_{\rm n}^{\rm iso}$.
This class is labeled with the letter $\mathbb{A}$ in \Fig{fig:abcdCartoon}, which illustrates the parameter space used in this procedure. 
The three other classes that are defined (labeled as $\mathbb{B}$, $\mathbb{C}$, and $\mathbb{D}$ in the figure) should dominantly contain background clusters. 
The contamination of the candidate sample $\mathbb{A}$ is then $C = B_{\rm n}^{\rm iso}/N_{\rm n}^{\rm iso}$ and the purity is $P \equiv 1 - C$.
Assuming that the proportion of background which is isolated is the same in the wide and narrow cluster areas 
and assuming that the proportion of signal in the control regions ($\mathbb{B}$, $\mathbb{C}$ and $\mathbb{D}$) is negligible compared to the background, the purity can be derived in a data-driven approach as
\begin{equation}
\label{eq:ABCDpurity}
P_{\rm dd}=1-\frac {B_{\rm n}^{\overline{\rm iso}}/N_{\rm n}^{\rm iso}} {B_{\rm w}^{\overline{\rm iso}}/B_{\rm w}^{\rm iso}} = 1-\frac {N_{\rm n}^{\overline{\rm iso}}/N_{\rm n}^{\rm iso}} {N_{\rm w}^{\overline{\rm iso}}/N_{\rm w}^{\rm iso}}.
\end{equation}
Unfortunately, both assumptions do not fully hold.
In part, this is due to the fact that single photons from meson decays can have a higher value of  $p_{\rm T}^{\rm iso}$ than merged decay photons at the same $\pt$, because of the presence of the second photon from the meson decay in the isolation cone.
Also, fluctuations in the cluster distributions, e.g.\ caused by overlapping showers from nearby particles originating from the same hard process, may lead to some energy contribution
either included in the cluster, which increases its width, or not included, which increases
the isolation momentum, causing an anti-correlation of the two parameters.\\
Since those are purely particle kinematics and detector effects, we assume the simulation reproduces them so that one can estimate the bias with the following equation~\cite{Acharya:2019jkx}:
\begin{equation}
\label{eq:ABCDpurityMC}
P = 1-\bigg(\frac {N_{\rm n}^{\overline{\rm iso}}/N_{\rm n}^{\rm iso}} {N_{\rm w}^{\overline{\rm iso}}/N_{\rm w}^{\rm iso}}\bigg)_{\rm data} \times \bigg(\frac {B_{\rm n}^{\rm iso}/N_{\rm n}^{\overline{\rm iso}}} {N_{\rm w}^{\rm iso}/N_{\rm w}^{\overline{\rm iso}}}\bigg)_{\rm MC} .
\end{equation}
The \gls{MC} corrections were determined from \gls{PYTHIA} simulations of pp
collisions. The signal events were generated with a high-energy
direct photon and a back-to-back jet and the background events were
generated with two high-energy back-to-back jets.

\begin{figure}[t]
\includegraphics[height=7.1cm]{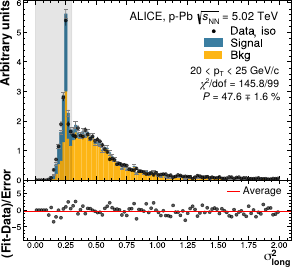} 
\includegraphics[height=7.1cm]{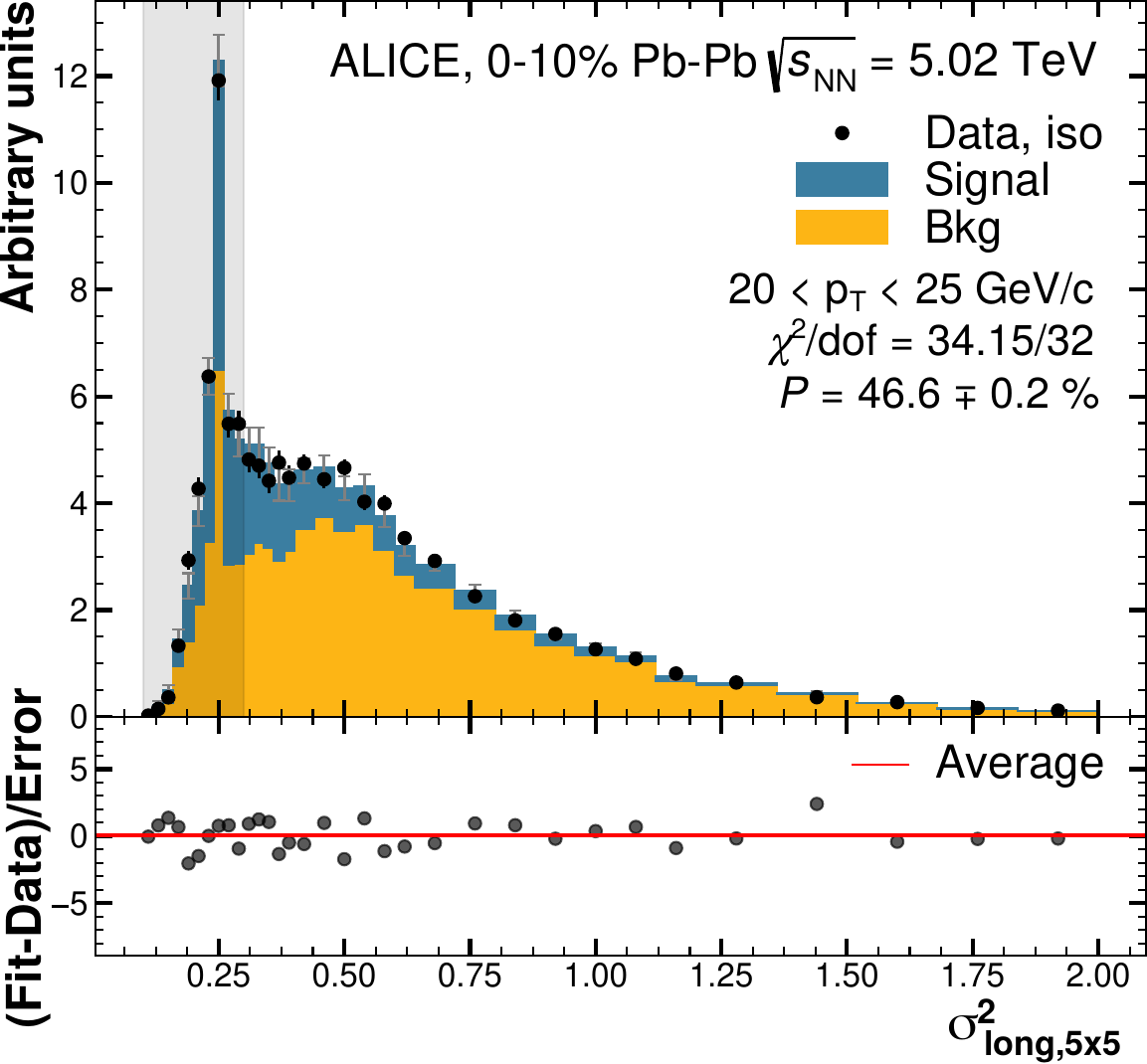}
\caption{(Color online) Example distribution of the template fit to the $\shshlo$ distribution in \pPb~\cite{Acharya:2020sxs} (left) and \PbPb\ (right) collisions at \sfivelead. }
\label{fig:templateFit}
\end{figure}

Alternatively, the purity can be estimated using a template fitting technique as described in Ref.~\cite{Acharya:2020sxs}.
This method is illustrated in \Fig{fig:templateFit}{ (left)} for \pPb\ collisions at \sfivelead. 
In \PbPb\ collisions the usage of the V1-clusterizer for the isolated photon analysis is not possible as too many particles would be overlapping within this cluster. 
Thus V2 clusters are used, but the shower shape is calculated around the leading cell of the V2 cluster in a 5x5 cell array, denoted as \shshloX\ for the isolated photon analysis. 
An example for \PbPb\ collisions at \sfivelead\ is shown in \Fig{fig:templateFit}{ (right)} yielding similar results as for \pPb\ collisions.\\
Using the template fit method, the $\shshlo$ distribution for the isolated cluster sample is fit with a linear combination of a signal \gls{PYTHIA} \gls{MC} distribution, and the background distribution is determined from data using an anti-isolated sideband ($5 < \pT^{\text{iso}}< 10$~\GeVc).
For the same reasons as described above, this method also needs \gls{MC} corrections based on a \gls{PYTHIA} simulation to take into account (anti-)correlations in the background between the regions.
For both methods, the correction based on the simulation results in an absolute correction on the purity ranging from 8\% to 14\% depending on the cluster \pT.

\begin{figure}[t]
\centering
\includegraphics[height=6.7cm]{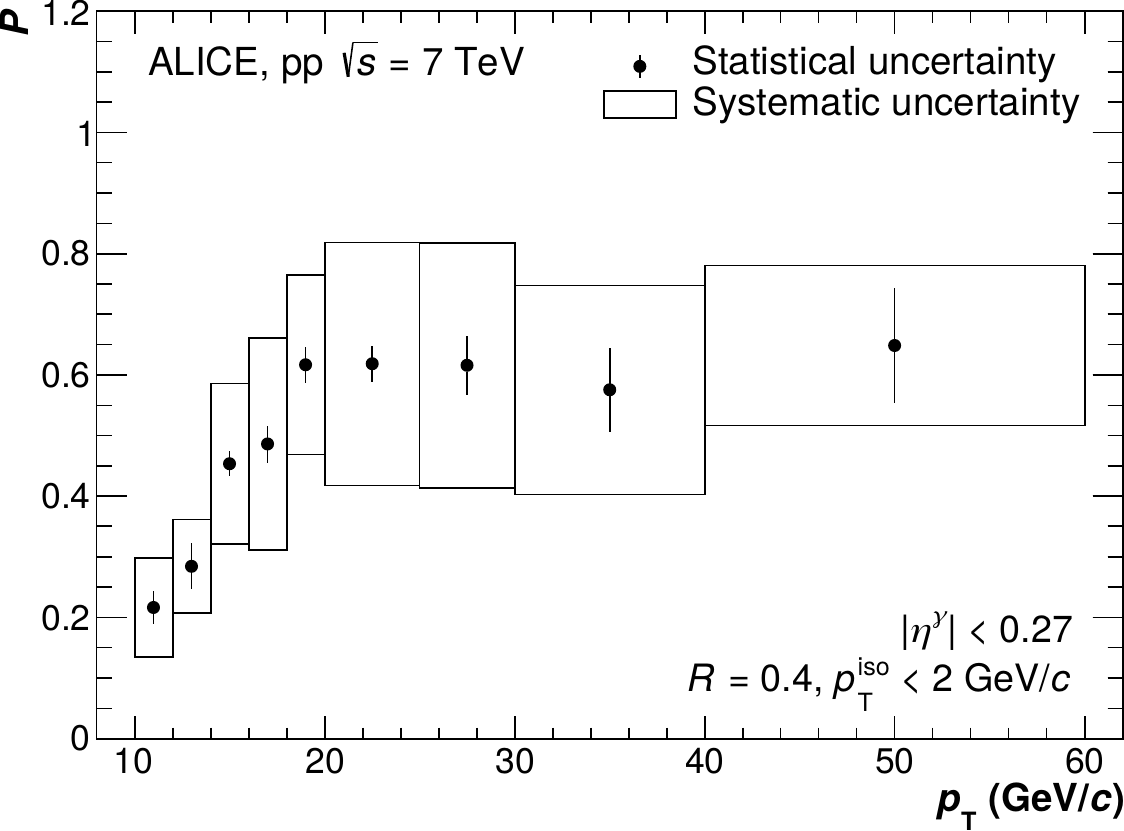}
\includegraphics[height=6.5cm]{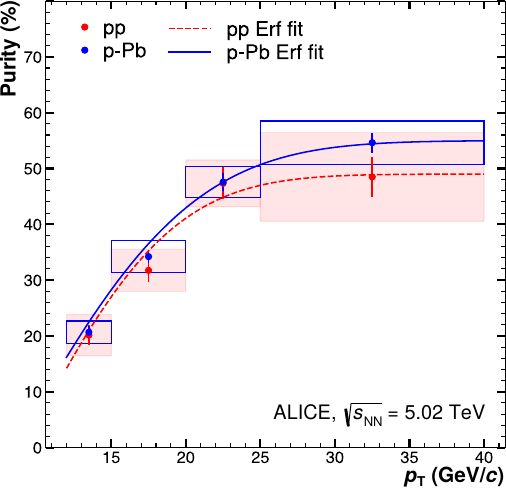}
\caption{(Color online) 
Isolated photon corrected purity in pp collisions at \sseven\ with $\pt^{\rm iso}<2$~\GeVc\ and $|\eta^{\gamma}| < 0.27$ 
calculated using \Eq{eq:ABCDpurityMC}, taken from~\cite{Acharya:2019jkx}~(left) and   
for pp and \pPb\ collisions at \sfivelead\ with $|\eta^{\gamma}| < 0.67$ using the template fit technique taken from~\cite{Acharya:2020sxs}~(right). 
The boxes indicate the systematic uncertainty, while the error bars reflect the statistical uncertainty. Figures are taken from the mentioned references. 
}
\label{fig:purityisolation}
\end{figure}

The left panel of \Fig{fig:purityisolation} shows the isolated photon purity calculated using \Eq{eq:ABCDpurityMC} in pp collisions at \sseven\ and the right panel shows the purity in pp and p--Pb collisions at \sfivelead\ determined using the template method.
The large 80\% contamination at $\pt^{\gamma} = 10$~\GeVc\ comes mainly from single decay photon clusters. 
The contamination decreases and saturates at $30-50\%$ for $\pt^{\gamma} > 18$~\GeVc, mainly from merged \piz\ mesons decay clusters. 

An interplay of physics and detector effects causes a $\pt^{\gamma}$ dependence of the purity.
On the one hand, the {\pt} spectra of prompt photons are harder than those of neutral pions, mainly because the latter are produced by fragmentation of a quark or gluon. 
Because of this, the purity, i.e. the ratio of direct photon and neutral pion \pt\ yields, increases with increasing \pt. 
In addition, the probability to find a photon as isolated varies with \pt; 
at higher $p_{\rm T}$, isolation of jet fragments is less probable for a fixed isolation momentum. 
On the other hand, due to the decreasing opening angle of meson decays at high \pt, the probability to obtain a narrow shower from the merged photons increases leading to a larger contamination from \piz\ mesons.
At $\pt = 20$~\GeVc, 5\% of the decay photons of the \piz\ mesons are found in the narrow shower shape region, and beyond {40~\GeVc} this contribution increases to more than 25\%. 
The combined effect of these mechanisms leads to the rise of the purity at low \pt\ and a saturation for $\pt > 18$~\GeVc. 

The purity obtained by both methods is similar, although in different colliding systems and energy, as shown in  \Fig{fig:purityisolation}, thus both techniques yield equivalent results for the isolated photon measurements. 
The \gls{EMCal} can be used in combination with the tracking detectors to select isolated photons 
with a reasonable purity in the \pt\ range from 10~\mom\ to at least 60~\mom\ in Run~1 for \pp\ and \pPb\ collision data.
The larger triggered data sets in \pp\ collisions at \sthirteen\ recorded in Run~2 will allow for measurements at even higher photon momenta.
 
\subsection{Neutral mesons}
\label{sec:mesons}
The $\piz$, $\eta$ and $\eta'$ mesons can be reconstructed with the \gls{EMCal} in their two-photon decay channel using their excess in the reconstructed invariant mass distribution (see \Eq{eq:KineIMgg}). 
The range in \pt\ for which this method can be used depends on the clusterizer and clusterization parameters used (see \Sec{sec:clusterization}): with the V1 clusterizer \piz\ mesons can be measured up to $E=15$~GeV, while clusterizers that are able to split the clusters (V2, $3\times3$ and V1+unfolding)  reach $E=22$~GeV, as shown in \Fig{fig:InvMassEBins}.\\
Considering this, neutral mesons can be identified in two ways depending on the energy range and clusterization: 
either from pairs of clusters using the invariant mass, called \gls{EMC} reconstruction (discussed in next section), or via a single merged cluster, called \gls{mEMC} reconstruction that relies on the cluster shower shape (discussed in \Sec{sec:MesonShowerShape}).  
Finally, in order to circumvent the cluster merging to a large extent, the neutral mesons can also be reconstructed via the invariant mass analysis using one photon reconstructed with the \gls{EMCal} and the other from a converted photon reconstructed with the \gls{TPC} and \gls{ITS} (\gls{PCM}, photon conversion method). 
This technique is called \gls{PCM-EMC} reconstruction and it is discussed in the next section.\\
In addition, heavier hadrons like $\omega$ and $\eta'$ mesons decay into $\piz$ or $\eta$ mesons. 
The decay channels $\omega \rightarrow \pi^0\pi^+\pi^-$ and $\eta' \rightarrow \eta\pi^+\pi^-$ can be reconstructed by choosing photon pairs in the corresponding invariant mass regions and pairing them with charged particles detected in the tracking detectors. 
The performance of these reconstruction channels will be discussed in \Sec{sec:heavyMeson}.

\subsubsection{Neutral meson reconstruction via two-photon invariant mass}
\label{sec:MesonInvMass}
The \piz, $\eta$ and $\eta$' mesons are reconstructed as excess yield around their nominal particle masses in the two-photon invariant mass spectrum on top of a combinatorial background. 
In this section, only the performance of the V2 clusterizer will be discussed as it proved to be the most suitable for a di-photon invariant mass analysis as discussed in \Sec{sec:clusterization}.
\begin{figure}
    \includegraphics[width=0.49\textwidth]{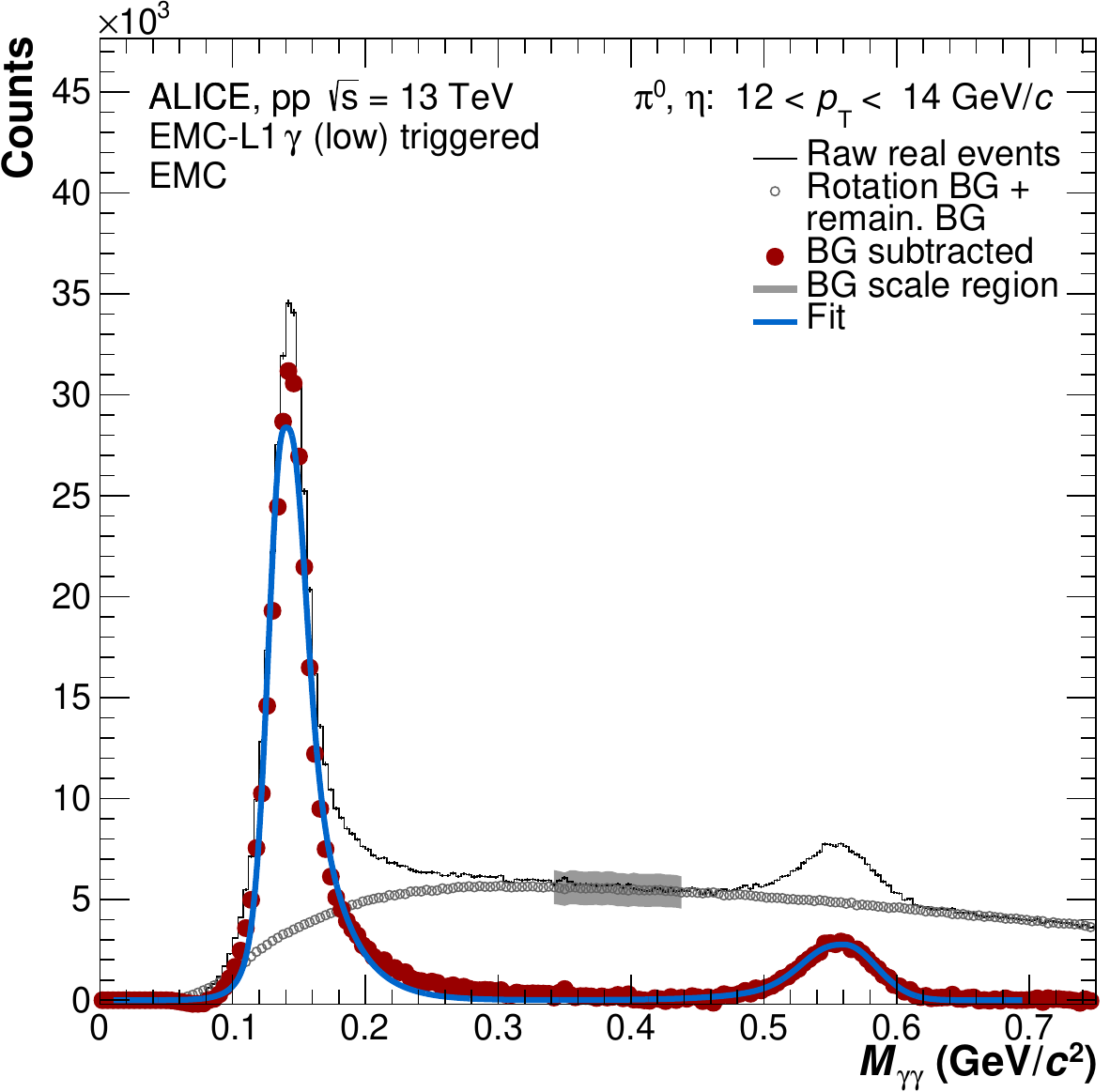}
    \includegraphics[width=0.49\textwidth]{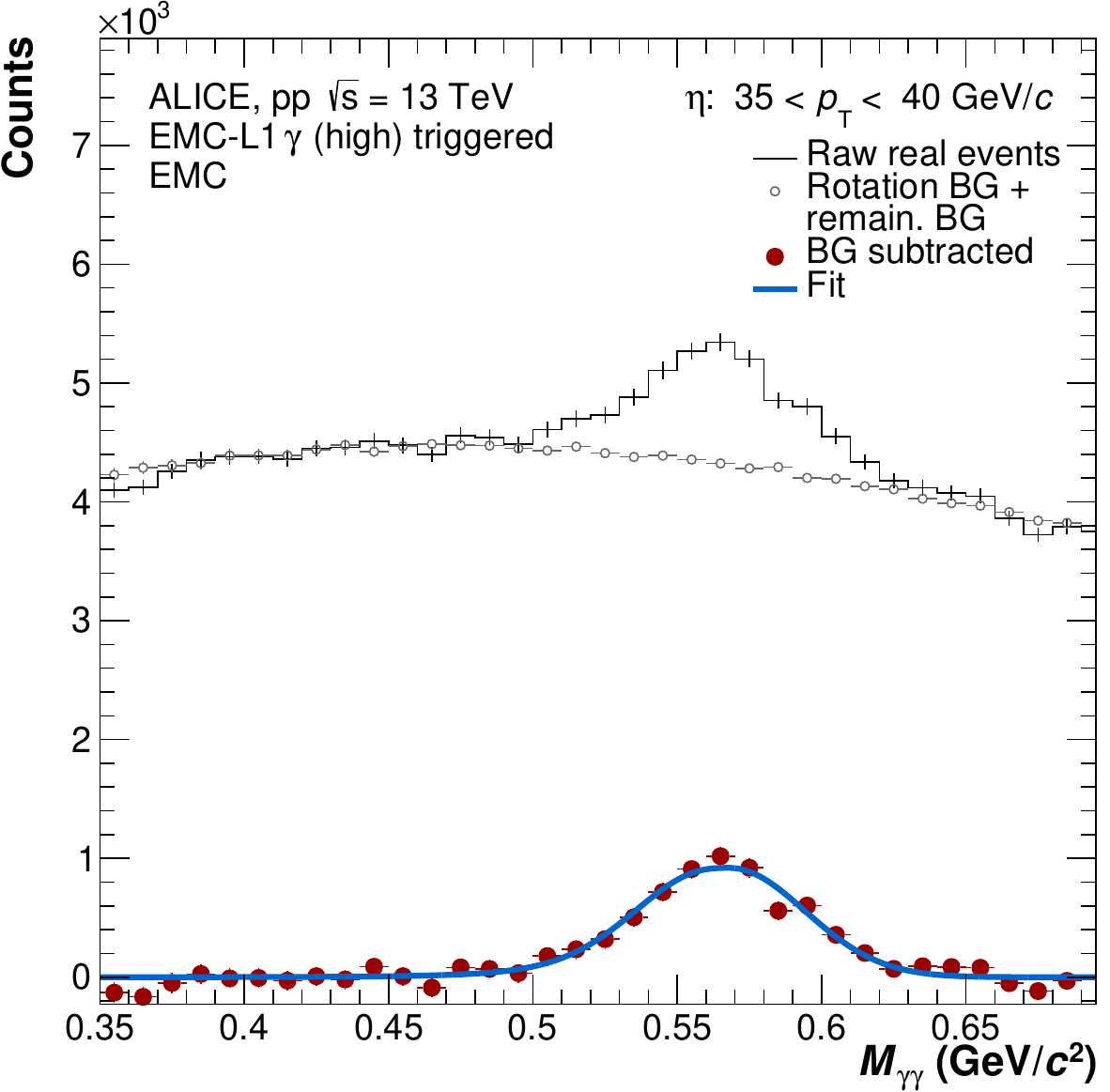}
    \caption{(Color online) Invariant mass distribution for neutral pion and $\eta$ meson candidates at intermediate (left) and high (right) transverse momenta reconstructed with both photons in the \gls{EMCal} in pp collisions at \sthirteen\ using the \gls{EMCal} \gls{L1} trigger sample. The combinatorial background is described using the rotation background technique. For the higher \pT\ slice only the $\eta$ meson invariant mass window is shown as the \piz\ meson cannot be reconstructed using the \gls{EMC} invariant mass technique at these momenta.}
    \label{fig:PeaksEMC}
    \includegraphics[width=0.49\textwidth]{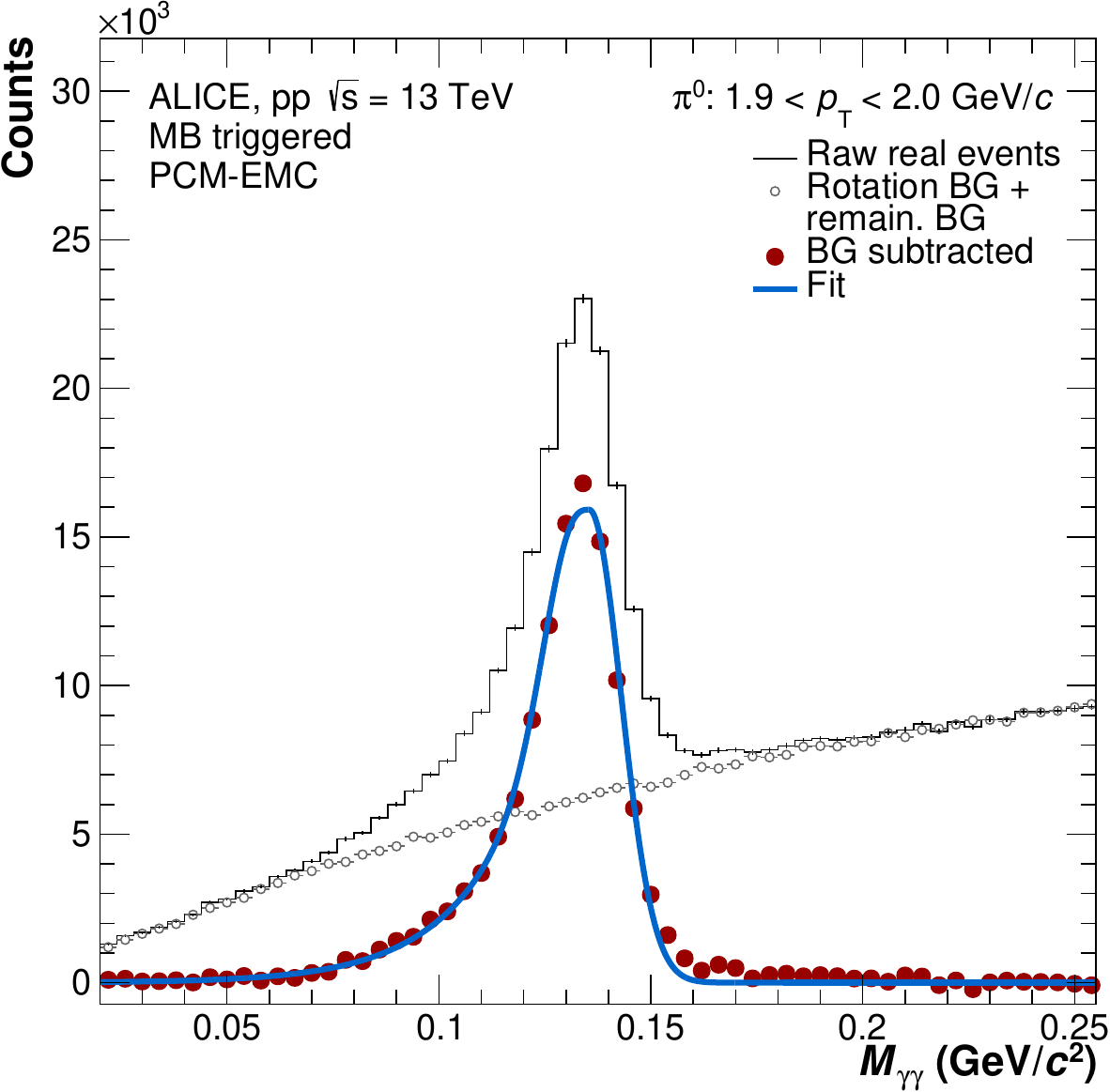}
    \includegraphics[width=0.49\textwidth]{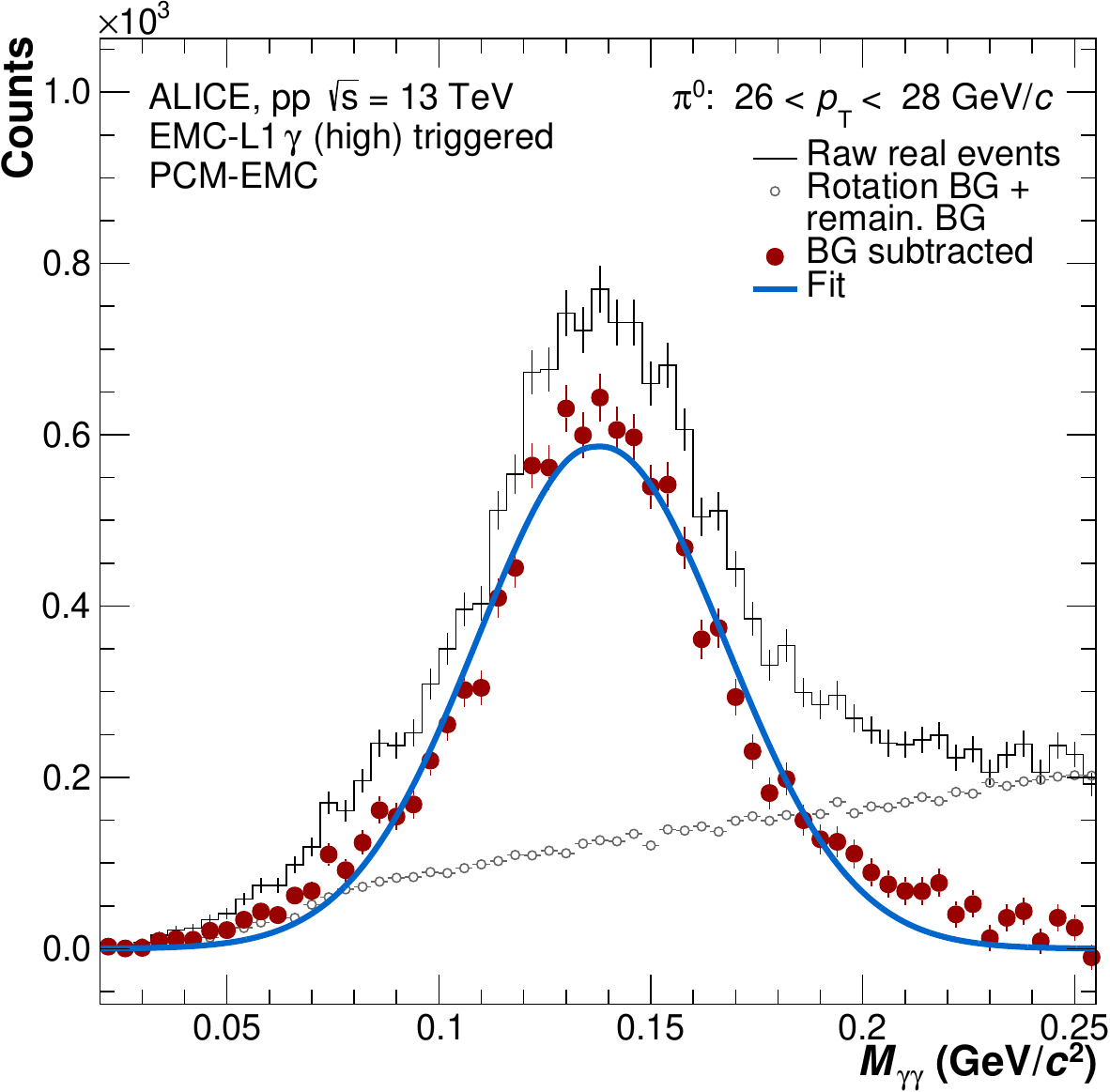}
    \caption{(Color online) Invariant mass distribution for neutral pion candidates at low (left) and high (right) transverse momenta reconstructed using the \gls{PCM-EMC} technique in pp collisions at \sthirteen\ using the minimum bias and \gls{EMCal} \gls{L1} trigger sample. The combinatorial background is described using the mixed-event background technique.}
    \label{fig:PeaksPCMEMC}
\end{figure}
\noindent \Figure{fig:PeaksEMC} shows the invariant mass distribution in two different transverse momentum intervals for neutral pion and $\eta$ meson candidates using the two-photon reconstruction based solely on the \gls{EMCal}.
\Figure{fig:PeaksPCMEMC} depicts the corresponding invariant mass distributions if one of the two photons was reconstructed with the \gls{PCM} instead.
The shape of the signal peaks is described using a Gaussian function with additional exponential tails on both sides of the peak.
The tail towards lower invariant masses mainly originates from Bremsstrahlung energy loss of the electron for the \gls{PCM} photon or missing energy for converted photons in front of the \gls{EMCal}.
On the higher invariant mass side and at higher cluster energies, cluster merging in the \gls{EMCal} contributes to the non-Gaussian tail at large invariant masses.

The combinatorial background below the peaks consists of two components: random combinations of photons, and a partially correlated background component arising from jets or particle decays with more than two photons in their final state, for instance, $\eta \rightarrow 3 \pi^0$ or K$^0_{\rm s} \rightarrow 2 \pi^0$. 
The correlated background is of particular importance for heavier mesons and at high transverse momenta, where the mesons are most likely accompanied by a jet. 
The random combinations of photons can be described by using mixed event techniques~\cite{Kopylov:1974th}.
The remaining background modulation due to correlations in the background can be parametrized by a polynomial of first or second order, adjusted to the measured distribution in a range close to the particle mass. 
For the mixed-event background, each photon from the signal event is paired with about 50--80 photons from different events with similar track or photon multiplicity and a primary vertex position along the beam line.
The absolute normalization of the mixed event invariant mass distribution is obtained by normalizing it to the corresponding same-event distribution in a signal free region. 
Then, the mixed-event background is subtracted and the remaining distribution is fitted with a polynomial added to the signal shape. 
\par
A more accurate description of the background including both the uncorrelated component and the correlated one due to the decays can be achieved by an adapted rotation technique developed within \gls{ALICE} based on Ref.~\cite{Adams:2004fm}.
In this technique, two arbitrary photons in the same event are paired and assumed to originate from a common mother particle. 
In order to get an approximation for the background an alternative decay of this mother particle is calculated. 
Each of the two newly created photons is then paired with all other photons in the event except the other newly created photon. 
In this process the direct correlation of the photons in the background calculation is removed because the two photons are not paired. 
All other correlations remain intact since the newly calculated decay could have also happened in the real event. 
Even if the photon pair does not originate from the same mother particle, the described process yields a good description of the background. 
The simplest, yet very effective approach to calculate the decay is to rotate the photons around the axis of their pair-momentum vector by 90 degrees. 
In this process the momenta and relative distance of the pair are kept the same.
Rotated photons are only considered in the pairing if they would still be reconstructable within the acceptance of the corresponding reconstruction technique.
This procedure is repeated for each possible photon pair combination. 
The invariant mass interval used to normalize the background estimated with the rotation technique can be significantly farther away from the $\pi^0$ and $\eta$ meson peak regions compared to the mixed event technique as the inherent shape of the in-event correlations is preserved.\\
%
The signal over background ratio of the $\piz$ ($\eta$) meson is a factor 2 to 3 (1 to 2) higher for the \gls{PCM-EMC} technique than for the \gls{EMC} reconstruction.
This can be attributed to two effects. 
Firstly to the better resolution and higher purity of the \gls{PCM} photon sample. 
Secondly, the worse reconstruction efficiency of the \gls{PCM} method results in less than one \gls{PCM} photon being reconstructed per event in pp collisions on average, while this average rises to 1.4 for the \gls{EMCal} reconstruction.
This results in a smaller pool of photon candidates that can be considered in a given event, which along with the meson candidate selection criteria results in a reduced combinatorial background below the true meson peaks.
The background below the \piz\ and $\eta$ meson peaks for the \gls{PCM-EMC} reconstruction can be approximated well using the mixed event technique in combination with an additional linear background as outlined in~\cite{Acharya:2017hyu,Acharya:2017tlv,Acharya:2018hzf}. 

The mixed-event background technique works well for the \gls{EMC} \piz\ meson reconstruction in the transverse momentum range from 2 to 15~GeV/$c$ in pp and \pPb\ collisions.
Beyond this momentum range, however, or in larger collision systems~\cite{Acharya:2018yhg}, the mixed-event background subtraction tends to lead to rather large systematic uncertainties as no signal-free region can be found where the distribution can be normalized.
This stems from the worse single cluster energy resolution at low energies and the smearing of cluster energies due to overlapping showers at higher energies, leading to a significant broadening of the meson invariant mass peaks.
In addition, rather strict selections on the minimum opening angle between the two photons have to be placed in the mixed event. 
This is necessary to mimic the minimal distance between measured photons arising from the finite cell size and the clusterization.
These selections significantly reduce the efficiency to reconstruct \piz\ meson beyond 15~\GeVc\ and completely remove the signal above 20~\GeVc. 
When using the rotation background, the distances between close pairs are kept.
Consequently, neutral pions can be reconstructed up to higher transverse momenta as long as a signal-free region can be found to which the rotation background can be normalized and as long as the decay photon showers do not fully overlap due to the decay kinematics. 
%

\begin{figure}[t]
  \centering
  \includegraphics[width=0.49\textwidth]{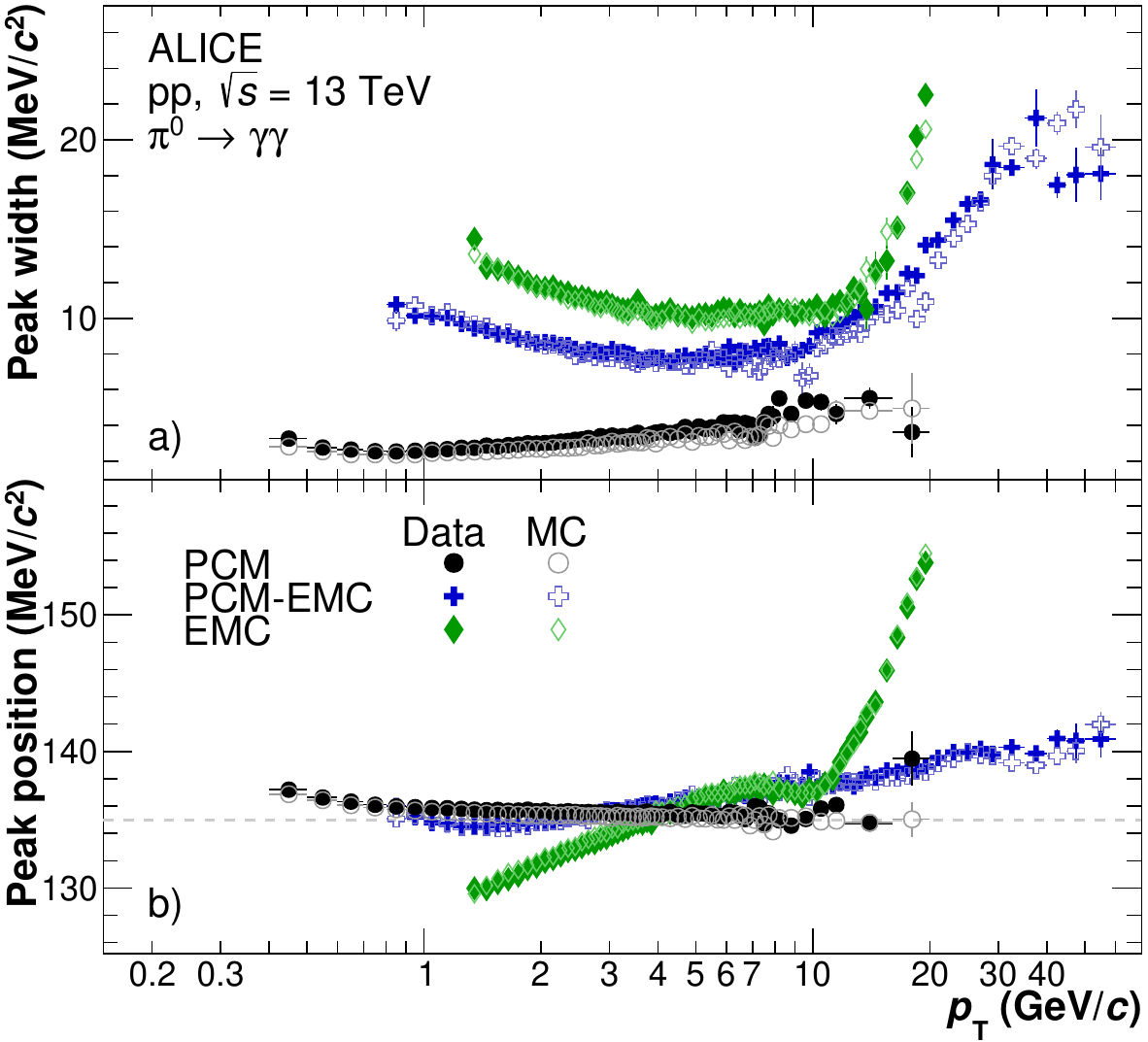}
  \includegraphics[width=0.49\textwidth]{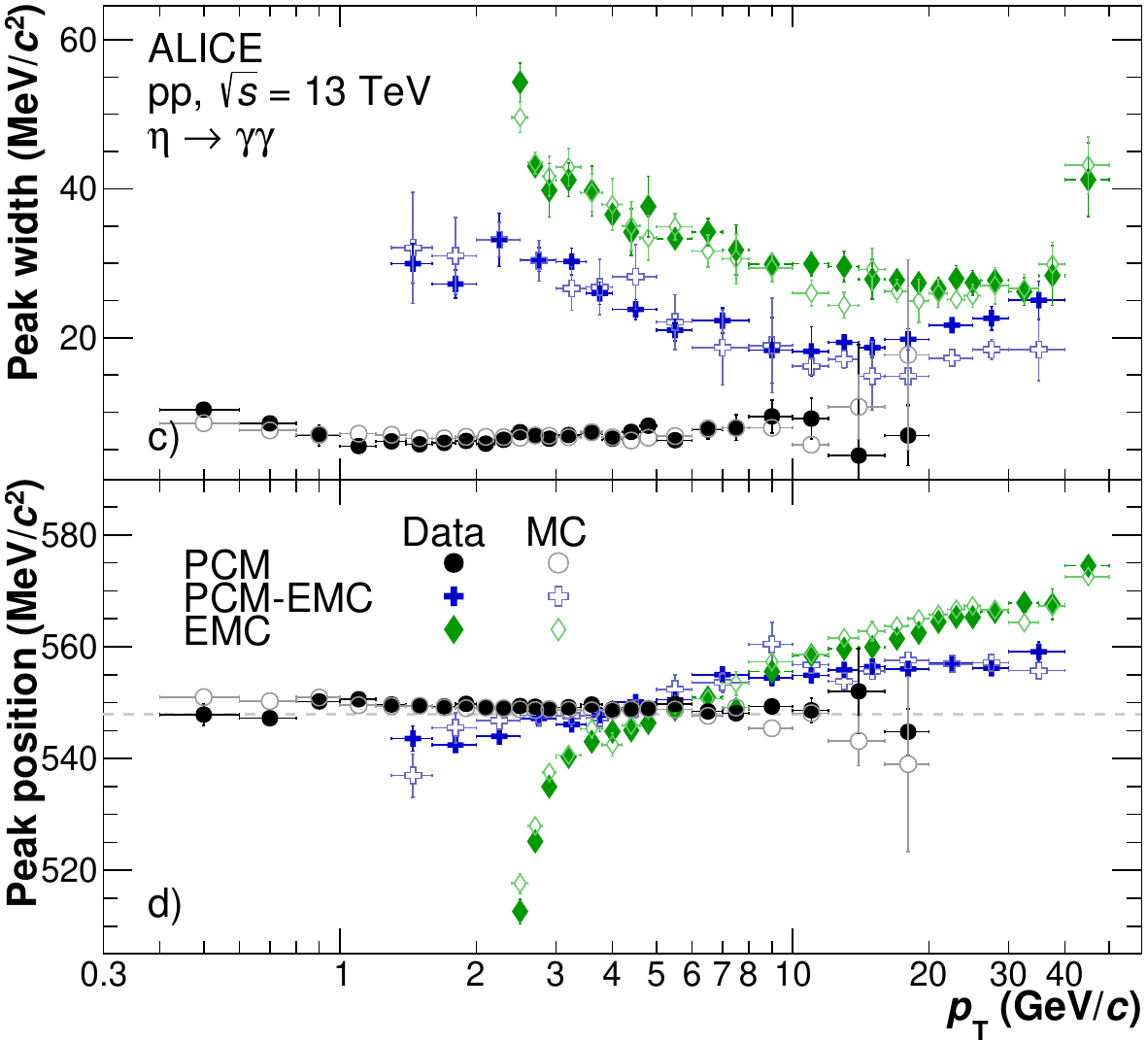}
  \caption{(Color online) \piz\ (left) and $\eta$ (right) meson peak position (b,d) and width (a,c) as a function of the meson momentum measured in pp collisions at \sthirteen\ combining two photons reconstructed with \gls{EMCal} or \gls{PCM}.} 
  \label{fig:pi0masswidth}
\end{figure}

\Figure{fig:pi0masswidth} shows the mean and width of the invariant mass peak for the \gls{EMC} and \gls{PCM-EMC} methods compared to the pure \gls{PCM} reconstruction for the neutral pion and $\eta$ meson. 
A significant broadening of the invariant mass peaks towards lower and higher \pt is observed for both mesons and both \gls{EMCal}-based reconstruction techniques.
As described in \Sec{sec:EMCalEnergyPositionCalib}, the \gls{EMCal} cluster energy was first corrected for the nonlinearity measured in a test beam (\Sec{sec:TestBeam}) and by a residual \gls{MC} correction (\Sec{sec:EMCalEnergyPositionCalib}).
The \gls{MC} correction appears to be system independent for pp and \pPb\ collisions but has to be adapted for semi-central and central \PbPb\ events, where the underlying event contributes a significant fraction of energy to the single cluster energies. 
The reconstruction of \piz \ mesons using \gls{EMC} experiences a peak position that increases with \pT~(lower left panel \Fig{fig:pi0masswidth}). 
At low \pT ($\pT < 4$~\GeVc), this is due to partial reconstruction of the photon energy due to conversions occurring in the material in front of the \gls{EMCal}. 
At higher \pT ($\pT > 8$~\GeVc), the showers merge into a single cluster and contribute to the inefficiency of reconstructing the left flank of the \piz\ meson peak. 
For the $\eta$ meson, on the other hand, only the loss of energy due to conversions has a significant effect on the peak position.
Both effects are very well reproduced in the simulation and thus allow for a neutral meson yield extraction in a wide transverse momentum region. 
The momentum reach of the neutral pion can be extended from $20$ to $50$ \GeVc\ when combining \gls{PCM} and \gls{EMCal} photon candidates using the \gls{PCM-EMC} reconstruction, until it reaches the limit where the opening angle for the neutral pion is too small and the electrons from the \gls{PCM} photon will point to the reconstructed \gls{EMCal} cluster.
Additionally, the conversion photon can be reconstructed down to momenta of 100 MeV$/c$, allowing \piz\ and $\eta$ mesons to be reconstructed starting from 0.8~\GeVc\ and 1.4~\GeVc, respectively. 
For the $\eta$ meson no decay photon merging is observed in the statistically accessible transverse momentum range.

\begin{figure}[t]
  \centering
  \includegraphics[width=0.5\textwidth]{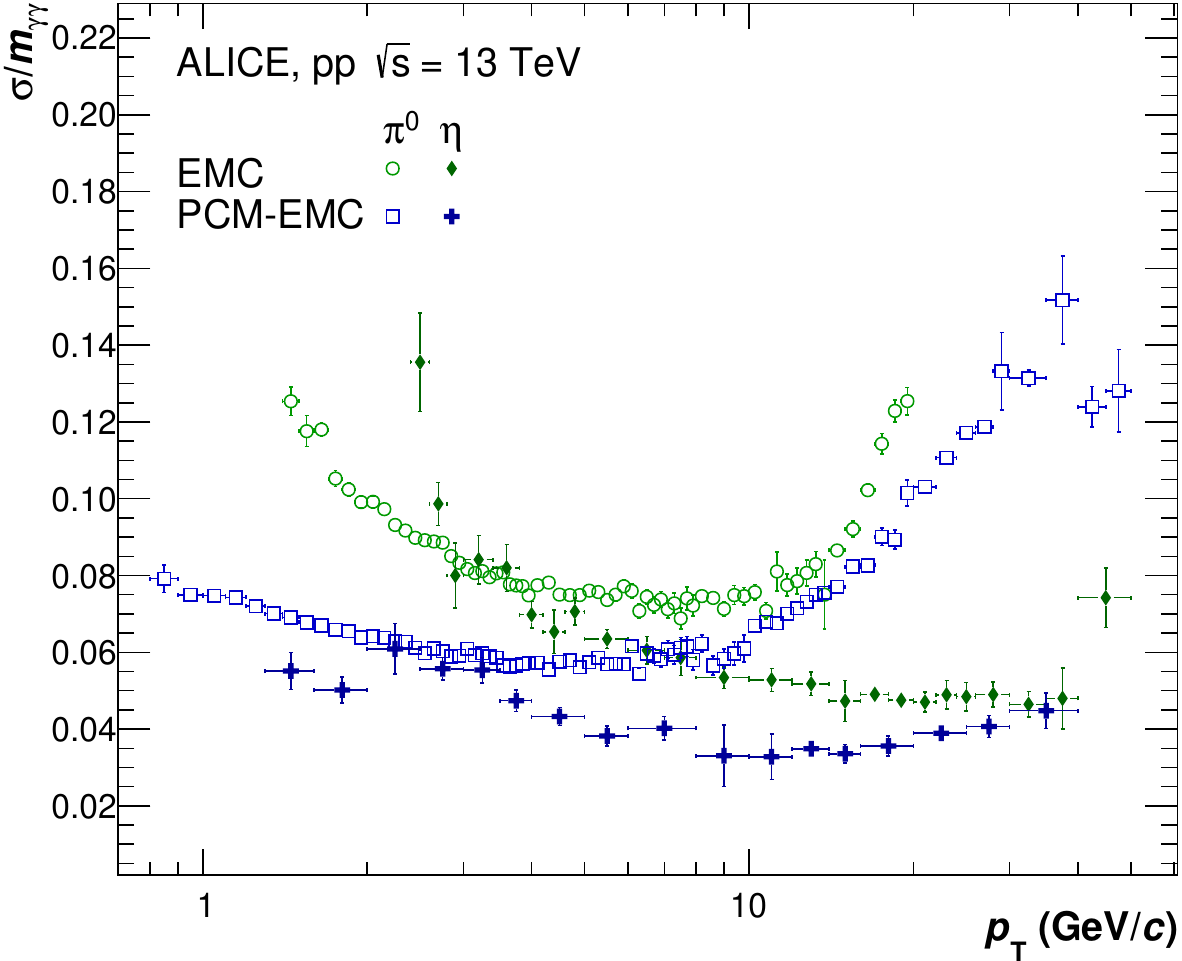} \hspace{0.5cm}
  \raisebox{+20mm}{%
  \begin{minipage}[b]{6.5cm}
    \caption{(Color online) Relative width of the \piz\ and $\eta$ meson as a function of transverse momentum in pp collisions at \sthirteen\ for the \gls{PCM-EMC} and \gls{EMC} reconstruction techniques.} 
    \label{fig:relwidth}
  \end{minipage}}
\end{figure}

\Figure{fig:relwidth} shows the peak width normalized to the nominal mass of the mesons as a function of transverse momentum.
By using a \gls{PCM} photon candidate for one of the decay photons, the neutral pion mass resolution significantly improves in the low \pT \ region, from about 9\% for the \gls{EMC} reconstruction to 6\% for the \gls{PCM-EMC} reconstruction.
The mass resolution appears to be even better for the $\eta$ meson, decreasing to about $5\%$ and $4\%$, respectively. 
While the relative width of the neutral pion peak with the \gls{EMC} reconstruction depends on the reconstructed momentum, the $\eta$ meson width, for the same reconstruction technique, exhibits only a mild \pT\ dependence.
The worsening of the \piz\ invariant mass resolution at lower momenta is driven by the energy resolution of the calorimeter, while the worsening at high \piz\ meson momenta is due to shower overlaps.
For the \gls{PCM-EMC} technique, the decreasing resolution with increasing \pT\ is mainly driven by the momentum resolution in the tracking.

The relative width of the mass peak for $\eta$ mesons can be used to calculate the expected width of heavier mesons decaying into two photons for the different reconstruction techniques at the nominal meson mass. 
For instance for the $\eta'$ meson with a mass of 957.78~MeV/$c^2$, the peaks would be about 40--100~MeV/$c^2$ wide, which can only be reconstructed given a very good understanding of the correlated background below the peak.
Considering the low branching ratio ($2.20 \pm 0.08 \%$) and the worse signal to background for the $\eta'$ with respect to the $\eta$ meson, the reconstruction of the $\eta'$ meson in its di-photon channel will not be possible with the present event sample.

\begin{figure}[t]
  \centering
  \includegraphics[height=7.3cm]{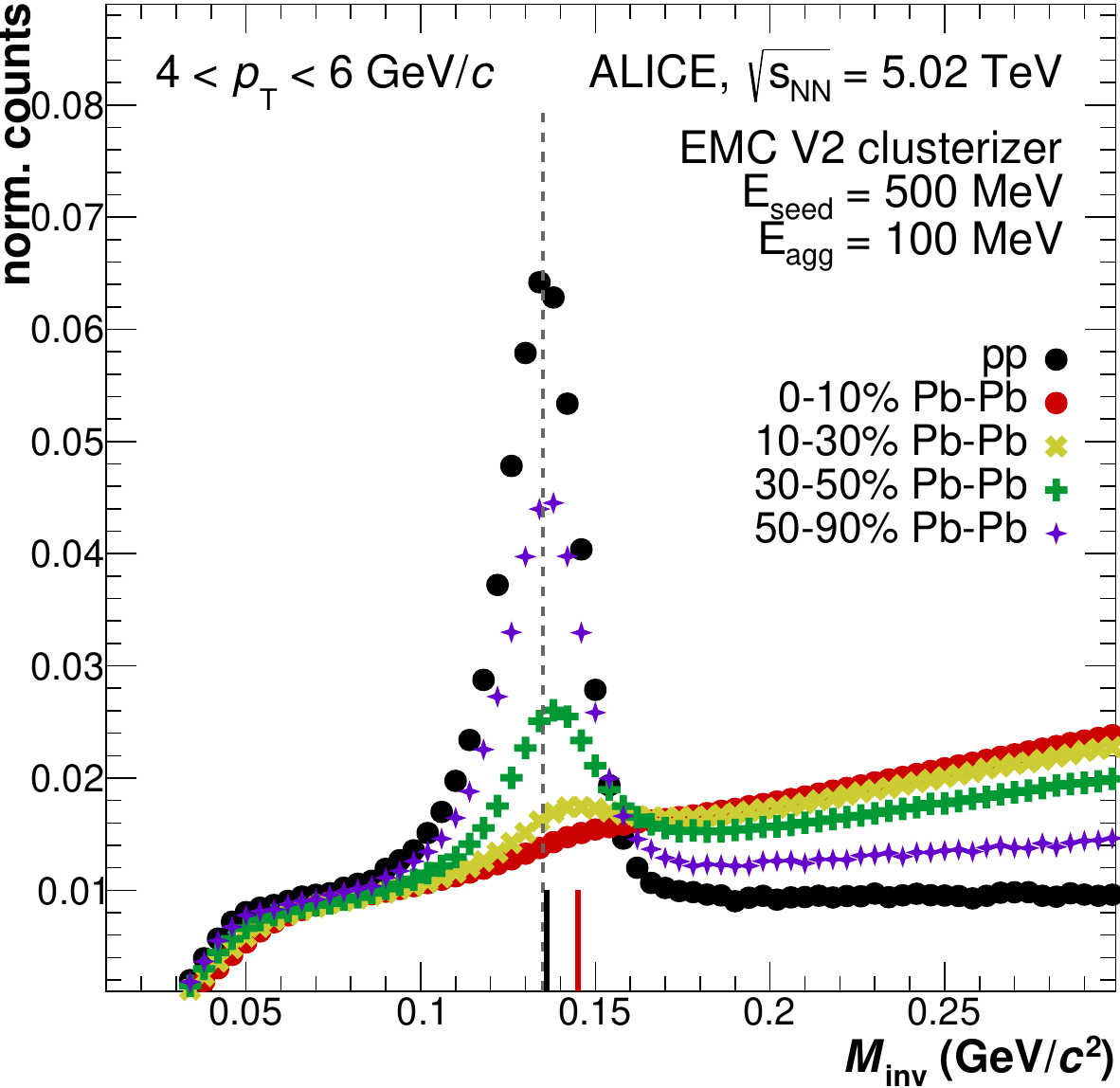} 
  \includegraphics[height=7.3cm]{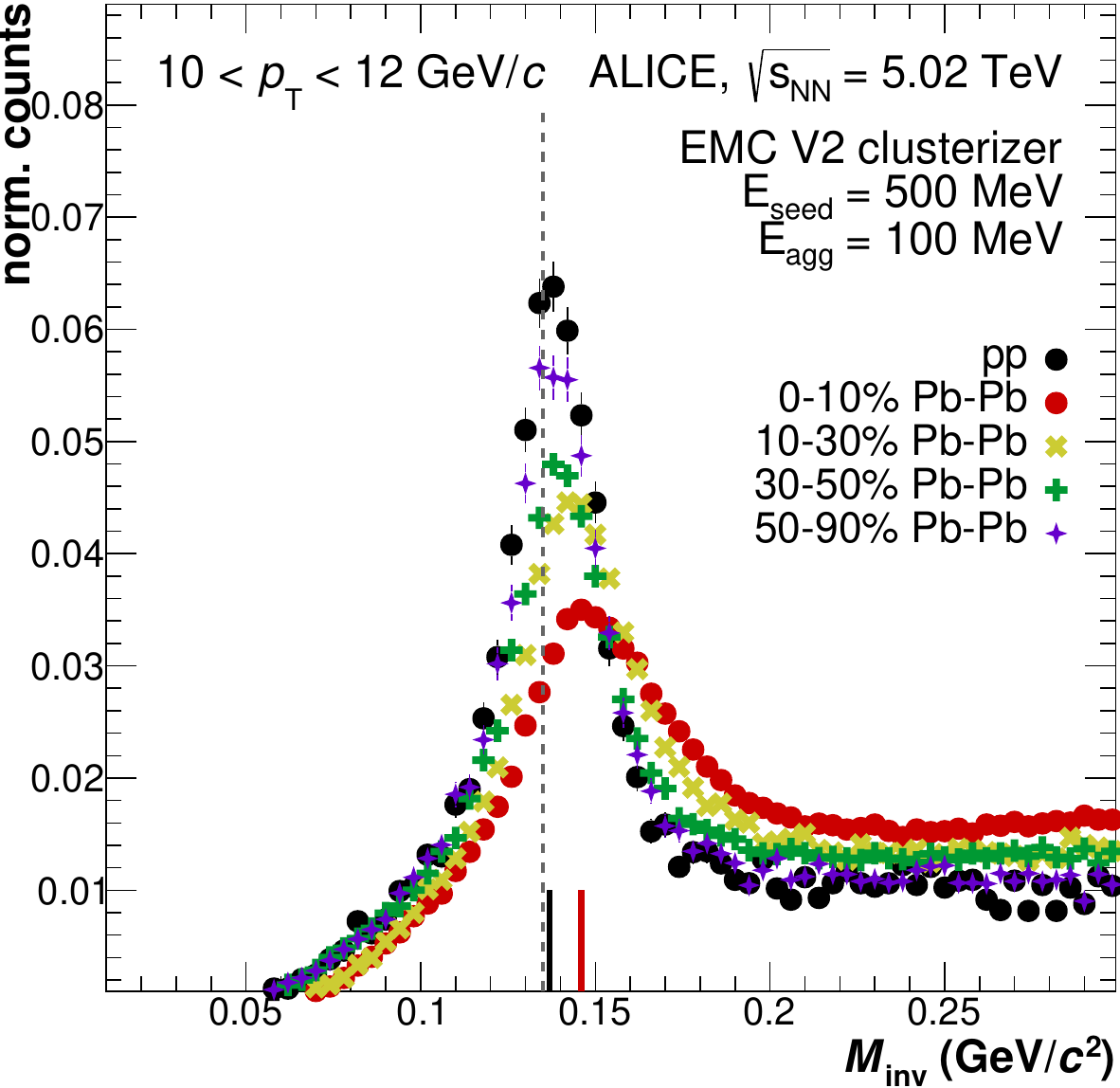} 
    \caption{(Color online) Invariant mass distribution for neutral pion candidates at intermediate (left) and high (right) transverse momenta reconstructed with both photons in the \gls{EMCal} for pp and \PbPb\ collisions in different centrality classes at \sfivelead. 
            The distributions are normalized to the integral in the displayed invariant mass region to be able to compare the shapes of the invariant mass distributions and their difference in the signal-to-background ratio. 
            The gray vertical line indicates the nominal pion mass, while vertical black and red lines indicate the reconstructed neutral pion mass in pp and central \PbPb\ collisions, respectively. }
    \label{fig:mesonPbPbV2}
\end{figure}

For larger collision systems and in particular most central \PbPb\ collisions, the performance of the \gls{EMCal} slightly deteriorates when using the same clusterization thresholds.
In particular at low transverse momentum ($\pT < 5$~\GeVc), the reconstructed neutral pion mass exhibits a significant shift which increases with increasing multiplicity as seen in \Fig{fig:mesonPbPbV2}.
This can be attributed to a substantial contribution of the underlying event to each cluster, which is not correlated to the energy deposition of the photon. 
While the simulations show a similar behavior, the details of the particle composition and transverse momentum dependence could, so far, not be reproduced for 0--10\% \PbPb\ collisions.
To overcome this problem, the minimum cluster energy was raised to 1.5(1.0) GeV in the analysis in central (semi-central) \PbPb\ collisions.
In addition, the \gls{MIP} energy was subtracted for each track that was matched to the cluster.
In Ref.~\cite{ALICE:2018mdl}, embedding simulated $\piz$ decays into real data was used to obtain the efficiency correction. 


\begin{figure}[t]
  \centering
  \includegraphics[width=0.49\textwidth]{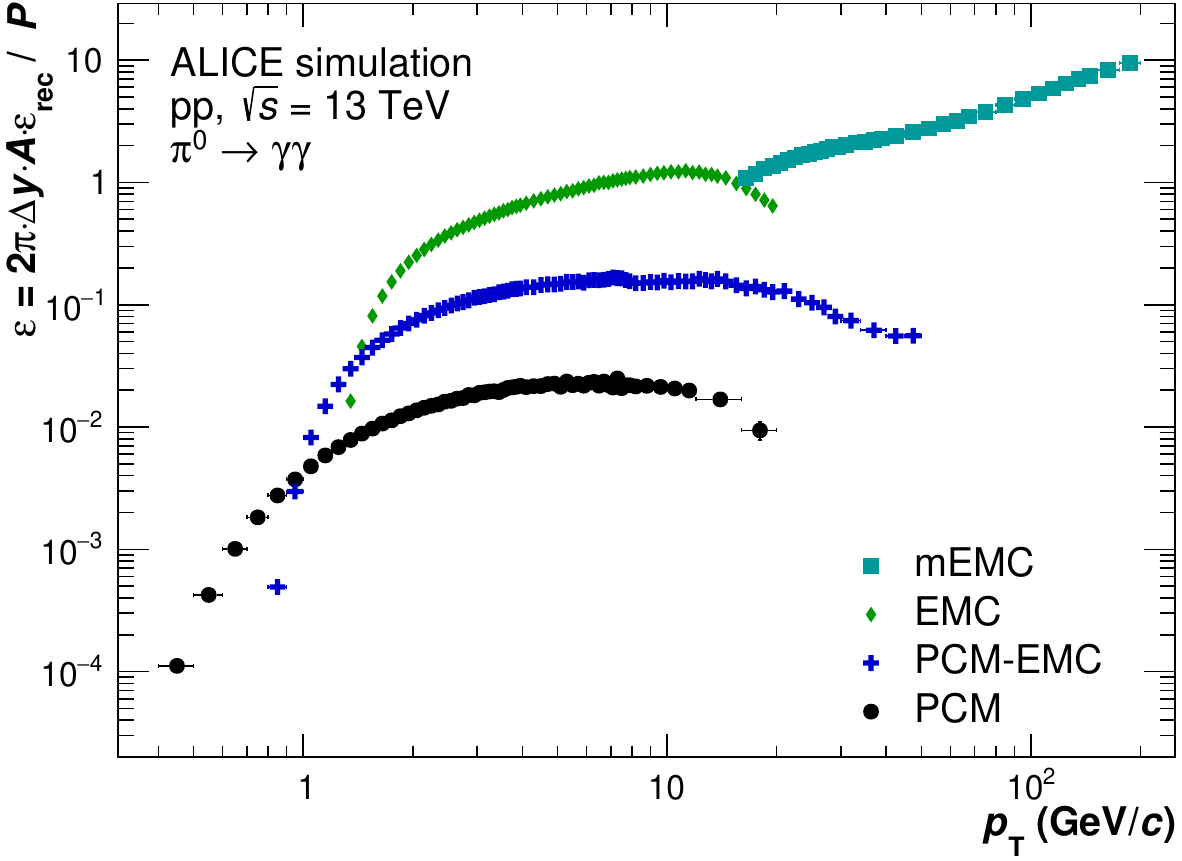} 
  \includegraphics[width=0.49\textwidth]{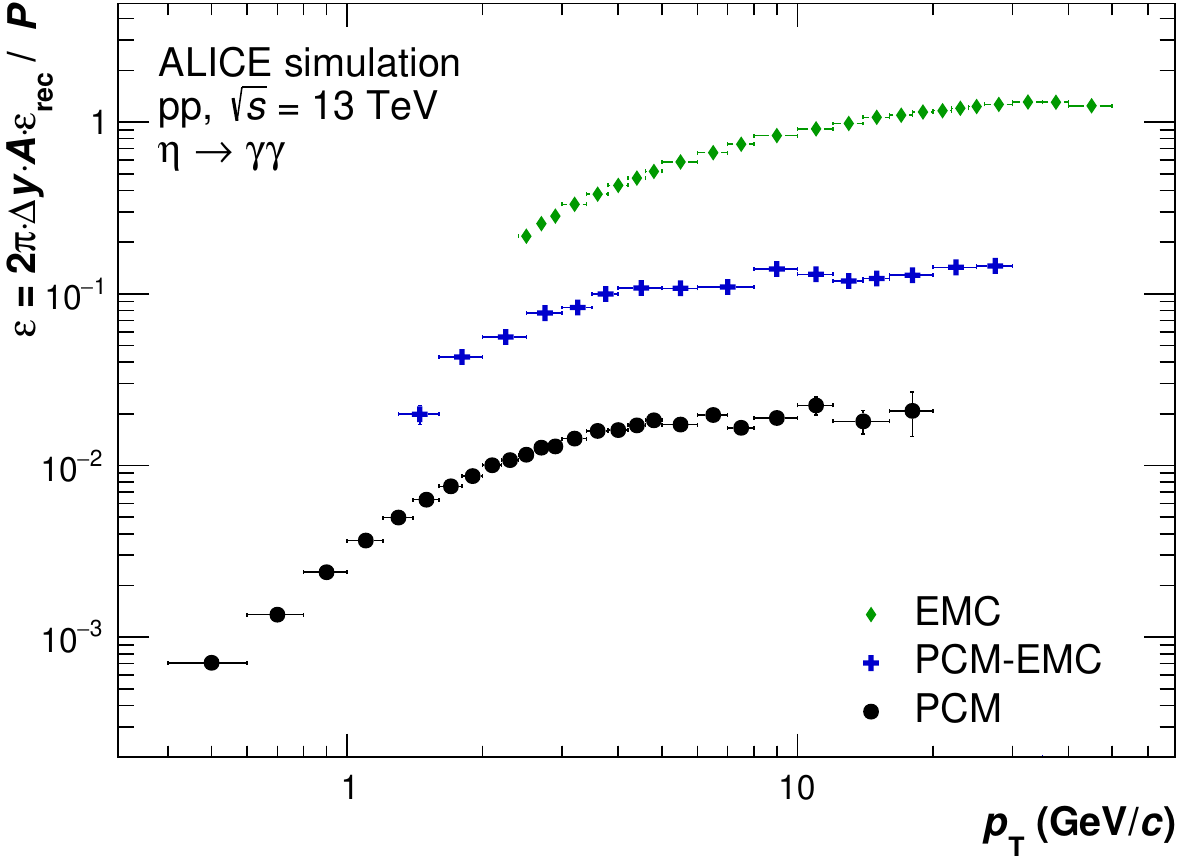} 
    \caption{(Color online) Combined correction factor for the neutral pion (left) and $\eta$ meson (right) for all \gls{EMCal} related reconstruction techniques and the pure \gls{PCM} reconstruction for comparison. 
              The correction factor contains the corrections for acceptance, efficiency and purity. }
    \label{fig:mesoneffi}
\end{figure}
The combined correction factor consisting of the meson reconstruction efficiency, acceptance, normalization constants, and meson purity is presented in \Fig{fig:mesoneffi} for \piz\ and $\eta$ mesons in \pp\ collisions at \sthirteen. 
For the invariant mass based techniques the meson purity is considered unity, as a clear peak can be used to extract the signal. 
The correction factors are shown for all available reconstruction techniques using the \gls{EMCal} for the respective mesons: \gls{EMC}, \gls{PCM-EMC} and \gls{mEMC}.
The total correction factor for the standalone \gls{PCM} reconstruction is added for comparison.
For the \gls{EMC} reconstruction technique the correction factor of the $\pi^0$ increases by several orders of magnitude from about $0.001$ at $1.2$ \GeVc\ to its maximum of about $1$ above $6$ \GeVc.
This characteristic rise is mainly driven by the increasing photon reconstruction efficiency seen in \Fig{fig:photonRecProp} together with the slightly increasing acceptance for mesons with larger opening angles. 
The decrease of the pion reconstruction efficiency above 10 \GeVc\ is due to cluster merging.
Similar features, slightly shifted in transverse momentum, are visible for the correction factor of the $\eta$ mesons for the \gls{EMC} reconstruction technique. 
The low momentum cut-off is caused by the lower signal to background ratio of the $\eta$ meson compared to the neutral pion. 
The maximum $\eta$ meson correction factor of about $1$ for the \gls{EMC} reconstruction is reached above $12$ \GeVc\ and no clear reduction at higher transverse momenta is observed within the covered momentum range.
Combining one photon from each reconstruction technique yields the expected reduction due to the conversion probability of about 9\%~\cite{Abelev:2012cn}. 
Furthermore, the correction factor appears to be closer in its \pt\ dependence to that of the \gls{PCM} technique and the effects from the reduction in the cluster efficiency for higher momenta due to merging are significantly reduced for the neutral pion.
This leads to stronger similarities between the correction factors of the neutral pion and $\eta$ meson up to 40 \GeVc, where both reach their current statistical limits.

\subsubsection{Neutral meson reconstruction based on single clusters}
\label{sec:MesonShowerShape}
\begin{figure}[t]
  \centering
  \includegraphics[height=2.5cm]{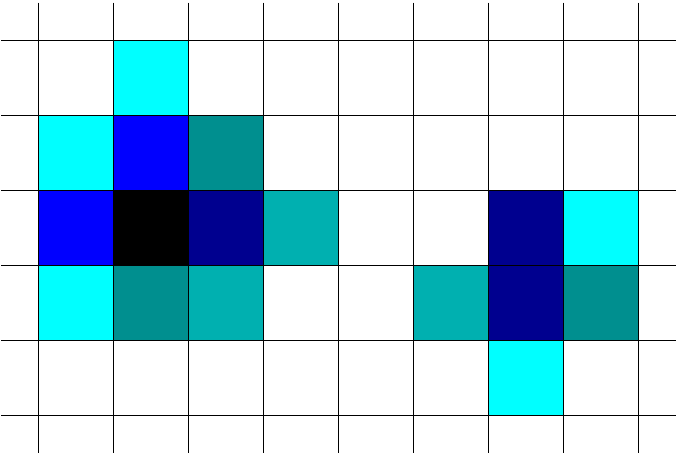} \hspace{0.2cm}
  \includegraphics[height=2.5cm]{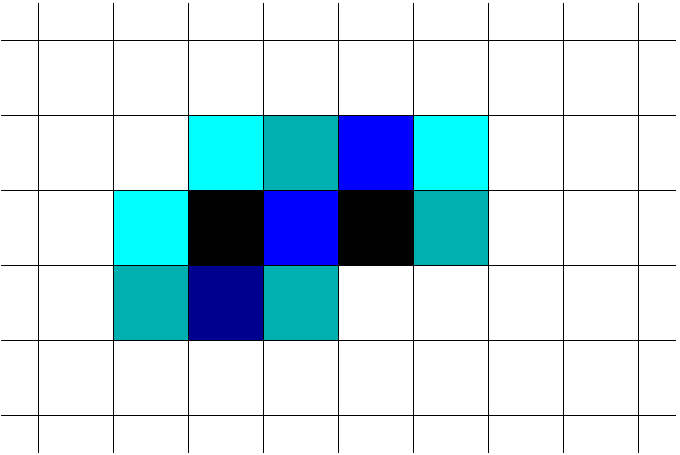} \hspace{0.2cm}
  \includegraphics[height=2.5cm]{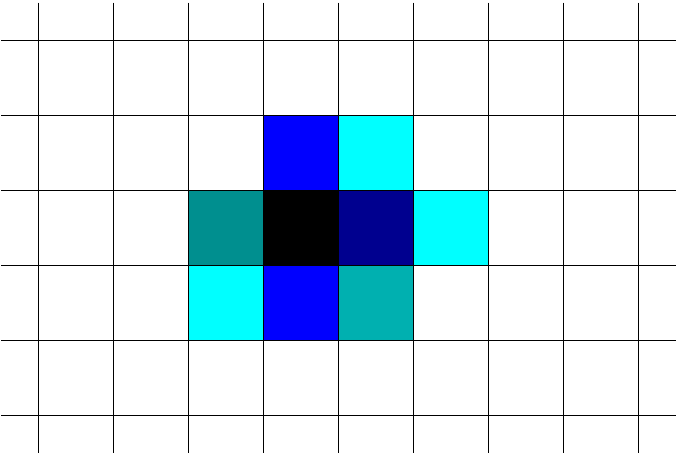}
  \caption{(Color online) Schematic view of cluster shower overlaps from \piz\ meson decays with $E_{\pi^{0}}=$~4, 10 and 20 GeV from left to right. 
  The cell color indicates the deposited energy; the darker, the more energy.}
\label{fig:decaymergeschema}
\end{figure}

For \piz\ mesons with $\pT>15$ \GeVc, the individual decay photons cannot be resolved by the clusterizer as their showers overlap on the \gls{EMCal} surface, as discussed in \Sec{sec:showershape}.
In this case, using distinguishing features of the profile and the spread of the energy distribution in the cells can help to distinguish between single photon clusters (symmetric deposition like in \Fig{fig:decaymergeschema} left) and merged particle clusters (\Fig{fig:decaymergeschema} middle and right). 
This \gls{mEMC} technique via the $\shshlo$ distribution becomes viable for the V1 clusterizer above $E\gtrsim6$~GeV and for the V2 clusterizer above $E\gtrsim15$~GeV.
The higher threshold in the latter case is chosen to avoid statistical overlap with the invariant mass-based techniques, where the V2 clusterizer can still split clusters between 6 and 15~GeV. 
Additionally, the V2 clusterizer absorbs less particles from the surrounding jet which otherwise distorts the $\shshlo$ distribution.
While photon clusters are rather round and peak at about 0.25 in the $\shshlo$ distribution, neutral pions appear to be elongated and predominately have values of $\shshlo$ larger than 0.27 for transverse momenta between $\sim15$ and $60$ \GeVc, see \Fig{fig:SSPhotonPi0MC}.
For $\pT > 60$ \GeVc, the $\shshlo$ distributions of photon and \piz\ clusters are more similar and the analysis must rely to a larger extent on the corrections from the simulation. \\
\begin{figure}[t]
  \centering
  \includegraphics[width=0.49\textwidth]{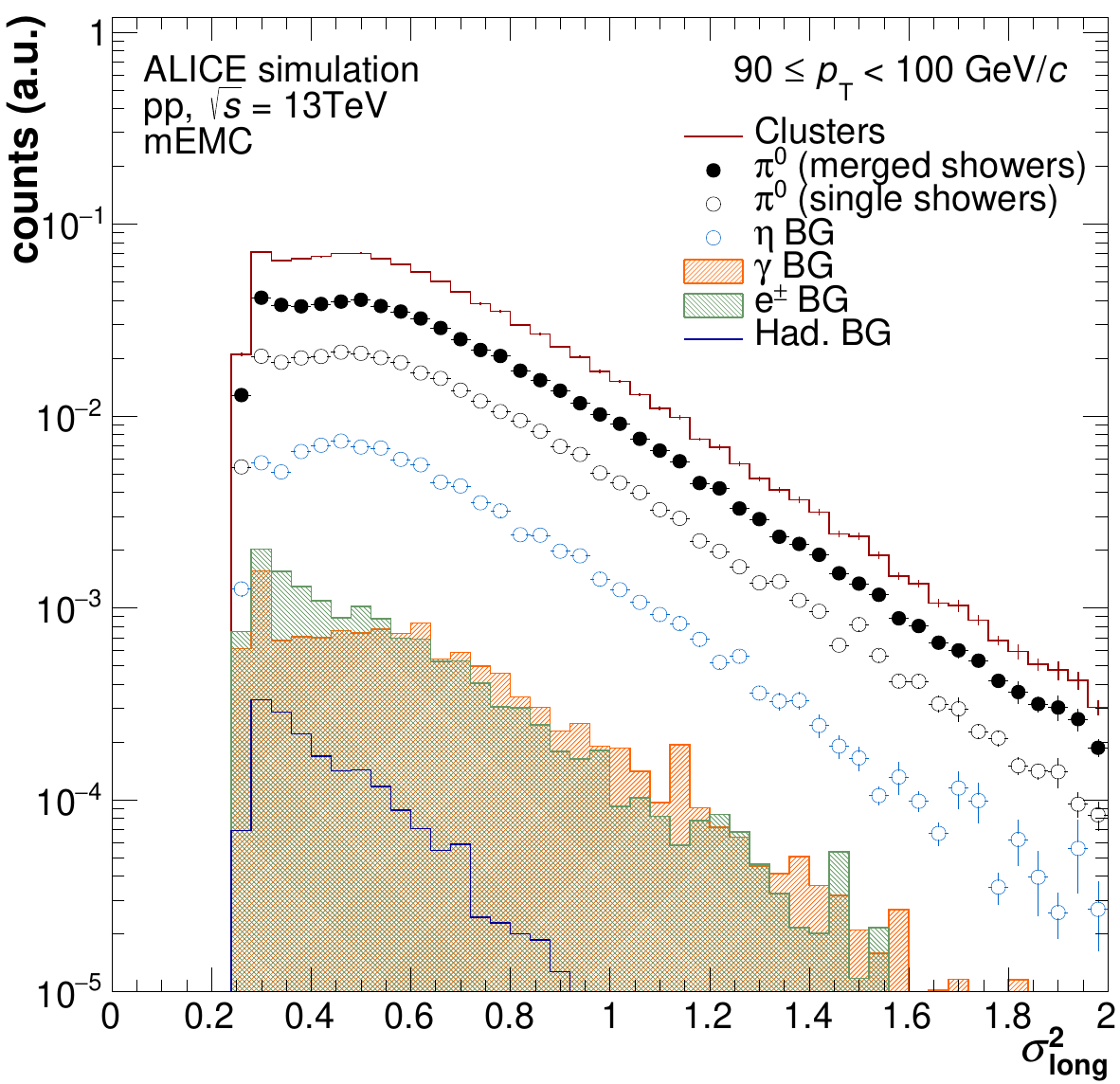}
  \includegraphics[width=0.49\textwidth]{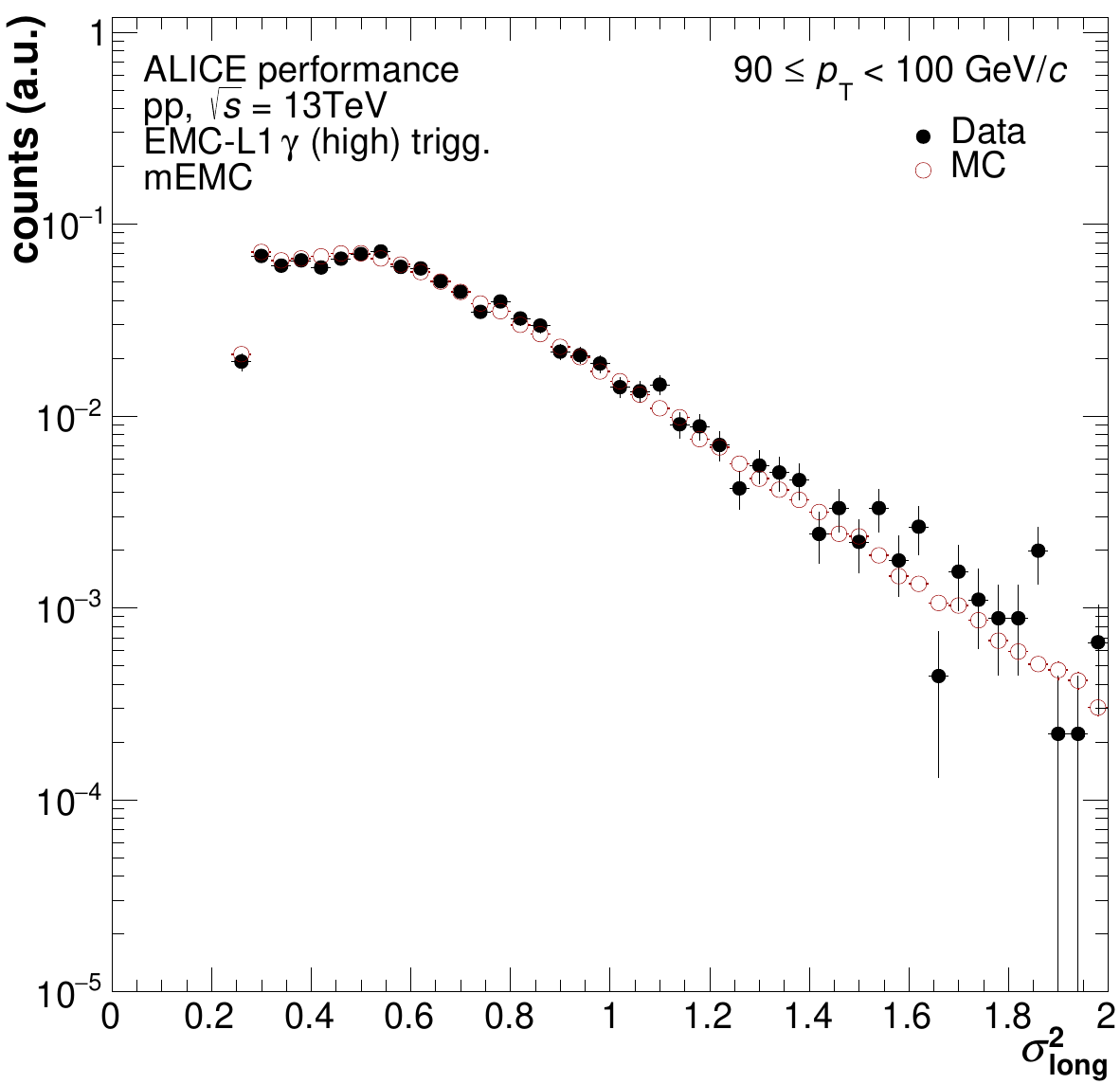}
  \caption{(Color online) Left: Decomposition of the $\shshlo$ distribution at $90 < \pT < 100$ \GeVc\ in its contributions from neutral pions reconstructed with both photons in one cluster (full black dots) or only one photon (open black dots) based on \gls{PYTHIA} 8 di-jet simulations. Additionally, the contributions from $\eta$ mesons (open blue dots), direct photons (orange histogram), primary electrons (green histogram) and other hadrons (blue histogram) are displayed. Right: Comparison of the $\shshlo$ distribution between data (black) and simulation (red). }
  \label{fig:M02Decomp}
\end{figure}
\begin{figure}[t]
  \centering
  \includegraphics[height=7cm]{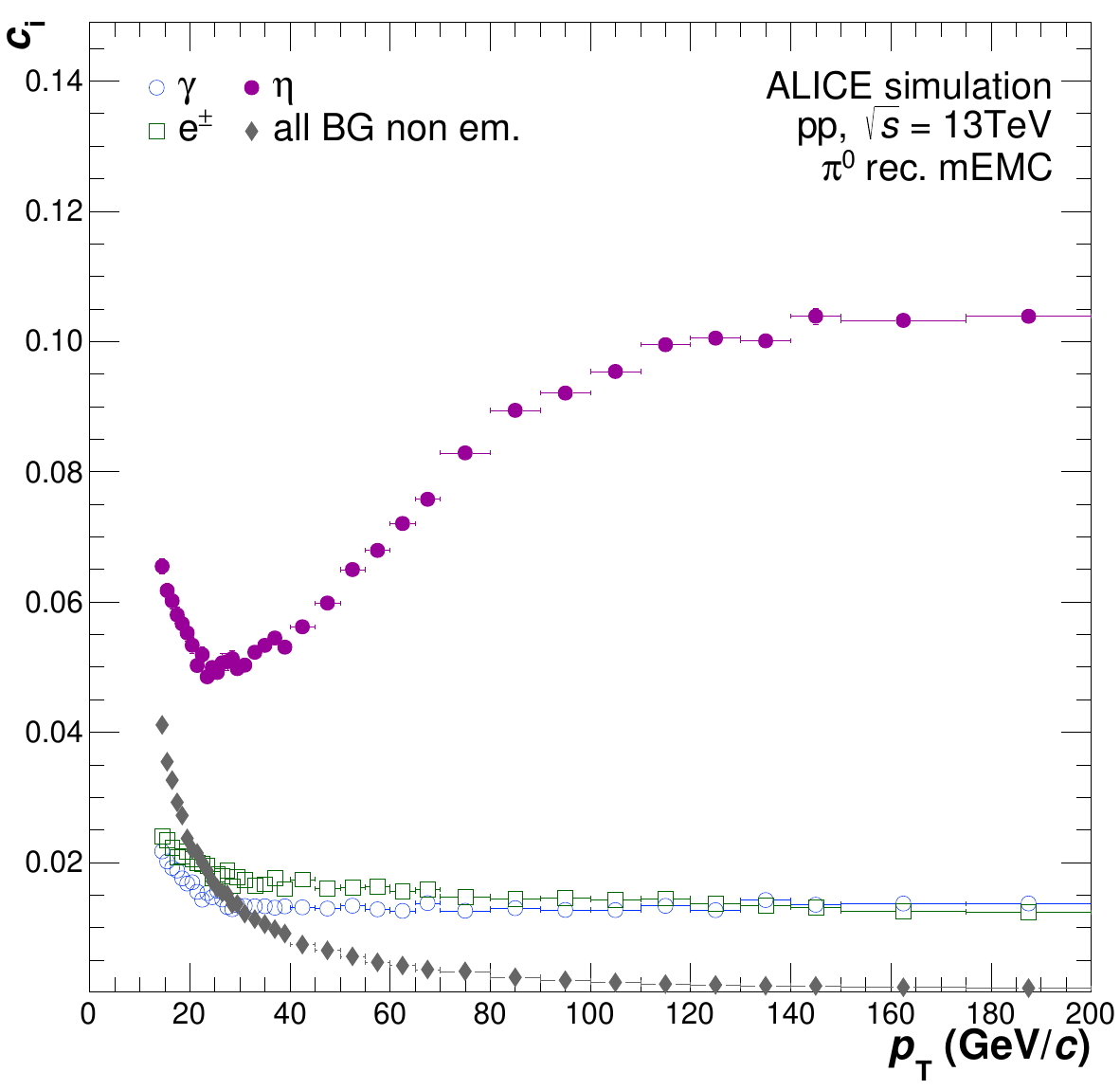}
  \includegraphics[height=7cm]{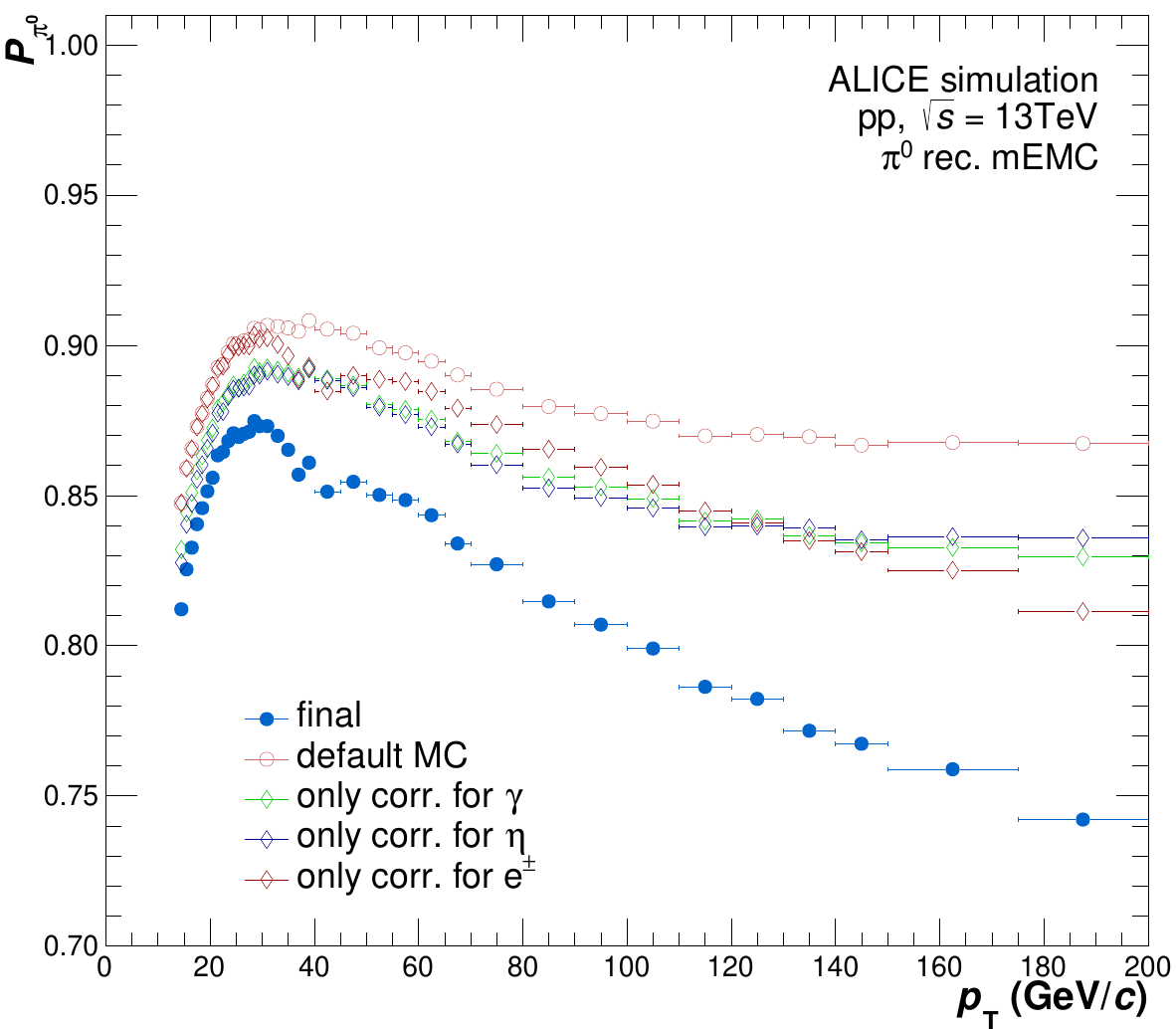}
  \caption{(Color online) Left: Contamination ($c_i$) of the neutral pion candidate sample split into the different contributions according to the \gls{PYTHIA} 8 di-jet simulations. Right: Purity of the neutral pion candidate sample without modifications (blue dots) and after additional corrections to adjust the simulation to the measured $\eta$ meson, expected direct photon and heavy-flavor-electron contributions (red open dots).}
\label{fig:PurityMerged}
\end{figure}
\begin{figure}[t]
 \centering
  \includegraphics[height=7.7cm]{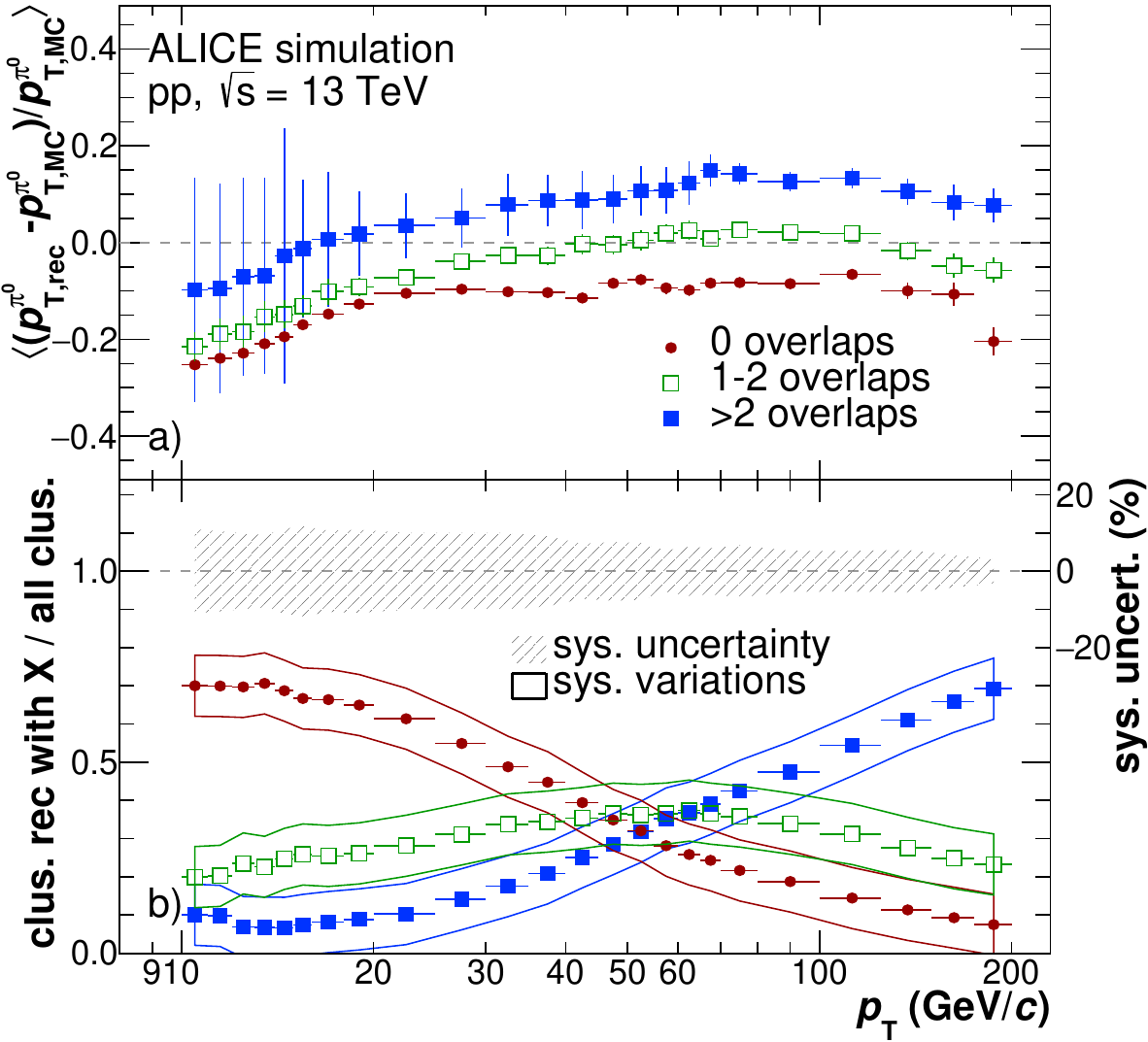}
    \caption{(Color online) \textit{a)} Mean cluster energy shift versus transverse momentum for three different neutral pion merged cluster types from \gls{PYTHIA} 8 simulations with two jets in the final state at $\sqrt{s}=13$ TeV. Clusters with no overlapping particles within $R<0.05$ around the momentum vector of the $\pi^0$ on Monte Carlo generator level are shown in red, while clusters with 1--2 overlapping particles are shown in green and clusters with more than two overlapping particles in blue. An increasing overlap of particles shifts the cluster energies to larger values. \textit{b)} Fractions of clusters from the three overlap types in the total cluster sample are shown in the same colors. The bands indicate the systematic variations on the fractions, which are applied in the toy simulation in order to obtain the final systematic uncertainty shown in the shaded gray band versus transverse momentum.} 
    \label{fig:mergedOverlap}
\end{figure}
\Figure{fig:M02Decomp} shows the $\shshlo$ distribution for clusters with a \pT\ between $90$ and $100$~\GeVc. 
The two dominant contributions arise from neutral pions, which are merged clusters containing both decay photons (full markers) and single decay photons without overlap from the other decay photon (open markers).
\Fig{fig:PurityMerged}~(left) illustrates that the $\eta$ meson contribution is the largest contamination to the neutral pion candidate sample and increases significantly with increasing pion momenta from around $5\%$ at low \pT\ to up to $11\%$ at 200~\GeVc.
The photon and electron contributions, on the other hand, can be significantly suppressed by the shower-shape cuts, in particular at low transverse momenta. 
However, the default \gls{PYTHIA} 8 Monte Carlo simulations used for the \gls{mEMC} analysis lack certain contributions~(prompt photons and electrons from weak decays) that need to be considered for the purity estimation. Moreover, the relative fraction of $\eta$ mesons is generally too small in these simulations compared to the measured ratio of $\eta$ to $\pi^0$ meson yields.
The effect of these additional contributions is shown in \Fig{fig:PurityMerged}~(right), resulting in a several percent lower final purity.
Additional primary track veto cuts as described in \Sec{sec:trackmatch} allow for an efficient reduction of the electron and charged hadron contributions.
As the track propagation beyond 20 \GeVc\ has rather large uncertainties and the neutral pions tend to appear within jets, the clusters for $\pT > 20$ \GeVc\ are only vetoed if their $E/p$ is larger than $1.7$.

The \gls{mEMC} technique was first used for a spectra analysis in Ref.~\cite{Acharya:2017hyu} for neutral pions between $16$ and $60$~\GeVc\ and was since improved to be able to reconstruct neutral pions up to 200~\GeVc~\cite{ALICE:2021est}. 
The main improvements in the analysis arose from a better understanding of the shower shape through the emulation of the cross talk as well as the nonlinearity of the energy response through measurements in the laboratory~\ref{sec:crosstalk}.
This allowed us to use the simulations even beyond the point where merged pions are clearly separated from photons, electrons, and single photons from $\eta$ mesons.

The correction for \gls{mEMC}, shown in \Fig{fig:mesoneffi}, consists of normalization constants, an acceptance component that is nearly constant in \pT, as well as the reconstruction efficiency.
The latter not only corrects for reconstruction losses but also for the \gls{EMCal} cluster energy resolution, which is strongly affected by particle overlaps within the same cluster, especially at high \pT, as shown in \Fig{fig:mergedOverlap}. 
The correction can therefore exceed unity due to the significant difference between reconstructed and true \pT\ of the neutral pion candidates. 
The energy resolution of the merged clusters is the main source of systematic uncertainty in the \gls{mEMC} measurement. 
It was estimated using a generator-level particle decay simulation where neutral pions are generated according to an input parametrization of the measured $\pi^0$ spectrum and were subsequently smeared according to the energy response matrices for three different event classes; events with no particle overlaps within a radial distance of $R<0.05$ around the generated neutral pions in the \gls{EMCal} acceptance, events with 1 to 2 allowed overlaps, and events with more than 2 overlaps in the same radial cone.
Each of these classes presents a different \pT-dependent reconstructed energy of the merged pion clusters as shown in \Fig{fig:mergedOverlap}~(top) with a visible energy loss in the zero overlap class and an up to 20\% higher reconstructed energy in the class with more than 2 particle overlaps.
The contribution of these event classes strongly changes as a function of \pT\ , thereby changing the final energy resolution correction.
The systematic uncertainty is determined by varying the composition of the three overlap classes, which effectively varies the spread of in-jet particle production within the $R<0.05$ cone. 
The contribution of a class is modified by up to  $\pm10\%$ of the total as indicated by the bands in \Fig{fig:mergedOverlap}~(bottom) and the generated $\pi^0$ spectrum based on the input parametrization is smeared according to the resolution matrices using the appropriate fractions of the yields. 
The procedure resulted in an uncertainty on the \gls{mEMC} $\pi^0$ spectrum of up to 10\% at \pT$<$40 \GeVc\ and about 5\% at $\pT=200$ \GeVc. \\
%
\begin{figure}[ht]
  \centering
  \includegraphics[width=0.49\textwidth]{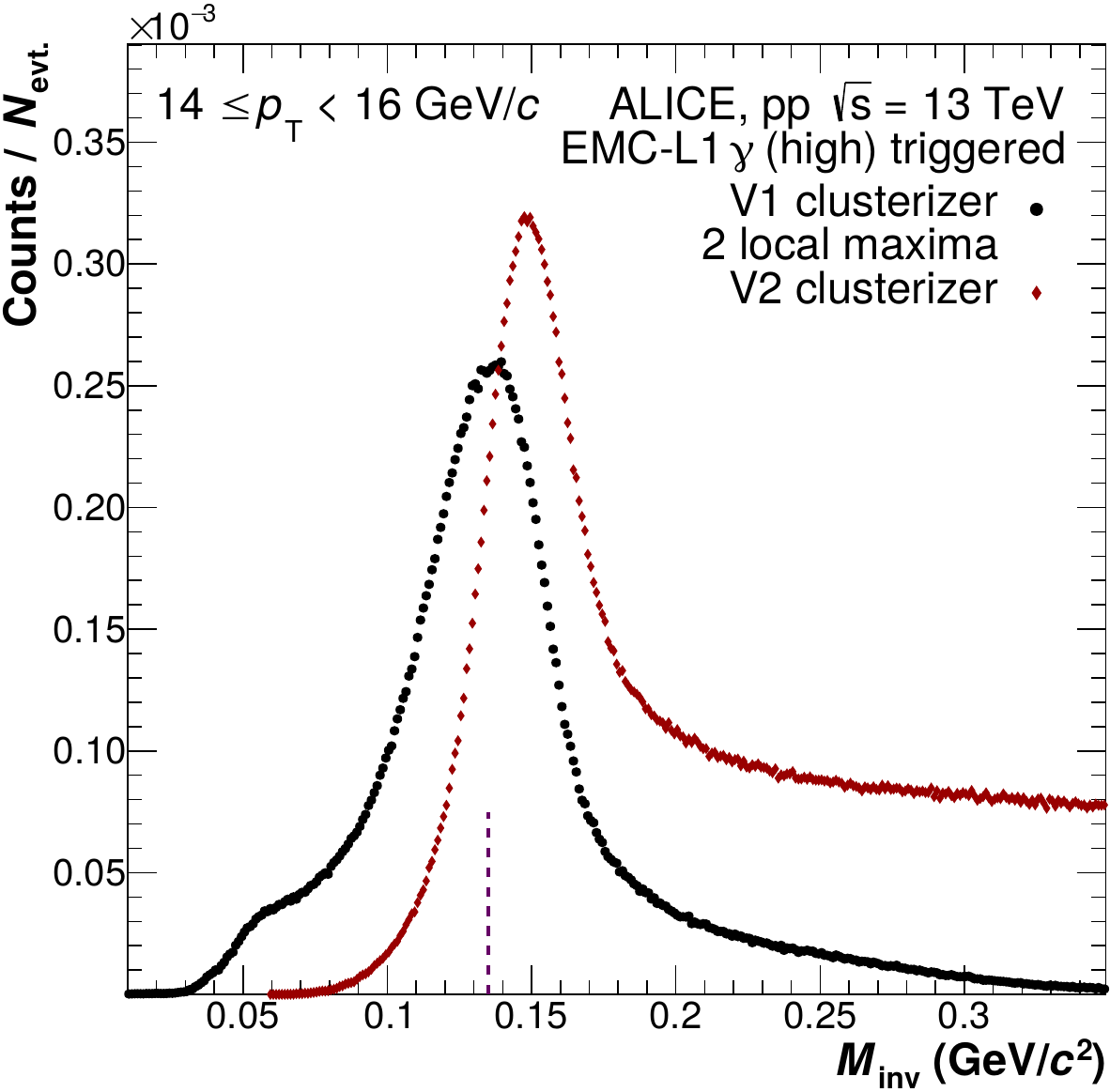} \hspace{0.2cm}
  \includegraphics[width=0.47\textwidth]{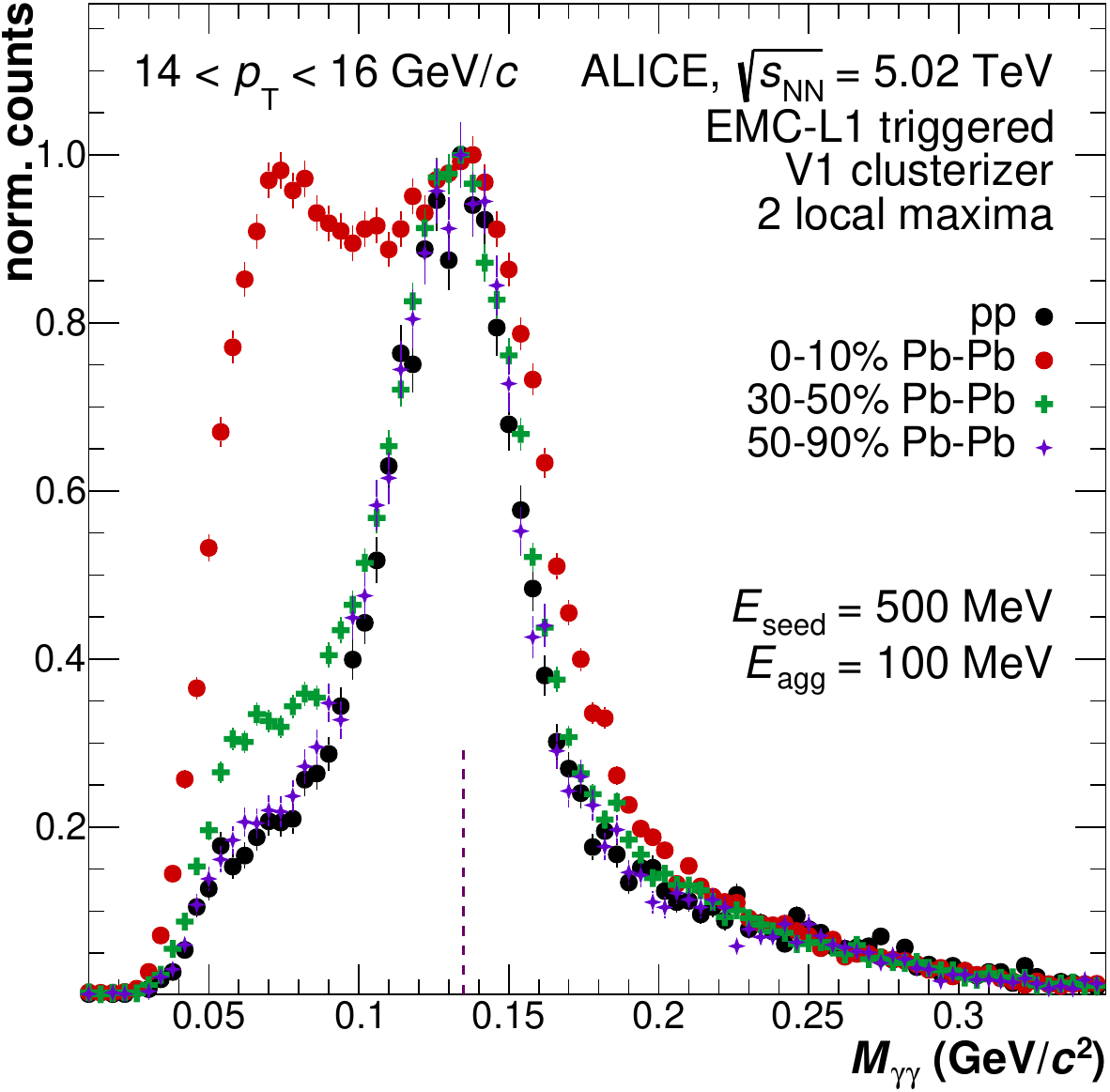}
  \caption{(Color online) Left: Mass of split clusters for pp collisions at \sthirteen\ for the V1 clusterizer with two local maxima compared to the invariant mass distribution obtained from pairs of V2 clusters for the neutral pion. 
            Right:  Mass of split clusters for pp and \PbPb\ collisions in different centrality classes at \sfivelead. 
            For the \PbPb\ comparison plot, the distributions are normalized to the maximum in the peak region to be able to compare the shapes of the invariant mass distributions. 
            The gray vertical line indicates the nominal pion mass.}
  \label{fig:Pi0SplitMass} 
\end{figure}
Besides extracting the neutral pion spectra, the \gls{mEMC} can also significantly simplify  particle correlation measurements as well as azimuthal anisotropy ($v_n$) measurements, as introduced in~\cite{Adam:2016xbp}. 
The main interest of these analyses lies in meson measurements in the range between $6$ and $20$ \GeVc.
For this, the V1 clusterizer has a particular advantage for clusters below $20$~GeV, where it allows to additionally make sure that the clusters in the \piz\ meson band are actually originating from \piz\ meson decays by splitting the cluster into two sub-clusters and calculating their invariant mass. 
The corresponding comparison of the invariant mass obtained from splitting V1 clusters with two local maxima and the pairing of two V2 clusters within the same sample are presented in \Fig{fig:Pi0SplitMass} (left).
For the split clusters the combinatorial background below the pion peak, even in \pp\ collisions, is significantly smaller than for the V2 clusterizer, as the probability of random overlaps is comparatively low.
This is even more apparent in high-multiplicity environments like in central \PbPb\ collisions, where this technique allows for a pre-selection of candidates without a significant loss in reconstruction efficiency.
Additionally, the precision of the mass peak position was improved by dividing the energy of cells, which are located between the two subclusters using the leading cell energies to calculate the fractions.
The invariant mass of splitted V1 clusters in \PbPb\ collisions in different centrality intervals is shown on the right side of \Fig{fig:Pi0SplitMass} together with the corresponding distribution in \pp\ collisions at the same center of mass energy. 
Keeping the clusterization thresholds the same in the different collision systems increases the correlated background below the pion mass peak, leaving the average peak position and width unchanged.
The increased background arises from a significantly larger underlying event contribution to each cluster and the larger cluster size in central \PbPb\ collisions.
A significant improvement of the signal-to-background ratio is observed in semi-central and peripheral collisions.
The background increase can be mitigated by raising the aggregation thresholds from about 100 MeV to 150 MeV or even 300 MeV, at the cost of a mild efficiency loss. \\
For clusters with two local maxima, a 2--3~$\sigma$ window around the fitted \piz\ meson invariant mass can be used to directly tag merged clusters as neutral pions, while an invariant mass window in between the \piz\ and $\eta$ mesons mass can simultaneously provide a background estimate. 
Additional constraints on \shshlo\ as described in Ref.~\cite{Adam:2016xbp} improve the purity even further for clusters with 2 local maxima and enable the pion identification for clusters with only one local maximum.
Clusters with more than two local maxima are normally not considered in this type of analysis.
These constraints allow to tag \piz\ meson clusters with purities of approximately 80\% in central \PbPb\ collisions and larger than 90\% in \pp\ collisions between $6<E_{\pi^{0}}<50$~GeV. 
In conclusion, with shower shape and intra-cluster splitting techniques, one can simplify event-by-event correlation analyses by directly tagging clusters stemming from \piz\ mesons up to very high energies with the \gls{EMCal}.

\subsubsection{Heavier meson reconstruction}
\label{sec:heavyMeson}
The neutral pions and $\eta$ mesons are the lightest mesons which are measured with the help of the \gls{EMCal}.
As many of these pions or $\eta$ mesons stem from decays of heavier mesons, they can be used to reconstruct their mother particles as well. 
A dominant source of decay pions is the $\omega(782)$ meson.

The $\omega$ meson predominately decays through the $\pi^0\pi^+\pi^-$ channel (\gls{BR} $89.2 \pm 0.7\%$~\cite{Tanabashi:2018oca}) .
Its charged decay products are reconstructed using the tracking detectors, while the $\pi^0$ can be reconstructed using the \gls{EMC}, \gls{PCM-EMC} or \gls{mEMC} technique up to very high momenta~\cite{Acharya:2020tjy}.
The statistics at high transverse momenta can be increased further by using the \gls{L0} or \gls{L1} \gls{EMCal} triggered samples as the momentum distributions of all three decay products are similar and thus an event trigger based on the $\pi^0$ part is possible without introducing any biases.
The neutral pion in the three-pion decay channel is measured by selecting a pair of clusters with an invariant mass within $3\sigma$ of the expected \gls{PDG} mass~\cite{Tanabashi:2018oca}. The selected neutral pion candidate is combined with two oppositely charged pions reconstructed in the \gls{TPC} and \gls{ITS} by selecting tracks of good quality which are in agreement with the expected energy loss for pions within $3 \sigma$.
Additionally, these tracks are constrained to originate from the primary vertex to suppress pileup and combinatorial background from other decays.
\begin{figure}[t]
  \centering
  \includegraphics[width=0.49\textwidth]{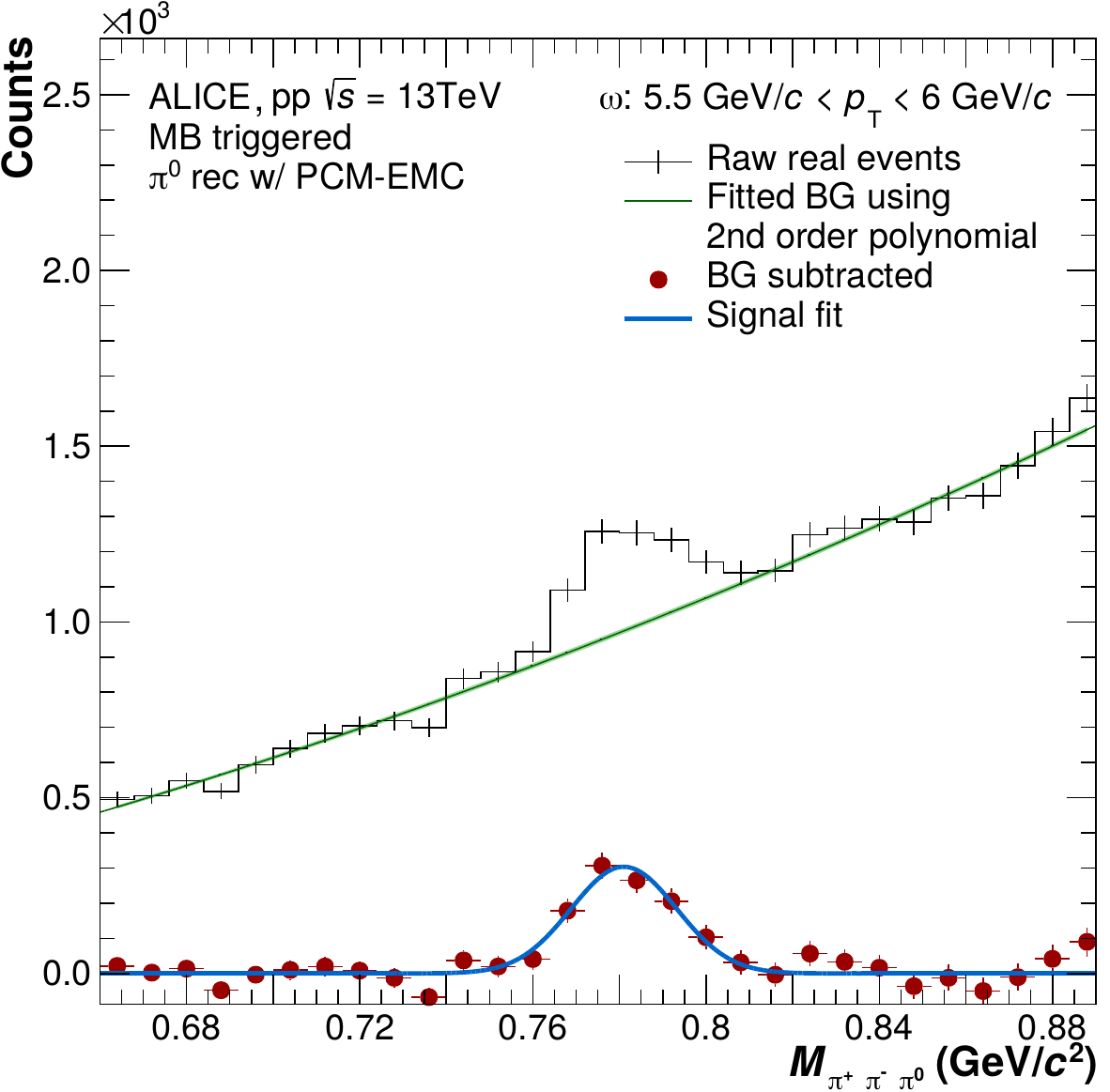}
  \includegraphics[width=0.49\textwidth]{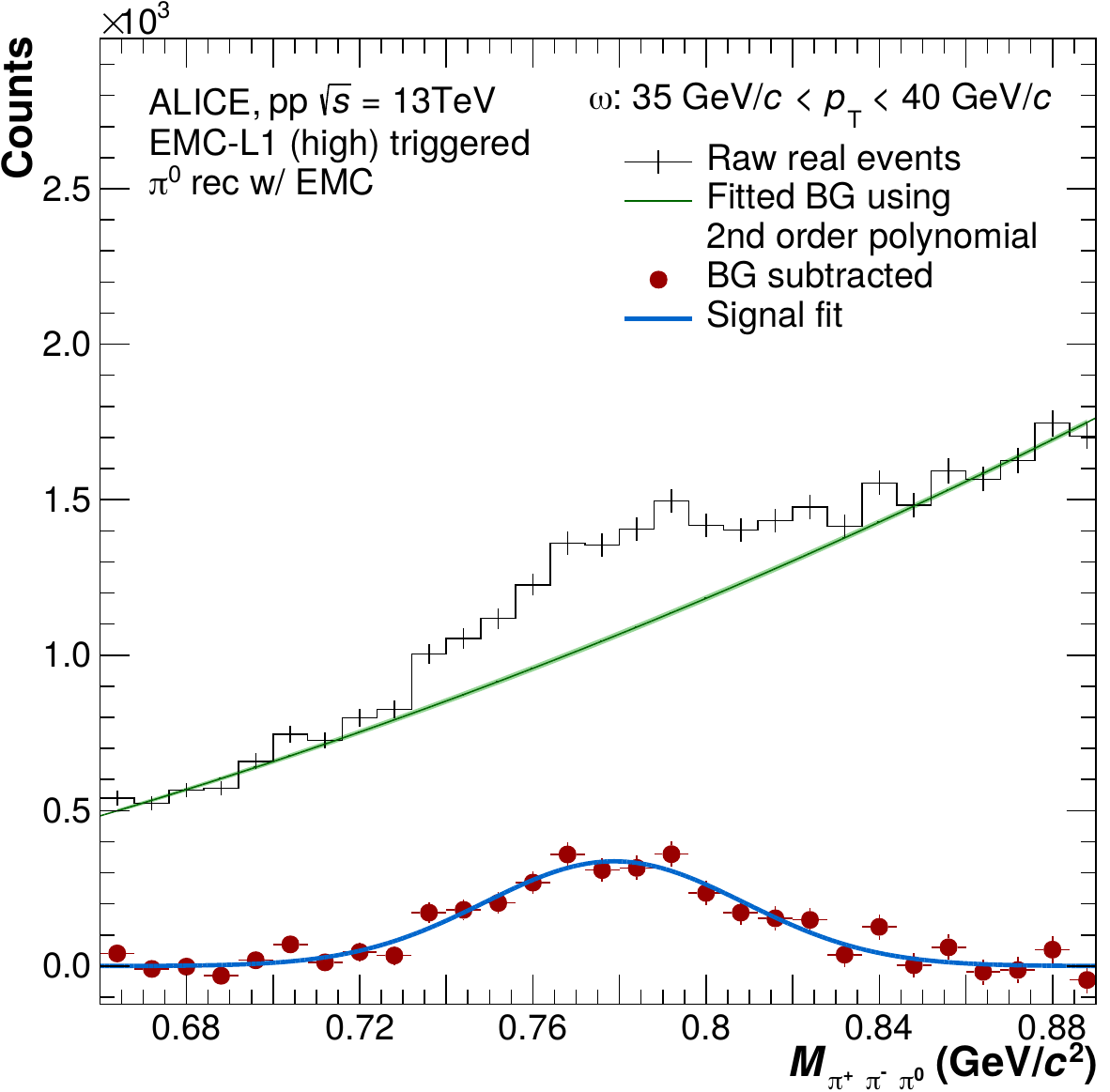}
  \caption{(Color online) Three-pion invariant mass distribution around the $\omega$ meson mass using neutral pion candidates reconstructed with the \gls{PCM-EMC} (left) and \gls{EMC} (right) reconstruction technique in two \pT{} intervals.}
  \label{fig:OmegaMass} 
\end{figure}

\begin{figure}[t]
  \centering
  \includegraphics[width=0.49\textwidth]{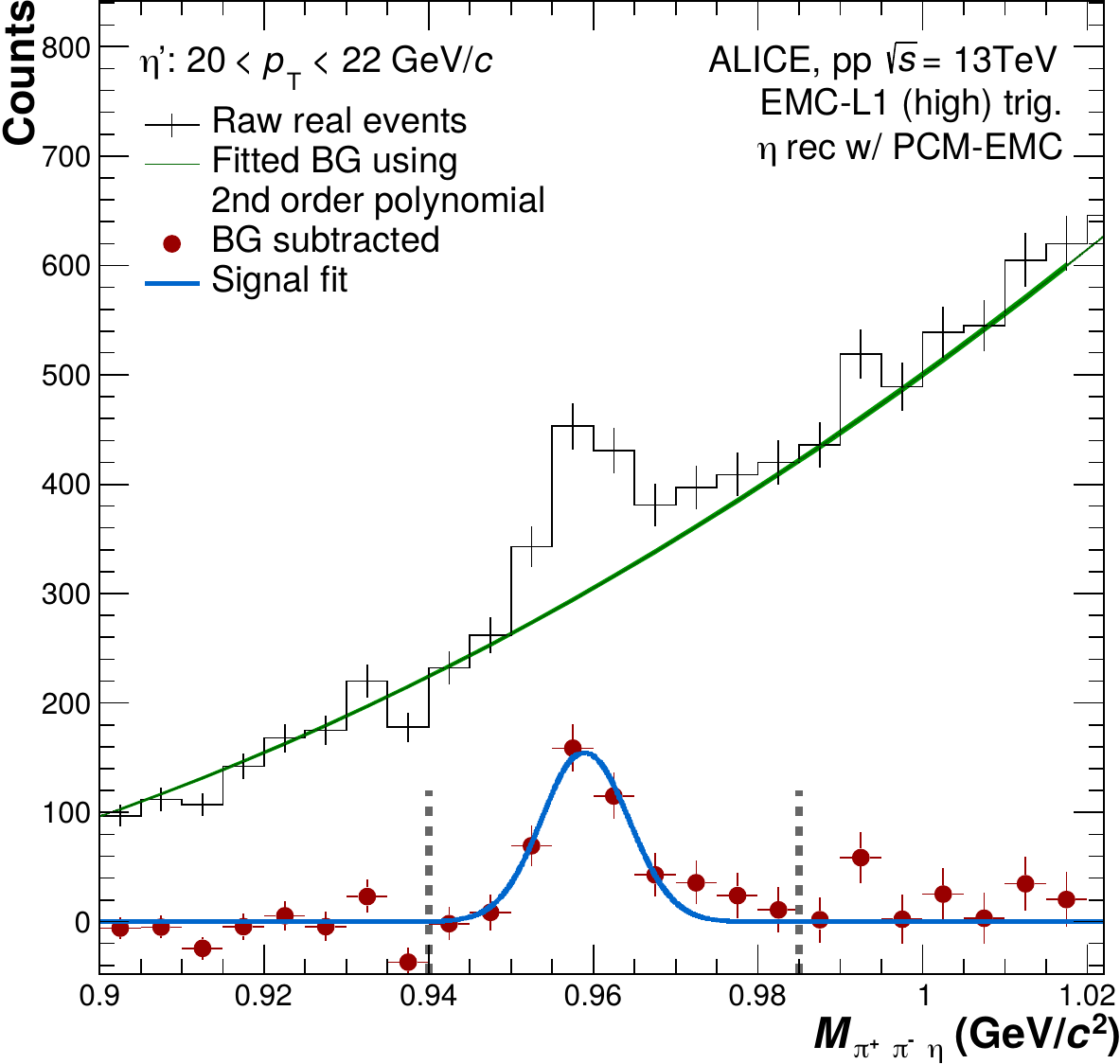}
  \includegraphics[width=0.49\textwidth]{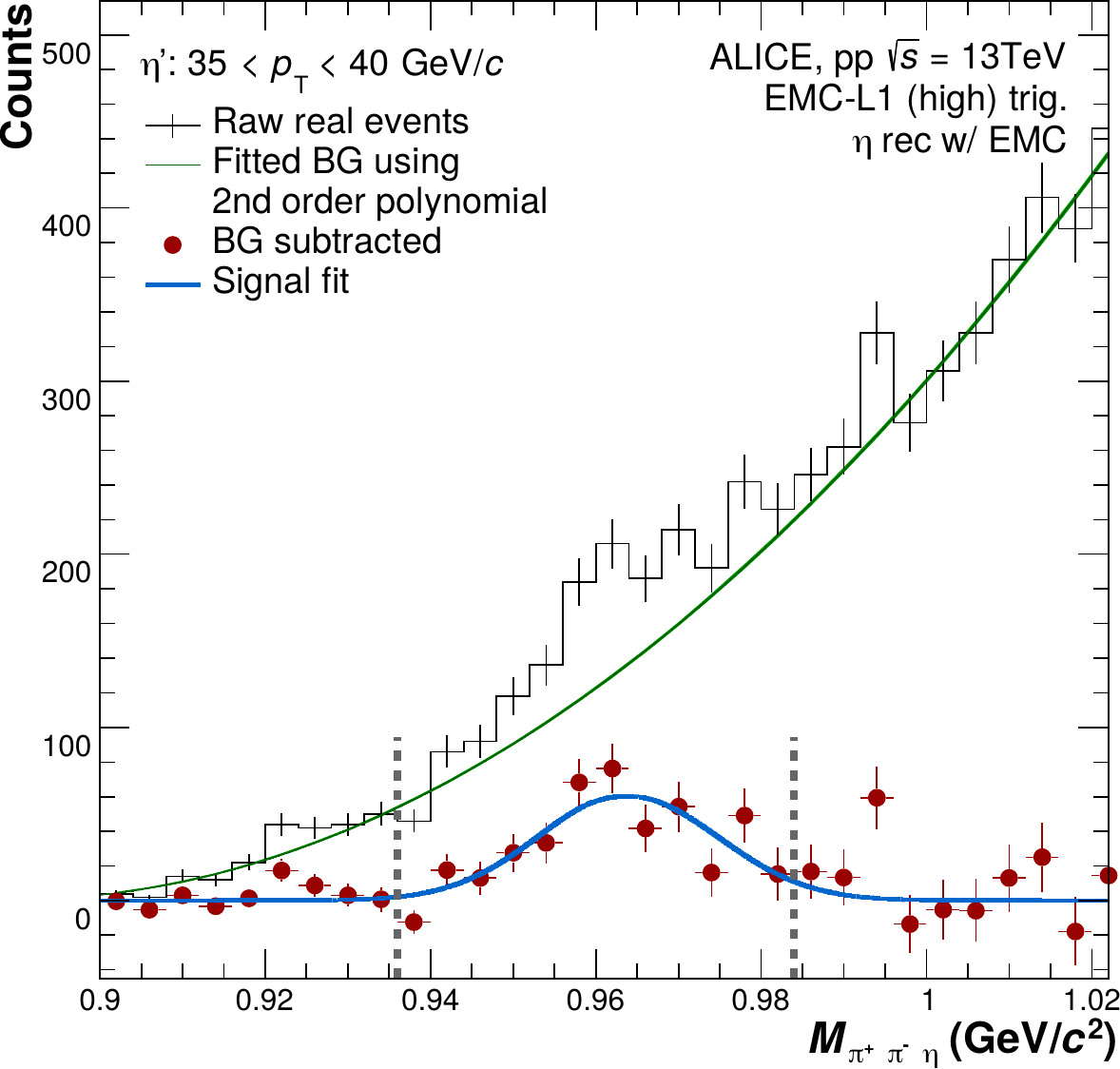}
  \caption{(Color online) Invariant mass distribution of $\pi^+\pi^-\eta$ around the $\eta'$ meson mass using $\eta$ meson candidates reconstructed with the \gls{PCM-EMC} (left) and \gls{EMC} (right) reconstruction technique in two \pT{} intervals.}
  \label{fig:EtaPrimeMass} 
\end{figure}

\Fig{fig:OmegaMass} shows example invariant mass distributions for which the neutral pions were reconstructed using either the \gls{PCM-EMC} or \gls{EMC} technique. 
The combinatorial background below the peak can be described and subtracted using a second order polynomial or exponential function.
This enables the determination of the signal yield of the $\omega$ meson for transverse momenta between $2.7$ and $28$ \GeVc\ when using pions reconstructed with the \gls{PCM-EMC} technique and $3.5$ and $40$ \GeVc\ when using the \gls{EMC} technique in pp collisions at \sthirteen.
Additionally, the $\omega$ meson could be reconstructed through the $\pi^0\gamma$ channel (\gls{BR} $8.28 \pm 0.28 \%$~\cite{Tanabashi:2018oca}) using the \gls{EMCal}. 
In this decay, reconstructing the $\gamma$ with the \gls{EMCal} is disfavored as the decay is rather asymmetric and the photon carries only little energy. 
Thus, the photon needs to be reconstructed with the \gls{PCM} technique, which introduces a reduction of the signal yield by another factor of ten due to the conversion probability of about 9\% in \gls{ALICE}. 
While the $\eta$ meson can be reconstructed in the three-pion decay channel as well, the precision of this measurements is significantly reduced as compared to the two-photon channel due to the higher combinatorial background arising from the charged pions and the smaller branching ratio.
For the $\eta'$ meson, on the other hand, the reconstruction in the two-photon channel is not favorable, but it can be reconstructed through its $\eta\pi^-\pi^+$ decay (\gls{BR} $42.9\pm0.7\%$~\cite{Tanabashi:2018oca}).
The $\eta$ meson is reconstructed in its two-photon decay channel within a $3\sigma$ window around its reconstructed mass as seen in \Figure{fig:pi0masswidth}.
Then, the mass is fixed to its \gls{PDG} value of $547.862 \pm 0.017$ MeV/$c^2$~\cite{Tanabashi:2018oca} for the combination with the oppositely charged pions. 
The corresponding examples of three-meson invariant mass distributions around the $\eta'$ meson mass reconstructed with the \gls{PCM-EMC} and \gls{EMC} techniques in two different \pT\ intervals  are shown in \Figure{fig:EtaPrimeMass} for pp collisions at \sthirteen. 
The combinatorial background is described using a polynomial or an exponential function and the signal yields can be extracted up to momenta of 40~\GeVc. \\
While so far only $\eta, \omega$  and $\eta'$ mesons reconstruction was explored, the results are promising to also use these techniques for even heavier mesons or baryons decaying to similar decay channels in particular at high momentum, where the \gls{EMCal} \gls{L0} and \gls{L1} triggers can be used to significantly enhance the sampled luminosity.
 
\subsection{Electrons}
\label{sec:electrons}
\begin{figure}[t]
  \centering
  \includegraphics[width=0.59\textwidth]{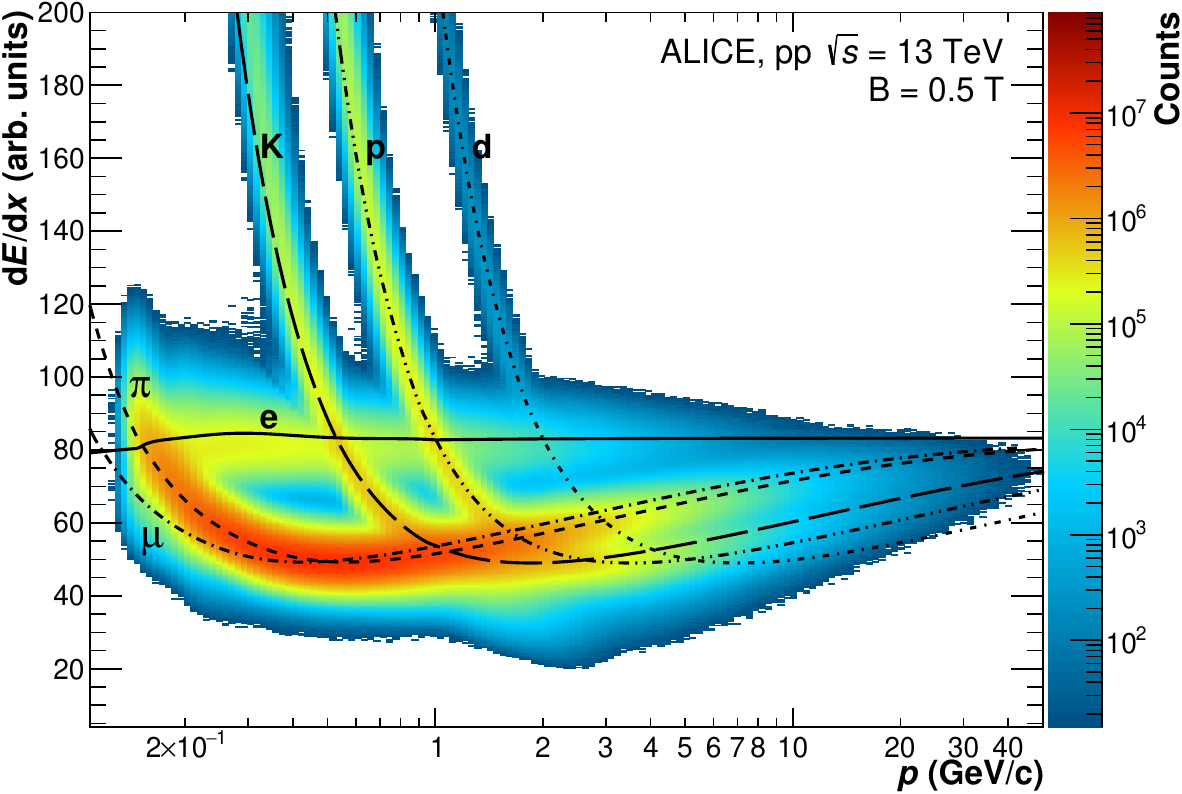}
  \caption{(Color online) d$E$/d$x$ distribution of tracks measured in the \gls{TPC} as function of the particle momentum.  }
  \label{fig:8-Elec-TPCnsig2D} 
\end{figure}

Heavy quarks (charm and beauty) are among the most important tools to study high-energy hadronic collisions. 
Due to their large masses, they are produced in hard-parton scattering processes and their production cross sections can be calculated in the framework of \gls{pQCD}~\cite{Cacciari:2012ny}.
In heavy-ion collisions, where a strongly coupled medium is formed, heavy quarks are used to study mass- and flavor-dependent interactions of partons in the hot and dense nuclear medium~\cite{Andronic:2015wma, Dokshitzer:2001zm}. 
With the ALICE central barrel detectors, heavy quarks are studied by measuring open heavy-flavor hadrons (D and B mesons) via their hadronic and semi-leptonic decay channels, and hidden heavy-flavor (J$/\psi$ and  $\Upsilon$) via their decay to e$^{+}$e$^{-}$ pairs.
The \gls{ALICE} \gls{TPC} has good electron/hadron discrimination in the region $\pT < 8~\GeVc$ as can be seen in \Fig{fig:8-Elec-TPCnsig2D}. 
At higher momenta, however, the electrons and charged pions cannot be separated efficiently as their d$E$/d$x$ signals become similar.
Using the \gls{EMCal} the electron identification capabilities were extended to up to $\pT \approx 40~\GeVc$ using the $E/p$ as discriminator,
where $E$ is the energy of the \gls{EMCal} ~cluster matched to the track and $p$ is the track momentum reconstructed with the \gls{ITS} and \gls{TPC}.
The discriminative power for electrons and positrons has proven to be identical and thus in the following discussion, when electrons are mentioned, the same consideration applies to positrons as well.
These extended \gls{PID} capabilities achieved via the inclusion of the \gls{EMCal} also improved the performance of the J$/\psi(\rightarrow$ e$^{+}$e$^{-})$ reconstruction at high $\pT$ by reducing the combinatorial background from misidentified pions.\\
A large improvement in the \pT\ reach for the respective measurements was also achieved by using various \gls{EMCal} single-shower triggers (\Sec{sec:trigger}) in \pp, \pPb\ and \PbPb\ collisions~\cite{Abelev:2012xe,Abelev:2014gla,Adam:2015qda,Acharya:2018upq,Acharya:2019hao,Adam:2016khe,Acharya:2019mom,Adam:2016ssk}.
By exploiting the full luminosity of these triggers, in particular for the \pp\ sample at \sthirteen, the \pT\ range could be nearly tripled compared to the corresponding minimum bias measurement.
The usage of the \gls{EMCal} L1 triggers also allowed for the full reconstruction of hadronic decays of the D*$^{+}$ meson for $\pT > 60$~\GeVc, by triggering on one of its decay products.
In the following section, the performance of heavy-flavor measurements using the \gls{EMCal} is demonstrated based on the full data set collected in \pp\ collisions at \sthirteen\ in the years 2016-2018 and the data collected in \PbPb\ collisions at \sfivelead\ in 2015~\cite{Adam:2016khe,Acharya:2019mom}.

\subsubsection{Electron identification} \label{sec:electronID}
\begin{figure}[t]
  \centering
  \includegraphics[width=0.32\textwidth]{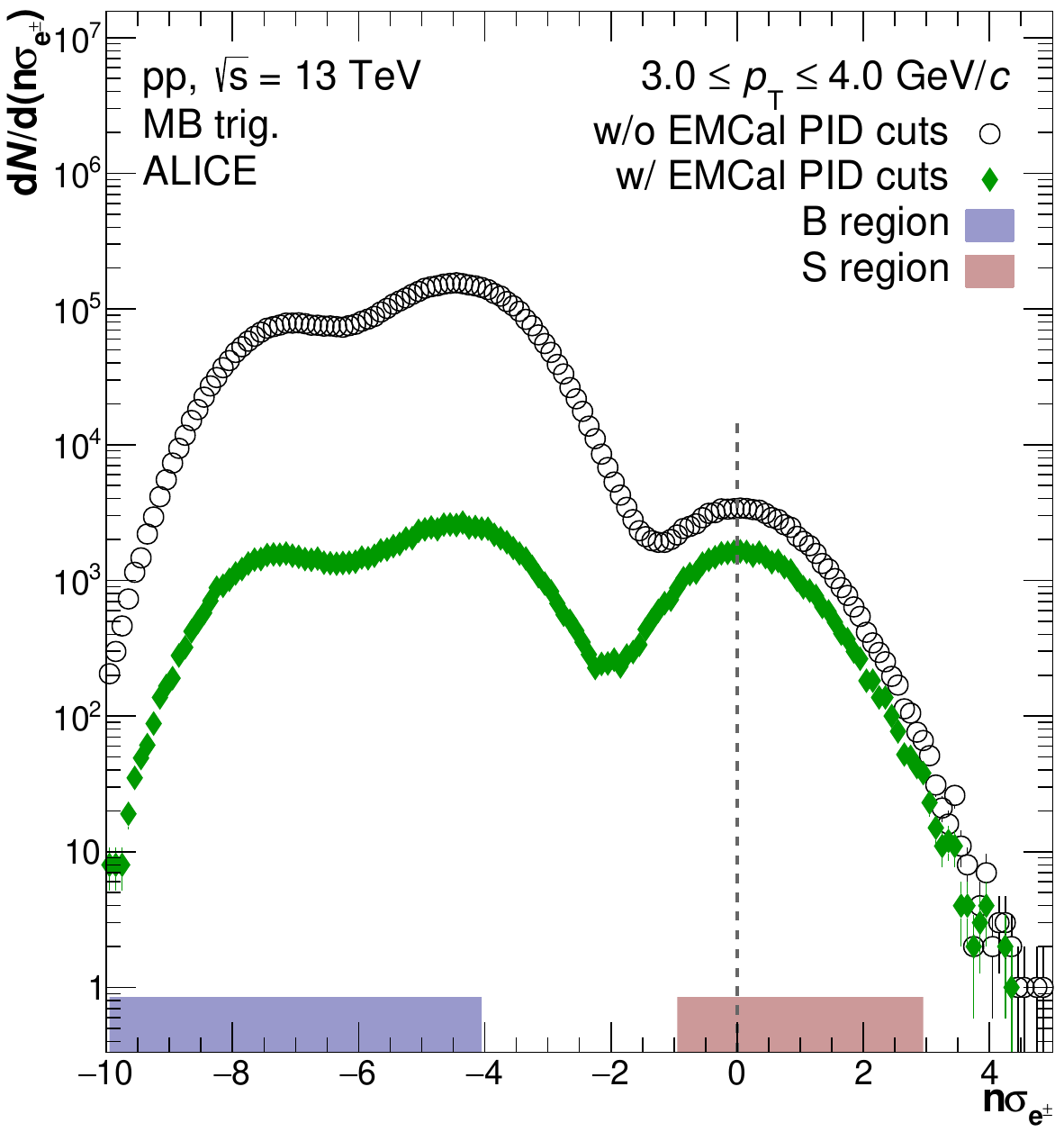} 
  \includegraphics[width=0.32\textwidth]{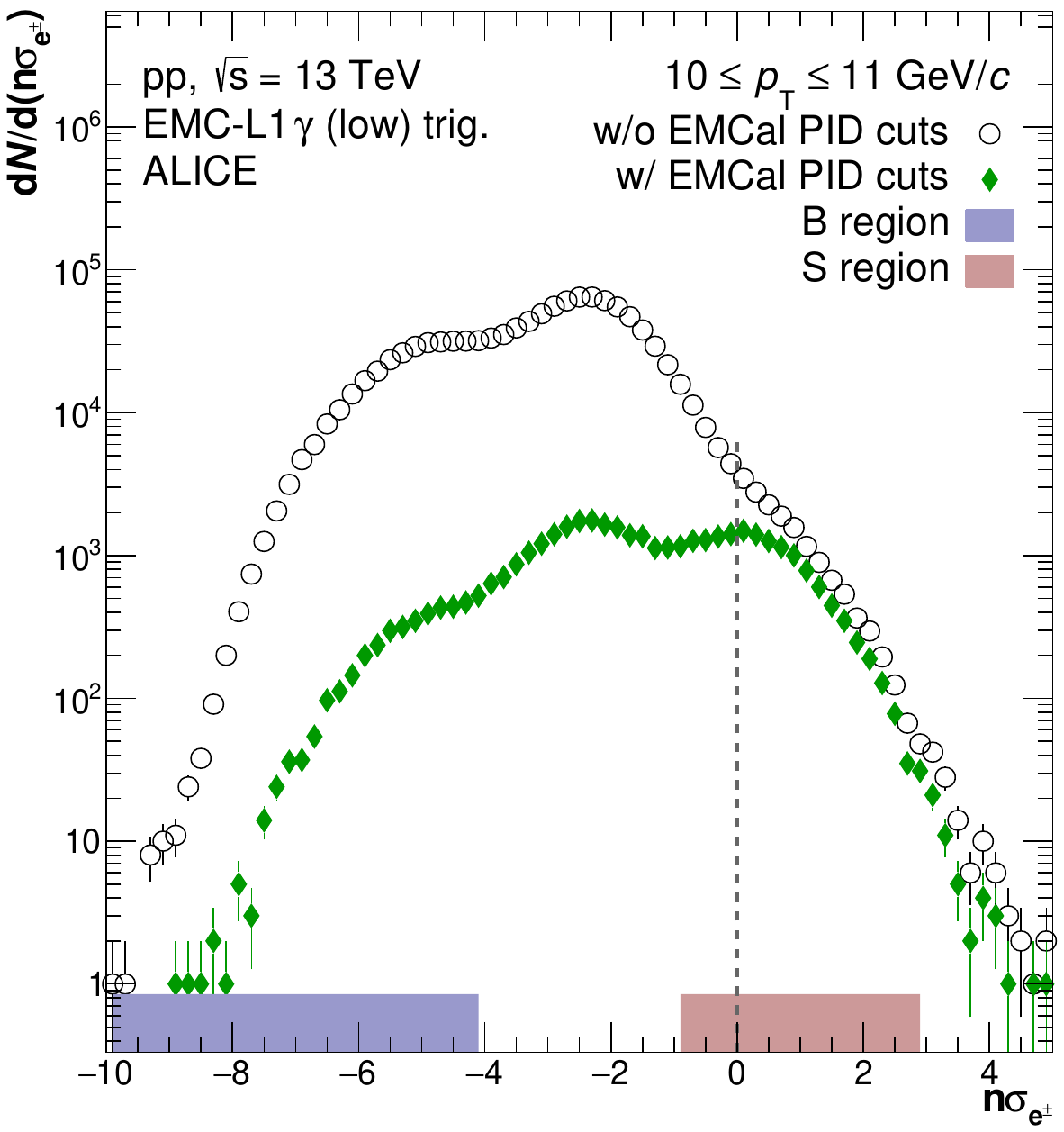} 
  \includegraphics[width=0.32\textwidth]{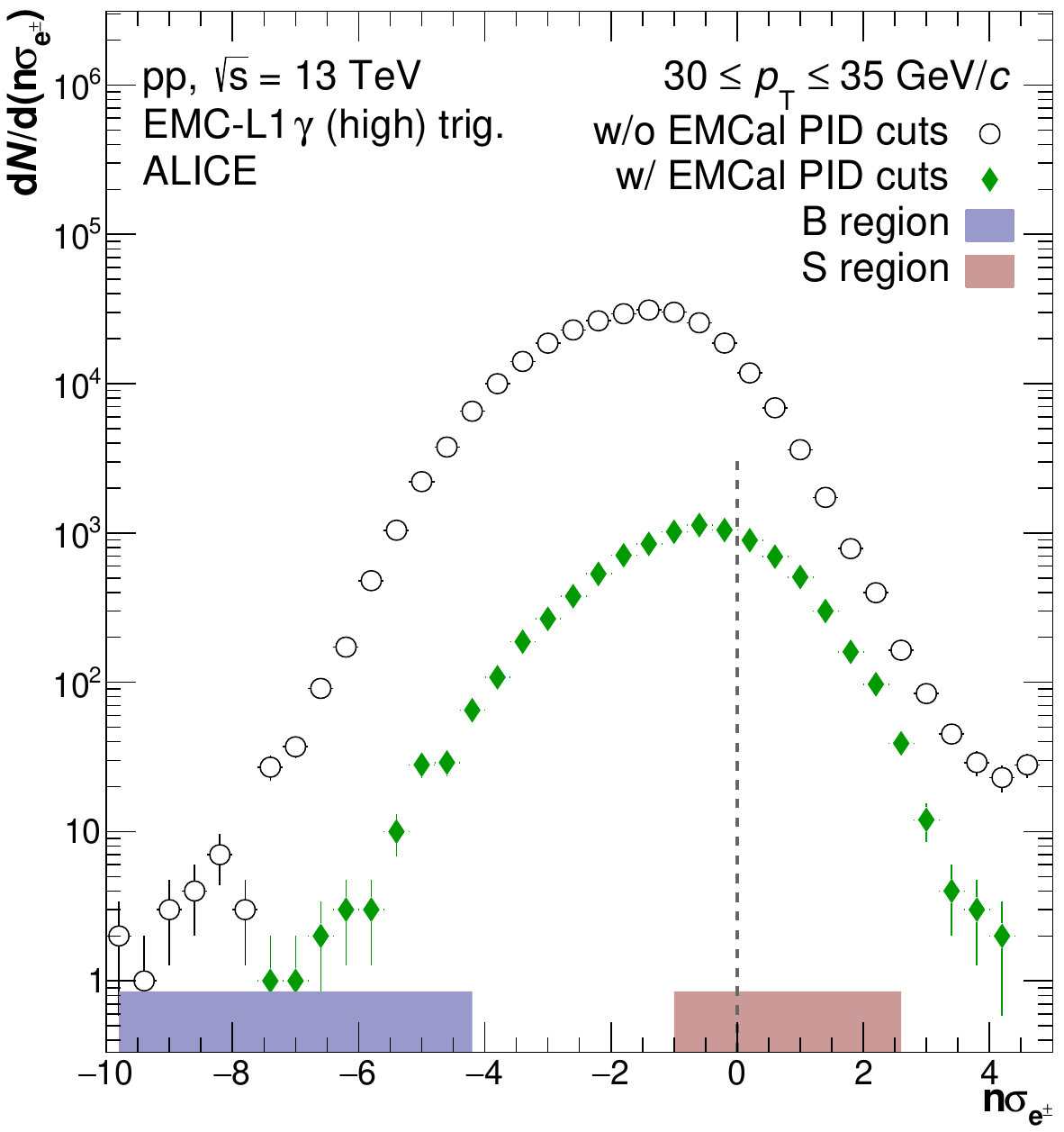} 
  \caption{$n\sigma^{\textrm{TPC}}_{e^\pm}$ distribution without (black) and with (green) \gls{EMCal} electron identification cuts of $E/p$ and $\shshlo$. Electrons form a Gaussian distribution centered around zero, indicated by the gray dashed line. The signal and background selection windows considered for the following $E/p$ plots are indicated by the red and blue shaded area, respectively. The distributions are shown for various event triggers in \pp\ collisions at \sthirteen\ in different \pT\ intervals.}
  \label{fig:8-Elec-TPCnsig1D} 
  \includegraphics[width=0.32\textwidth]{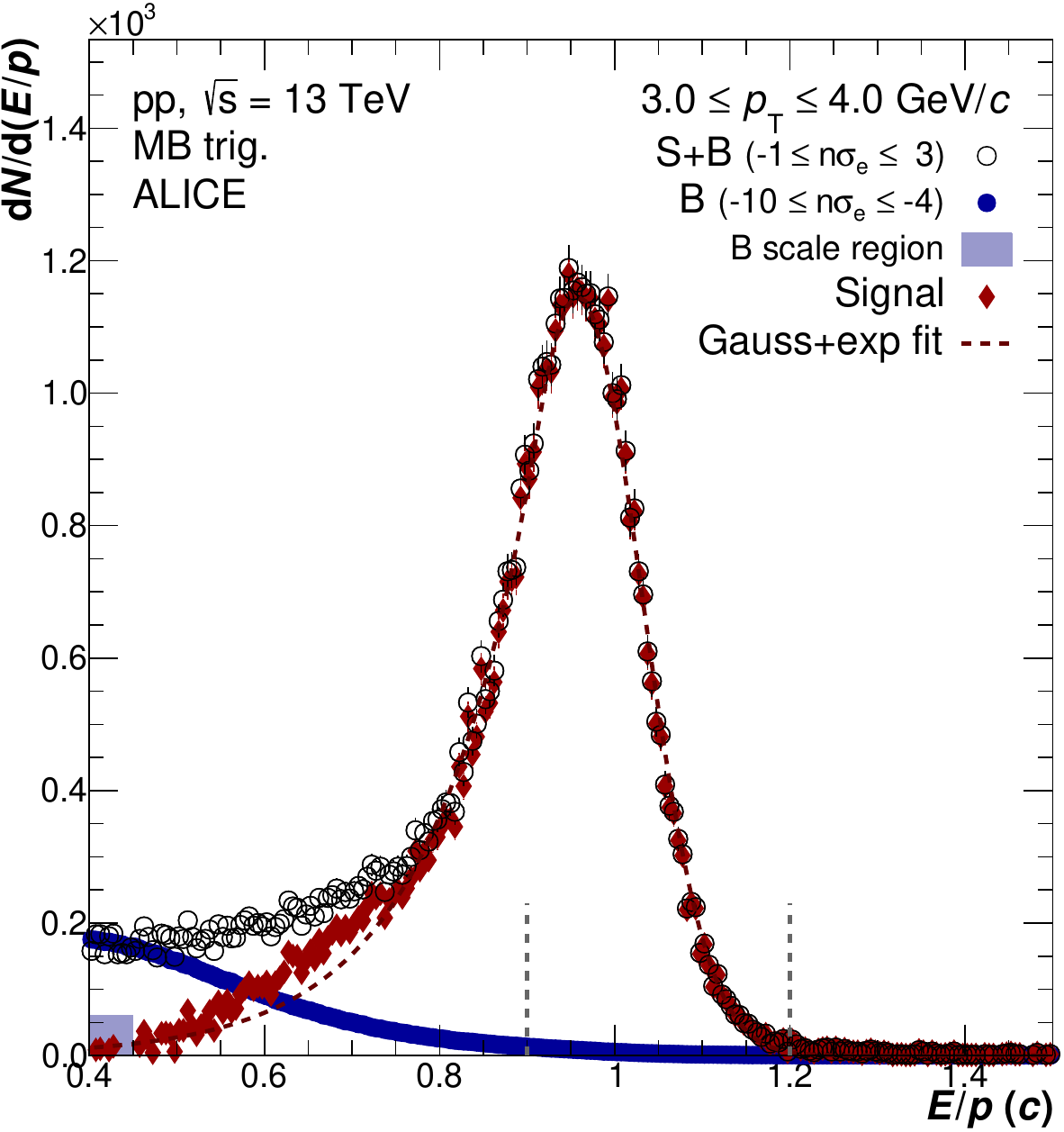} 
  \includegraphics[width=0.32\textwidth]{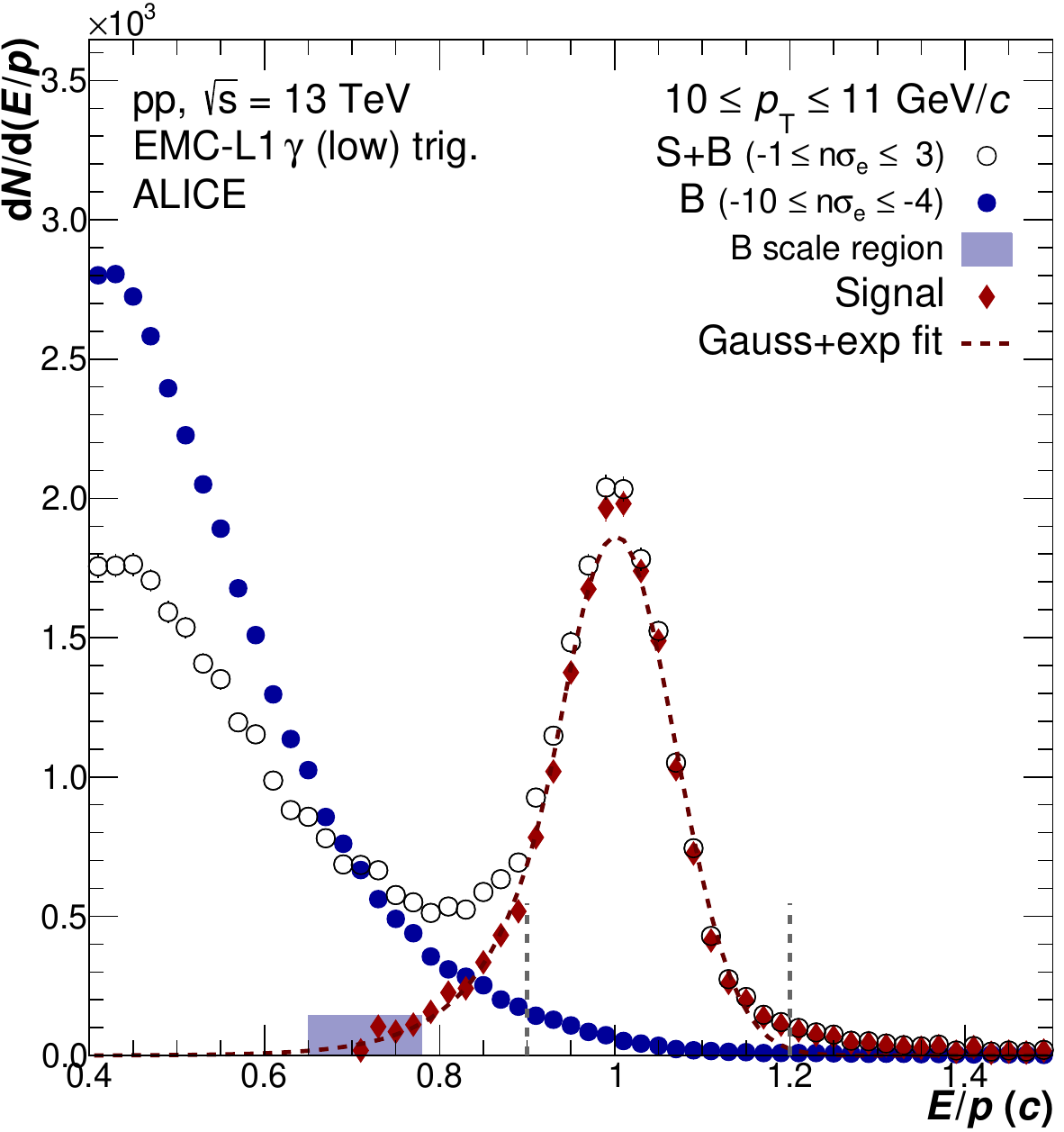} 
  \includegraphics[width=0.32\textwidth]{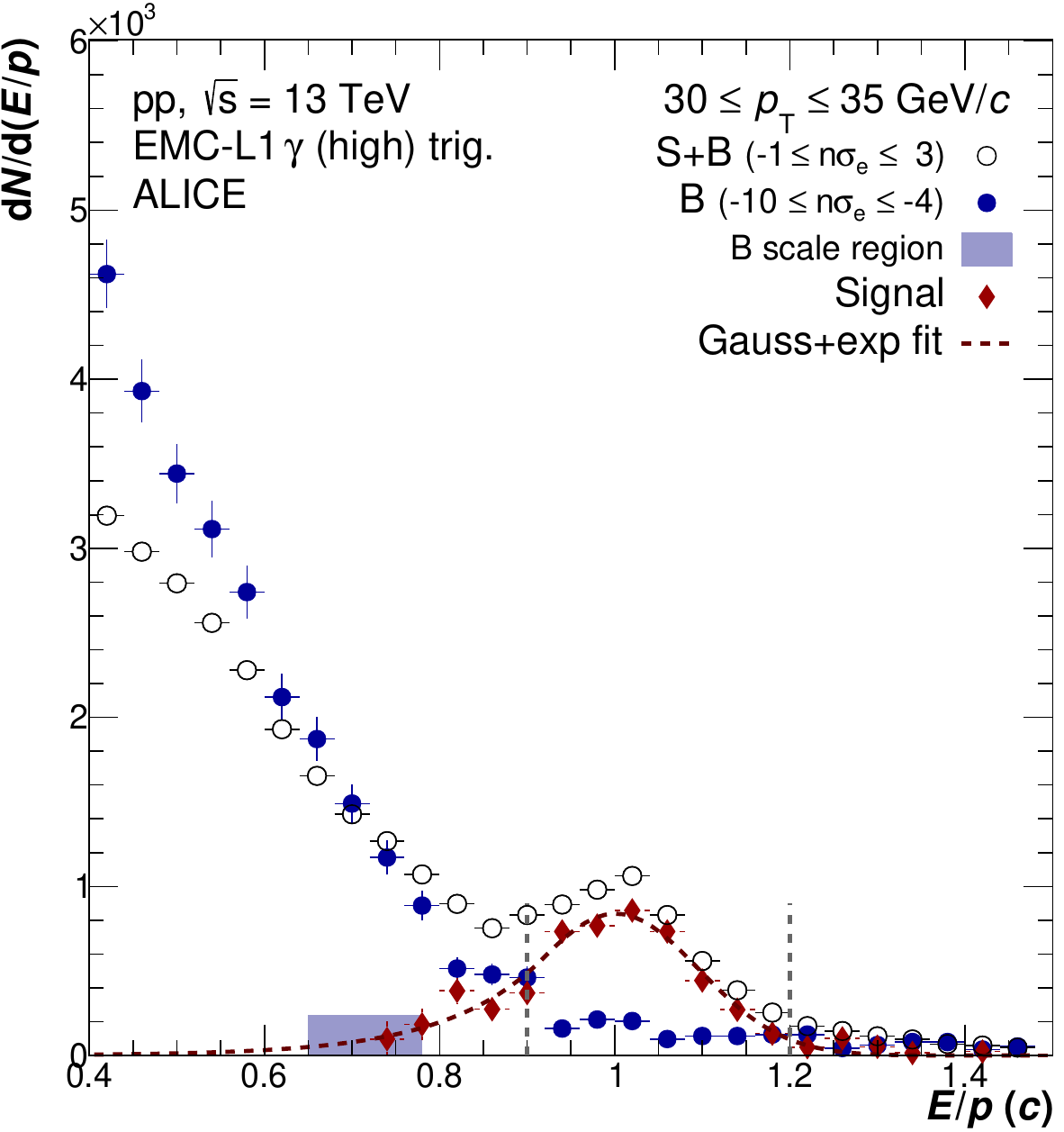} 
  \caption{(Color online) $E/p$ distribution for electron candidates selected by applying $-1 \leq n\sigma^{\textrm{TPC}}_{e^\pm} \leq 3$ (black open circles), and for hadrons with $-10 \leq n\sigma^{\textrm{TPC}}_{e^\pm} \leq -4$ (blue dots) scaled to match the electron distribution in the range indicated by the blue shaded box. The red diamonds reflect the remaining signal distribution after background subtraction and the corresponding signal fit using a Gaussian with an exponential tail is overlaid in dark red. The distributions are shown for various event triggers in \pp\ collisions at \sthirteen\ in different \pT\ intervals. }
  \label{fig:8-Elec-EovP} 
  
\end{figure}

To reconstruct electrons originating close to the primary vertex, fully reconstructed charged tracks based on \gls{ITS}-\gls{TPC} tracking are selected.
These tracks have to pass standard quality selections that furthermore ensure they do not originate from a conversion or other secondary interactions~\cite{Adam:2015qda}.
Moreover, a loose electron identification is employed by selecting tracks with the specific ionisation energy loss inside the \gls{TPC} of $-1 \leq n\sigma^{\textrm{TPC}}_{e^\pm} \leq 3$, where $n\sigma^{\textrm{TPC}}_{e^\pm}$ is the difference between the measured and expected detector response signals (d$E/$d$x$) for electrons normalised to the response resolution. 
Tracks are geometrically matched to the clusters reconstructed in the \gls{EMCal} ~along $\eta$ and $\varphi$~(see \Sec{sec:trackmatch}). 
Tracks that are matched to clusters within $\Delta\eta < 0.01$ and $\Delta\varphi < 0.01$ rad are selected. 
\Figure{fig:8-Elec-TPCnsig1D} shows the $n\sigma^{\textrm{TPC}}_{e^\pm}$ distribution for the selected track sample with an associated cluster in the \gls{EMCal} for different transverse momentum intervals and triggers in \pp\ collisions at \sthirteen.
The black points represent the distribution for tracks without further identification criteria based on the \gls{EMCal} cluster, and the signal and background regions considered for the further \gls{EMCal} measurements are indicated by the shaded red and blue bands, repectively.
For low momenta, electrons appear to be well separated in the $n\sigma^{\textrm{TPC}}_{e^\pm}$ distribution and centered around 0, as indicated by the gray dashed line.
A multi-Gaussian fit allows the statistical extraction of the corresponding electron yield. 
Above $\pT = 8$~\GeVc, the pion and electron responses are close to each other and the electron signal extraction becomes increasingly difficult until it fully breaks down at about 15 \GeVc.
In this critical region and above these transverse momenta, the electron identification based on the \gls{EMCal} $E/p$ distribution significantly improves the signal extraction.
Tracks with $E/p$ around 1 are identified as electrons using the \gls{EMCal} response. 
Hadrons have a lower $E/p$ ratio as they deposit only a fraction of their initial energy in the \gls{EMCal}. 
In addition to the $E/p$ ratio, the \gls{EMCal} cluster shape (see \Sec{sec:showershape}) is used to further improve the purity of the electron sample. 
The dispersion along the long axis of the cluster is required to be in the range $0.05 < \shshlo < 0.9$ for low \pt\ ($<15$ \GeVc) and $0.05 < \shshlo< 0.6$ for higher \pt. 
\Figure{fig:8-Elec-EovP} shows the $E/p$ distribution of electron candidates after applying the $n\sigma^{\textrm{TPC}}_{e^\pm}$ and $\shshlo$ selections for \pp ~collisions at \sthirteen. 
The distribution for $3 < \pT< 4$ \GeVc ~is obtained using the minimum bias triggered sample, while the \gls{EMCal} triggered data sample with low and high trigger threshold is used for $10 < \pT< 11$ \GeVc ~and $30 < \pT< 35$ \GeVc, respectively. 
The electron peak at $E/p$ around 1 is clearly visible up to 35 \GeVc. 
Extending the electron identification capabilities beyond 8--10~\GeVc\ is only possible by using the \gls{EMCal} triggered data sample and by applying the \gls{EMCal} electron identification criteria to better discriminate electron and hadron candidates. \\
We estimate the hadron contamination of the electron sample in the data by measuring the $E/p$ for hadrons with $-10 \leq n\sigma^{\textrm{TPC}}_{e^\pm} \leq -4$ (blue shaded regions in \Fig{fig:8-Elec-TPCnsig2D}). 
The hadron $E/p$ distribution is scaled to match the electron candidate's $E/p$ distribution in the blue shaded regions shown in \Fig{fig:8-Elec-EovP}. 
Subsequently, we obtain the electron yield by integrating the $E/p$ distribution for $0.9 \leq E/p \leq 1.2$ and subtracting the hadron contamination statistically. 
The improvement in the discrimination power between electrons and hadrons using the \gls{EMCal} \gls{PID} cuts based on the shower shape and $E/p$ selection is demonstrated in \Fig{fig:8-Elec-TPCnsig1D} by the green markers.
In the lowest transverse momentum bin, where the separation based on the $n\sigma^{\textrm{TPC}}_{e^\pm}$ is performing well, the additional constraints provided by the \gls{EMCal} can further improve the signal-to-background in the vicinity of the Gaussian for the electrons centered around 0. 
Furthermore, it can be seen that at lower transverse momenta also the hadron contribution at negative $n\sigma^{\textrm{TPC}}_{e^\pm}$ is significantly suppressed.
At intermediate \pT\ ($10 < \pT < 11$ \GeVc), where the electron peak in $n\sigma^{\textrm{TPC}}_{e^\pm}$ is not clearly visible due to the large hadron background and the \gls{TPC} reaches its separation limit, the electron identification based on the \gls{EMCal} starts to outperform the identification purely based on the \gls{TPC} signal.
For even higher transverse momenta, electrons can no longer be identified using the \gls{TPC} d$E/$d$x$ and only a single broad distribution centered at $n\sigma^{\textrm{TPC}}_{e^\pm} \approx -2$ is visible.
After applying the additional \gls{EMCal} selection criteria, the mean shifts to about 0.5 implying that a large fraction of the contamination could be rejected.
Thus, the joint \gls{PID} using the \gls{TPC} and \gls{EMCal} capabilities enables the measurement of electrons over a wide transverse momentum range.
The usage of the \gls{L1} single shower triggers further aids the extension of the transverse momentum reach.

\begin{figure}[t]
  \centering
  \begin{tabular}{p{0.45\textwidth} p{0.5\textwidth}}
    \vspace{0pt} \includegraphics[height=6.7cm]{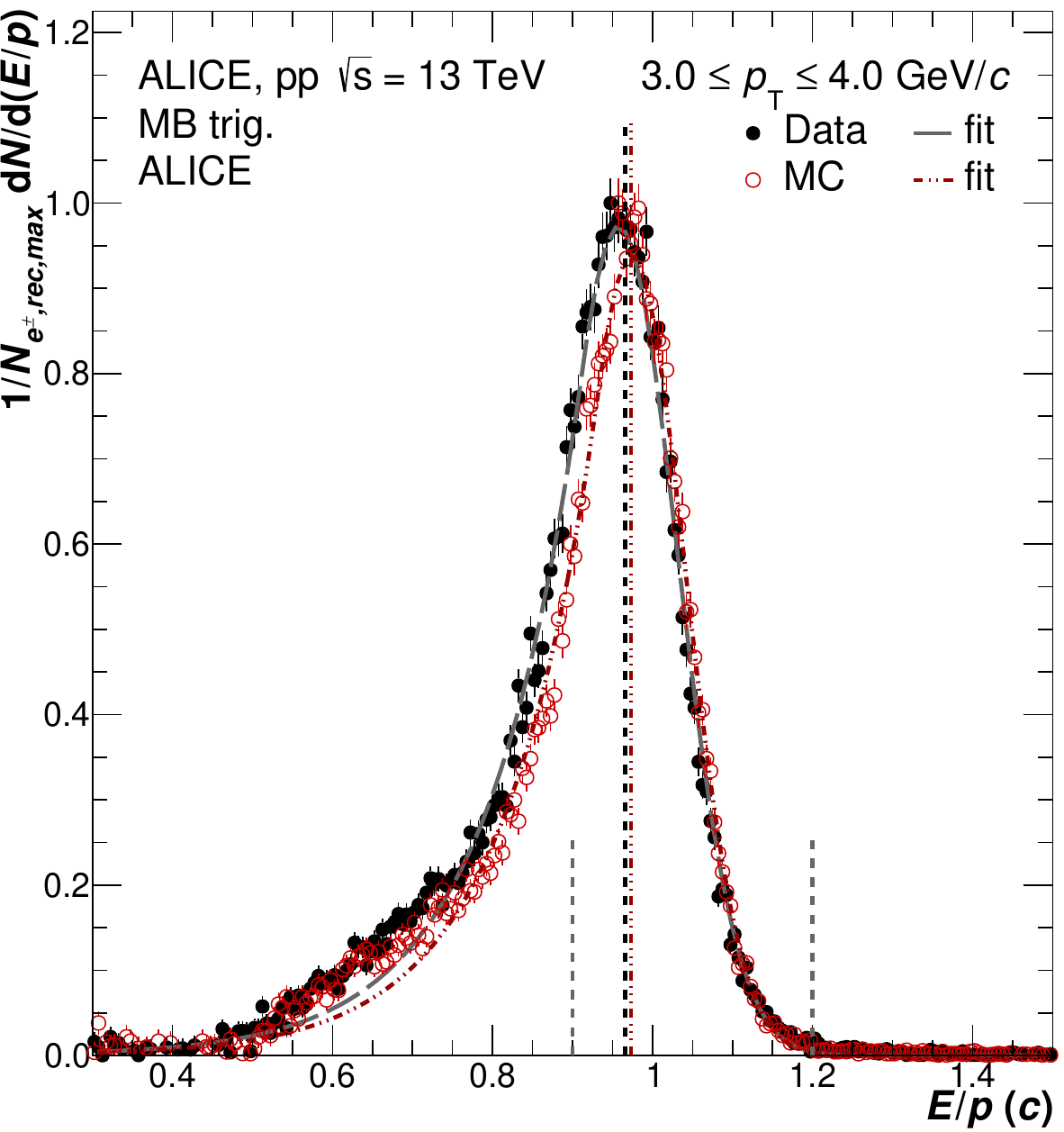} & 
    \vspace{0pt} \includegraphics[height=6.9cm]{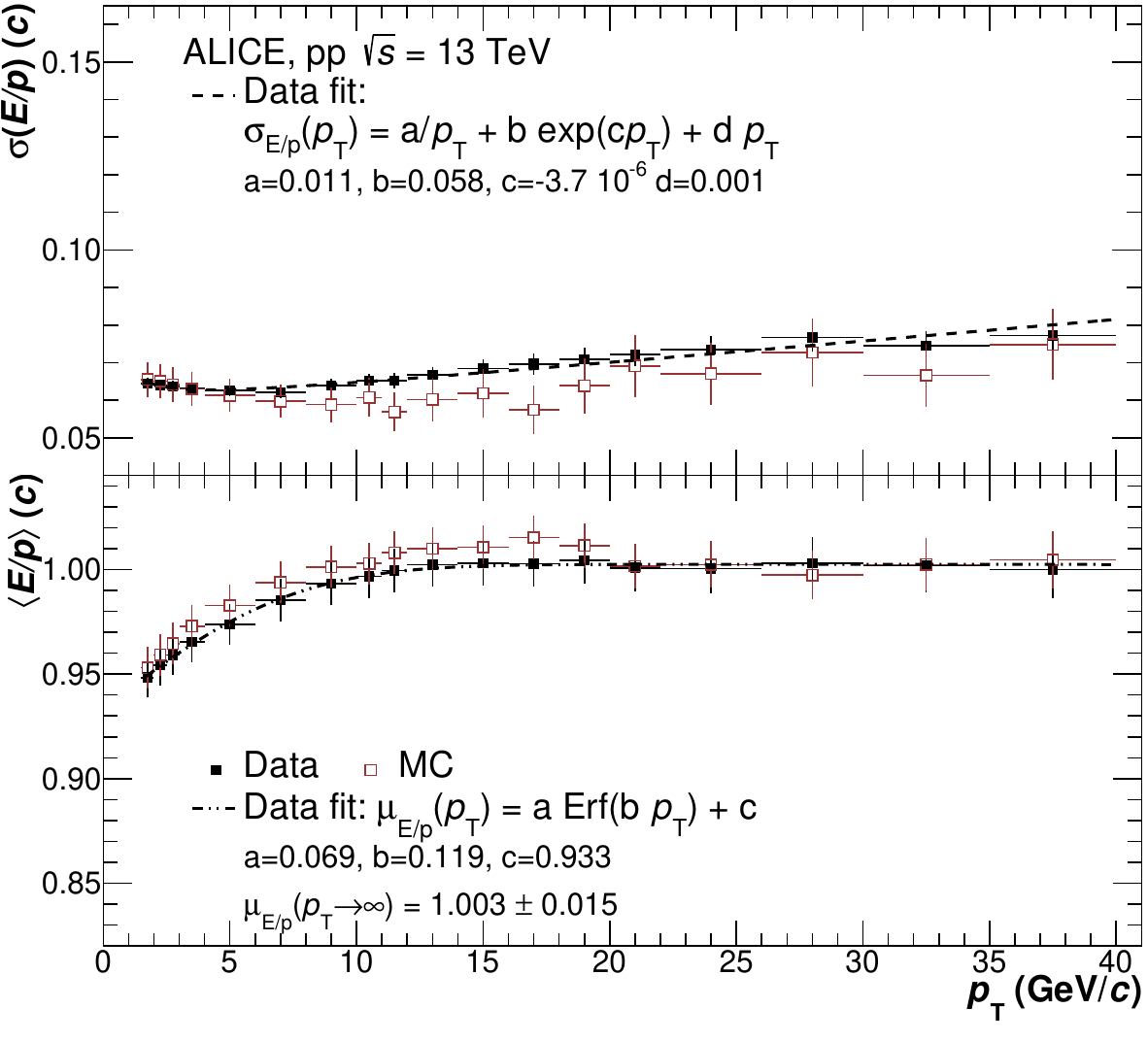} 
  \end{tabular}

  \caption{(Color online) Left: Comparison of the $E/p$ distribution between real (black) and simulated (red) data for electron candidates with $3.0 < \pT < 4$ \GeVc\ in pp collisions at \sthirteen.
            Gaussian fits with exponential tails to both sides are superimposed for both distributions in the corresponding color. 
            The truncated means for data and \gls{MC} are indicated by vertical colored lines.
            The truncation and signal integration window is indicated by vertical dashed gray lines.
            Right: Data and simulation comparison of width (top) and mean (bottom) of the $E/p$ distributions of electrons as a function of transverse momentum in \pp\ collisions at \sthirteen. 
            Fit functions are shown as dashed lines, where their functional form is given in the respective legends. 
            }
  \label{fig:8-Elec-EovPdataMCvsPt} 
\end{figure}

The \gls{MC} description of the \gls{EMCal} performance for electron identification is demonstrated by comparing the $E/p$ distribution of electrons after subtracting the estimated hadron contributions in data and simulations.
\Figure{fig:8-Elec-EovPdataMCvsPt}{ (left)} shows the corresponding comparison for electron candidates in data and simulation taken from the minimum bias sample in pp collisions at \sthirteen. 
In addition, Gaussian fits with exponential tails on both sides are superimposed for data and MC.
A slight difference between the fits and the distributions can be observed, mainly at low $E/p$ below the peak.
This is partially caused by the fact that the distributions are obtained using the track momentum evaluated at the first track point or the \gls{PCA} to the primary vertex. 
Further energy losses, e.g.~due to Bremsstrahlung when passing through the \gls{TRD}, are not accounted for.
Instead of using the fits to evaluate mean and $\sigma$ of the $E/p$ distributions, the use of truncated mean and the standard deviation of the distributions was found to be more accurate and stable, especially at  high \pT. 
A comparison of the truncated mean and width of electron $E/p$ peaks is displayed in \Fig{fig:8-Elec-EovPdataMCvsPt}{ (right)}.
The uncertainties shown in the figure reflect the statistical uncertainties as well as the systematic ones arising from variations of the truncation window.
The mean of the $E/p$ distribution in data is reproduced in the simulations within less than $1\%$. 
The \pT-dependence of the mean can be parametrized using an error function.
It is found to converge to $E/p\approx 1$ at high transverse momenta, as expected for electrons.
Even though the width in data and \gls{MC} agrees within uncertainties, the peaks appear to be systematically narrower in the simulations, due to the more expressed tail at lower $E/p$ observed in the data.
This tail can be attributed to the aforementioned energy loss in the detector material in front of the \gls{EMCal} ~detector which is not fully reproduced in the simulations.\\
The increase in the $E/p$ resolution at high momenta is driven by the momentum resolution of the tracks.
It can be concluded that the electron $E/p$ is reasonably well described by the \gls{MC}, and the residual differences below 1\% were found to be negligible on the level of electron reconstruction efficiencies.

\begin{figure}[t!]
  \centering
  \includegraphics[width=0.49\textwidth]{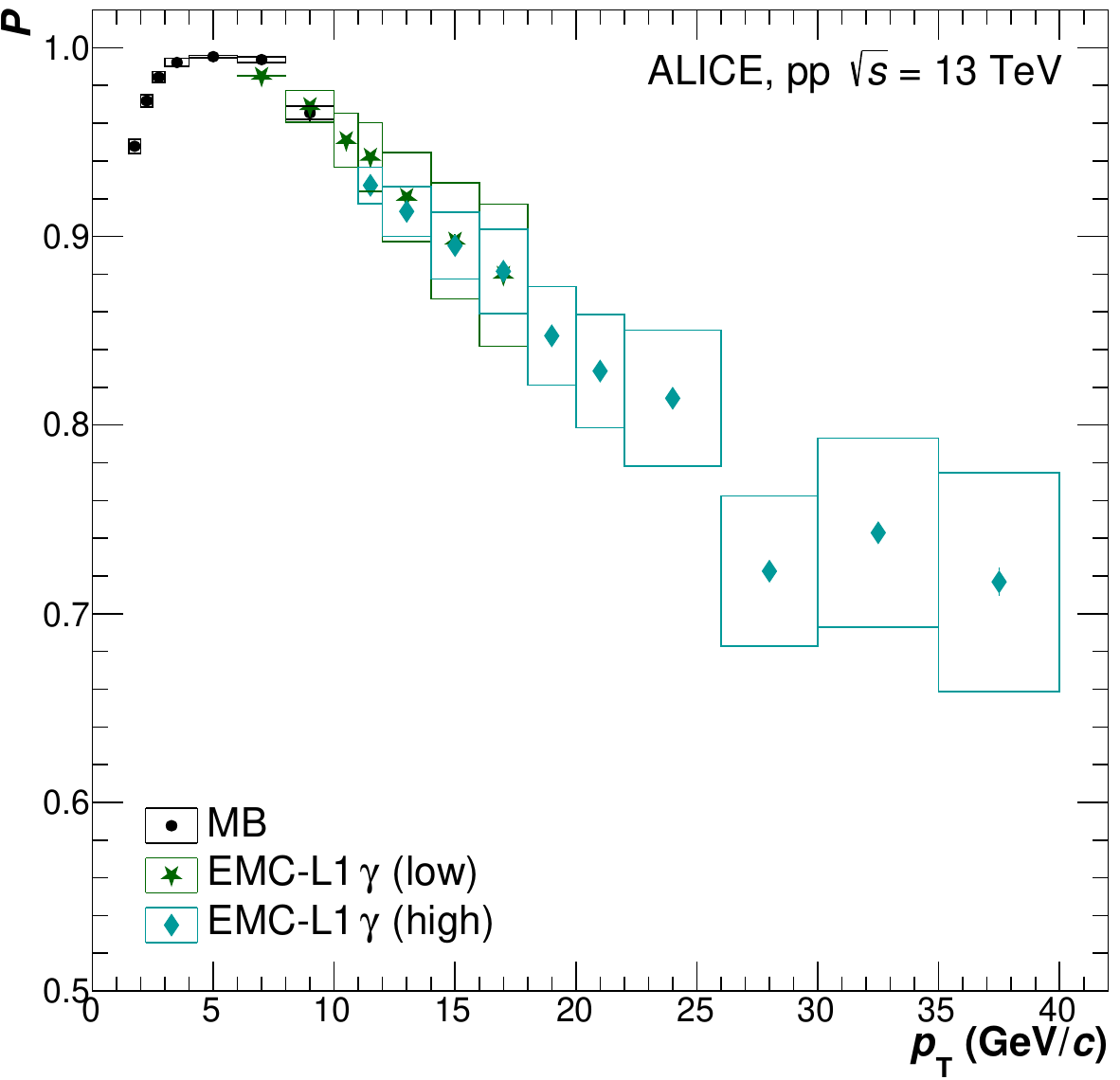} 
  \includegraphics[width=0.49\textwidth]{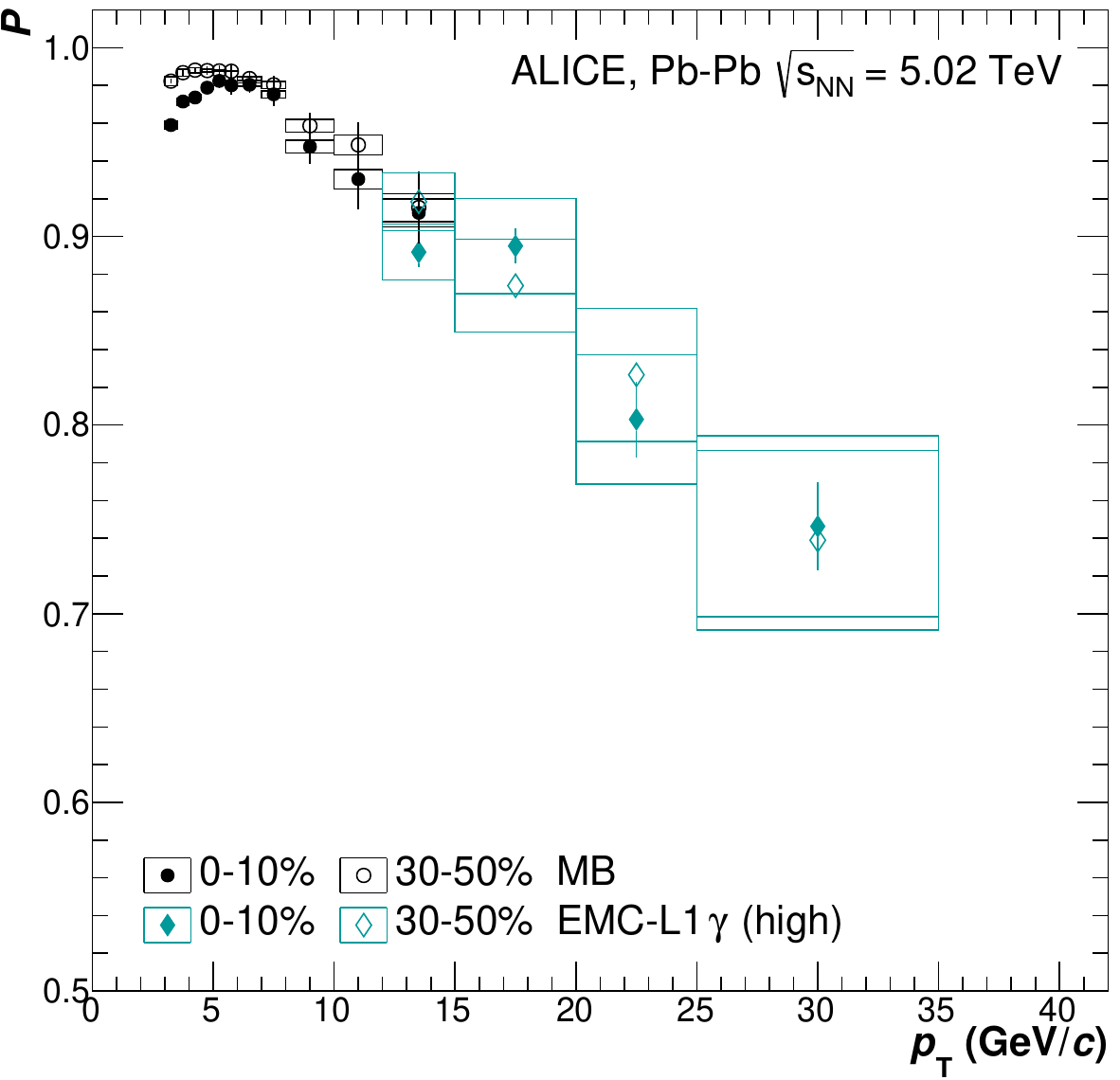} 
  \caption{(Color online) Data-driven purity ($P$) estimates of the electron sample after applying \gls{TPC} and \gls{EMCal} electron identification criteria as a function of \pT\ in \pp\ collisions at \sthirteen\ (left) and in \PbPb\ collisions at \sfivelead\ (right). The boxes indicate the systematic uncertainty arising from the use of different scaling ranges for the hadronic background.}
  \label{fig:8-Elec-purity} 
\end{figure}
The data-driven purity estimate of the electron sample obtained after applying the $n\sigma^{\textrm{TPC}}_{e^\pm}$, $\shshlo$ and $E/p$ selection criteria in \pp ~collisions at \sthirteen\ is shown  as a function of \pT\ in \Fig{fig:8-Elec-purity}.
It is obtained via the ratio of the signal distribution shown in red in \Fig{fig:8-Elec-EovP} over all candidates in the $n\sigma^{\textrm{TPC}}$ signal region (open black), calculated within the indicated $E/p$ range.
The purity in \PbPb\ collisions at \sfivelead\ was obtained with a loosened $E/p$ selection of $0.9\leq E/p\leq 1.3$, in order to account for shower overlaps which were found to shift the $E/p$ signal to higher values, especially for most-central collisions.
With the information from the \gls{EMCal} detector, an electron sample with purity $> 90\%$ is obtained up to $\pT \sim 15$ \GeVc ~in pp collisions. 
The validity of the data-driven method was tested using \gls{MC} treated as data, where one finds agreement of the extracted purity with the true purity of the simulated sample within uncertainties. 
The boxes indicate the systematic uncertainty arising from the use of different scaling ranges for the hadronic background.
In \PbPb\ collisions, with a higher multiplicity and hence higher hadron contamination, the purity of the electron sample is $> 90\%$ up to $\pT \sim 10$ \GeVc ~and $> 80\%$ up to $\pT \sim 20$ \GeVc, as shown in \Fig{fig:8-Elec-purity}~(right).  
In addition to improving the purity at high transverse momenta with respect to an analysis purely based on the tracking, the \gls{EMCal} triggered data in parallel allows to extend the \pT\ range in all collision systems. 
The corresponding extension of the \pT\ range for \pp\ collisions at \sthirteen\ and \PbPb\ \sfivelead\ is shown in \Fig{fig:8-Elec-purity} by comparing the transverse momentum range for the minimum bias and \gls{EMCal} \gls{L1} triggered data. 

\begin{figure}[t!]
\centering
\includegraphics[width=0.49\textwidth]{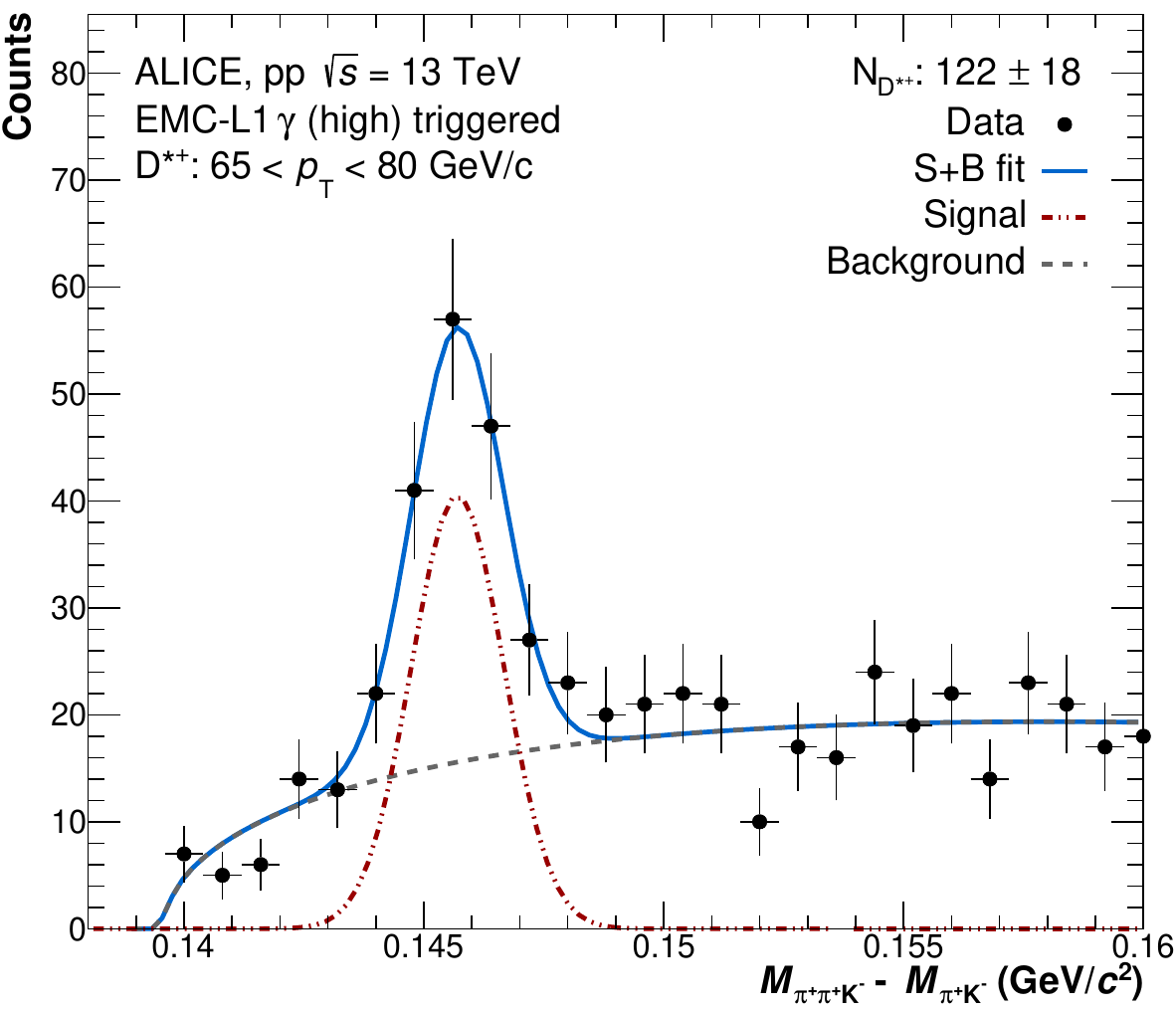}
\includegraphics[width=0.49\textwidth]{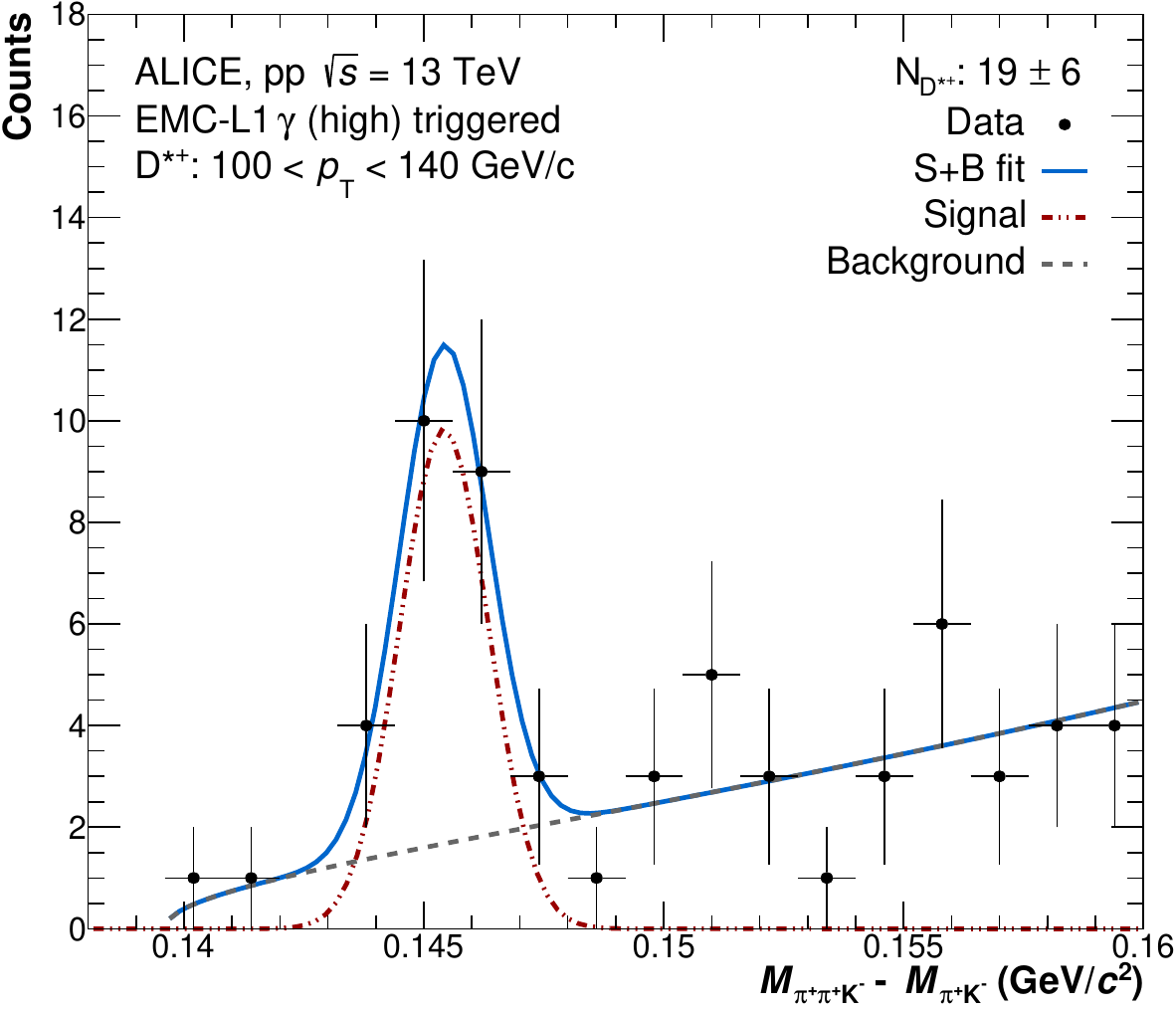}
\caption{(Color online) Invariant mass distribution of D*$^{+}$ candidates in \pp~collisions at \sthirteen\ for $65 < \pT < 80 $ \GeVc~(left) and $100< \pT < 140 $ \GeVc~(right) using the \gls{EMCal} \gls{L1} triggered data. The raw data distribution is shown in black, while the combined signal and background fit is overlayed as a blue line. The separated components of the signal and background contribution to the fit are displayed as red and gray lines, respectively.}
\label{fig:8-Elec-DStarPlus} 
\end{figure}

By making use of the triggered samples, the \pT\ range can be nearly tripled for the \pp\ data and doubled for the most central 0-10\% \PbPb\ data set.
Furthermore, the combined \gls{TPC} and \gls{EMCal} \gls{PID} capabilities and their high purity can be used to tag electron candidates on a track-by-track basis in order to measure correlations of these electrons with other hadrons in the event~\cite{Adam:2016ssk}.

The data samples with the \gls{EMCal} trigger can also be used to enrich the sample of events containing heavy flavor (charm and beauty) hadrons, opening the possibility of reconstructing D-mesons in their hadronic decay channels up to higher transverse momenta as compared to the minimum-bias triggered samples.
An example for the D*$^{+}$ reconstruction in its D*$^{+} \rightarrow $D$^0(\pi^+$K$^-) \pi^+$ decay channel is shown in \Fig{fig:8-Elec-DStarPlus}, where the transverse momentum coverage could be extended from 80~\GeVc{} up to 140~\GeVc.
The D*$^{+}$ signal is extracted through the difference of the three-particle ($\pi^+\pi^+$K$^-$) invariant mass and the reconstructed D$^0$ mass for D$^0 \rightarrow$ K$\pi$ decay candidates having an invariant mass within $3 \sigma$ of the nominal D$^0$ mass.
For about $40\%$ of the events with a D*$^{+}$ at high \pT, the trigger is fired by at least one of its decay products hitting the \gls{EMCal} and creating a hadronic shower exceeding the threshold energy.
The remaining enhancement can be attributed to the electromagnetic component of the jet accompanying the D-meson or the recoiling jet coming from the other charm quark and its energy deposit in the calorimeter.

\begin{figure}[t!]
\centering
\includegraphics[width=0.49\textwidth]{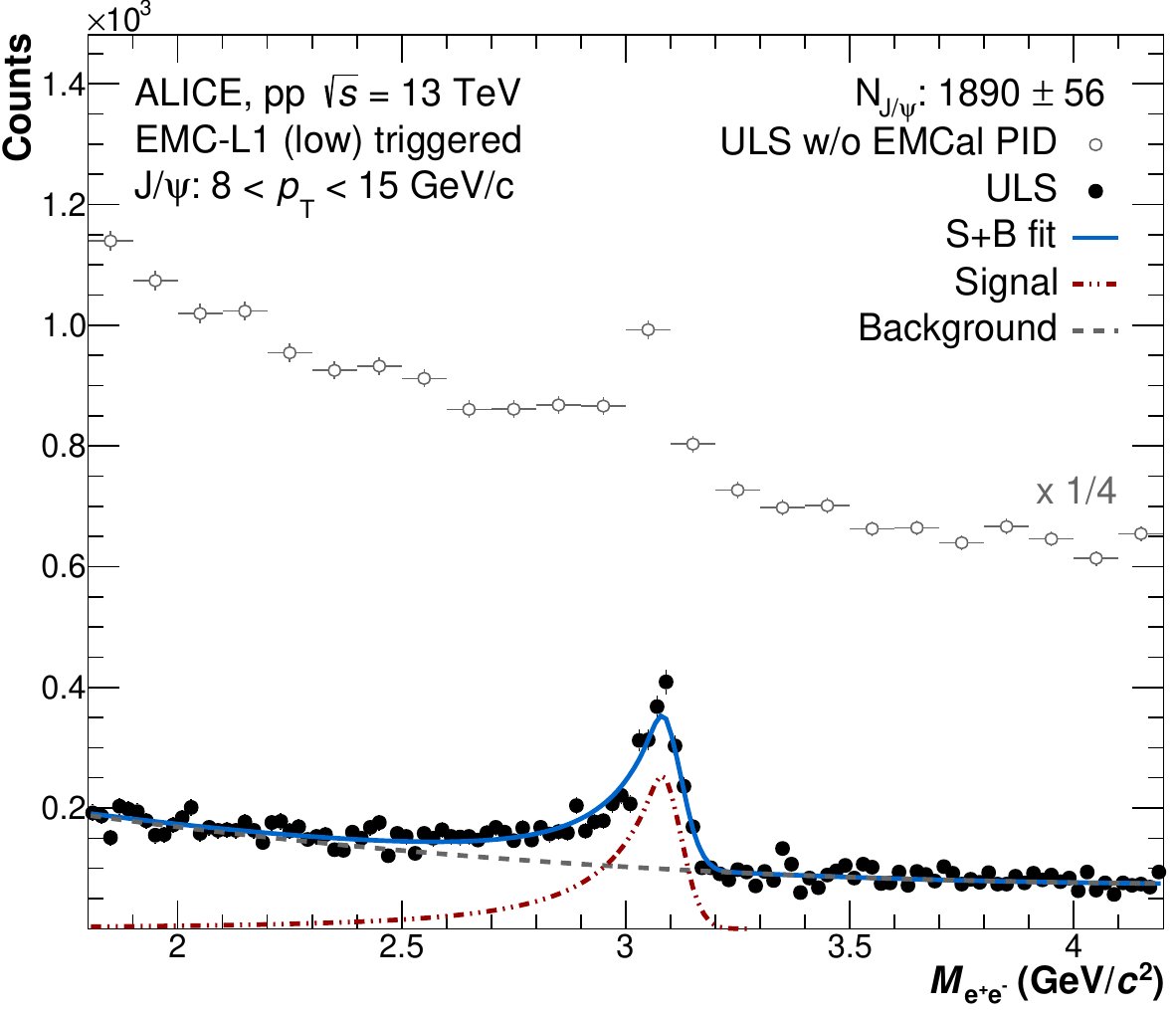}
\includegraphics[width=0.49\textwidth]{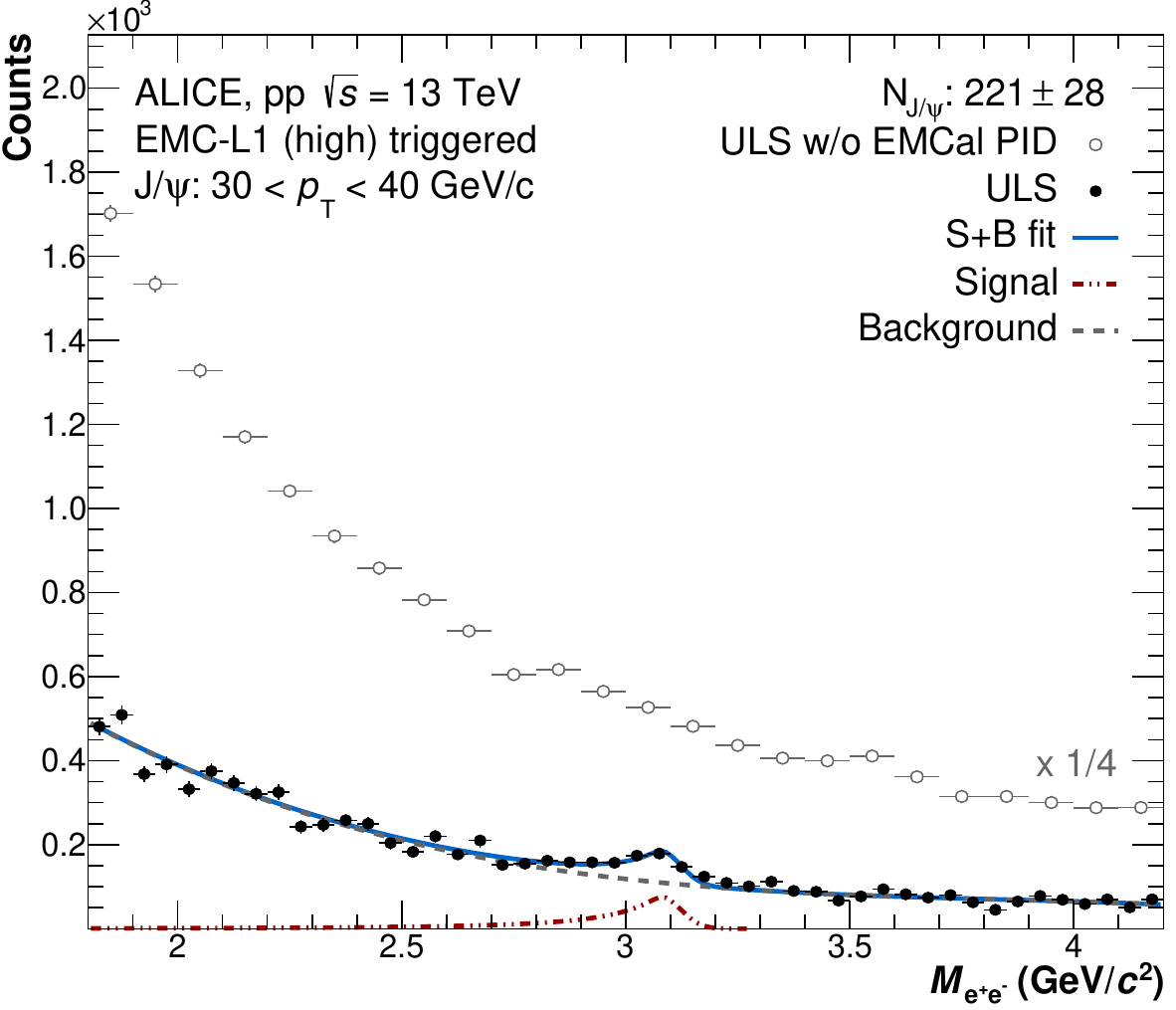}
\caption{(Color online) Invariant mass distribution of J/$\psi$ candidates in \pp~collisions at \sthirteen\ for $8 < \pT < 15 $ \GeVc~(left) and $30< \pT < 40 $ \GeVc~(right). 
                        The gray open markers depict the distribution for e$^+$e$^-$ pairs where at least one track could be matched to an \gls{EMCal} cluster. 
                        It is scaled by $1/2$ and $1/4$, respectively, for the different \pT\ intervals, to enhance the visibility. 
                        The black closed markers represent the distribution after applying the \gls{EMCal} \gls{PID} selections on at least one of the J/$\psi$ decay products.
                        The combined signal and background fit is shown by the blue line, while the polynomial background and pure signal fit are shown as gray and red lines respectively~\cite{ALICE:2021dtt}. }
\label{fig:8-Elec-Jpsi} 
\end{figure}

\subsubsection{J$/\psi$ meson reconstruction}
The production of charmonium (bound state of c and $\bar{\textrm{c}}$ quarks) at \gls{RHIC} and \gls{LHC} energies is not yet fully understood and can give important information on perturbative and non-perturbative \gls{QCD}.
Furthermore, quarkonium production in heavy-ion collisions provides important information on the nature and properties of the produced medium. 
The \gls{EMCal} can be used to extend the measurements of the J/$\psi$ mesons, reconstructed in their e$^+$e$^-$ decay channel, up to higher \pT\ values as compared to those that can be reliably identified and reconstructed relying on the barrel tracking and the minimum bias triggered samples~\cite{Acharya:2020pit,ALICE:2021dtt}.
This is achieved by exploiting the identification of electrons in the \gls{EMCal} and the largely enhanced luminosity that is sampled with the high-\pt\ single shower triggers.
Electrons are first identified using the \gls{TPC}, and then at least one of the J/$\psi$ decay products is required to be in the \gls{EMCal}, with a cluster energy above the trigger threshold and in the range $0.8 < E/p < 1.2$. 
\Figure{fig:8-Elec-Jpsi} shows an example of the invariant mass of di-electrons for $8 < \pt < 15 $ GeV/$c$ and $30< \pt < 40 $ GeV/$c$. 
The gray open markers represent the invariant mass distribution for pairs where at least one of the tracks could be matched to an \gls{EMCal} cluster, while the distribution depicted with closed black markers also has the $E/p$ selection applied on at least one of the J/$\psi$ decay products.
Additionally, the corresponding cluster is required to be above the trigger energy threshold.
The improvement in the signal to background ratio is clearly visible at the cost of a minimal efficiency reduction.
Once the additional \gls{EMCal} \gls{PID} criteria are applied, the J/$\psi$ peak emerges around $M_{\text{e}^+\text{e}^-} \approx$ 3.09 \GeVcs\ for both \pT\ intervals and the combinatorial background can be described by a second or third order polynomial fit performed excluding the peak region. 
The production yield of J/$\psi$ is then calculated in the mass range $2.92 < M_{\text{e}^+\text{e}^-} < 3.16$~\GeVcs.

 
\subsection{Jets}
\label{sec:jets}
\FloatBarrier
The \gls{EMCal} \com{in \gls{ALICE} }can also be used to reconstruct larger objects, namely ``jets'', which consist of a set of correlated particles emerging from the fragmentation and hadronization of partons produced in partonic scatterings with large momentum transfer.

Many jet analyses in \gls{ALICE} used only the charged-track information to reconstruct jet properties~\cite{ALICE:2014dla,Adam:2015hoa,Adam:2016jfp,Acharya:2018eat,Acharya:2019tku,Abelev:2013kqa, Adam:2015doa, Acharya:2018uvf,Adam:2015mda, Acharya:2017goa, Acharya:2019zup,Acharya:2019djg}.  
The central barrel of the \gls{ALICE} detector has unique tracking capabilities, which  enable a measurement of charged particles down to transverse momenta as low as 150~MeV/$c$. 
While charged-particle jets provide improved angular precision with respect to fully reconstructed jets when studying their substructure and are experimentally simpler, they inherently violate \gls{IRC} safety since jet fragmentation does not generally conserve the charged component. 
Analytical calculations of charged-particle jet observables therefore require the introduction of additional non-perturbative functions~\cite{Chang:2013rca}.

By additionally measuring the neutral-jet constituents with the \gls{EMCal}, \gls{IRC}-safe jet observables can be constructed and standard perturbative calculations can be directly compared to experimental measurements.
Including the neutral component in jet reconstruction also enables the use of the trigger to significantly enhance the number of reconstructed jets at high $\pT$ providing access to momentum ranges that cannot be covered by using only the tracking information~(see \Sec{sec:trigger}).
Jets reconstructed from the combination of information from the tracking system and the \gls{EMCal} are referred to as ``full jets''. 
\gls{ALICE} has measured inclusive full jet invariant transverse momentum spectra in \pp\ and \PbPb\ collisions at \stwolead~\cite{Abelev:2013fn,Adam:2015ewa} and \sfivelead ~\cite{Acharya:2019jyg}, as well as correlations involving full jets~\cite{Adam:2015xea, Acharya:2019psw}.
These measurements demonstrate significant jet quenching~\cite{Cunqueiro:2021wls} effects in heavy-ion collisions, such as strong suppression of jet yields in \PbPb\ collisions compared to appropriately-normalized jet transverse momentum spectra measured in \pp\ collisions. 
Moreover, comparisons of inclusive full jet transverse momentum spectra in \pp\ collisions to analytical \gls{pQCD} calculations demonstrated the importance of \gls{NNLO} and \gls{NLL} contributions to the jet cross section calculation.

\subsubsection{Full jet reconstruction}
The \gls{EMCal} measures inclusive photons and electrons with high efficiency, which in combination with charged-particle tracks includes the vast majority of directly-measurable jet constituents. 
Neutral long-lived hadrons (n, K$^{0}_{L}$) are not reliably measured in the \gls{EMCal}~(which has a hadronic scattering length of $\lambda\approx1$), however these comprise only a small fraction of the jet energy, typically of the order of 3-6\%~\cite{Abelev:2013fn}, and its effects are corrected for with \gls{MC} simulations at the analysis level.\\
In order to combine charged-particle tracks and \gls{EMCal} clusters (built with the V2 clusterizer described in \Sec{sec:clusterization}), a simple procedure, called the {\it hadronic} correction, is employed to account for the double counting introduced by charged tracks depositing energy in the \gls{EMCal}.
After the standard energy calibrations are applied to the clusters~(energy nonlinearity and exotic cluster removal, see \Tab{tab:photonbasiccuts}), all charged tracks are extrapolated and matched to clusters, as described in \Sec{sec:trackmatch}.
The possible hadronic energy constribution to clusters with one or more associated tracks is subtracted using 
\begin{equation}
    E_{\rm sub}  = E_{\rm clus} - f\, \sum_{i}p_{i}^{\rm track}
    \label{eq:hadcorr}
\end{equation}
where $f$ is the fraction of the subtracted energy, $p_{\mathrm{i}}^{\mathrm{track}}$ is the momentum of the i-th track matched to the cluster and $E_{\rm clus}$ is the cluster energy after energy correction. 
Clusters for which the energy after subtraction, $E_{\rm sub}$, is negative are discarded.
The entire track momentum is usually subtracted from the cluster energy~($f=1$), which is correct for electrons, but leads to an oversubtraction for hadrons; this effect is however is compensated for in the detector response obtained using simulations, where the same hadronc correction procedure is applied.

The clusters are converted into four-momentum vectors assuming they are massless particles originating from the center of the collision. 
Jets are usually reconstructed using the anti-$k_{\mathrm{T}}$ clustering algorithm~\cite{Cacciari:2008gp}, using only clusters for which the energy is larger than $300$~MeV/$c$ (see \Sec{sec:clusterization}) and charged-particle tracks with $\pt > 150$~MeV/$c$. 
The requirement on the energy of the cluster $E$ was selected to be at the minimal limit allowed for by the energy resolution of the detector in order to maximize the efficiency of the jet energy reconstruction. 
This is particularly important for jets with large resolution parameters which otherwise would exhibit a significant shift in the jet energy scale for stringent cluster selections implying larger corrections.
Jets are reconstructed with different choices for the jet resolution parameter $R$ (the analog of the jet radius parameter for sequential recombination algorithms such as the anti-$k_{\mathrm{T}}$ algorithm).
To ensure that the entire jet energy is deposited in the \gls{EMCal}, the jets are required to have their axis at a distance larger than $R$ from the border of the \gls{EMCal} to fully fit into the fiducial acceptance of the \gls{EMCal}.
Therefore jets reconstructed in the EMCal are limited to $|\eta_{\rm jet}| < 0.7- R$, making $R = 0.6$ the maximum possible resolution parameter.
On the contrary, with the \gls{DCal}, only jets with about $R=0.1$ can be reconstructed due to its comparatively smaller acceptance. 

\subsubsection{Performance of full jet reconstruction}
To quantify the performance of the jet reconstruction in the detector, two quantities are used: the \gls{JES} and the \gls{JER}. 
The \gls{JES} describes the mean energy difference between the generated jet at particle level and the reconstructed jet at detector level, defined as
\begin{equation}
  {\rm JES} = \mu (\Delta \pT) \equiv \mu \left[ \frac{\pT^{\rm det} - \pT^{\rm part}}{\pT^{\rm part}} \right]\,,
    \label{eq:JES}
\end{equation}
where $\pT^{\rm det}$ and $\pT^{\rm part}$ are the jet $\pT$ at reconstructed~(detector) and truth~(particle) level obtained in simulations.
It describes the fraction of energy that is on average missed when reconstructing jets. 
For an ideal detector all energy of the jet is captured, thus the shift on the jet energy scale would be zero.

The \gls{JER} describes the variance of the $\Delta \pT$ distribution, defined as 
\begin{equation}
{\rm JER} \equiv \sigma \left[ \frac{\pT^{\rm det} - \pT^{\rm part}}{\pT^{\rm part}} \right]\,,
\label{eq:JER}
\end{equation}
and hence characterizes the degree to which stochastic effects add up to the difference in $\Delta p_{T}$ between detector and particle-level jets. 

\begin{figure}[t!]
    \centering
    \includegraphics[width=1\textwidth]{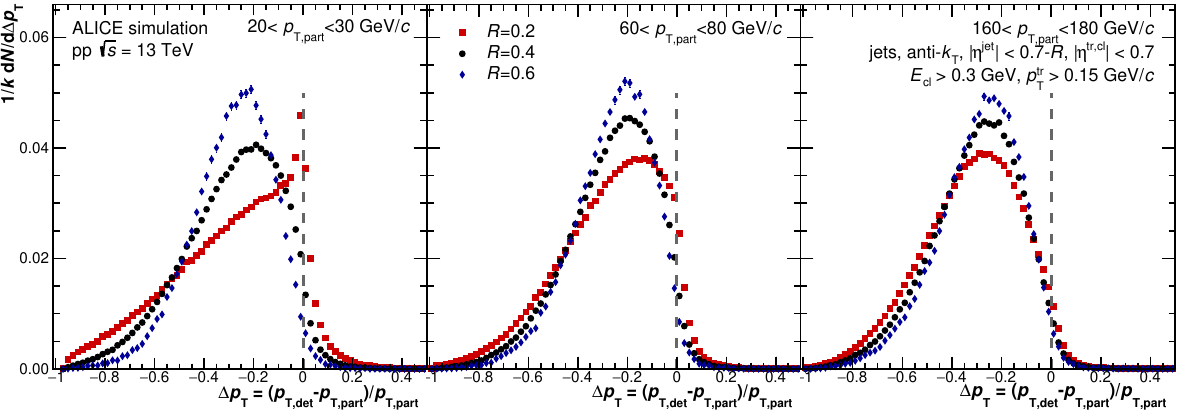}\\ \vspace{0.2cm}
    \includegraphics[width=0.49\textwidth]{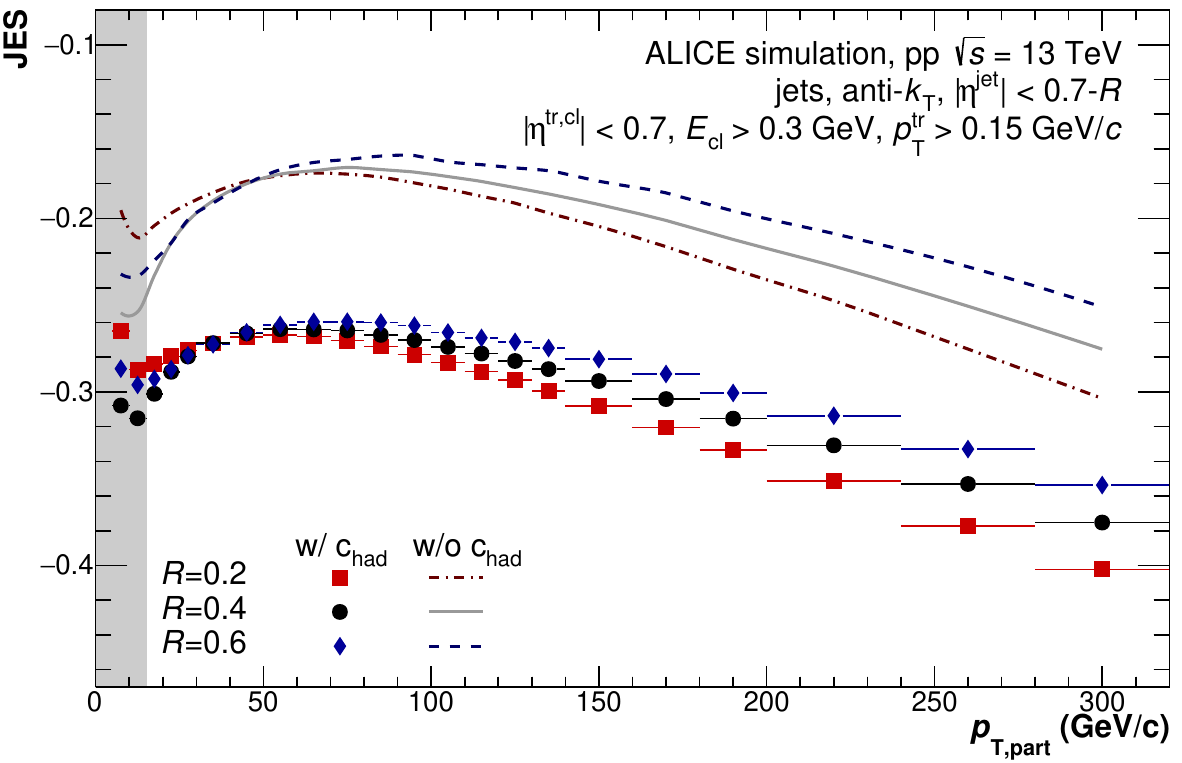}
    \includegraphics[width=0.49\textwidth]{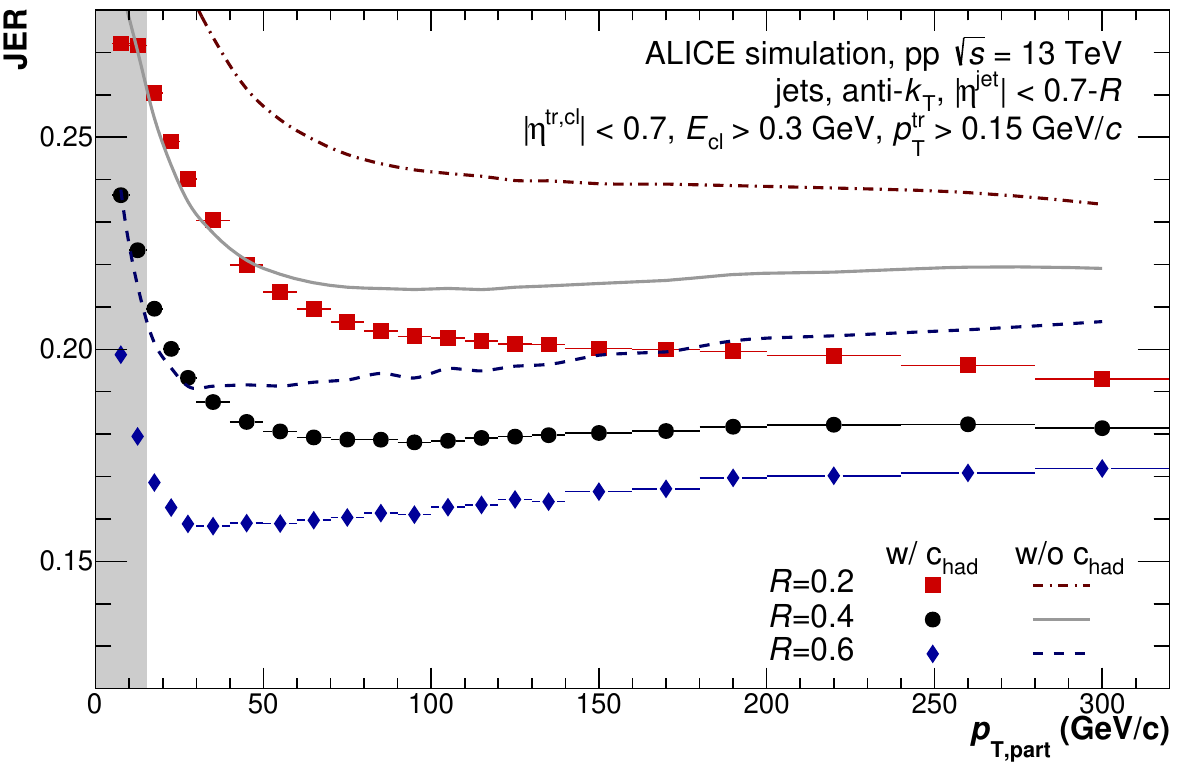}
    \caption{(Color online) Instrumental effects on the jet energy measurement at \sthirteen\ in \pp\ collisions as a function of the jet resolution parameter ($R=0.2$, $R=0.4$ and $R=0.6$).
            Upper panel: jet-by-jet distribution for various intervals in jet \pt. 
            Lower panels: \gls{JES} as mean~(left) and \gls{JER} as standard deviation~(right) of these distributions with ($f=1$) and without ($f=0$) the hadronic correction~($c_{\rm had}$), shown as points and lines, respectively.
            The gray bands indicate the \pT\ regions not taken into account for the final measurements.} 
    \label{fig:6-Perf-jesjervsRpp13}
\end{figure}

\Figure{fig:6-Perf-jesjervsRpp13} illustrates the performance of the full jet reconstruction for $R=0.2$, $0.4$ and $0.6$ using detector simulations in \pp\ collisions at \sthirteen. 
The upper panels show the distributions of $\Delta$\pt\ for three jet \pt\ intervals, which 
exhibit large jet-by-jet fluctuations of reconstructed jet \pT, as well as a significant tail of the distributions to negative values. 
These instrumental effects are mostly caused by the tracking inefficiency, with additional contributions from \gls{EMCal} resolution effects, and unreconstructed neutral energy.
The lower panels show the \gls{JES}~(left) and the \gls{JER}~(right) of the $\Delta$\pt\ distributions, as a function of particle-level jet \pt.
The \gls{JES} exhibits a deviation from zero towards negative values, which becomes larger with increasing jet \pt, and which is larger for smaller jet radii than larger jet radii.
The \gls{JER} depends weakly on the jet \pt, and decreases modestly with increasing jet radius.
Additionally, \Figure{fig:6-Perf-jesjervsRpp13}~(bottom) shows the effect of the correction for the hadronic contribution ($c_{\text{had}}$) to the cluster energy on the \gls{JES} (left) and \gls{JER} (right) for jets. 
A \pt\ independent shift of the \gls{JES} in the positive direction can be observed in case the correction is not applied, indicating that the energy lost due to detector inefficiency is partially compensated by the double counting of energy deposited by charged particles in the \gls{EMCal}. 
The \gls{JER} improves by about 5\% when applying the correction since it reduces fluctuations induced by hadronic energy deposits.

\begin{figure}
    \centering
    \includegraphics[width=0.45\textwidth]{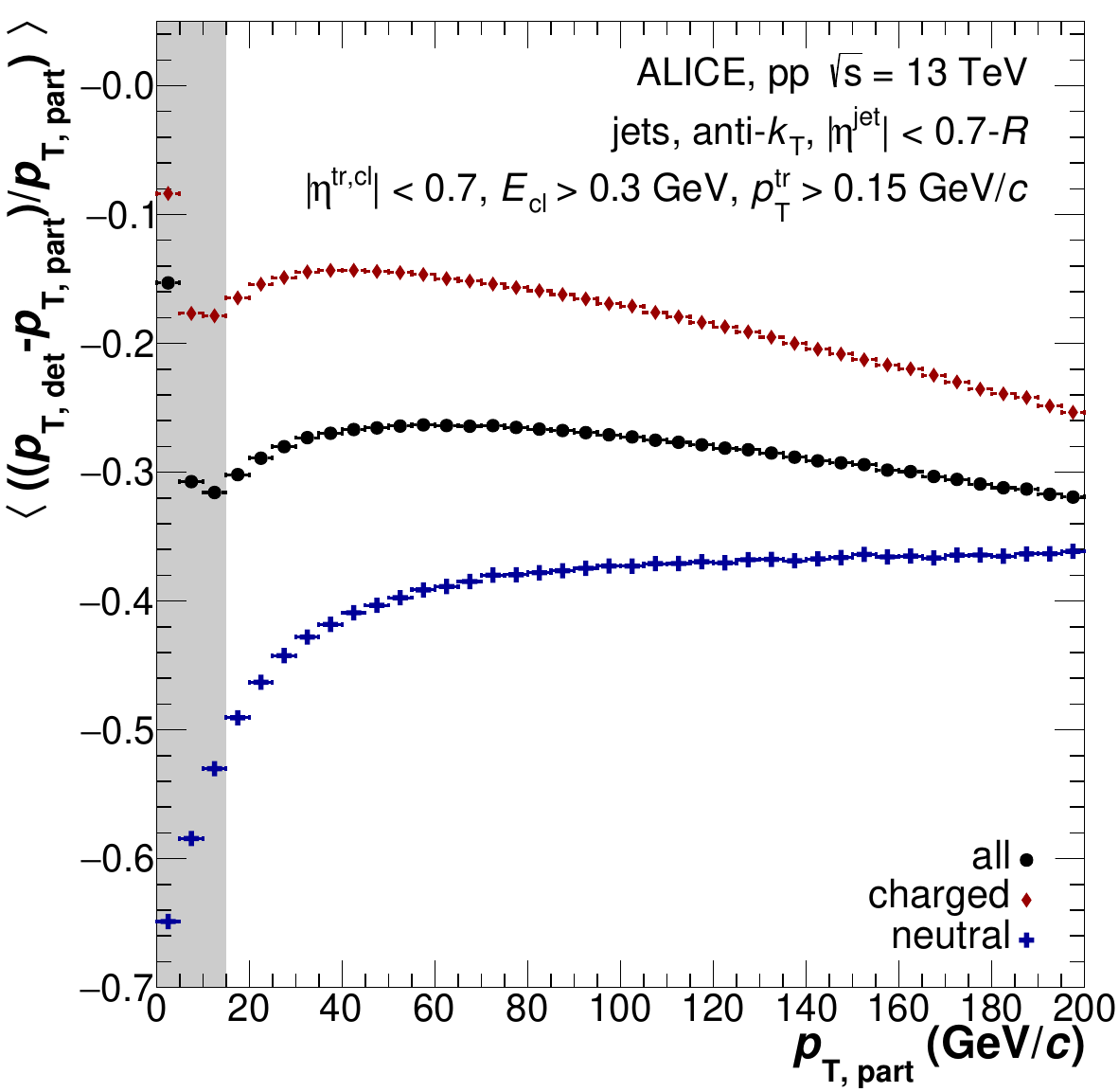}
    \includegraphics[width=0.45\textwidth]{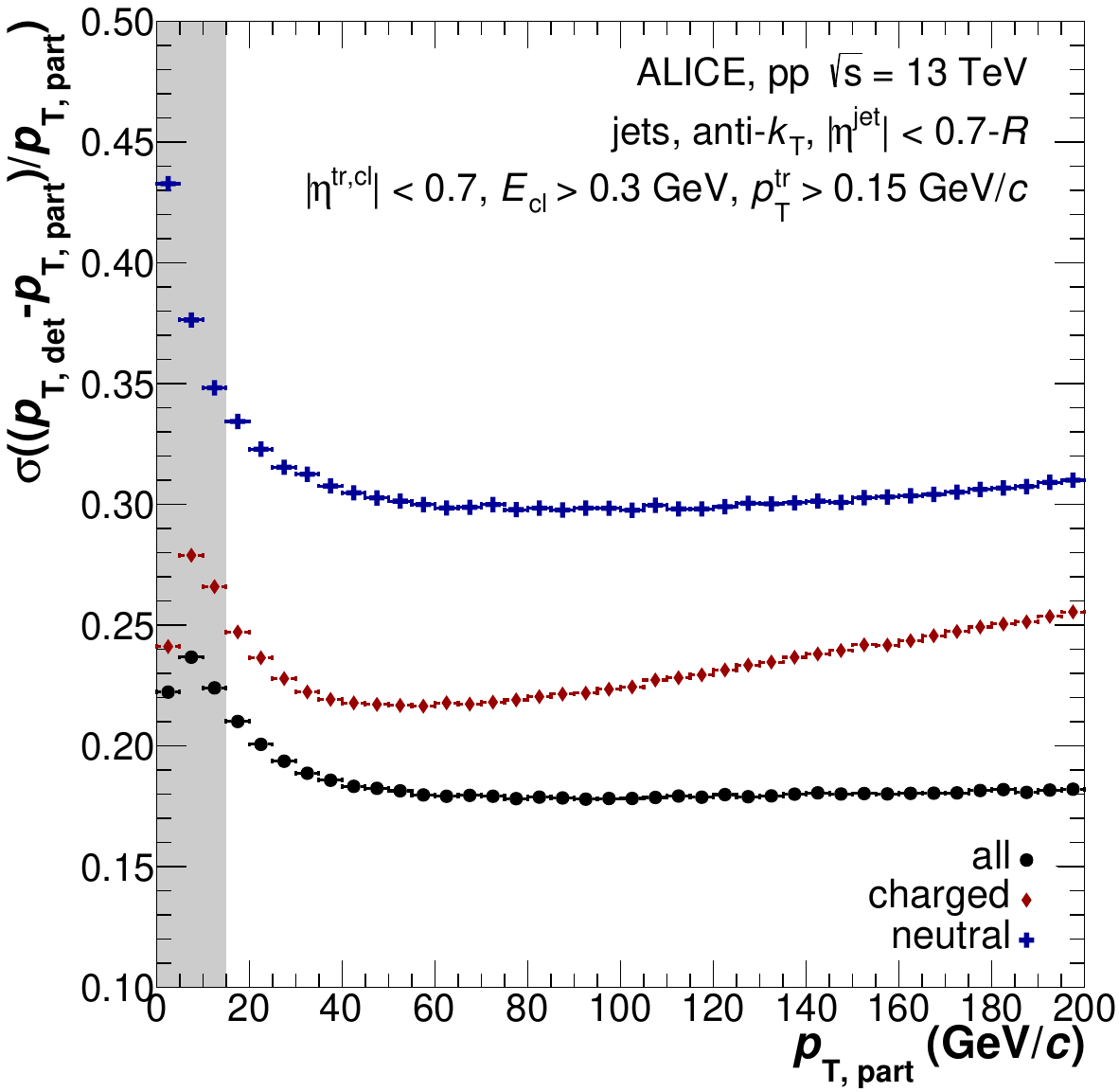}
    \caption{Jet energy scale (left) and resolution (right) for jets with $R$~=~0.4 using all, charged and neutral constituents.}
    \label{fig::JES::charged_neutral}
\end{figure}
In \Fig{fig::JES::charged_neutral} the \gls{JES} for charged and neutral particles is shown separately.
A smaller shift can be observed for charged constituents, which is approximately $-0.15$ for jets with $R=0.4$ and $\pT=40$~GeV/$c$.
Its magnitude increases with momentum up to a shift of $-0.26$ at $\pT=200$~GeV/$c$, consistent with measurements of the jet-energy scale for track-based jets. 
Considering only neutral constituents the JES is approximately constant at $-0.4$ for $\pT>60$~GeV/$c$ and $R=0.4$. 
The increase of the \gls{JES} shift with increasing \pT\ for charged constituents results from a reduction of the tracking efficiency in environments with a large local track density of high $\pT$ tracks due to limitations in the two-track resolution in the central barrel detectors and the low magnetic field of $B=0.5$~T. 
The \pT-dependence of the \gls{JES} shift for charged constituents translates into the scale shift for full jets. 
The \gls{JES} shift considering only neutral particles is independent of $\pT$ for sufficiently high $\pt$ as the measurement of the neutral energy does not depend on the two-particle resolution. 
The \gls{JER} for individual charged or neutral jets is larger than for full jets, with a \gls{JER} for charged constituents of $\approx0.23$ at $\pT=40$~GeV/$c$ for jets with $R=0.4$, increasing with $\pt$ to $0.25$ at $\pT=200$~GeV/$c$, while it is approximately constant at $0.3$ for neutral constituents for jets with $\pT>60$~GeV/$c$. 

\begin{figure}[t!]
    \centering
    \includegraphics[width=.49\textwidth]{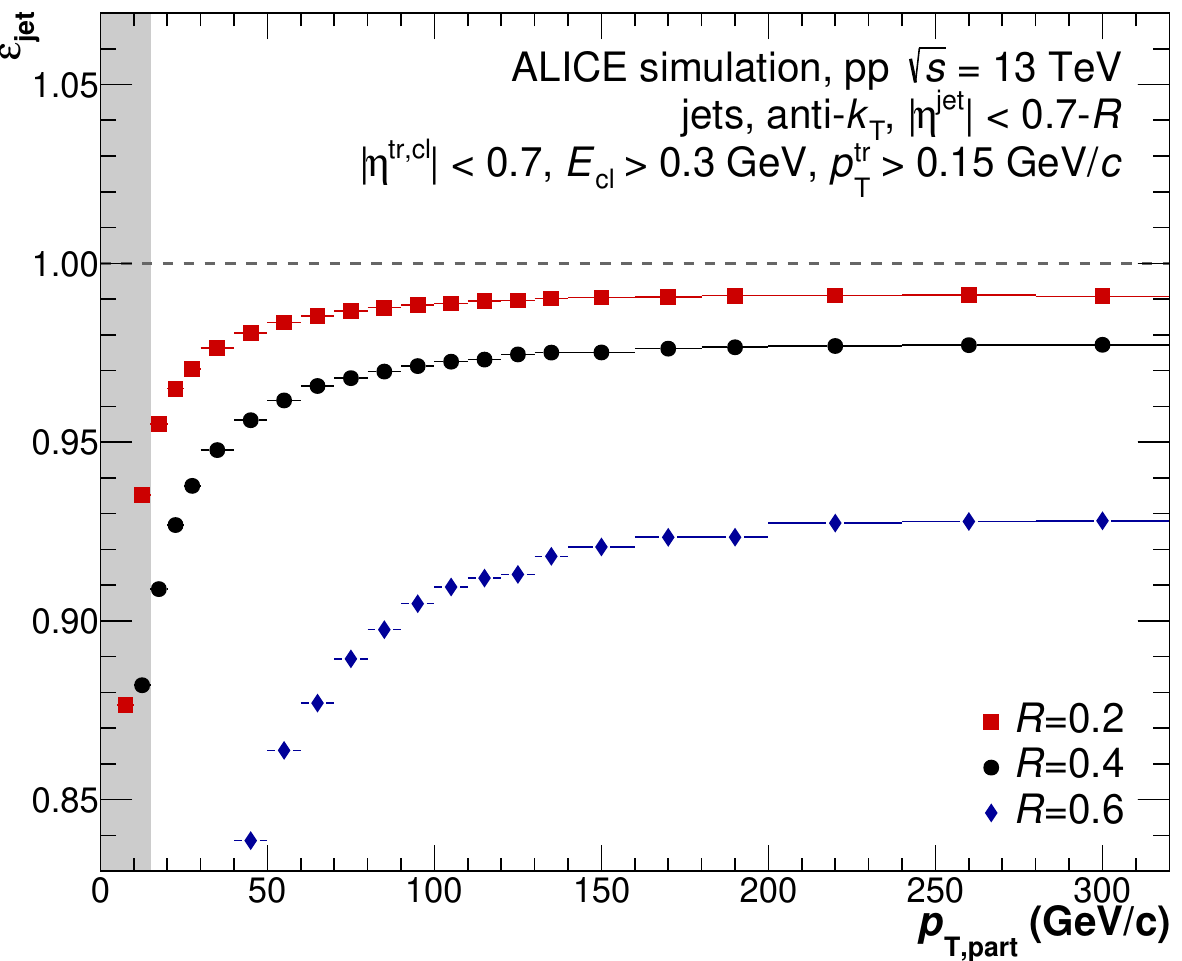}
    \includegraphics[width=.49\textwidth]{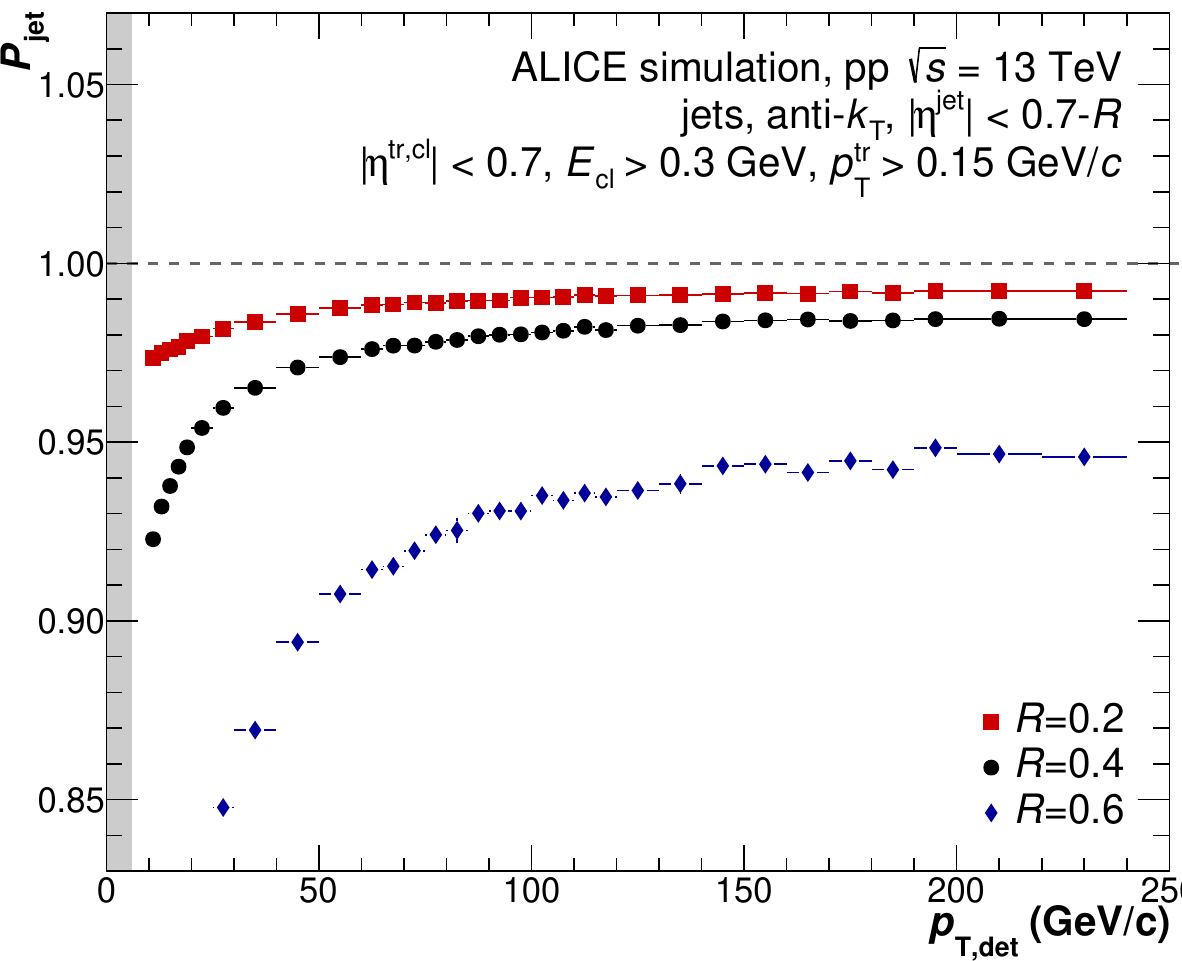}
    \caption{(Color online) Jet finding efficiency $\varepsilon_{\rm jet}$ (left) and purity $P_{\rm jet}$ (right) for jets with different jet resolution parameter measured in \pp{} collisions at \sthirteen. 
              The gray bands indicate the \pT\ regions not taken into account for the final measurements.}
    \label{fig:6-Jet-jetfindingeffpp13TeV}
\end{figure}

\subsubsection{Jet finding efficiency}
The jet finding efficiency is defined as the fraction of particle-level jets in the \gls{EMCal} fiducial acceptance at a given \pt{} for which a detector-level jet at any \pt{} was reconstructed and matched to the particle-level jet via a matching criterion that typically depends on the distance between the true and reconstructed jet axes.
\Figure{fig:6-Jet-jetfindingeffpp13TeV}~(left) shows the jet finding efficiency for various jet resolution parameters in \pp{} collisions at \sthirteen{}. 
In order to match particle and detector-level jets, the acceptance was restricted at the detector level to the \gls{EMCal} fiducial acceptance, while at particle level the acceptance was extended by $R$ in both $\eta$ and $\varphi$. 
In the determination of the jet finding efficiency only those jet pairs are considered for which both particle and detector-level jets are within the acceptance.
Furthermore, matched jets are required to have their jet axis separated by a distance smaller than $R$. 
Consequently, the detector-level jet sample contains jets for which the closest particle-level jet might contain areas outside the \gls{EMCal} fiducial acceptance. 
In order to address the contamination of the jet sample by jets originating from particle jets not fully contained in the \gls{EMCal} fiducial acceptance, we define the jet finding purity as the fraction of detector-level jets which are matched to the corresponding particle-level jet at any \pt{} inside the \gls{EMCal} fiducial acceptance. 
The contamination is then estimated and subtracted from the measured jet sample. The jet finding efficiency corrects for the fraction of particle-level jets in the acceptance which cannot be matched to a detector-level jet. 
At sufficient high \pt, the jet finding efficiency is approximately constant for a given jet resolution parameter. 
For jets with larger resolution parameter, the probability to match with a jet partially outside the \gls{EMCal} fiducial acceptance increases. 
This is reflected in a decrease in both the jet finding efficiency and purity with increasing $R$. 
Towards lower \pt, a drop in both the jet finding efficiency and purity is observed. 
The drop in the efficiency originates from jets with a corresponding detector-level jet in a \pt{} range inaccessible with the apparatus. 
Furthermore, towards lower \pt{}, contributions from the underlying event affect the jet reconstruction to a larger extent, leading to an increase in matched jets with one partner partially outside the acceptance. 
The latter is also reflected in a decrease of the jet finding purity towards lower \pt{}.

\begin{figure}[t!]
    \centering
    \includegraphics[width=0.49\textwidth]{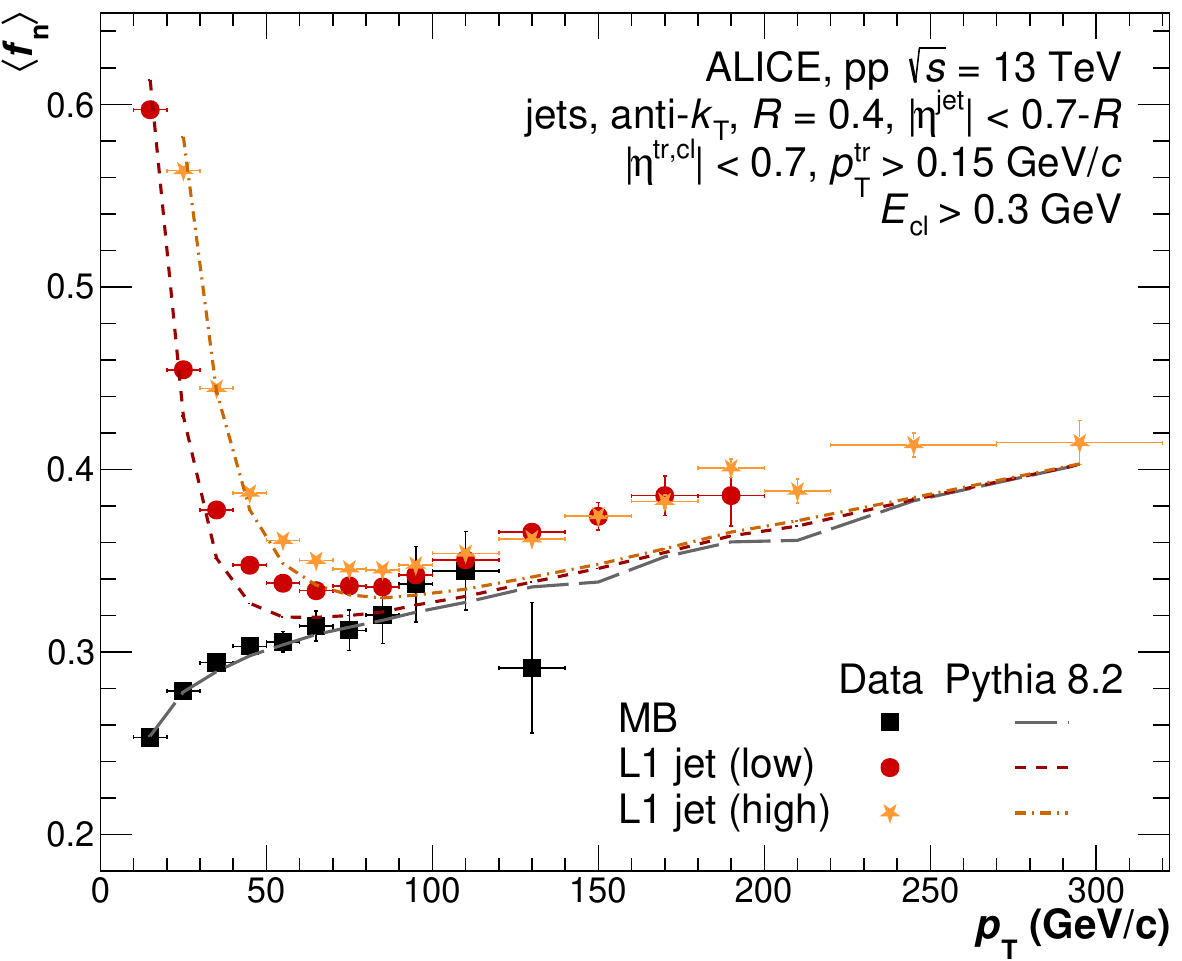}
    \includegraphics[width=0.49\textwidth]{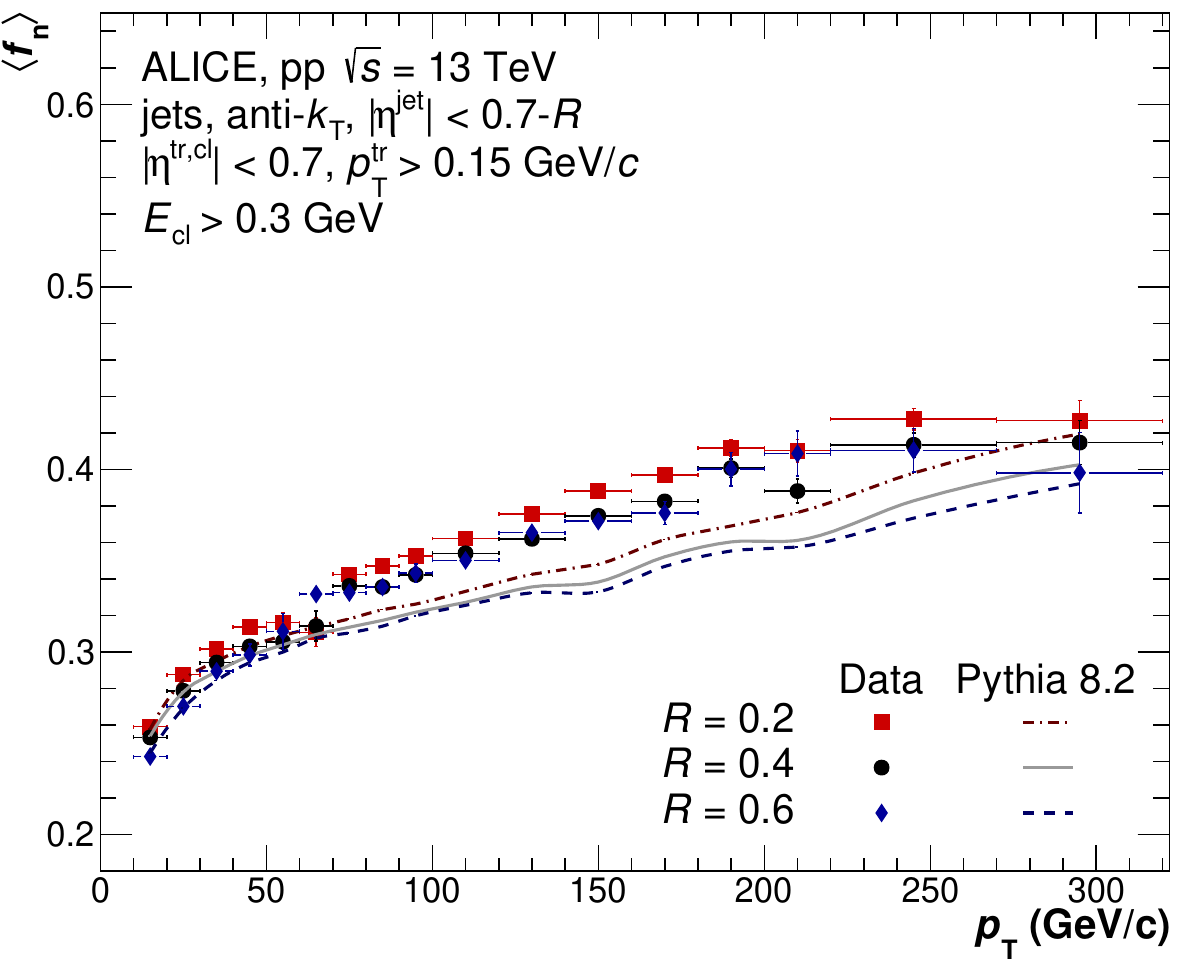}
    \caption{(Color online) Mean neutral energy fraction~($\left<f_{n}\right>$) as function of the jet $p_{\rm{T}}$ for \gls{MB} and \gls{L1} triggers for jets with $R = 0.4$ (left) and for various jet resolution parameters~(right).}
    \label{fig:6-Pro-meanNEF}
\end{figure}
\begin{figure}[t!]
    \centering
    \includegraphics[width=\textwidth]{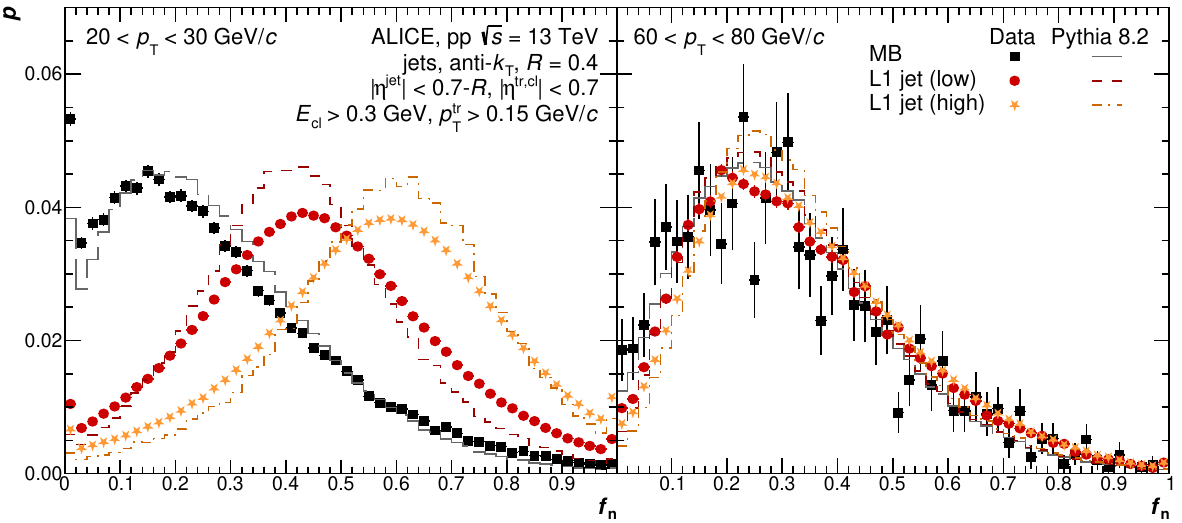}
    \caption{(Color online) Probability distribution of the neutral energy fractions~($f_n$) for jets with $R$ = 0.4 for different triggers for $20<\pt<30$~\GeVc~(left) and $60<\pt <80$~\GeVc~(right).}
    \label{fig:6-Pro-NEFdistPtBin}
\end{figure}

\subsubsection{Properties of reconstructed jets}
To characterize the fraction of the jet energy deposited in the \gls{EMCal}, the mean \gls{NEF} or the ratio of the neutral energy in a jet to the total energy of the jet, are shown in \Fig{fig:6-Pro-meanNEF}~(left) as function of jet \pT{} for jets with $R=0.4$ for minimum bias events and \gls{L1}-jet triggered events. 
The mean \gls{NEF} is also presented  for various jet resolution parameters~ in \Fig{fig:6-Pro-meanNEF}~(right). 
The results reported in \Fig{fig:6-Pro-meanNEF}{ (right)} were obtained using minimum bias-triggered events for $20<\pT<70$~\GeVc, the low-threshold jet trigger for $70<\pT<100$~\GeVc, and the high-threshold jet trigger for $100<\pT<320$~\GeVc. 
The \gls{NEF} increases from $\approx 0.3$ to $\approx 0.4$ with increasing $\pT$ in the range $60<\pT<320$~GeV/$c$.
No dependence on the jet resolution parameter is observed. 
The \gls{NEF} agrees among the different triggers in the \pT - intervals where the triggers are maximally efficient, where the minimum \pT\ of the intervals are approximately $60$~\GeVc\ for the low threshold trigger and approximately $80~\GeVc$ for the low threshold trigger . 
Low \pT\ jets with a low \gls{NEF} do not deposit enough energy in the \gls{EMCal} to pass the trigger threshold, so the trigger enhances jets with a higher \gls{NEF}. 
To illustrate this, \Figure{fig:6-Pro-NEFdistPtBin} shows the \gls{NEF} distributions for two different \pt{} intervals for different triggers. 
It can be seen that the trigger becomes unbiased for \pT\ above approximately $60$~\GeVc, while a strong bias towards jets with higher neutral energy fraction is observed at low \pt{}. 
The \gls{NEF} distributions are qualitatively described by \gls{PYTHIA} simulations. 
At high \pt{} we observe a mild difference in the mean \gls{NEF} between data and simulations, originating from a larger contribution in the tail towards larger \gls{NEF}. 
The remaining difference between data and simulation can originate from noise contributions in data, not taken into account in simulation and leading to an underestimation of the energy resolution, as well as differences in the particle composition between data and simulation. 

\subsubsection{Subtraction of background contributions in \PbPb\ collisions}
In heavy-ion collisions, the large \gls{UE} activity and its local fluctuations can make up a significant contribution to the reconstructed jet \pt.
The reconstructed jet spectrum is obtained by subtracting the average \gls{UE} contribution from the raw jet spectrum, using $\ptj^{\rm rec}=\ptj^{\rm raw} - \rho A$, where $\rho$ is estimated by the median of the jet momentum density distribution~(excluding the two leading jets in the estimate) and $A$ is the area of the jet~\cite{Abelev:2013kqa}.
To estimate the background of full jets in \gls{ALICE} we rely on two inputs~(see details of reconstruction in Ref.~\cite{Adam:2015ewa}):
i)~The typical momentum density of charged jets in each event $\rho_{\rm ch}$ estimated by the median and~ii) the ratio of charged energy in the \gls{TPC} to neutral energy in  the \gls{EMCal}, normalized by their respective acceptances.
This scale factor denoted as $S_{\rm{EMCal}}$ is first calculated event-by-event and then the mean of these scale factors as a function of centrality is parameterized with a second order polynomial. 
The reason for this hybrid approach is that the significantly larger acceptance of the \gls{TPC} leads to a larger sample of unbiased background jets that can be used to determine the average $\rho_{\rm ch}$. 
The total event-by-event and centrality-dependent \gls{UE} contribution that is subtracted from full jets is:
\begin{equation}
\rho(C) = \rho_{\rm ch} \times \left<S_{\text{EMCal}}(C)\right>\,.
\label{Eq:6-Est-FullRho}
\end{equation}
\par
\begin{figure}[t!]
  \centering
  \includegraphics[height=5.7cm]{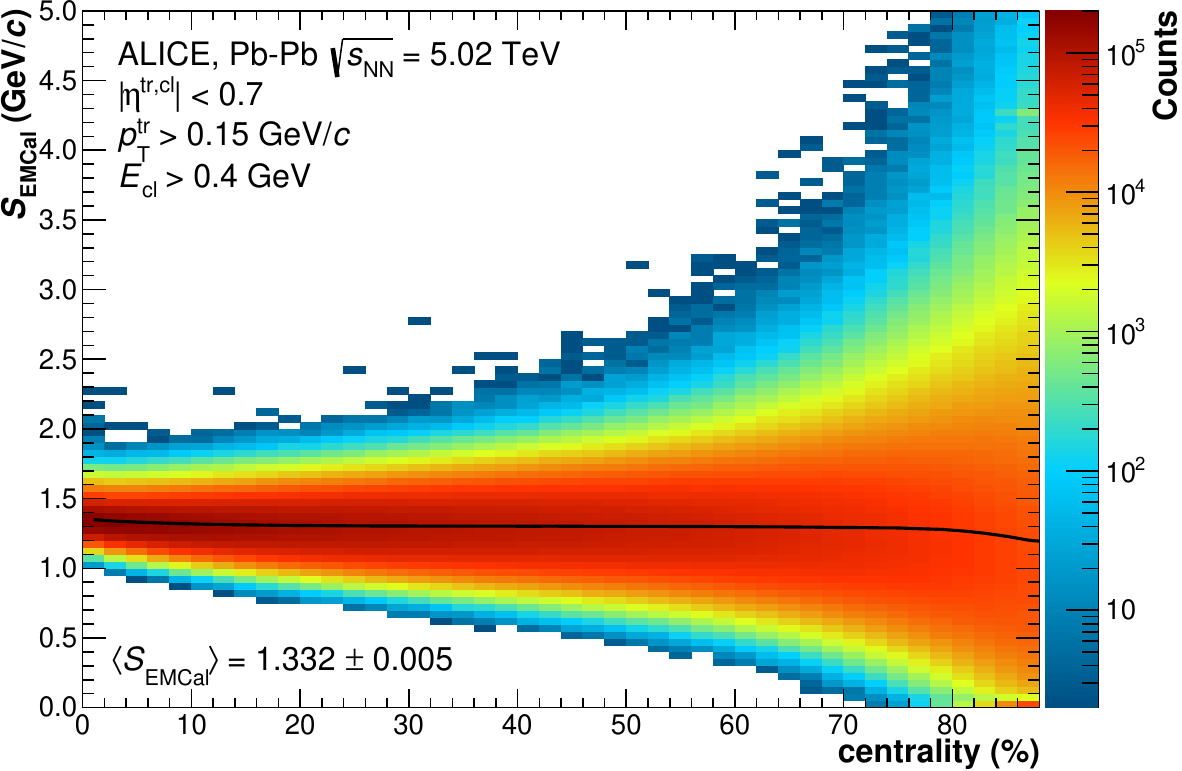}
  \includegraphics[height=5.7cm]{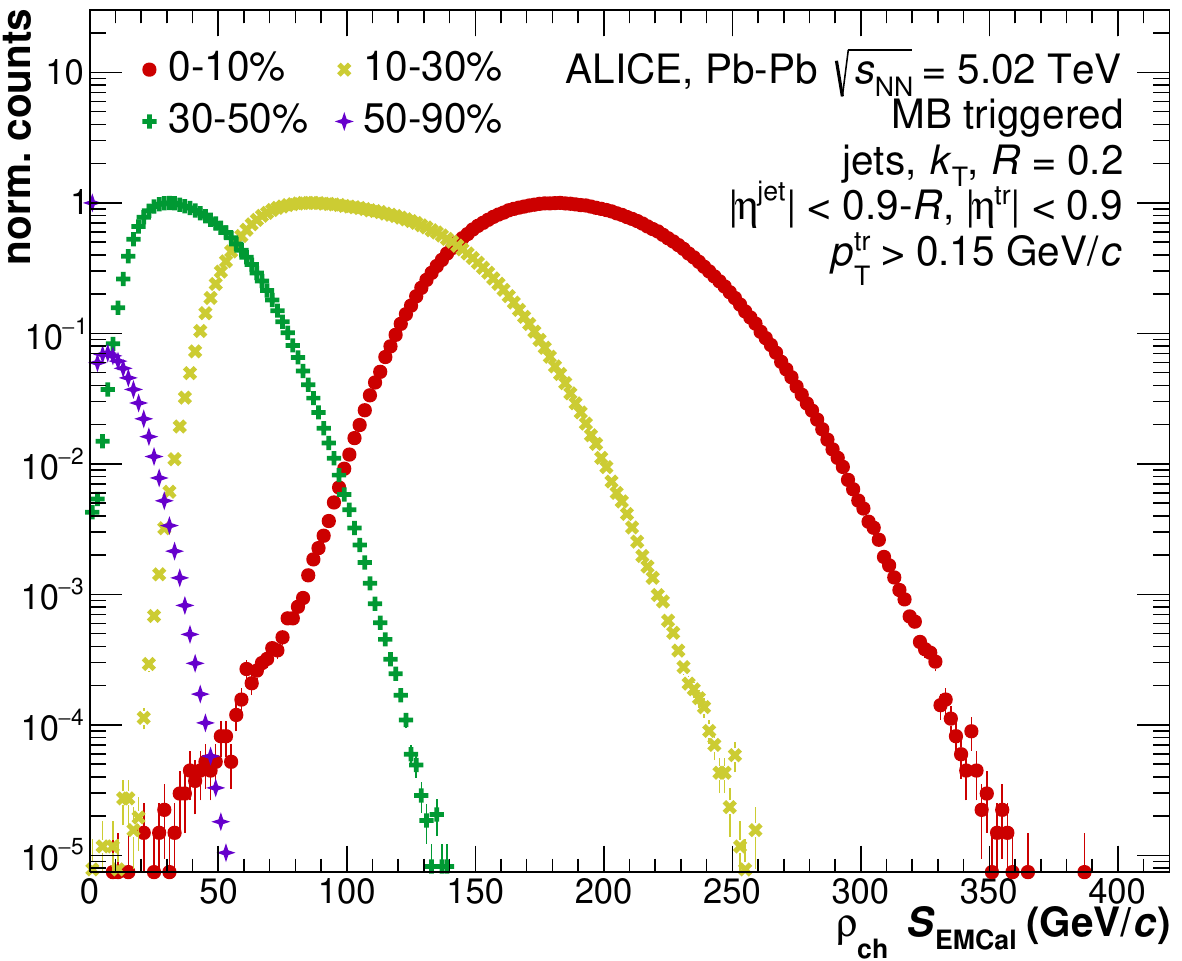}
  \caption{(Color online) Left: Background scale factor as a function of centrality for various events~(with the mean shown in black).
  Right: Comparison of the scaled $\rho_{\rm ch} \times S_{\text{EMCal}}$ for different centrality intervals.}
  \label{fig:6-Est-scaleFactor}
\end{figure}

The value of $S_{\text{EMCal}}$ is significantly dependent on the cell thresholds and hadronic correction procedure used in the specific analysis. 
\Figure{fig:6-Est-scaleFactor}~(left) shows the ratio of charged to neutral energy as a function of centrality for \PbPb\ collisions at \sfivelead. 
The spread in values originates from a variation of this ratio event-by-event at a given centrality. 
The mean value of $S_{\rm{EMCal}}$ is also shown in \Fig{fig:6-Est-scaleFactor}{ (left)} and its centrality dependence was parametrized by a second-order polynomial, as reported above. 
In case detector conditions and/or analysis selections change, $S_{\text{EMCal}}$ has to be determined again, since it is specific for a given analysis.  
\Figure{fig:6-Est-scaleFactor}~(right) shows the  distributions of the \GLS{UE} momentum density $\rho$ obtained using the hybrid approach for several centrality intervals.

\begin{figure}[t!]
    \centering
    \includegraphics[width=0.49\textwidth]{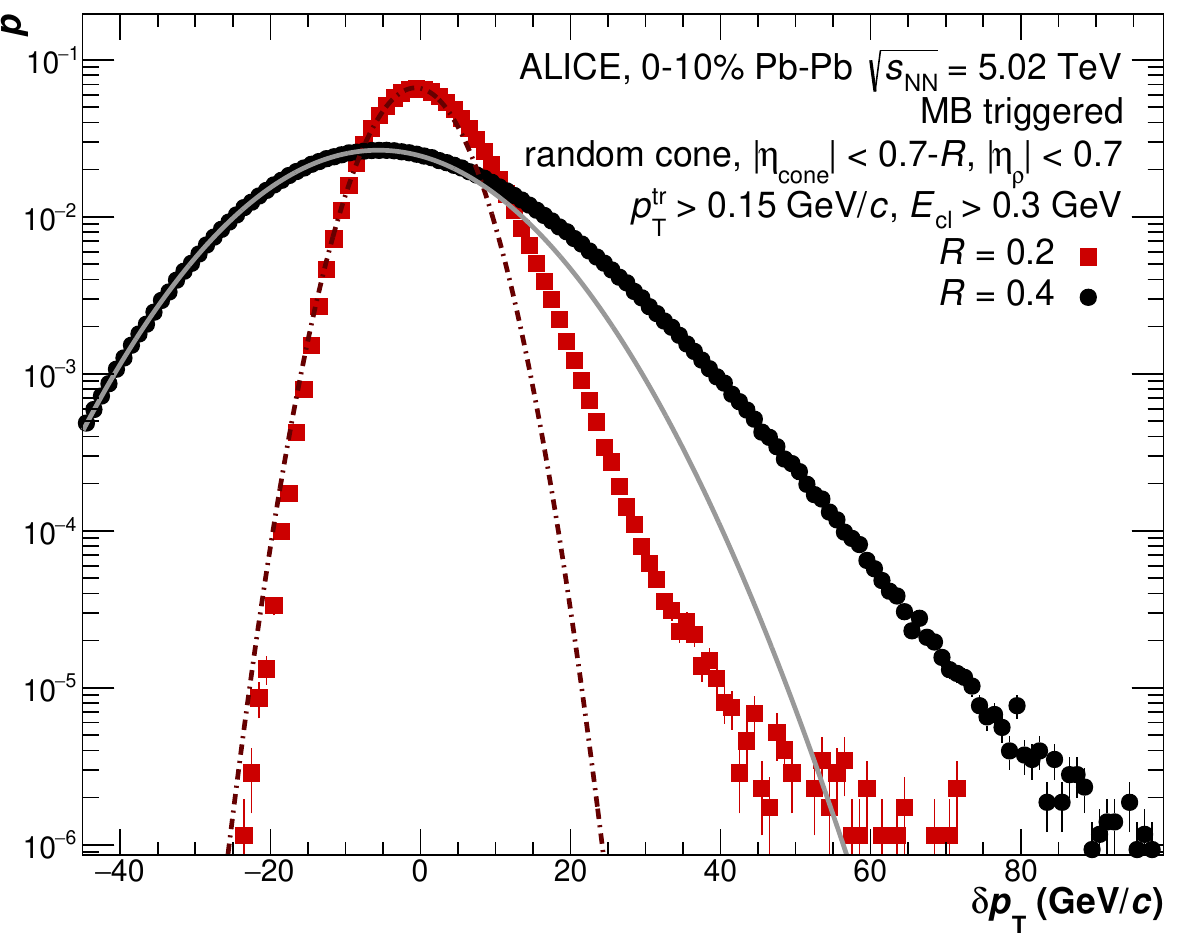}
    \includegraphics[width=0.49\textwidth]{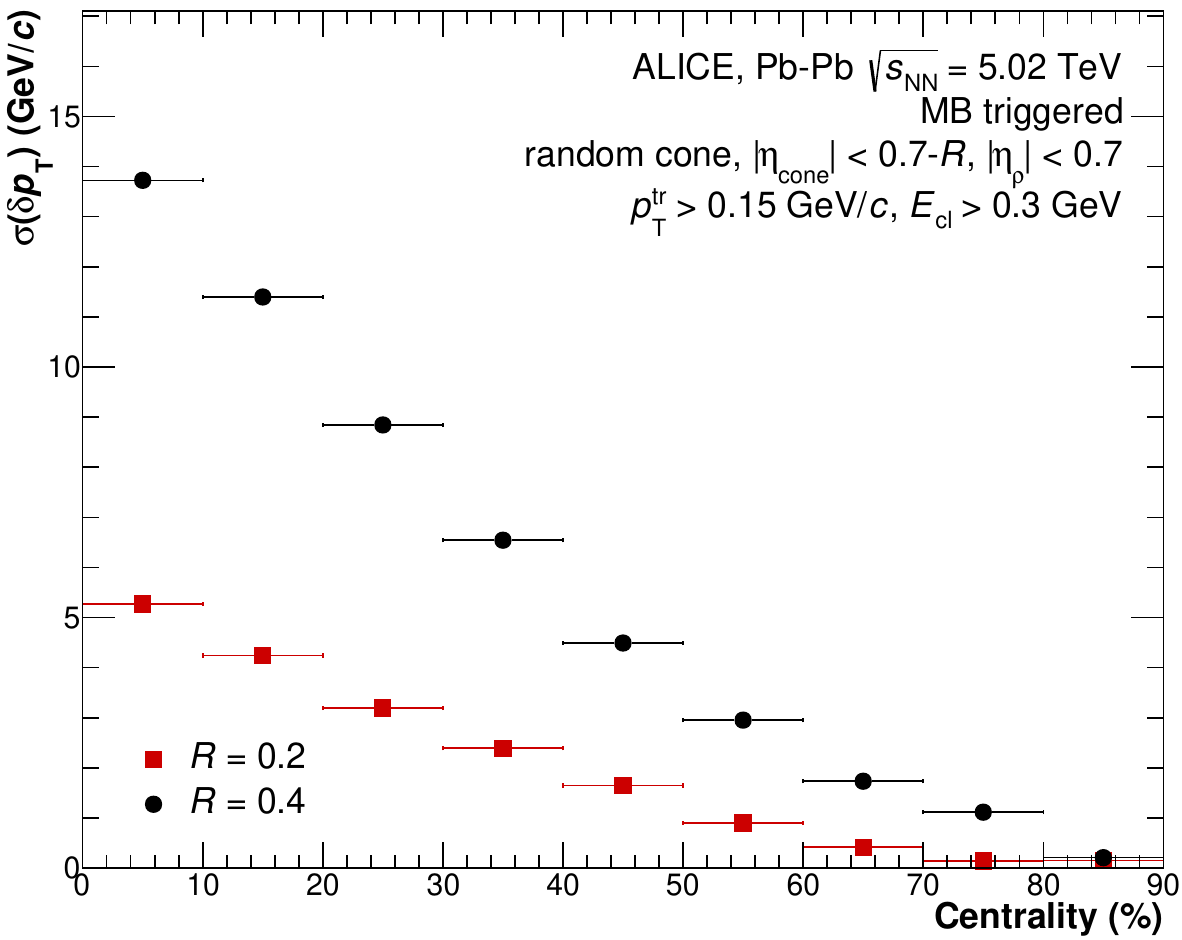}
    \caption{(Color online) Left: Probability distribution of the $\delta$\pT\ distribution for random cones with radii of $R = 0.2$ and $R = 0.4$ excluding the 2 leading jets in the \gls{EMCal} for the $10\%$ most central \PbPb\ collisions at \sfivelead. 
    On top of the distributions, the corresponding Gaussian fits for $\delta\pT < 0$ are displayed as dashed and dotted lines.
    Right: Comparison of the Gaussian width of the $\delta$\pT\ distribution as a function of centrality for $R = 0.2$ and $R = 0.4$ in \PbPb\ collisions at \sfivelead.}
    \label{fig:deltaPtPbPb}
\end{figure}
\begin{figure}[t!]
    \centering
    \includegraphics[width=1\textwidth]{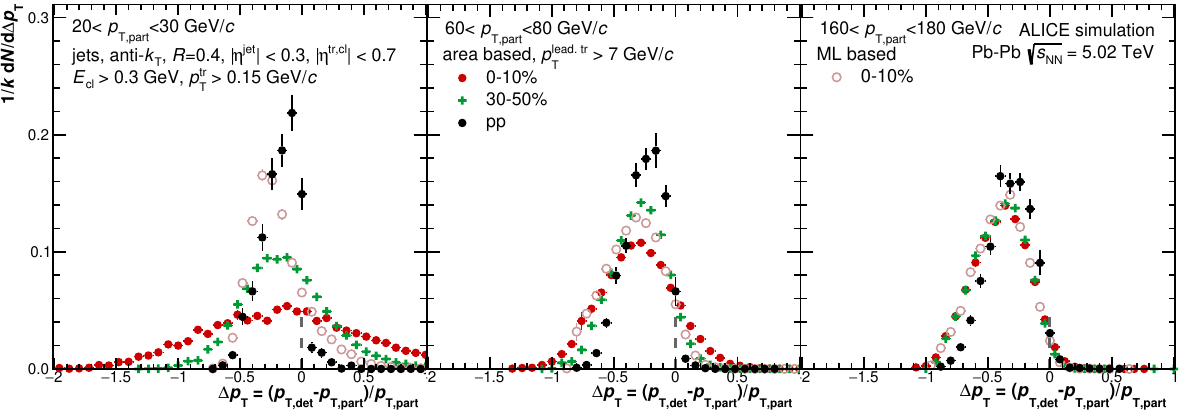}\\ \vspace{0.2cm}
    \includegraphics[width=0.49\textwidth]{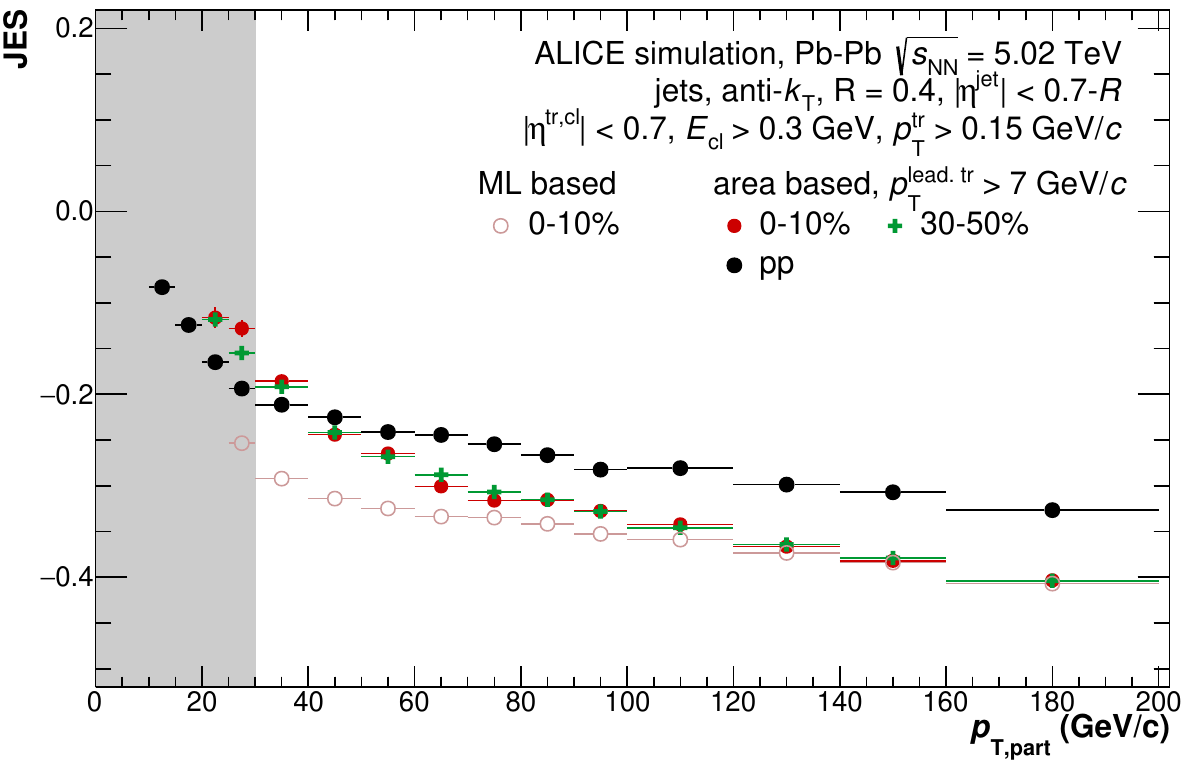}
    \includegraphics[width=0.49\textwidth]{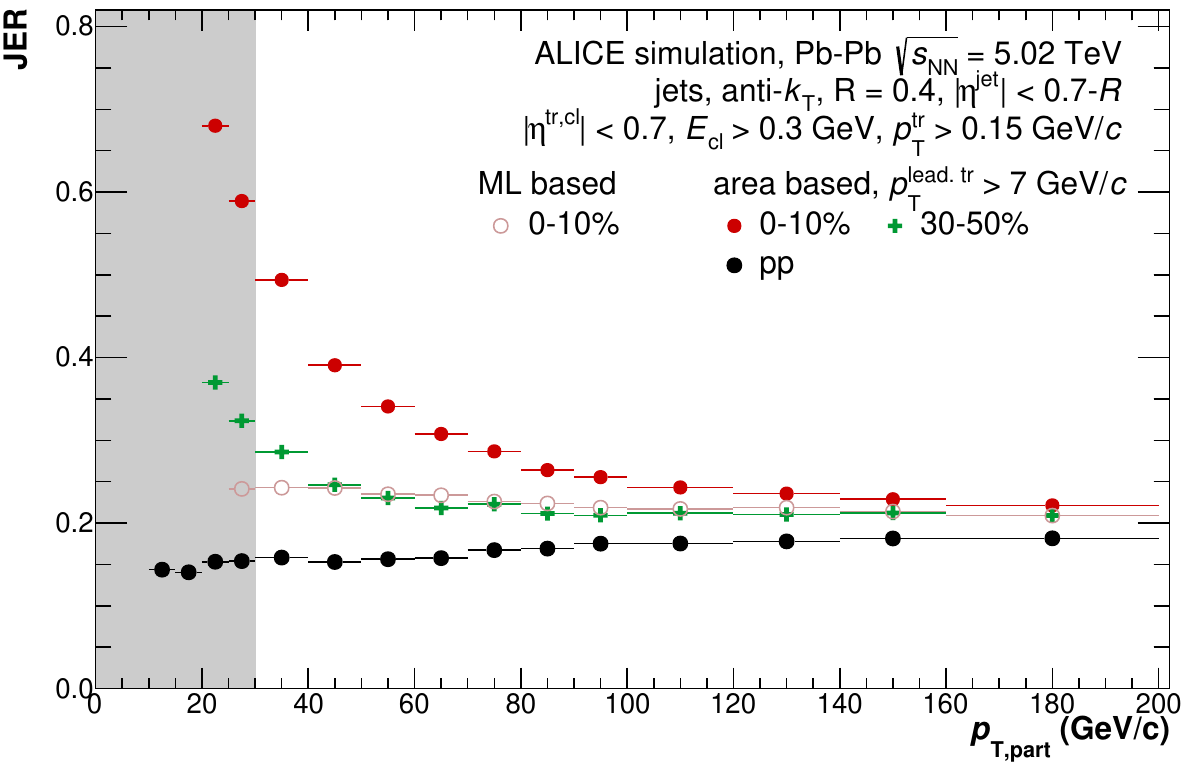}
    \caption{(Color online) Instrumental effects on the jet energy measurement in the 0-10\% (red) and 30-50\% (green) central \PbPb\ collisions at \sfivelead\ for the jet resolution parameter $R=0.4$ for jets corrected with the area-based method as well as the ML-based background description for the 10\% most central events (red open circle).
    When using the area based correction method the jet reconstruction is done with a leading track bias of $\pT=7$~\GeVc, while this is not the case for the machine learning based background description. 
    For comparison also the \pp\ results at \sfive\ are shown with a leading track bias of $\pT = 7$ \GeVc\ are shown in black.
            Upper panel: jet-by-jet distribution for various intervals in jet \pt. 
            Lower panels: the \gls{JES} is the mean~(left) and \gls{JER} is the standard deviation~(right) of these distributions.
            The gray bands indicate the \pT\ regions not taken into account for the final measurements.} 
    \label{fig:6-Perf-jesjerPbPb}
\end{figure}

The \gls{UE} contribution that is subtracted from the jets does not contain region-to-region fluctuations in $\rho_{\rm ch}$ and event-by-event fluctuations of $S_{\text{EMCal}}$, either of which can originate from statistical or dynamical fluctuations. 
Fluctuations arise from both the difference between the average $\rho_{\rm ch}$ in the event and the local $\rho$ background fluctuations in ($\eta$, $\varphi$), and the difference of the average $S_{\text{EMCal}}$ at a given centrality and the actual neutral to charged energy ratio in a specific event.
The size of these background fluctuations was studied by comparing the average to the local $\rho$. 
To do so, random cones of a given size, $A$, were placed in the event and the difference between the summed \pT\ of tracks and clusters in the cone~(local background) and $\rho A$~(average background) was used to obtain $\delta_{\pt}$. 
\Figure{fig:deltaPtPbPb}~(left) presents the distribution of the background fluctuations for two different jet resolution parameters, $R=0.2$ and $R=0.4$. 
The size of the fluctuations increases with larger $R$, making jet measurements with larger-radius jets challenging. 
The background fluctuations also increase strongly with increasing centrality, as shown in \Fig{fig:deltaPtPbPb}~(right), where the width of the $\delta_{\pt}$-distribution is displayed as a function of the centrality percentile for two considered values of $R$.
These widths reach values of 5 and 14~\GeVc\ in central collisions for $R=0.2$ and $0.4$, respectively.

\Figure{fig:6-Perf-jesjerPbPb} illustrates the performance of the full jet reconstruction for $R=0.4$ using detector simulations in \PbPb\ collisions at \sfive for different centrality classes and \pp\ collisions at the same energy. 
The upper panels show the distributions of $\Delta$\pt\ for three jet \pt\ intervals, which exhibit large jet-by-jet fluctuations of reconstructed jet \pT, as well as a significant tail of the distributions to negative values.
These fluctuations decrease for more peripheral events.
For the \PbPb\ case, the large uncorrelated background in heavy-ion collisions adds an additional effect besides the detector effects described for the \pp\ case.
The lower panels show the \gls{JES}~(left) and the \gls{JER}~(right) of the $\Delta$\pt\ distributions, as a function of particle-level jet \pt.
The \gls{JER} additionally increases for more central events due to the additional background contributions for jets with $R = 0.4$.
One approach to decrease the residual fluctuations remaining after background subtraction is to use machine learning~(ML) as described in Ref.~\cite{Haake:2018hqn}. 
Using regression techniques to create a mapping for jet properties and properties of the constituents of the jet to the corrected jet $\pT$, achieves a significant performance improvement as seen it can be seen from the open red markers in  \Fig{fig:6-Perf-jesjerPbPb}. 
Such performance improvements allow for an extension of the kinematic region of the measurement to larger jet radii and lower transverse momentum than previously possible. 

\subsubsection{Jet substructure}
Measurements of jet substructure in \pp\ and heavy-ion collisions are an essential tool to further study \gls{pQCD} and jet quenching in hot and dense \gls{QCD} matter (see \eg~\cite{Larkoski:2017jix, Andrews:2018jcm, Aad:2019vyi, Sirunyan:2017bsd, Acharya:2019djg, Adam:2020kug, Kauder:2017cvz}).
In order to measure the fine substructure of jets, the angular resolution of the detector must be good enough to distinguish nearly collinear jet constituents. 
Typically, this is best achieved with tracking detectors~\cite{Aad:2019vyi}, however charged-particle jets cannot be easily compared to theoretical calculations. 
The \gls{EMCal} is a relatively fine-grained calorimeter, which enables the possibility to measure jet substructure observables using full jets while maintaining a fairly small angular cutoff. 

\begin{figure}[t!]
    \centering
    \includegraphics[width=.49\textwidth]{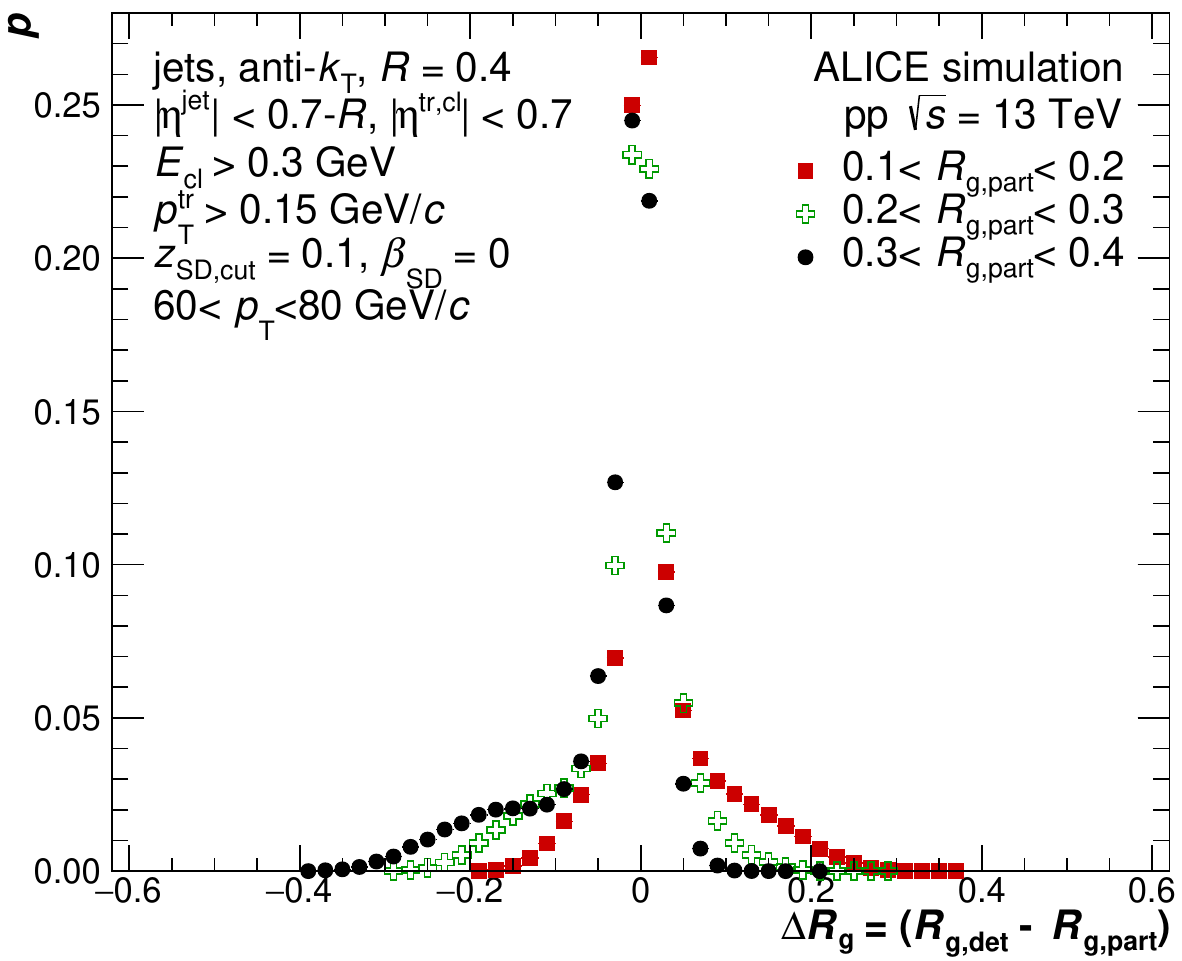}
    \includegraphics[width=.49\textwidth]{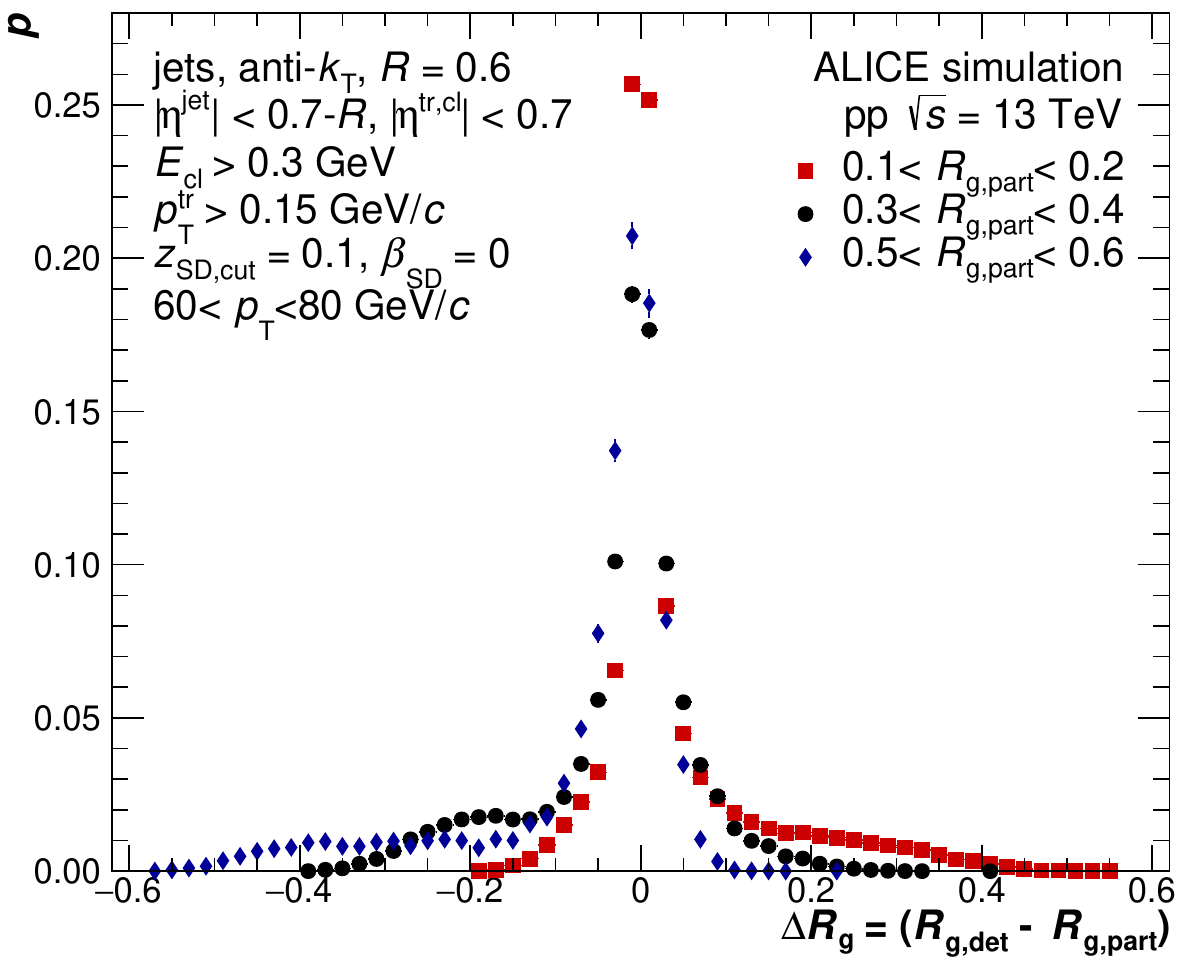}\\
    \includegraphics[width=.49\textwidth]{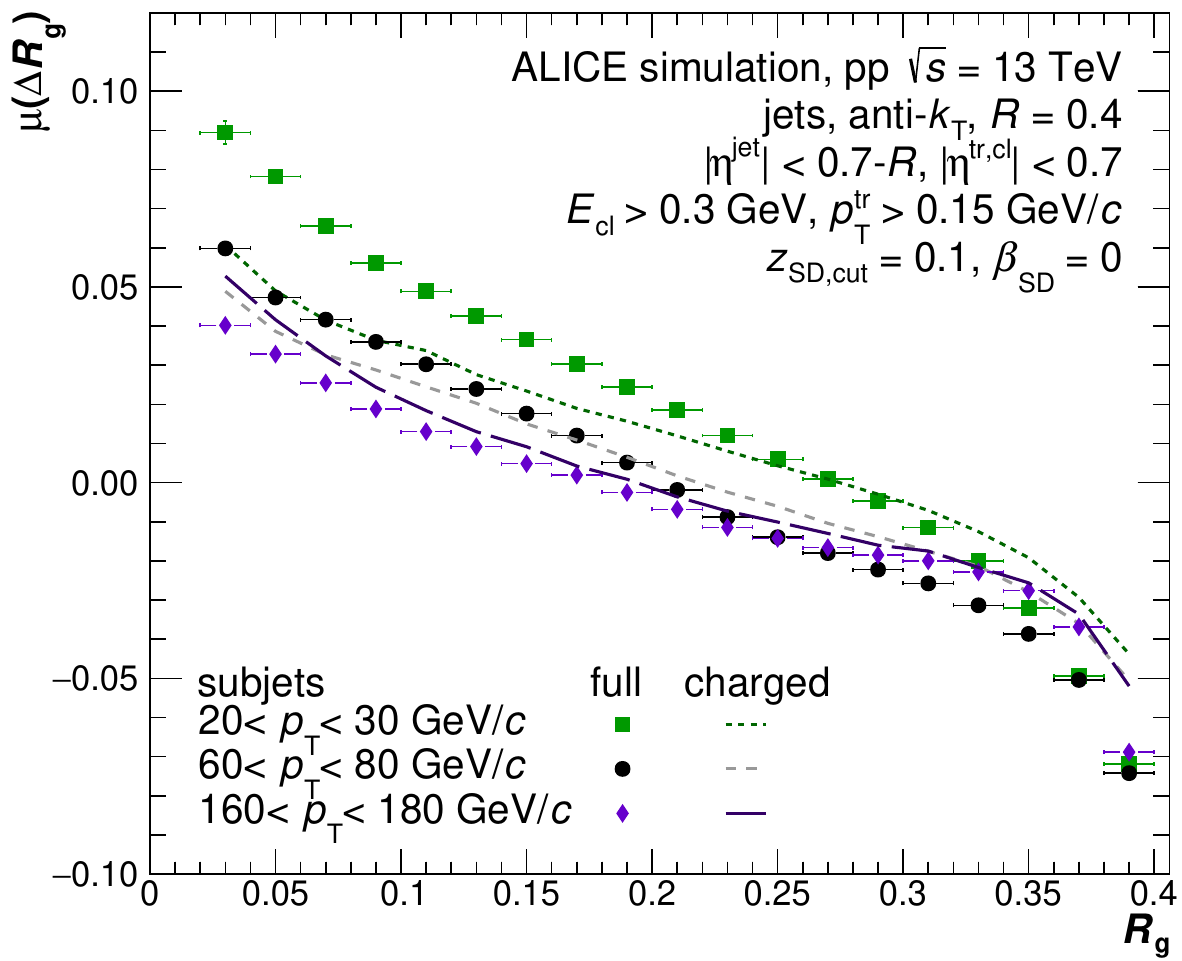}
    \includegraphics[width=.49\textwidth]{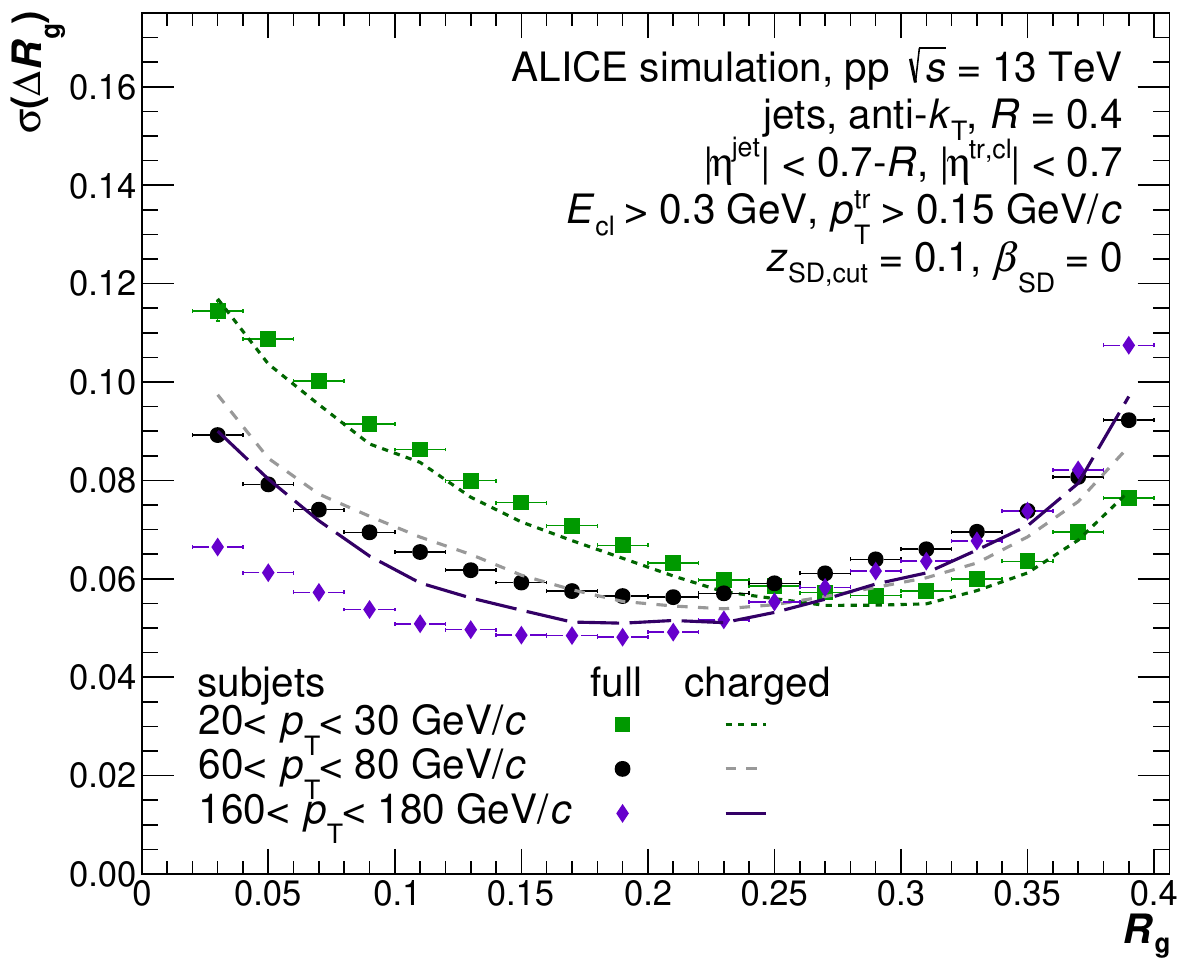}
    \caption{(Color online) Top: Probability distribution of the $R_{g}$ residuals for jets with $60<\pT<80$~\GeVc\ for a jet resolution parameter of $R=0.4$~(left) and $R=0.6$~(right). 
    Bottom: Mean~(left) and width~(right) of the $\Delta R_{g}$ distribution versus $R_{g}$ for jets with $R=0.4$ for different bins in $\pt$. Lines denote the case where subjets were reclustered with charged constituents only.}
    \label{fig:6-Jet-JetRgRes13TeV}
\end{figure}

\Figure{fig:6-Jet-JetRgRes13TeV} shows an example of the residuals between the value measured from the reconstructed jet and the generated value from the particle-level jet for one typical substructure observable, the Soft Drop groomed jet radius \Rg~\cite{Larkoski:2014wba, Kang:2019prh}, which represents the radial distance between the two hardest subjets, resulting from reclustering of the constituents of the original jet with the C/A algorithm~\cite{Dokshitzer:1997in, Wobisch:1998wt}.
The distributions of the residuals~($\Delta R_{\rm g}$) are displayed for $R=0.4$~(left) and $R=0.6$~(right) for $60<\pT<80$~\GeVc\ and they show a peak at $\Delta R_{\rm{g}}\approx0$. 
Tails can be observed in direction of positive $\Delta$\Rg\ at small \Rg\, indicating that the angular separation is overestimated in the detector for small radial distances between the two hardest subjets, and towards negative $\Delta$\Rg\ at large \Rg\, indicating an underestimation of the angular separation in this \Rg\ interval. 
At intermediate \Rg\ the distribution is almost symmetric.
The \Rg\ dependence of the mean and the width of the distribution, which characterizes the resolution, are shown in the bottom panels of \Fig{fig:6-Jet-JetRgRes13TeV} for three different intervals of jet \pT{} for $R=0.4$. 
Lines indicate the case where subjets are reclustered with charged constituents only. 
At small \Rg\ the mean is shifted towards larger \Rg\ ($\langle \Delta\Rg \rangle > 0$). 
The shift increases  with decreasing \pT{}. 
In this region, the sensitivity to constituents picked up from the underlying event, which can be clustered at different angles, is enhanced. 
This leads to an  overestimation of \Rg\ at detector level, which is strongest in the tails of the distribution. 
As \Rg\ increases, the mean of the $\Delta$\Rg\ distribution shifts to negative values at all \pT, indicating an underestimation of \Rg\ at detector level due to geometrical constraints. 
The resolution is affected by the tails in the distribution and it is optimal in the \Rg\ region where the distribution is most symmetric. 
At low \pT, the performance is best for subjets reclustered with charged constituents only profiting from the position and momentum resolution of tracks, resulting in a smaller shift of the mean as compared to full jets. 
Towards higher \pT{}, where charged-particle reconstruction in \gls{ALICE} becomes more challenging due to the moderate magnetic field strength, the resolution improves when including neutral constituents, because they help constrain the orientation of the main axes of the subjets.

\ifflush
\clearpage
\fi
\section{Summary and prospects} 
\label{sec:summary}
An overview of the performance of the \gls{EMCal} during operation in the years 2010--2018 in \gls{ALICE} at the \gls{LHC} was presented.
Details regarding the design and readout of the detector are given in \Sec{sec:hardware}.
The procedures for data taking, reconstruction and validation are documented in \Sec{sec:dat}. 
The setup and analysis of the electron and muon test beam data are discussed in \Sec{sec:testeam}. 
The calibration procedures are described in \Sec{sec:calibration}. 
The capabilities of the \gls{EMCal} to reconstruct and measure photons, light mesons, electrons and jets, are discussed in \Sec{sec:physics}.
A summary of key properties is provided in \Tab{tab:KeyQuantities}.
The design goals of the \gls{EMCal} were achieved, and the \gls{EMCal} is crucial for numerous analyses in \gls{ALICE}.

This report was made during the \gls{LHC} long shutdown~2. 
No hardware modifications are planned for the future operations of \gls{EMCal} during Run~3/4. 
However, for the future operations two measures have been taken: the upgrade of the Front-End Electronics firmware and production of spare hardware.

{\it Firmware upgrade:} The \gls{EMCal} continues to be operated as a trigger detector also during Run~3/4.
The firmware of the \gls{SRU} and Summary Trigger Units (\gls{STU}) are already upgraded according to the Run~3 trigger and \gls{DAQ} protocols~\cite{Antonioli:2013ppp}.
In addition, the readout rate is increased and currently $\sim$35~kHz readout rate is expected for minimum bias \PbPb\ collisions at 50~kHz, which is close to the design value. 
During the future operations both \gls{EMCal} and \gls{DCal} will continue providing \gls{L0}, \gls{L1}-$\gamma$ and \gls{L1}-jet triggers. 

{\it Spare production:} For a smooth operation through Run~3/4, new \gls{FEE} boards have been produced, which are identical to the ones used during Run~1 and 2. 
This accounts for 15\% of the units used in the experimental cavern: 100 \glspl{FEC} and 6 \glspl{TRU}.
In addition, 2 \glspl{STU} have been produced as spares for \gls{EMCal} and \gls{DCal}.

\begin{table}[t!]
  \caption{Summary of key characteristics of the \gls{EMCal}. If not otherwise indicated, energy~($E$) is given in GeV and transverse momentum (\pT) in \GeVc.}
  \centering
 \begin{tabular}{lll}
    Key quantity & Value & Section \\
    \toprule
    High gain range             & $15  $ MeV $\lesssim E  \lesssim 15.6$ GeV, & \ref{sec:readout}\\
                                & 1 ADC $\approx$  16 MeV  \\ 
    Low gain range              & $250  $ MeV $\lesssim E \lesssim 250$ GeV, & \ref{sec:readout} \\
                                & 1 ADC $\approx$  250 MeV  \\  
    \midrule    
    MIP energy (MeV)            & $235.6 \pm 0.9$                                                           & \ref{sec:TBanalysis} \\
    \midrule
    Energy resolution (\%)      & $\sigma_{E}(E)/E = 1.4 \oplus 9.5/\sqrt{E} \oplus 2.9/E$                    & \ref{sec:TBNonLin}\\
    \midrule
    Channel-by-channel miscalib. & $< 1\%$                                                                & \ref{sec:e-calibchapter}\\
    \midrule
    Nonlinearity               &  $ f(E) = \frac{4.3 + 0.06~\ln(E)}{1+3.5~\exp(E/4172)} $
                                                                                                                              & \ref{sec:TBNonLin}\\
    \midrule                                                                                                                          
    MC-cluster-fine-tuning       &                                                                           & \ref{sec:EMCalEnergyPositionCalib}\\
    \hspace{0.2cm} \piz $(M_{\text{data}}-M_{\text{MC}})/M_{\text{MC}}$& $< 0.3\%$                                                            & \\ 
    \hspace{0.2cm} $\eta$ $(M_{\text{data}}-M_{\text{MC}})/M_{\text{MC}}$ & $< 0.6\%$                                                            & \\ 
    \midrule
    $E/p$ &  & \ref{sec:electronID}\\
    \hspace{0.2cm} resolution                & $\sigma_{E/p} (\pT) = 0.011/\pT + 0.001 \pT  + $\\
                                & \hspace{2cm} $0.058~\exp( -3.7\times10^{-6} \pT)$                 & \\
    \hspace{0.2cm} calibration               & $\mu_{E/p} (\pT)= 0.069 \text{~Erf}(0.119 \pT) + 0.933 $       & \\
    \hspace{0.2cm} $\mu_{E/p,\text{ data}}/\mu_{E/p,\text{ MC}}$  & $< 1.5\%$                                                                          & \\
    \midrule
    Position resolution (cm)    & $\sigma_{x,y} (E) = 0.27 \oplus 1.04/\sqrt{E}$                            & \ref{sec:PosResol} \\
    \midrule
    Time resolution (ns)        & $\sigma_t = 2.4$                                                          & \ref{sec:timeCalib}\\ 
    \bottomrule
 \end{tabular}
  \label{tab:KeyQuantities}
\end{table}

\ifflush
\clearpage
\fi
\newenvironment{acknowledgement}{\relax}{\relax}
\begin{acknowledgement}
\section*{Acknowledgments}

The ALICE Collaboration would like to thank all its engineers and technicians for their invaluable contributions to the construction of the experiment and the CERN accelerator teams for the outstanding performance of the LHC complex.
The ALICE Collaboration gratefully acknowledges the resources and support provided by all Grid centres and the Worldwide LHC Computing Grid (WLCG) collaboration.
The ALICE Collaboration acknowledges the following funding agencies for their support in building and running the ALICE detector:
A. I. Alikhanyan National Science Laboratory (Yerevan Physics Institute) Foundation (ANSL), State Committee of Science and World Federation of Scientists (WFS), Armenia;
Austrian Academy of Sciences, Austrian Science Fund (FWF): [M 2467-N36] and Nationalstiftung f\"{u}r Forschung, Technologie und Entwicklung, Austria;
Ministry of Communications and High Technologies, National Nuclear Research Center, Azerbaijan;
Conselho Nacional de Desenvolvimento Cient\'{\i}fico e Tecnol\'{o}gico (CNPq), Financiadora de Estudos e Projetos (Finep), Funda\c{c}\~{a}o de Amparo \`{a} Pesquisa do Estado de S\~{a}o Paulo (FAPESP) and Universidade Federal do Rio Grande do Sul (UFRGS), Brazil;
Bulgarian Ministry of Education and Science, within the National Roadmap for Research Infrastructures 2020-2027 (object CERN), Bulgaria;
Ministry of Education of China (MOEC) , Ministry of Science \& Technology of China (MSTC) and National Natural Science Foundation of China (NSFC), China;
Ministry of Science and Education and Croatian Science Foundation, Croatia;
Centro de Aplicaciones Tecnol\'{o}gicas y Desarrollo Nuclear (CEADEN), Cubaenerg\'{\i}a, Cuba;
Ministry of Education, Youth and Sports of the Czech Republic, Czech Republic;
The Danish Council for Independent Research | Natural Sciences, the VILLUM FONDEN and Danish National Research Foundation (DNRF), Denmark;
Helsinki Institute of Physics (HIP), Finland;
Commissariat \`{a} l'Energie Atomique (CEA) and Institut National de Physique Nucl\'{e}aire et de Physique des Particules (IN2P3) and Centre National de la Recherche Scientifique (CNRS), France;
Bundesministerium f\"{u}r Bildung und Forschung (BMBF) and GSI Helmholtzzentrum f\"{u}r Schwerionenforschung GmbH, Germany;
General Secretariat for Research and Technology, Ministry of Education, Research and Religions, Greece;
National Research, Development and Innovation Office, Hungary;
Department of Atomic Energy Government of India (DAE), Department of Science and Technology, Government of India (DST), University Grants Commission, Government of India (UGC) and Council of Scientific and Industrial Research (CSIR), India;
National Research and Innovation Agency - BRIN, Indonesia;
Istituto Nazionale di Fisica Nucleare (INFN), Italy;
Japanese Ministry of Education, Culture, Sports, Science and Technology (MEXT) and Japan Society for the Promotion of Science (JSPS) KAKENHI, Japan;
Consejo Nacional de Ciencia (CONACYT) y Tecnolog\'{i}a, through Fondo de Cooperaci\'{o}n Internacional en Ciencia y Tecnolog\'{i}a (FONCICYT) and Direcci\'{o}n General de Asuntos del Personal Academico (DGAPA), Mexico;
Nederlandse Organisatie voor Wetenschappelijk Onderzoek (NWO), Netherlands;
The Research Council of Norway, Norway;
Commission on Science and Technology for Sustainable Development in the South (COMSATS), Pakistan;
Pontificia Universidad Cat\'{o}lica del Per\'{u}, Peru;
Ministry of Education and Science, National Science Centre and WUT ID-UB, Poland;
Korea Institute of Science and Technology Information and National Research Foundation of Korea (NRF), Republic of Korea;
Ministry of Education and Scientific Research, Institute of Atomic Physics, Ministry of Research and Innovation and Institute of Atomic Physics and University Politehnica of Bucharest, Romania;
Ministry of Education, Science, Research and Sport of the Slovak Republic, Slovakia;
National Research Foundation of South Africa, South Africa;
Swedish Research Council (VR) and Knut \& Alice Wallenberg Foundation (KAW), Sweden;
European Organization for Nuclear Research, Switzerland;
Suranaree University of Technology (SUT), National Science and Technology Development Agency (NSTDA), Thailand Science Research and Innovation (TSRI) and National Science, Research and Innovation Fund (NSRF), Thailand;
Turkish Energy, Nuclear and Mineral Research Agency (TENMAK), Turkey;
National Academy of  Sciences of Ukraine, Ukraine;
Science and Technology Facilities Council (STFC), United Kingdom;
National Science Foundation of the United States of America (NSF) and United States Department of Energy, Office of Nuclear Physics (DOE NP), United States of America.
In addition, individual groups or members have received support from:
Marie Sk\l{}odowska Curie, European Research Council, Strong 2020 - Horizon 2020 (grant nos. 950692, 824093, 896850), European Union;
Academy of Finland (Center of Excellence in Quark Matter) (grant nos. 346327, 346328), Finland;
Programa de Apoyos para la Superaci\'{o}n del Personal Acad\'{e}mico, UNAM, Mexico.
\end{acknowledgement}
\clearpage
\printnoidxglossaries
\bibliographystyle{utphys}
\bibliography{biblio}
\newpage
\appendix

\section{GEANT configuration}
\label{sec:GEANTConf}


\begin{longtable}{p{0.12\textwidth}p{0.1\textwidth}p{0.78\textwidth}}
\caption{ \label{table:GEANT3Switchs} GEANT3 physics process flags. 
          These flags can be set on a material by material basis, from the \gls{GEANT}3 documentation PHYS001-3~\cite{geant3doc}. }\\
  \multicolumn{3}{l}{}\\
       \midrule
      Switch   & \small ALICE Default values & Description \\ \toprule
  \endfirsthead
      \multicolumn{3}{l}{\emph{Table~\ref{table:GEANT3Switchs} continued}}\\
      \midrule
      Switch   & \small ALICE Default value & Description \\
      \toprule
   \endhead
      \midrule
       \multicolumn{3}{r}{\emph{Table~\ref{table:GEANT3Switchs} continued 
                          on next page.}}
   \endfoot
    \endlastfoot
    \footnotesize
    13 ANNI & 1 & Positron annihilation. The e$^+$ is stopped.\newline
               0 No position annihilation.\newline
               1 Positron annihilation with generation of $\gamma$.\newline
               2 Positron annihilation without generation of $\gamma$.\\
    \footnotesize
    14 BREM & 1 & bremsstrahlung. The interaction particle (e$^-$, e$^+$, 
                $\mu^-$, $\mu^+$) is stopped.\newline
               0 No bremsstrahlung. \newline
               1 bremsstrahlung with generation of $\gamma$.\newline
               2 bremsstrahlung without generation of $\gamma$.\\
    \footnotesize
    15 COMP & 1 & Compton scattering.\newline
               0 No Compton scattering.\newline
               1 Compton scattering with generation of e$^-$. \newline
               2 Compton scattering without generation of e$^-$.\\
    \footnotesize
    16 DCAY & 1 & Decay in flight. The decaying particles stops. \newline
               0 No decay in flight \newline 
               1 Decay in flight with generation of secondaries \newline
               2 Decay in flight without generation of secondaries \\
    \footnotesize
    17 DRAY & 0 & $\delta$-ray production.\newline
               0 No $\delta$-ray production.\newline
               1 $\delta$-ray production with generation of e$^-$.\newline
               2 $\delta$-ray production without generation of e$^-$.\\
    \footnotesize
    18 HADR & 1 & Hadronic interactions. The particle is stopped in case 
                of inelastic interactions, while it is not stopped in case 
                of elastic interactions.\newline
               0 No hadronic interactions.\newline
               1 Hadronic interactions with generation of secondaries.\newline
               2 Hadronic interactions without generation of 
                 secondaries.\newline
               $>2$ can be used in the user code \texttt{GUPHAD} and 
                 \texttt{GUHADR} to choose 
                 a hadronic package. These values have no effect on the 
                 hadronic packages themselves. Not supported in AliRoot.\\
    \footnotesize
    19 LOSS & 2 & Continuous energy loss.\newline
               0 No continuous energy loss, DRAY is forced to 0.\newline
               1 Continuous energy loss with generation of $\delta$-rays 
                  which have an energy above DCUTE and restricted 
                  Landau-fluctuations 
                  for $\delta$-rays which have an 
                  energy below DCUTE (no $\delta$-ray produced).\newline
               2 Continuous energy loss without generation of $\delta$-rays 
                  and full Landau-Vavilov-Gauss 
                  fluctuations. In this case 
                  DRAY is forced to 0 to avoid double counting of 
                  fluctuations.\newline
               3 Same as 1, kept for backwards compatibility.\newline
               4 Energy loss without fluctuations. The value obtained 
                 from the tables is used directly.\\
    \footnotesize
    20 MULS & 1 & Multiple scattering.\newline
               0 No multiple scattering.\newline
               1 Multiple scattering according to Moli\`ere.
                 theory.\newline
               2 Same as 1. Kept for backwards compatibility.\newline
               3 Pure Gaussian scattering according to the Rossi 
                 formula.\\ 
    \footnotesize
    21 PAIR & 1 & Pair production. The interacting $\gamma$ is 
                  stopped. \newline
               0 No pair production. \newline
               1 Pair production with generation of e$^+/$e$^-$.\newline
               2 Pair production without generation of e$^+/$e$^-$.\\
    \footnotesize
    22 PHOT & 1 & Photoelectric effect. The interacting photon is 
                  stopped.\newline
               0 No photo-electric effect. \newline
               1 Photo-electric effect with generation of e$^-$.\newline
               2 Photo-electric effect without generation of e$^-$.\\
    \footnotesize
    23 RAYL & 1 & Rayliegh effect. 
                  The interacting $\gamma$ is not stopped.\newline
               0 No Raylieght effect.\newline
               1 Rayliegh effect.\\
    \footnotesize
    24 STRA & 0 & Turns on the collision sampling method to simulate 
                  energy loss in thin materials, particularly gasses.\newline
               0 Collision sampling is off.\newline
               1 Collision sampling is on. \\
    \footnotesize
    PFIS & 0 & Nuclear fission induced by a photon The photon stops.\newline
               0 No photo-fission.\newline
               1 Photo-fission with generation of secondaries.\newline
               2 Photo-fission without generation of secondaries.\\
    \footnotesize
    MUNU & 1 & Muon-nucleus interactions. The muon is not stopped.\newline
               0 No muon-nucleus interactions.\newline
               1 Muon-nucleus interactions with generation of 
                 secondaries.\newline
               2 Muon-nucleus interactions without generation of secondaries.\\
    \footnotesize
    CKOV & 1 & Light absorption. This process is the absorption of light 
                photons in dielectric materials. It is turned on by default 
                when the generation of ${\rm \check{C}}$erenkov 
                light is requested (in GEANT manual it is LABS).\newline
                0 No absorption of photons.\newline
                1 Absorption of photons with possible detection.\\
    \footnotesize
    SYNC & 0 & Synchrotron radiation in magnetic fields.\newline
               0 Synchrotron radiation is not simulated.\newline
               1 Synchrotron photon are generated, at the end of the 
                 tracking step.\newline
               2 Photons are not generated, the energy is deposited 
                 locally.\newline
               3 Synchrotron photons are generated, distributed along the 
                 curved path of their particle. \\
   \normalsize
   \label{table:GEANT3PhysicsFlags}
\end{longtable}
\pagebreak

\begin{table}
  \caption{\label{table:GEANT3PhysicsLimits} \gls{GEANT}3 physics process limits. 
              These ``cuts'' can be set on a 
              material by material basis, from the 
              GEANT3 documentation ZZZZ010-2~\cite{geant3doc}.
    }
  \centering
  \begin{tabular}{lll}   
    \footnotesize
      Parameter   & \small ALICE Default value & Description \\ \toprule
    \footnotesize
    3 CUTGAM & $1.\times 10^{-3}$ GeV & Threshold for gamma transport.\\
    \footnotesize
    4 CUTELE & $1.\times 10^{-3}$ GeV & Threshold for electron and positron 
                                   transport.\\
    \footnotesize
    5 CUTNEU & $1.\times 10^{-3}$ GeV & Threshold for neutral hadron 
                                         transport.\\
    \footnotesize
    6 CUTHAD & $1.\times 10^{-3}$ GeV & Threshold for charged hadron 
                                        and ion transport.\\
    \footnotesize
    7 CUTMUO & $1.\times 10^{-3}$ GeV & Threshold for muon transport.\\
    \footnotesize
    8 BCUTE &  $1.\times 10^{-3}$ GeV & Threshold for photons produced by \\
            &                         & electron bremsstrahlung.\\ 
    \footnotesize
    9 BCUTM &  $1.\times 10^{-3}$ GeV & Threshold for photons produced by \\
            &                         & muon bremsstrahlung.\\ 
    \footnotesize
    10 DCUTE &  $1.\times 10^{-3}$ GeV & Threshold for electrons produced by \\
            &                         &electron $\delta$-rays.\\ 
    \footnotesize
    11 DCUTM &  $1.\times 10^{-3}$ GeV & Threshold for electrons produced by \\
            &                         &muon or hadron $\delta$-rays.\\ 
    \footnotesize
    12 PPCUTM & $1.\times 10^{-3}$ GeV & Threshold for e$^{\pm}$ direct pair \\
            &                         &production by muons.\\
    \footnotesize
    TOFMAX & $1.\times 10^{10}$ s & Threshold on time of flight counted \\
            &                         & from primary interactions time.\\ \bottomrule

  \end{tabular}
  
\end{table}




\ifflush
\clearpage
\newpage
\fi
\section{The ALICE Collaboration}
\label{app:collab}
\ifdraft
\else
\begin{flushleft} 
\small

S.~Acharya\,\orcidlink{0000-0002-9213-5329}\,$^{\rm 126}$, 
D.~Adamov\'{a}\,\orcidlink{0000-0002-0504-7428}\,$^{\rm 87}$, 
A.~Adler$^{\rm 70}$, 
G.~Aglieri Rinella\,\orcidlink{0000-0002-9611-3696}\,$^{\rm 32}$, 
M.~Agnello\,\orcidlink{0000-0002-0760-5075}\,$^{\rm 29}$, 
N.~Agrawal\,\orcidlink{0000-0003-0348-9836}\,$^{\rm 51}$, 
Z.~Ahammed\,\orcidlink{0000-0001-5241-7412}\,$^{\rm 133}$, 
S.~Ahmad\,\orcidlink{0000-0003-0497-5705}\,$^{\rm 15}$, 
S.U.~Ahn\,\orcidlink{0000-0001-8847-489X}\,$^{\rm 71}$, 
I.~Ahuja\,\orcidlink{0000-0002-4417-1392}\,$^{\rm 37}$
% ...

\section*{Affiliation Notes}

$^{\rm I}$ Deceased\\
$^{\rm II}$ Also at: Max-Planck-Institut f\"{u}r Physik, Munich, Germany\\
$^{\rm III}$ Also at: Italian National Agency for New Technologies, Energy and Sustainable Economic Development (ENEA), Bologna, Italy\\
$^{\rm IV}$ Also at: Dipartimento DET del Politecnico di Torino, Turin, Italy\\
$^{\rm V}$ Also at: Department of Applied Physics, Aligarh Muslim University, Aligarh, India\\
$^{\rm VI}$ Also at: Institute of Theoretical Physics, University of Wroclaw, Poland\\
$^{\rm VII}$ Also at: An institution covered by a cooperation agreement with CERN\\

\section*{Collaboration Institutes}

$^{1}$ A.I. Alikhanyan National Science Laboratory (Yerevan Physics Institute) Foundation, Yerevan, Armenia\\
$^{2}$ AGH University of Science and Technology, Cracow, Poland\\
$^{3}$ Bogolyubov Institute for Theoretical Physics, National Academy of Sciences of Ukraine, Kiev, Ukraine\\
$^{4}$ Bose Institute, Department of Physics  and Centre for Astroparticle Physics and Space Science (CAPSS), Kolkata, India\\
$^{5}$ California Polytechnic State University, San Luis Obispo, California, United States\\
$^{6}$ Central China Normal University, Wuhan, China\\
$^{7}$ Centro de Aplicaciones Tecnol\'{o}gicas y Desarrollo Nuclear (CEADEN), Havana, Cuba\\
$^{8}$ Centro de Investigaci\'{o}n y de Estudios Avanzados (CINVESTAV), Mexico City and M\'{e}rida, Mexico\\
$^{9}$ Chicago State University, Chicago, Illinois, United States\\
$^{10}$ China Institute of Atomic Energy, Beijing, China\\
$^{11}$ Chungbuk National University, Cheongju, Republic of Korea\\
$^{12}$ Comenius University Bratislava, Faculty of Mathematics, Physics and Informatics, Bratislava, Slovak Republic\\
$^{13}$ COMSATS University Islamabad, Islamabad, Pakistan\\
$^{14}$ Creighton University, Omaha, Nebraska, United States\\
$^{15}$ Department of Physics, Aligarh Muslim University, Aligarh, India\\
$^{16}$ Department of Physics, Pusan National University, Pusan, Republic of Korea\\
$^{17}$ Department of Physics, Sejong University, Seoul, Republic of Korea\\
$^{18}$ Department of Physics, University of California, Berkeley, California, United States\\
$^{19}$ Department of Physics, University of Oslo, Oslo, Norway\\
$^{20}$ Department of Physics and Technology, University of Bergen, Bergen, Norway\\
$^{21}$ Dipartimento di Fisica, Universit\`{a} di Pavia, Pavia, Italy\\
$^{22}$ Dipartimento di Fisica dell'Universit\`{a} and Sezione INFN, Cagliari, Italy\\
$^{23}$ Dipartimento di Fisica dell'Universit\`{a} and Sezione INFN, Trieste, Italy\\
$^{24}$ Dipartimento di Fisica dell'Universit\`{a} and Sezione INFN, Turin, Italy\\
$^{25}$ Dipartimento di Fisica e Astronomia dell'Universit\`{a} and Sezione INFN, Bologna, Italy\\
$^{26}$ Dipartimento di Fisica e Astronomia dell'Universit\`{a} and Sezione INFN, Catania, Italy\\
$^{27}$ Dipartimento di Fisica e Astronomia dell'Universit\`{a} and Sezione INFN, Padova, Italy\\
$^{28}$ Dipartimento di Fisica `E.R.~Caianiello' dell'Universit\`{a} and Gruppo Collegato INFN, Salerno, Italy\\
$^{29}$ Dipartimento DISAT del Politecnico and Sezione INFN, Turin, Italy\\
$^{30}$ Dipartimento di Scienze MIFT, Universit\`{a} di Messina, Messina, Italy\\
$^{31}$ Dipartimento Interateneo di Fisica `M.~Merlin' and Sezione INFN, Bari, Italy\\
$^{32}$ European Organization for Nuclear Research (CERN), Geneva, Switzerland\\
$^{33}$ Faculty of Electrical Engineering, Mechanical Engineering and Naval Architecture, University of Split, Split, Croatia\\
$^{34}$ Faculty of Engineering and Science, Western Norway University of Applied Sciences, Bergen, Norway\\
$^{35}$ Faculty of Nuclear Sciences and Physical Engineering, Czech Technical University in Prague, Prague, Czech Republic\\
$^{36}$ Faculty of Physics, Sofia University, Sofia, Bulgaria\\
$^{37}$ Faculty of Science, P.J.~\v{S}af\'{a}rik University, Ko\v{s}ice, Slovak Republic\\
$^{38}$ Frankfurt Institute for Advanced Studies, Johann Wolfgang Goethe-Universit\"{a}t Frankfurt, Frankfurt, Germany\\
$^{39}$ Fudan University, Shanghai, China\\
$^{40}$ Gangneung-Wonju National University, Gangneung, Republic of Korea\\
$^{41}$ Gauhati University, Department of Physics, Guwahati, India\\
$^{42}$ Helmholtz-Institut f\"{u}r Strahlen- und Kernphysik, Rheinische Friedrich-Wilhelms-Universit\"{a}t Bonn, Bonn, Germany\\
$^{43}$ Helsinki Institute of Physics (HIP), Helsinki, Finland\\
$^{44}$ High Energy Physics Group,  Universidad Aut\'{o}noma de Puebla, Puebla, Mexico\\
$^{45}$ Horia Hulubei National Institute of Physics and Nuclear Engineering, Bucharest, Romania\\
$^{46}$ Hubei University of Technology, Wuhan, China\\
$^{47}$ Indian Institute of Technology Bombay (IIT), Mumbai, India\\
$^{48}$ Indian Institute of Technology Indore, Indore, India\\
$^{49}$ INFN, Laboratori Nazionali di Frascati, Frascati, Italy\\
$^{50}$ INFN, Sezione di Bari, Bari, Italy\\
$^{51}$ INFN, Sezione di Bologna, Bologna, Italy\\
$^{52}$ INFN, Sezione di Cagliari, Cagliari, Italy\\
$^{53}$ INFN, Sezione di Catania, Catania, Italy\\
$^{54}$ INFN, Sezione di Padova, Padova, Italy\\
$^{55}$ INFN, Sezione di Pavia, Pavia, Italy\\
$^{56}$ INFN, Sezione di Torino, Turin, Italy\\
$^{57}$ INFN, Sezione di Trieste, Trieste, Italy\\
$^{58}$ Inha University, Incheon, Republic of Korea\\
$^{59}$ Institute for Gravitational and Subatomic Physics (GRASP), Utrecht University/Nikhef, Utrecht, Netherlands\\
$^{60}$ Institute of Experimental Physics, Slovak Academy of Sciences, Ko\v{s}ice, Slovak Republic\\
$^{61}$ Institute of Physics, Homi Bhabha National Institute, Bhubaneswar, India\\
$^{62}$ Institute of Physics of the Czech Academy of Sciences, Prague, Czech Republic\\
$^{63}$ Institute of Space Science (ISS), Bucharest, Romania\\
$^{64}$ Institut f\"{u}r Kernphysik, Johann Wolfgang Goethe-Universit\"{a}t Frankfurt, Frankfurt, Germany\\
$^{65}$ Instituto de Ciencias Nucleares, Universidad Nacional Aut\'{o}noma de M\'{e}xico, Mexico City, Mexico\\
$^{66}$ Instituto de F\'{i}sica, Universidade Federal do Rio Grande do Sul (UFRGS), Porto Alegre, Brazil\\
$^{67}$ Instituto de F\'{\i}sica, Universidad Nacional Aut\'{o}noma de M\'{e}xico, Mexico City, Mexico\\
$^{68}$ iThemba LABS, National Research Foundation, Somerset West, South Africa\\
$^{69}$ Jeonbuk National University, Jeonju, Republic of Korea\\
$^{70}$ Johann-Wolfgang-Goethe Universit\"{a}t Frankfurt Institut f\"{u}r Informatik, Fachbereich Informatik und Mathematik, Frankfurt, Germany\\
$^{71}$ Korea Institute of Science and Technology Information, Daejeon, Republic of Korea\\
$^{72}$ KTO Karatay University, Konya, Turkey\\
$^{73}$ Laboratoire de Physique des 2 Infinis, Ir\`{e}ne Joliot-Curie, Orsay, France\\
$^{74}$ Laboratoire de Physique Subatomique et de Cosmologie, Universit\'{e} Grenoble-Alpes, CNRS-IN2P3, Grenoble, France\\
$^{75}$ Lawrence Berkeley National Laboratory, Berkeley, California, United States\\
$^{76}$ Lund University Department of Physics, Division of Particle Physics, Lund, Sweden\\
$^{77}$ Nagasaki Institute of Applied Science, Nagasaki, Japan\\
$^{78}$ Nara Women{'}s University (NWU), Nara, Japan\\
$^{79}$ National and Kapodistrian University of Athens, School of Science, Department of Physics , Athens, Greece\\
$^{80}$ National Centre for Nuclear Research, Warsaw, Poland\\
$^{81}$ National Institute of Science Education and Research, Homi Bhabha National Institute, Jatni, India\\
$^{82}$ National Nuclear Research Center, Baku, Azerbaijan\\
$^{83}$ National Research and Innovation Agency - BRIN, Jakarta, Indonesia\\
$^{84}$ Niels Bohr Institute, University of Copenhagen, Copenhagen, Denmark\\
$^{85}$ Nikhef, National institute for subatomic physics, Amsterdam, Netherlands\\
$^{86}$ Nuclear Physics Group, STFC Daresbury Laboratory, Daresbury, United Kingdom\\
$^{87}$ Nuclear Physics Institute of the Czech Academy of Sciences, Husinec-\v{R}e\v{z}, Czech Republic\\
$^{88}$ Oak Ridge National Laboratory, Oak Ridge, Tennessee, United States\\
$^{89}$ Ohio State University, Columbus, Ohio, United States\\
$^{90}$ Physics department, Faculty of science, University of Zagreb, Zagreb, Croatia\\
$^{91}$ Physics Department, Panjab University, Chandigarh, India\\
$^{92}$ Physics Department, University of Jammu, Jammu, India\\
$^{93}$ Physics Department, University of Rajasthan, Jaipur, India\\
$^{94}$ Physics Program and International Institute for Sustainability with Knotted Chiral Meta Matter (SKCM2), Hiroshima University, Hiroshima, Japan\\
$^{95}$ Physikalisches Institut, Eberhard-Karls-Universit\"{a}t T\"{u}bingen, T\"{u}bingen, Germany\\
$^{96}$ Physikalisches Institut, Ruprecht-Karls-Universit\"{a}t Heidelberg, Heidelberg, Germany\\
$^{97}$ Physik Department, Technische Universit\"{a}t M\"{u}nchen, Munich, Germany\\
$^{98}$ Politecnico di Bari and Sezione INFN, Bari, Italy\\
$^{99}$ Research Division and ExtreMe Matter Institute EMMI, GSI Helmholtzzentrum f\"ur Schwerionenforschung GmbH, Darmstadt, Germany\\
$^{100}$ Saga University, Saga, Japan\\
$^{101}$ Saha Institute of Nuclear Physics, Homi Bhabha National Institute, Kolkata, India\\
$^{102}$ School of Physics and Astronomy, University of Birmingham, Birmingham, United Kingdom\\
$^{103}$ Secci\'{o}n F\'{\i}sica, Departamento de Ciencias, Pontificia Universidad Cat\'{o}lica del Per\'{u}, Lima, Peru\\
$^{104}$ Stefan Meyer Institut f\"{u}r Subatomare Physik (SMI), Vienna, Austria\\
$^{105}$ SUBATECH, IMT Atlantique, Nantes Universit\'{e}, CNRS-IN2P3, Nantes, France\\
$^{106}$ Suranaree University of Technology, Nakhon Ratchasima, Thailand\\
$^{107}$ Technical University of Ko\v{s}ice, Ko\v{s}ice, Slovak Republic\\
$^{108}$ The Henryk Niewodniczanski Institute of Nuclear Physics, Polish Academy of Sciences, Cracow, Poland\\
$^{109}$ The University of Texas at Austin, Austin, Texas, United States\\
$^{110}$ Universidad Aut\'{o}noma de Sinaloa, Culiac\'{a}n, Mexico\\
$^{111}$ Universidade de S\~{a}o Paulo (USP), S\~{a}o Paulo, Brazil\\
$^{112}$ Universidade Estadual de Campinas (UNICAMP), Campinas, Brazil\\
$^{113}$ Universidade Federal do ABC, Santo Andre, Brazil\\
$^{114}$ University of Cape Town, Cape Town, South Africa\\
$^{115}$ University of Houston, Houston, Texas, United States\\
$^{116}$ University of Jyv\"{a}skyl\"{a}, Jyv\"{a}skyl\"{a}, Finland\\
$^{117}$ University of Kansas, Lawrence, Kansas, United States\\
$^{118}$ University of Liverpool, Liverpool, United Kingdom\\
$^{119}$ University of Science and Technology of China, Hefei, China\\
$^{120}$ University of South-Eastern Norway, Kongsberg, Norway\\
$^{121}$ University of Tennessee, Knoxville, Tennessee, United States\\
$^{122}$ University of the Witwatersrand, Johannesburg, South Africa\\
$^{123}$ University of Tokyo, Tokyo, Japan\\
$^{124}$ University of Tsukuba, Tsukuba, Japan\\
$^{125}$ University Politehnica of Bucharest, Bucharest, Romania\\
$^{126}$ Universit\'{e} Clermont Auvergne, CNRS/IN2P3, LPC, Clermont-Ferrand, France\\
$^{127}$ Universit\'{e} de Lyon, CNRS/IN2P3, Institut de Physique des 2 Infinis de Lyon, Lyon, France\\
$^{128}$ Universit\'{e} de Strasbourg, CNRS, IPHC UMR 7178, F-67000 Strasbourg, France, Strasbourg, France\\
$^{129}$ Universit\'{e} Paris-Saclay Centre d'Etudes de Saclay (CEA), IRFU, D\'{e}partment de Physique Nucl\'{e}aire (DPhN), Saclay, France\\
$^{130}$ Universit\`{a} degli Studi di Foggia, Foggia, Italy\\
$^{131}$ Universit\`{a} del Piemonte Orientale, Vercelli, Italy\\
$^{132}$ Universit\`{a} di Brescia, Brescia, Italy\\
$^{133}$ Variable Energy Cyclotron Centre, Homi Bhabha National Institute, Kolkata, India\\
$^{134}$ Warsaw University of Technology, Warsaw, Poland\\
$^{135}$ Wayne State University, Detroit, Michigan, United States\\
$^{136}$ Westf\"{a}lische Wilhelms-Universit\"{a}t M\"{u}nster, Institut f\"{u}r Kernphysik, M\"{u}nster, Germany\\
$^{137}$ Wigner Research Centre for Physics, Budapest, Hungary\\
$^{138}$ Yale University, New Haven, Connecticut, United States\\
$^{139}$ Yonsei University, Seoul, Republic of Korea\\
$^{140}$  Zentrum  f\"{u}r Technologie und Transfer (ZTT), Worms, Germany\\
$^{141}$ Affiliated with an institute covered by a cooperation agreement with CERN\\
$^{142}$ Affiliated with an international laboratory covered by a cooperation agreement with CERN.\\

\end{flushleft} 
\fi
\end{document}